\documentclass{umthesis}
\usepackage{amsmath,amssymb,amsfonts,bbold}
\usepackage{latexcad}
\usepackage{epsfig}
\usepackage{float}
\usepackage[export]{adjustbox}
\usepackage{graphicx,subcaption}
\usepackage{multirow}
\usepackage[table,xcdraw]{xcolor}
\usepackage{caption}
\usepackage[font={scriptsize}]{subcaption}
\usepackage[utf8x]{inputenc}

\newcommand{\N}{\mathcal{N}}
\newcommand{\D}{\mathcal{D}}
\newcommand{\M}{\mathcal{M}}
\newcommand{\RNum}[1]{\uppercase\expandafter{\romannumeral #1\relax}}
\def\be{\begin{equation}}
\def\ee{\end{equation}}
\newcommand{\bea}{\begin{eqnarray}}
\newcommand{\eea}{\end{eqnarray}}

\usepackage{hyperref}
\hypersetup{
     colorlinks   = false, 
     urlcolor     = blue, 
    linkbordercolor    = {green}, 
    citebordercolor = {blue}
}

\begin{document}
\pdfstringdefDisableCommands{%
\let\MakeUppercase\relax
}

\title{Investigations In Higher Dimensional Gravity Theory}
\author{Safinaz Ramadan Abdel-Rahman Farag Salem}
\date{August 2022}
\copyrightyear{2022}

\bachelors{B.Sc.}{Al-Azhar University} %
\masters{M.Sc.}{Al-Azhar University}   %

\committeechair{Doctor Moataz Hassan Mahmoud Emam}       %
\firstreader{Doctor Hala Hashem Mohamed Salah}       %
\departmentchair{}   %
\departmentname{Physics}

\degree{Doctor of Philosophy}{Ph.D.}

\frontmatter                        %
\maketitle                          %
\copyrightpage                      %
\signaturepage

\begin{dedication}
  \begin{center}

\emph{Yet nature is made better by no mean.\\
But nature makes that mean: so, over that art\\
which you say adds to nature, is an art\\
that nature makes.\vskip6pt
Shakespeare, The Winter’s Tale. }
  \end{center}
\end{dedication}

\begin{dedication}
  \begin{center}
      \centerline{\epsfxsize=12.0cm\epsfysize=6.0cm\epsfbox{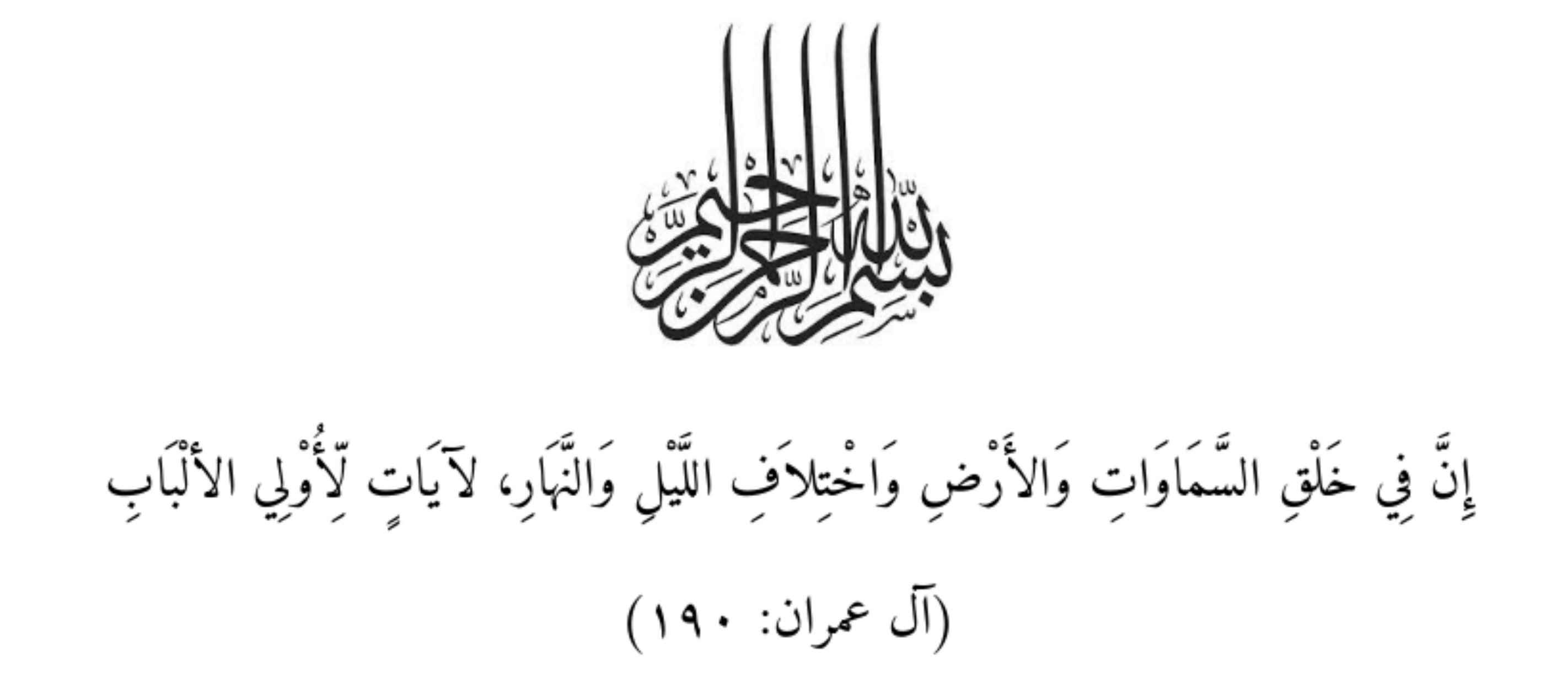}}\vskip6pt
        \centerline{``Indeed, in the creation of the heavens and the earth and} 
\centerline{the alternation of the night and the day are signs for those of understanding '' }\vskip6pt
        \centerline{``En vérité, dans la création des cieux et de la terre, et dans} 
\centerline{ l'alternance de la nuit et du jour, il y a certes des signes pour les doués d'intelligence '' }\vskip6pt
\centerline{[Quran 190:Al Imran]}
  \end{center}
\end{dedication}

\chapter{Acknowledgments}

I write these words while I'm close to discussing my proposal for the doctorate, indeed I'm so overwhelmed by many emotions and aspirations and feels so much gratitude.
Praise to All-mighty God who created all that beauty and all that vast magnificent universe. 

I'm grateful for all my professors who made the way for theoretical physics easy for me
and who made all that possible. I'm so grateful because I was so lucky to have a plentiful time before starting the Ph.D. to learn vital courses in theoretical physics. It was a precious chance to study, make research, and know-how the scientific research can be. 

It's from my pleasure to meet Prof. Emam. At the beginning of the Ph.D., the field of gravitational physics was almost new for me, but with his consistent support and encouragement, the field of gravitational physics and supergravity became facile for me. 
Your guidance and your continuous advice to take each research step in
a deeper viewpoint sharpen my experiences and make this work completed as it has been today. So I'm grateful for your concern, and patience.

I'm so thankful for Prof. Hala Hashem, for her valuable support, and assistance. 
I would like to thank the head of the physics department at the faculty of science, Al-Azhar University. I would like to thank my colleagues and my professors at Al-Azhar University.

I'm so grateful for my father and my mother, love you so much, rest in peace. 

Hope this dissertation is useful for anyone who reaches for it. Hope our work includes answers for bizarre questions about our universe and even raises many other inquiries. 

\begin{abstract}
In this thesis we study BPS 3-brane embedded in five dimensional $\N=2$ supergravity. We have found that when modeling our universe as a brane world filled with matter, radiation and/or pure energy, the time evolution of the brane world is strongly correlated to the complex structure moduli of the underlying Calabi-Yau manifold. 
Moreover most solutions exhibit an early time inflation and late time acceleration, which means our universe’s whole cosmological history may be explained by only the bulk effects, without introducing an inflation field or even cosmological constant term in some cases. The constructed solutions are numeric and have been manifested analytically as well. Finally, since our solutions to the field equations are constrained by how far we know about the Calabi-Yau manifold itself, we have studied the solely time dependence of the complex structure moduli and the metric of the complex structure space in case the dimension of the complex structure space of the Calabi-Yau manifold $h_{(2,1)}=1$ . 

\end{abstract}

\setcounter{tocdepth}{2}        %

\tableofcontents                %
\listoffigures
\mainmatter                     %

\unnumberedchapter{Introduction}
The trials to extend general relativity (GR) emerged very early, indeed just after its discovery by Einstein in 1915 by a couple of years. The questions at those times rose about if GR is a unique theory of gravity or not. Later on, in the beginning of the sixties, extending GR became indispensable because it is not renormalizable and, therefore, can not be conventionally quantized.

On the other hand, supersymmery (SUSY) proved itself as a theory that can alleviate some of the UV divergences of quantum field theory, via cancellations between bosonic and fermionic loops, hence the UV divergences of quantum gravity become milder in supergravity. 

Supergravity (SUGRA) is a supersymmetric extension of GR, or a theory of local supersymmetry. It involves the graviton, a spin 2 elementary particle described by Einstein gravity, and extra matter, in particular a fermionic partner of the graviton called the gravitino (in case of $\N=1$ SUGRA).

Supergravity manifested itself as a realistic theory that describes nature including the four types of interactions, and also it has been found that SUGRA in ten spacetime dimensions is the low energy limit of superstring theories. Moreover, there exist a unique $\N=1$  supergravity theory in eleven spacetime dimensions which is the low energy limit of a more general nonperturbative quantum field theory in $\D=11$ called M-theory, from which the known superstring theories in ten dimensions arise as perturbative limits. 

In contemporary fundamental physics M- theory is known as the leading candidate for the theory of everything (TOE). Nowadays the search for the complete formulation of TOE or any empirical evidence for it, namely the compactification of its extra dimensions to construct candidate models of our four- dimensional world is one of the most vital quests of the new century, if not the millennium.   

There are many works have been done before to compactify $\N=1~ \D=11$ SUGRA over Calabi- Yau (CY)  3-fold to $\N=2~ \D=5$ SUGRA. We use these results to find different BPS (Bogomol’nyi-Prasad-Sommerfield) 3-brane solutions for the reduced theory. 
These solutions couple to new fields and moduli that depend on the particular topology of CY. 
BPS solutions conserve supersymmetry by setting the SUSY transformations of the model fermions to zero, 
so they are bosonic solutions. Worthy to mention here that SUSY is half broken in supergravity as we will see later.

In this framework, we show that when our universe is modeled as a 3-brane embedded in five dimensional spacetime, 
one of the cosmological ambiguity of the universe can be interpreted, typically what happens in the inflationary epoch when 
adding an extra spatial dimension in case of matter, radiation and/or pure energy dominant brane-universes ?

Also the solutions of the field equations does not interpret only the early time inflation of the brane-world, but when adding cosmological constants for the brane and the bulk, they show late time acceleration which means the recent accelerated time expansion of the universe can be explained as well. Further they introduce possible scenarios about the future of the universe time evolution. 

We manifest on the problem of the initial conditions and the multiverses. That we solve the field equations for different initial conditions after the Big Bang leading to a variety of possible universes including several universes that follow the accepted history of our own. 

Last but not least, these solutions depend on the geometery of the Calabi–Yau manifold because the time dependence of the brane-worlds is found to be strongly correlated to the hypermultiplet fields, particularly the complex structure moduli of the underlying Calabi-Yau submanifold. So that we study the  moduli and the metric of the complex structure space of the Calabi-Yau manifold in case $h_{(2,1)} =1$. 
 
So this work has two main aims, first the deeper understanding of the topology of the Calabi-Yau manifold, 
and the dimensional reduction techniques, that gets us closer to the structure of superstring theories 
and the peculiar M- theory, and second to find possible applications of our results to our universe’s cosmological evolution. 
The dissertation is organized as follows:
\begin{description}
    \item[Chapter One] We introduce a brief review about string theory and supergravity in eleven dimensions.
    \item[Chapter Two] We give a review about the different types of manifold, their developments and properties, 
then we will talk about Calabi-Yau manifold and its complex structure space.
    \item[Chapter Three] We will introduce the dimensional reduction of $\D = 11$ SUGRA to $\D = 5$ SUGRA over Calabi-Yau manifold , 
ending up by the theory in its symplectic form.
    \item[Chapter Four] We numerically find solutions for $\N=2~ \D= 5$ SUGRA by solving Einstein field equations in case of radiation only, and dust only filled branes. We have found that when the universe is modeled
as a 3-brane world, in both cases the time evolution of the Calabi–Yau complex structure moduli strongly correlated to the expansion of the brane-world, showing an early short period of rapid expansion, that can interpret the inflationary epoch.
    \item[Chapter Five]  We study a brane- world filled with all matter, radiation and energy (represented by a brane cosmological constant term) to interpret the time being era of the universe's evolution with time. We also added a bulk cosmological constant, to see the whole effect of the bulk and the CY moduli on the brane- universe scale factor. 
Also we find out that the solutions can predict the future time evolution of our universe.
We study the moduli's variation with respect to the fifth extra dimension, where the cosmological constant problem can be resolved
analytically.
\item[Chapter Six] We use our solutions to the field equations in case of a world brane filled by dust, radiation and energy with the field equation of the moduli and its complex conjugate to get analytic expressions to the time dependence of the moduli and the metric of the CY complex structure space in case the dimension of the complex structure space $h_{(2,1)}=1$. We show the moduli and the metric component G in the complex plane. Then in this case we can find analytic solutions to the the potential of the complex structure space and the volume of the Calabi–Yau manifold which are both related to the metric. Finally we have solved
Einstein equations analytically to manifest all the numerical results introduced. 
    \item[Conclusion] We summerize the main results of our study and point out to possible streams of future research.
\end{description}

\setcounter{equation}{0}

\chapter{Theory of Everything Towards Unification }

The standard model (SM) agrees with all
the existing particle physics experiments, and the general relativity is consistent with most of the astrophysical and cosmological observations. However both the models have some drawbacks as the following:
\begin{itemize}
	\item The general relativity is a classical theory, not valid at the quantum level.
	\item The standard model involves a large number of parameters, some of which are unnaturally small.
	\item Neutrino oscillations and the convergence of gauge coupling constants point to the existence 
of a heavy mass scale, the Grand Unification scale, $M_U \sim 10^{16}$ GeV, and/or the string
scale $M_{str} \sim 10^{17}$ GeV, not explained by the SM. 
\item Many features of the standard cosmological model such as the very nature of dark matter
and dark energy, the asymmetry between matter and antimatter, the initial Big Bang 
singularity (still present in inflationary models), the quantum structure of black holes and of
space- time itself, are not yet well understood.
\end{itemize}

Supersymmetry, in its spontaneously broken phase, is the leading proposal for new physics
beyond the Standard Model, subject to experimental tests in modern colliders \cite{Endo:2021}, and is the
leading paradigm for the early universe and inflationary cosmology. Supersymmetry is naturally
realized in the framework of superstrings, which is the only known candidate for
a consistent quantum theory of gravity. 

Furthermore, superstring theory (1984-1994) has successfully
passed a series of nontrivial tests \cite{Kounnas}:
\begin{itemize}
	\item Gravity and the basic ingredients of the Standard Model emerge very naturally.
	\item Superstring theory is free of ultraviolet divergences and anomalies.
	\item It incorporates in a very beautiful and natural way many of the major theoretical ideas
such as supersymmetry, Grand Unification and the possible existence of large extra dimensions.
\item It has explained the microscopic origin of the entropy and other thermodynamical properties of black holes.
\end{itemize}

In 1995, the physicist Edward Witten discovered the mother of all string theories. He found various indications that the perturbative string theories fit together into a coherent nonperturbative theory, which he dubbed M-theory. M-theory looks like each of the string theories in different physical contexts but does not itself have limits on its regime of validity - a major requirement for the theory of everything.
So though Witten didn't write down complete equations of M-theory, why, then, string/M-theory have given the edge of theoretical physics searches for this decade and most probably for the next decade?

\begin{figure}[!t]
\centering
\includegraphics[scale=0.3]{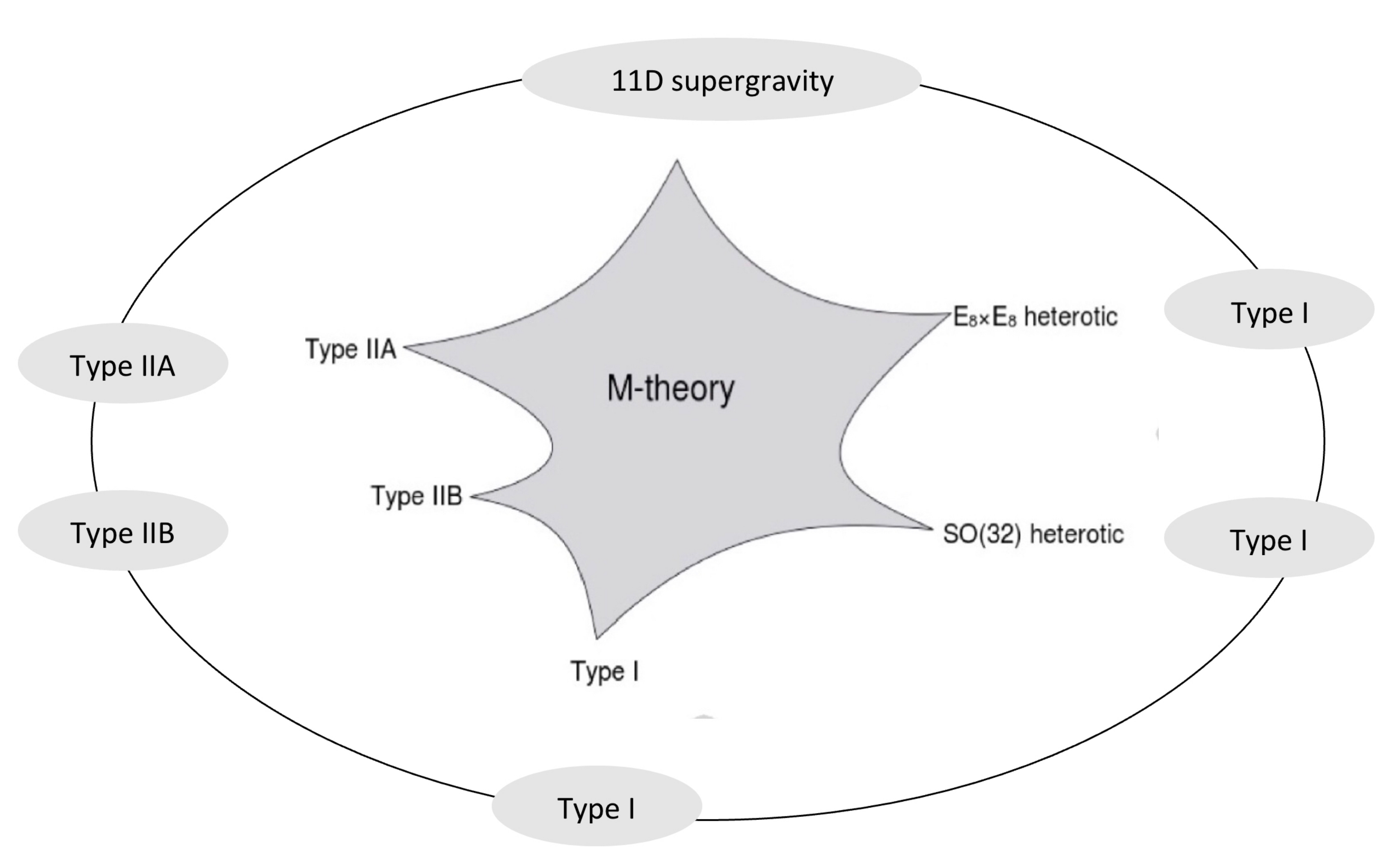}
\caption{The current state of string theory: The inner star gives the five known
superstring theories in ten dimensions. The outer circuit represents the three known
supergravity theories in ten dimensions in addition to the eleven dimensional theory. }
\label{Mom}
\end{figure}

Fig. (\ref{Mom}) shows how M- theory unifies all the higher dimensional theories in a beautiful and an elegant way.
In this chapter we will introduce the basics of superstring theories, also we will give a general overview on supergravity, especially  the eleven dimensional theory, in such way we can directly deal with in chapter 3.


\section{String Basics}  

\begin{figure}[!t]
\centering
\includegraphics[scale=0.35]{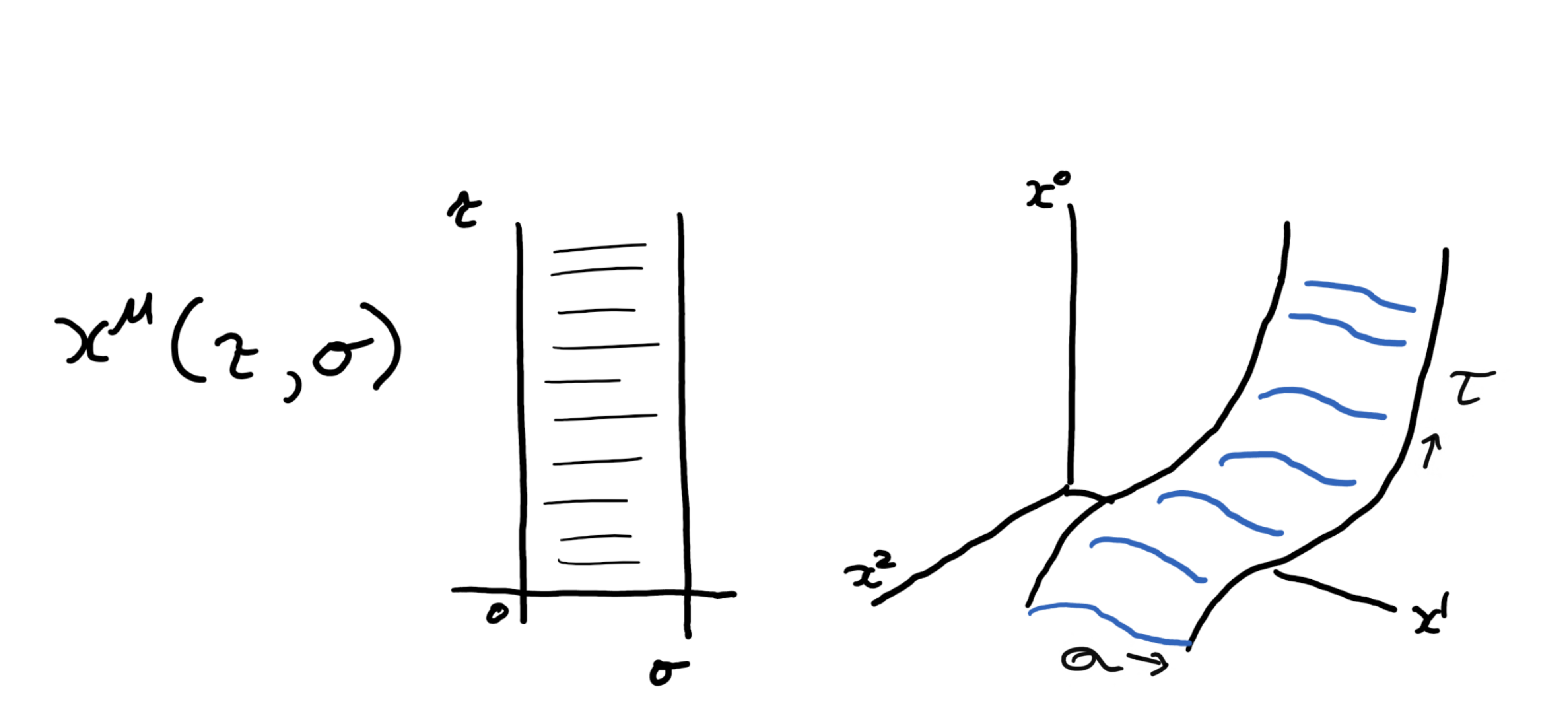}
\caption
{A string's world sheet. The function $X^\mu(\tau,\sigma)$ embeds the world sheet,
parameterized by $(\tau,\sigma)$ into space-time coordinates $X^\mu$}
\label{sheet}
\end{figure}

The string is one dimensional object, as it moves, it sweeps out a two dimensional surface in space-time called a worldsheet. To describe the dynamics of the string we use the simplest Poincar\'e invariant action, the Nambu-Goto action 
\bea
\nonumber S_{NG} &=&- \frac{1}{2\pi\alpha'}~ \int d\tau d\sigma~ \sqrt{- det~ h_{ab}} \\
& =& - \frac{1}{2\pi\alpha'}~ \int d\tau d\sigma~ \sqrt{- |(\partial_a X^\mu) (\partial_b X^\nu) \eta_{\mu\nu}|},
\label{ng}
\eea
this generalize the relativistic action of a point particle. It describes a string worldsheet of a proper time $\tau$ and a proper length $\sigma$, propagating in flat D- dimensional space-time background $X^\mu(\tau,\sigma)$, $\mu=0,...,D-1$,
Fig. (\ref{sheet}) \cite{Nambuac}. $h_{ab}= \partial_a  X^\mu \partial_b X^\nu ~ \eta_{\mu\nu} $ 
\footnote{\em Where our convention for the signature of the Lorentzian metrices is $(-,+,+,+)$.} is the induced metric on the world sheet, defining the pullback of space-time on the world sheet as an object embedded in space-time. By analogue to the point particle case, the constant $\frac{1}{2 \pi \alpha’}$ is the tension of the string or the mass per unit length.$\alpha’$ is called Regge slope and it’s 1 for bosonic string \footnote{ \em Look for instance at Refs. \cite{BRegge, Kir:2007}.}.
It’s useful however to write the action in a form which removes the square root \footnote{\em For quantization.}. Polyakov showed that this can be achieved by using the formalism of general relativity. Introducing a two dimensional metric $\gamma^{ab}$ . Then the invariant action is
\be
S_P =  - \frac{1}{4\pi\alpha'}~ \int d\tau d\sigma~ \sqrt{-\gamma} \gamma^{ab} ~(\partial_a X^\mu) (\partial_b X^\nu) \eta_{\mu\nu}, 
\label{sp} 
\ee
besides Poincar\'e symmetry, this action has diffeomorphism invariance under the worldsheet coordinates transformation and Weyl invariant under the rescaling of the worldsheet metric  $ \gamma_{ab} \to \gamma'_{ab} = e^{2 \omega(\tau,\sigma)} \gamma_a$ .

Varying the action with respect to $\gamma_{ab}$ yields a two-dimensional energy momentum tensor
\be
T^{ab}(\tau,\sigma)=-\frac{4\pi}{\sqrt{-\gamma}}\frac{\delta S}{\delta \gamma_{ab}} =
- \frac{1}{\alpha'} \{\partial^a X_\mu \partial^b X^\mu - \frac{1}{2} \gamma^{ab} \gamma_{cd}
\partial^c X_\mu \partial^d X^\mu  \}.
\ee
While varying the action with respect to $X^\mu$ yields the string equation of motion
\be
\nabla^2 X^\mu =0,
\label{eom} 
\ee
where the Laplacian is with respect to the worldsheet coordinates. This is a single
string propagating without sources. Since the canonical momentum to $X^\mu$ is $(\partial_a X^\mu)$,
we find for an open string we are forced to impose Neumann boundary conditions Equ. (\ref{bco}) in order to avoid momentum flow off the ends of the string:
\be 
\left.\begin{aligned}
    X^{' \mu}(\tau,0) &= 0\\
    X^{' \mu}(\tau,\pi)&= 0
  \end{aligned} \right\} \mbox{ Open String Neumann BCS}
\label{bco}
\ee
While for a closed string the following periodic boundary condition have been chosen:
\be
  \left.
  \begin{aligned}
   X^{' \mu}(\tau,0) &=  X^{' \mu}(\tau,\pi)\\
    X^{\mu}(\tau,0)&= X^{\mu}(\tau,\pi)\\
 \gamma_{ab}(\tau,0)& = \gamma_{ab} (\tau,\pi)\\
  \end{aligned} \right\}   \text{ Closed String Periodic BCS}
  \label{bcc}
\ee

\subsection{Spectrum}
Solving the string wave equation (\ref{eom}) and imposing the boundary conditions Equ. (\ref{bco}), gives the most general solutions for the open string
\be
X^\mu (\sigma,\tau)=x^\mu + 2 \alpha' p^\mu \tau + i \sqrt{2\alpha'} \sum^{n=\infty}_{\substack{n=-\infty\\
 n\neq0~~~}} \frac{1}{n} \alpha^\mu_n e^{-in\tau}~ cos (n\sigma). \label{FM}
\ee
While imposing the boundary conditions Equ. (\ref{bcc}) gives the most general solutions for
the closed string, wherein it sweeps out a cylinder in space-time as it propagates. A closed string has a couple of solutions corresponding to right moving and left moving waves 
\be 
X^\mu(\tau, \sigma)= X^\mu_R(\sigma^-)+ X^\mu_L(\sigma^+)  
\ee
\be
\begin{aligned}
X^\mu_R(\sigma^-) &= \frac{x^\mu}{2} +  \alpha' p^\mu \sigma^- + i \sqrt{\frac{\alpha'}{2}} \sum^{n=\infty}_{\substack{n=-\infty\\
 n\neq0~~~}} \frac{1}{n} \alpha^\mu_n e^{-2in\sigma^-}\\
X^\mu_L(\sigma^+) &= \frac{x^\mu}{2} +  \alpha' p^\mu \sigma^+ + i \sqrt{\frac{\alpha'}{2}} \sum^{n=\infty}_{\substack{n=-\infty\\
 n\neq0~~~}} \frac{1}{n} \alpha^\mu_n e^{-2in\sigma^+}.
\end{aligned}
\ee
$x^\mu$ and $p^\mu$ are the position and momentum vectors of the center of mass of
the string respectively, and $\alpha^\mu_n$ are arbitrary Fourier modes, which in the quantum
theory are interpreted as annihilation and creation operators in the Hilbert space of
the string’s oscillations. The masses of open string states are
\be
m^2=\frac{1}{\alpha'} (N-1),
\ee
this implies a condition on the space-time dimension where the string theory lives that the dimension should equal 26. 
The ground state  (N = 0) is tachyonic, having negative mass-squared. Since the
potential for a scalar field is $\frac{1}{2} m^2 \phi^2$, this means that the ground state is unstable. One way
to think about this is that we are using the bosonic string only as a toy model, and are indeed expanding around an unstable
state. Tachyonic state can be removed by extending space-time symmetry further than Poincar\'e symmetry 
and going to the superstring theory.
The masses of closed strings are
\be
m^2=\frac{2}{\alpha'} (N-\tilde{N}-2).
\ee

The string states presumed to yield the observed elementary particles in lower
dimensions are the ones corresponding to $N = \tilde{N} = 1$, representing massless particles.
An important such state is a particle with spin two. This, naturally, is interpreted as
the graviton, confirming that string theory necessarily contains gravity as a solution.
Since the harmonics on the string can go arbitrarily high, we find that there is an
infinite tower of massive states allowed on the string. The masses of these ‘particles’
are of the order of the Planck mass, making them highly inaccessible to us. It is
assumed that the particles we observe today are the result of the massless string
states, acquiring mass via spontaneous symmetry breaking.

\subsection{Interactions}
In string theory interactions arise very elegantly because they are described by processes
in which strings join and split. In these processes, the worldsheets of free strings combine
to form a single worldsheet. The worldsheets of interacting strings have the 
construction of Riemann surfaces, that are some of the most interesting two-dimensional surfaces. They are,
roughly speaking, surfaces where the two coordinates make up a complex variable. In equivalent Riemann surfaces can be distinguished by parameters called moduli
\footnote{\em We will see later how these moduli play a vital rule in the topology of the theory.}.
Finding a way to construct all Riemann surfaces with their associated moduli is a
difficult mathematical problem.

\begin{figure}[!t]
\centering
\includegraphics[scale=0.25]{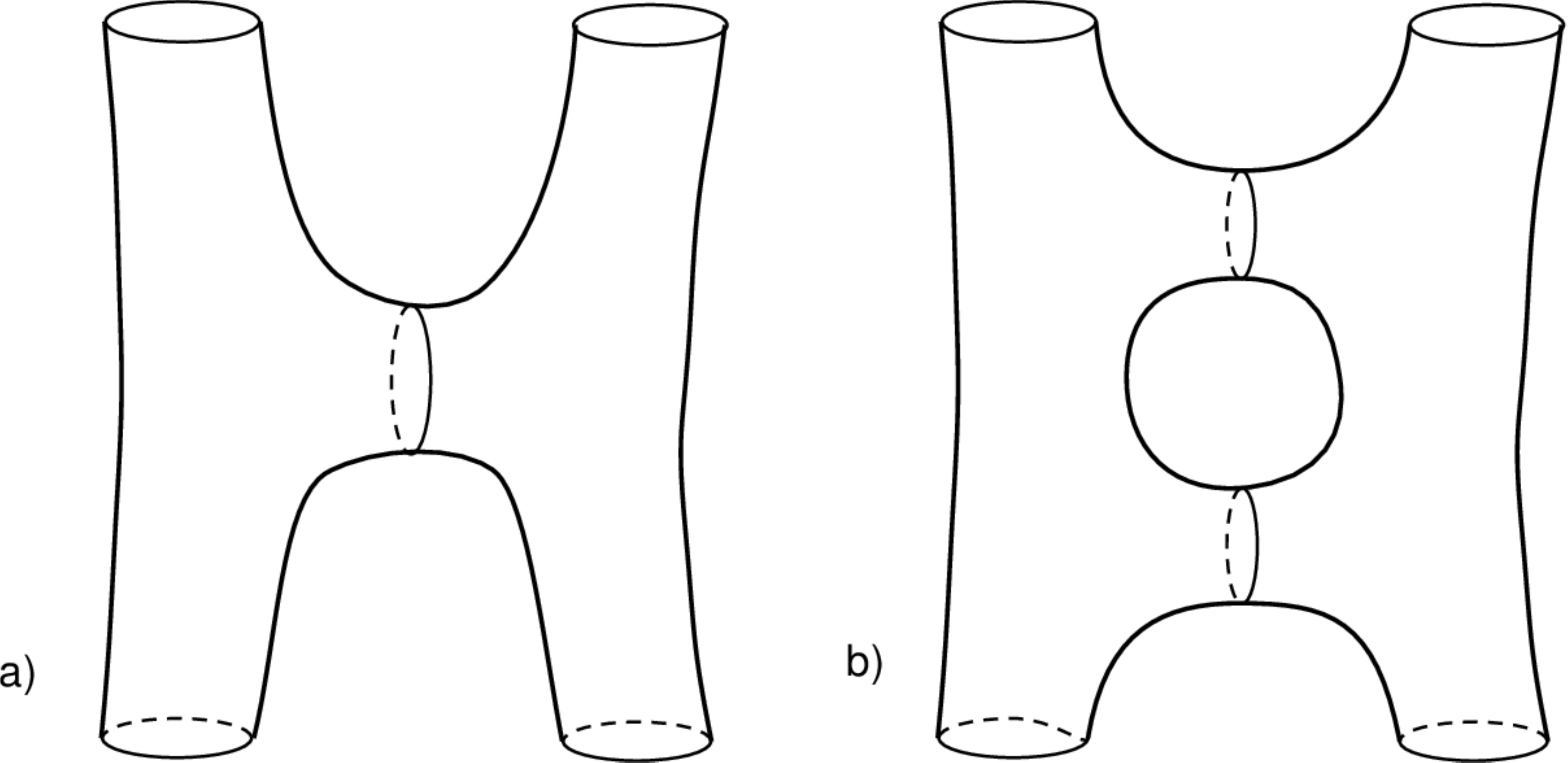}
\caption{Closed strings scattering by the exchange of (a) one string, and (b) two strings. }
\label{int}
\end{figure}

To include these interaction, another terms are added to Polyakov action Equ. (\ref{sp}). Indeed, that action is the simplest version for D-dimensional bosonic fields $X^\mu(\tau, \sigma)$. By the analogy with GR, one can add the Einstein-Hilbert action  
\be 
\chi = \frac{1}{4\pi} \int \sqrt{g} \mathcal{R},
\ee
$\mathcal{R}$ again being the worldsheet curvature. This is not merely diff $\times$ Weyl invariant, it is
topologically invariant. For worldsheets with boundaries, as
in open string theory, (diff $\cdot$ Weyl) invariance requires also a boundrey term:
\be 
\chi = \frac{1}{4\pi} \int d\sigma d\tau  \sqrt{\gamma}~ \mathcal{R} +
\frac{1}{2\pi} \int_{\mbox{boundary}} ds~~  k,
\ee
where k is the extrinsic curvature on the boundary \cite{polchinski1994}.
The quantity $\chi$ is called Eular characteristic and it depends only on the topology of the world sheet and so will only matter when comparing world sheets of different topology. Now, the full string action
that resembles two-dimensional gravity coupled to D bosonic "matter" fields $X^\mu$ is given by
\be 
S= S_P + \lambda~ \chi,
\ee 
where $\lambda$ is an arbitrary parameter \footnote{ \em It will turn out that $\lambda$ is not a free parameter. In the full string theory, it has dynamical meaning, and will be equivalent to the expectation value of one of the massless fields - the dilaton - described by the string.}.
So what is $\lambda$'s role $?$  The  path integral defining the string theory:
\be 
Z_P = \int dX d\gamma ~ e^{-S}.
\ee
For a closed surface with h bundles $\chi= 2 - 2 \mbox{h}$. At closed string emission or reabsorbtion, 
as in Fig. (\ref{int}), the resulting amplitudes will be weighted by a factor $e^{-\lambda \chi}$. So while this term does not affect anything local like the worldsheet equations of motion, it does affect the relative weights of surfaces of
different topologies. The quantity $g_c = e^\lambda$ is called the closed string coupling.
For open string amplitudes analogue to Fig. (\ref{int}) the Euler number is $2 - 2\mbox{h} - \mbox{b}$, where b are
the boundaries. Adding a strip increases b by one and so the path integral weight changes by $e^{\lambda}$. So that 
$g_c= g_o^2$, where $g_o$ is the open string coupling.

Treated as a two dimensional conformal field theory, string
theory in a flat 26-dimensional background can be reformulated as if it were in a
curved background. Roughly speaking, a theory of strings on a curved background is
equivalent to one embedded in a flat metric with a background teaming with coherent
gravitonic string states. This enables us to rewrite the Polyakov action in terms of
a curved background metric $G_{\mu\nu}(X)$. We further generalize this by including other
massless background fields; namely a totally antisymmetric tensor field $B_{\mu\nu}(X)$ and
a scalar field $\phi$ known as the dilaton. So, even though these are stringy states, one
can treat them as independent fields that couple to the string. The Polyakov action
then becomes
\be
S_\sigma =\frac{1}{4\pi\alpha'}~ \int d\tau d\sigma ~\sqrt{\gamma} [(\gamma^{ab} G_{\mu\nu}+i\varepsilon^{ab}B_{\mu\nu}) (\partial_a X^\mu)
(\partial_b X^\nu) +\alpha' R \phi],
\ee
this action is known as the sigma model \cite{polchinski1994} . In addition
to these fields, we will see in the next section that certain p-form gauge fields arise in
superstring theory. Again, these are really massless string states which we organize in
terms of classical background fields. In other words, the spectrum of massless states
in string theory is identical to that of a free field theory with such fields.

What about the free parameters in string theory ? Indeed , the string theory has no free parameters.
The strength of interactions is parameterized by the string coupling g which is a dimensionless number determined by the expectation value of the dilaton field. The Regge slope $\alpha'$ is dimensionful, but the string coupling g, together with the
Regge slope $\alpha'$, determined by the value of Newton’s constant and determines the tension of D-branes.
So that $\alpha'$ is not a free parameter and even can simply be absorbed in the definition of $X^\mu$. 

\section{Superstrings}
The bosonic string theory that have been previewed so far contains important bosonic particles, such as the photon and the graviton. Non- Abelian gauge bosons, needed to transit the strong and weak forces , also arise in string theory as we will see later. However this theory suffers from many drawbacks. First, to be consistent and anomaly free, it has to live in 26 space-time dimensions. Second, the spectrum of both closed and open strings does not contain fermions. Third, the theory ground state is a tachyon. Tachyons are unphysical states because they assemble an instability of the vaccum. Indeed all these obstacles can be resolved on the way of  making the theory a realistic one that describing the nature by including the fermions in its spectrum. 

In string theory, the incorporation of fermions turns out to require symmetry farther than Poincar\'e symmetry. It requires supersymmetry, a symmetry that relates fermions and bosons, and the resulting theory is called superstring. In superstring theory, there is an equal number of fermions and bosons at every mass level. If supersymmetry exists in nature it must be spontaneously broken because we do not observe degeneracies between fermions and bosons. 
Anyway supersymmetry is considerd one of most presumably scenarios beyond  the Standard Model of particle physics.        

To extend the Polyakov action to be supersymmetric invariant, the world sheet bosonic fields $X^\mu(\sigma,\tau)$  are paired up with internal degrees of freedom describing fermionic partener $\psi^\mu(\sigma,\tau)$ . The new fields  $\psi^\mu(\sigma,\tau)$ are two component Majorana spinors and vectors under Lorentz rotations $SO(D-1,1)$ . The final worldsheet action is obtained by adding the standard Dirac of $D$ freely massless fermions to the free theory of $D$ massless bosons 
\be S= - \frac{1}{2\pi} \int d^2 \sigma (\partial_\alpha X^\mu \partial^\alpha X^\nu + \bar{\psi}^\mu \partial_\alpha \gamma^\alpha \psi^\nu  ) \eta_{\mu\nu},
\label{susyA}
\ee
here $\gamma^\alpha$ , with $\alpha =0,1$. In two dimensions, a particular simple choice for the $\gamma$- matrices is:
\be
\gamma^0 = \sigma_2 , ~~~~~~ \gamma^1 = i \sigma_1,
\label{gamma}
\ee
Dirac matrices obey the Clifford algebra $\{ \gamma_\alpha,\gamma_\beta \} =2 \eta_{\alpha\beta}$. The fermionic worldsheet $\psi^\mu$ are made of Grassmann variables , so they obey the anticommutation relations
\be \{ \psi^\mu ,\psi^\nu \}= 0. \ee

Since the Dirac matrices are purely real , Equ. (\ref{gamma})  is a Majorana representation, with eigenfunctions 
\be 
\psi = \begin{pmatrix}
\psi_- \\
\psi_+ 
\end{pmatrix}.
\ee 
The Dirac conjugate of a spinor is $\bar{\psi} = \psi^\dagger \beta$, $\beta=i\gamma_0$, which for a Majorana spinor $\psi^T \beta$ . In this notation the fermionic part of the action is \cite{POLCHINSKI:2005, Dine2007} 
\be 
S_f = \frac{i}{\pi} \int d^2 \sigma (\psi_- \partial_+ \psi_- + \psi_+ \partial_- \psi_+),
\label{sf}
\ee
where $\psi_\pm = \frac{1}{2} (\partial_\tau\pm \partial_\sigma)$ refer to the worldsheet coordinates defined by $\sigma^\pm = \tau\pm \sigma$. The equation of motion for the two spinor components is the Dirac equation , which now take the form 
\be
\partial_+ \psi_- = 0,  ~~~~~~~~~~~ \partial_- \psi_+ = 0,
\label{eom2}
\ee  
these equations describe left-moving and right-moving waves.  Thus in two dimensions,  
the fields $\psi_\pm$ are Majorana- Weyl spinors.

\section{Model expansions and types of superstring theories}
The possible boundary conditions for the fermionic fields $\psi^\mu$ are exactly the same as for the case of the bosonic string theory $X^\mu$ . The equations of motion (\ref{eom2}) guarantee that the action Equ. (\ref{sf}) is stationry . Requiring the vanishing of the action variation at the boundaries is fulfilled by admitting that 
\be 
\begin{aligned}
\psi(\tau,\pi) &= + \psi (\tau, -\pi) ~~~~~ \mbox{Ramond boundary condition}\\
\psi(\tau,\pi) &= - \psi (\tau, -\pi) ~~~~~ \mbox{Neveu-Schwarz boundary condition}.
\end{aligned}
\ee
So as the superstring moves in space , if we considered it moves through an interval $\sigma \in [-\pi,\pi]$ \footnote{\em Equivalently $\sigma \in [0,2\pi]$ in the sense of closed string .}, the full superstring state space breaks into two subspaces or, two sectors : a Ramond (R) sector which contains periodic fermions , and 
Neveu–Schwarz (NS) sector which contains antiperiodic fermions. 
Correspondingly, the fields mode expansion: 

$\bullet$ For open strings: In the R sector
\be 
\psi_-^\mu(\tau,\sigma) = \frac{1}{\sqrt{2}} \sum_{n\in \mathbb{Z}} d^\mu_n e^{-in(\tau-\sigma)} , ~~~~~~~~~ \psi_+^\mu(\tau,\sigma) = \frac{1}{\sqrt{2}} \sum_{n\in \mathbb{Z}} d^\mu_n e^{-in(\tau+\sigma)}
\ee 

In the NS sector
\be 
\psi_-^\mu(\tau,\sigma) = \frac{1}{\sqrt{2}} \sum_{r\in \mathbb{Z}+ 1/2} b^\mu_r e^{-i r (\tau-\sigma)} , ~~~~~~~~~ \psi_+^\mu(\tau,\sigma) = \frac{1}{\sqrt{2}} \sum_{r\in \mathbb{Z}+1/2} b^\mu_r e^{-ir(\tau+\sigma)}
\ee 
The Majorana condition requires these expansions to be real, and hence $d^\mu_{-n}= d^{\mu \dagger}_{n}$.

$\bullet$ For closed strings: In the R sector
\be 
\psi_-^\mu(\tau,\sigma) =  \sum_{n\in \mathbb{Z}} d^\mu_n e^{-2 i n (\tau-\sigma)} , ~~~~~~~~~ \psi_+^\mu(\tau,\sigma) =  \sum_{n\in \mathbb{Z}} \tilde{d}^\mu_n e^{- 2 i n (\tau+\sigma)},
\ee

In the NS sector
\be 
\psi_-^\mu(\tau,\sigma) =  \sum_{r\in \mathbb{Z}+ 1/2} b^\mu_r e^{-2 i r (\tau-\sigma)} , ~~~~~~~~~ \psi_+^\mu(\tau,\sigma) = \sum_{r\in \mathbb{Z}+1/2} \tilde{b}^\mu_r e^{-2 i r (\tau+\sigma)}.
\ee 

So that an open superstring has two sectors (NS and R), while closed
string sectors corresponding to the different pairings of the left- and right-movers there are four distinct closed-string sectors: (NS, NS), (NS, R), (R, NS), (R,R). 

In closed superstring theories spacetime bosons arise from the (NS,NS) sector and also from the (R,R) sector, since this sector is “doubly” fermionic. The spacetime fermions arise from the (NS,R) and (R, NS)
sectors. In order to get a closed string theory with supersymmetry we must truncate the four sectors above. A consistent truncation arises if we use truncated left and right sectors to begin with \cite{Zwi:2009}. Suppose we take, for example, 
\be 
 \mbox{left sector}:~~
\begin{Bmatrix}
 \mbox{NS+}& \\
 \mbox{R-}& 
\end{Bmatrix}, ~~~~~~~~
 \mbox{right sector}:~~
\begin{Bmatrix}
 \mbox{NS+}& \\
 \mbox{R+}&  
\end{Bmatrix}.
\ee
Combining these sectors multiplicatively we find the four sectors of the type \RNum{2}A superstring:

\begin{center}
Type \RNum{2}A : (NS+, NS+), (NS+,R+), (R-, NS+), (R-,R+).
\end{center}

The  (NS+, NS+) sector is the one responsible for the graviton $G_{\mu\nu}$ , 
the antisymmetric $B_{\mu\nu}$ field (called Kalb- Ramond) and the dilaton $\phi$.  
A different theory, a type \RNum{2}B superstring, arises if the chosen Ramond sectors are of the same type:
\be 
 \mbox{left }:~~
\begin{Bmatrix}
\mbox{NS+}& \\
 \mbox{R-}&  
\end{Bmatrix}, ~~~~~~~~
 \mbox{right sector}:~~
\begin{Bmatrix}
 \mbox{NS+}& \\
 \mbox{R-}&  
\end{Bmatrix}.
\ee
We then get

\begin{center}
Type \RNum{2}B : (NS+, NS+), (NS+,R-), (R-, NS+), (R-,R-).
\end{center}

While the (NS+, NS+) bosons of type \RNum{2}A and type \RNum{2}B theories are the same 
\footnote{\em This sector is universal in all five superstring theories.}, the R-R bosons are rather different:

\begin{center}
R–R massless fields in type \RNum{2}A : $A_\mu, A_{\mu\nu\rho},$
\end{center}

\begin{center}
R–R massless fields in type \RNum{2}B: $A$, $A_{\mu\nu}$, $A_{\mu\nu\rho\sigma}$ ,
\end{center}
where $A_\mu$ is Maxwell field, $A_{\mu\nu\rho}$  is a 3-form gauge field, $A$ is a scalar field and $A_{\mu\nu}$  is a Kalb–Ramond field. The above R–R fields are deeply related to the existence of stable
solutions of string theory, known as branes. The index \RNum{2} refers to the theories being 
$\mathcal{N}=2$ supersymmetric. 
Actually the previous spectrum truncating (or projecting) is made in a very specific way that eliminates tachyons
appearance and  leads to a supersymmetric theory in ten-dimensional space-time  
\footnote{ \em This projection is called the
GSO projection, since it was introduced by Gliozzi, Scherk and Olive. For more details look for example at \cite{Katrin}.}.

In addition to the type \RNum{2} theories, there are also two heterotic superstring theories. They
 are also closed string theories, but while a type \RNum{2} closed superstring arises by combining
together left moving and right moving copies of open superstrings, in the heterotic
string we combine a left moving open bosonic string with a right moving open superstring.

Out of the 26 left moving bosonic coordinates of the bosonic fields only ten of them are
matched by the right moving bosonic coordinates of the superstrings. As a result, this
theory effectively lives in ten-dimensional spacetime. 

Heterotic strings come in two versions:
$E_8 \times E_8$ type and $SO(32)$ type. These labels characterize the groups of symmetries
that exist in the theories. $E_8$ is a group, in fact, it is the largest exceptional group (the E is
for exceptional). The group $SO(32)$ is the group generated by 32-by-32 matrices that are
orthogonal and have unit determinant. 
Finally, in addition to both type \RNum{2} and heterotic theories, there is the type \RNum{1} theory.
This is a supersymmetric theory of open and closed unoriented strings. A string theory is
unoriented, if the states of the theory are invariant under
an operation that reverses the orientation of the strings. Whereas  type \RNum{2} theories and the
heterotic theories are theories of oriented closed strings.
 Thus the complete list of ten-dimensional supersymmetric string theories is therefore

\begin{itemize}
\item Type \RNum{2} A,
\item Type \RNum{2} B,
\item  $E_8 \times  E_8$ heterotic,
\item $SO(32)$ heterotic,
\item Type \RNum{1}.
\end{itemize}

These five theories have all been known since the middle of the 1980s. Some relationships (known as dualities )
between them were found soon after their discovery, but a clearer picture emerged
only in the late 1990s. The limit of type \RNum{2}A theory as the string coupling is taken to infinity
was shown to give a theory in eleven dimensions. This theory is called M-theory. 
\footnote{\em With the meaning of M to be decided when the nature of the theory becomes clear. Also sometimes it is refered by
the mother of all string theories.}

It is known, however, that M-theory is not a string theory. M-theory contains membranes (2-branes) and
5-branes. However, the discovery of many other relationships   between the five
string theories and M-theory has made it clear that we really have just one
theory. In conclusion: M-theory is a unique theory, and the five superstrings and
are different limits of this theory. For now days the edges of this physics subject have not been reached yet.




\section{Supergravity}
The first development of supergravity theory was in the seventies and constructed quite separately from string theory. 
The $\mathcal{N}=1$ SUGRA in four spacetime dimensions as the minimal possible supersymmetric extension of GR. Today it is well known supergravity theories are the low
energy limits of string theory as we will show later. Basically one expands the string action in $\alpha'$ and drops
all terms of $\mathcal{O}(\alpha'^2)$ and higher. Yet that is was not of course the case when SUGRA has been 
discovered. Historically, it was expected since in $\mathcal{N}=1~ D=4$ SUGRA there is more symmetry than in pure gravity, 
the high–energy (short distance) behavior will improve. Although this is true, still the (super)symmetry is not enough to
cancel all divergences in the theory. That led to extend the supersymmetry itself for ($\mathcal{N} > 1$) in higher dimensions ($D > 4$) to get for the time being vast realm of supergravity theories \cite{theME}. 
However $\mathcal{N}=1~ D=11$ has earned its salient features because it is a unique theory, and its soltions are the solutions of M-theory as well. In the following we will present a whole figure about SUGRA, then we will speak solely about $\mathcal{N}$=1~ D=11 SUGRA, starting from how its action is built, a brief review about Calab- Yau 3 fold, and then how it is reduced over CY to 
$\mathcal{N}=2~ D=5$ SUGRA, the theory that we introduce solutions for it after that.   

Generally speaking there are several ways to build a supergravity theory:
\begin{itemize}
\item By Brute force. By supersymmetrizing the Einstein-Hilbert action, i.e. adding fermionc degrees of freedom (D.O.F.) as the superpartner of the graviton $g_{\mu\nu}$, then finding out the SUSY transformations that keep the new action Lorentz invariant. Unfortunately, on quantization this theory is not renormalizable exactly as GR. Hence this way is not so accurate or, at best, the theory is low energy effective limit of a better idea, which is indeed what SUGRA turned out to be.
\item To begin with a supersymmetric Yang-Mills action without gravity and then make the supersymmetric
transformations local, i.e. become spacetime dependent. This directly implies the existence of gravity in the theory. 
In te next section we will give more details about this procedure, but keeping in mind this minimal SUGRA is not perturbative. 
\item Pertubative SUGRA in ten spacetime dimensions is a low energy limit of perturbative superstring theories in $D=10$. As illustrated in Fig. (\ref{Mom}), type  \RNum{2}A string theory reduces to type \RNum{2}A SUGRA, type \RNum{2}B string theory reduces to type \RNum {2}B SUGRA, type \RNum{1} string theory reduces to type \RNum{1} SUGRA and the heterotic string theories both reduce to type \RNum{1} SUGRA.  
\item Different SUGRA models in lower spacetime dimensions is obtained form the compactifiction of higher dimensional SUGRA models over various types of manifolds. Indeed this way is very important because all the trials now to detect any signal for the higher dimensional theories within our four dimensions world. we will talk bout the dimensional reduction in next pages.     
\end{itemize}

\subsection{Generating SUGRA from SUSY}
In general new fields should be added when a global symmetry becomes local. Consider the gauge symmetry, for example the Dirac action \cite{Freedman1976, Cerdeno:1998hs}
\be S= i \int d^4 x ~ \bar{\psi} \gamma^\mu \partial_\mu \psi  \ee
is invariant under the global symmetry $\psi'= e^{i\alpha} \psi$. If we make this symmetry local $\psi' = e^{i\alpha(x)} \psi$, the action becomes
\be S'= i \int d^4 x ~ e^{i\zeta(x)} \bar{\psi} \gamma^\mu \partial_\mu (e^{-i\zeta(x)} \psi),  \ee
\be S'=  S+ 
i \int d^4 x \bar{\psi} \gamma^\mu  \psi (-i) (\partial_\mu \zeta),  \ee
adding a new term:
\be
i A_\mu \bar{\psi} \gamma^\mu  \psi e^{i\zeta(x)},
\ee
now; 
\be S'=  S+ 
i \int d^4 x \bar{\psi} \gamma^\mu  \psi (A_\mu +\partial_\mu \zeta) e^{i\zeta(x)}, \ee
\be A'_\mu=A_\mu+\partial_\mu \zeta. \ee

Second example, consider the action of a scalar field in flat space- time
\be 
S= \int d^4 x \Big[ \frac{1}{2} \eta^{\mu\nu} \partial_\mu \phi \partial_\nu \phi - V(\phi)  \Big],
\ee
where $\eta^{\mu\nu}$ is the Minkowski metric. This action is invariant under Poincar\'e symmetry
\be 
x^\mu \to \Lambda^\mu_\nu x^\nu + a^\mu,
\ee
to make this transformation local, we replace the Lorentz four vector $\mathbf{\Lambda}$ by "some thing" depends on space- time $\Lambda^\mu_\nu=\frac{\partial x^\mu}{\partial x^\nu}$ which is the generalized coordinate transformations in special relativity. Now to keep the action invariant we have to change to
\be
S= \int d^4 x \sqrt{-|g|} ~\Big[ \frac{1}{2} g^{\mu\nu} \partial_\mu \phi \partial_\nu \phi - V(\phi) + \mathcal{R}  \Big] ,
\ee
where $g^{\mu\nu}$ is the metric. From the view point of the particle physics it is a spin- 2 particle called the graviton. $\mathcal{R} $ is the Ricci scalar. In the case of supersymmetry, consider the free Wess–Zumino Lagrangian
\be \mathcal{L}  = \partial_\mu \phi^* \partial^\mu\phi + \chi^\dagger i \bar{\sigma}^\mu (\partial_\mu \chi). \ee  
The corresponding action is invariant under the SUSY transformations
\be \delta \phi = \zeta. \chi= \zeta^a \chi_a= \zeta^\dagger (-i \sigma_2) \chi ,~~~~~~~~ \delta \chi=-i \sigma^\mu (i\sigma_2 \zeta^*) \partial_\mu \phi, \ee
together with their hermitian conjugates. When $\zeta \to \zeta(x)$
\be \delta \mathcal{L} = \mbox{terms seen before} + \chi^\dagger (i\sigma_2) (\partial^\mu \phi ) (\partial_\mu \eta^*) 
- \chi^\dagger~ i\bar{\sigma}^\mu~ i \sigma^\nu~ (\partial_\mu \phi) (\partial_\nu \bar{\chi}), \ee
where
\be \delta \mathcal{L}_{\mbox{Extra}} \sim K_\mu (\partial^\mu \zeta). \ee
To cancel these extra terms, we add to the Lagrangian
\be \mathcal{L}_{\mbox{Extra}}= K_\mu \Psi^\mu_a,\ee
provided that:
\be \Psi'^\mu = \Psi^\mu + \partial^\mu \zeta,\ee
$\Psi^\mu$ is a spin $\frac{3}{2}$ particle called the gravitino. So that the Lagrangian becomes
\be \mathcal{L} = \partial_\mu \phi^* \partial^\mu\phi - \frac{1}{2} \bar{\psi} \gamma^\mu  \partial_\mu \psi +K_\mu \Psi^\mu, \ee  
where in the second term we used Dirac 4-spinor formula. The variation of $\Psi^\mu $ cancels $\delta \mathcal{L}_{\mbox{Extra}}$. However it has been found that there are still extra terms
\be
\delta \mathcal{L} \sim - K_\mu (\partial^\mu \zeta) - \bar{\Psi} \gamma^\nu \zeta(x) \left( (\partial_\mu \phi^*) (\partial_\nu \phi) + \frac{1}{2} \bar{\psi} \gamma_\mu \gamma_\nu \psi \right).
\ee
The last two terms are defined as the energy-momentum tensor which is given by
\be 
T_{\mu\nu} = \frac{1}{\sqrt{-|g|}} \frac{\partial \mathcal{L}}{\partial \eta^{\mu\nu}},
\ee
so to be canceled , another extra term is added which might be now contracted with $T_{\mu\nu}$
\be \mathcal{L}_{Extra~\RNum{2}} \sim g_{\mu\nu} T^{\mu\nu}, \ee
provided that
\be \delta g_{\mu\nu} = \bar{\psi}_\mu \gamma_\nu \zeta(x), \ee
where $g_{\mu\nu}$ is the graviton. In $\mathcal{N}=1$ supersymmetry the gravitino $\Psi_\mu$ is the superpartener of the graviton. 
Some thing can be concluded when gravity arises naturally in a locally supersymmetric theory. That is
supersymmetry is not just an arbitrary restriction on nature, but rather a maximal symmetry of spacetime with far reaching implications.
\subsection{SUGRA in $11$- dimensions}
This theory is the most general SUGRA and it is completely unique if we required that we can not never have  particles have higher spin more than 2. Then 11 dimension is the highest possible number of spacetime dimension that we can construct supergravity theory \cite{freedman:2012}.

Note that, since D=11 is the highest dimension where supersymmetry can live, it was expected when D=11 SUGRA dimensionaly reduced, SUSY and Einstein's relativity are obtained. But unfortunately this is not the case. At dimensional reduction, there are many extra fields produced not exist in the original theories, that makes the theoreticians till nowadays keep wondering what is the right relation between D=11 SUGRA and SUSY and GR.   
Consider for instance D=3  $\mathcal{N}$=1 supergravity, on shell, $e^a_\mu$ has $\Big (\frac{(d-1)(d-2)}{2}-1 \Big)$ D.O.F. and $\psi^a_\mu$ has $ \Big ( (d-3) 2^{\frac{d-1}{2}}/2 \Big) $. So on shell $\{e^a_\mu,\psi^a_\mu\}$ has no D.O.F. Off-shell however, we have for $e^a_\mu$, $\frac{d(d-1)}{4}=3$ D.O.F. and for $\psi^a_\mu$ , $(d-1) 2^{\frac{d-1}{2}}=4$. So we need to add an auxiliary scalar S \cite{Intro1}. 

Now consider D=11 $\mathcal{N}$=1 supegravity. On sell $e^{\hat{\alpha}}_\mu$ has 44 D.O.F. and $\psi^a_\mu$ has 128 D.O.F. ($\mu=0,...,10$). So we need extra (84) bosonic D.O.F. That will be a 3-form gauge field 
\be A_3=\frac{1}{3!}~ A_{\mu\nu\rho}~ dx^\mu \wedge dx^\nu \wedge  dx^\rho. \ee 
The field strength will be a 4-form defined by
\be
F=dA=\frac{1}{4!}~ F_{\mu\nu\rho\lambda}~ dx^\mu \wedge dx^\nu \wedge  dx^\rho \wedge  dx^\lambda,
\ee
where
\be
F_{\mu\nu\rho\lambda} = \partial_\mu A_{\nu\rho\lambda} - \partial_\nu A_{\mu\rho\lambda} + \partial_\rho A_{\mu\nu\lambda}
- \partial_\lambda A_{\mu\nu\rho}.
\ee
The Lagrangian 
\be
2~ k_{11}^2~ \mathcal{L} = |e|~ \mathcal{R} -2~ i ~|e|~ \bar{\psi}_\mu \gamma^{\mu\nu\rho} (\bar{\nabla}_\nu \psi_\rho)- \frac{1}{2}~ \frac{|e|}{4!}~
F_{\mu\nu\rho\lambda} F^{\mu\nu\rho\lambda} + \mbox{additional Ferminoic terms}, \label{11lag}
\ee
where the greek symbols run over $0,...,10$, $|e|=\sqrt{-|g|}$, $g= \mbox{det}~  g_{\mu\nu}$ , $\gamma^{\mu\nu\rho}\sim \gamma^{[\mu\nu}~~ \gamma^{\rho]}$, $\gamma^{\mu\nu}\sim[\gamma^\mu,\gamma^\nu]$, and $\gamma^\mu=e^\mu_{\hat{\alpha}} \gamma^{\hat{\alpha}}$.  The Lagrngian is invariant under the following transformations:
\be
\delta e^{\hat{\alpha}}_\mu = i \bar{\zeta} \gamma^{\hat{\alpha}} \psi_\mu,
\ee
\be 
\delta \psi_\mu = \bar{\nabla}_\mu \zeta.
\ee

To find out the transformation of $A_{\mu\nu\rho}$ which keep the Lagrangian invariant, we should add extra term 
$\frac{1}{6~ 3! (4!)^2}~ \varepsilon_{\mu_1,..\mu_{11}} A^{\mu_1,..\mu_3} F^{\mu_4,..\mu_7} F^{\mu_7,..\mu_{11}}$. Focusing only on the bosonic terms, and write the Lagrangian in the vielbein formula \footnote{See Appendix (\ref{Df}).}:
\be
2~ k_{11}^2~ \mathcal{L} = |e|~ \Big[ \mathcal{R} - \frac{1}{48} F^2\Big] - \frac{1}{6} A \wedge F \wedge F  ,
\label{Lag111}
\ee
which now is invariant under
\be
\delta e^{\hat{\alpha}}_\mu =- \frac{i}{2} \bar{\zeta} \gamma^{\hat{\alpha}} \psi_\mu,
\ee
\be 
\delta \psi_\mu = 2 \bar{\nabla}_\mu \zeta + \frac{1}{144} F_{\nu\rho\lambda\sigma} \Big[ \gamma_\mu^{\nu\rho\lambda\sigma}- 
8 \delta^\nu_\mu \gamma^{\rho\lambda\sigma}\Big]\zeta,
\ee
\be 
\delta A_{\mu\nu\rho} = \frac{3}{2} \bar{\zeta} \gamma_{[\mu\nu} ~ \psi_{\rho]}.
\ee
The next step to write the equation of motions of the gauge field strength . Returning to Maxwell's equations in 4- dimensions, $\partial_\mu F^{\mu\nu}= j^\nu$. They can be written as $ d * F = J$, which gives Gauss’s law and Ampere’s law  and
$ d F =0$, Bianchi identity that gives Faraday's law. Similarly in 11-dimensions, the source free E.O.M of the 4- form field strength are
\be
\begin{aligned} 
d * F + \frac{1}{2} ~ F \wedge F =0 ,
~~~~~~d F =0.
\end{aligned}
\ee
If there is a source
\be 
\begin{aligned} 
\int (d * F + \frac{1}{2} F \wedge F) = Q_E,
~~~~\int d F = Q_M,
\end{aligned} 
\ee
where $Q_E$ and $Q_B$ are constants and calling them with these names has nothing to do with the electric and the magnetic charges, but that’s just an analogy with the Maxwell’s equations. 

The question now, in eleven dimensions, what are these constants ? and what is the source of these charges ? 

For instance , in Maxwell’s equations , the source of electric charges is the electrons. Can we have here point- like particles with $Q_E$ and $Q_M$? 

In 4- dimensions: ~~~~~~ $\mathcal{L}_{int} \propto A_\mu J^\mu, $ 
where $J^\mu$ is the current of a point- like particle and the index $\mu$ describes the motion of a particle that has a world line in 4 dimensions space- time. However

In 11- dimensions : ~~~~~ \be \nonumber \mathcal{L}_{int} \propto A_{\mu\nu\rho} J^{\mu\nu\rho}, \ee
$J^{\mu\nu\rho}$ describes a hyperplane ( some thing with higher dimensions ) propagates in time, but it has two space dimensions , i.e., it’s a 3-dimensional hyperplane , two spatial propagate in time , imagine it as “ a paper”. 
It’s a generalization of the world line and it’s called a supermembrane or (2Brane). Two here refers it’s a two dimensional brane propagating in time .

When the 2Brane lives in 11- dimensions is wrapped a round 1- dimension of radius $R$, then $R \to 0$
limit is taken, this mean we go from 11-dimensions lower to 10-dimensions . This’s called klein kaluza compactification . Indeed it has been found that superstrings in 10- dimensions come from membranes in 11- dimensions when wrapped around a circle . The membrane has an electric $Q_E$, exactly like a point particle in 4- dimensions.  

So we can see how the extra term added to the Lagrangian that has a gauge field is strongly related to the presence of a super 2Brane in 11- dimensions. So what do we have now? 
a gravitational object carries an electric charge ( you can call the coupling constant of Lagrangian term that has $A$ ), has a mass and propagates gravitationaly in space- time ( so it can be described by a metric ). 

These discoveries were incredible in supergravity, because it were discovered separately in superstigns in the same time.

Turns now to the magnetic charges in 11- dimensions. The first question we start by, do the electric and magnetic charges couple to the same branes or no ? 

In Maxwell’s equations $Q_E=\int d * F$ which means $Q_E$ is related to $d*F$, while $Q_M=\int d F$ which means $Q_M$ is related to $d F$. $F$ is a two form, while $* F$ is a $(D-2)$ form, i.e., $*F $ is a two form . So that in Electromagnetism (EM) $F_{\mu\nu}$ is a $4 \times 4$ matrix, and its dual $F^{\mu\nu}$ is a $4 \times 4 $ too, so in 4- dimensions the point particle can carry a magnetic and an electric charges .

In 11- dimensions. however $F$ is a 4 form , while $* F$ is a 7- form. $*F$ has a seven indices $F^{\mu\nu\rho\lambda\sigma\delta\zeta}$ and it’s the field strength of a six form gauge field, $F=d A$, that is responsible for the magnetic charge. And as the 3- form gauge field couples to two dimensional brane (called M2-brane) , the 6- form brane couples to 5- dimensional brane ( 5- dimensional surface propagating in time that’s called M5- brane).

M2-brane and M5-brand are considered the simplest solutions of supergravity in 11- dimensions, and there are proofs that you can not construct any other branes dimensions. However a combinations of these two branes can be constructed because they are gravitational objects, hence there can be multiple branes. But any ways M2 and M5 are the fundamental branes. When the M5- brane is wrapped over a 1- dimension, we get a 4- dimensions brane just like when we wrap the M2-brane over 1- dimension. and get a string, and it is found there are 4- dimensions branes in string theory. Indeed that is the goal from the beginning to start by the most possible high dimension and then compictify to lower dimensions and see what we will have. So as bold lines, there are no point particles in 11- dimensions, they can not exist , you can consider Maxwell’s equations in EM as a special case of higher gauge field theories, and finally M-branes ( M2- M5) solutions of supergravity are non perturbative.

The anti-commutations relations of the SUSY generators now become
\be 
\{Q,\bar{Q}\} = \gamma^\mu P_\mu + \gamma^{\mu\nu} Z_{\mu\nu} + \gamma^{\mu\nu\rho\lambda\sigma} Z_{\mu\nu\rho\lambda\sigma}.
\ee

It has been found that these ``Zs`` are topological charges of the M-branes, $Z_{\mu\nu}$ is the topological charge of M2-brane and $Z_{\mu\nu\rho\lambda\sigma}$ is the topological charge of M5-brane.

\subsection{Branes in string theory}
To manifest that the origin of branes in string theory is completely different than in supergravity. 
As previously pointed out in string theory branes raised to conserve momentum at the end points boundary
conditions for the open strings. So open strings always bounded to dynamical objects that have extended dimensions, i.e., 
strings end points always attached to some hyperplanes which is called p-branes. Where if $p=0$ it is a point particle, if $p=1$ it is a string, if $p=2$ it’s a membrane and so on.
However, although branes in string theory are in 10- dimensions and branes in supergravity are in 11- dimensions , theoreticians found how they relate to each other via Kaluza Klein reduction as will be mentioned next.
\subsection{Branes as solution for SUGRA}
The question now how these branes are written as solution for SUGRA?
We have admitted before that they are gravitational objects, i.e., they have an energy density and so they gravitate in addition they have a coupling with the form gauge \cite{theME}. Let us write M2-brane. First of all as these branes  exist "if they exist" , they break SUSY because they define a preferred direction of space. So Poincar\'e symmetry is broken, i.e., supersymmetry is broken. In the case of M2-brane , it breaks the space into two pieces , i.e., M2-brane breaks SUSY to two halfs.


The simplest form of M2-branes happens when the fermions vanish, and we get pure bosonic solutions. 
The M2-brane metric is then 
\be 
ds^2_{11}= ds^2_{M2} + ds^2_{\mbox{transverse}},
\ee
it is  11- dimensional metric, but 11- dimensions written in two terms , a term explains the world volume of the brane and the other term explains the remaining dimensions. So $ds^2_{\mbox{transverse}}$  is the metric of all dimensions that perpendicular to my brane. Writing general matrices for each piece
\be 
ds^2_{11}= g_{ab}~ dx^a~ dx^b + g_{\mu\nu}~ dx^\mu~ dx^\nu,
\label{M2}
\ee
where $a,b=0,1,2$ and $\mu,\nu=3,...,10$. Take into account that we are assuming a symmetry in space, like a rotation symmetry for these dimensions. For instance , I need to describe the 8- dimensions by $SO(8)$ , so the metrices $g_{ab}$ and $g_{\mu\nu}$ most likely will depend on a radius r where $r^2 = x^2_3+ ...+ x^2_{10}$ is a radial distance in 8-dimensions. 

Take all of these and apply on the equations of motion, and in the Lagrangian Equ.(\ref{Lag111}) that yielded from 
Einstein field equations (EFE) \footnote{ \em Appendix (\ref{GR}).} including the Energy-moment tensor. Also $\delta \psi_\mu=0$ can be checked (there is a theorem states that this condition implies EFE). At the end we get the metric of M2-brane as 
\be
ds^2_{11} = H^{-3/2}(r) ~\eta_{ab}~ dx^a dx^b + H^{3/2}(r)~ \delta_{\mu\nu} ~dx^\mu dx^\nu,
\label{M232}
\ee
where $H(r)$ is a radial function , $\eta_{ab}$ is the Minkowski metric. $H^{3/2}$ is a radial direction of the transverse space. Remember that the M2-brane has an electric charge under the gauge field A
\be
A_{012}= \pm H^{-1},
\ee
or in general
\be
A_{\mu\nu\rho}= \pm \varepsilon{\mu\nu\rho} H^{-1}.
\ee

From EFE, H(r) satisfies the Laplace equation 
\be
\nabla_8^2 H=0 ~~\to \delta^{\mu\nu} \partial_\mu \partial_\nu H=0,
\ee
that yields
\be 
H= 1 +\frac{M_2}{r^6}, \label{GP}
\ee
this is the equation of the gravitational potential. Where $M_2$ is an arbitrary integration constant proportional to the energy density of the brane. To figure out this formula of $H$, Equ. (\ref{GP}), let us return to the Laplace equation in 3- dimensions $\nabla^2 \phi =0$ with the spherically symmetric solution $\phi \sim 1 + \frac{q}{r}$, with the Electric field $E\sim \frac{1}{r^2}$. So we say the point particle is surrounded by a two dimensional spherical charge with volume $r^3$ (Gauss law). 

Now in 11- dimensions what is the surface should surround the M2-brane ? it’s a surface having a volume $\sim r^8$,  and an Electric field $\frac{1}{r^7}$ and so that the potential $\propto \frac{1}{r^6}$. Another case, if you make similar calculus for instance for a black hole (BH) , you will get $H \sim \frac{1}{r}$ ( similar for Schwarzschild radius). 
\bigskip

\textbf{M5-brane}:
In the same manner, let us construct a metric that has a symmetry in the transverse directions
\be 
dS^2_{11} = g_{ab}~ dx^a dx^b + g_{\mu\nu}~ dx^\mu dx^\nu,
\ee 
where $a,b=(0,1,2,3,4,5)$ and $\mu,\nu=(6,7,8,9,10)$. 
\be
ds^2_{11} = H^{-3/2} ~ \eta_{ab} ~ dx^a dx^b + H^{3/2} ~ \eta_{\mu\nu} ~ dx^\mu dx^\nu ,
\label{M5}
\ee
where the first part is the world volume of the M5-brane and the second part is transverse direction. The 6- form gauge filed is given by 
\be 
A_{012345} = \pm H^{-1}(r), 
\ee
where $H(r)$ is given by $\nabla^2_5 H(r)=0 ~ \to ~ H(r)= 1 + \frac{M_5}{r^3}.$

Now to figure out what are $M_2$ and $M_5$ and how they are related to the "electric" and the "magnetic" charges , literally the 3- form and the 6- form gauge fields, let us recall the familiar GR in 4- dimensions with an electric field. So we will have the Einstein equation with the Maxwell’s equations as
\be 
\mathcal{L} = |e| [ \mathcal{R} - \frac{1}{4}~ F_{\mu\nu} F^{\mu\nu} ] ,
\ee
take the Reissner–Nordström solution which represents electrically charged black hole with a metric 
\be
ds^2= - f(r) dt^2 + f^{-1} (r) dr^2+ r^2 d \Omega^2,
\ee
where $d\Omega^2= d\theta^2 + \sin^2 \theta d\phi^2$ and
\be 
f(r) = 1 - \frac{2M}{r} + \frac{Q^2}{r^2} ,
\ee
where Q is the ordinary electric charge of the black hole. Note that can be the Schwarzschild black hole solution except for the last term which has Q. Here the event horizon of the black hole depends on two parameters, it’s charge as well as its mass, so simply if $Q \to 0 $ we will get Schwarzschild BH. The electric field specified by $A_0 = \frac{Q}{r}$ ( just a Coulomb field). To get the BH’s event horizon   we set $f(r)=0$ and solve for r and since the equation is quadratic in $r$, we get two event horizons 
\be 
r_\pm = M \pm \sqrt{M^2 - Q^2} ,
\ee
which means we have certain conditions. 

Case : $M=Q \to $ there is a single horizon (maximal case) \\

Case: $M^2 < |Q| \to $ no horizon ( a naked singularity). 

There are hypothesis that called  "cosmic censorship" states that the naked singularities are not allowed by nature . So whatever a nature try to make a naked singularity , it doesn't work. So physically speaking, M should be greater than $|Q|$, $M \geq |Q|$. This’s called BPS (Bogomol’nyi-Prasad-Sommerfield) condition and when $M=|Q|$ it’s called saturated BPS condition. 

Now let us rewrite the RN solution at the maximal case $M=|Q|$, change the coordinates $r \to (r+M)$, then the metric becomes
\be 
ds^2_{M=|Q|} = -H^{-2} dt^2 + H^2 \delta_{ij}~ dx^i dx^j ,
\ee
where $i,j=1,2,3$ and $H=1+\frac{M}{r}$  as a solution of $\nabla^2_3 H(r)=0$. And the gauge field $A_0 = H^{-1}$. Now notice the similarity between this case and the M2 or M5 branes case Equ. (\ref{M232}), and Equ. (\ref{M5}). Therefore M2 and M5 are interpreted as generalized Maxwell's BPS black holes (or black branes) with one event horizon . They behave as just multi-dimensions black holes.
Also we can add more branes
\be
H = 1 + \sum_{i=1}^k \frac{M_i}{r_i},
\ee
so we have $k$ 2M-branes at $r=({\textbf r}-{\textbf r_I})$. This is a stable configuration, i.e., these BHs are static, they do not move because the electric repulsion exactly cancels the gravitational attraction. But this is only when the BPS condition is realized, so if BPS didn't saturated, there will be instability , the BHs start to move. 
\section{Compactificaton; An overview}
\label{compact}
Consider Kaluza- Klein compactificaton. It is have been made in 1922, that there was a hope to unify GR and EM. It assumed that we live  in a 5- dimensional space- time with only gravity. 

But what about the fifth dimension that we can not see. It was assumed that this dimension is very small, it can be wrapped it over a circle 
\be
 0 \leq x_5 \leq 2\pi R,
\ee
and assume we live where R is very small, so we can divide the metric as
the extra 4 degrees of freedom will be a gauge field $A_\mu$ , that can be the gauge field of EM. But what about thee extra diagonal scalar term that doesn't belong to the GR nor the EM. Nowadays this scalar is called the dilaton. But this model has been ruled out because it predicted a scalar field which was not known at these times.

However upon the discovery of D=11 supergravity and trying to dimensionaly reduced, Kaluza- Klein reduction has been bring to light again. But every time we go lower in dimension , we get extra fields. For instance down to 10- dimensions, the 11- dimensions metric will be 
divided as:
\be
ds^2_{11} = (e^{2\alpha} g_{\mu\nu} + e^{2\alpha} A_\mu A_\nu ) dx^\mu dx^\nu + 2 e^{2\alpha} A_\mu dx^\mu~ dy+ e^{2\alpha} dy^2,
\ee
where $\mu,\nu=0,...,9$, the first term is the solution of the EFE in 10 dimensions. So the graviton breaks down to a new gauge field $A_\mu$ and the dilaton $e^{2\phi}$. The 11- dimensions gauge field and the field strength also break down
\be 
A_{LMN}^{(3)} \to A_{\mu\nu\rho}^{(3)} + A_{\mu\nu y} \equiv B_{\mu\nu}^{(2)},
\ee
\be 
F_{LMNP} \to F_{\mu\nu\rho\lambda} + G_{\mu\nu\kappa} ~~~ (G=dB),
\ee
where $y$ is the extra dimension we wrapping $0 \leq y \leq 2\pi R$. Also the spinors will split into two. So we get $\N=2$ supergravity in 10- dimensions, which called type \RNum{2}A SUGRA.

This wrapping over a circle can be made many times down to 4- dimensions.
In each time we reduce a dimension, we add new field breaks SUSY into half's . So this way in reduction is not preferable because SUSY will be broken too much than we phenomenologicaly need. 

We can make dimension reduction over manifolds. But what are the type of manifolds can be used to reduce from higher to lower dimensions while SUSY is preserved. For instance we can reduce over Calabi-Yau manifold to go from D=10 to D=4 with preserving SUSY. The 10- dimensions metric will then be
\be
ds^2_{10}= g_{\mu\nu} ~ dx^\mu dx^\nu + g_{ab} ~ dx^a dx^b,
\ee
where the first part is the 4- dimensions we want $\mu,\nu=0,...,3$ and the second part is the 6- dimensions manifold $a,b=4,...,10$ . There is also what is called G-Calibrated manifold which is 7- dimensional, so we can reduce over it form D=11 directly to D=4.

Calabi-Yau manifolds has many characteristics \footnote{\em As we will see in the next chapter.}, like they always have even dimensions, and usually written in complex number. For instance the 6- dimensions Calabi-Yau can be written as
\bea
\nonumber z_1&=&x_1+ix_2, \\ 
\nonumber z_2&=&x_3+ix_4, \\ 
z_3 &=&x_5+ix_6, 
\eea 
so instead of 6- dimensions real manifold, now we have only 3- dimensions complex manifold. Anyhow , although the Calabi- Yau manifold's properties are well known, till these days no body knows what it looks like. That because Calabi- Yau manifold has to be closed manifold not opened and unfortunately no one could construct yet a closed Calabi- Yau manifold, but we know it just exist due to Yau's theorem, which proves the existence of a metric , even though we do not know yet how this metric looks like.

\section{Low energy supergravity }

Let us see how \RNum{2}A superstring theory arises from the dimensional reduction of the eleven-dimensional supergravity 
\cite{POLCHINSKI:2005}

As shown before, the eleven-dimensional supergravity has two bosonic fields, the metric $G_{MN}$ and a 3-form potential $A_{MNP} \equiv A_3$ with field strength $F_4$. The bosonic part of the action  is given by
\be
2 \kappa_{11}^2 S_{11} = \int d^{11} x ~ ( - G )^{1/2} ~ \Big ( R -  \frac{1}{2} |F_4|^2 \Big ) - \frac{1}{6} \int A_3 \wedge F_4 \wedge F_4, 
\label{s11}
\ee
now dimensionaly reduce . The general metric that is invariant under translations in the 10-direction is
\bea
\nonumber ds^2 &= & G_{MN}^{11} (x^\mu ) ~ dx^M ~ d x^N\\ 
&=& G_{\mu\nu}^{10} (x^\mu ) ~ dx^\mu ~ d x^\nu + \text{exp} (2 \sigma (x^\mu )) [ dx^{10} + A_\nu (x^\mu) dx^\nu]^2 , 
\eea 
where $M,~ N$ run from 0 to 11 and $\mu,\nu$ from 0 to 9. The eleven-dimensional metric reduces to a ten-dimensional metric, a gauge field $A_1$, and a scalar $\sigma$. The potential $A_3$ reduces to two $A_3$ and $A_2$ . The three terms in Equ.(\ref{s11}) become
\bea 
\nonumber S_1 &=& \frac{1}{2\kappa_{10}^2} \int ~ d^{10} x ~ (-G)^{1/2} ~\Big ( e^\sigma R - \frac{1}{2} e^{3\sigma} |F_2|^2 \Big ), \\ \nonumber
S_2 &=& - \frac{1}{4\kappa_{10}^2} \int ~ d^{10} x ~ (-G)^{1/2} ~\Big ( e^{-\sigma} |F_3|^2 + e^\sigma |\tilde{F_4}|^2 \Big) , \\ 
S_3 &=& - \frac{1}{4\kappa_{10}^2} \int A_2 \wedge F_4 \wedge F_4 = 
- \frac{1}{4\kappa_{10}^2} \int A_3 \wedge F_3 \wedge F_4, \label{s10}
\eea 
where
\be
\tilde{F_4}= d A_3 - A_1 \wedge F_3,
\ee
the fields of the reduced theory are the same as the bosonic fields of the \RNum{2}A string, as they must be. In particular the scalar $\sigma$ must be the dilaton $\phi$ up to some field redefinition. The terms in the action have a variety of $\sigma$- dependency.

Since we have arrived at the action Equ. ( \ref{s10}) without reference to string theory, we have no idea as yet how these fields are related to those in the world-sheet action. We will proceed by guesswork, and then explain the result in world-sheet terms. First redefine
\be 
G_{\mu\nu} = e^{-\sigma} G_{\mu\nu} ~ (\text{new}), ~~~ \sigma =\frac{2\phi}{3}, \label{dil}
\ee
substitute by the new metric. Then
\bea 
\nonumber S_{\RNum{2}A} &=& S_{NS} + S_R + S_{CS} , \\ \nonumber 
S_{NS} &=& \frac{1}{2\kappa_{10}^2} \int ~ d^{10} x ~ (-G)^{1/2} 
e^{-2 \phi} ~\Big ( R - 4 \partial_\mu \Phi \partial^\mu \Phi - \frac{1}{2} |H_3|^2 \Big) , \\ \nonumber
S_R&=& - \frac{1}{4\kappa_{10}^2} \int ~ d^{10} x ~ (-G)^{1/2} ~\Big (  |F_2|^2 +  |\tilde{F_4}|^2 \Big) , \\ 
S_{CS} &=& - \frac{1}{4\kappa_{10}^2} \int B_2 \wedge F_4 \wedge F_4 , \label{ss}
\eea 
note that $R \to e^\sigma R + ....$,  that $(-G)^{1/2} \to e^{-5\sigma} ~ (- G)^{1/2}$. 

The NS action now involves the dilaton in standard form. Equ. (\ref{dil}) is the unique redefinition that does this. The R action does not have the expected factor of $e^{-2\phi}$, but can be brought to this form by the further redefinition 
\be
C_1 = e^{-\Phi} C’_1,
\ee
after which 
\be 
\int d^{10} x ~ (-G)^{1/2}~ |F_2|^2 = \int d^{10} x ~ (-G)^{1/2}~ e^{-2\Phi} ~|F’_2|^2,
\ee
\be 
F’_2 \equiv dC’_1- d\Phi \wedge C’_1, 
\ee
and similarly for $F_3$ and $C_3$. With the Bianchi identity and gauge transformations
\be
dF'_2=d\Phi\wedge F'_2,~~~~~~~~~~~ \delta C'_I = d\lambda'_0 - \lambda'_0 d\Phi.
\label{B-gt}
\ee
Let us now make contact with string theory and see why the background R–R fields appearing in the world-sheet action have the more complicated properties Equ. (\ref{B-gt}). We work at the linearized level, in terms of the vertex operators
\be 
\eta_\alpha \tilde{\eta}_\beta ( C \Gamma^{\mu_1...\mu_p} )_{\alpha \beta} e_{\mu_1...\mu_p} (X),
\label{vo}
\ee
here $\eta_\alpha$ is the R ground state vertex operator and $\Gamma^{\mu_1...\mu_p}= \Gamma^{[\mu_1...\Gamma^{\mu_p}]}$ . 

Amount to two Dirac equations, one acting on the left spinor index and one on the right:
\be
\Gamma^\nu \Gamma^{\mu_1...\mu_p} \partial_\nu e_{\mu_1....\mu_p} (X) = 
\Gamma^{\mu_1...\mu_p} \Gamma^\nu \partial_\nu e_{\mu_1....\mu_p} (X) =0,
\ee
this Dirac equation is equivalent to: 
\be
d e_p =d \star e_p =0,
\ee
these equations have the same form as the field equation and Bianchi identity for a $p$-form field strength. Thus the function $e_{\mu_1...\mu_p} (X) $ appearing in the vertex operator as the R–R field strength rather than potential. To confirm this, observe that in the \RNum{2}A theory the spinors in the R–R vertex operator, Equ. (\ref{vo}) have opposite chirality and so their product contains forms of even rank, the same as the    \RNum{2}A  R–R field strengths.

\section{Experimental signatures for string theory} 

Detecting superstring theory has many perspective and can be in many different ways, as:

\begin{itemize}

\item Finding new supersymmetric particles by the high energy accelerators like the Large Hadron Collider (LHC) at CERN will prove supersymmetry and enhance string theory situation . On the other side, finding standard model- like particles but at a greater mass will be a discovery of what’s called Kaluza-Klein recurrences which may suggest the existence of extra dimensions, one of the  predominant predictions of string theories. 

\item As gravity is thought to be a force able to probe these extra dimensions, another way in which we may be find evidence for string theory is through the disappearance of gravitons, the hypothesized carrier of gravity, into these other dimensions. The particle might be carried away without a trace, but it would leave behind an imbalance in momentum and energy. So careful analysis would be needed to decide if a graviton has traveled to another dimension or we have instead produced a SUSY particle, 
whose decay products would also be invisible to our detector.

\item 	Through its impact on the earliest, most extreme moments of the universe, the physics of string theory may have left faint cosmological signatures—for example, in the form of gravitational waves or a particular pattern of temperature variations in the cosmic microwave background radiation \cite{Erik2011}.

\end{itemize}

\setcounter{equation}{0}
\chapter{Manifolds}

The space-time in string theory and though in supergravity is often described by means of a mathematical object called manifold.
So in order to understand the dimensional reduction of $D=11$ SUGRA to  $D=5$ SUGRA over Calabi-Yau manifold
as we will show in the next chapter, we should to understand pretty well what are CY manifolds as we will illustrate in this chapter. CY 3-fold is a complex special K$\ddot{\mbox{a}}$hler manifold. Before talking about this class of manifolds, we will give first a brief overview on the different types of manifolds, starting with elementary definitions and properties \cite{Ooguri2015}.  
\section{Riemannian geometry}
The $n$- dimensional manifold $\M$ is a topological space that each point $x \in \M$ has an open set $U$ $\subseteq$ $\M$ such that $U$ has a mapping $\phi$ to $\mathbb{R}^n$. The pair $(U, \phi)$ is called coordinate chart, while $\{(U_i,\phi_i)\}$ is called atlas, where
$n \in \mathbb {Z}$. A smooth manifold is called  \textbf{A Riemannian manifold} if it admits a symmetric positive-definite metric 
$ g= g_{\mu\nu} (x) dx^\mu dx^\nu$. A manifold is called \textbf{Lorentzian} if its metric
is the Minkowski metric $\eta= \mbox{diag}(-1,1,...,1)$, i.e. behaves locally like
$\mathbb{R}^{1,n-1}$ . A Levi-Civita connection may be chosen, leading to the usual
expressions for the Christoffel symbols, the Riemann and Ricci tensors and the Ricci scalar:
\bea
\Gamma^\lambda_{\mu\nu} &=& \frac{1}{2} g^{\lambda\kappa} [(\partial_\mu g_{\nu\kappa}) + (\partial_\nu g_{\mu\kappa})-
(\partial_\kappa g_{\mu\nu}) ],\\ 
R_{\mu\rho\nu}^{~~~\sigma} &=& (\partial_\rho \Gamma^\sigma_{\mu\nu})  - (\partial_\mu \Gamma^\sigma_{\rho\nu})
+  \Gamma^\alpha_{\mu\nu} \Gamma^\sigma_{\alpha\rho} - \Gamma^\alpha_{\rho\nu} \Gamma^\sigma_{\alpha\mu},\\
R_{\mu\nu} &=& R_{\mu\rho\nu}^{~~~\rho} = (\partial_\rho \Gamma^\rho_{\mu\nu})  - (\partial_\mu \Gamma^\rho_{\rho\nu})
+  \Gamma^\alpha_{\mu\nu} \Gamma^\rho_{\alpha\rho} - \Gamma^\alpha_{\rho\nu} \Gamma^\rho_{\alpha\mu},\\
R &=& g^{\mu\nu} R_{\mu\nu}  = R^\mu_\mu.
\eea
\section{Complex and K$\ddot{\mbox{a}}$hler manifolds}
Similarly, a complex $n$-dimensional manifold defined as a set of points that behaves locally like $\mathbb{C}^p$, 
where $\mathbb{C}^p$ is the space of differential forms of degree $p$, and $n \in \mathbb{Z}$, i.e., they are parameterized by complex coordinates. So that the dimension of $n$ should be even.
Like the Riemannian manifold are manifested by a metric 
$g_{\mu\nu} (x)$, complex manifolds are endowed by the so- called complex structure, a mixed matrix
$J^{~\nu}_\mu (x)$ defined on the manifold, where $\mu,\nu= 1,..,2m$, such that
\be
J^{~\nu}_\mu ~ J^{~\rho}_\nu = - \delta^\rho_\mu, ~~~~ J^2=-\mathbb{1},  
\ee                                                                                                                                      
this's called integrability condition for almost complex coordinates.    
To say the manifold has a complex structure the so- called Nijenhuis tensor
\be
N^\rho_{\mu\nu} = J^\alpha_\mu~
[ (\partial_\alpha J^\rho_\nu ) - 
(\partial_\nu J^\rho_\alpha)] -
J^\alpha_\nu [ (\partial_\alpha J^\rho_\mu ) - (\partial_\mu J^\rho_\alpha)] 
\ee
should vanish everywhere. Note that this's not trivial for any manifold, for instance $S^6$ (6- dimensional sphere) has almost complex structure, while $S^4$ (4- dimensional sphere) has no any complex structure.  
     
Now if the manifold has both Riemannian and complex structure, $g_{\mu\nu}$ and $J^{~\nu}_\mu$, respectively, the manifold is called 
K$\ddot{\mbox{a}}$hler manifold. The compatibility conditions between Riemannian and complex structures are given by 
\begin{enumerate}
  \item $\nabla_\mu ~ J^\nu_{~\rho} =0$,
  \item $g_{\mu\nu} J^\mu_{~\rho} J^\nu_{\sigma} = g_{\rho\sigma}$,
\end{enumerate}
the second condition leads to the metric $ ds^2= g_{\mu\nu} dx^\mu dx^\nu$ can be written as 
\be 
ds^2= 2 g_{i\bar{j}} d\omega^i d\omega^{\bar{j}}, 
\ee
the holomorphic coordinates $i,\bar{j}= 1,.., m$, where $2m$ is the dimension of the manifold. Introduce a 2- form ( anti-symmetric tensor)
\begin{equation} 
\begin{split}
\kappa & = \frac{1}{2} g_{\mu\nu} J^\mu_{~\rho} ~ dx^\rho \wedge dx^\nu \\
 & = i g_{i\bar{j}} d\omega^i \wedge d\omega^{\bar{j}}, 
\end{split}
\end{equation}
$\kappa$ is called the K$\ddot{\mbox{a}}$hler form. The first condition leads to :
\be 
d \kappa =0,
\label{eq1}
\ee
means the metric $g_{i\bar{j}}$ satisfying ; 
\begin{equation} 
\begin{split}
\partial_i g_{j\bar{k}} &= \partial_j g_{i\bar{k}},\\
\partial_{\bar{j}} g_{i\bar{k}} &= \partial_{\bar{k}} g_{i\bar{j}},
\label{eq2}
\end{split}
\end{equation}
plus the complex conjugate equations. So “ locally “ on each coordinate patch on the manifold. 
\be 
g_{i\bar{k}} = \partial_i f_{\bar{k}} ,
\ee
$ f_{\bar{k}}$ can also written as a derivative of another function 
\be
g_{i\bar{k}} = \partial_i \partial_{\bar{j}} K,
\label{pot22}
\ee
$K$ is called the K$\ddot{\mbox{a}}$hler potential , then
\be
\kappa = i  \partial_i \partial_{\bar{j}} K ~ d\omega^i \wedge d\omega^{\bar{j}}.
\ee
Note that $K$ can not globally defined on the manifold, otherwise it can be defined locally for every coordinate patch, and if the coordinates had transformed, $K$ transforms in away makes all its relations work. Consequently, if $d\kappa =0$, and $ \kappa = d\Lambda$ , $\Lambda$ can only locally defined.

This’s obvious when we know that the volume of the 
K$\ddot{\mbox{a}}$hler manifold is proportional to  $\kappa^m$ , so if $\kappa$ is exact and  assuming $M$ is a non-compact manifold, with no boundaries, then by Stoke’s theorem
\be
\int_M d \Lambda = \int_{\partial M} \Lambda =0, 
\ee
which is a disaster because the volume of $M$ should be positive and definite.
Also the condition (\ref{eq1} ) or equivalently (\ref{eq2}) simplifies the properties of the manifold considerably, for example one finds that
\be 
\begin{aligned}
\Gamma^k_{ij} &= g^{k\bar{l}} (\partial_i g_{j\bar{l}}), \\
\Gamma^{\bar{k}}_{\bar{i}\bar{j}} &= g^{l\bar{k}} (\partial_{\bar{i}} g_{\bar{j}l}),
\end{aligned}
\ee
are the only non-vanishing Christoffel symbols, indicating that parallel transport does not mix the holomorphic and the antiholomorphic components of a vector. Also the non-vanishing components of the Ricci tensor are found to be
\be 
R_{i\bar{j}} = \partial_i \partial_{\bar{j}} ~\mbox{log (det}~ g)
\ee
\section{De Rham cohomology groups}
Consider a $p$- form $\omega$ if it closed form, i.e., $d \omega=0$, can we say $\omega$ is exact, i.e., $\omega=d\lambda$, where $\lambda$ is a $(p-1)$ form ? clearly if 
$\omega$ is an exact form it is closed, but what about vise versa ? we will try to answer this question \cite{Nakahara2003}.

$\textbf{Poincar\'e's ~ Lemma}$:
On $n$- dimensional Euclidean space $\mathbb{R}^n$ a closed form is exact. This theorem can be generalized if we have a manifold $M$ that can be continuously deformed to a point. So if you have a closed form on any coordinate batch $U_i$ on $M$, this form is always exact, because $U_i$ can be contracted to a point in $\mathbb{R}^n$.

So now what kind of closed forms that are not exact ? 

Define the space of closed forms:
\be 
Z^p \mathcal{M} = \{ \omega \in C^p \mathcal{M} ; d\omega =0 \},
\ee   
$Z^p$ is infinite dimension space because you can choose any $(p-1)$ form then take its exterior derivative. Also
\be
B^p \mathcal{M} = \{ d\lambda : \lambda \in C^{p-1} \mathcal{M} \}.
\ee
The cohomolgy is defined by
\be
H^p \mathcal{M} = Z^p \mathcal{M}/ B^p \mathcal{M},
\ee    
the space of closed forms modulo by exact forms. In Mankowski flat space, cohomology is trivial by Poincar\'e's Lemma. The dimension of this space
\be
b_k = \mbox{dim}~ H^k \mathcal{M},
\ee   
is called Betti number and it is a topological invariant, which means if you continuously deformed the manifold, this quantity keeps invariant. Also from the topological invariants, Euler characteristic which defined by
\be
\chi = b_0-b_1 +b_2-b_3+ ...
\ee 
\section{Homology groups}
They are Abelian groups which partially count the number of holes in a topological space. In particular, homology groups form a measure of the hole structure of a space, but they are one particular measure and they do not always pick up everything.
Before going further, we need first introduce the following notation.
\subsection{Triangulation}
It means to approximate any topological space by a generalizaton of triangles. 
When we talk about tringle, it will include its interior, its 3 vertices, 3 edges and one face. While a 3 dimentional version of a tringle is called tetrahedron. Generalize this to define $p$- simplex $\subset R^p$. They consists of ($p+1$) vertices, so the triangle of is 2- simplex, while the tetrahedron is 3- simplex.

$p$- simplex itself is a topological space becasue it is a subspace of the Euclidean space which's a topological space. Also they can be combined to form more complicated spaces. So we can have a continuous map from a topological space to another and we can say when two topological spaces are equivalent to each other. Consider two topological spaces $M_1$ and $M_2$, where 
\be
f ~~:~~ M_1 \to M_2, ~~~~~~~~~~~~~~~~ g ~~:~~ M_2 \to M_1
\ee        
If these mapping functions are inverse to each other 
\be
f \circ g = id, ~~~~~~~ g \circ f = id 
\ee 
we say $M_1$ and $M_2$ are equivalent ($\sim$) to each other, consequently if $ k \sim M $, where $k$ is the simplicial complex, we say $k$ is a triangulation for $M$. In summary, given an arbitrary topological space $M$, to be traingulated , we cover it up by $k$ which is $\sim$ for $M$.

Of course traingulation is not unique, because $k$ can be more complicated, but this won't change the topology of space. One also can consider different $\sim$ relations between different triangulation. 

Theorem: If you have smooth differential manifolds, you always have triangulation. 

\subsection{Homology}
It is a dual space of cohomology space. Consider $k$ which contains $\sigma$ a $p$-simplex with $(p =0,..,n)$ and $n = dim M$. $\sigma$ has a rotation direction around its vertices , so $\sigma$ denoted by its vertices as 
\be 
\sigma = \langle v_0,v_1,...,v_p \rangle
\ee
Define $C_p$ a vector space generated by these simplexes. We can add or substract simplexes (as elements of vectors) and reverse their orintation
\be 
\langle v_0,v_1,...,v_p \rangle = - \langle v_1,v_0,...,v_p \rangle,  
\ee
Similar for the extrior operator in the differential forms, define 
here $\partial$ a boundry operator
\be
\partial \langle v_0 ....v_p \rangle = \sum^p_{i=0} (-1)^i~ 
\langle v_0,...,\hat{v_i}, ...,v_p \rangle, 
\ee     
removing the $ith$ component and yields $(p-1)$ simplex and so that 
here $\partial$ is called a boundry operator. Consider more complicated example of two triangles
\be
\partial : C_p \to C_{p-1} 
\ee 
and $\partial^2=0$. Define
\be
Z_p(k)= \{ c \in C_p(k) : \partial c =0 \}
\ee
i.e., elements of $C_p(k)$ which vanish by $\partial$. These means there are kind of simplical complex that have no boundries, because when they acted on by the boundry operator they vanish. So that the elements of $Z_p (k)$ are called cycles.   
Now define
\be
B_p(k)= \{ \partial c : c \in C_{p+1}(k) \}
\ee
The boundry itself has no boundry. So $ B_p(k) \subset Z_p (k)$ . The homology defined as 
\be
H_p(k) = Z_p(k) / B_p(k),
\ee
the cycles modulo by boundries. So if you want to compute the homology of any topoligcal space, what you do is to consider any trigulation of that space, then apply the above definition. Homology is a topological invariant (dose not depend on the way of triangulation). 
Consider homology with integer or with real coefficients $H_p(M,Z)$ and $H_p(M,R)$ respectivly. The first detects more topolgy than the second. Consider for example SO(3) rotational group, this's a manifold because you can introduce coordintes on it ( so it has a topological structure ). Consider $S^3$ (3- dimentional sphere) with  anti- polar points identified, where once you identify a point on the upper hemisphere, you identify a point on the lower hemisphere, so the whole sphere can be drawn just as a hemisphere, take 
\be 
H_1 (SO(3), Z) = Z_2 = \{ 0,1 \}
\ee    
$Z_2$ is a space consists of 2 numbers and it is binary (1+1=0), which means you have one cycle when repeated you return to the same point. While $H_1 (SO(3), R)=0$ because $H_1 (SO(3), Z)$ is very discrete. $Z_2$ is called torsion and it is an element of homology which corresponds to some finite group. While $H_0 (SO(3), Z)= Z$ or $H_1 (SO(3), R)= R$ are trivial. 
\subsection{Fundamental theorem of de Rham}        
There are dual features between homology and cohomolgy. Consider
\be
H^p \mathcal{M} \in \omega,
\ee
where $\omega$ is a closed $p$-form, but not exact and
\be
H_p \mathcal{M} \in c,
\ee 
where $c$ cycles in $p$- dimensions. Now integrate $p$-form over $p$- dimensional cycle
\be 
(c,w) = \int_c w.
\ee
Due to Stoke's theorem this pair is independent of the change of $\omega$ by exact form or changing $c$ by a boundry
\be 
\int_\gamma d \lambda = \int_{\partial \gamma } \lambda ,
\ee
where $\partial \gamma$ is the boundry of some orintable manifold $\gamma$. So
\be
\int_c  (\omega+d\lambda) = \int_c \omega + \int_{\partial c} \lambda = \int_c \omega,  
\ee
\be
\int_{c+\partial \gamma} \omega =  \int_c \omega + \int_{\partial \gamma} \omega = \int_c \omega,  
\ee
because the boundry of $c$ is absent. Also $(c,\omega)$ dose not affected by trivial shifts
\be 
(c+ \partial \gamma, \omega) = (c,\omega) ,
\ee
\be 
(c, \omega + d\lambda) = (c,\omega),
\ee
this manifest the pairing between $H_p\mathcal{M}$ (cycles or homology) and $H^p\mathcal{M}$ (closed forms or cohomology). dim $H_p = b_b$ 
(Batti number) is a topoligcal invariant and $H_p$ has basis of $\{c_1,...,c_{b_p}\}$. The two theorems of de Rham state that:

Theorem 1:

For any integers $v_i$, there are a closed $p$-form $\omega$, where is  
\be
\int_{c_i} \omega = v_i,
\ee 
$c_i$ is the period of integral that gives you this integer. 

Theorem 2 : 

If $\int_{c_i} \omega =0$ for all $c_i$, then $\omega=0$. This means $c_i$ basis are not degenerate. Note that
dim $H^p =$ dim $H_p = b_p $, where the basis of $H^p$ are $\{ \omega^1, \omega^2,...., \omega^{b_p}\}$, so that we can define a square matrix
\be
(c^i, \omega^j) = \int_{c^i} \omega^j = \Delta_{ij},
\ee
called a period matrix, and the de Rham theorems say: det $\Delta \neq 0$.

\section{Hodge- K$\ddot{\mbox{a}}$hler cohomology}
Rembeber a $k$-form on a manifold is defined by:
\be 
C^k\mathcal{M} \ni \omega = \frac{1}{k!} \omega_{\mu_1,.......\mu_k} ~~~~ dx^{\mu_1} \wedge ...... dx^{\mu_k},
\ee
in de Rham cohomology we take the exterior derivative of this form and consider the form closed, then the space of the closed forms modulo by exact forms. 

Now if we have complex coordinates $\omega^i$ and
$d\omega^{\bar{j}}$ that do not mix under holomorphic coodinate transfomation. So that the space of $k$-forms $C^k\mathcal{M}$ decomposing to 
$C^{p.q}\mathcal{M}$ and $(p-q)$ form is defined instead by;
\be 
\omega= \frac{1}{p!~q!}~ \omega_{i_1,....., i_p, \bar{j}_1, ....., \bar{j}_q} ~~~ d\omega^{i_1} \wedge ..... d\omega^{i_p} ~~ \wedge  d\omega^{\bar{j}_1} \wedge ...... \wedge d\omega^{\bar{j}_q},
\ee
where $p$ is the number of $d\omega$’s and $q$ is the number of $d\bar{\omega}$’s. $H^{p.q}\mathcal{M}$ is called Hodge- de Rhame comology group, where the  total degrees of freedom is given by $k=p+q$. Let’s consider now an arbitrary 2- form 
\be
C^{1.1} \mathcal{M} : \sigma_{i\bar{j}}~~ d\omega^i \wedge d\omega^{\bar{j}},
\ee
if $d\sigma=0$ it will be a member of the de Rham class $H^{1.1}$. The K$\ddot{\mbox{a}}$hler form is a non trivial member of $H^{1.1}$, so that (locally) the dimension of $H^{1.1}$ is 1. The question now does $H^{1.1}$ contain any other known members? what about the Ricci tensor? it’s a 2-form ;
\be 
\mathcal{R}_{i\bar{j}}= \partial_i \partial_{\bar{j}}~ (\mbox{log}~ g),
\ee
\be
\mathcal{R} = \mathcal{R}_{i\bar{j}} d\omega^i \wedge d\omega^{\bar{j}},
\ee
but it’s not clear if $d \mathcal{R}=0$ or not. May be a nice remark here is that Einstein field equations in the vaccum $\mathcal{R}_{\mu\nu}$ is trivial with $\mathcal{R}=d\Lambda=0$. 

So in general is $R$ a trivial element of $H^{1.1}$? it has been found this question is so important and carries many interesting topology data about the manifold. $R$ is called the $1^{\mbox{st}}$ Chern- Class, and it was deduced that like when $\kappa$ is a member of $H^{1.1}$, the manifold contains a k$\ddot{\mbox{a}}$hler structure, similarly if $R$, can a metric $g_{i\bar{j}}$ found on the manifold with the same complex structure and with the same k$\ddot{\mbox{a}}$hler structure? that what was proved by Yau. K$\ddot{\mbox{a}}$hler manifolds satisfy the Ricci flatness condition are called Calabi- Yau manifolds. 
\section{Calabi- Yau manifold}
\begin{figure}[!t]
\centering
\includegraphics[scale=0.4]{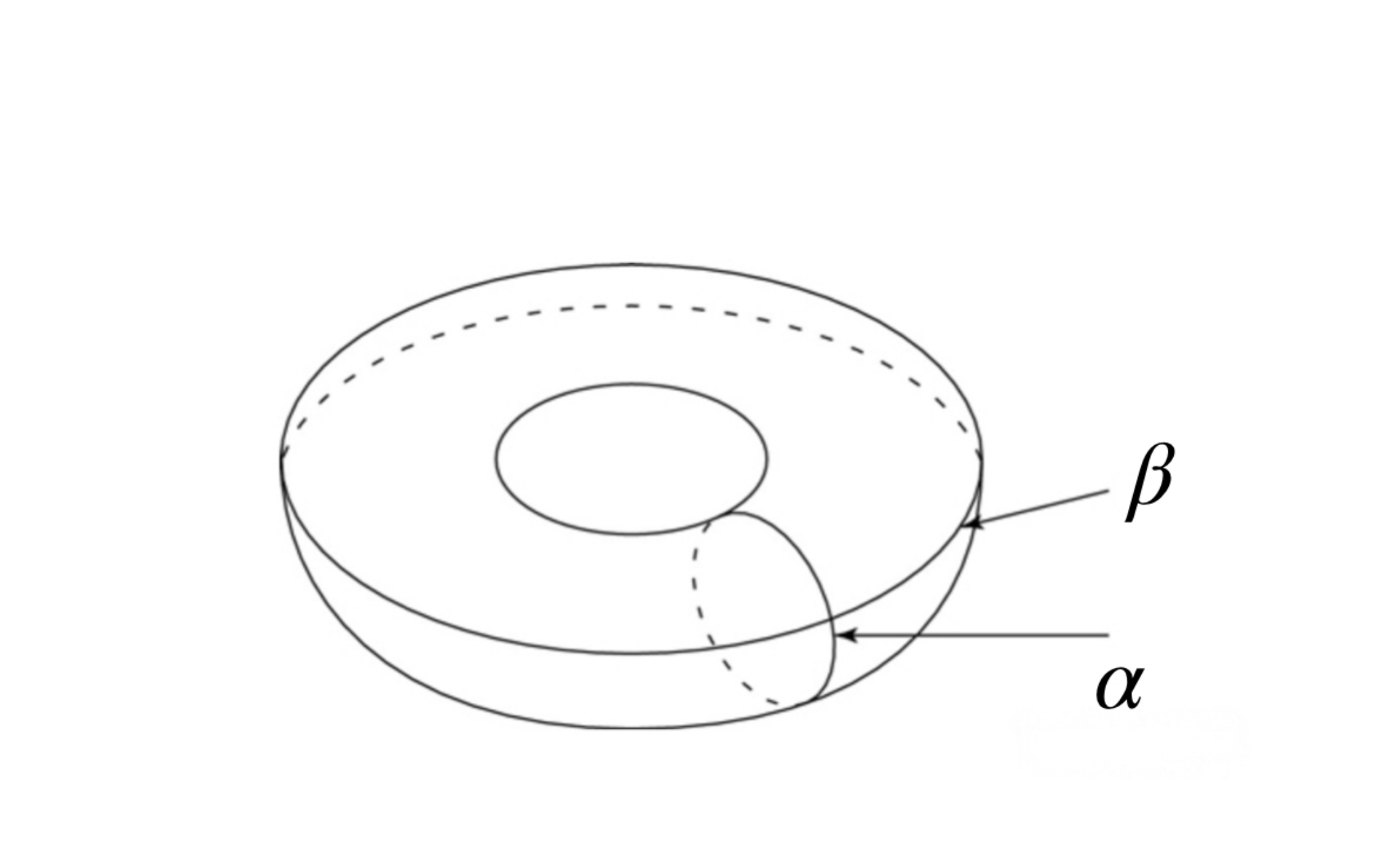}
\caption{The torus as Calabi- Yau manifold in one complex dimension.}
\label{to}
\end{figure}

In 1954 Calabi proposed the following conjecture: If $\M$ is a complex manifold with a K$\ddot{a}$hler metric and vanishing first Chern class, then there exists a unique Ricci flat metric for each K$\ddot{a}$hler class on $\M$. In 1976, Calabi's conjecture was proven by Yau \cite{CYM2-2}. Consider:
\be 
\partial_i \partial_{\bar{j}}~ log (\mbox{det}~ g)=0,
\ee
then:
\be 
log (\mbox{det}~ g) = f(w) + \bar{f} (\bar{w}),
\ee
so that $det~ g$ transforms as a determinant of ($i,\bar{j}$) matrix, where $i$ and $\bar{j}$ are the holomorphic and anti-holomorphic coordinates, respectively.
\be
\mbox{det}~ g = \Omega(w) \bar{\Omega} (\bar{w}),
\label{volume}
\ee
where $\Omega(w)$ is $(i,0)$ form and $\bar{\Omega}(\bar{w})$ is $(0,\bar{j})$ form. So that $\mathcal{R}_{i\bar{j}}=0 ~ \implies \exists ~ \Omega(i,0)$ form:
\begin{itemize}
\item $\bar{\partial} \Omega =0$, ~~~~~ (holomorphic)
\item $\Omega \neq 0.$ ~~~~~~~ any where on $\M$,
\end{itemize}
that what was proven by Yau. Examples of Calabi-Yau manifold in one complex dimension:
\begin{itemize}
  \item A complex plane $\mathbb{C}$, with a flat metric
  $dw~ d\bar{w}$, $w \in \mathbb{C}$.
  \item $ \mathbb{C}^w =  \mathbb{C} \backslash \{0\}$, which is a cylinder.
  \item $T^2 = \mathbb{C} / \mathbb{Z} + \tau~ \mathbb{Z}$,
\end{itemize}
where $\tau$ is an integer $\in \mathbb{Z}$. $w=w^1+\tau~ w^2$, where $\tau$ is called the complex structure moduli, it parameterizes the shape of the torus. Since $T^2$ is a CY manifold, it should has a holomorphic form $\Omega$, $(i,0)$, $i=1$, because the complex dimension is 1, $\Omega= d~w$.
We can extract the complex strucure moduli by integrating $\Omega$ over a homology cycle of the CY manifold, 
Fig. (\ref{to})

\begin{figure}[H]
\centering
\includegraphics[scale=0.3]{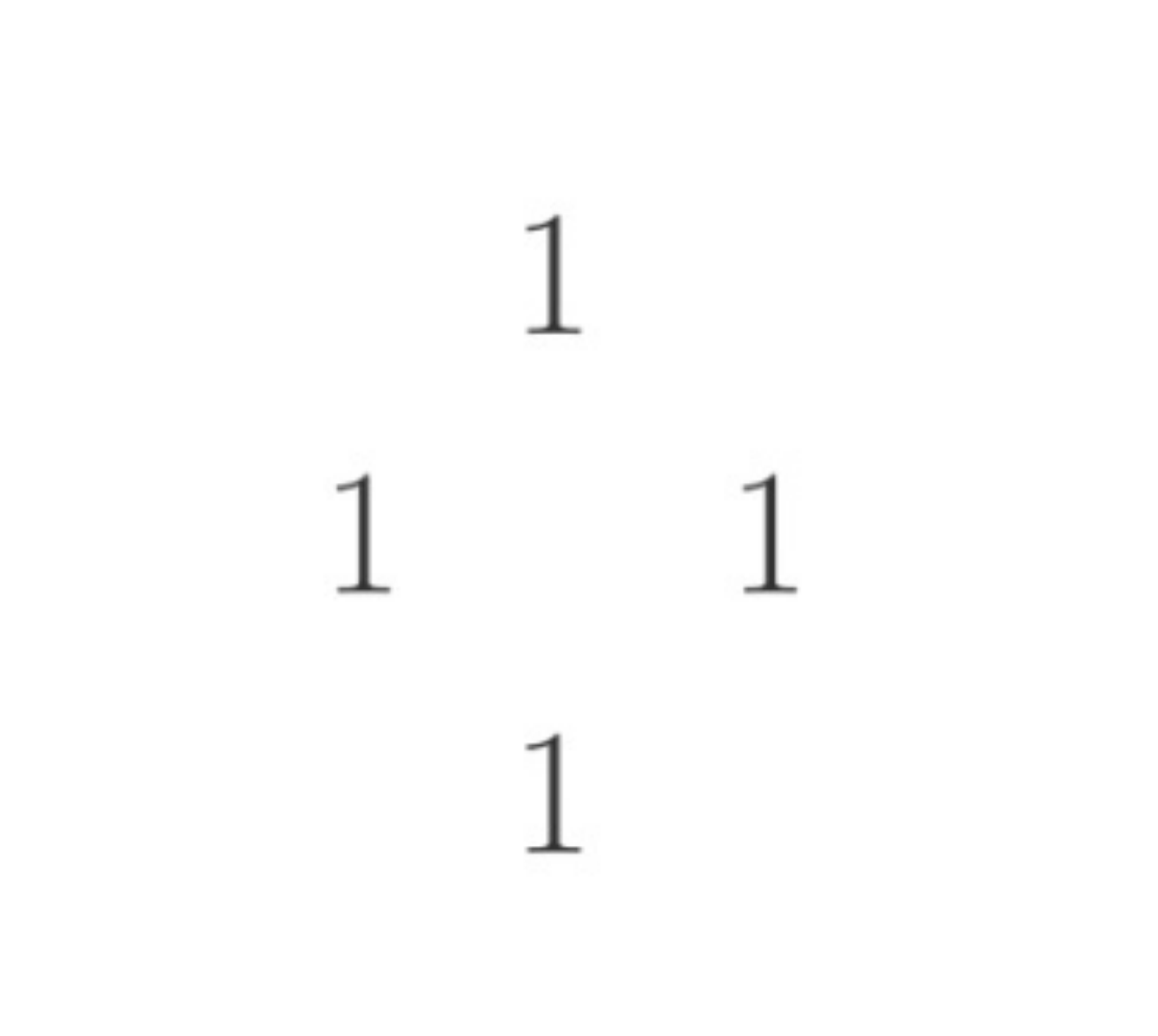}
\caption{The hodge diamond of $T^2$.}
\label{hod1}
\end{figure}

\begin{figure}[t]
\centering
\includegraphics[scale=0.3]{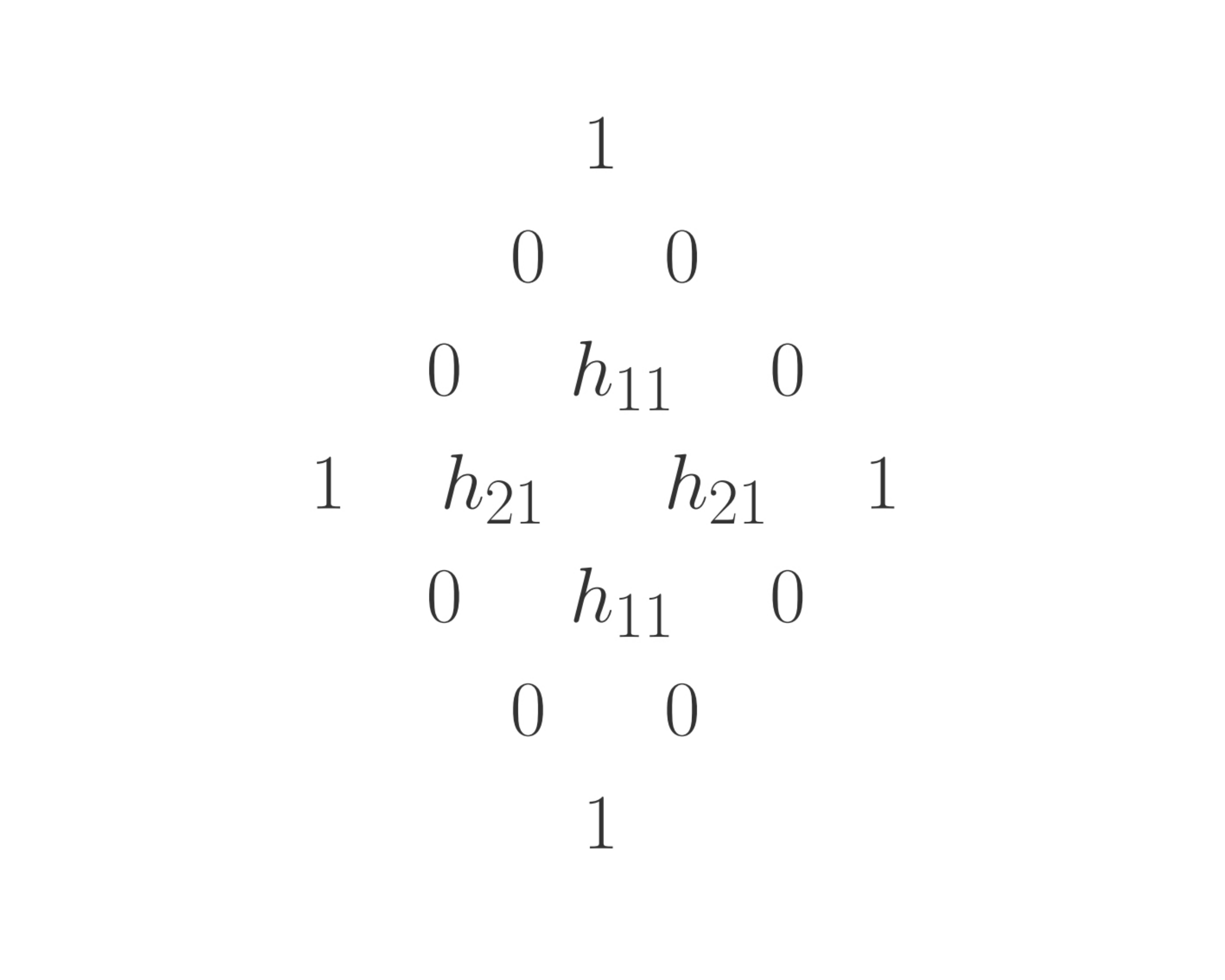}
\caption{The hodge diamond of the 3-fold CY.}
\label{hod3}
\end{figure}

\be
\int_\alpha \Omega =1 , ~~~~~~~~~ \int_\beta \Omega = \tau,
\ee
in this case the Hodge- de Rham cohomology will be given by: 
\be 
H^{p,q} = \{ w \in \Omega^{p,q}, d w =0\}/ d\lambda.
\ee
The dimension of $H^{p,q}$ is called the hodge number $h^{p,q}$. The hodge diamond of $T^2$ is given by Fig. (\ref{hod1}),
where:
\begin{equation} 
\begin{split}
h^{1,0} &=1, ~~~~~~~~~~ \mbox{only}~ \Omega= dx ~ (1,0)~ \mbox{form}, \\
h^{1,1} &=1, ~~~~~~~~~~ dw~ d\bar{w} ~ \mbox{the volume.}
\end{split}
\end{equation}
While the hodge diamond of the 3- complex dimensions CY manifold is given by Fig. (\ref{hod3}),
where $h^{11} = h^{22}$ and $h^{12}= h^{21}$. The Eular characterstic is given by $\chi= 2 (h^{11}- h^{12})$. $h^{2,1}$ determines the number of complex structure moduli on the complex strucutre space $M_C$ and $h^{1,1}$ determines the number of K$\ddot{\mbox{a}}$hler forms on  $M_K$, where $M_K$ being a complexification of the parameters of the K$\ddot{\mbox{a}}$hler class. In addition, there exists a symmetry in the structures of $M_C$ and $M_K$ which lends support to the so-called mirror symmetry hypothesis of CY 3-folds \cite{CYM2-1}. In terms of homology groups, Calabi-Yau 3-folds admit a non-trivial $H^3$ that can be Hodge decomposed as follows:
\be 
H^3= H^{3,0} \oplus H^{2,1} \oplus H^{1,2} \oplus H^{0,3}.
\ee
So that CY 3-folds have a single $(3,0)$ cohomology form; 
$h_{3,0} =dim  (H^{3,0})= 1$, which we will call $\Omega$ (holomorphic: $\bar{\partial} \Omega=0$ ).  
\subsection{Deformation of complex structure}
$\mathbf {The~line~ sub- bundle}$:
in different complex structures, the decompositions of the tangent bundle into holomorphic and
anti-holomorphic parts are different \cite{Hori:2003ic}.  
Therefore, what was a closed and holomorphic $(n, 0)$ form in one complex structure will no longer be of type $(n, 0)$, (nor holomorphic) in another complex structure. However, the form will still
be closed, as the exterior derivative is independent of complex structure.
In this description it is easy to see that, to linear order, a (3, 0) form
can only change into a linear combination of (3, 0) and (2, 1) forms. The
change can be measured by $H^{2,1}$.
The cohomology class in $H^3$ representing the holomorphic
$(n, 0)$ form must change over the moduli space of complex structures $M_C$ .
In fact, $H^3$ forms a bundle over the moduli space, and the Calabi-Yau form
is a complex line in this bundle. We can define a natural metric on this line bundle
\be
h = ||\Omega||^2 = (\Omega,\Omega) = i \int \Omega \wedge \bar{\Omega},
\ee
with
\be
K= - log ||\Omega||^2 = - log \int \Omega \wedge \bar{\Omega},
\ee
transforms as a K$\ddot{\mbox{a}}$hler potential,
$K \to K - f - f$. We therefore can define a metric on moduli space by
\be
G_{m\bar{n}}= \partial_m \bar{\partial}_{\bar{n}} K, 
\ee
and this is well defined no matter the gauge choice of $\Omega$, since $f + \bar{f}$ is killed
by $\partial \bar{\partial}$.

We already decided that the tangent space to
moduli space is $H^{2,1}$ . To see the above metric in more detail,
let us write the variation of $\Omega$ with
respect to a coordinate direction $z^a$ as
\be
\begin{aligned}
\partial_a \Omega &= (3,0) \mbox{piece} + (2,1) \mbox{piece}\\
&= k_a \Omega + \chi_a,
\label{q1}
\end{aligned}
\ee
where there are no other terms since the variation of a holomorphic (1,0)
form has a (1, 0) and a (0, 1) piece. Then, using Eq. (\ref{q1}), one finds
\be
\begin{aligned}
\partial_a \bar{\partial}_{\bar{b}} K &= \partial_a \big[ \frac{-1}{\int \Omega \wedge \bar{\Omega}} \int \Omega \wedge \bar{\partial}_{\bar{b}} \bar{\Omega} \big] \\
&=  - \frac{1}{(\int \Omega \wedge \bar{\Omega})^2} \int \partial_a \Omega \wedge \Omega \int \Omega \wedge \bar{\partial}_{\bar{b}} \bar{\Omega} + \frac{1}{\int \Omega \wedge \bar{\Omega}} \int \partial_a \Omega \wedge \bar{\partial}_{\bar{b}} \bar{\Omega}\\
 &= \frac{\int \chi_a \wedge \chi_{\bar{b}}}{(\int \Omega \wedge \bar{\Omega})}.
\end{aligned}
\ee
While the variation of the Ricci-flat
metric corresponding to the ath direction 
\be
\delta_a g_{\bar{i}\bar{j}} = \frac{\partial g_{\bar{i}\bar{j}}}{\partial z^a} = - \frac{1}{||\Omega||^2} \bar{\Omega}^{~kl}_{\bar{i}} (\chi_{a~ k l \bar{j}}), 
\ee
Equivalent to
\be
\delta g_{\bar{i}\bar{j}} = \frac{\partial g_{\bar{i}\bar{j}}}{\partial z^a} = - \frac{1}{||\Omega||^2} \bar{\Omega}^{~kl}_{\bar{i}} \chi_{a|~ k l \bar{j}}~~ z^a, 
\ee
$\chi_{a|~ kl \bar{i}} = - \frac{1}{2} \Omega_{kl}^{~~\bar{j}} ~\frac{\partial g_{\bar{i}\bar{j}}}{\partial z^a}.$ Also $z^a$ define the parameters, or moduli, of the complex structure ($z^a: a=1,...,h_{2,1}$) . 
And each $\chi$ defines a $(2,1)$ cohomology class. The important observation here is that when the moduli treated as complex coordinates, it can define a special K$\ddot{\mbox{a}}$hler metric $G_{a\bar{b}} $ on $\mathcal{M}_C$ as follows \cite{Candelas:355}:
\be
V_{CY}~ G_{a\bar{b}} (\delta z^a ) (\delta z^{\bar{b}} ) = \frac{1}{4} \int_{\mathcal{M}} g^{i\bar{j}} g^{k\bar{l}} (\delta g_{ik})~ (\delta g_{\bar{j}\bar{l}} ) , \label{41-ch2}
\ee
where $V_{CY}$ is the volume of the Calabi-Yau.  
From $(\ref{volume})$, the volume of the CY 3-fold is given by
\be
Vol(\mathcal{M}) = i \int \Omega \wedge \bar{\Omega},
\ee
which simply relates the K$\ddot{\mbox{a}}$hler potential of $\mathcal{M}_C$ to the volume of $\mathcal{M}$ by
\be
Vol(\mathcal{M}) = e^{-K}.
\label{veve}
\ee
\subsection{Periods and coordinates on moduli space}
Let ($\alpha_a, \beta^b$) be the dual cohomology basis of the homology 3-cycles $H^3$ on $\mathcal{M}_C$, where $ a, b, c = 0, . . , h_{2,1}$. Such that 
\be 
\begin{aligned}
& \int_\mathcal{M} \alpha_a \wedge \beta^b = \int_{A^b} \alpha_a = \delta^b_a, \\
& \int_\mathcal{M} \beta^a \wedge \alpha_b = \int_{\beta_b} \beta^a = -\delta^a_b, \\
& \int_\mathcal{M} \alpha_a \wedge \alpha_b = \int_\mathcal{M} \beta^a \wedge \beta^b =0, \label{dual-coh}
\end{aligned}
\ee
and let  ($A^a , B_b$)  be canonical homology basis, such that 
\be 
\begin{aligned}
& A^a \cap B_b = \delta^a_b, \\ 
& B_a \cap A^b = - \delta^b_a, \\
& A^a \cap A^b = B_a \cap B_b=0. 
\end{aligned}
\ee

\be 
\mbox{dim}~ H^3 = h^{3,0} + h^{2,1} + h^{1,2} +h^{0,3} = 2 (1+ h^{2,1}) 
\ee
So that we can expand the Calabi–Yau form $\Omega$ as \cite{CYM2-1} 
\be 
\Omega = z^a \alpha_a - f_b \beta^b,
\ee
for some $z^a$ , $f_b$ , $a, b = 1, . . , h^3, \mathcal{M}/2 = h^{2,1} (\mathcal{M}) + 1$. 
The coordinates $z^a$ and $f_b$ will change as we move in Calabi–Yau moduli space,
since $\Omega$ will change. In fact, as we have mentioned, since the location of
$\Omega$ in $H^3$ $\mathcal{M}$ determines the complex structure, the $z^a$ and $f_b$ determine the
point in moduli space – even over determine it, as can be seen by counting
parameters (moduli space is $h^{2,1}$ ($\mathcal{M}$) dimensional).
It is immediate from the dual basis relations that $z^a$ and $f_b$ can be
expressed in terms of the “period integrals”
\be 
z^a = \int_{A^a} \Omega, ~~~f_b = \int_{B_b}      \Omega.\label{ZZ}
\ee
Therefore we can express the complex structure in terms of
periods $\int_C \Omega$ of the Calabi–Yau form.
In fact, it can be shown that the $z^a$ alone locally determine the complex
structure . We can therefore imagine solving for
the $f_b$ in terms of the $z^a$ . Then the $z^a$ are only redundant by one extra
variable, but there is also an overall scale of $\Omega$ that is arbitrary, and it is
often convenient to keep the $z^a$ as homogeneous coordinates on $\mathcal{M}$ . 
Some of the ingredients of this structure require the knowledge of the Hodge duality relations
(with respect to $\mathcal{M}$) of the forms ($\alpha$, $\beta$) \cite{CYM2-2}:
\be
\begin{aligned}
\star \alpha_a &= (\gamma_{ab}+ \gamma^{cd} \theta_{ac} \theta_{bd}) \beta^b - \gamma^{cb} \theta_{ca} \alpha_b \\
\star \beta^a &= \gamma^{ac} \theta_{cb} \beta^b - \gamma^{ab} \alpha_b, \label{hodge-dual}
\end{aligned}
\ee
where $\theta_{ab}$ and $\gamma_{ab}$ are real matrices defined by
\be
\begin{aligned}
\mathcal{N}_{ab} &= \bar{f}_{ab} + 2i \frac{N_{ac} z^c _{bd} z^d}{z^p N_{pq} z^q}\\
&= \theta_{ab}-i \gamma_{ab}
\end{aligned}
\ee
where $f_{ab}=\partial_a f_b$  (the derivative is with respect to $z^a$ ), $\mathcal{N}_{ab}=Im(f_{ab})$, $\gamma^{ab} \gamma_{bc}=\delta^a_c$ and 
$\mathcal{N}_{ab}$ is known as the {\textbf periods matrix}.

\setcounter{equation}{0}

\chapter{$\D = 5~~ \mathcal{N}= 2$ supergravity with hypermultiplets}
%
%
\section{Dimensional reduction}

The unique supersymmetric gravity theory in eleven dimensions has the following bosonic action:
\be
S_{11}= \frac{1}{2\kappa_{11}^2} \int~ d^{11} x~~~~ \Big[ \sqrt{-G}~ \Big( \mathcal{R} - 
\frac{1}{48} \mathcal{F}^2 \Big) + \frac{1}{6~ 3! (4!)^2}~ \epsilon_{\mu_1,..\mu_{11}} \mathcal{A}^{\mu_1,..\mu_3} \mathcal{F}^{\mu_4,..\mu_7} \mathcal{F}^{\mu_7,..\mu_{11}} \Big].
\ee
Write this in the differential forms language, \footnote{Appendix (\ref{Df}).}, 
the Einstein-Hilbert
term
\be 
 d^{11} x~ |e|~ \mathcal{R} \to  ~ \mathcal{R} \star 1,
\ee 
the Maxwell's terms
\be
\begin{aligned}
 d^{11} x~ \frac{|e|}{48} \mathcal{F}^2 \to  ~~ \frac{1}{2} \mathcal{F} \wedge \star \mathcal{F}, \\
\mbox{and the third term} \to - \frac{1}{6} \mathcal{A} \wedge \mathcal{F} \wedge \mathcal{F},
\end{aligned}
\ee
So that;
\be
S_{11}=  \int_{11} \Big( \mathcal{R} \star 1 - \frac{1}{2} \mathcal{F} \wedge \star \mathcal{F} - \frac{1}{6} \mathcal{A} \wedge \mathcal{F} \wedge \mathcal{F} \Big),
\ee
$\mathcal{R}$ is the $D = 11$ Ricci scalar, $\mathcal{A}$ is the 3-form gauge potential and 
$\mathcal{F} = d \mathcal{A}$. Make the dimensional reduction:
\be
\begin{aligned}
ds^2 &= g_{MN} ~ dx^M dx^N \\
 & = e^{\frac{2}{3}\sigma}~g_{\mu\nu} ~ dx^\mu dx^\nu + e^{\frac{\sigma}{3}}~ k_{ab} ~ dx^a dx^b, \\
& ~~~ M,N = 0,..., 10 ~~~ \mu,\nu = 0,..., 4 ~~~ a,b = 1,...,6, \label{metric}
\end{aligned}
\ee
where the metric $g_{\mu\nu}$ is the five dimensional metric, which will eventually contain our 
2-brane solutions. $k_{ab}$ is a metric on the six dimensional compact subspace $\mathcal{M}$, the dilaton $\sigma$ is a function in $x^\mu$ only and the warp factors are chosen to give the conventional coefficients in five dimensions, guaranteeing that the gravitational term in the action
will have the standard Einstein-Hilbert form. We choose a complex structure on $\mathcal{M}$ such that
\be 
k_{ab} dx^a dx^b = k_{mn} dx^m dx^n +  k_{\bar{m}\bar{n}} dx^{\bar{m}} dx^{\bar{n}} + 2 k_{m\bar{n}} dx^{m} dx^{\bar{n}},
\ee
where the holomorphic and antiholomorphic indices ($m,n; \bar{m}, \bar{n}$) are three dimensional on $\mathcal{M}$. The
Hermiticity condition demands that $k_{mn} = k_{\bar{m}\bar{n}}=0$, while the Ricci tensor for 
$\mathcal{M}$ is set to zero
as dictated by Yau’s theorem. Furthermore, we consider the case where only the complex structure
is deformed, which requires $\delta k_{m\bar{n}}=0$, and $(\delta k_{mn},\delta k_{\bar{m}\bar{n}} \neq 0 )$.
Now, the flux compactification of the gauge field is done by expanding $\mathcal{A}$ into two forms, one
is the five dimensional gauge field A while the other contains the components of $\mathcal{A}$
on $\mathcal{M}$ written
in terms of the cohomology forms $(\alpha_I, \beta^I)$ as follows:
\bea
\nonumber
\mathcal{A} &=& A + \sqrt{2} (\zeta^I \alpha_I + \tilde{\zeta}_I \beta^I), ~~~ ( I=0,..., h_{2,1}) \\ \nonumber
\mathcal{F} &=& d \mathcal{A} = \mathcal{F} + \mathcal{F}_\mu dx^\mu, \\ \nonumber
F_\mu dx^\mu &=& \sqrt{2} [ (d\zeta^I) \wedge \alpha_I + (d\tilde{\zeta}_I) \wedge \beta^I ] \\
&=&  \sqrt{2} [ (\partial_\mu \zeta^I) \alpha_I + (\partial_\mu \tilde{\zeta}_I)  \beta^I ] \wedge dx^\mu,
\label{af}
\eea
where $A\equiv \frac{1}{3!} A_{\mu\nu\rho} dx^\mu dx^\nu dx^\rho$. The coefficients $\zeta^I$ and $\tilde{\zeta}_I$ appear as axial scalar fields in the lower dimensional theory. We also note that $A$ in five dimensions is dual to a scalar
field which we will call $a$ (also known as the universal axion). The set ($a$ , $\sigma,~ \zeta^0,~ \tilde{\zeta}_0$ ) as the universal hypermultiplet \cite{CYM2-2,theME}. 
The rest of the hypermultiplets are ($ z^i,~ z^{\bar{i}},~ \zeta^i,~ \tilde{\zeta}_i:~ i=1,...,h_{2,1}$). Also included in the hypermultiplets are the fermionic partners of the hypermultiplet scalars known as the hyperini
(singular: hyperino). However, in what follows, we will only discuss the bosonic part of the action.
The hyperini, as well as the gravitini, will make their appearance in the SUSY variation equations
later. The parts one needs for the dimensional reduction are as follows:

1. The metric components
\be 
\begin{aligned}
G_{\mu\nu} &= e^{\frac{2}{3}\sigma} g_{\mu\nu}, ~~~ G^{\mu\nu} = e^{-\frac{2}{3}\sigma} g^{\mu\nu} \\
G_{ab} &= e^{-\frac{\sigma}{3}} k_{ab}, ~~~ G^{ab} = e^{\frac{\sigma}{3}} k^{ab} \\
G &= \mbox{det} G_{MN} = e^{\frac{4}{3}\sigma} gk, \\
g &= \mbox{det} g_{\mu\nu}, ~~~ k= \mbox{det} k_{ab}.
\end{aligned}
\ee

2. The Christoffel symbols
\be 
\begin{aligned}
\Gamma^\mu_{\nu\rho} &= \tilde{\Gamma}^\mu_{\nu\rho} [g] + \frac{1}{3} [ \delta^\mu_\nu (\delta_\rho \sigma) + 
\delta^\mu_\rho (\delta_\nu \sigma) - \delta_{\nu\rho}  \delta^{\mu \kappa} (\partial_\kappa \sigma)] \\
\Gamma^\mu_{ab} &=  \frac{1}{6} e^{-\sigma} k_{ab} (\partial^\mu \sigma)  - \frac{1}{2} e^{-\sigma} (\partial^\mu k_{ab}) \\
\Gamma^a_{\mu  b} &= \frac{1}{2} k^{ac}  (\partial_\mu k_{cb}) - \frac{1}{6} \delta^a_b  (\partial_\mu \sigma) \\
\Gamma^a_{bc} &= \hat{\Gamma}^a_{bc} [k],   
\end{aligned}
\ee
where the (\~{}) and the ($\verb!^!$) refer to the purely five and six dimensional components respectively. Now, calculating the eleven dimensional Ricci scalar based on this gives:
\be 
\sqrt{|G|} \mathcal{R} = \sqrt{|gk|} \Big [ \mathcal{R}[g] - \frac{1}{2} (\partial_\mu \sigma) (\partial^\mu \sigma) 
-  \frac{1}{4} k^{m\bar{n}}  k^{r\bar{p}} (\partial_\mu k_{mr})  (\partial_\mu k_{\bar{n}\bar{p}})\Big],
\ee
where we have used $\hat{\mathcal{R}}_{ab}=0$ since $\mathcal{M}$ is Ricci-flat, 
as well as dropped all total derivatives and terms
containing $k_{mn}$,  $k_{\bar{m}\bar{n}}$, and $\delta k_{m\bar{n}}$. Using Equ. (\ref{41-ch2}) 
\be
\int_{11}~~ \mathcal{R} \star 1 = \int d^5 x~ \sqrt{|g|}~ \Big[ R -\frac{1}{2} (\partial_\mu \sigma) (\partial^\mu \sigma)
- G_{i\bar{j}} (\partial_\mu z^i)  (\partial^\mu z^{\bar{j}}) \Big],
\ee
where we have normalized the volume of the compact space to $V_{CY}= \sqrt{dw}^6 |k| = 1$. Next, the Maxwell term is:
\be 
- \frac{1}{2} \mathcal{F} \wedge \star \mathcal{F} \to - \frac{1}{2} \frac{1}{4!} \mathcal{F}_{LMNP} \mathcal{F}^{LMNP} = - \frac{1}{48} \Big(e^{-2\sigma} F_{\mu\nu\rho\sigma} F^{\mu\nu\rho\sigma} 
+ e^{\frac{2}{3}\sigma} F_\mu F^\mu \Big).
\ee
Substituting, we get:
\be
\begin{aligned}
- \frac{1}{2} \int_{11} \mathcal{F} \wedge \star \mathcal{F} &= - \frac{1}{48}~ \int d^5 x~ \sqrt{|g|} e^{-2\sigma}~ F_{\mu\nu\rho\sigma} F^{\mu\nu\rho\sigma}  \\
&- \int d^5 x ~  \sqrt{|g|} e^{\sigma}~ \Big [ (\partial_\mu \zeta^I) (\partial^\mu \zeta^J) \int_\mathcal{M} \alpha_I \wedge \star \alpha_J \\
&+ (\partial_\mu \zeta^I) (\partial^\mu \tilde{\zeta}_J) \int_\mathcal{M} \alpha_I \wedge \star \beta^J + 
+(\partial_\mu \tilde{\zeta}_I) (\partial^\mu \zeta^J) \int_\mathcal{M} \beta^I \wedge \star \alpha_J \\
&+ (\partial_\mu \tilde{\zeta}_I) (\partial^\mu \tilde{\zeta}_J) \int_\mathcal{M} \beta^I \wedge \star \beta^J \Big ],  
\end{aligned}
\ee
where the Hodge star on the right hand side is with respect to $\mathcal{M}$. Now, using Equ. (\ref{dual-coh}), and Equ. (\ref{hodge-dual})
we end up with:
\be
\begin{aligned}
- \frac{1}{2} \int_{11} \mathcal{F} \wedge \star \mathcal{F} = &- \int d^5 x~ \sqrt{|g|} \Big\{ \frac{1}{48}~ e^{-2\sigma}~ F_{\mu\nu\rho\sigma} F^{\mu\nu\rho\sigma}  \\
& + e^\sigma \Big[(\gamma_{IJ} + \gamma^{KL} \theta_{IK} \theta_{JL})(\partial_\mu \zeta^I) (\partial^\mu \zeta^J) +
 \gamma^{IJ} (\partial_\mu \tilde{\zeta}_I) (\partial^\mu \tilde{\zeta}_J)  \\
&+ 2 \gamma^{IK} \theta_{JK} (\partial_\mu \zeta^J)  (\partial^\mu \tilde{\zeta}_J) \Big] \Big\}
\end{aligned}
\ee
Finally, the Chern-Simons term gives:
\bea
\nonumber
- \frac{1}{6} \int_{11} \mathcal{A} \wedge \mathcal{F} \wedge \mathcal{F} &=& - \int_5 \Big[ \zeta^I F \wedge d \tilde{\zeta}_J \int_{\mathcal{M}} \alpha_I \wedge \beta^J
- \tilde{\zeta}_I F \wedge d \zeta^J \int_{\mathcal{M}} \alpha_J \wedge \beta^I \Big] \\
&=& - \int_5 F \wedge (\zeta^I d \tilde{\zeta}_I - \tilde{\zeta}_I d\zeta^I).  
\eea
To sum up, the ungauged five dimensional $\mathcal{N} = 2$ supergravity bosonic action with vanishing
vector multiplets is \cite{CYM2-2} :
\be
\begin{aligned}
S_5 =& \int~ d^5 x |g|~ \Big\{ R -\frac{1}{2} (\partial_\mu \sigma) (\partial^\mu \sigma) - 
G_{i\bar{j}} \partial_\mu z^i \partial^\mu z^{\bar{j}} - \frac{1}{2} e^{-2\sigma} F^2 \\
-& e^\sigma \Big[(\gamma_{IJ} + \gamma^{KL} \theta_{IK} \theta_{JL})(\partial_\mu \zeta^I) (\partial^\mu \zeta^J) +
 \gamma^{IJ} (\partial_\mu \tilde{\zeta}_I) (\partial^\mu \tilde{\zeta}_J) 
+ 2 \gamma^{IK} \theta_{JK} (\partial_\mu \zeta^J)  (\partial^\mu \tilde{\zeta}_J) \Big] \Big\} \\
-& \int_5 F ~(\zeta^I \partial_\mu \tilde{\zeta}_I - \tilde{\zeta}_I \partial_\mu \zeta^I) \wedge dx^\mu,
\end{aligned}
\ee
Or:
\bea
\nonumber 
S_5 &=& \int_5 \Big[ \mathcal{R} \star 1 - \frac{1}{2} d\sigma \wedge \star d\sigma - G_{i\bar{j}} dz^i \wedge \star dz^{\bar{j}} -
F \wedge (\zeta^I d \tilde{\zeta}_I - \tilde{\zeta}_I d\zeta^I) 
\\ &-& \frac{1}{2} e^{-2\sigma} F \wedge \star F  -e^\sigma ~ X  \Big],
\label{act}
\eea
where
\be
X = \Big[ (\gamma_{IJ} + \gamma^{KL} \theta_{IK} \theta_{JL}) d \zeta^I \wedge \star d \zeta^J
+ \gamma^{IJ} d \tilde{\zeta}_I \wedge \star d\tilde{\zeta}_J +2 \gamma^{IK} \theta_{JK} d \zeta^J \wedge \star d\tilde{\zeta}_I \Big],
\label{xx}\ee
To complete the picture, we vary the action and present the field equations of $\sigma$, ($z^i, z^{\bar{i}}$), A and  
($\zeta^I, \tilde{\zeta}_I$) respectively:
\be 
(\Delta \sigma) \star 1 - e^\sigma X + e^{-2\sigma} F \wedge \star F = 0
\ee
\be
\begin{aligned}
(\Delta z^i) \star 1 + \Gamma^i_{jk} dz^j \wedge \star dz^k - \frac{1}{2} e^\sigma G^{i\bar{j}} (\partial_{\bar{j}} X) \star 1 &=0 \\
(\Delta z^{\bar{i}}) \star 1 + \Gamma^{\bar{i}}_{\bar{j}\bar{k}} dz^{\bar{j}} \wedge \star dz^{\bar{k}} - \frac{1}{2} e^\sigma G^{\bar{i}j} (\partial_{\bar{j}} X) \star 1 &=0 
\end{aligned}
\label{zEOM}
\ee
\be
d^\dagger \Big[ e^{-2\sigma} F + \star (\zeta^I d\tilde{\zeta}_I - \tilde{\zeta}_I d \zeta^I ) \Big] =0 
\ee \label{AA}
\be
\begin{aligned}
d^\dagger \Big[ e^{\sigma} \gamma^{IK} \theta_{JK} d\zeta^J + e^\sigma \gamma^{IJ} d\tilde{\zeta}_J + \zeta^I \star F \Big] &=0 \\
d^\dagger \Big[ e^{\sigma} ( \gamma_{IJ} + \gamma^{KL} \theta_{IK} \theta_{JL}) d\zeta^J + e^\sigma \gamma^{JK}
\theta_{IK} d\tilde{\zeta}_J - \tilde{\zeta}_I \star F \Big] &=0 \label{zeta}
\end{aligned}
\ee
From a five dimensional perspective, the moduli ($z^i$ , $z^{\bar{i}}$) behave as scalar fields. 
We recall, however, that the behavior of the other fields is dependent on the moduli, i.e. they are functions in them. 
Hence it is possible to treat Equ. (\ref{zEOM}) as constraints that can be used to reduce the degrees of freedom of the other field
equations.

Equations (\ref{AA}), and (\ref{zeta}) are clearly the statements that the forms:
\be
\begin{aligned}
\jmath_2 &= e^{-2\sigma} F + \star (\zeta^I d \tilde{\zeta}_I - \tilde{\zeta}_I d \zeta^I) \\
\jmath_5 &= e^\sigma \gamma^{IK} \theta_{JK} d\zeta^J + e^\sigma \gamma^{IJ} d\tilde{\zeta}_J + \zeta^I \star F \\
\tilde{\jmath}_{5|I} &= e^\sigma (\gamma_{IJ} + \gamma^{KL} \theta_{IK} \theta_{JL} ) d\zeta^J + e^\sigma \gamma^{JK} \theta_{IK} 
d\tilde{\zeta}_J - \tilde{\zeta}_I \star F
\end{aligned}
\ee
From a five dimensional perspective, they can be thought of as the result of the invariance of the action
under particular infinitesimal shifts of $A$ and ($\zeta$, $\tilde{\zeta}$).  
The charge densities corresponding to them can then be found in the usual way by:
\be 
\wp_2 = \int \jmath_2, ~~~~~ \wp_5^I = \int \jmath_5^I, ~~~~~ \tilde{\wp}_{5|I} = \int \tilde{\jmath}_{5|I}. 
\ee
The geometric way of understanding these charges is noting that they descend from the eleven
dimensional electric and magnetic M-brane charges, hence the ($2, 5$) labels. M2-branes wrapping
special Lagrangian cycles of $\mathcal{M}$ generate $\wp_2$ while the wrapping of M5-branes excite 
($\wp_5^I$, $\tilde{\wp}_{5|I}$).

Finally, for completeness sake we also give $da$, where $a$ is the universal axion dual to $A$. Since
Equ. (\ref{AA}) is equivalent to $d^2 a = 0$, we conclude that
\be 
da = e^{-2\sigma} \star F - (\zeta^I d\tilde{\zeta}_I - \tilde{\zeta}_I d\zeta^I), \label{aa1}
\ee
where $a$ is governed by the field equation
\be
d^\dagger \Big[ e^{2\sigma} da + e^{2\sigma} \Big(\zeta^I d \tilde{\zeta}_I - \tilde{\zeta}_I d\zeta^I  \Big)  \Big] = 0;
\label{aa2} \ee
as a consequence of $dF = 0$. Both terms involving $F$ in Equ. (\ref{act}) could then be replaced by the single expression 
\be
S_a = \frac{1}{2} \int e^{2\sigma} \Big[ da + \Big(\zeta^I d \tilde{\zeta}_I - \tilde{\zeta}_I d\zeta^I  \Big) \Big] 
\wedge \star \Big[ da + \Big(\zeta^I d \tilde{\zeta}_I - \tilde{\zeta}_I d\zeta^I  \Big) \Big].
\label{aa3}\ee

\section{The theory in manifestly symplectic form}
For the sake of completeness, we also give a recently proposed form of the $\mathcal{N} = 2$ theory,
clearly highlighting its symplectic structure \cite{CYM2-2}. Since the action is invariant under rotations in $\mathbf{Sp}$,
then it is clear that $R,~ d\sigma,~ dz$ and $F$ are themselves symplectic invariants. The axion fields ($\zeta,\tilde{\zeta}$),
however, can be thought of as components of an $\mathbf{Sp}$ ‘axions vector’. If we define:
\be 
|Xi \rangle = \begin{pmatrix} \zeta^I \\ -\tilde{\zeta}_I \end{pmatrix} , ~~~~
|d Xi \rangle = \begin{pmatrix} d \zeta^I  \\ -d \tilde{\zeta}_I \end{pmatrix}
\ee
then
\be
\langle \Xi | d \Xi \rangle = \zeta^I d \tilde{\zeta}_I - \tilde{\zeta}_I d \zeta^I,
\ee
as well as:
\be
\begin{aligned}
& \langle \partial_\mu \Xi | \Lambda | \partial^\mu \Xi \rangle = \\ 
& - (\gamma_{IJ} + \gamma^{KL} \theta_{IK} \theta_{JL}) (\partial_\mu \zeta^I)  (\partial^\mu \zeta^J) -
\gamma^{IJ} (\partial_\mu \tilde{\zeta}_I)  (\partial^\mu \tilde{\zeta}_J) 
- 2 \gamma^{IK} \theta_{JK} (\partial_\mu \zeta^J)  (\partial^\mu \tilde{\zeta}_I),
\end{aligned}
\ee
such that Equ. (\ref{xx}) becomes
\be
\begin{aligned}
X &=  (\gamma_{IJ} + \gamma^{KL} \theta_{IK} \theta_{JL}) d \zeta^I \wedge \star d \zeta^J
+ \gamma^{IJ} d \tilde{\zeta}_I \wedge \star d\tilde{\zeta}_J +2 \gamma^{IK} \theta_{JK} d \zeta^J \wedge \star d\tilde{\zeta}_I,\\
&= - \langle \partial_\mu \Xi | \Lambda | \partial^\mu \Xi \rangle \star 1.
\end{aligned}
\ee
As a consequence of this language, the field expansion Equ. (\ref{af}) could be rewritten
\be
\begin{aligned}
\mathcal{A} &= A + \sqrt{2} \langle \Theta | \rangle, \\
\mathcal{F} &= d \mathcal{A} = F + \sqrt{2} \langle \Theta \mathop{|}_\Lambda d \Xi \rangle.
\end{aligned}
\ee
The bosonic action in manifest symplectic covariance is hence:
\be
\begin{aligned}
S_5 = \int_5 &\Big[ \mathcal{R} \star 1 - \frac{1}{2} d\sigma \wedge \star d\sigma - G_{i\bar{j}} dz^i \wedge \star dz^{\bar{j}} 
- F \wedge \langle  \Xi | d \Xi \rangle 
\\&- \frac{1}{2} e^{-2\sigma} F \wedge \star F  +e^\sigma ~ \langle \partial_\mu \Xi | \Lambda | \partial^\mu \Xi \rangle \star 1 \Big].
\end{aligned} \label{act-f}
\ee
The equations of motion are now
\be 
(\Delta \sigma) \star 1 + e^\sigma \langle \partial_\mu \Xi | \Lambda | \partial^\mu \Xi \rangle \star 1  + e^{-2\sigma} F \wedge \star F = 0
\label{sigma}
\ee
\be
\begin{aligned}
(\Delta z^i) \star 1 + \Gamma^i_{jk} dz^j \wedge \star dz^k + \frac{1}{2} e^\sigma G^{i\bar{j}} \partial_{\bar{j}}  \langle \partial_\mu \Xi | \Lambda | \partial^\mu \Xi \rangle \star 1 &=0 \\
(\Delta z^{\bar{i}}) \star 1 + \Gamma^{\bar{i}}_{\bar{j}\bar{k}} dz^{\bar{j}} \wedge \star dz^{\bar{k}} + \frac{1}{2} e^\sigma G^{\bar{i}j} \partial_{\bar{j}}  \langle \partial_\mu \Xi | \Lambda | \partial^\mu \Xi \rangle \star 1  &=0 \label{zzz}
\end{aligned}
\ee
\be
d^\dagger \Big[ e^{-2\sigma} F + \star \langle \Xi | d \Xi \rangle \Big] =0 \label{AA-f}
\ee 
\be
d^\dagger \Big[ e^{\sigma} | \Lambda d\Xi \rangle + \star F | \Xi \rangle \Big] =0. \label{zeta-ff} 
\ee
Note that, as is usual for Chern-Simons actions, the explicit appearance of the gauge potential $| \Xi \rangle$
in Equ. (\ref{AA-f}), and Equ. (\ref{zeta-ff}) does not have an effect on the physics since:
\be
\begin{aligned}
d^\dagger \star \langle\Xi|d\Xi\rangle &\to d \langle\Xi|d\Xi\rangle = \langle d \Xi \mathop{|}_\Lambda d\Xi\rangle \\  
d^\dagger \star F|\Xi\rangle &\to d[F|\Xi\rangle]= F \wedge |d\Xi\rangle.
\end{aligned}
\ee
The Noether currents and charges become
\be
\begin{aligned}
\jmath_2 &= e^{-2\sigma} F + \star \langle \Xi | d \Xi \rangle \\
|\jmath_5 \rangle &= e^\sigma | \Lambda d\Xi \rangle + \star F |\Xi \rangle
\wp_2 = \int \jmath_2, ~~~~~ |\wp_5\rangle = \int |\jmath_5 \rangle.
\end{aligned}
\ee
The equations of the universal axion Equ. (\ref{aa1}), Equ. (\ref{aa2}), and Equ. (\ref{aa3}) are now
\be 
da = e^{-2\sigma} \star F - \langle \Xi | d \Xi \rangle, 
\ee
\be
d^\dagger \Big[ e^{2\sigma} da + e^{2\sigma} \langle \Xi | d \Xi \rangle  \Big] = 0 ~~~~~~~~ \mbox{and}
\ee
\be
S_a = \frac{1}{2} \int e^{2\sigma} \Big[ da + \langle \Xi | d \Xi \rangle \Big] 
\wedge \star \Big[ da + \langle \Xi | d \Xi \rangle \Big].
\ee
So Equ. (\ref{AA-f}), and Equ. (\ref{zeta-ff}) can also written as:
\be
d^\dagger \{ e^\sigma | \mathbf{\Lambda} d\Xi \rangle - e^{2\sigma} [ d a + \langle \Xi | d \Xi \rangle] |\Xi \rangle \} =0, 
\label{v-f}
\ee
\be
d^\dagger [ e^{2\sigma} d a + e^{2\sigma} \langle \Xi | d \Xi \rangle] =0. 
\label{k-f}
\ee
The full action is symmetric under the following SUSY transformations:
\bea
 \delta _\epsilon  \psi ^1  &=& D \epsilon _1  + \frac{1}{4}\left\{ {i {e^{\sigma } \left[ {d\phi + \left\langle {\Xi }
 \mathrel{\left | {\vphantom {\Xi  {d\Xi }}}
 \right. \kern-\nulldelimiterspace} {{d\Xi }} \right\rangle } \right]}- Y} \right\}\epsilon _1  - e^{\frac{\sigma }{2}} \left\langle {{\bar V}}
 \mathrel{\left | {\vphantom {{\bar V} {d\Xi }}} \right. \kern-\nulldelimiterspace} {{d\Xi }} \right\rangle\epsilon _2  \nonumber\\
 \delta _\epsilon  \psi ^2  &=& D \epsilon _2  - \frac{1}{4}\left\{ {i {e^{\sigma } \left[ {d\phi + \left\langle {\Xi }
 \mathrel{\left | {\vphantom {\Xi  {d\Xi }}} \right. \kern-\nulldelimiterspace}
 {{d\Xi }} \right\rangle } \right]}- Y} \right\}\epsilon _2  + e^{\frac{\sigma }{2}} \left\langle {V}
 \mathrel{\left | {\vphantom {V {d\Xi }}} \right. \kern-\nulldelimiterspace} {{d\Xi }} \right\rangle \epsilon _1,  \label{SUSYGraviton}
\eea
\bea
  \delta _\epsilon  \xi _1^0  &=& e^{\frac{\sigma }{2}} \left\langle {V}
    \mathrel{\left | {\vphantom {V {\partial _\mu  \Xi }}} \right. \kern-\nulldelimiterspace} {{\partial _\mu  \Xi }} \right\rangle  \Gamma ^\mu  \epsilon _1  - \left\{ {\frac{1}{2}\left( {\partial _\mu  \sigma } \right) - \frac{i}{2} e^{\sigma } \left[ {\left(\partial _\mu \phi\right) + \left\langle {\Xi }
    \mathrel{\left | {\vphantom {\Xi  {\partial _\mu \Xi }}} \right. \kern-\nulldelimiterspace}
    {{\partial _\mu \Xi }} \right\rangle } \right]} \right\}\Gamma ^\mu  \epsilon _2  \nonumber\\
     \delta _\epsilon  \xi _2^0  \nonumber &=& e^{\frac{\sigma }{2}} \left\langle {{\bar V}}
    \mathrel{\left | {\vphantom {{\bar V} {\partial _\mu  \Xi }}} \right. \kern-\nulldelimiterspace} {{\partial _\mu  \Xi }} \right\rangle \Gamma ^\mu  \epsilon _2  + \left\{ {\frac{1}{2}\left( {\partial _\mu  \sigma } \right) + \frac{i}{2} e^{\sigma } \left[ {\left(\partial _\mu \phi\right) + \left\langle {\Xi }
    \mathrel{\left | {\vphantom {\Xi  {\partial _\mu \Xi }}} \right. \kern-\nulldelimiterspace}
    {{\partial _\mu \Xi }} \right\rangle } \right]} \right\}\Gamma ^\mu  \epsilon
     _1, \\ \label{SUSYHyperon1}
\eea
and
\bea
     \delta _\epsilon  \xi _1^{\hat i}  &=& e^{\frac{\sigma }{2}} e^{\hat ij} \left\langle {{U_j }}
    \mathrel{\left | {\vphantom {{U_j } {\partial _\mu  \Xi }}} \right. \kern-\nulldelimiterspace} {{\partial _\mu  \Xi }} \right\rangle \Gamma ^\mu  \epsilon _1  - e_{\,\,\,\bar j}^{\hat i} \left( {\partial _\mu  z^{\bar j} } \right)\Gamma ^\mu  \epsilon _2  \\
     \delta _\epsilon  \xi _2^{\hat i}  &=& e^{\frac{\sigma }{2}} e^{\hat i\bar j} \left\langle {{U_{\bar j} }}
    \mathrel{\left | {\vphantom {{U_{\bar j} } {\partial _\mu  \Xi }}} \right. \kern-\nulldelimiterspace} {{\partial _\mu  \Xi }} \right\rangle \Gamma ^\mu  \epsilon _2  + e_{\,\,\,j}^{\hat i} \left( {\partial _\mu  z^j } \right)\Gamma ^\mu  \epsilon    _1,\label{SUSYHyperon2} 
\eea
where $\left(\psi ^1, \psi ^2\right)$ are the two gravitini and $\left(\xi _1^I, \xi _2^I\right)$ are the hyperini. The quantity $Y$ is defined by:
\begin{equation}
    Y   = \frac{{\bar Z^I N_{IJ}  {d  Z^J }  -
    Z^I N_{IJ}  {d  \bar Z^J } }}{{\bar Z^I N_{IJ} Z^J
    }},\label{DefOfY}
\end{equation}
where $N_{IJ}  = \mathfrak{Im} \left({\partial_IF_J } \right)$. The $e$'s are the beins of the special K\"{a}hler metric $G_{i\bar j}$, the $\epsilon$'s are the five-dimensional $\N=2$ SUSY spinors and the $\Gamma$'s are the usual Dirac matrices. The covariant derivative $D$ is given by $D=dx^\mu\left( \partial _\mu   + \frac{1}{4}\omega _\mu^{\,\,\,\,\hat \mu\hat \nu} \Gamma _{\hat \mu\hat \nu}\right)\label{DefOfCovDerivative}$ as usual, where the $\omega$'s are the spin connections and the hatted indices are frame indices in a flat tangent space.

\setcounter{equation}{0}
\chapter{Inflationary Brane-Worlds Coupled to the Calabi-Yau 
Complex Structure Moduli}
\label{ch4}

Still there are many quests have not been resolved yet about the Cosmological evolution. For instance, around the first seconds of
our universe birth. Namely, what happened before the Big Bang ? 
What do drive the universe as its way till now times? What were the universe initial conditions (ICs) ? 
And how it rapidly exapanded " in size " 
in a very short period of time during the inflation era ? Or what does the cosmological constant really represent ? 
The term has been added to Einstein field equations to account for the accelerated expansion on the time being epoch.
\footnote{See Appendix (\ref{GR}) for more information about the cosmological constant.}   
Finally how the universe will end up ? 
What are the most probable scenarios for our universe future? 

There are many theories try to interpret all these mystries, like quantum gravity, see for example \cite{Hertog:2013, James:2018},
or higher curvature theories of gravity
\footnote{Including string theory itself.}, as \cite{Baumann:2009, McAllister:2010}. 
However, we try to look at these from other perspective through modeling the universe as a 3-brane imbedded in a higher five dimensional spacetime called the bulk. 

Indeed modeling the universe as a 3-brane embedded in a higher dimensional bulk, initially proposed by \cite{Randall:1999ee} and further developed by many authors such as \cite{Brax:2003fv}, \cite{Maartens:2010ar}, and \cite{Roane:2007zz}. While our work 
is an extension to a previous studies have been initiated before \cite{Emam:2015laa}.
In this chapter we manifest the correlation has been deduced between 
the moduli of the complex structure of the underlying Calabi-Yau manifold and the dynamic behavior of the brane-world . We numerically solve the field equations in case of dust filled and radiation filled branes. 
Also we study the time dependence of the $\mathcal{N} = 2$ hypermultiplets scalars. Although these fields are restricted to the bulk, the dilaton, for instance, is related to the volume of CY manifold, so it’s important to see how it behaves at different universes (ICs). 
\section{Brane dynamics}
We consider a metric of the form 
\be%
ds^2= - e^{2\alpha(t,y)} dt^2 + e^{2\beta(t,y)} (dr^2 + r^2 d \Omega^2 ) + e^{2\gamma(t,y)} dy^2,
\label{metricE}
\ee%
where $d\Omega^2= d\theta^2+\sin^2(\theta) d\phi^2.$ This metric may be interpreted as representing a single 3-brane located at $y = 0$ in the transverse space, it may also represent a stack of N branes located at various values of $y = y_I (I = 1, ...., N \in \mathbb{Z} ) $ where the warp functions $\alpha$, $\beta$ and $\gamma$ are rewritten such that
$y \to \sum_{I=1}^N |y-y_I|$.  Either way, we will eventually focus on the four dimensional $(t,  r, \theta, \phi)$ dynamics, effectively evaluating the warp functions at a specific, but arbitrary, $y$ value. We also
note that a metric of the form Equ. (\ref{metricE}) was shown in
\cite{Kallosh:2001} to be exactly the type needed for a consistent BPS cosmology . 
The components of the Einstein tensor are
\bea \nonumber G_{tt} &=&3 (\dot\beta^2+\dot\beta \dot\gamma)-3 e^{2(\alpha-\gamma)} (\beta^{\prime \prime} + 2 \beta^{\prime 2}-\beta^{\prime} \gamma^{\prime}) \\
\nonumber G_{rr} &=& - e^{2(\beta-\alpha)}[ 2 \ddot\beta+3\dot\beta^2+\ddot\gamma+ \dot\gamma^2 + 2\dot\beta(\dot\gamma-\dot\alpha) -\dot\alpha \dot\gamma)] 
\\ \nonumber &+& e^{2(\beta-\gamma)}[ 2\beta^{\prime \prime}+3\beta^{\prime 2}+\alpha^{\prime \prime} + \alpha^{\prime 2} + 2\beta^{\prime} (\alpha^{\prime}-\gamma^{\prime}) -\alpha^{\prime} \gamma^{\prime})] 
\eea
\newpage
\bea 
\nonumber G_{(\theta\theta,~  \phi\phi)} &=& e^{2(\beta-\gamma)} ~ (r^2,~ r^2 sin^2\theta)~
[3\beta^{\prime 2} - 2 \beta^{\prime} (\gamma^{\prime}-\alpha^{\prime})- \gamma^{\prime}\alpha^{\prime}+\alpha^{\prime 2}+2\beta^{\prime \prime}+\alpha^{\prime \prime}] \\ \nonumber &-& e^{2(\beta-\alpha)} ~ ( r^2, ~ r^2 \sin^2\theta)
[3 \dot\beta^2+ 2 \dot\beta( \dot\gamma- \dot\alpha )+ \dot\gamma^2 - \dot\gamma \dot\alpha+ 2  \ddot\beta+\ddot\gamma], \\
\nonumber G_{yy} &=&  3 (\beta^{ \prime 2} + \beta' \alpha')-3 e^{2(\gamma-\alpha)} (\ddot\beta + 2 \dot\beta^2-\dot\beta \dot\alpha) \\
G_{ty} &=& 3( \alpha^{\prime} \dot\beta-\beta^{\prime} \dot\beta+ \beta^{\prime} \dot\gamma - \dot\beta^{\prime}) ,
\label{GG}
\eea
where a prime is a derivative with respect to $y$ and a dot is a derivative with respect to $t$. We consider only the fields' dependence on time, so that $G_{ty}$ totally vanishes.  
The energy momentum tensor is given by \cite{BookME} :
\be%
T_{\mu\nu} = -2 \frac{\partial \mathcal{L}}{\partial g^{\mu\nu}}+ g_{\mu\nu} \mathcal{L}.
\ee%
ln addition to the usual perfect fluid stress tensor on the brane:
\be%
T_{\mu\nu} =( \rho + p ) U_\mu U_\nu + g_{\mu\nu} p 
\ee%
From (\ref{act-f});
\begin{multline}
 \mathcal{L}= - R + \frac{g^{\alpha\beta}}{4} \partial_\alpha \sigma \partial_\beta \sigma + \frac{g^{\alpha\beta}}{2} G_{i\bar{j}} \partial_\alpha z^i \partial_\beta z^{\bar{j}} - \frac{g^{\alpha\beta}}{2}e^\sigma [\partial_\alpha \Xi|\Lambda|\partial_\beta \Xi ] \\+
\frac{e^{2\sigma}}{2} [\partial_\alpha \phi +\langle \Xi|\partial_\alpha \Xi\rangle ] ~ [ \partial_\beta \phi+ \langle \Xi|\partial_\beta \Xi\rangle ]~~~~~~~~~~~~~~~~~~~~~~~~~~~~
\end{multline}
\be%
T_{\mu\nu} = T_{\mu\nu}^{Bulk} + T_{\mu\nu}^{3b},
\ee%
\be%
T_{yy} = T_{yy}^{Bulk}, ~~~~~~~~~~ T_{yt}=0.
\ee%
\begin{multline}
-2\frac{\partial \mathcal{L}}{\partial g^{\mu\nu}}
= -\frac{1}{2} \partial_\mu \sigma \partial_\nu \sigma  - G_{i\bar{j}} \partial_\mu z^i \partial_\nu z^{\bar{j}} + e^\sigma [\partial_\mu \Xi|\Lambda|\partial_\nu \Xi ] \\ - \frac{e^{2\sigma}}{2} [\partial_\mu \phi + \langle \Xi|\partial_\mu \Xi\rangle ] ~ [ \partial_\nu \phi+ \langle \Xi|\partial_\nu \Xi\rangle ],
~~~~~~~~~~~~~~~~~~~
\end{multline}
so that:
\bea \nonumber
T_{\mu\nu}^{Bulk} &=&  -\frac{1}{2} \partial_\mu \sigma \partial_\nu \sigma  - G_{i\bar{j}} \partial_\mu z^i \partial_\nu z^{\bar{j}} + e^\sigma [\partial_\mu \Xi|\Lambda|\partial_\nu \Xi ] 
 \\\nonumber &-& \frac{e^{2\sigma}}{2} [\partial_\mu \phi + \langle \Xi|\partial_\mu \Xi\rangle ] ~ [ \partial_\nu \phi+ \langle \Xi|\partial_\nu \Xi\rangle ] +  \frac{g_{\mu\nu}}{4} \partial_\alpha \sigma \partial^\alpha \sigma 
+ \frac{g_{\mu\nu}}{2} G_{i\bar{j}} \partial_\alpha z^i \partial^\alpha z^{\bar{j}}
\\\nonumber &-& \frac{g_{\mu\nu}}{2}  e^\sigma[\partial_\alpha \Xi|\Lambda|\partial^\alpha \Xi ]  
+ g_{\mu\nu} \frac{e^{2\sigma}}{4} [\partial_\alpha \phi + \langle \Xi|\partial_\alpha \Xi\rangle ] ~ [ \partial^\alpha \phi+ \langle \Xi|\partial^\alpha \Xi\rangle ]
,\\
\label{munutensor}
\eea
and:
\bea
\nonumber T_{yy} &=&  -\frac{1}{2} \partial_y \sigma \partial_y \sigma  - G_{i\bar{j}} \partial_y z^i \partial_y z^{\bar{j}} + e^\sigma [\partial_y \Xi|\Lambda|\partial_y \Xi ] 
\\\nonumber
 &-& \frac{e^{2\sigma}}{2} [\partial_y \phi + \langle \Xi|\partial_y \Xi \rangle ] ~ [ \partial_y \phi+ \langle \Xi|\partial_y \Xi\rangle ] 
+  \frac{g_{yy}}{4} \partial_y \sigma \partial^y \sigma 
 +\frac{g_{yy}}{2} G_{i\bar{j}} \partial_y z^i \partial^y z^{\bar{j}}\\\nonumber
&-& \frac{g_{yy}}{2}  e^\sigma [\partial_y \Xi|\Lambda|\partial^y \Xi ]  + g_{yy} \frac{e^{2\sigma}}{4} [\partial_y \phi + \langle \Xi|\partial_y \Xi\rangle ] ~ [ \partial^y \phi+ \langle \Xi|\partial^y \Xi \rangle ]
.\\
\eea
Then:
\be%
T^{3b}_{tt} = \rho(t), ~~~~~~~
T^{3b}_{rr} = e^{2\beta} p(t), ~~~~~~~
T^{3b}_{\theta\theta} = e^{2\beta} r^2 p(t), ~~~~~
T^{3b}_{\phi\phi} = e^{2\beta} \sin{\theta}^2 p(t), ~~~~~~~
\ee%
where $\rho$ and $p$ depend on time.  We are interested in bosonic configurations that preserve some supersymmetry, 
so the stress tensor (\ref{munutensor})
can be considerably simplified by considering the vanishing of the supersymmetric variations
(\ref{SUSYHyperon1}, \ref{SUSYHyperon2}), that leads to the BPS condition \cite{Emam:2015laa, Emam2013}
\be e^\sigma \langle \Xi|\mathbf {\Lambda}| \star d\Xi \rangle
 = \frac{1}{2} d\sigma \wedge \star d\sigma + \frac{1}{2} e^{2\sigma} [d\phi + \langle\Xi|~d\Xi \rangle] \wedge \star [d\phi + \langle \Xi|~d\Xi\rangle] + 2 |\dot z|^2,
\label{bps}
\ee
Substitute by (\ref{bps}) in (\ref{munutensor}), leaving the dynamics to depend only on the complex
structure moduli ($z^i,z^{\bar{i}}$) :
\be%
T_{\mu\nu}= G_{i\bar{j}} \partial_\mu z^i \partial_\nu z^{\bar{j}} - \frac{1}{2}  g_{\mu\nu} G_{i\bar{j}} \partial_\alpha z^i \partial^\alpha z^{\bar{j}}.
\ee%
Then:
\bea 
\label{EFE1}
\nonumber G_{tt} &=& \frac{1}{2}G_{i\bar{j}} {\dot z^i}  {\dot z^{\bar{j}}}+ \frac{1}{2} e^{2(\alpha-\gamma)} G_{i\bar{j}} z^{`i}   z^{`\bar{j}}+e^{2\alpha}\rho,  \\\nonumber
G_{rr} &=& \frac{1}{2}G_{i\bar{j}} e^{2(\beta-\alpha)} {\dot z^i}  {\dot z^{\bar{j}}}- \frac{1}{2} e^{2(\beta-\gamma)} G_{i\bar{j}} z^{`i} z^{`\bar{j}}+e^{2\beta} p, \\\nonumber
 G_{\theta\theta} &=& \frac{1}{2} G_{i\bar{j}} e^{2(\beta-\alpha)} r^2 {\dot z^i}  {\dot z^{\bar{j}}} - \frac{1}{2}G_{i\bar{j}}r^2 e^{2(\beta-\gamma)}  z^{`i} z^{`\bar{j}} + e^{2\beta} r^2 p,  \\\nonumber
 G_{\phi\phi} &=& \frac{1}{2} G_{i\bar{j}} e^{2(\beta-\alpha)} r^2 \sin{\theta}^2 {\dot z^i}  {\dot z^{\bar{j}}} - \frac{r^2 \sin{\theta}^2}{2}G_{i\bar{j}} e^{2(\beta-\gamma)}  z^{`i} z^{`\bar{j}} + e^{2\beta} r^2 \sin{\theta}^2 p, \\
\eea
and; 
\be%
\label{EFE2} 
G_{yy} = \frac{1}{2}  G_{i\bar{j}} z^{'i}   z^{\bar{'j}}+\frac{1}{2} e^{2(\gamma-\alpha)}  G_{i\bar{j}} \dot z^{i} \dot z^{\bar{j}},
\ee
The components of the Einstein tensor are given by Equ. (\ref{GG}).
The universal axion equations (\ref{v-f}) and (\ref{k-f}) can be integrated to give
\be 
d\phi + \langle\Xi|d\Xi\rangle = n e^{-2\sigma} dh,
\label{phiEOM} 
\ee
where $h$ is harmonic in $(t,y)$, i.e. satisfies $\Delta h=0$, and n $\in \mathbb{R}$. Similarly:
\be e^\sigma|\Lambda d\Xi\rangle- n dh |\Xi\rangle = s|dK\rangle ~~~~~
\mbox{where} ~~~~~ |\Delta K \rangle =0, ~~~
s \in \mathbb{R}.
\label{uniAx}
\ee
Using the hyperini transformations (\ref{SUSYHyperon1}) and (\ref{SUSYHyperon2}) with (\ref{uniAx}) and make the assumption that 
$\epsilon_1 = \pm \epsilon_2$ we find the following for the axions \cite{Emam:2015laa}:
\be 
|d\Xi\rangle = e^{-\frac{\sigma}{2}}~ Re ~ [
( e^{-\sigma} dh-id\sigma) |V\rangle + 2i |U_i\rangle dz^i
],
\label{xi1}
\ee
\be 
|\Xi\rangle dh = e^{-\frac{\sigma}{2}}~ Re ~ [
( d\sigma+i e^{-\sigma} dh) |V\rangle + 2 |U_i\rangle dz^i],
\label{xi2}
\ee
at $n=1$ and $s=0$. On the other hand, the equation of the harmonic function $h$, that arises from $\Delta h=0$ is given by:
\be 
e^{\alpha-\gamma} [h'' + (\alpha' + 3 \beta' - \gamma') h' ] =
e^{\gamma-\alpha} [\ddot{h} - (\dot{\alpha} - 3 \dot{\beta} - \dot{\gamma}) \dot{h} ] 
\label{harmo}
\ee
\section{The cosmology of a single brane}
The warp functions as well as $h$ can be considered as separable as follows:
\be
\begin{split}
e^{\beta(t,y)} &= a(t)~F(y) \\
e^{\gamma(t,y)} &= b(t)~ K(y) \\
e^{\alpha(t,y)} &= c(t)~ N(y) \\
h(t,y) &= k(t)~ M(y)
\end{split}
\ee
Our interest is the dynamics of a single brane out of an infinite number of possible 3-branes along $y$, so we will evalutate the functions $F (y),~ K (y),~ N (y)$, and $M (y)$ near the brane of interestand normalize the result to unity, i.e. $F (0) = 1$  and so on, where the brane under study is located at $y = 0$.
So that:
\be
\begin{aligned}
& e^{2\alpha} = c^2(t) \to 1, ~~~~~~ \mbox{(RW-like metric)}, ~~~~~ \dot\alpha = 0, \\
& e^{2\gamma} = b^2(t), ~~~~~~~~~\dot\gamma = \frac{\dot b}{b}, \\
& e^{2\beta} = a^2(t), ~~~~~~~~ \dot\beta= \frac{\dot a}{a}.
\end{aligned} 
\ee
Equations (\ref{GG}) and (\ref{harmo}) are the basic
equations governing the dynamics of the multi-brane spacetime. Einstein equations 
reduce to the Friedmann-like form 
\bea
\label{eee}
 3\left[ {\left( {\frac{{\dot a}}{a}} \right)^2  + \left( {\frac{{\dot a}}{a}} \right)\left( {\frac{{\dot b}}{b}} \right)} \right] &=& G_{i\bar j} \dot z^i \dot z^{\bar j}  + \rho(t) \nonumber\\
 2\frac{{\ddot a}}{a} + \left( {\frac{{\dot a}}{a}} \right)^2  + \frac{{\ddot b}}{b} + 2\left( {\frac{{\dot a}}{a}} \right)\left( {\frac{{\dot b}}{b}} \right) &=&  - G_{i\bar j} \dot z^i \dot z^{\bar j} - p(t) \nonumber\\
 3\left[ {\frac{{\ddot a}}{a} + \left( {\frac{{\dot a}}{a}} \right)^2 } \right] &=&  - G_{i\bar j} \dot z^i \dot z^{\bar j}.
\eea
It is clear that the quantity $G_{i\bar j} \dot z^i \dot z^{\bar j}$  plays an important role in the solution since 
as it will be shown to be the source of the dynamics of the brane. It is interpreted as the norm of the moduli's flow velocity.
Now the equation of the harmonic function (\ref{harmo}) becomes:
\be 
\ddot k + \Big[3 \frac{\dot a}{a} +\frac{\dot b}{b} \Big] \dot k= 0
\label{keq}
\ee
We tend to make $k$ and all the hypermultiplet bulk fields depend only on the scale factor $a$. So by using  
(\ref{EFE1}) and (\ref{GG}) in (\ref{keq}) we get for $k$:
\be
\ddot k - \Big(3 \frac{\dot a}{a} +3\frac{\ddot a}{\dot a} - \frac{1}{a^2 \dot a} \Big) \dot k= 0
\ee 
For the dilaton, from (\ref{sigma}), (\ref{AA-f}), and (\ref{zeta-ff}) we get 
the time dependence equation of $\sigma$:
\be
\ddot \sigma + \frac{1}{2} \dot \sigma^2 -
\frac{e^{-2\sigma}}{2}  \dot k^2 - 6 \Big(  \frac{\ddot a}{a} + \frac{\dot a^2}{a^2} \Big)=0.
\ee
To get the time dependence of $\phi$ (\ref{phiEOM}) we should to get $\langle\Xi|\dot \Xi\rangle$ first. From
(\ref{xi1}) and (\ref{xi2}) 
\footnote{Within the text some times we use $|\dot{z}|^2 \equiv G_{i\bar{j}} d\dot{z}^i d\dot{z}^{\bar{j}}$.} 
\be
\langle \Xi|\dot {\Xi}\rangle = \frac{e^{-\sigma}}{2 \dot k} \Big( e^{-2\sigma} \dot k^2 + \dot \sigma^2 + 4 |\dot z|^2 \Big),
\ee
so that 
\be 
\dot \phi =e^{-2\sigma} {\dot k} -\frac{e^{-\sigma}}{2 \dot k} 
\Big[ e^{-2\sigma} \dot k^2 + \dot \sigma^2 - 12 \Big(  \frac{\ddot a}{a} + \frac{\dot a^2}{a^2} \Big) \Big]
\ee
Finally, while again using the assumption $\epsilon_1=\pm\epsilon_2$, the vanishing of the gravitini equations (\ref{SUSYGraviton}) gives the following for the near-brane spinors
\be\label{Spinor}
    \epsilon\left( t\right)  = e^{\frac{\sigma }{2} + \frac{3}{4}in\Omega k - \Upsilon } \hat \epsilon,
\ee
where $\hat \epsilon$ is an arbitrary constant spinor and the functions $\Upsilon$ and $\Omega$ are solutions of $\dot \Upsilon  = Y$ and
\be\label{Omega}
    \frac{d}{{dt}}\left( {\Omega k} \right) = e^{ - \sigma } \dot k
\ee
respectively. This and all the field equations are the same in case of dust- filled brane, radiation- filled brane and
radiation, dust with energy filled brane.
The equations derived here are much too complicated to solve analytically. 
We then present a numerical solution for the time dependence of the quantities involved.
\section{A dust-filled brane}
\label{DD}
\begin{figure}[H]
  \begin{subfigure}[t]{.5\linewidth}
    \centering
    \includegraphics[width=0.7\columnwidth]{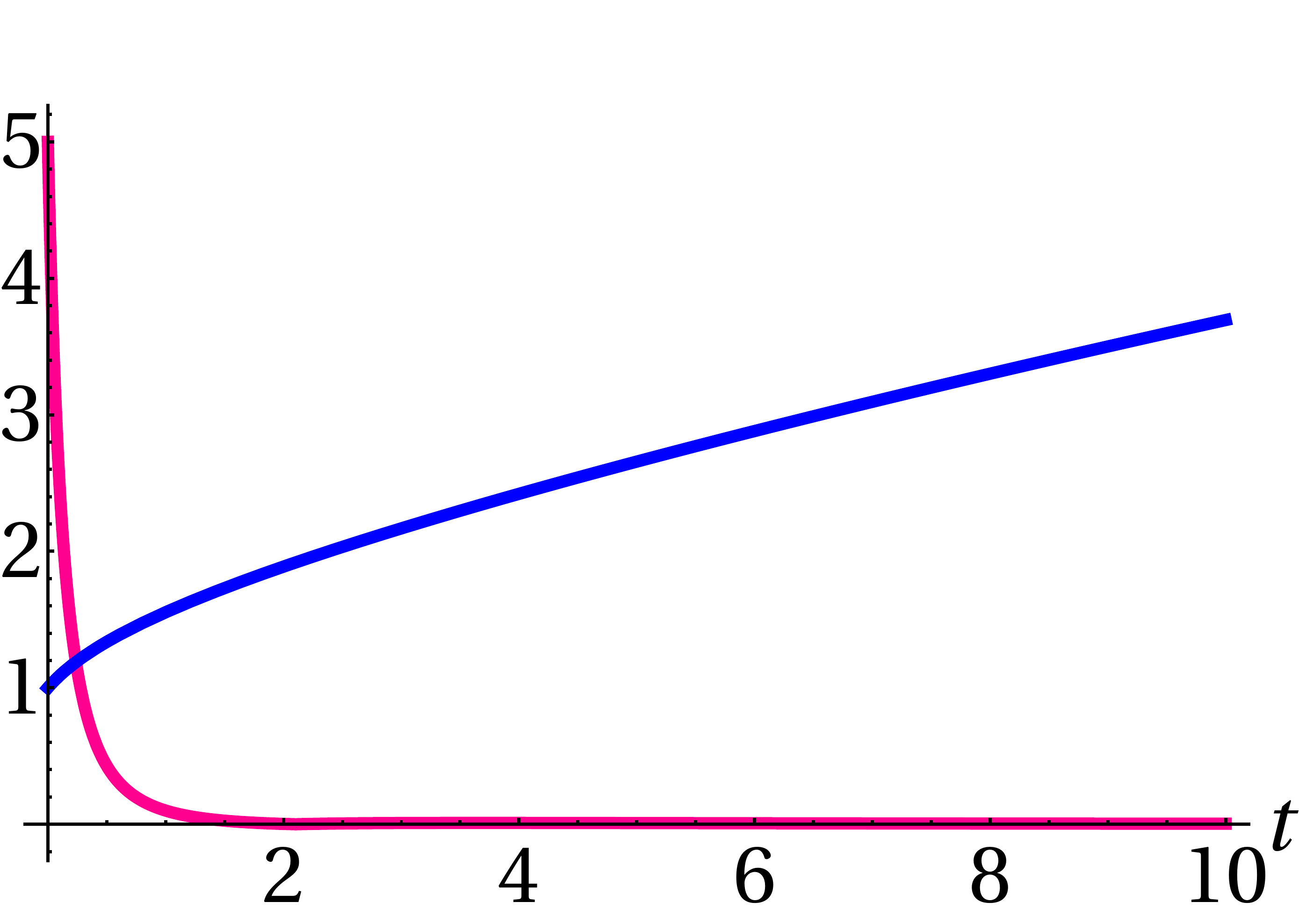}
    \caption{The scale factors $a$ and $b$; represented by the blue curve, while ${G_{i\bar j} \dot z^i \dot z^{\bar j}}$ (already positive) represented by the red curve.}
    \label{33}
  \end{subfigure}
\qquad
  \begin{subfigure}[t]{.5\linewidth}
    \centering
    \includegraphics[width=0.7\columnwidth]{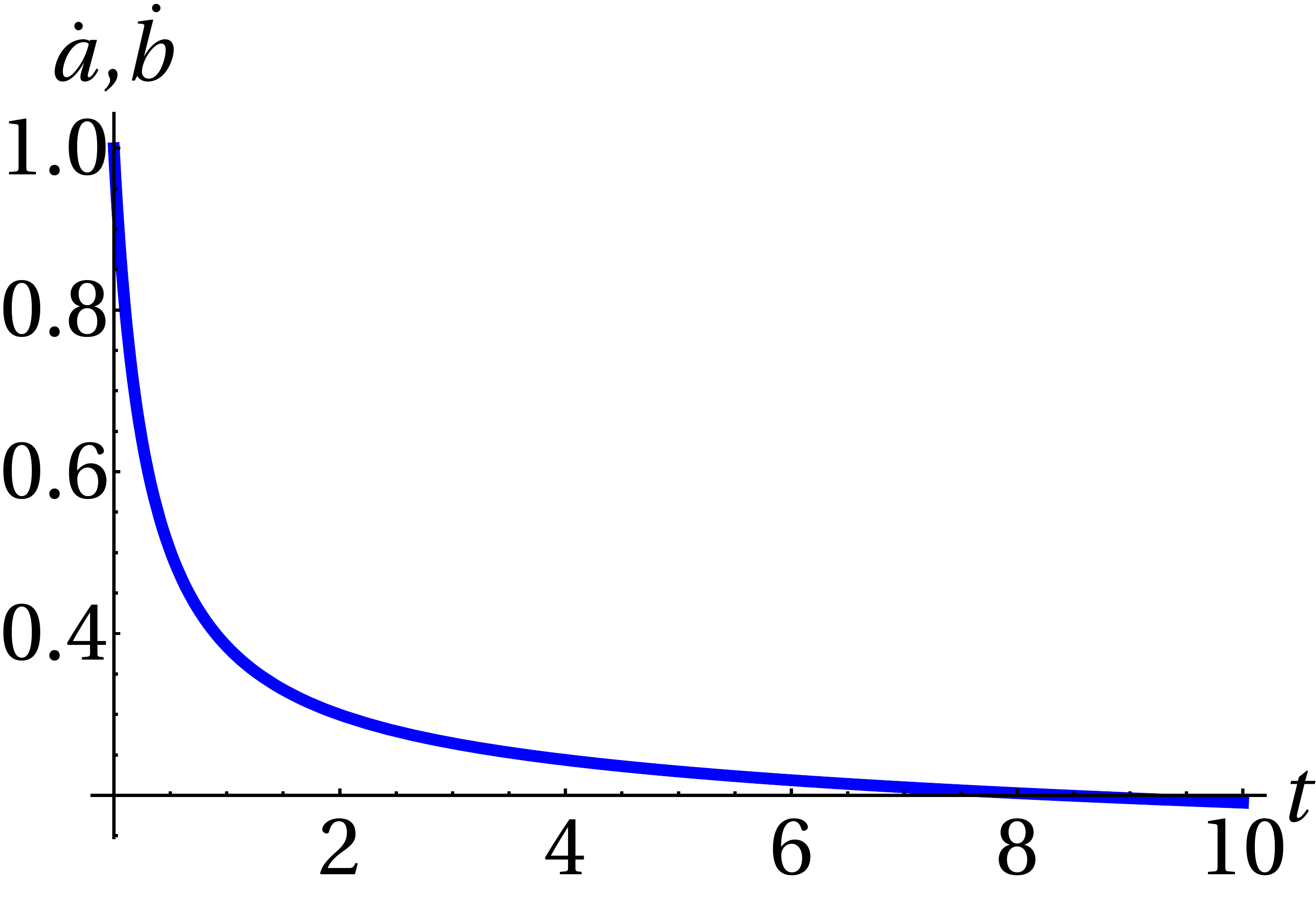}
    \caption{The expansion rates of the scale factors. Both $\dot a$ and $\dot b$ are represented by the shown curve.}
    \label{34}
  \end{subfigure}
\\[9em]
  \begin{subfigure}[t]{.5\linewidth}
    \centering
    \includegraphics[width=0.7\columnwidth]{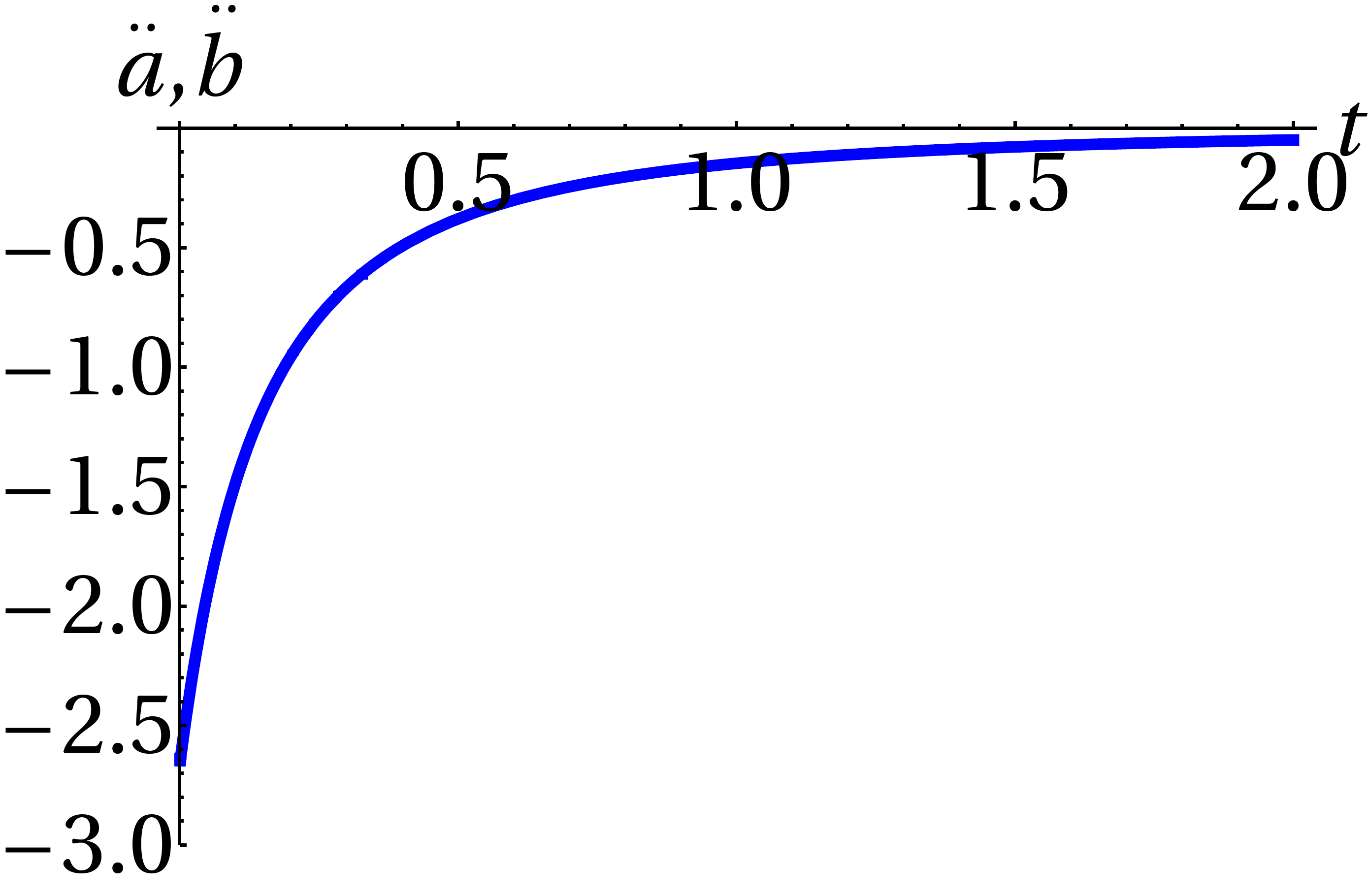}
    \caption{The accelerations of the scale factors. Both $\ddot a$ and $\ddot b$ are represented by the shown curve.}
    \label{35}
  \end{subfigure}
\qquad
  \begin{subfigure}[t]{.5\linewidth}
    \centering
    \includegraphics[width=0.7\columnwidth]{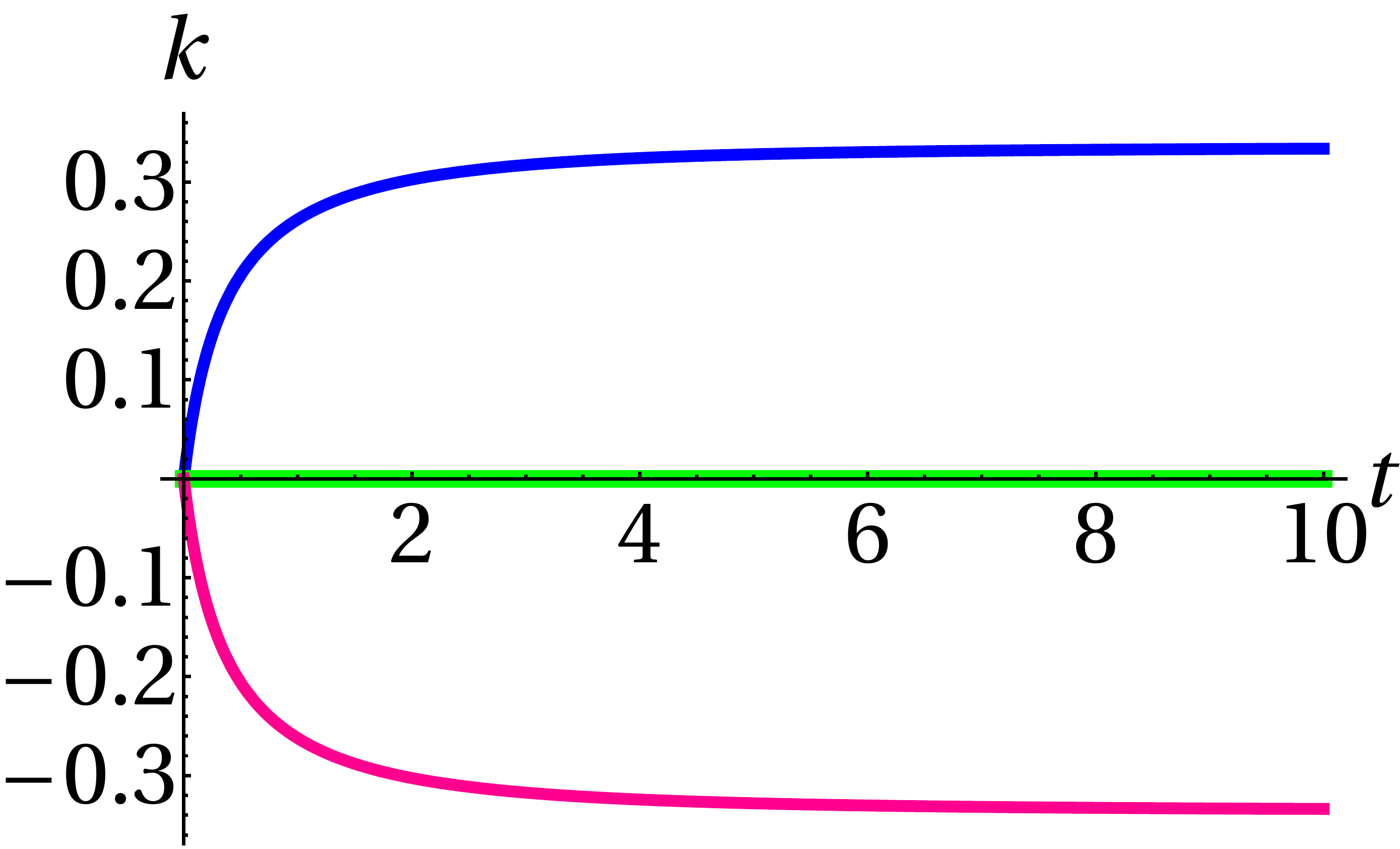}
    \caption{The harmonic function $k$ using: $\dot k\left(0\right)=1$ (blue curve), $\dot k\left(0\right)=0$ (green line), and $\dot k\left(0\right)=-1$ (red curve).}
    \label{36}
  \end{subfigure}
  \caption{Dust-filled brane world with initial conditions set number 4.}
  \label{Fig8}
\end{figure}

For a dust- filled brane 
\be\label{Density1}
    T_{tt}^{\textit {Brane}}  = \rho\left(t\right) = \frac{1}{{a^3 }}, ~~~~~~~~~~ p(t) =0,
\ee
so that the Einstein equations (\ref{eee}) become
\bea
 3\left[ {\left( {\frac{{\dot a}}{a}} \right)^2  + \left( {\frac{{\dot a}}{a}} \right)\left( {\frac{{\dot b}}{b}} \right)} \right] &=& G_{i\bar j} \dot z^i \dot z^{\bar j}  + \frac{1}{{a^3 }} \nonumber\\
 2\frac{{\ddot a}}{a} + \left( {\frac{{\dot a}}{a}} \right)^2  + \frac{{\ddot b}}{b} + 2\left( {\frac{{\dot a}}{a}} \right)\left( {\frac{{\dot b}}{b}} \right) &=&  - G_{i\bar j} \dot z^i \dot z^{\bar j}  \nonumber\\
 3\left[ {\frac{{\ddot a}}{a} + \left( {\frac{{\dot a}}{a}} \right)^2 } \right] &=&  - G_{i\bar j} \dot z^i \dot z^{\bar j}.
\eea
We numerically solve the given equations for the scale factors $a$ and $b$. We also find $G_{i\bar j} \dot z^i \dot z^{\bar j}$ and note that all three quantities are clearly related; confirming \cite{Emam:2015laa}. In fact this work takes this relation further than was found in the previous work; wherein the form of $a$ had to be assumed to fit the \emph{assumption} of an expanding universe. In our case we find all three results \emph{without} having to make assumptions about any of them, and show that the expansion of the brane-world is inevitable in the presence of decaying moduli. The fact that we find this without any imposed assumptions is ascribed to the presence of the constraints (\ref{Density1}). To complete the picture, we also solve for $k$, $\sigma$, $\phi$, $\left\langle {\Xi } \mathrel{\left | {\vphantom {\Xi  {\dot \Xi }}} \right. \kern-\nulldelimiterspace} {{\dot \Xi }} \right\rangle $, as well as $\Omega$. Clearly we do not, and in fact \emph{cannot}, have a complete solution because the field equations for the moduli themselves cannot be solved without explicit knowledge of a metric on the Calabi-Yau submanifold and its exact topology. In fact, we take this work to be one step closer a possible deeper understanding of these topics; namely what the Calabi-Yau submanifold exactly looks like. Finally, let's note that because of the nature of numerical analysis, we find and present all results using a variety of initial conditions, some of which pertain to the possible use of this result to model our universe (\emph{i.e.} a big bang-like singularity as an initial condition), while the remaining ones are given for completeness and additional confirmation of the causal relation between the moduli and the scale factors. These ICs are listed in tabel (\ref{112-4}), where 
for the first five initial conditions we find that $b\propto a$ up to an arbitrary scaling constant determined by the initial conditions. As such and without loss of generality we choose a scale of unity.
Consequently $a$, $b$, and their derivatives exactly coincide in the dust plots of IC1 through IC5. In the sixth case different initial conditions are chosen for $a$ and $b$, and thus they no longer coincide. In most solutions in this subsection the norm $G_{i\bar j} \dot z^i \dot z^{\bar j}$ is negative; except IC4 Fig. (\ref{33}), however we plot its absolute value for ease of reference.

\begin{table}[H]
\centering
\caption{The six sets of initial conditions (IC) used in the dust and radiation filled branes computations.}
\label{112-4}
\vspace{1cm}
\begin{adjustbox}{max width=\textwidth}
\begin{tabular}{|c|c|c|c|c|l|}
\hline \rule{0pt}{1cm}
\large{\textbf{IC Set Number}}  & $a$ & $b$ & $\dot a$ & $\dot b$ & \large{\textbf{Description}}                                        \\ \hline \rule{0pt}{1cm}
\textbf{1} & 0          & 0          & 0                     & 0                     & \large{Big bang-like IC with vanishing initial velocities} \\\hline \rule{0pt}{1cm}
\textbf{2} & 1          & 0          & 1                     & 0                     & \large{Non-singular IC with vanishing initial velocities}  \\ \hline \rule{0pt}{1cm}
\textbf{3} & 0          & 1          & 0                     & 1                     & \large{Big bang-like IC with initial positive velocities}    \\ \hline \rule{0pt}{1cm}
\textbf{4} & 1          & 1          & 1                     & 1                     & \large{Non-singular IC with initial positive velocities}     \\ \hline \rule{0pt}{1cm}
\textbf{5} & 1          & -0.2       & 1                     & -0.2                  & \large{Non-singular IC with initial negative velocities}     \\ \hline \rule{0pt}{1cm}
\textbf{6} & 0          & 0          & 1                     & 0                     & \large{Initial singularity in $a$ only with vanishing initial velocities}    \\  \hline 
\end{tabular}
\end{adjustbox}
\end{table}


\begin{figure}[H]
  \begin{subfigure}[t]{.5\linewidth}
    \centering
    \includegraphics[width=0.7\columnwidth]{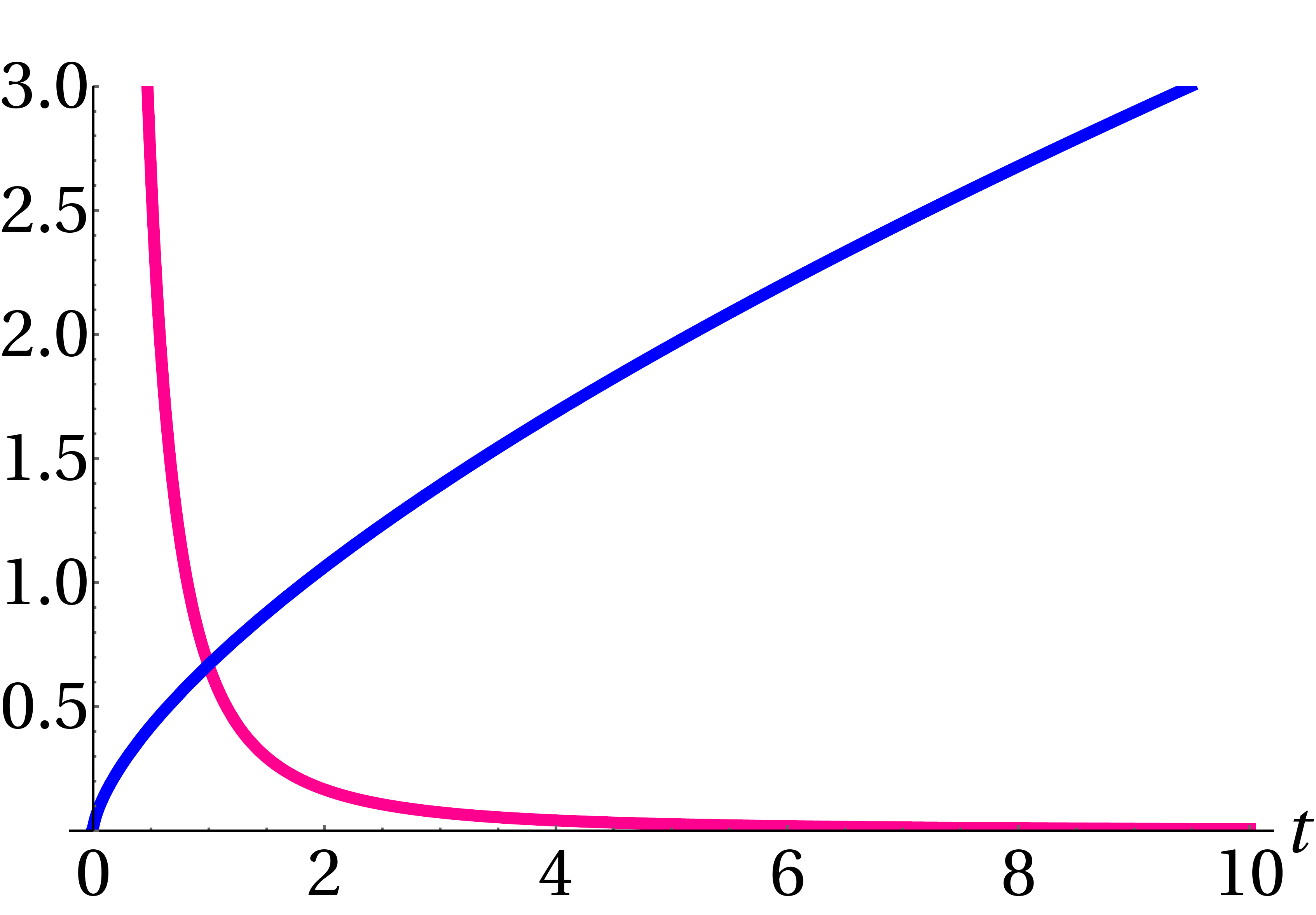}
    \caption{The scale factors $a$ and $b$; represented by the blue curve, while $\left| {G_{i\bar j} \dot z^i \dot z^{\bar j}} \right|$ represented by the red curve.}
    \label{1-4}
  \end{subfigure}
\qquad 
  \begin{subfigure}[t]{.5\linewidth}
    \centering
    \includegraphics[width=0.7\columnwidth]{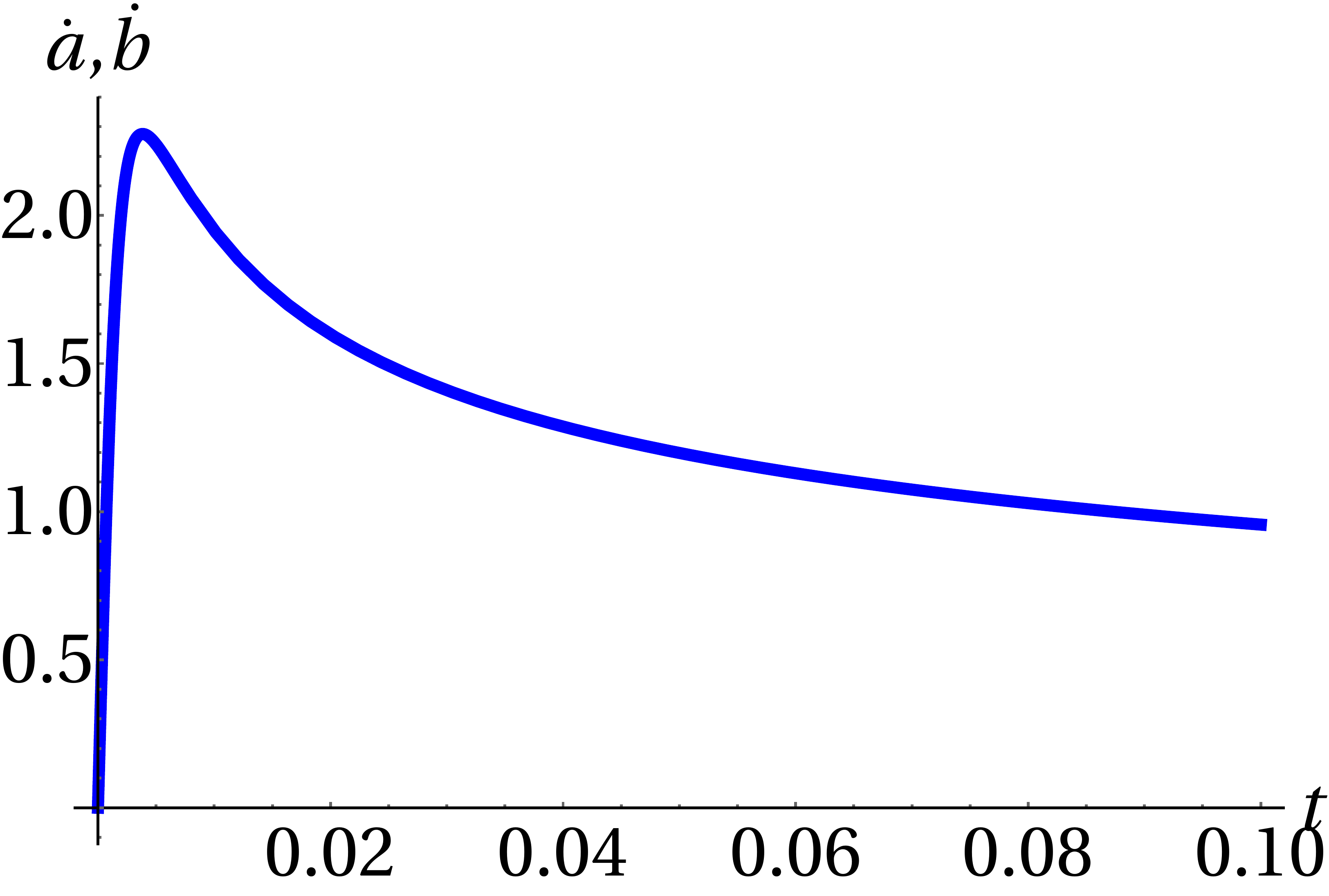}
     \caption{The expansion rates of the scale factors. Both $\dot a$ and $\dot b$ are represented by the shown curve.}
    \label{2}
  \end{subfigure}
\\[9em]
\begin{subfigure}[t]{.5\linewidth}
    \centering
    \includegraphics[width=0.7\columnwidth]{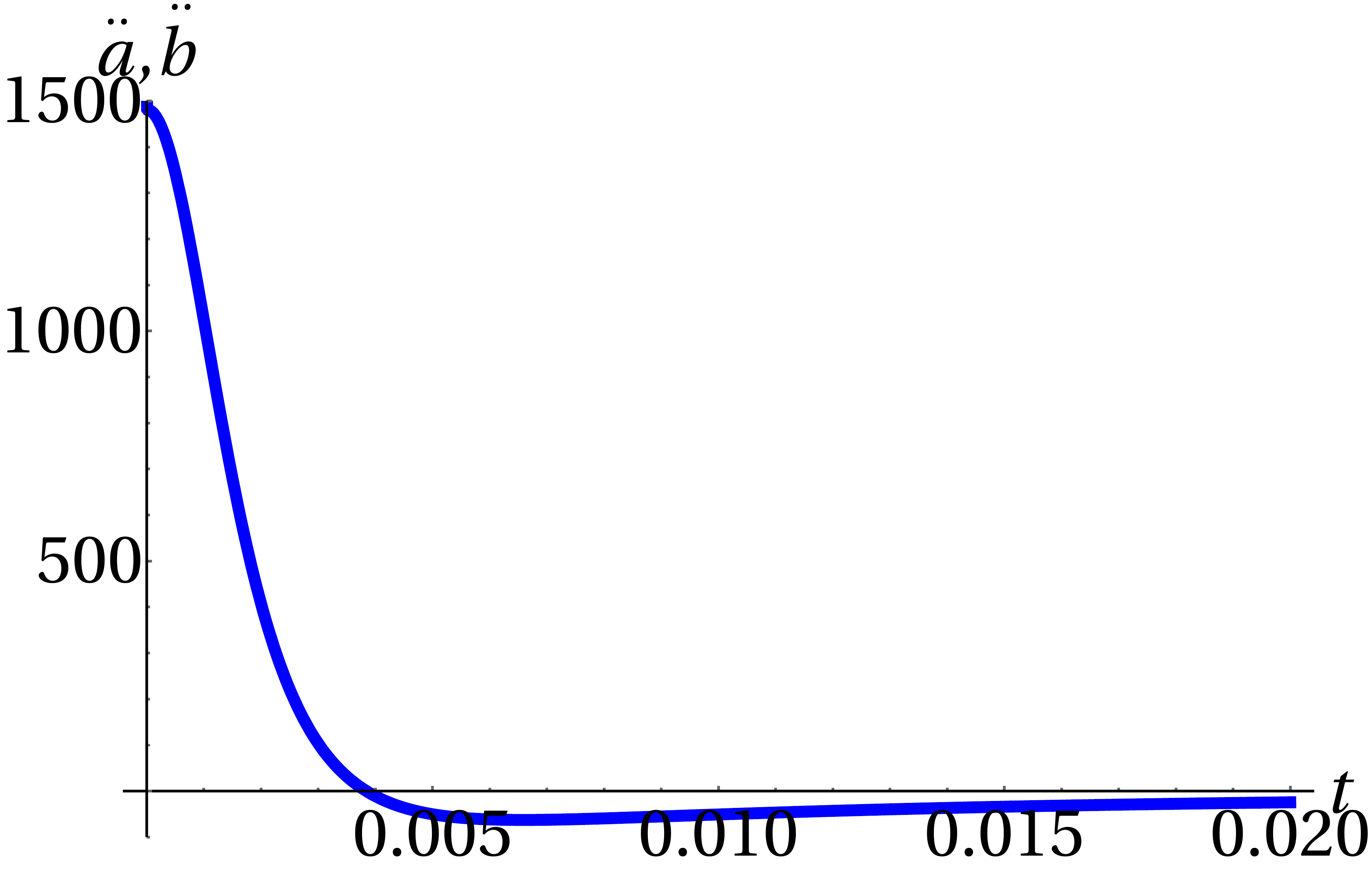}
    \caption{The accelerations of the scale factors. Both $\ddot a$ and $\ddot b$ are represented by the shown curve.}
    \label{3}
  \end{subfigure}
\qquad
  \begin{subfigure}[t]{.5\linewidth}
    \centering
    \includegraphics[width=0.7\columnwidth]{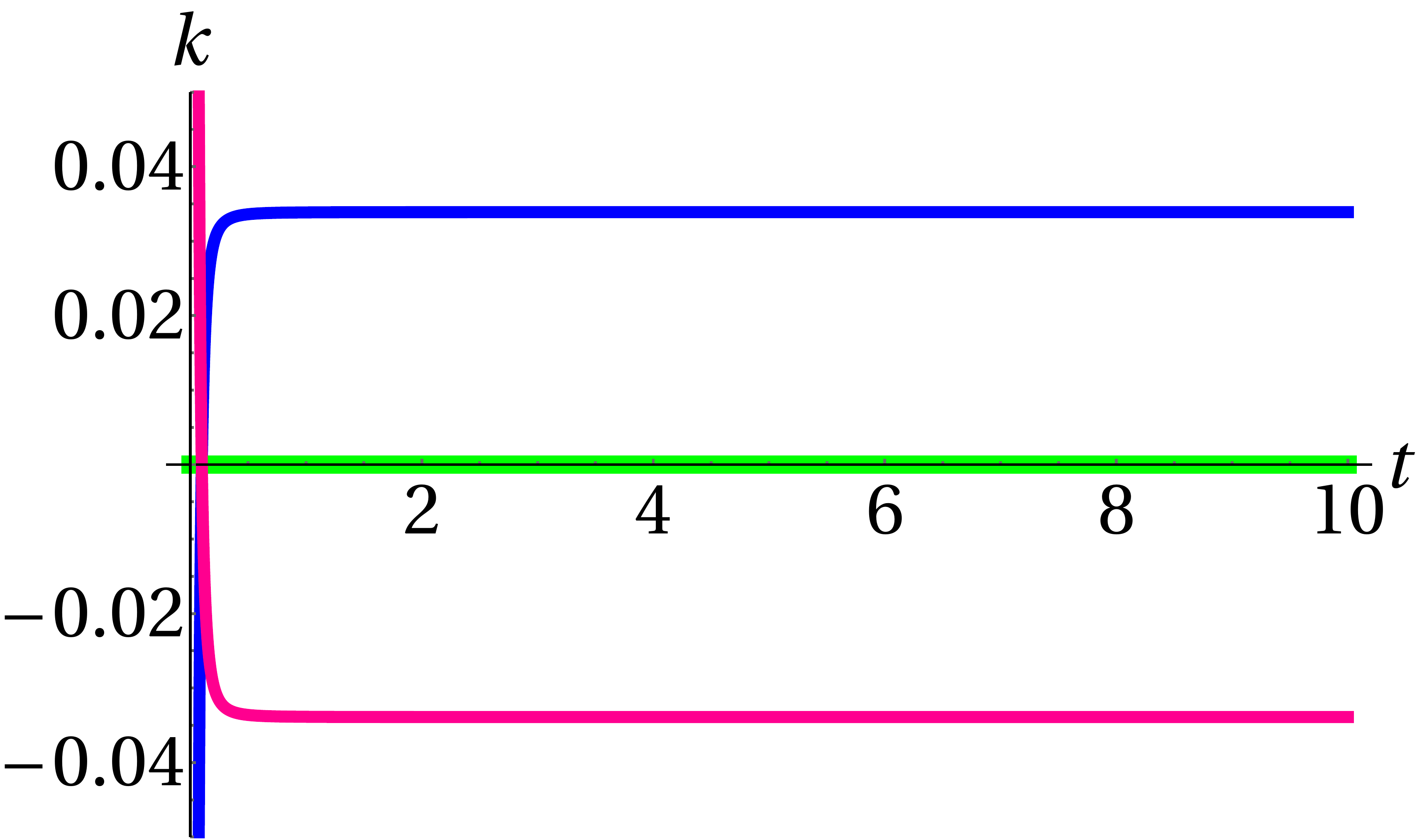}
    \caption{The harmonic function $k$ using: $\dot k\left(0\right)=1$ (blue curve), $\dot k\left(0\right)=0$ (green line), and $\dot k\left(0\right)=-1$ (red curve).}
    \label{4}
  \end{subfigure}
   \caption{Dust-filled brane world with initial conditions set number 1 .}
\label{Fig1}
 \end{figure}

\begin{figure}[H]  
\begin{subfigure}[t]{.5\linewidth}
    \centering
    \includegraphics[width=0.7\columnwidth]{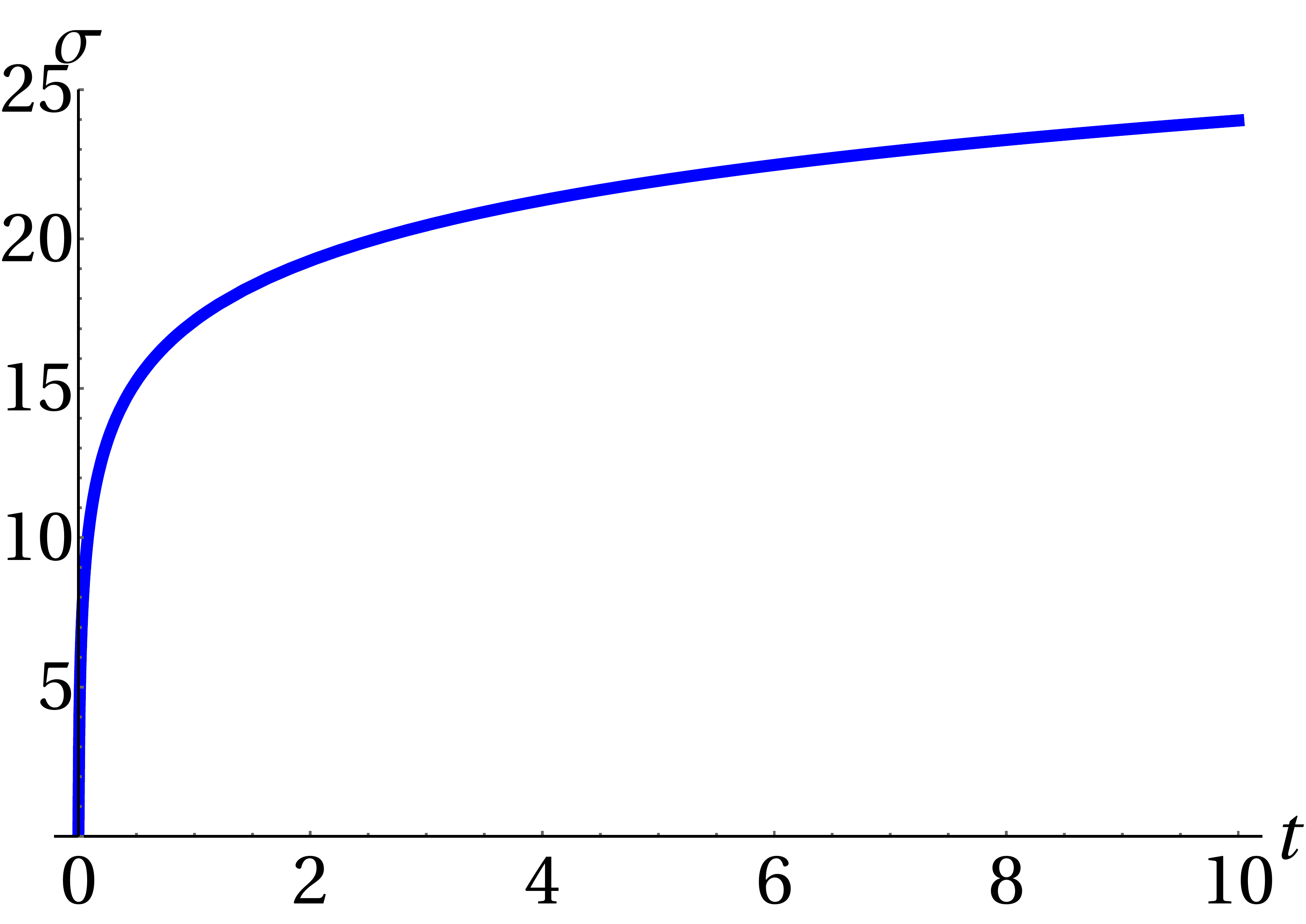}
    \caption{The dilaton $\sigma$; same for all three $\dot k\left(0\right)$.}
    \label{5}
  \end{subfigure}
\qquad%
 \begin{subfigure}[t]{.5\linewidth}
    \centering
    \includegraphics[width=0.7\columnwidth]{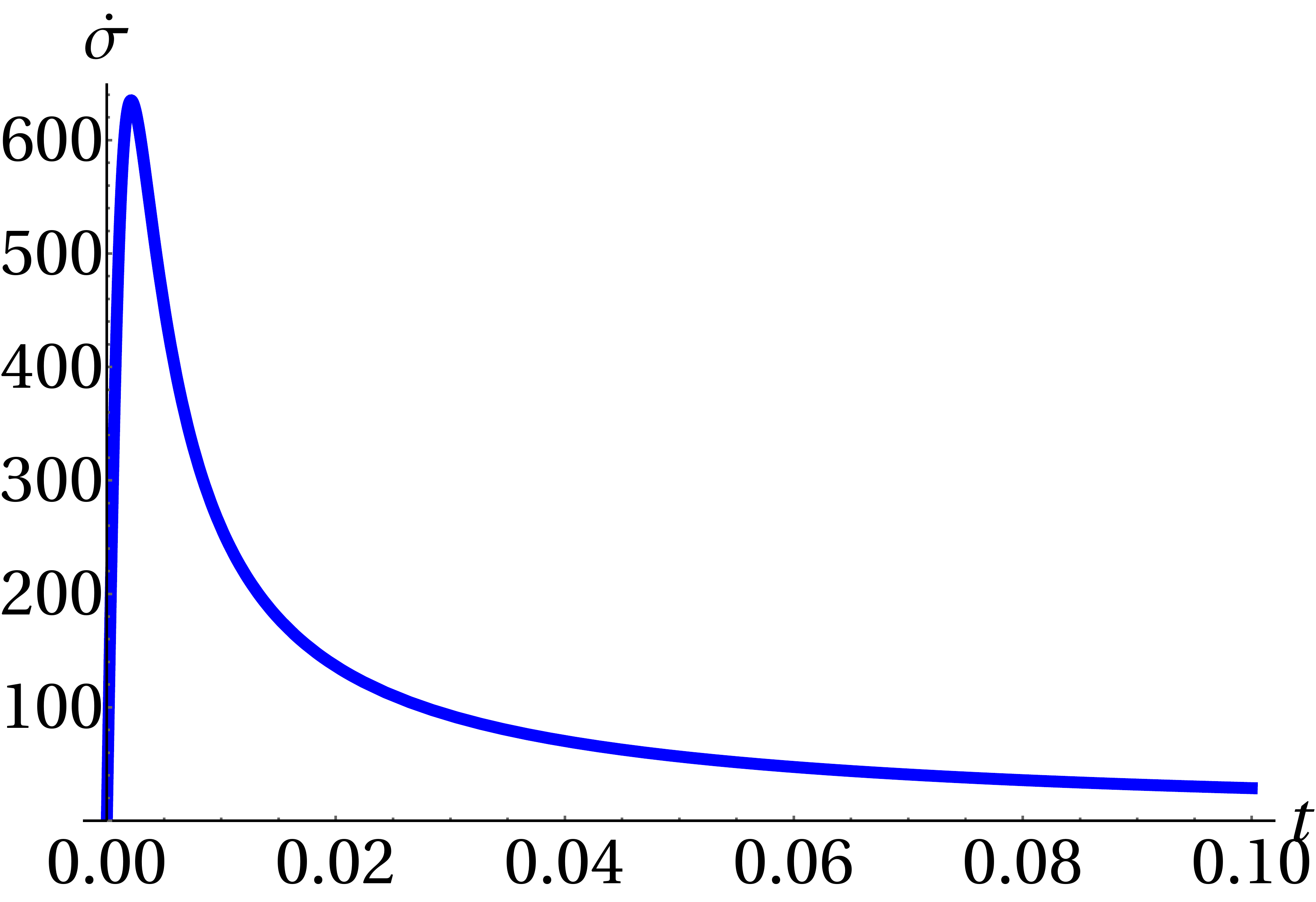}
    \caption{The dilatonic field strength $\dot\sigma$.}
    \label{6}
  \end{subfigure}
\qquad
\\[9em]  
\begin{subfigure}[t]{.5\linewidth}
    \centering
    \includegraphics[width=0.7\columnwidth]{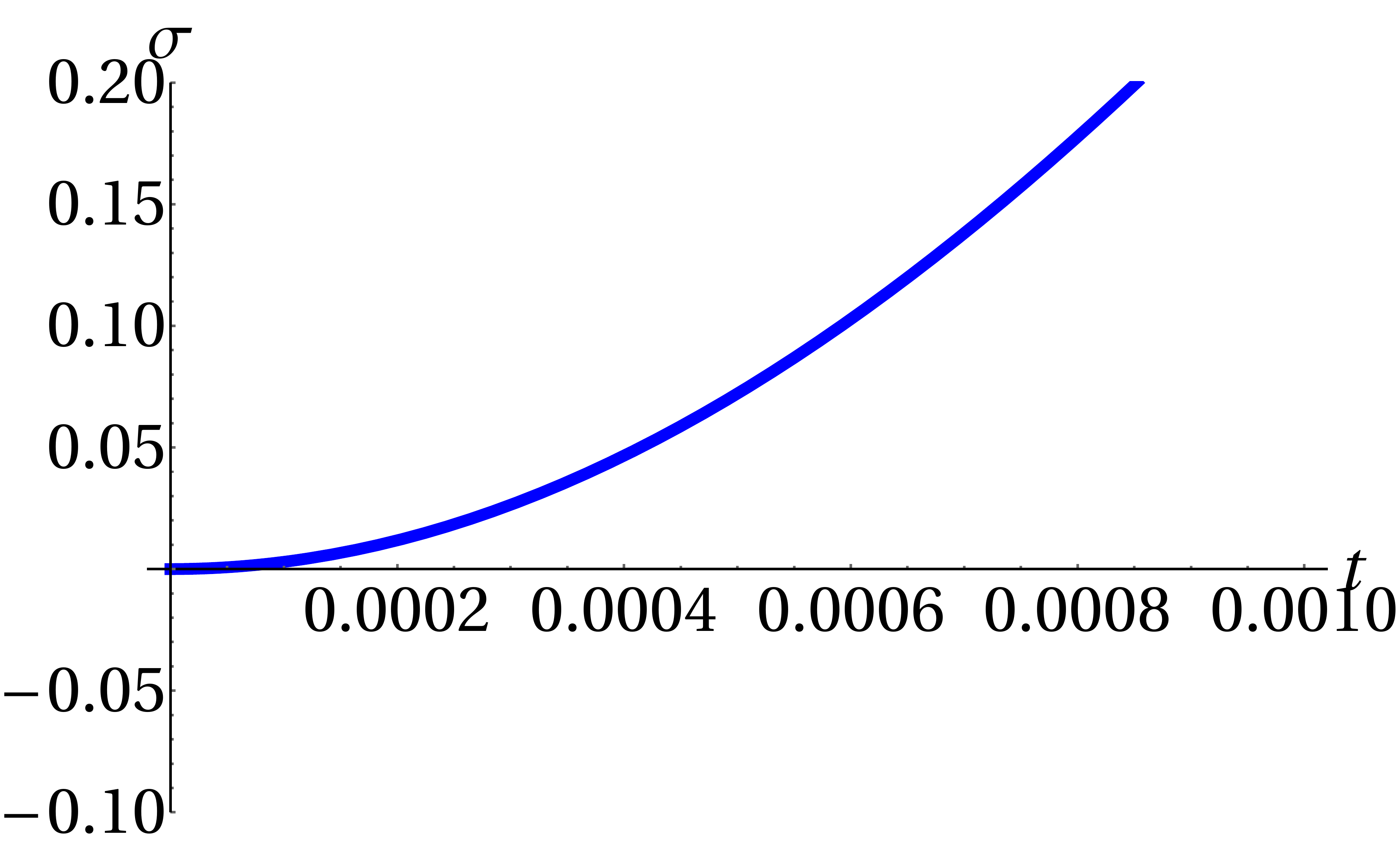}
    \caption{A zoom-in on the dilaton for $\dot k\left(0\right)=0$.}
    \label{7}
  \end{subfigure}
\qquad 
 \begin{subfigure}[t]{.5\linewidth}
    \centering
    \includegraphics[width=0.7\columnwidth]{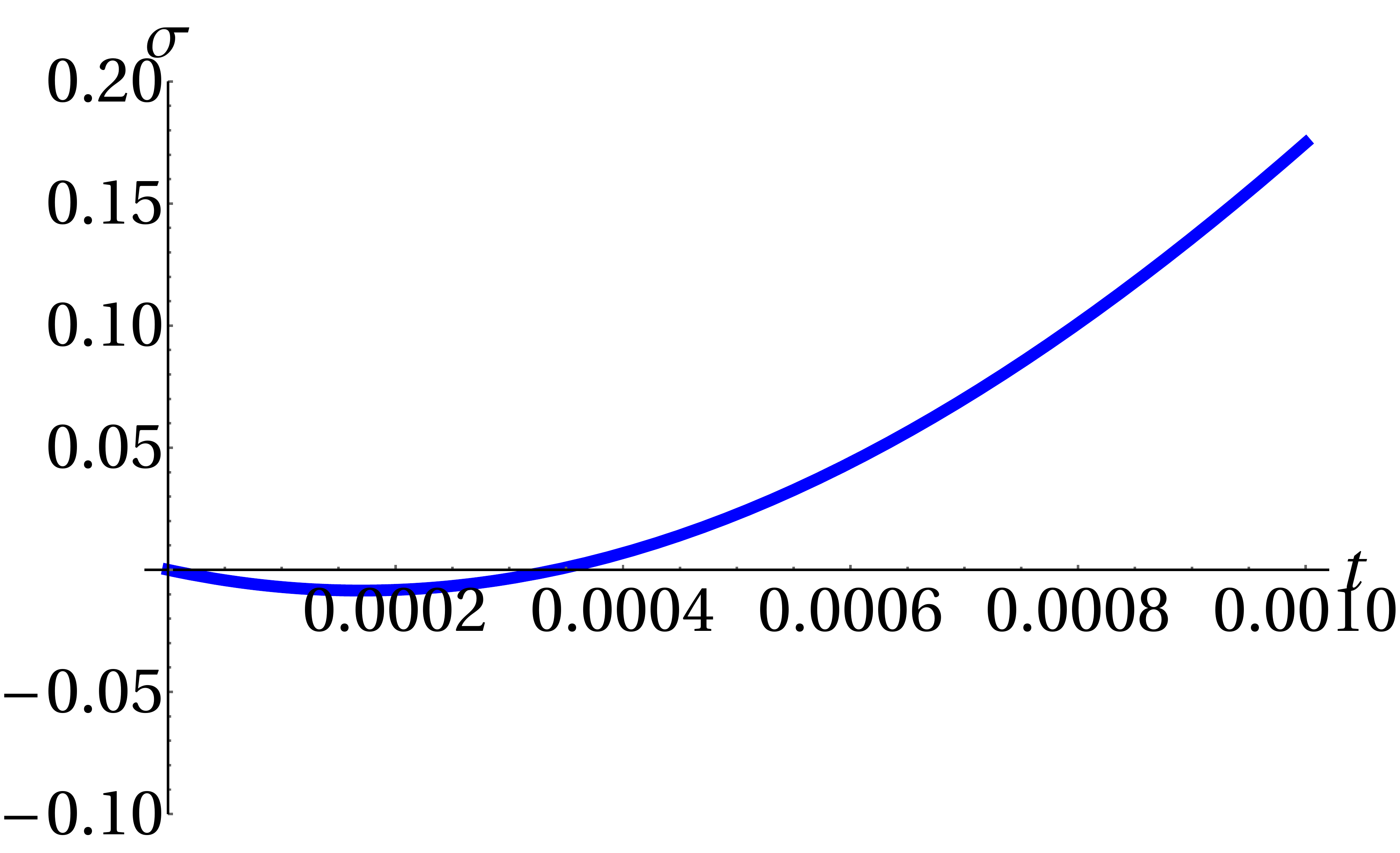}
    \caption{A zoom-in on the dilaton for $\dot k\left(0\right)=-1$.}
    \label{8}
  \end{subfigure}
  \caption{Dust-filled brane world with initial conditions set number 1 (continued).}
  \label{Fig2}
\end{figure}
%
\begin{figure}[H]
  \begin{subfigure}[t]{.5\linewidth}
    \centering
    \includegraphics[width=0.7\columnwidth]{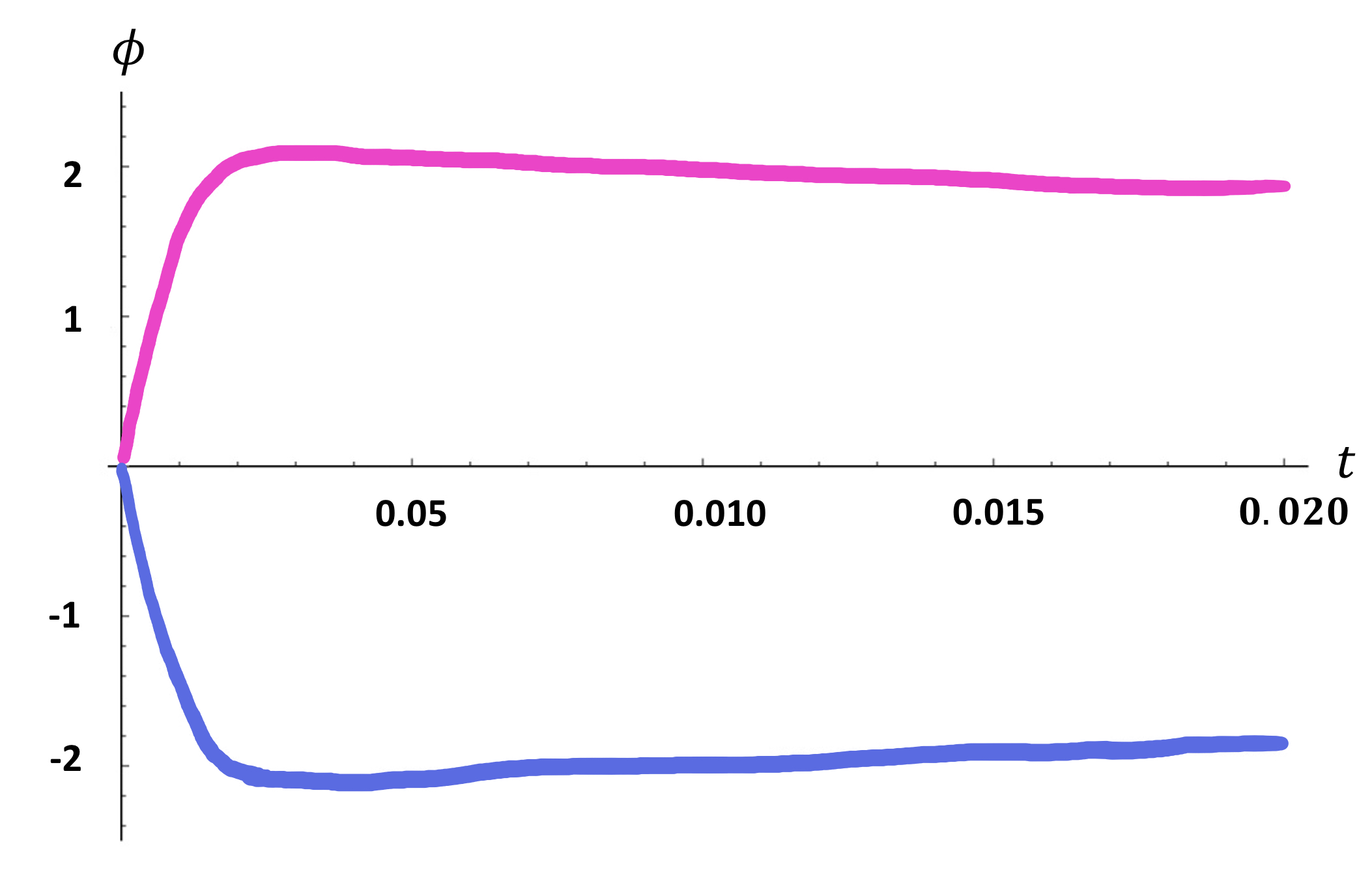}
    \caption{The universal axion $\phi$ for $\dot k\left(0\right) = 1$ (blue curve), and $\dot k\left(0\right) = -1$ (red curve). The solution diverges for $\dot k\left(0\right)=0$.}
    \label{9}
  \end{subfigure}
\qquad
  \begin{subfigure}[t]{.5\linewidth}
    \centering
    \includegraphics[width=0.7\columnwidth]{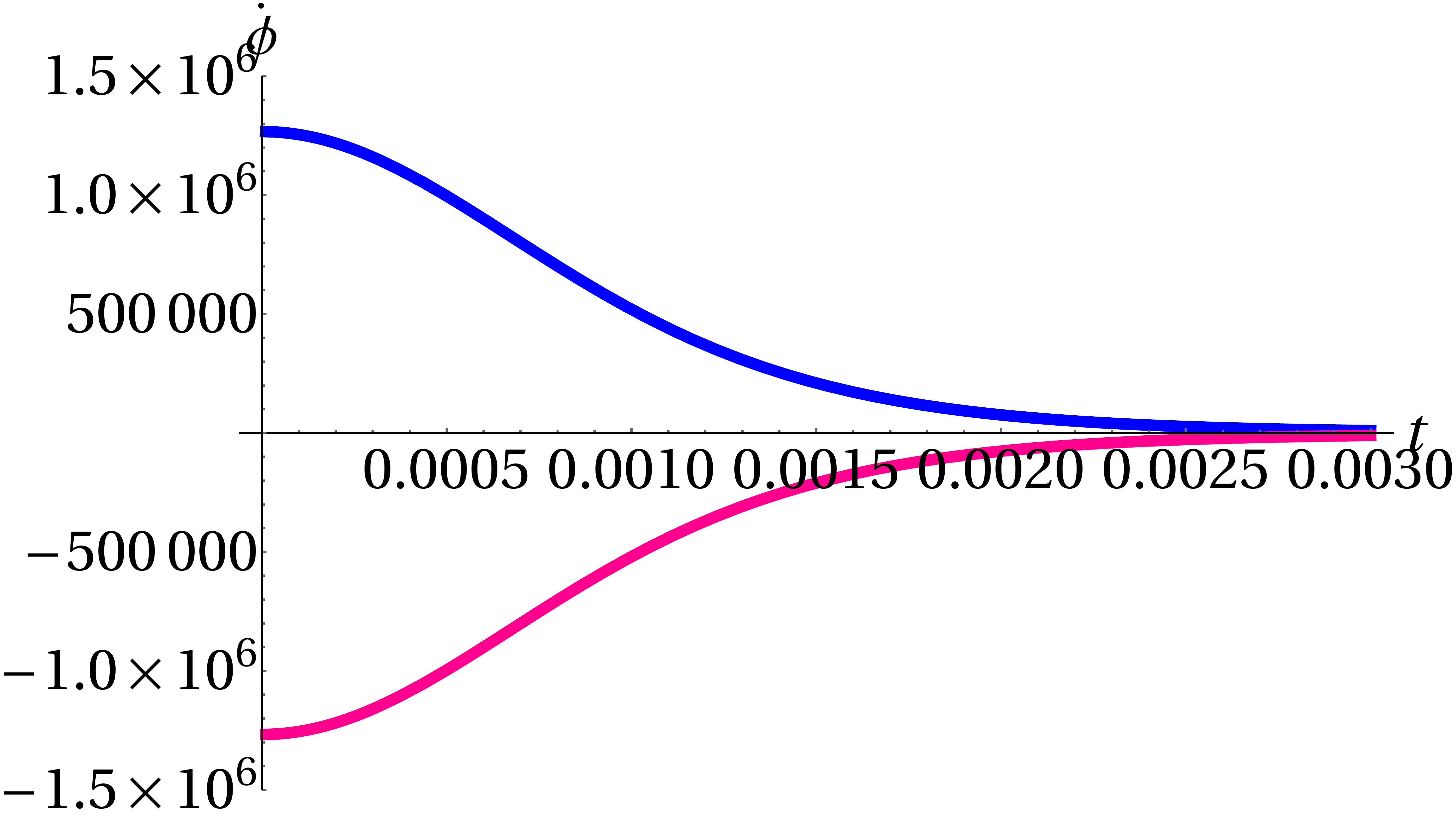}
    \caption{The axionic field strength $\dot\phi$ for $\dot k\left(0\right) = 1$ (blue curve), and $\dot k\left(0\right) = -1$ (red curve).}
    \label{10}
  \end{subfigure}
\qquad 
\\[9em] 
\begin{subfigure}[t]{.5\linewidth}
    \centering
    \includegraphics[width=0.7\columnwidth]{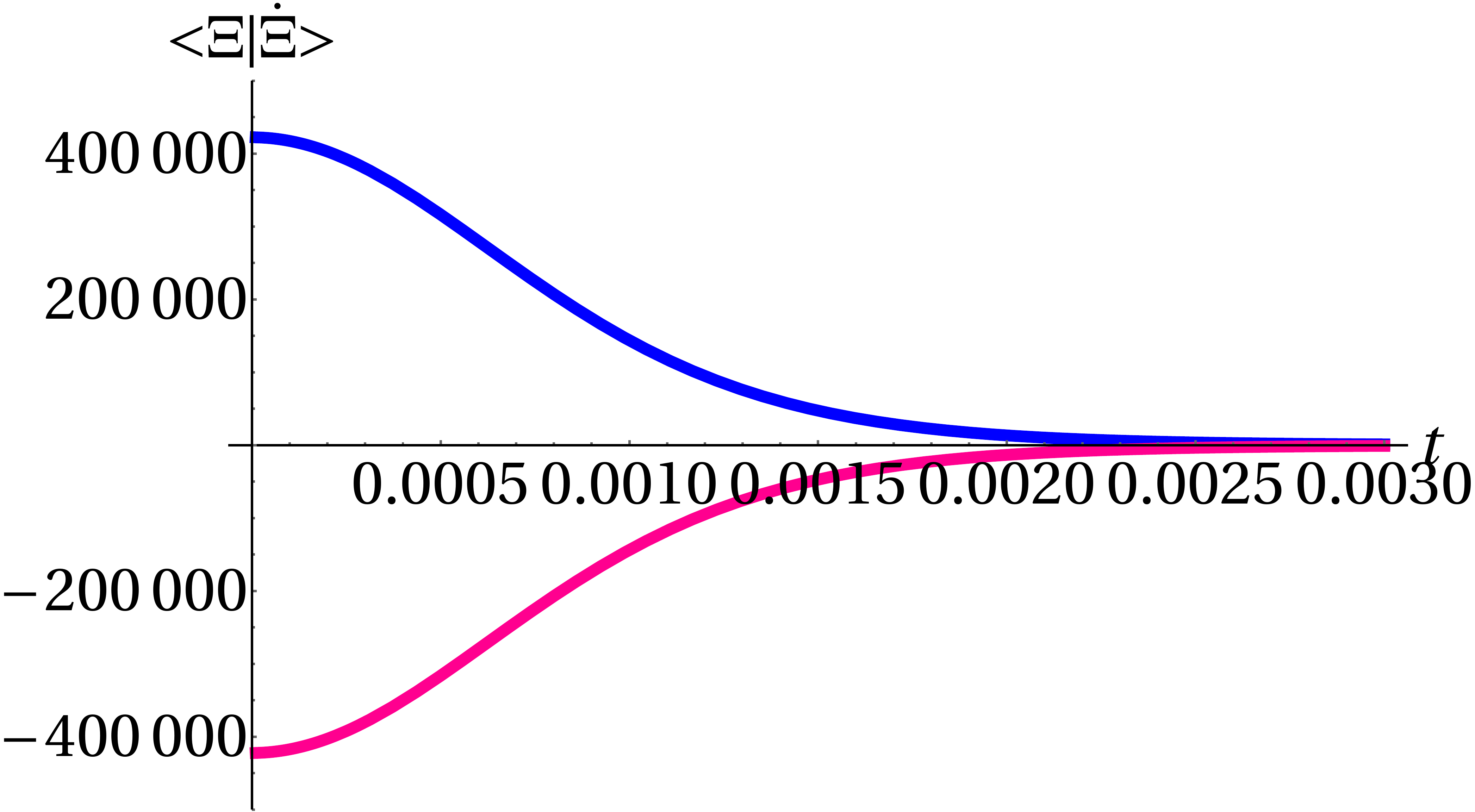}
    \caption{$\langle \Xi | \dot{\Xi} \rangle$ for $\dot{k}(0)= 1$  (blue), and $\dot{k}(0)  = -1$ (red).}
    \label{11}
  \end{subfigure}
\qquad
  \begin{subfigure}[t]{.5\linewidth}
    \centering
    \includegraphics[width=0.7\columnwidth]{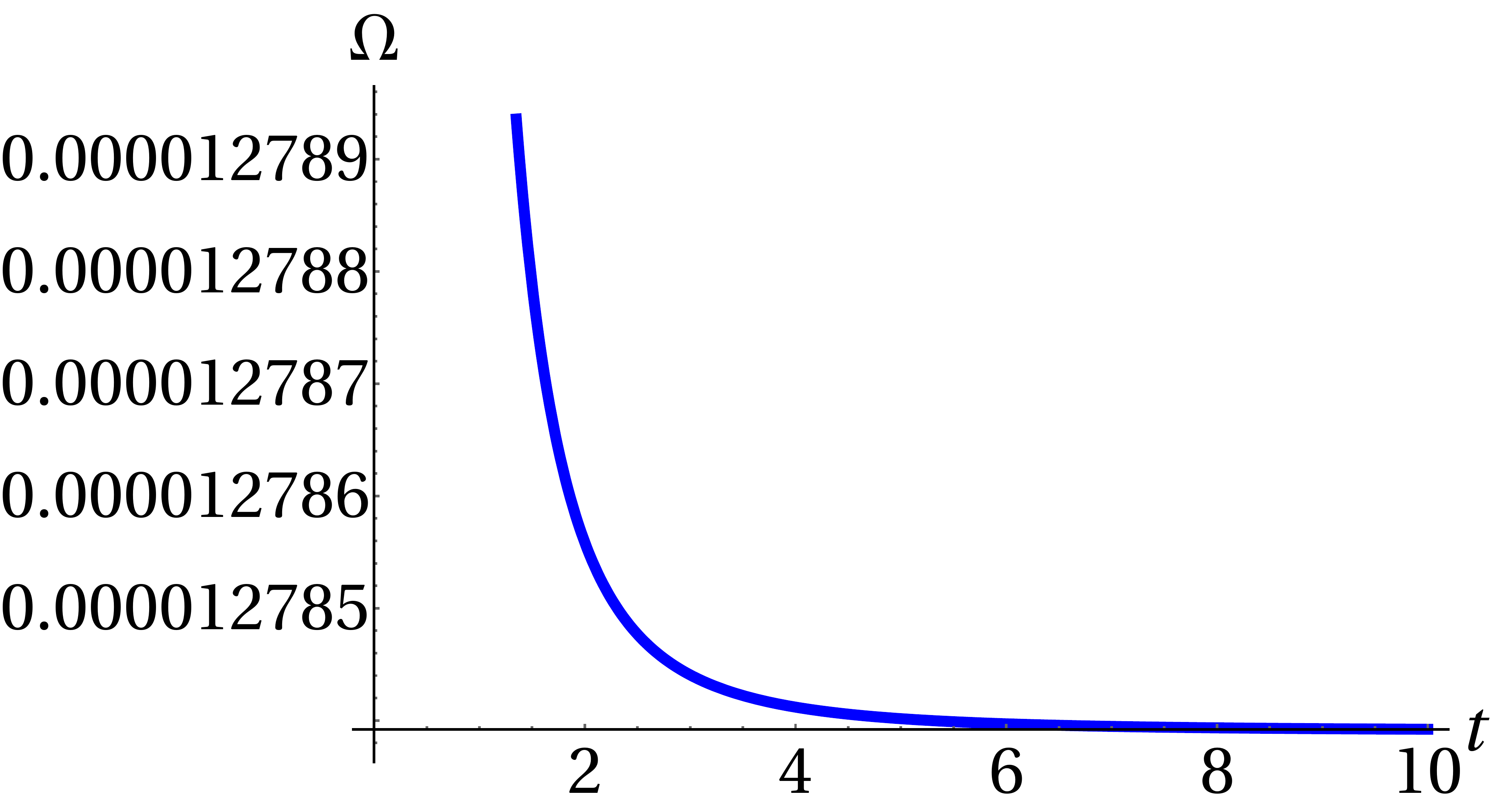}
    \caption{$\Omega$ at $\dot k\left(0\right) = 1$ and  $ \dot\sigma \left(0\right) =0 $.}
    \label{12}
  \end{subfigure}%
  \caption{Dust-filled brane world with initial conditions set number 1 (continued).}
  \label{Fig22}
\end{figure}


\begin{figure}[H]
  \begin{subfigure}[t]{.5\linewidth}
    \centering
    \includegraphics[width=0.7\columnwidth]{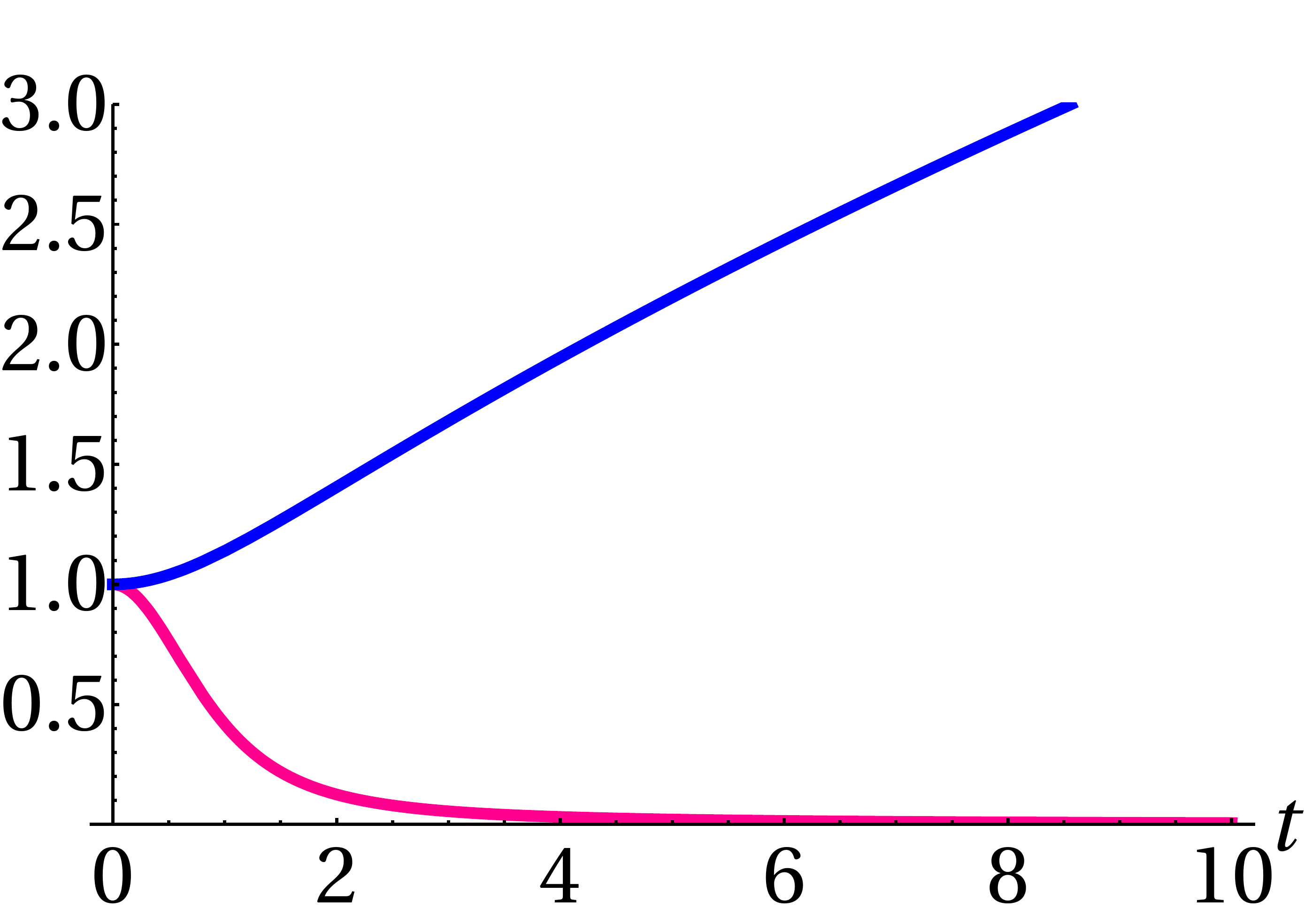}
    \caption{The scale factors $a$ and $b$; represented by the blue curve, while $\left| {G_{i\bar j} \dot z^i \dot z^{\bar j}} \right|$ represented by the red curve.}
    \label{13}
  \end{subfigure}
\qquad
  \begin{subfigure}[t]{.5\linewidth}
    \centering
    \includegraphics[width=0.7\columnwidth]{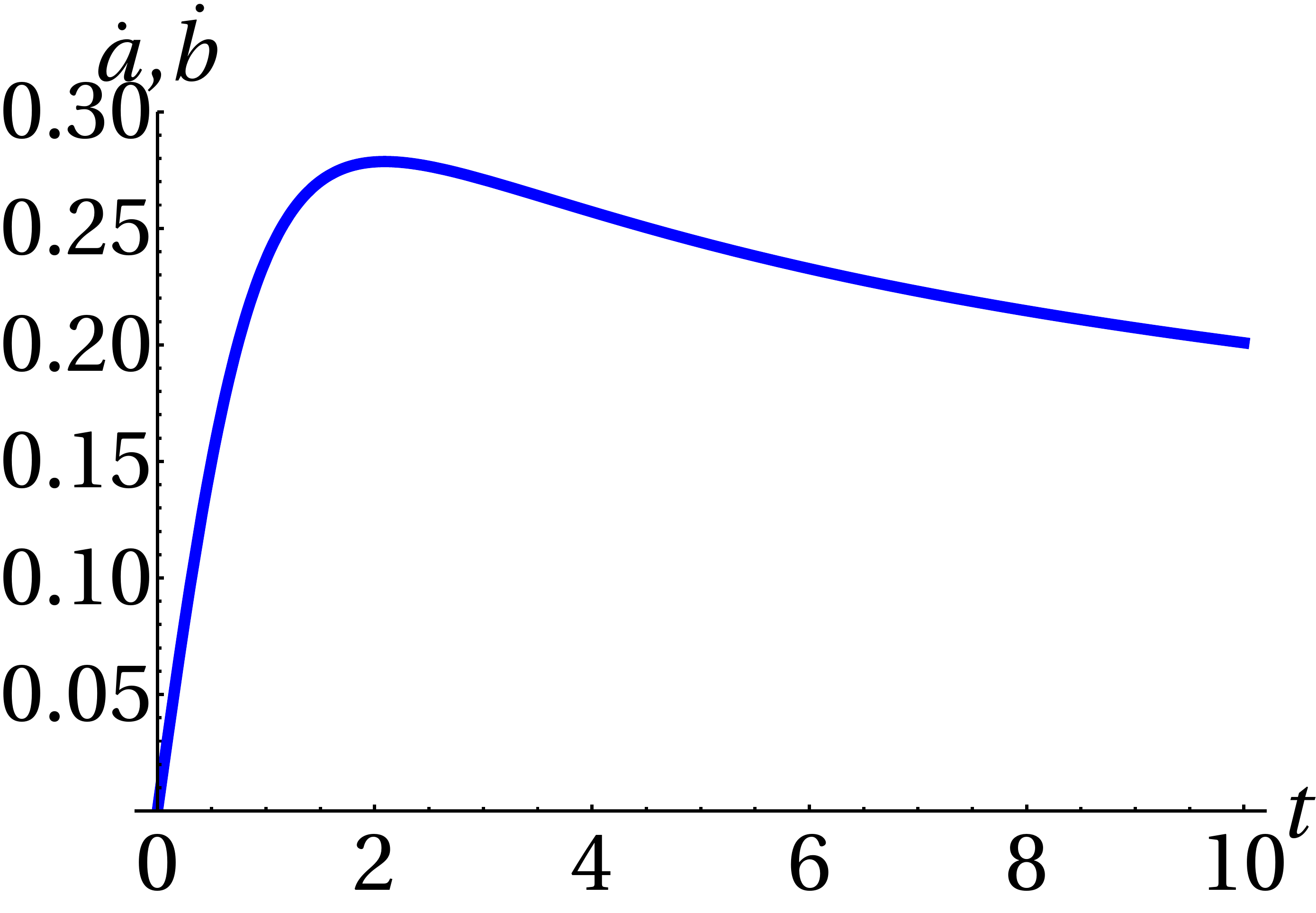}
    \caption{The expansion rates of the scale factors. Both $\dot a$ and $\dot b$ are represented by the shown curve.}
    \label{14}
  \end{subfigure}
\\[9em]
\begin{subfigure}[t]{.5\linewidth}
    \centering
    \includegraphics[width=0.7\columnwidth]{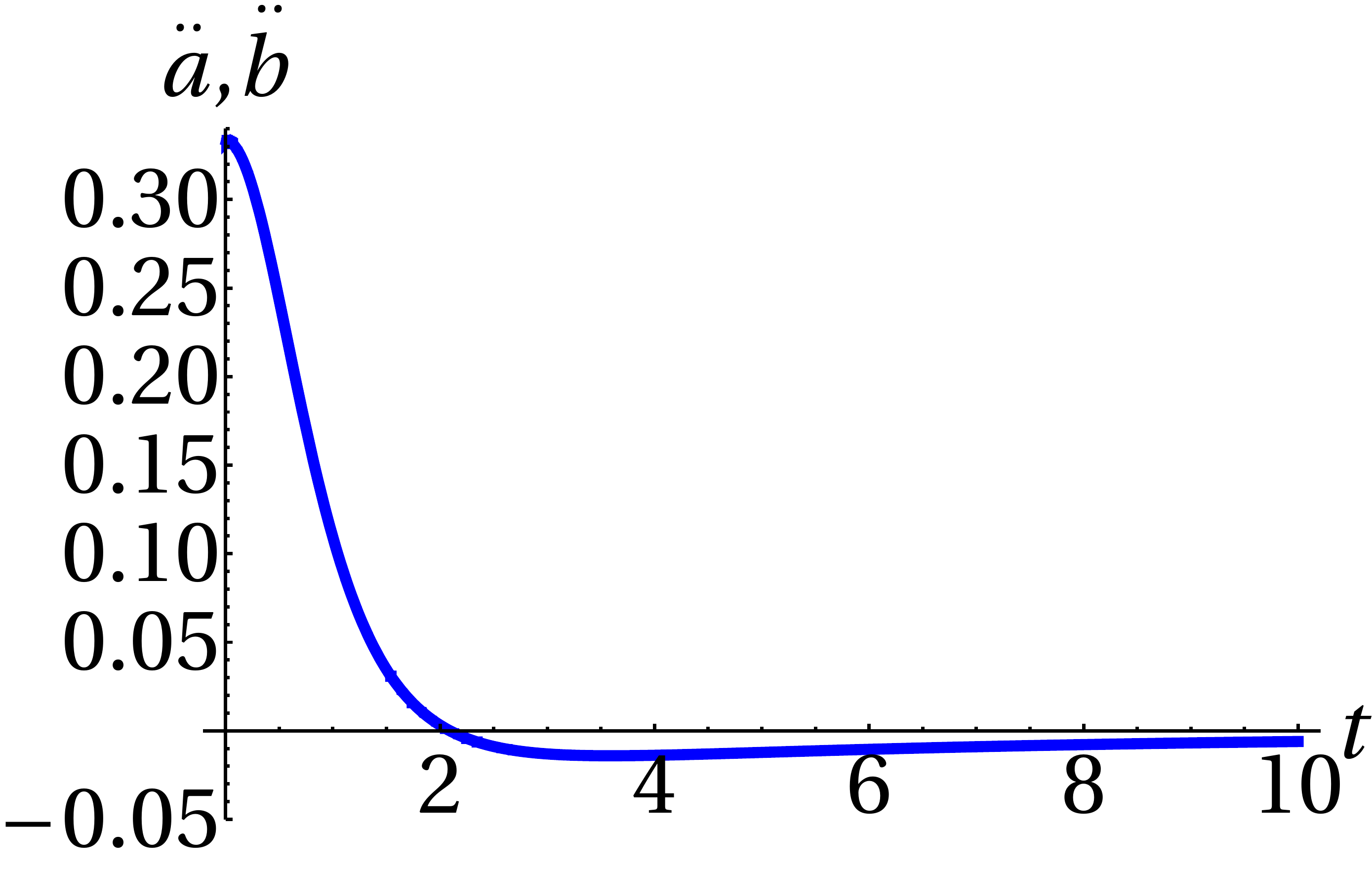}
    \caption{The accelerations of the scale factors. Both $\ddot a$ and $\ddot b$ are represented by the shown curve.}
    \label{15}
  \end{subfigure}
\qquad
  \begin{subfigure}[t]{.5\linewidth}
    \centering
    \includegraphics[width=0.7\columnwidth]{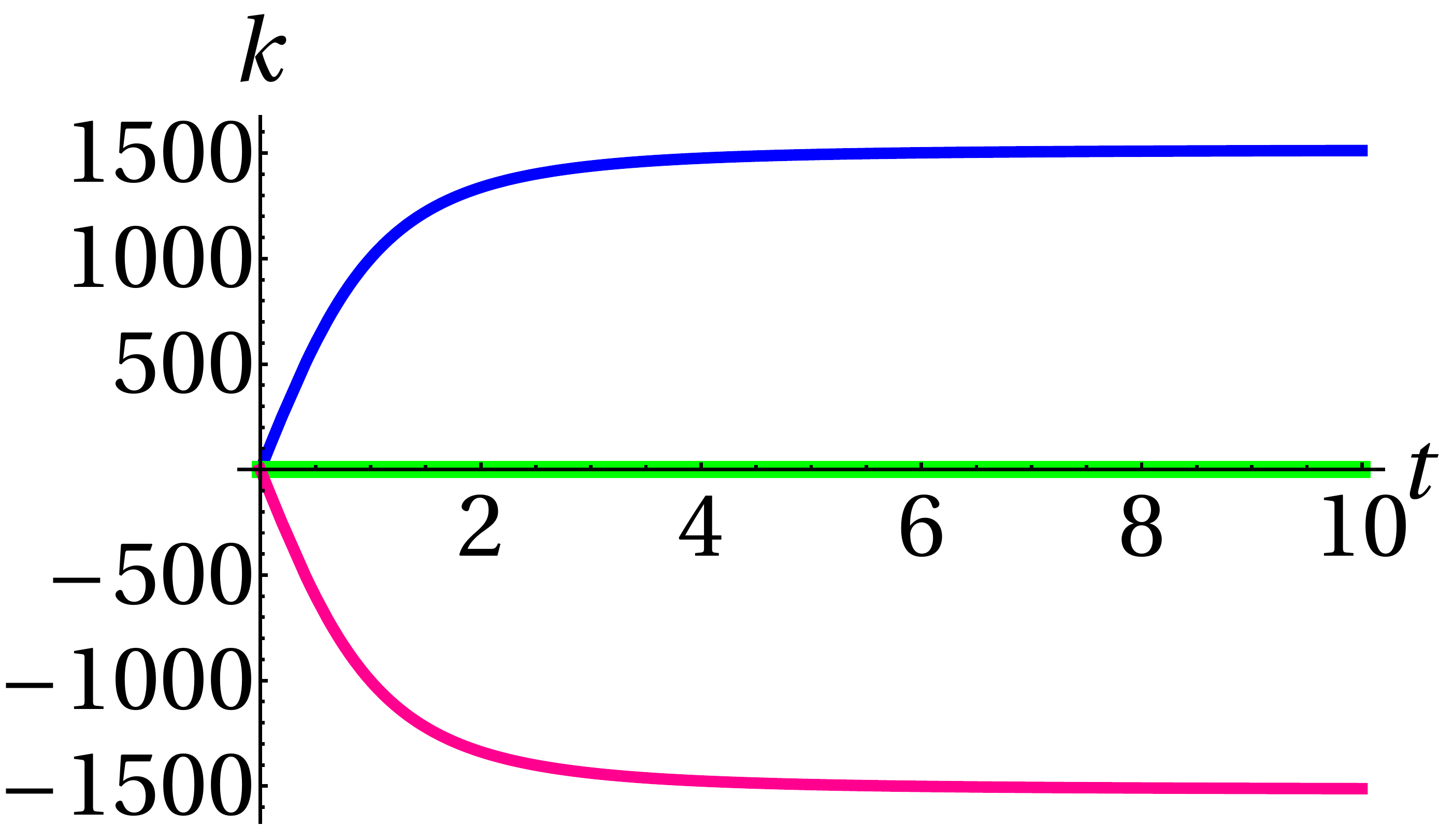}
    \caption{The harmonic function $k$ using: $\dot k\left(0\right)=1$ (blue curve), $\dot k\left(0\right)=0$ (green line), and $\dot k\left(0\right)=-1$ (red curve).}
    \label{16}
  \end{subfigure}
  \caption{Dust-filled brane world with initial conditions set number 2.}
  \label{Fig4}
\end{figure}

  \begin{figure}[H]
    \begin{subfigure}[t]{.5\linewidth}
    \centering
    \includegraphics[width=0.7\columnwidth]{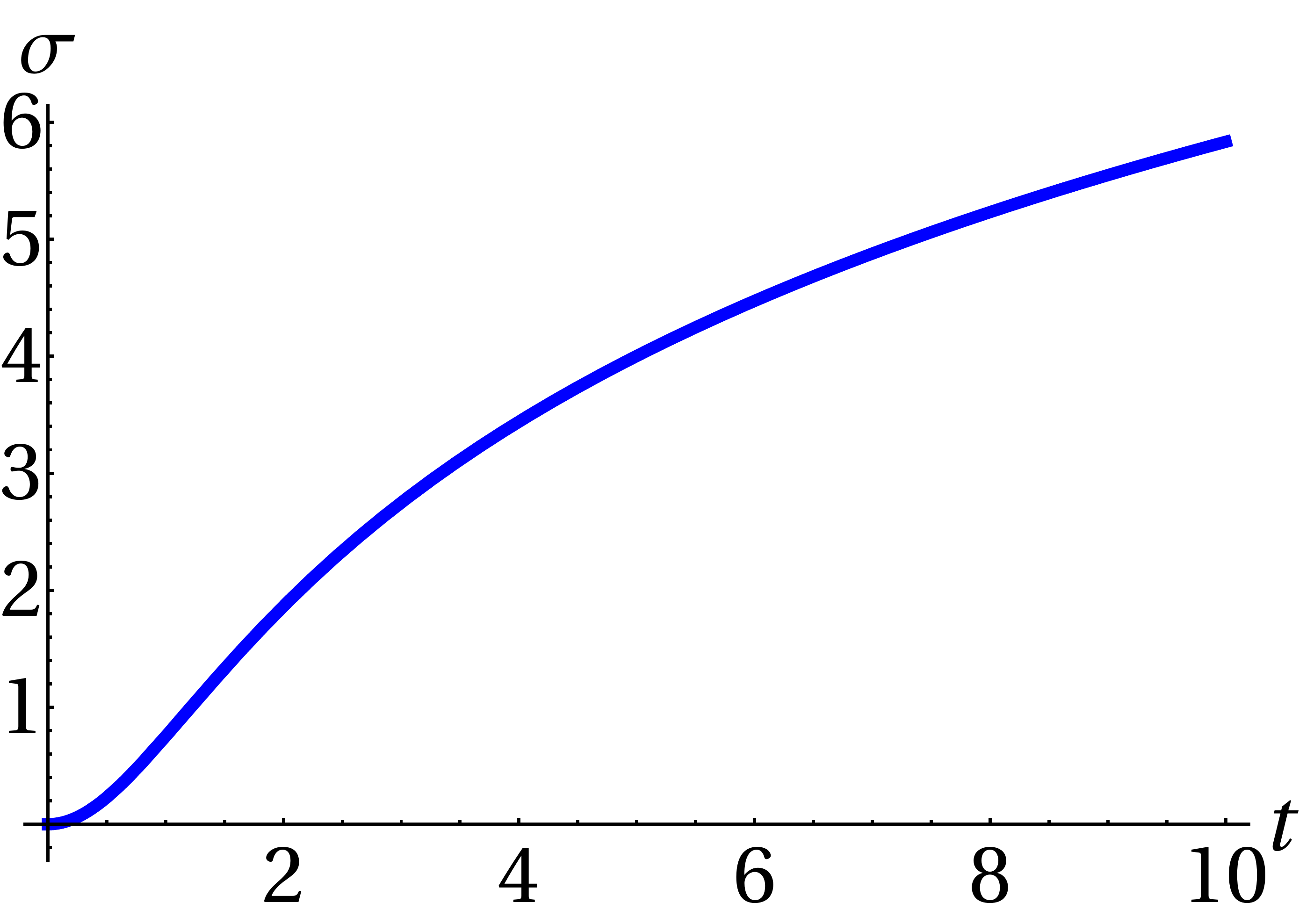}
    \caption{The dilaton $\sigma$; same for all three $\dot k\left(0\right)$.}
    \label{17}
  \end{subfigure}
\qquad
  \begin{subfigure}[t]{.5\linewidth}
    \centering
    \includegraphics[width=0.7\columnwidth]{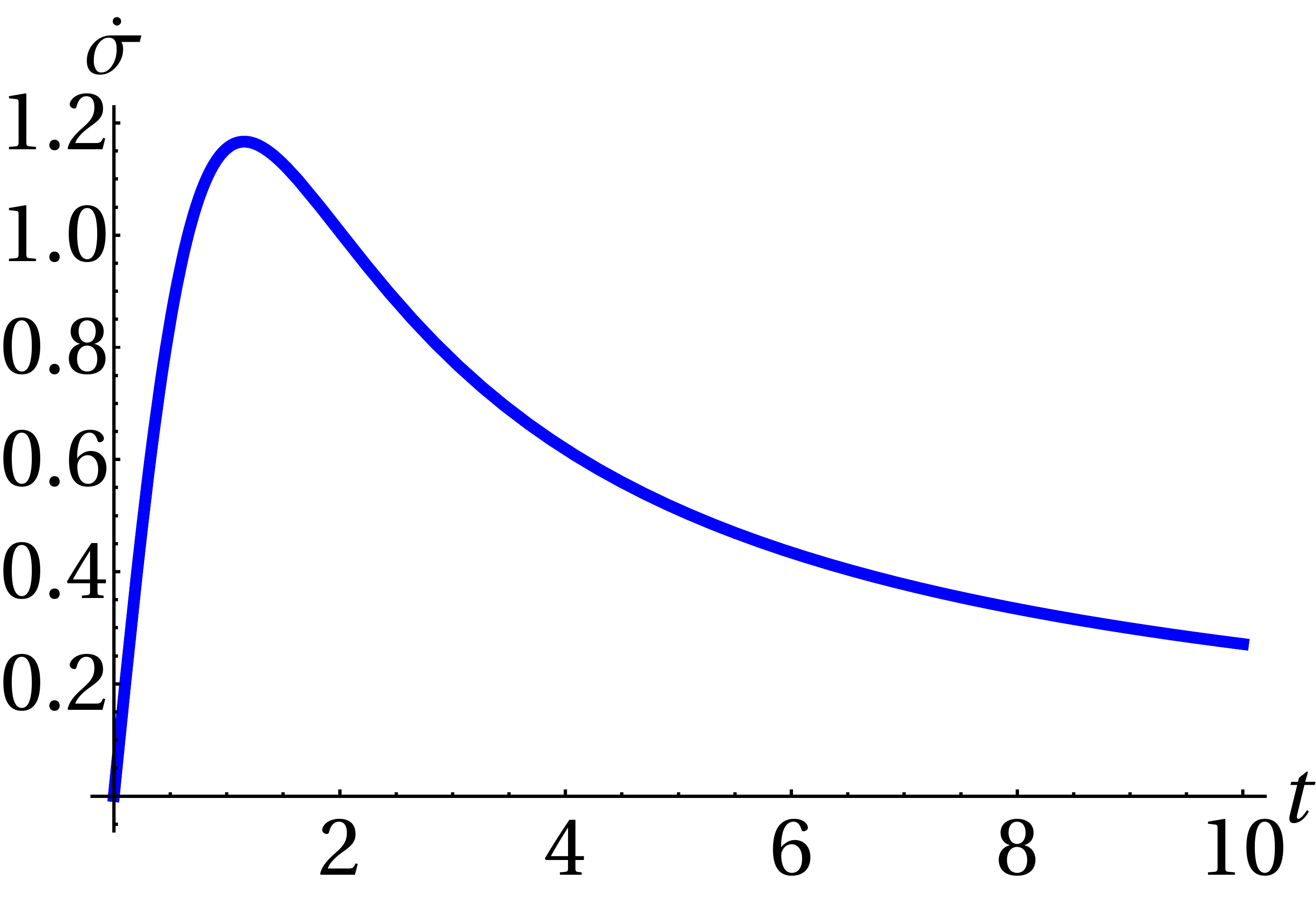}
    \caption{The dilatonic field strength $\dot\sigma$.}
    \label{18}
  \end{subfigure}  
\\[4em]
\begin{subfigure}[t]{.5\linewidth}
    \centering
    \includegraphics[width=0.7\columnwidth]{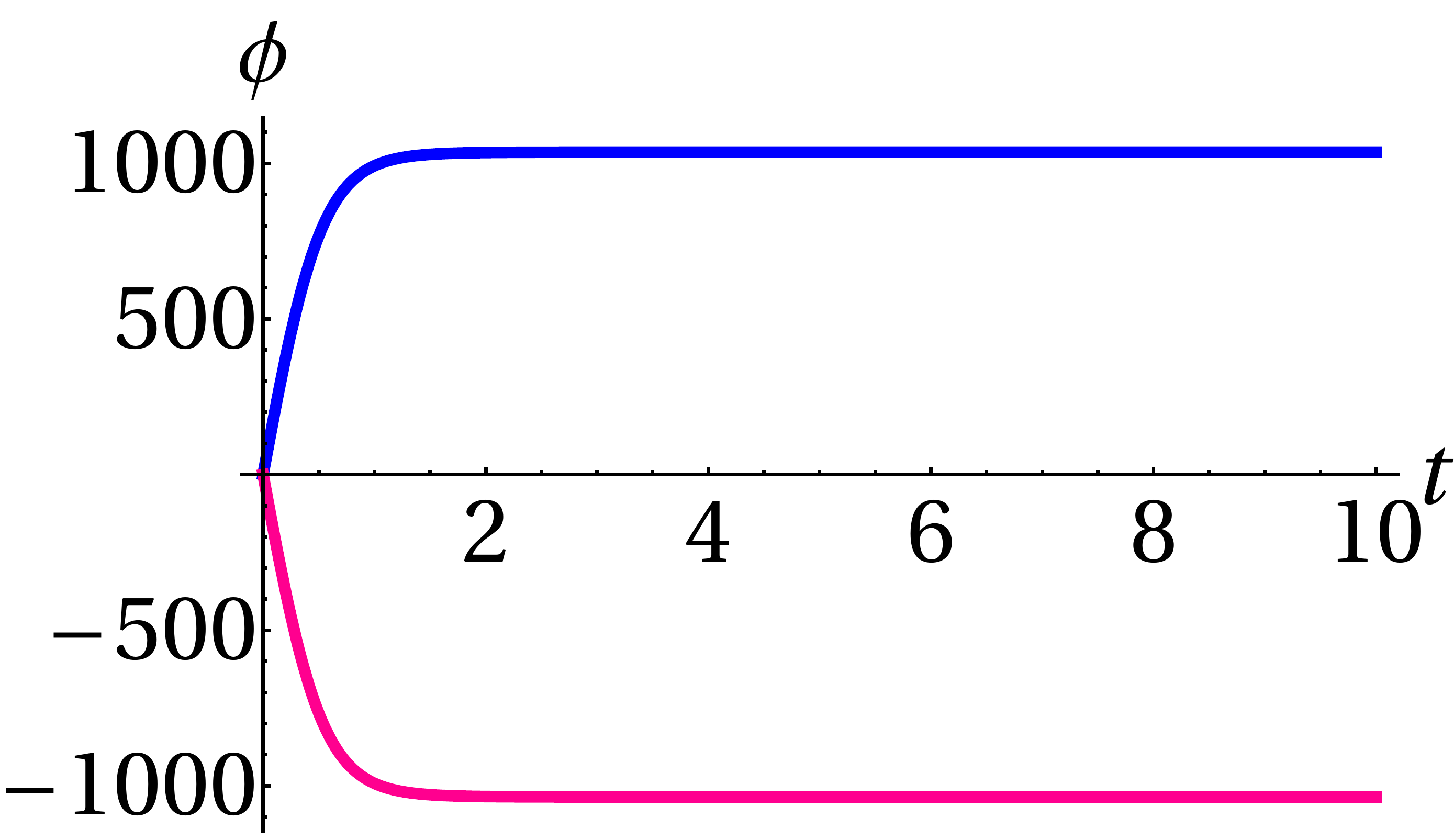}
    \caption{The universal axion $\phi$ for $\dot k\left(0\right) = 1$ (blue curve), and $\dot k\left(0\right) = -1$ (red curve). The solution diverges for $\dot k\left(0\right)=0$.}
    \label{19}
  \end{subfigure}
\qquad
  \begin{subfigure}[t]{.5\linewidth}
    \centering
    \includegraphics[width=0.7\columnwidth]{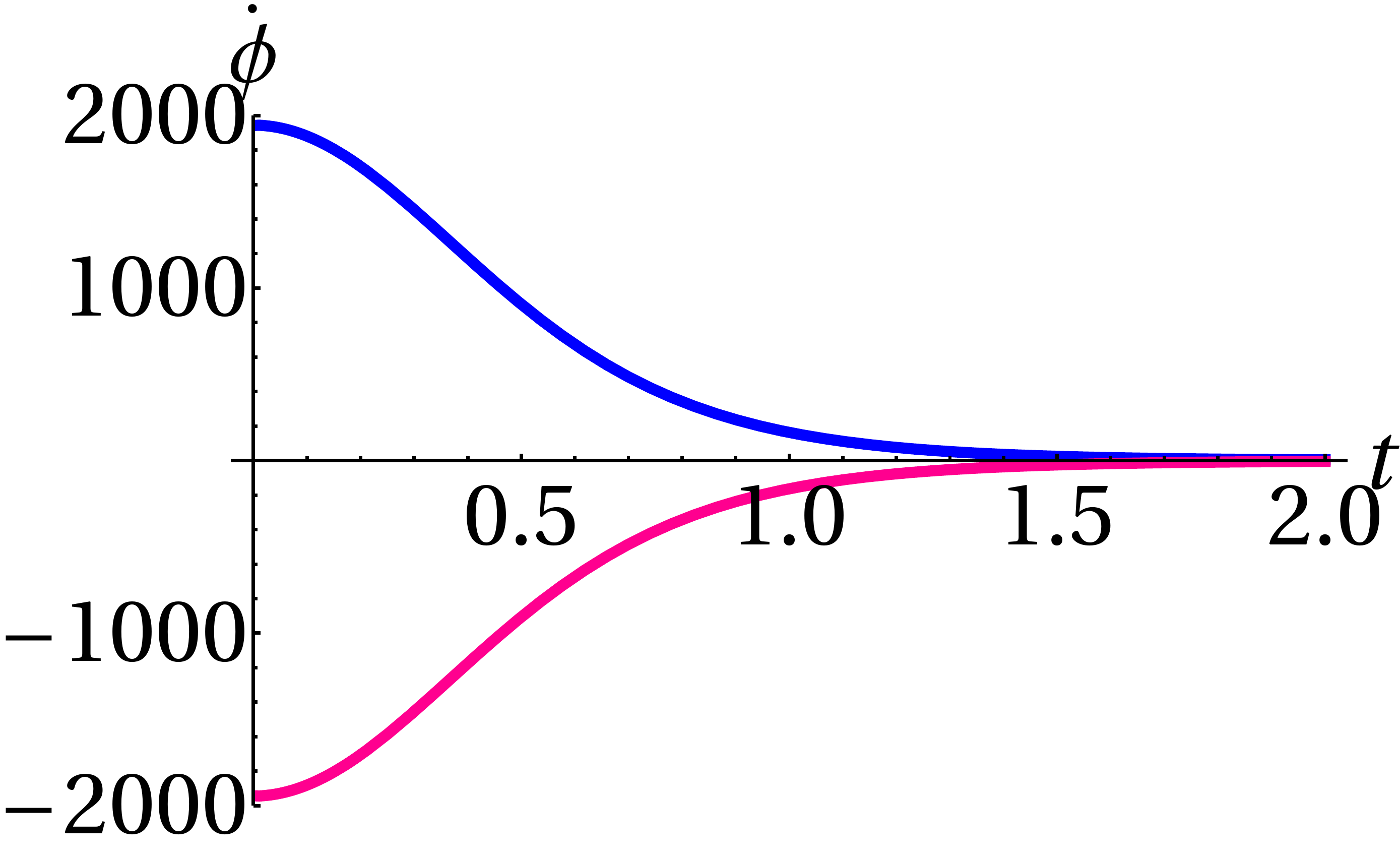}
    \caption{The axionic field strength $\dot\phi$ for $\dot k\left(0\right) = 1$ (blue curve), and $\dot k\left(0\right) = -1$ (red curve).}
    \label{20}
  \end{subfigure}
\\[4em]
  \begin{subfigure}[t]{.5\linewidth}
    \centering
    \includegraphics[width=0.7\columnwidth]{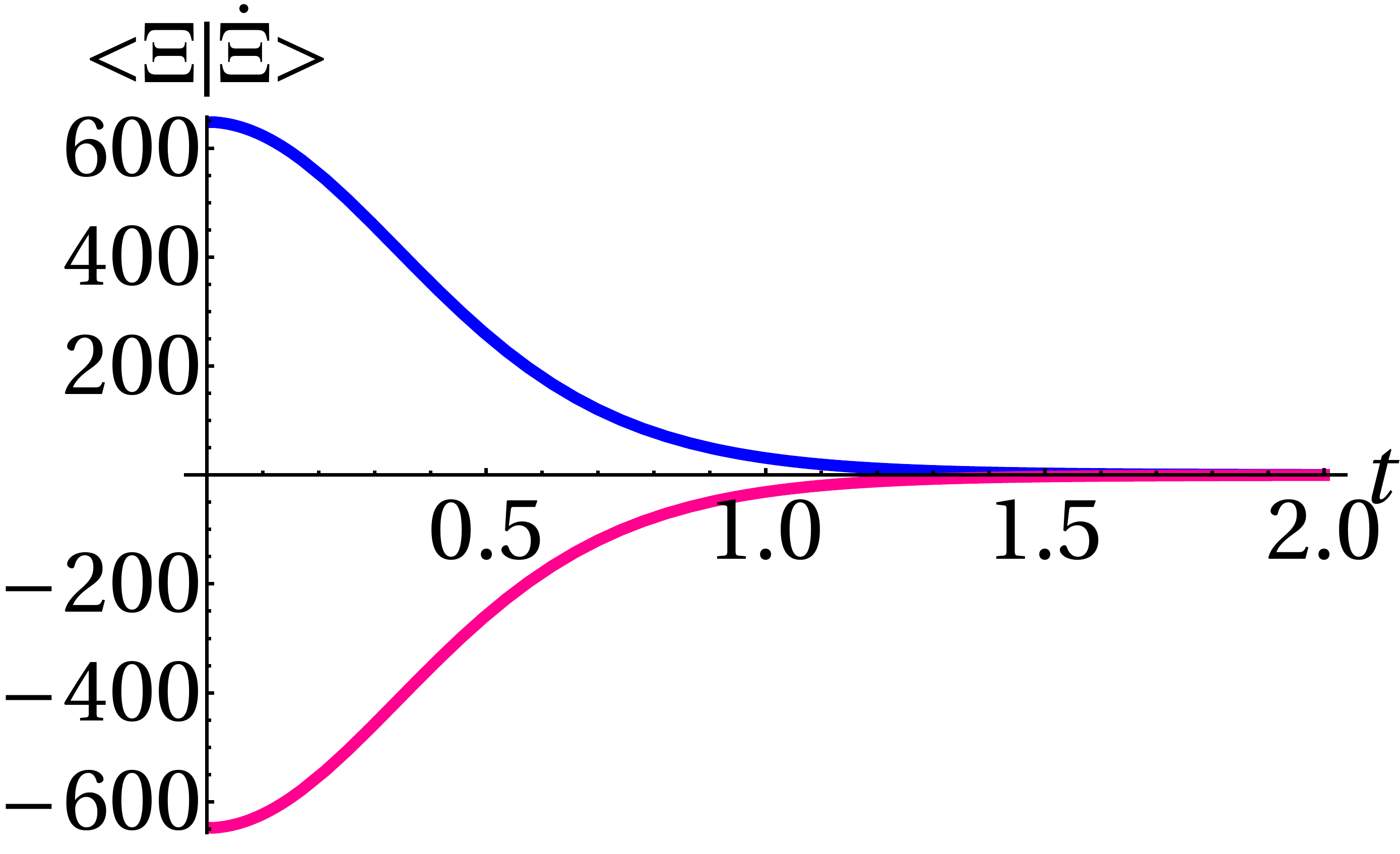}
    \caption{$\langle \Xi | \dot{\Xi} \rangle$ for $\dot{k}(0)= 1$  (blue), and $\dot{k}(0)  = -1$ (red).}
    \label{21}
  \end{subfigure}
  \qquad
\begin{subfigure}[t]{.5\linewidth}
    \centering
    \includegraphics[width=0.7\columnwidth]{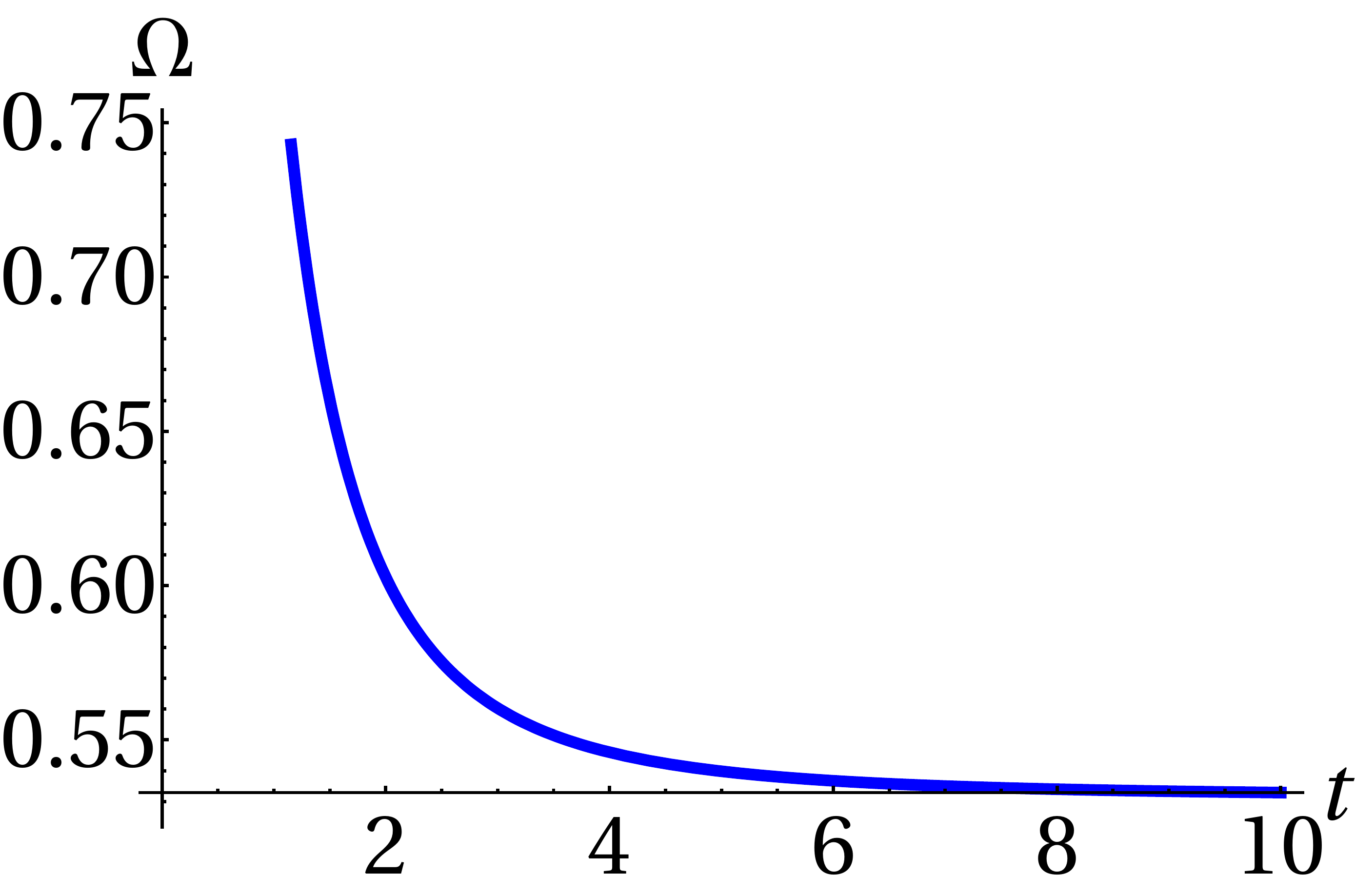}
    \caption{$\Omega$ at $\dot k\left(0\right) = 1$ and  $ \dot\sigma \left(0\right) =0 $.}
    \label{22}
  \end{subfigure}
  \caption{Dust-filled brane world with initial conditions set number 2 (continued).}
  \label{Fig44}
\end{figure}


\begin{figure}[H]
  \begin{subfigure}[t]{.5\linewidth}
    \centering
    \includegraphics[width=0.7\columnwidth]{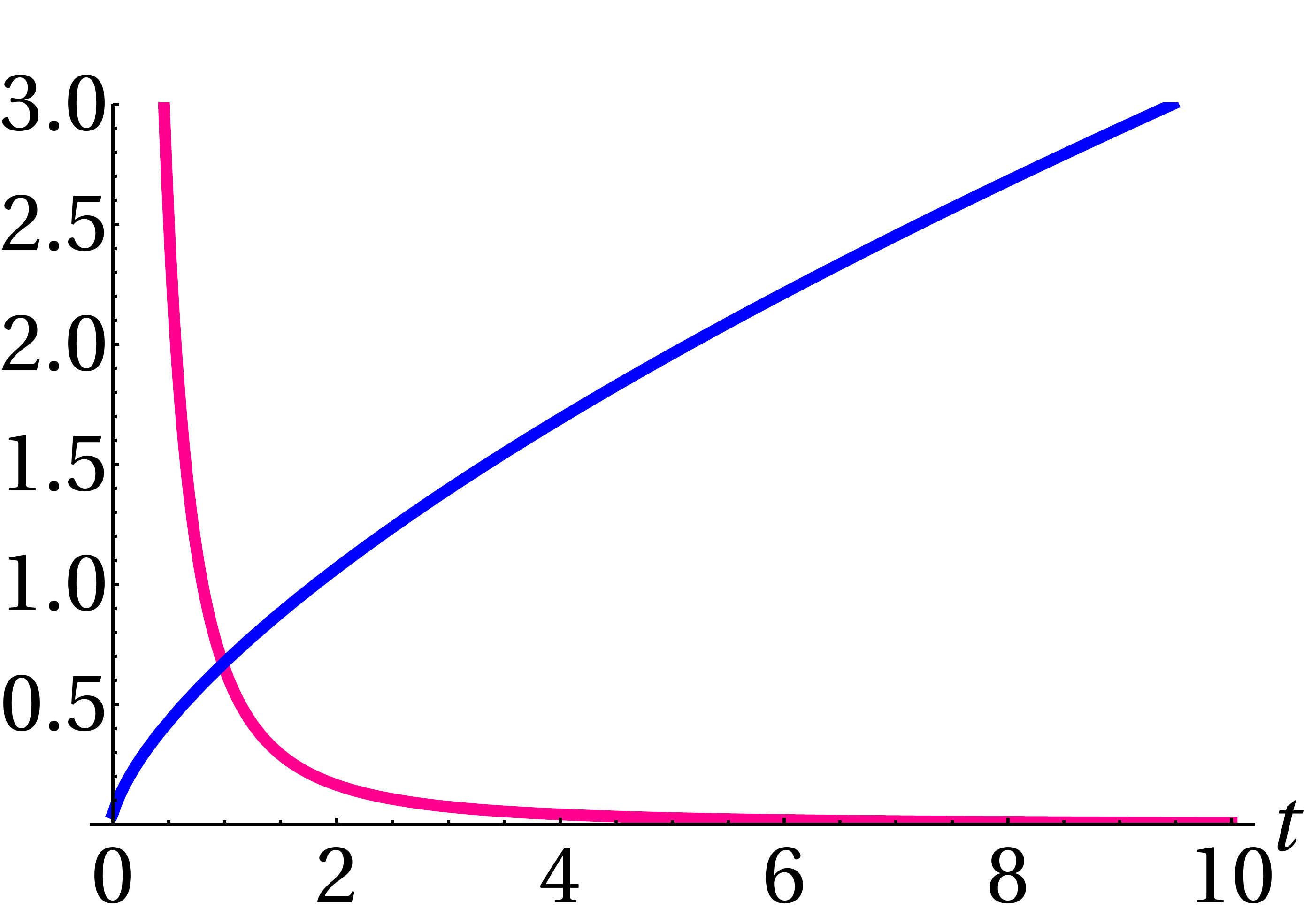}
    \caption{The scale factors $a$ and $b$; represented by the blue curve, while $\left| {G_{i\bar j} \dot z^i \dot z^{\bar j}} \right|$ represented by the red curve.}
    \label{23}
  \end{subfigure}
\qquad
  \begin{subfigure}[t]{.5\linewidth}
    \centering
    \includegraphics[width=0.7\columnwidth]{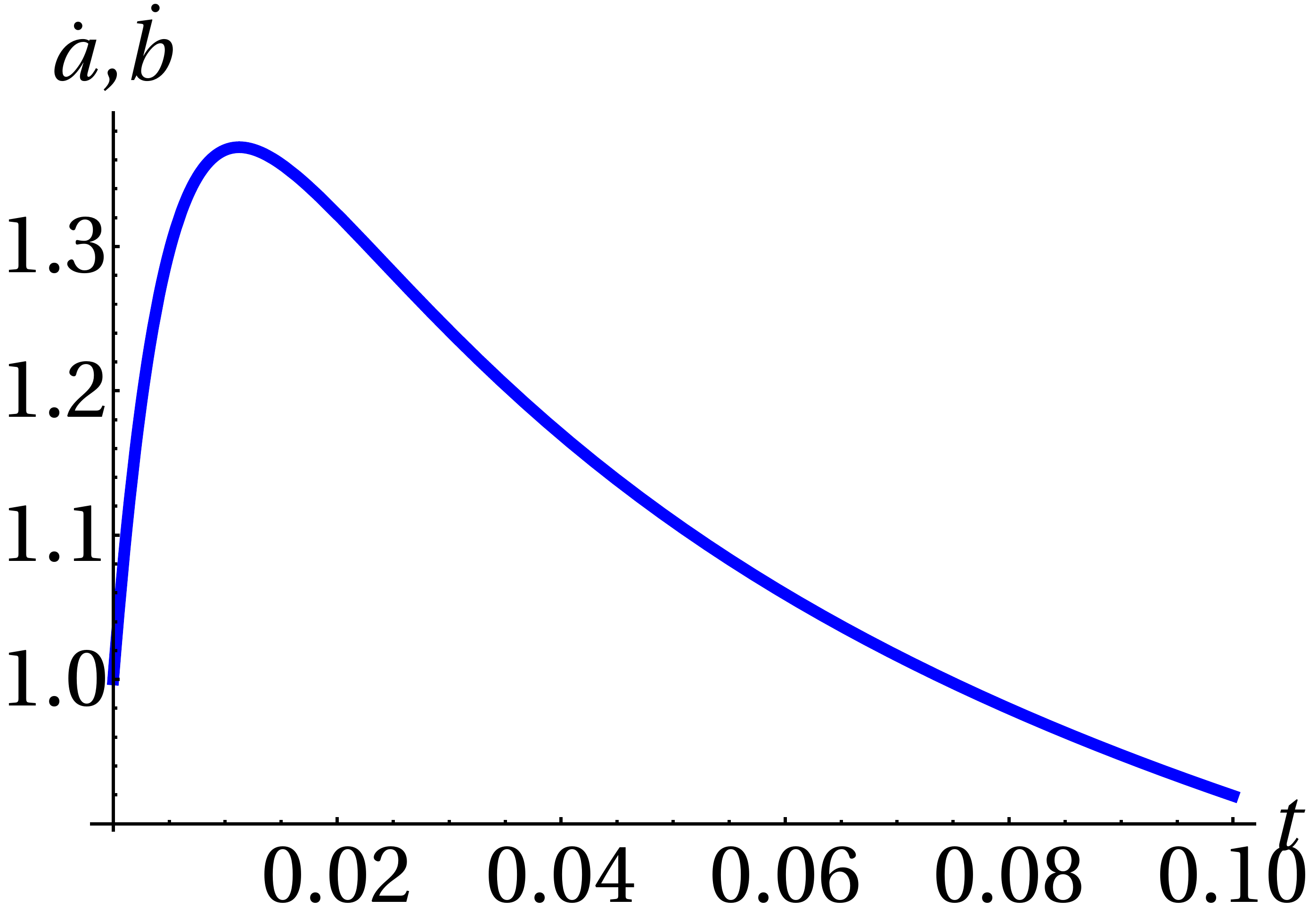}
    \caption{The expansion rates of the scale factors. Both $\dot a$ and $\dot b$ are represented by the shown curve.}
    \label{24}
  \end{subfigure}
\\[9em]
      \begin{subfigure}[t]{.5\linewidth}
    \centering
    \includegraphics[width=0.7\columnwidth]{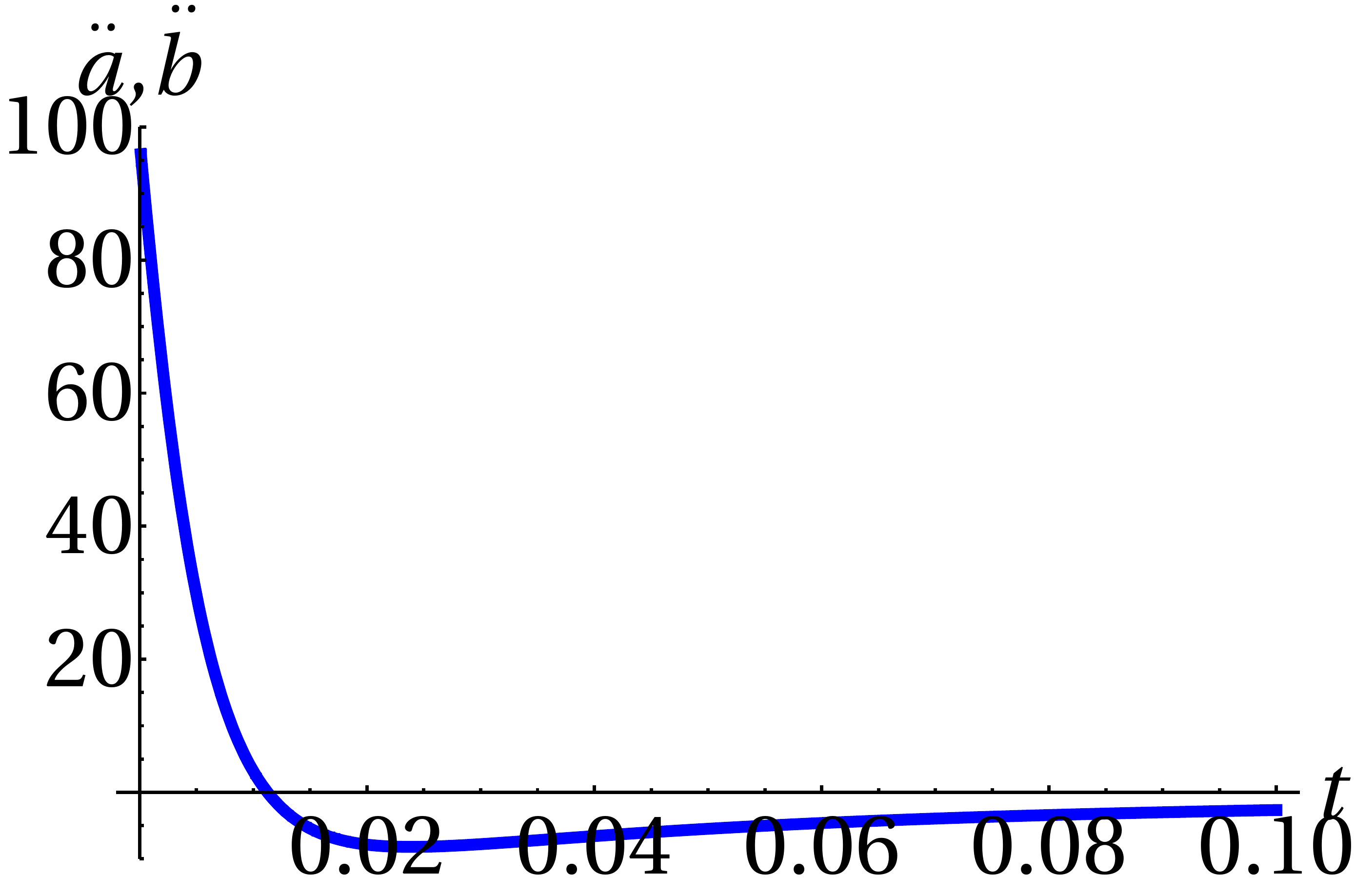}
    \caption{The accelerations of the scale factors. Both $\ddot a$ and $\ddot b$ are represented by the shown curve.}
    \label{25}
  \end{subfigure}
\qquad
  \begin{subfigure}[t]{.5\linewidth}
    \centering
    \includegraphics[width=0.7\columnwidth]{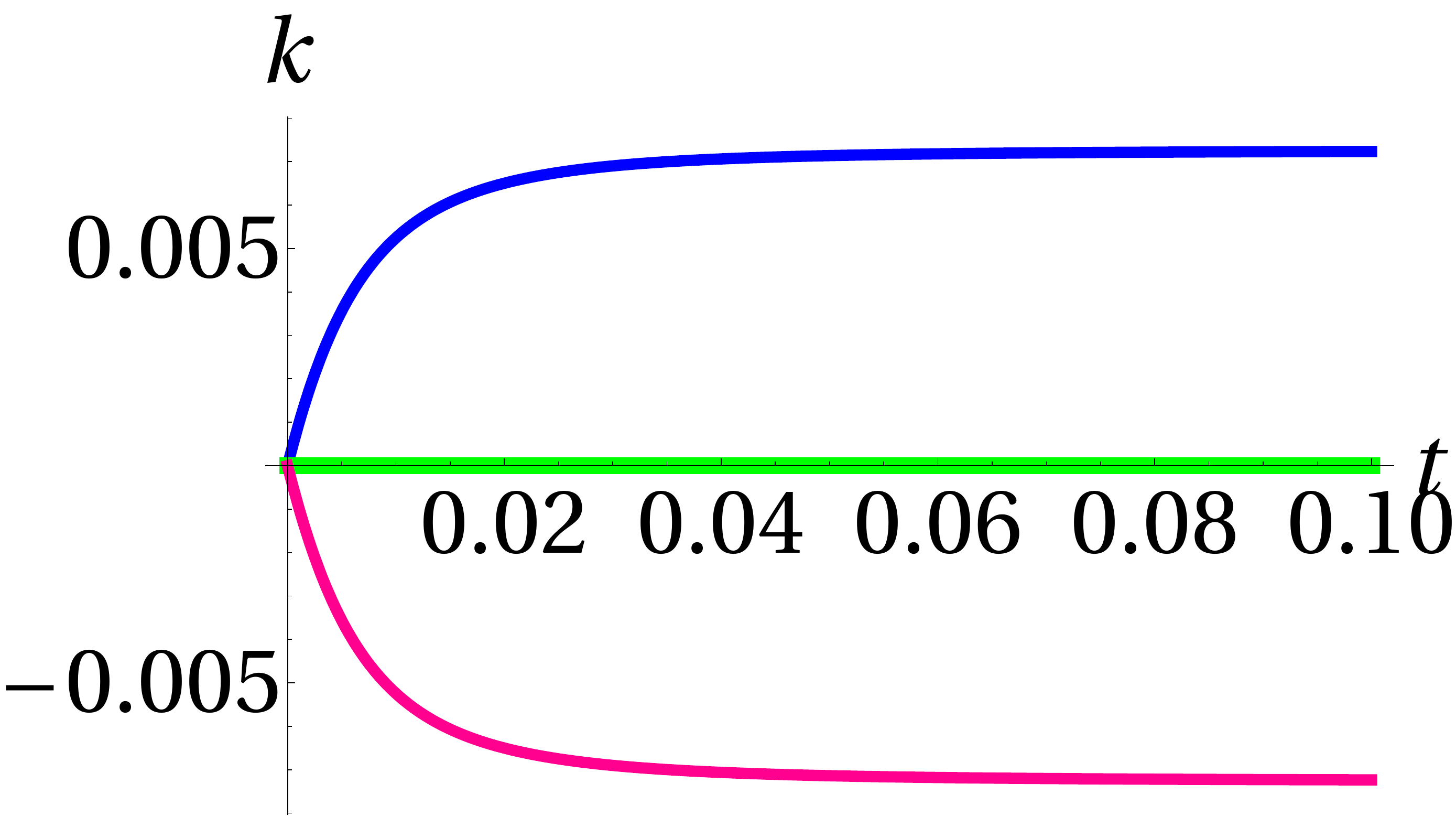}
    \caption{The harmonic function $k$ using: $\dot k\left(0\right)=1$ (blue curve), $\dot k\left(0\right)=0$ (green line), and $\dot k\left(0\right)=-1$ (red curve).}
    \label{26}
  \end{subfigure}
  \caption{Dust-filled brane world with initial conditions set number 3.}
  \label{Fig6}
\end{figure}

\begin{figure}[H]
\begin{subfigure}[t]{.5\linewidth}
    \centering
    \includegraphics[width=0.7\columnwidth]{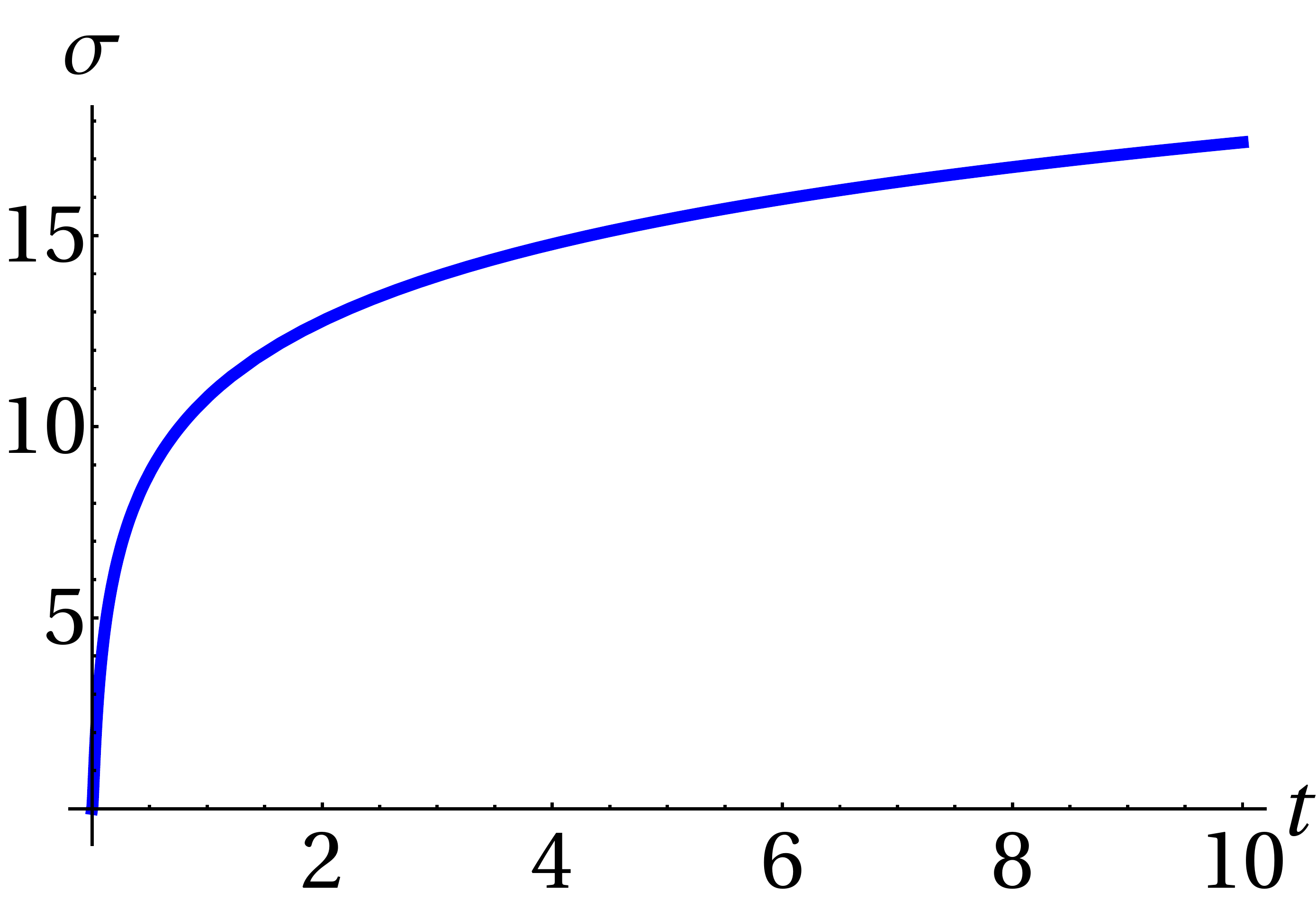}
    \caption{The dilaton $\sigma$; same for all three $\dot k\left(0\right)$.}
    \label{27}
  \end{subfigure}
\qquad  
\begin{subfigure}[t]{.5\linewidth}
    \centering
    \includegraphics[width=0.7\columnwidth]{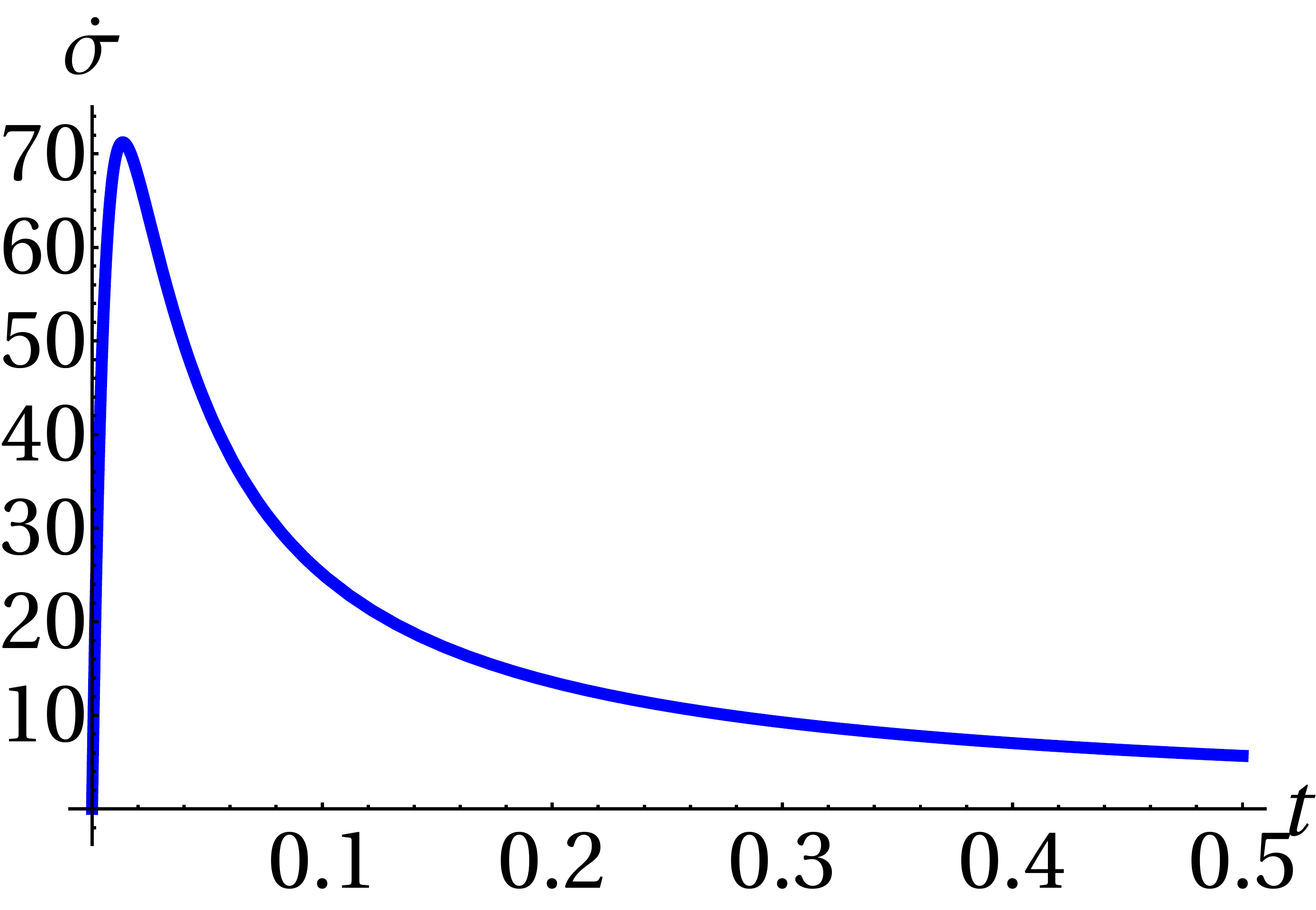}
    \caption{The dilatonic field strength $\dot\sigma$.}
    \label{28}
  \end{subfigure}
\\[4em]  
\begin{subfigure}[t]{.5\linewidth}
    \centering
    \includegraphics[width=0.7\columnwidth]{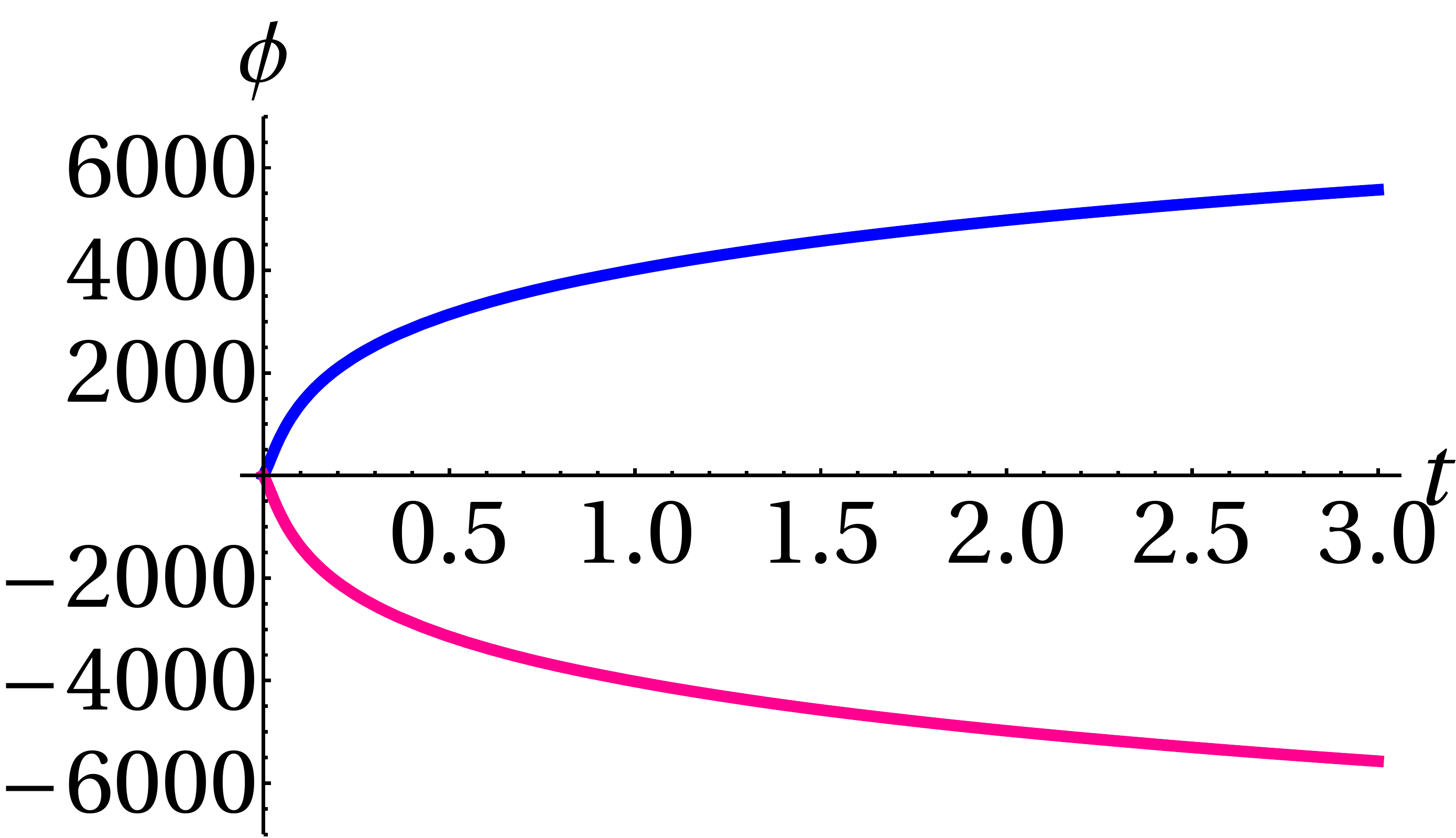}
    \caption{The universal axion $\phi$ for $\dot k\left(0\right) = 1$ (blue curve), and $\dot k\left(0\right) = -1$ (red curve). The solution diverges for $\dot k\left(0\right)=0$.}
    \label{29}
  \end{subfigure}
\qquad
  \begin{subfigure}[t]{.5\linewidth}
    \centering
    \includegraphics[width=0.7\columnwidth]{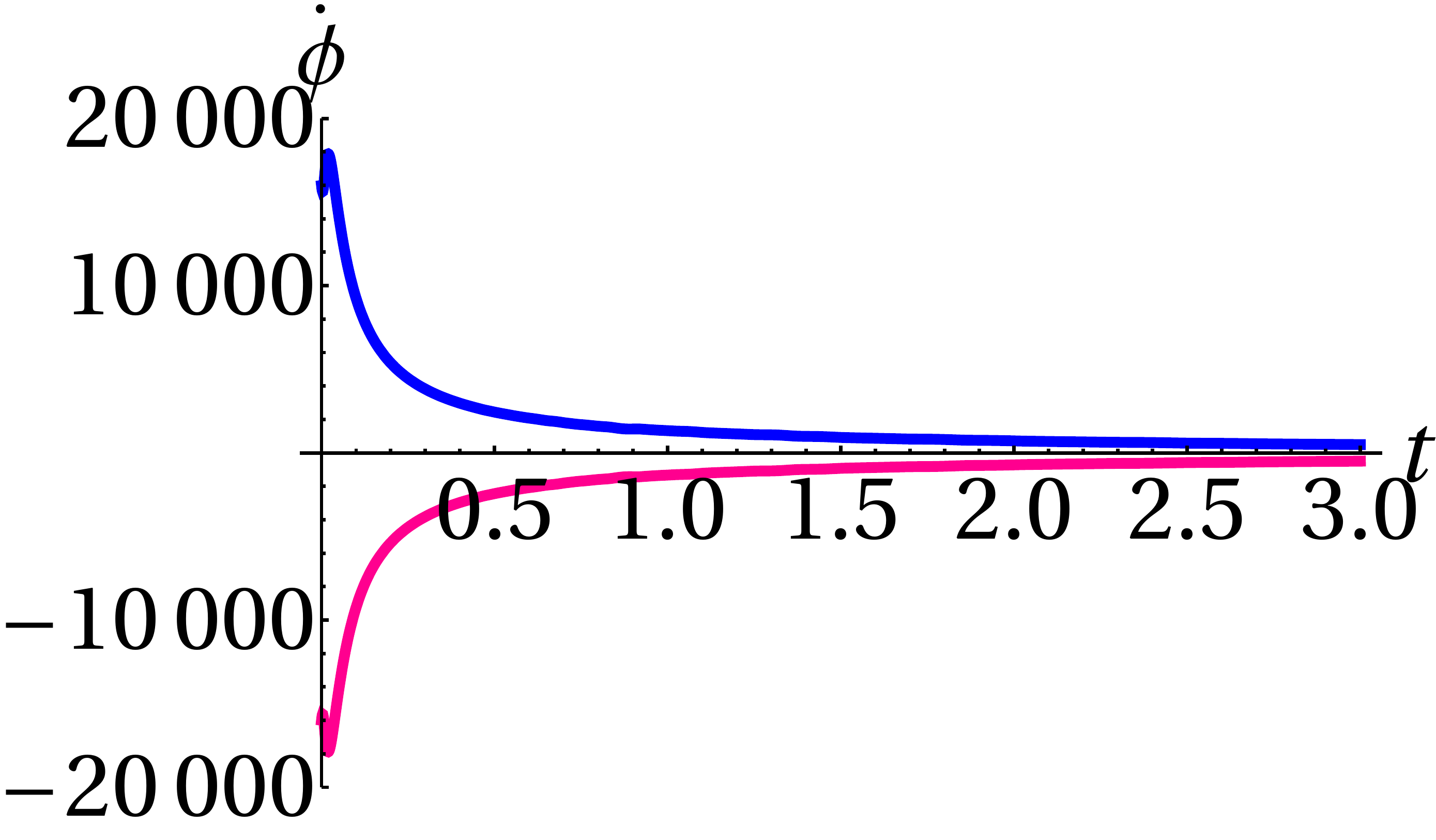}
    \caption{The axionic field strength $\dot\phi$ for $\dot k\left(0\right) = 1$ (blue curve), and $\dot k\left(0\right) = -1$ (red curve).}
    \label{30}
  \end{subfigure}
\\[4em]
  \begin{subfigure}[t]{.5\linewidth}
    \centering
    \includegraphics[width=0.7\columnwidth]{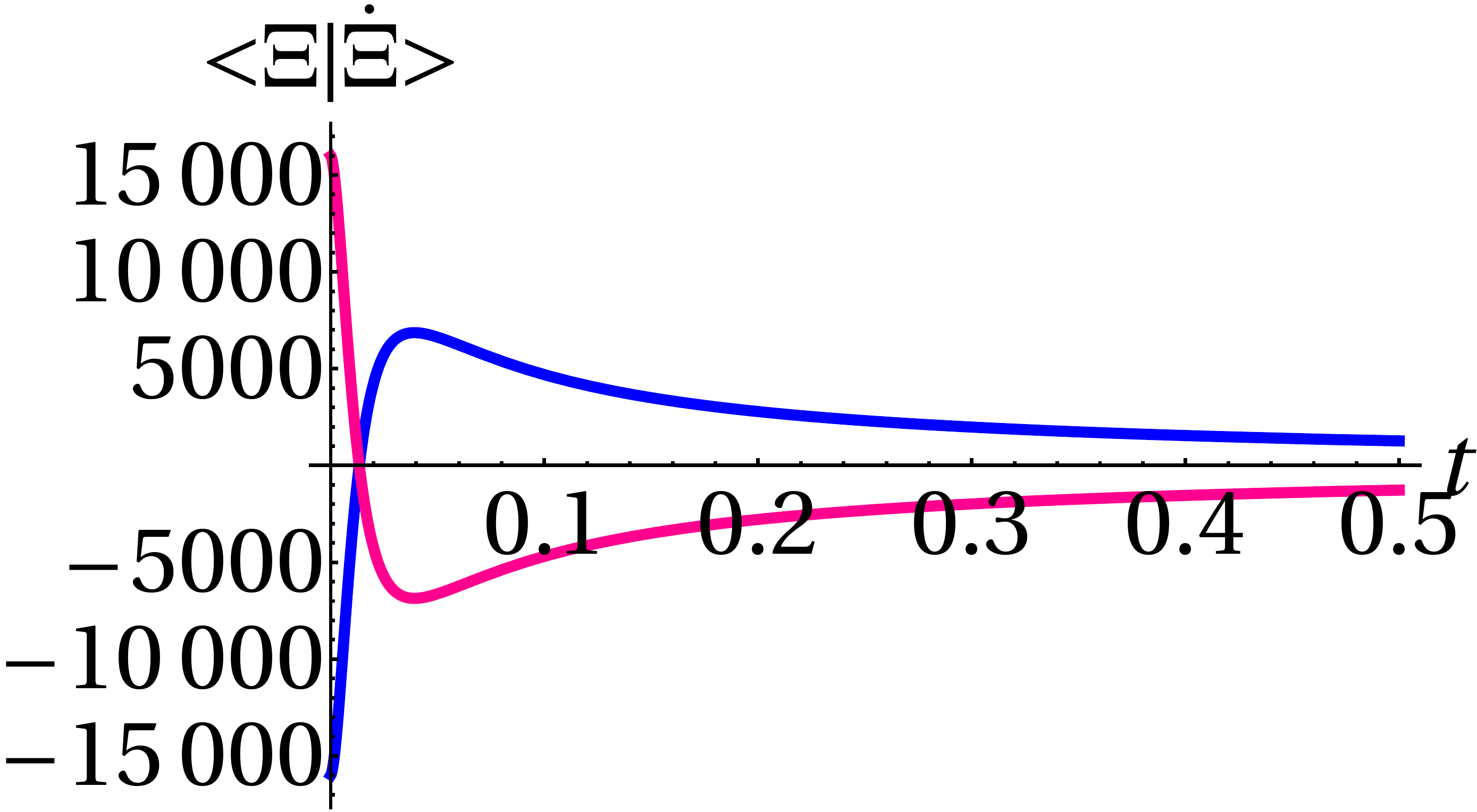}
    \caption{$ \langle \Xi | \dot{\Xi} \rangle$ for $\dot{k}(0)= 1$  (blue), and $\dot{k}(0)  = -1$ (red).}
    \label{31}
  \end{subfigure}
\qquad
  \begin{subfigure}[t]{.5\linewidth}
    \centering
    \includegraphics[width=0.7\columnwidth]{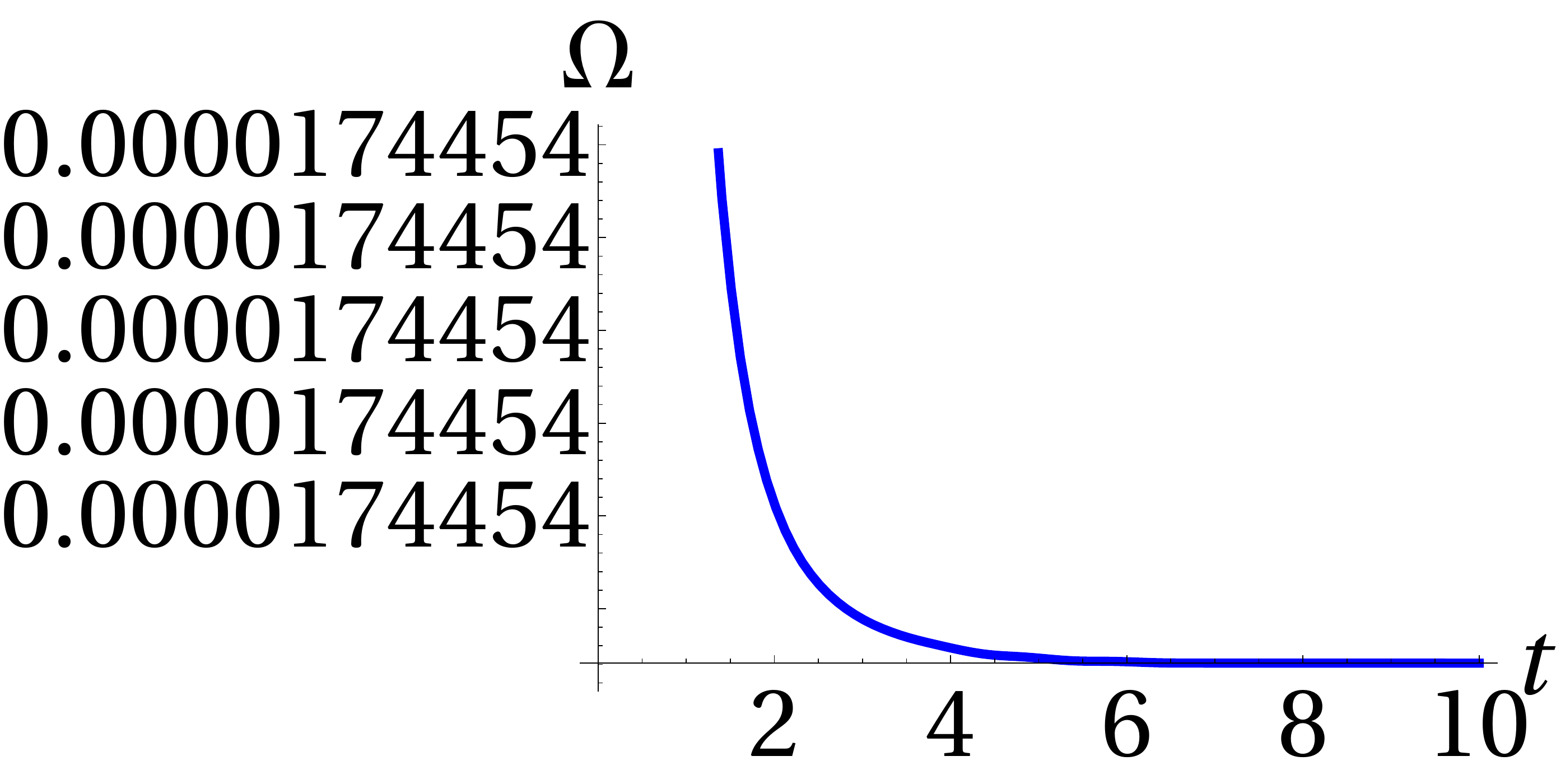}
    \caption{$\Omega$ at $\dot k\left(0\right) = 1$ and  $ \dot\sigma \left(0\right) =0 $.}
    \label{32}
  \end{subfigure}
  \caption{Dust-filled brane world with initial conditions set number 3 (continued).}
  \label{Fig66}
\end{figure}


\begin{figure}[H]
 \begin{subfigure}[t]{.5\linewidth}
    \centering
    \includegraphics[width=0.7\columnwidth]{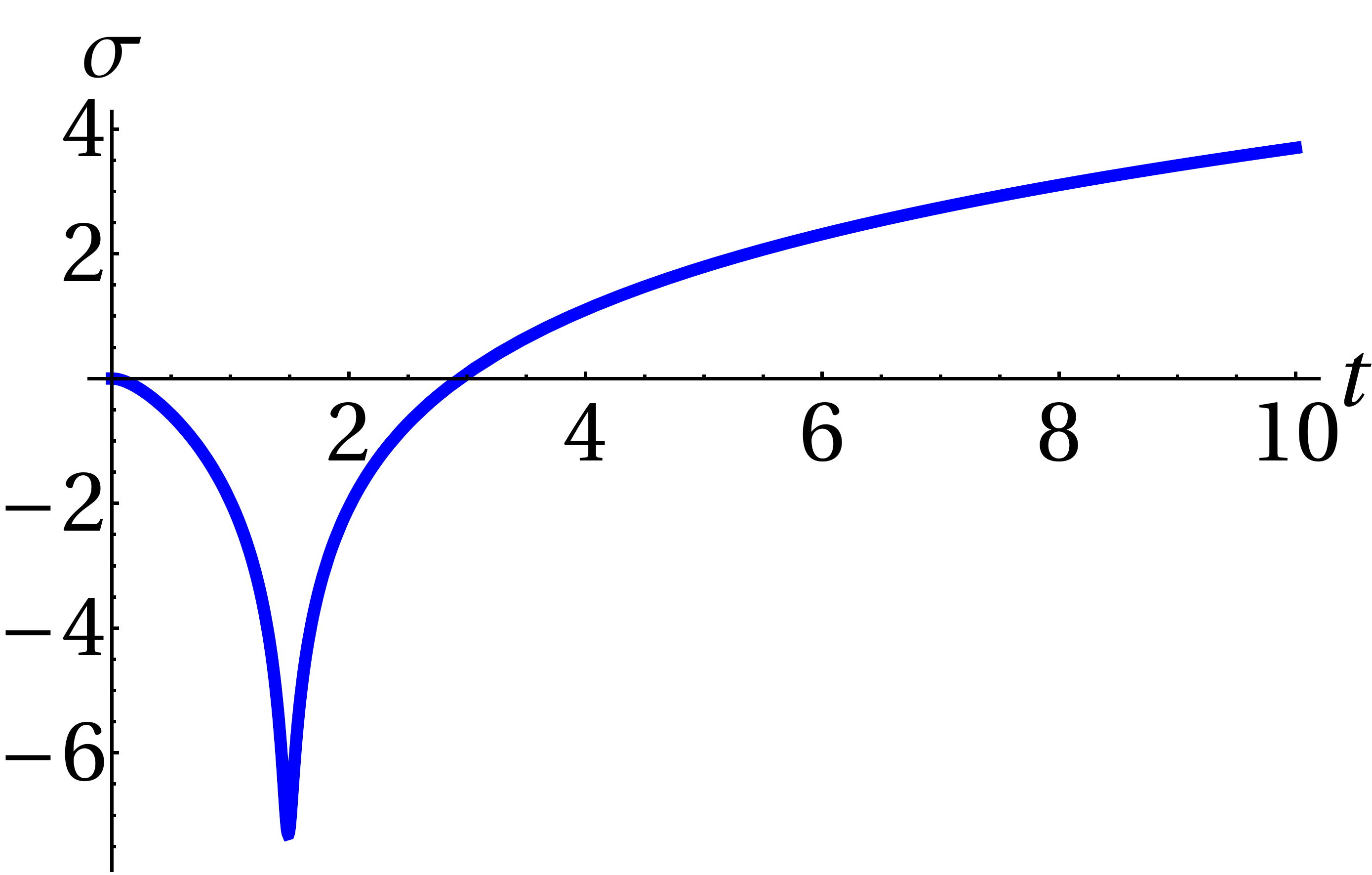}
    \caption{The dilaton $\sigma$; same for all three $\dot k\left(0\right)$.}
    \label{37}
  \end{subfigure}
\qquad
  \begin{subfigure}[t]{.5\linewidth}
    \centering
    \includegraphics[width=0.7\columnwidth]{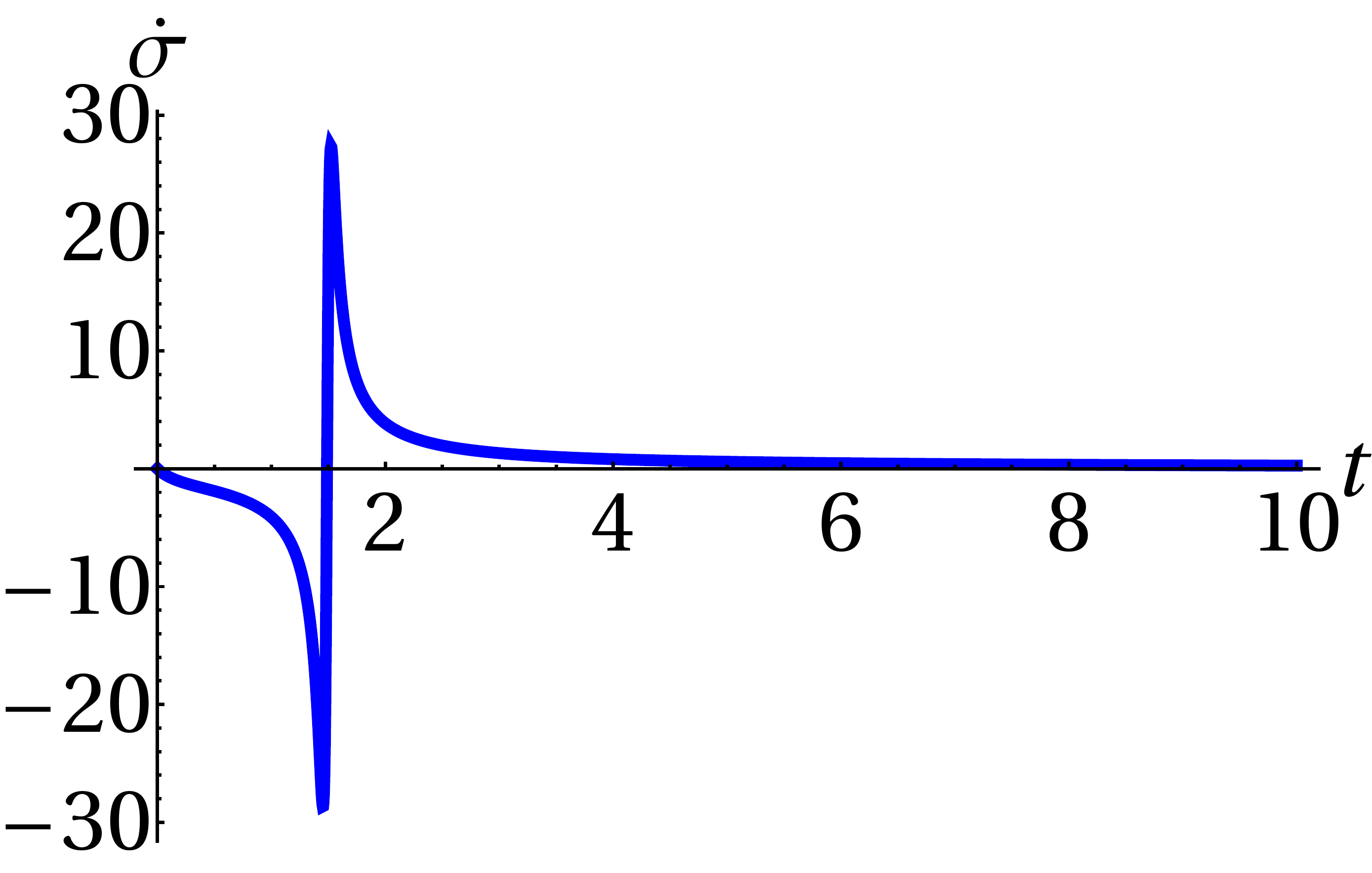}
    \caption{The dilatonic field strength $\dot\sigma$.}
    \label{38}
  \end{subfigure}
\\[4em]
\begin{subfigure}[t]{.5\linewidth}
    \centering
    \includegraphics[width=0.7\columnwidth]{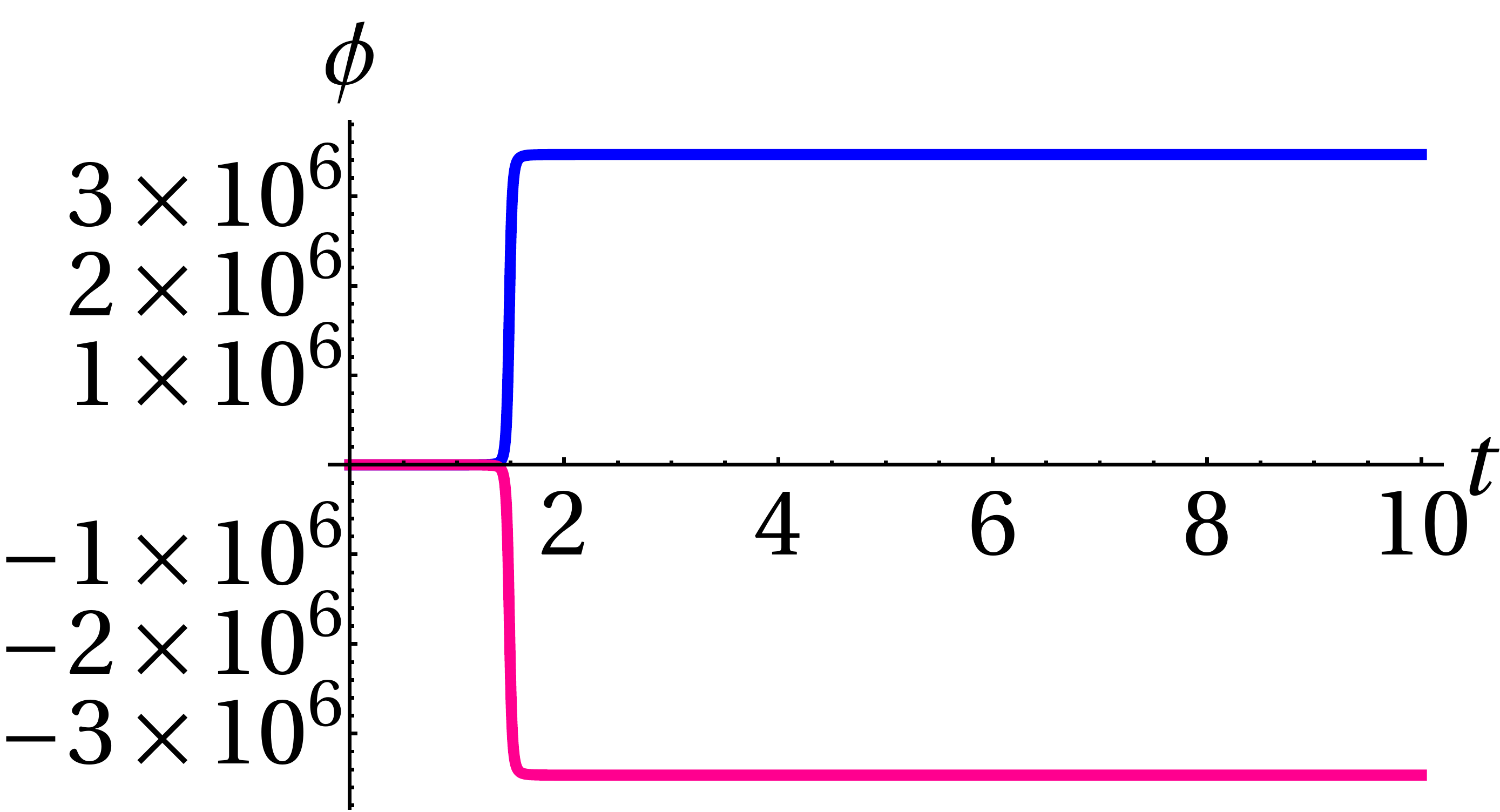}
    \caption{The universal axion $\phi$ for $\dot k\left(0\right) = 1$ (blue curve), and $\dot k\left(0\right) = -1$ (red curve). The solution diverges for $\dot k\left(0\right)=0$.}
    \label{39}
  \end{subfigure}
\qquad
  \begin{subfigure}[t]{.5\linewidth}
    \centering
    \includegraphics[width=0.7\columnwidth]{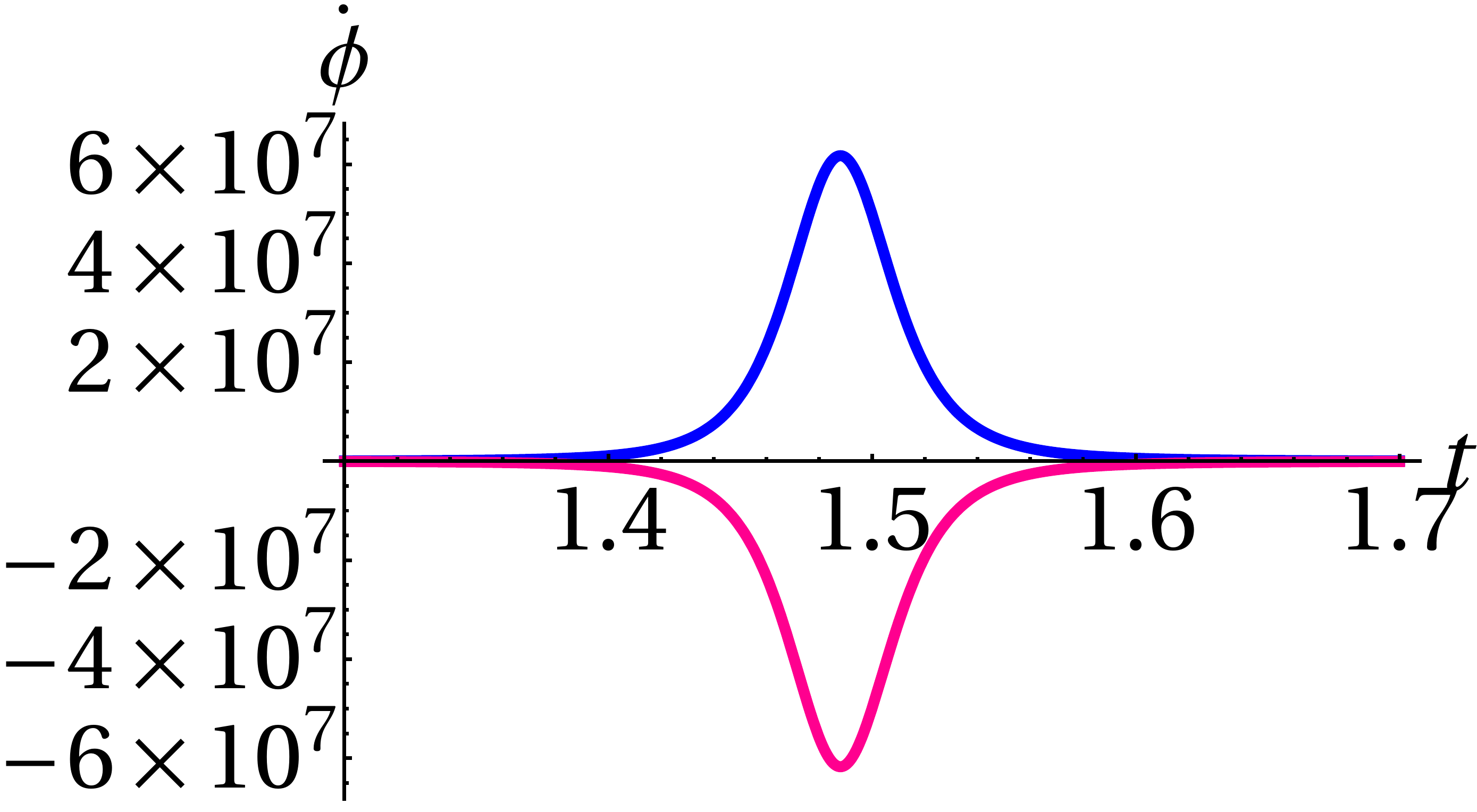}
    \caption{The axionic field strength $\dot\phi$ for $\dot k\left(0\right) = 1$ (blue curve), and $\dot k\left(0\right) = -1$ (red curve).}
    \label{40}
  \end{subfigure}
\\[4em]
    \begin{subfigure}[t]{.5\linewidth}
    \centering
    \includegraphics[width=0.7\columnwidth]{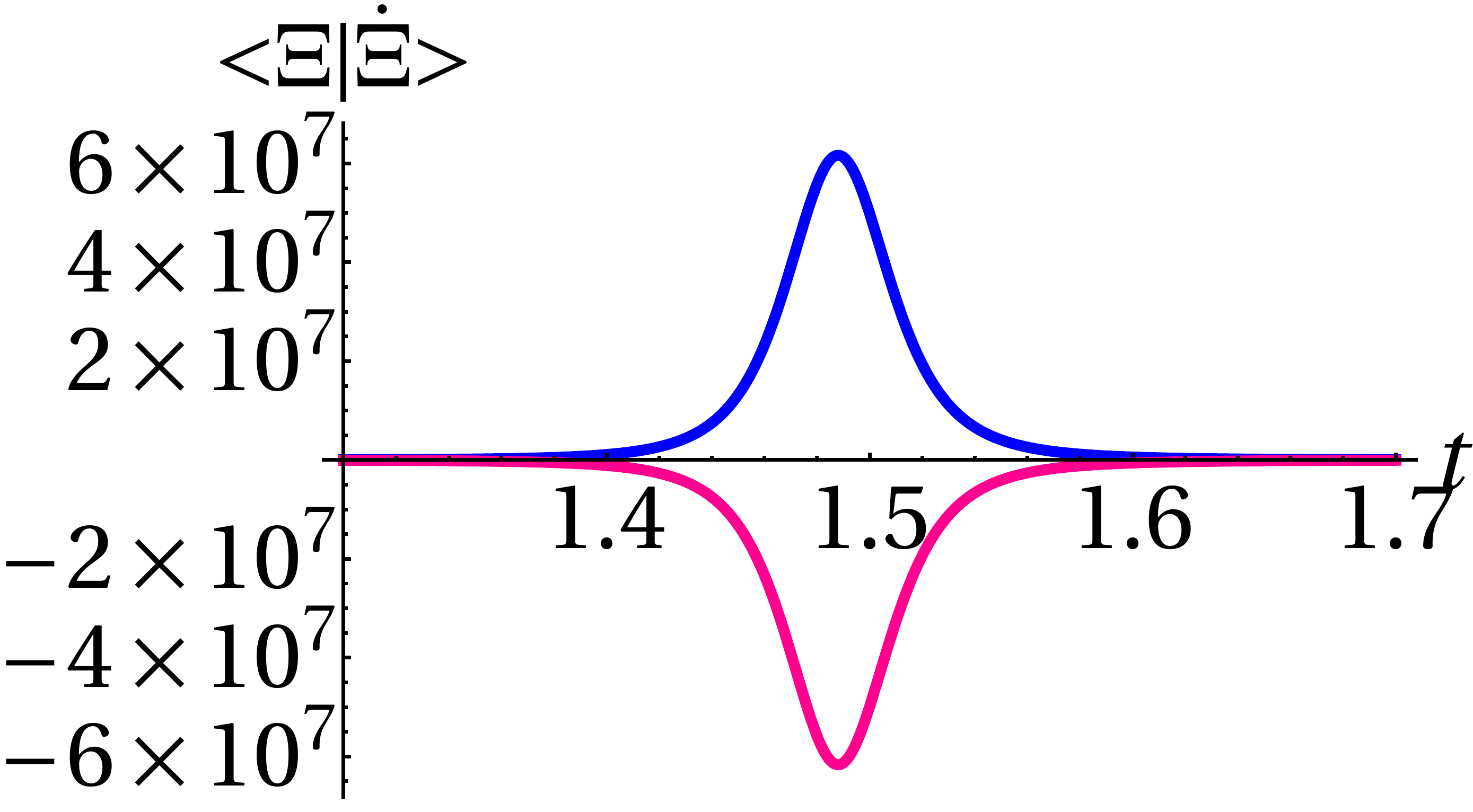}
    \caption{$ \langle \Xi | \dot{\Xi} \rangle$ for $\dot{k}(0)= 1$  (blue), and $\dot{k}(0)  = -1$ (red).}
    \label{41}
  \end{subfigure}
\qquad
  \begin{subfigure}[t]{.5\linewidth}
    \centering
    \includegraphics[width=0.7\columnwidth]{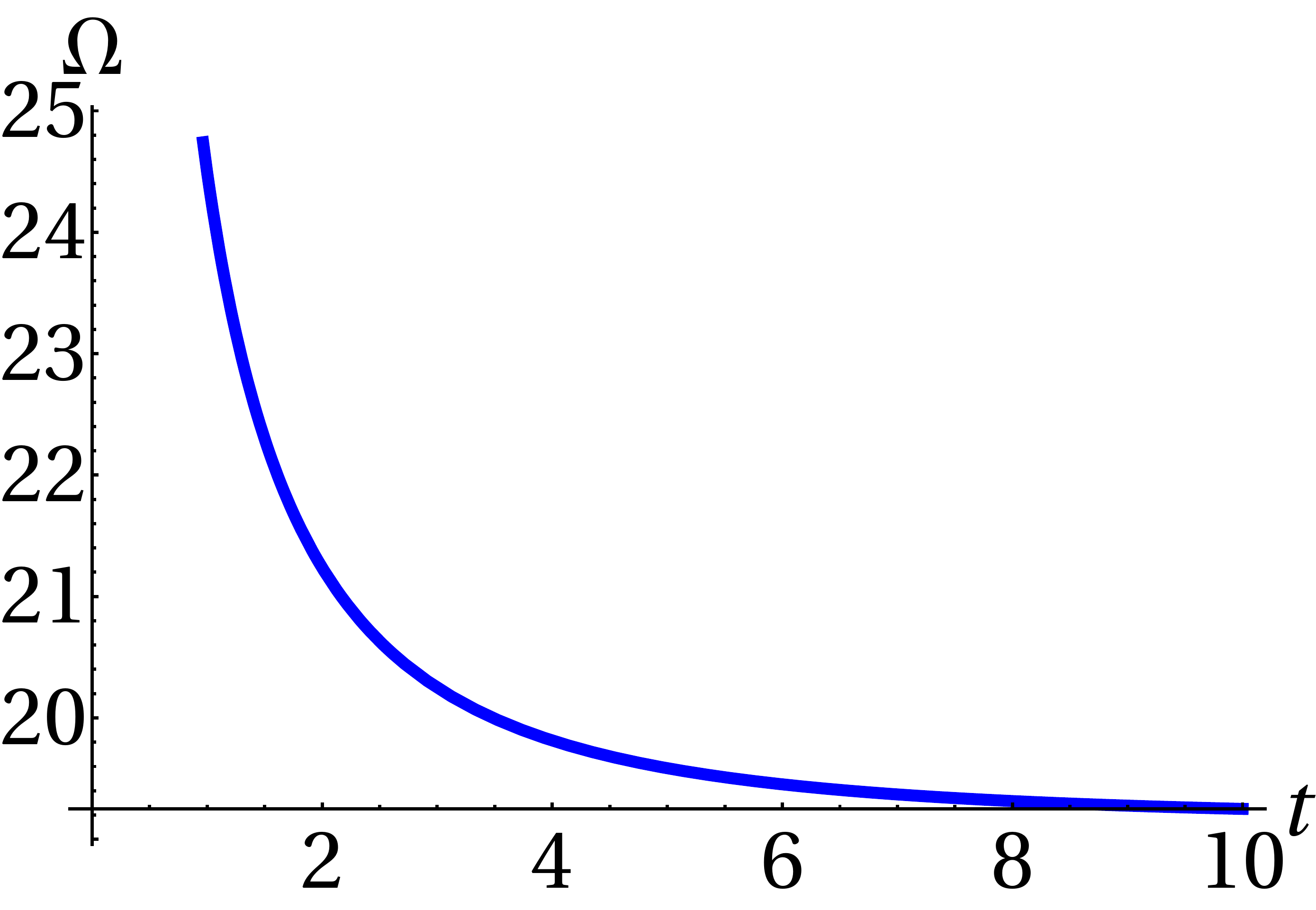}
    \caption{$\Omega$ at $\dot k\left(0\right) = 1$ and  $ \dot\sigma \left(0\right) =0 $.}
    \label{42}
  \end{subfigure}
 \caption{Dust-filled brane world with initial conditions set number 4 (continued).}
  \label{Fig88}
\end{figure}


\begin{figure}[H]
  \begin{subfigure}[t]{.5\linewidth}
    \centering
    \includegraphics[width=0.7\columnwidth]{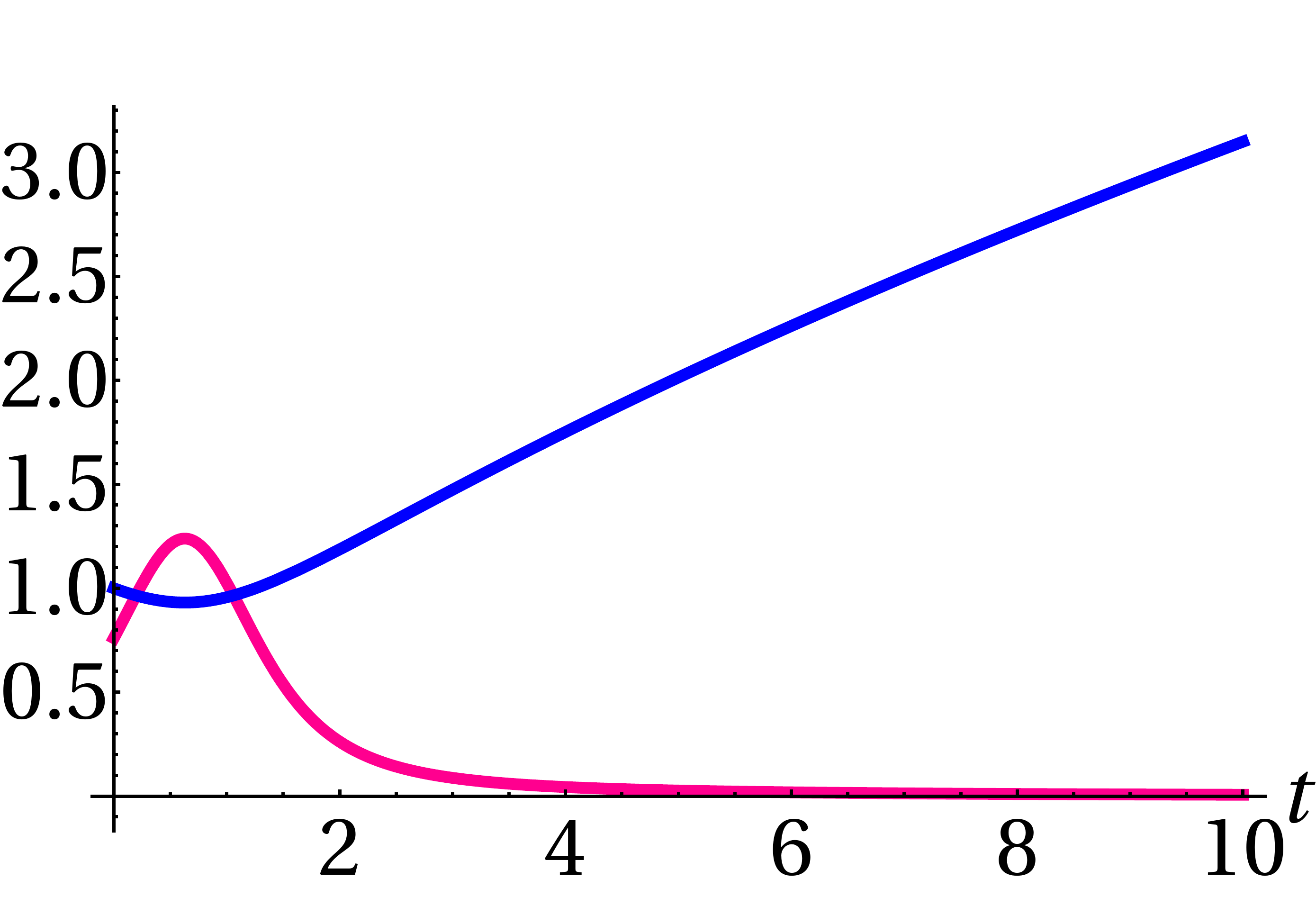}
    \caption{The scale factors $a$ and $b$; represented by the blue curve, while $\left| {G_{i\bar j} \dot z^i \dot z^{\bar j}} \right|$ represented by the red curve.}
    \label{43}
  \end{subfigure}
\qquad
  \begin{subfigure}[t]{.5\linewidth}
    \centering
    \includegraphics[width=0.7\columnwidth]{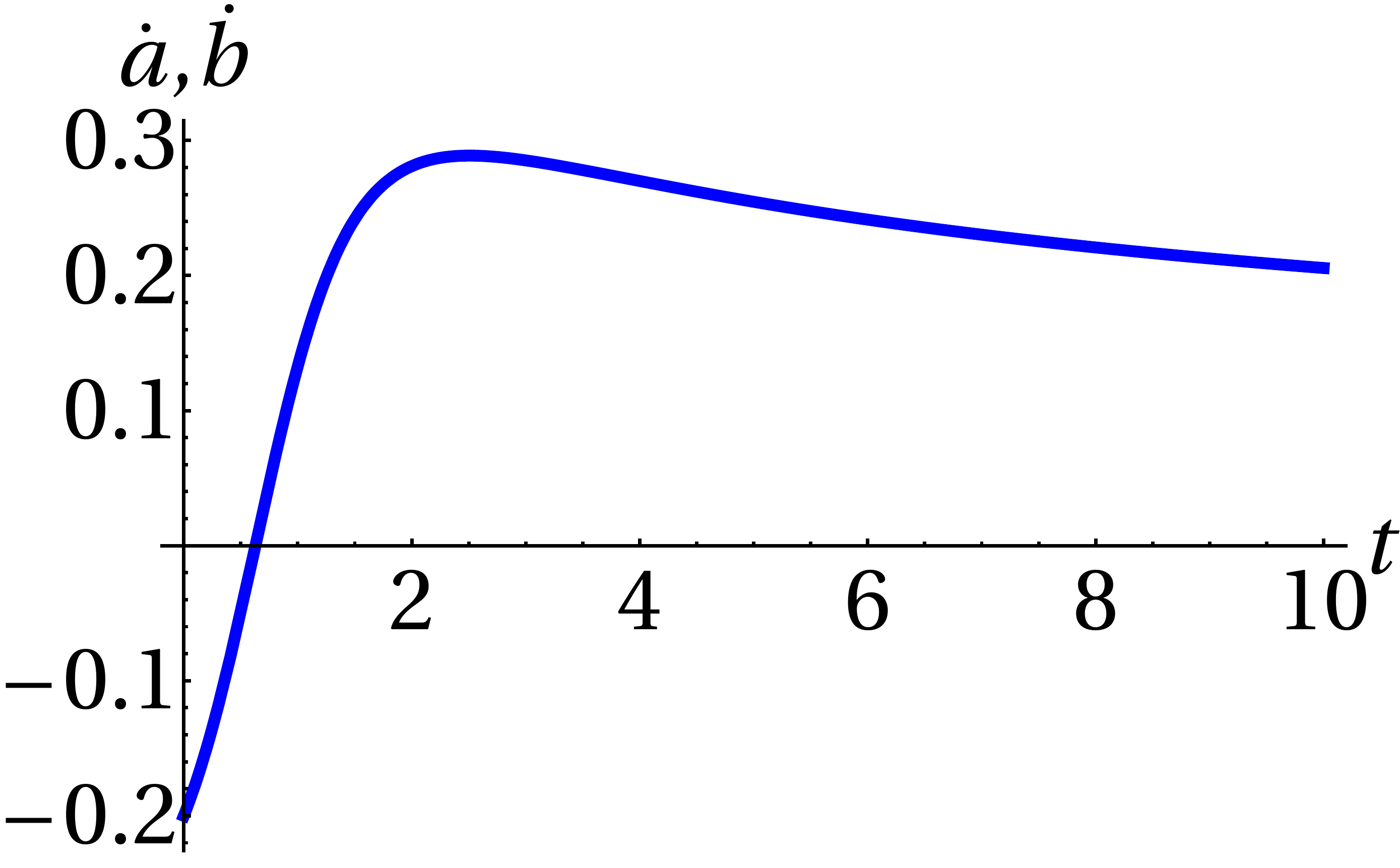}
    \caption{The expansion rates of the scale factors. Both $\dot a$ and $\dot b$ are represented by the shown curve.}
    \label{44}
  \end{subfigure}
\\[9em]
  \begin{subfigure}[t]{.5\linewidth}
    \centering
    \includegraphics[width=0.7\columnwidth]{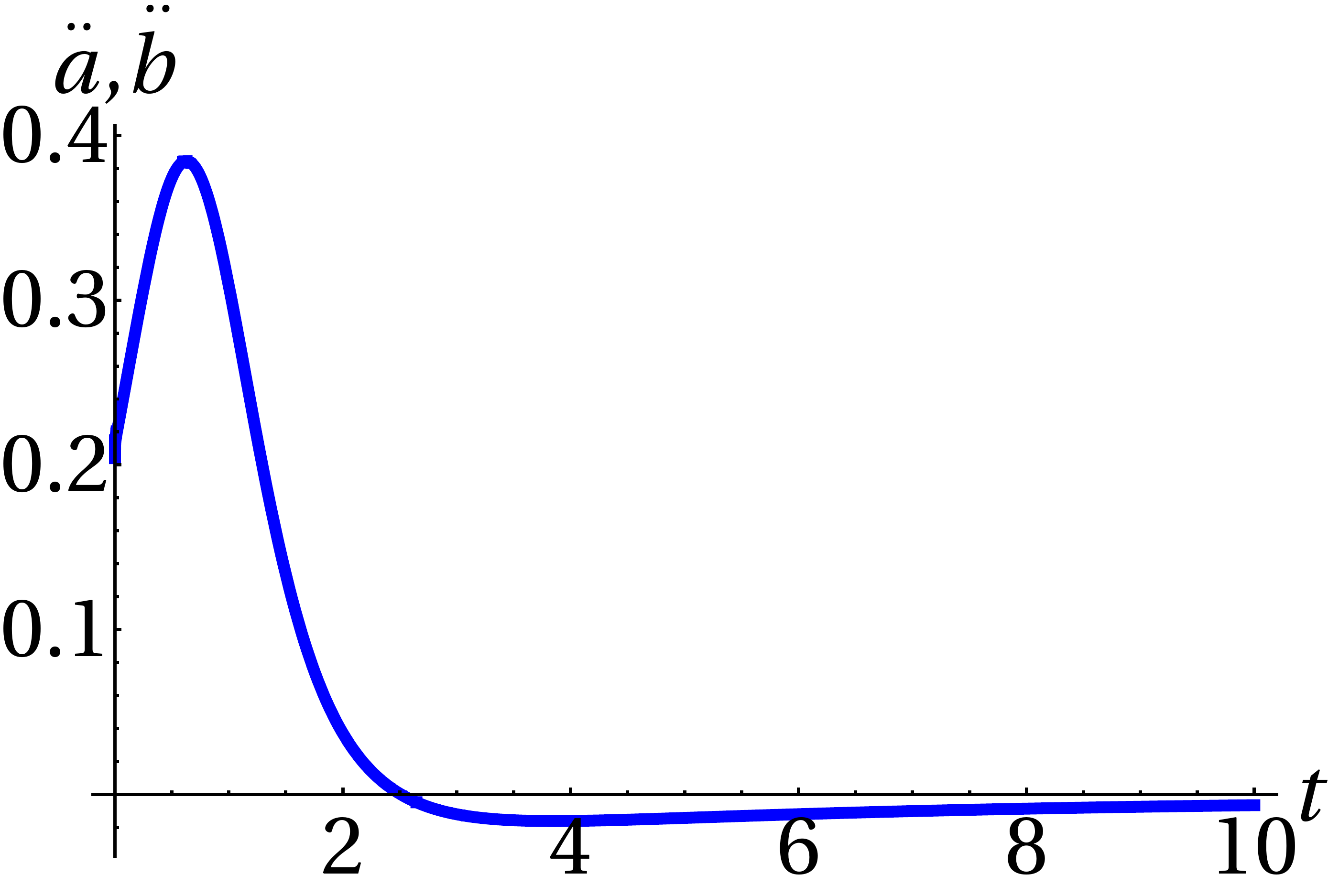}
    \caption{The accelerations of the scale factors. Both $\ddot a$ and $\ddot b$ are represented by the shown curve.}
    \label{45}
  \end{subfigure}
\qquad
  \begin{subfigure}[t]{.5\linewidth}
    \centering
    \includegraphics[width=0.7\columnwidth]{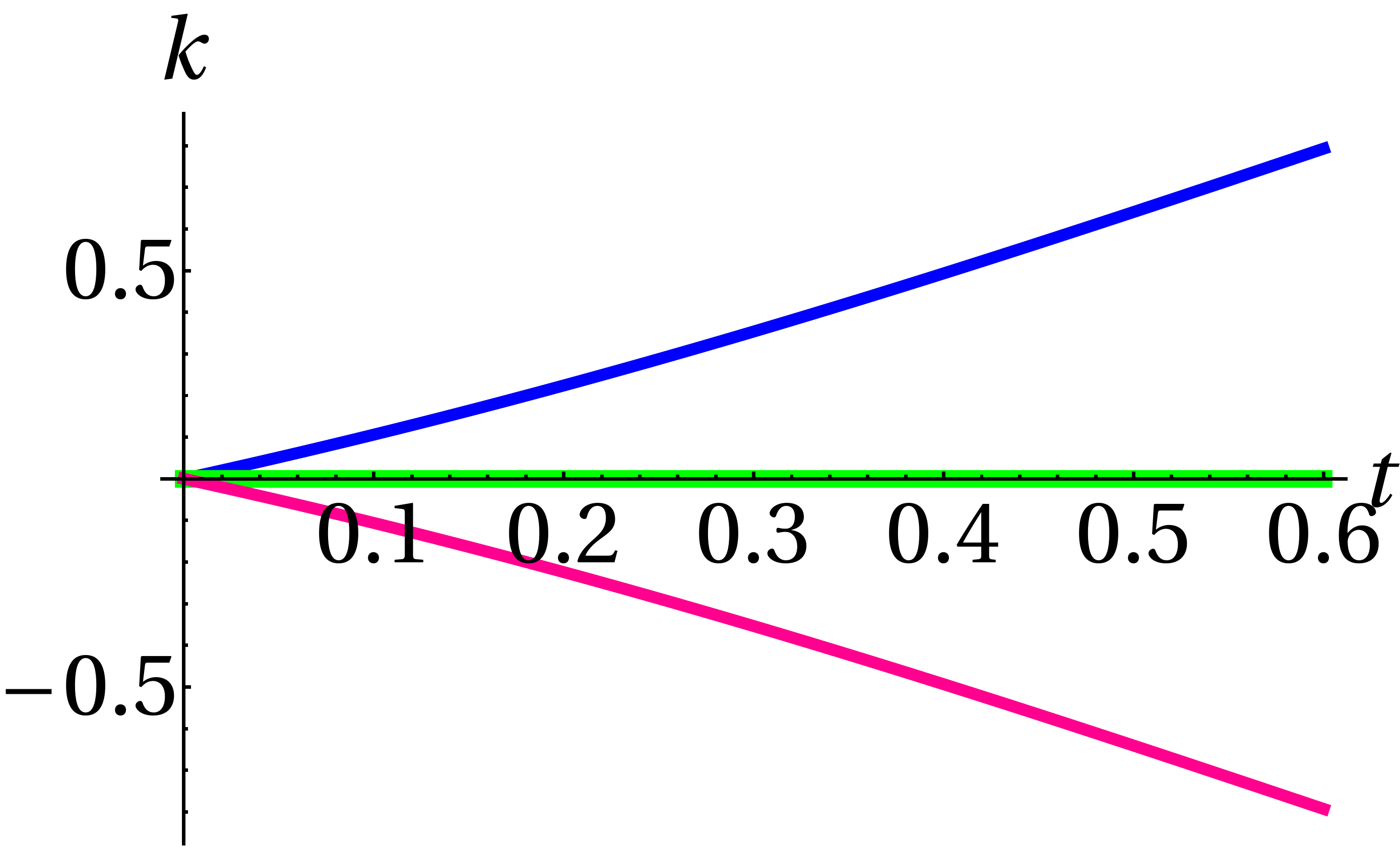}
    \caption{The harmonic function $k$ using: $\dot k\left(0\right)=1$ (blue curve), $\dot k\left(0\right)=0$ (green line), and $\dot k\left(0\right)=-1$ (red curve).}
    \label{46}
  \end{subfigure}
  \caption{Dust-filled brane world with initial conditions set number 5.}
  \label{Fig10}
\end{figure}
\begin{figure}[H]
 \begin{subfigure}[t]{.5\linewidth}
    \centering
    \includegraphics[width=0.7\columnwidth]{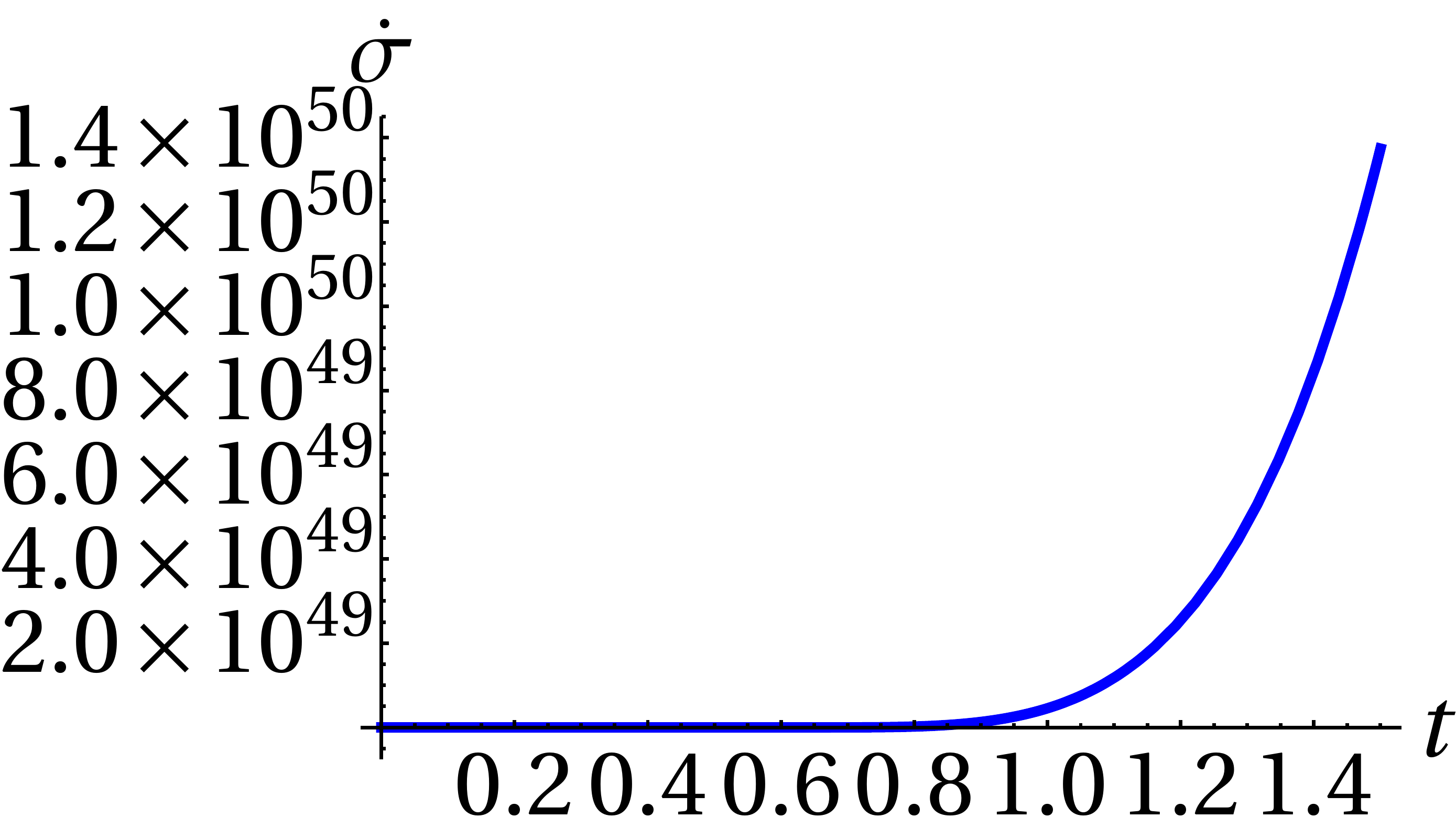}
    \caption{The dilaton $\sigma$; same for all three $\dot k\left(0\right)$.}
    \label{47}
  \end{subfigure}
\qquad
  \begin{subfigure}[t]{.5\linewidth}
    \centering
    \includegraphics[width=0.7\columnwidth]{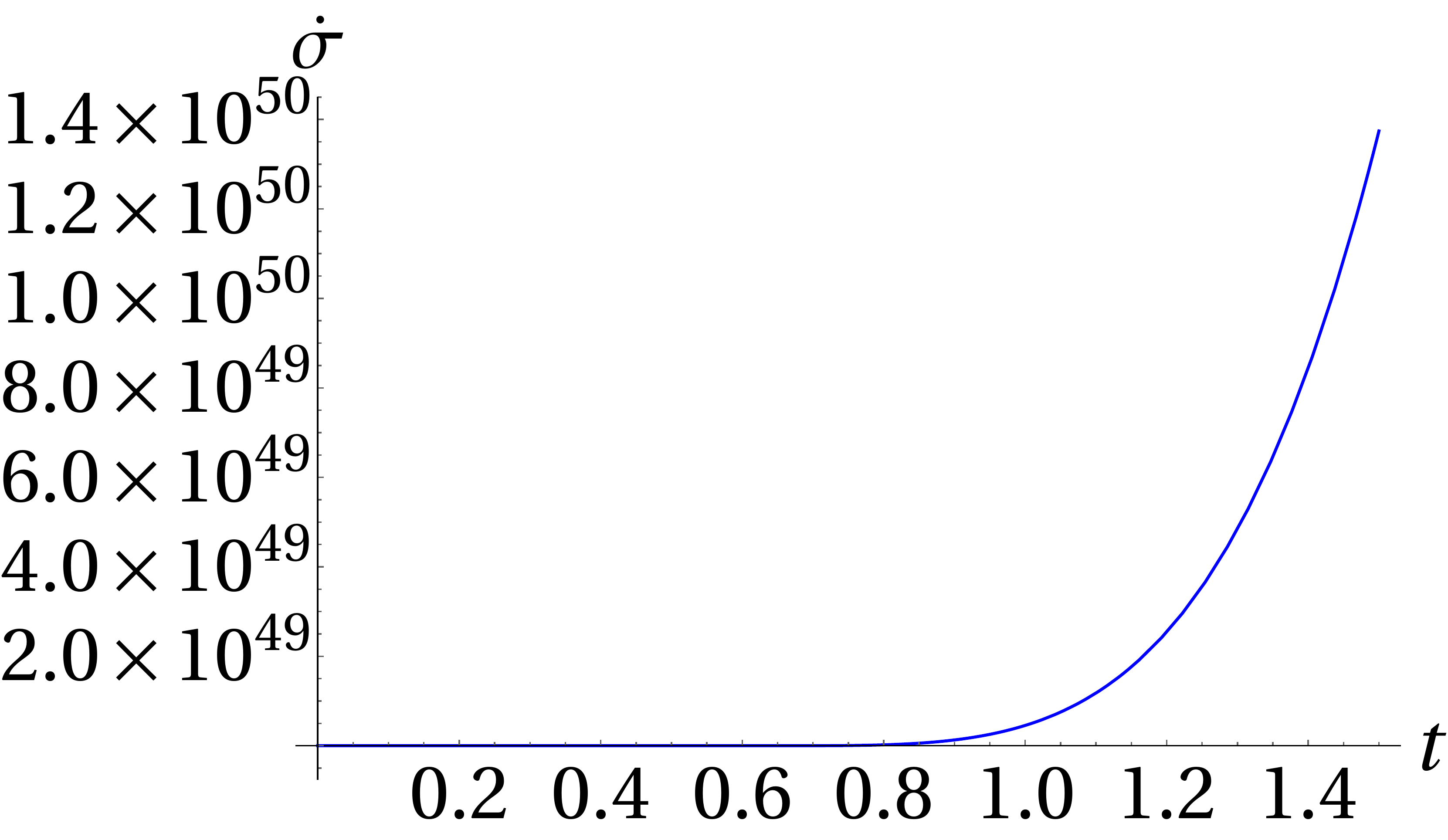}
    \caption{The dilatonic field strength $\dot\sigma$.}
    \label{48}
  \end{subfigure}
  \\[4em]
 \begin{subfigure}[t]{.5\linewidth}
    \centering
    \includegraphics[width=0.7\columnwidth]{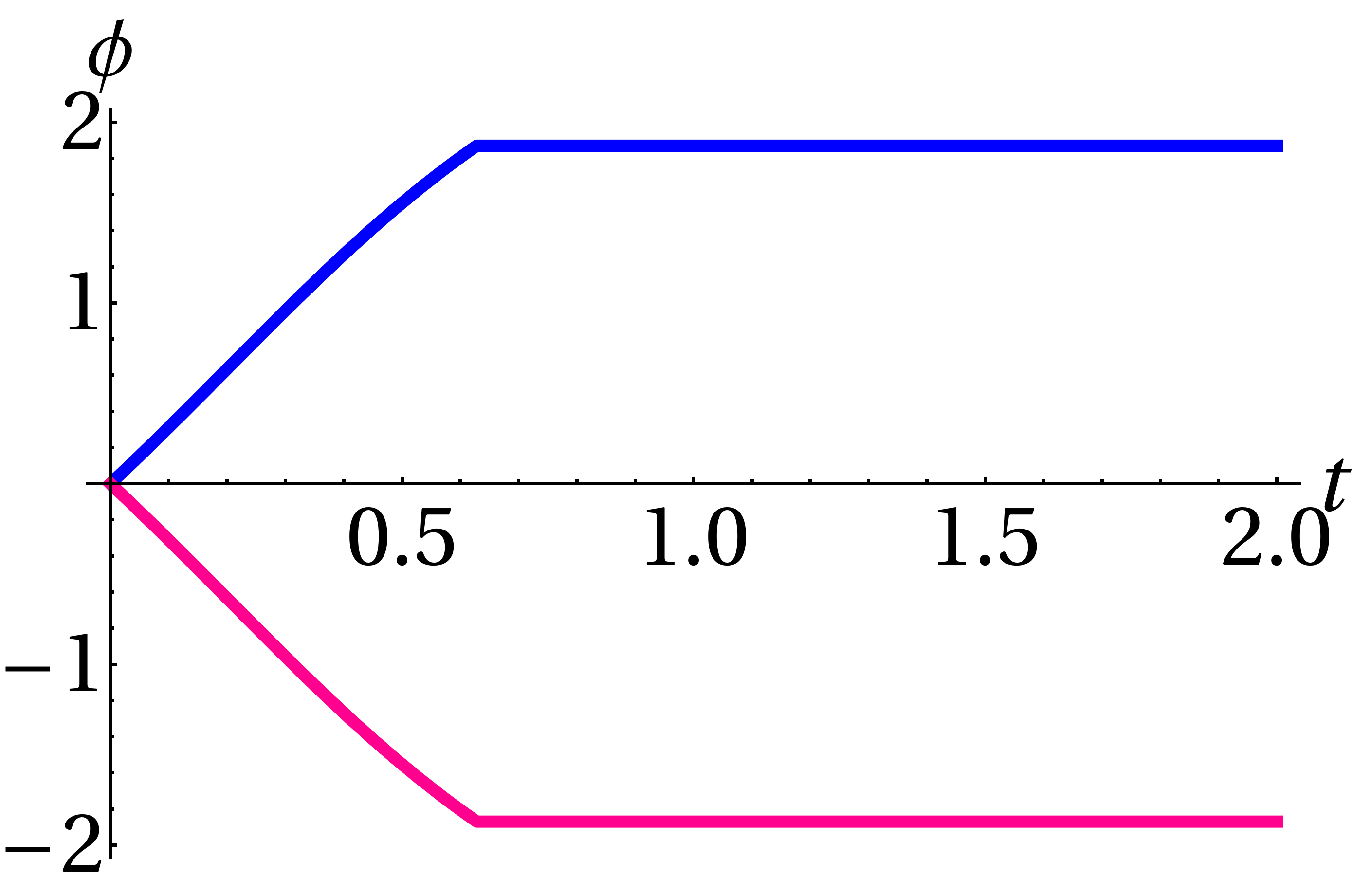}
    \caption{The universal axion $\phi$ for $\dot k\left(0\right) = 1$ (blue curve), and $\dot k\left(0\right) = -1$ (red curve). The solution diverges for $\dot k\left(0\right)=0$.}
    \label{49}
  \end{subfigure}
\qquad
  \begin{subfigure}[t]{.5\linewidth}
    \centering
    \includegraphics[width=0.7\columnwidth]{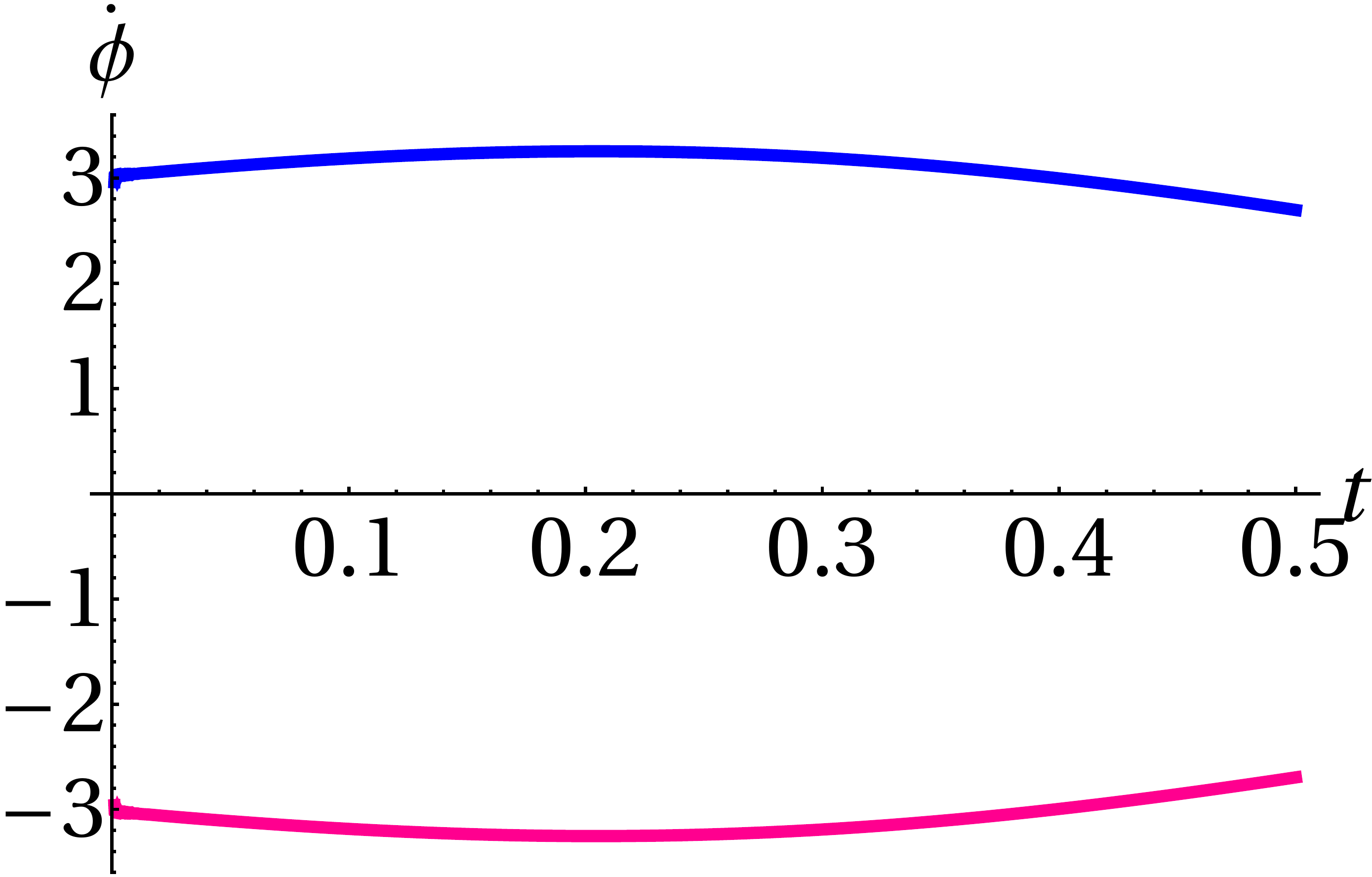}
    \caption{The axionic field strength $\dot\phi$ for $\dot k\left(0\right) = 1$ (blue curve), and $\dot k\left(0\right) = -1$ (red curve).}
    \label{50}
  \end{subfigure}
\\[4em]  
\begin{subfigure}[t]{.5\linewidth}
    \centering
    \includegraphics[width=0.7\columnwidth]{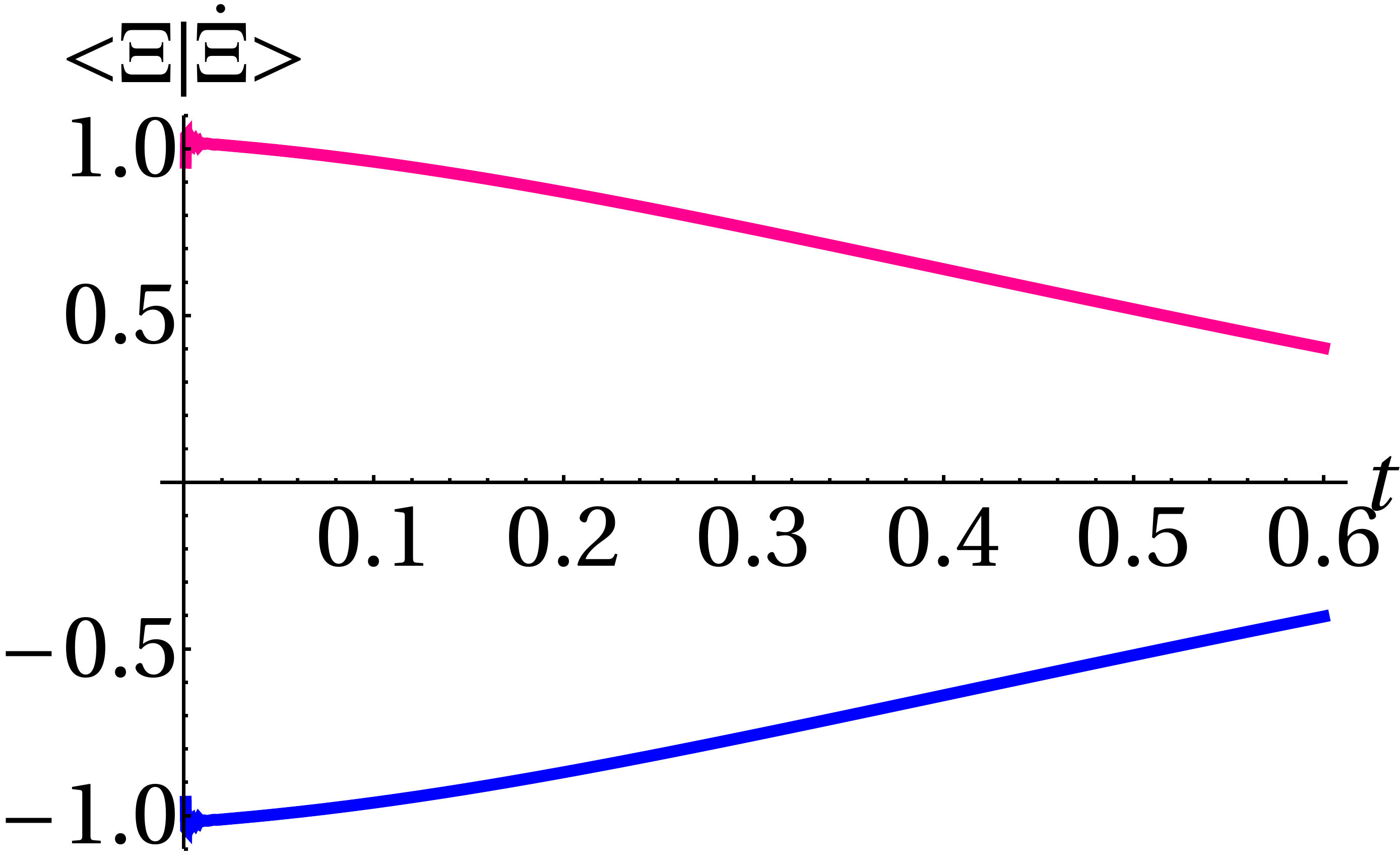}
    \caption{$ \langle \Xi | \dot{\Xi} \rangle$ for $\dot{k}(0)= 1$  (blue), and $\dot{k}(0)  = -1$ (red).}
    \label{51}
  \end{subfigure}
\qquad
  \begin{subfigure}[t]{.5\linewidth}
    \centering
    \includegraphics[width=0.7\columnwidth]{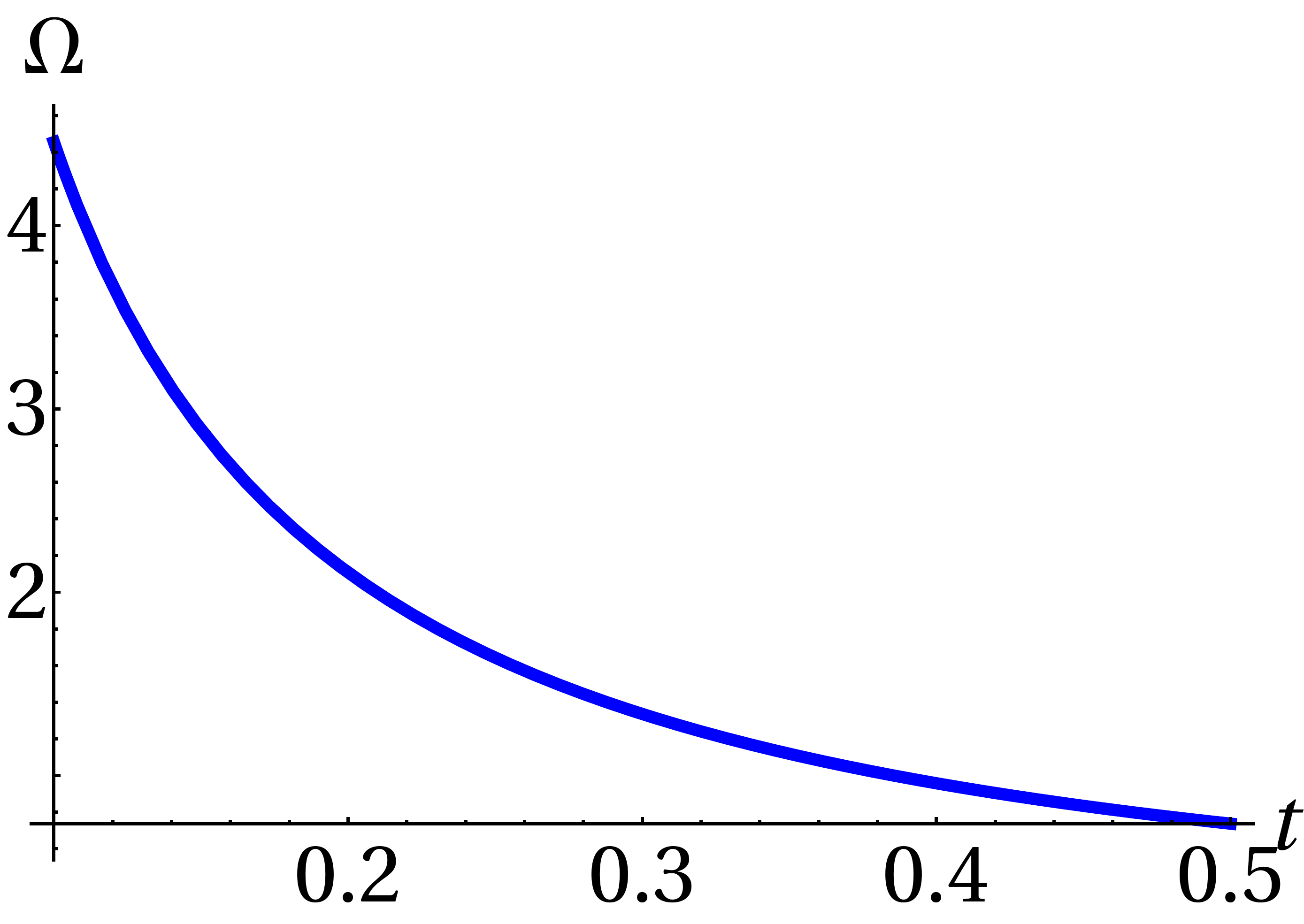}
    \caption{$\Omega$ at $\dot k\left(0\right) = 1$ and  $ \dot\sigma \left(0\right) =0 $.}
    \label{52}
  \end{subfigure}
   \caption{Dust-filled brane world with initial conditions set number 5 (continued).}
  \label{Fig101}
\end{figure}


\begin{figure}[H]
  \begin{subfigure}[t]{.5\linewidth}
    \centering
    \includegraphics[width=0.7\columnwidth]{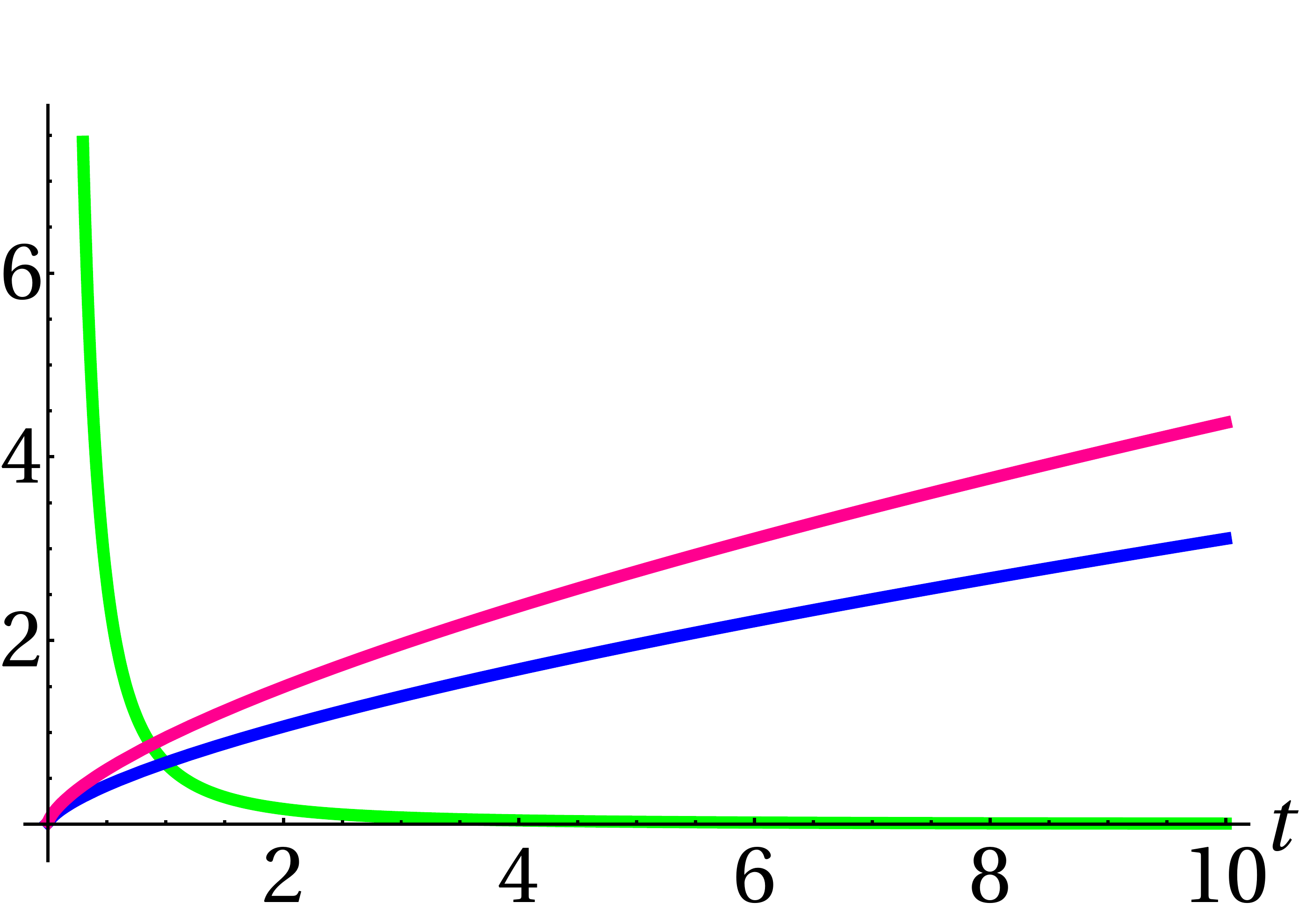}
    \caption{The scale factor $a$ is represented by the blue curve, $b$ by the red curve, while $\left| {G_{i\bar j} \dot z^i \dot z^{\bar j}} \right|$ by the green curve. The curve for $b$ is scaled down by a factor of 40 to fit in the graph.}
    \label{103}
  \end{subfigure}
\qquad
  \begin{subfigure}[t]{.5\linewidth}
    \centering
    \includegraphics[width=0.7\columnwidth]{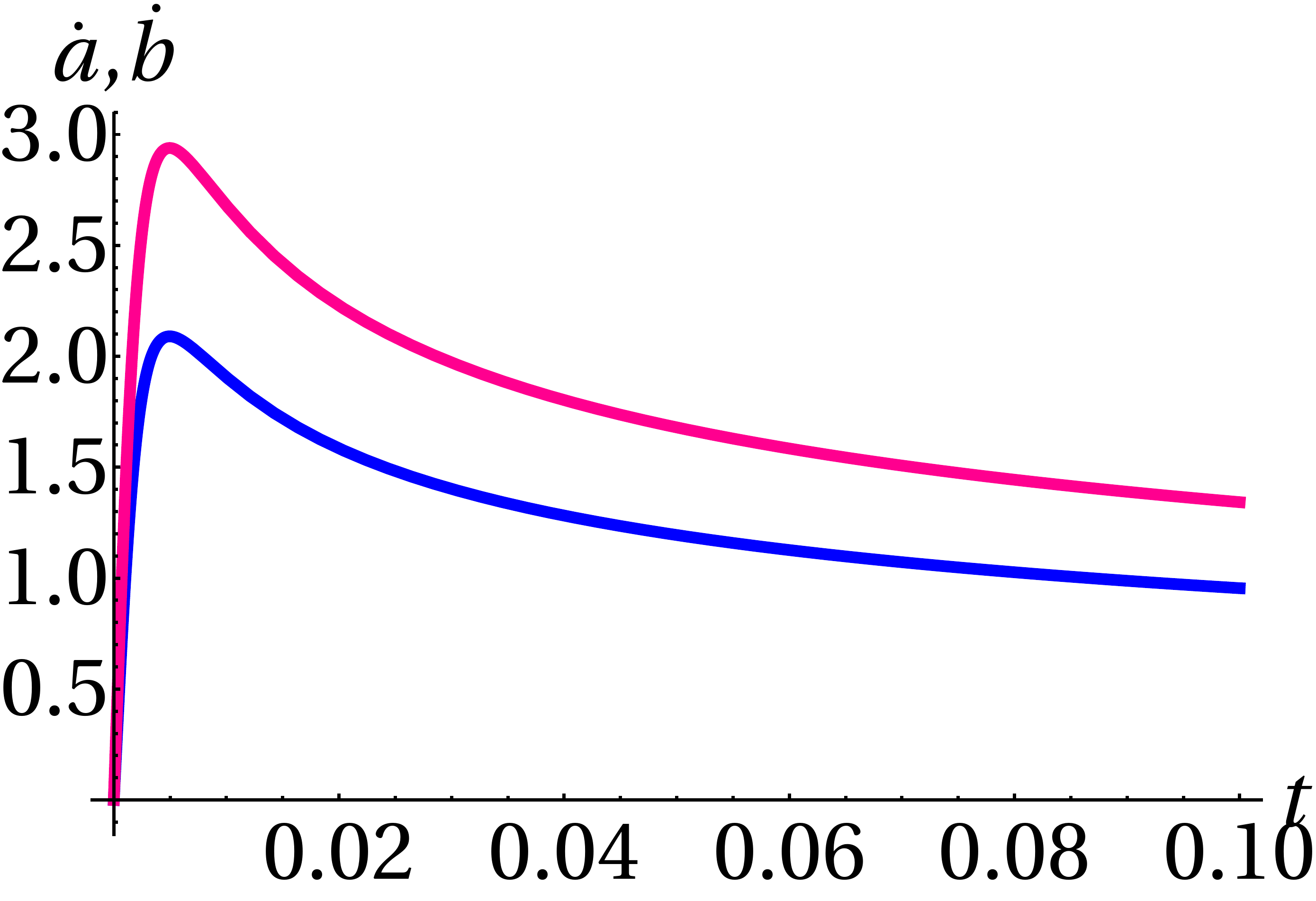}
    \caption{The expansion rates of the scale factors: $\dot a$ is represented by the blue curve, and $\dot b$ by the red curve. The curve for $\dot b$ is scaled down by a factor of 40 to fit in the graph.}
    \label{104}
  \end{subfigure}
\\[9em]
  \begin{subfigure}[b]{.5\linewidth}
    \centering
    \includegraphics[width=0.7\columnwidth]{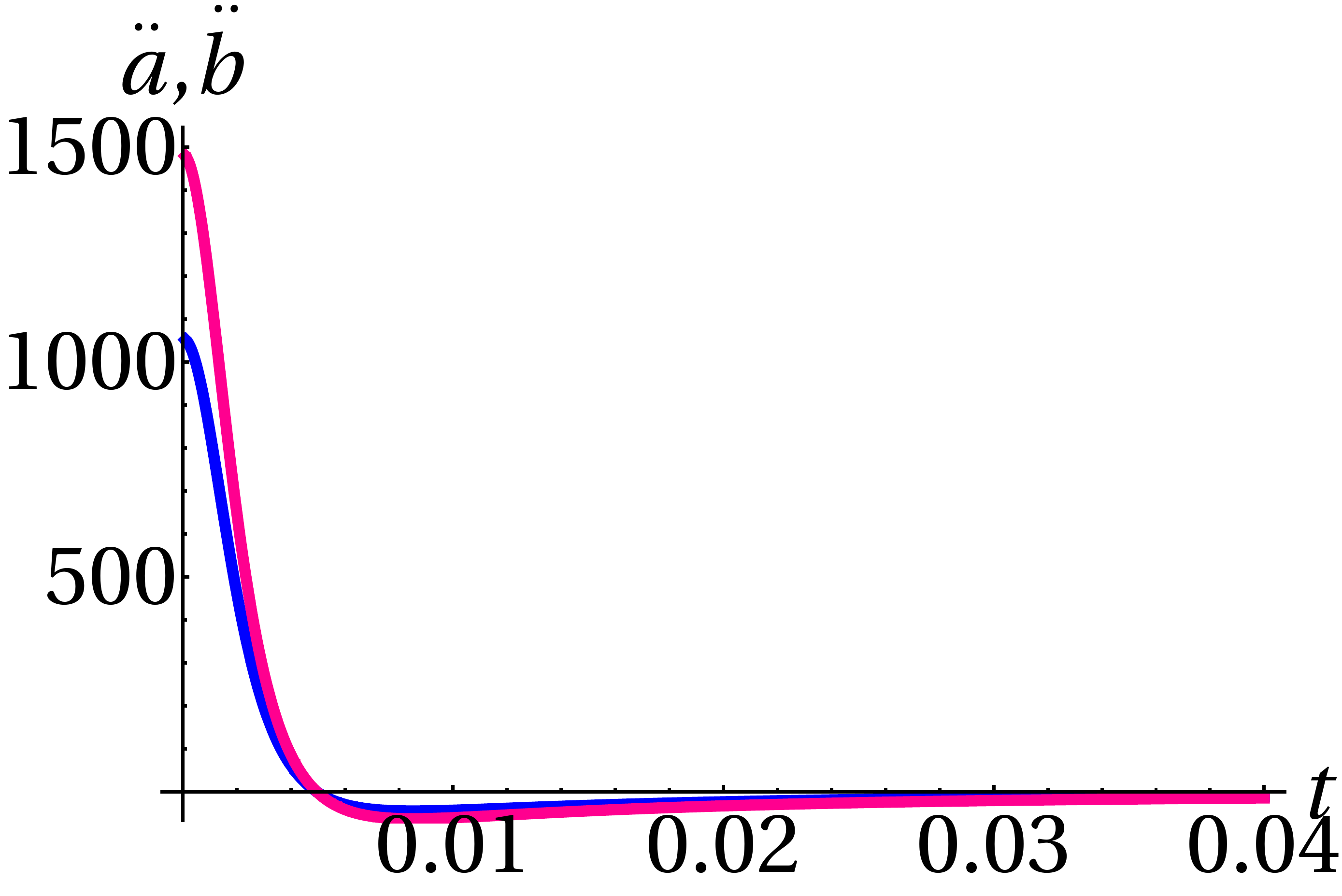}
    \caption{The accelerations of the scale factors: $\ddot a$ is represented by the blue curve, and $\ddot b$ by the red curve (scaled down 40-fold).}
    \label{105}
  \end{subfigure}
\qquad
  \begin{subfigure}[b]{.5\linewidth}
    \centering
    \includegraphics[width=0.7\columnwidth]{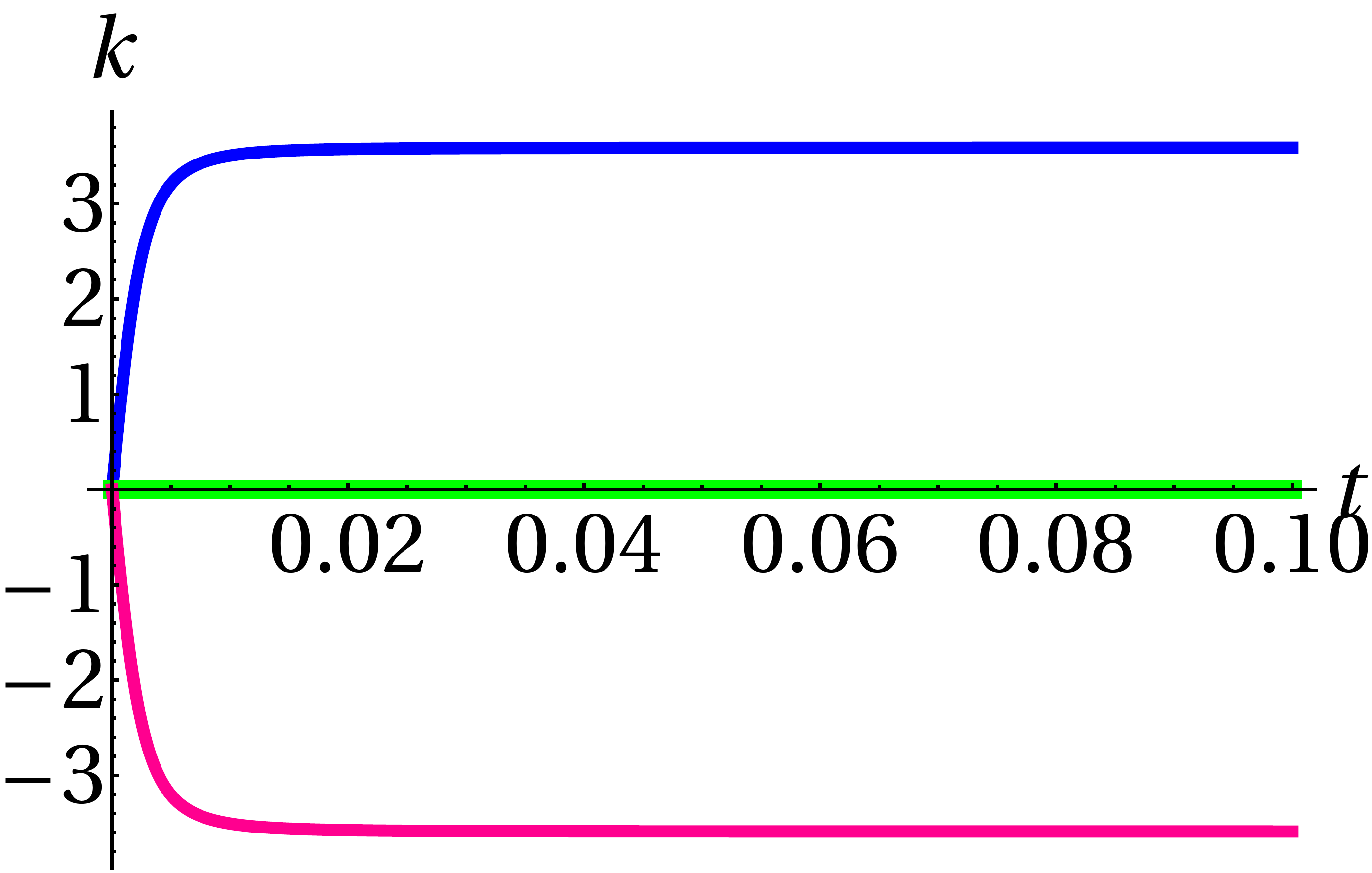}
    \caption{The harmonic function $k$ using: $\dot k\left(0\right)=1$ (blue curve), $\dot k\left(0\right)=0$ (green line), and $\dot k\left(0\right)=-1$ (red curve).}
    \label{106}
  \end{subfigure}
 \caption{Dust-filled brane world with initial conditions set number 6.}
  \label{Fig21}
\end{figure}
\begin{figure}[H]
\begin{subfigure}[t]{.5\linewidth}
    \centering
    \includegraphics[width=0.7\columnwidth]{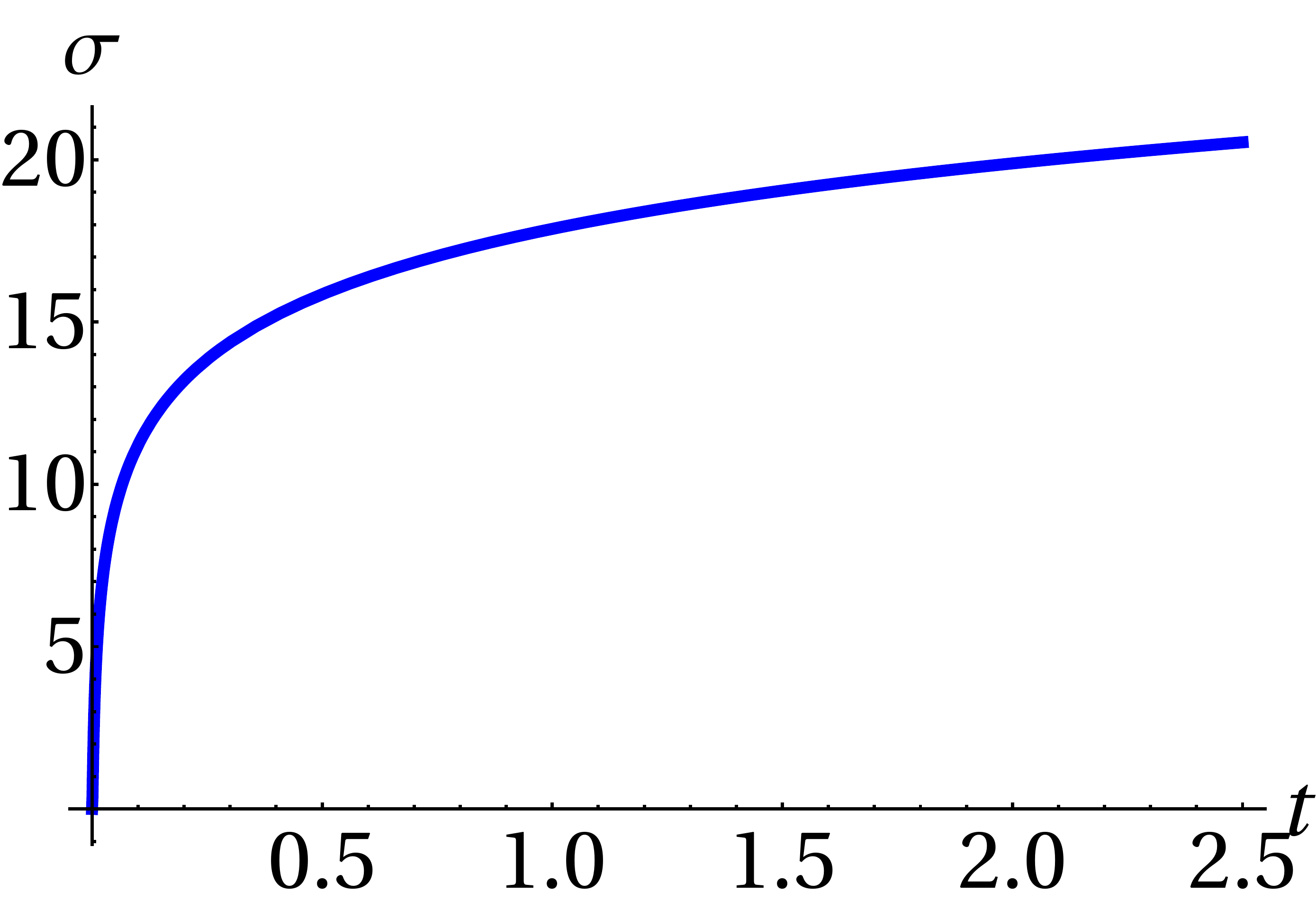}
    \caption{The dilaton $\sigma$; same for all three $\dot k\left(0\right)$.}
    \label{107}
  \end{subfigure}
\qquad
  \begin{subfigure}[t]{.5\linewidth}
    \centering
    \includegraphics[width=0.7\columnwidth]{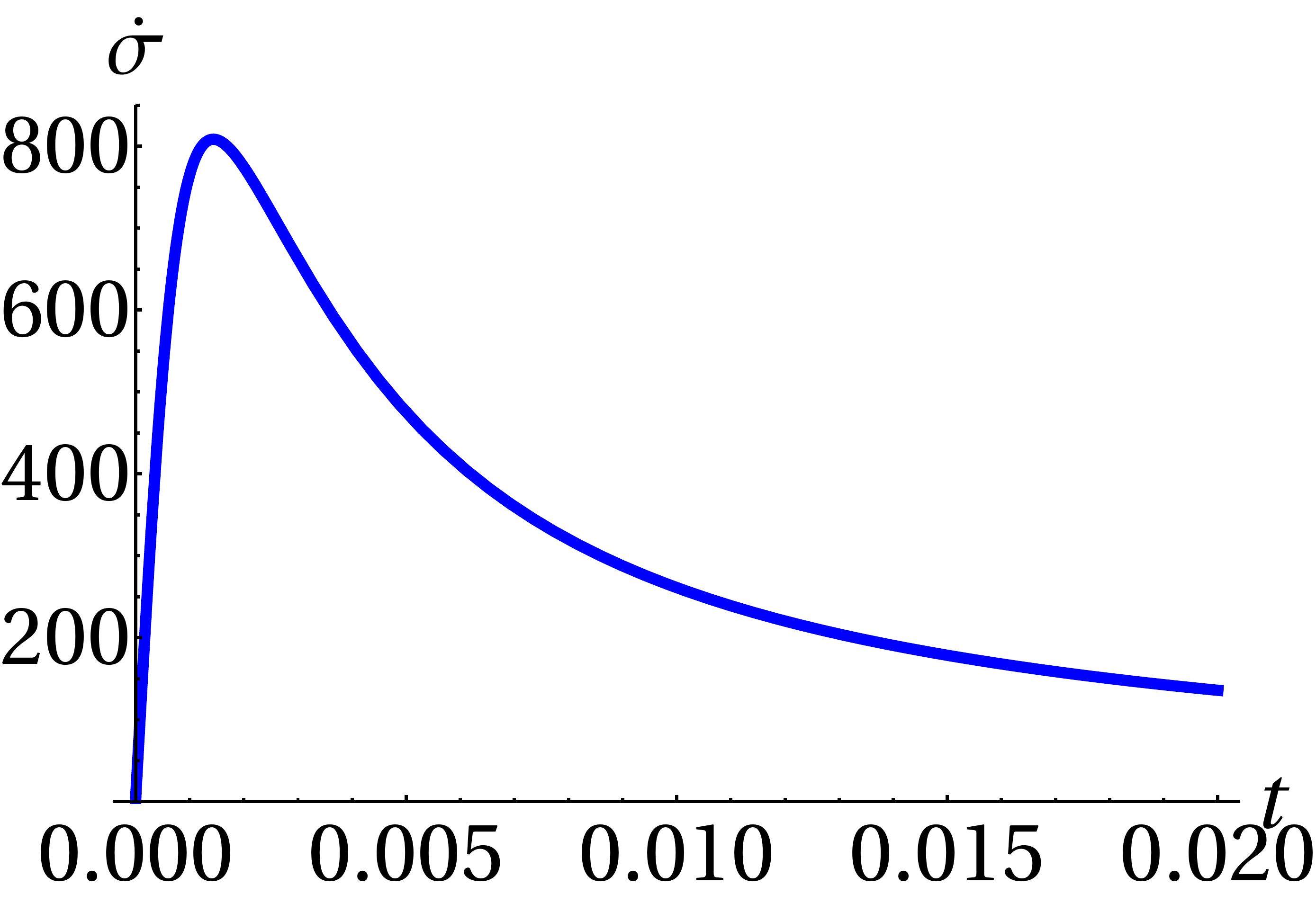}
    \caption{The dilatonic field strength $\dot\sigma$.}
    \label{108}
  \end{subfigure}  
\\[4em]
\begin{subfigure}[t]{.5\linewidth}
    \centering
    \includegraphics[width=0.7\columnwidth]{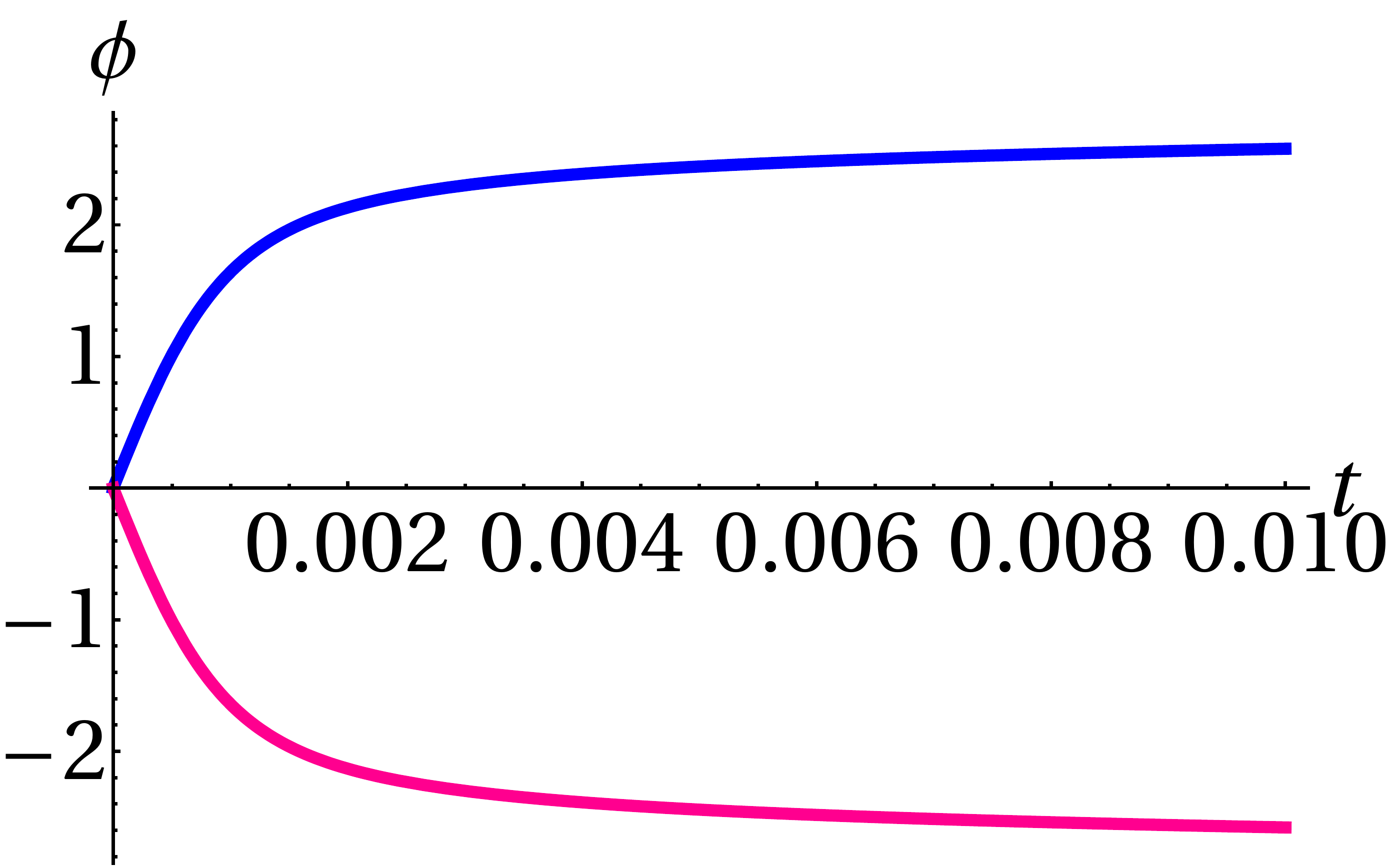}
    \caption{The universal axion $\phi$ for $\dot k\left(0\right) = 1$ (blue curve), and $\dot k\left(0\right) = -1$ (red curve). The solution diverges for $\dot k\left(0\right)=0$.}
    \label{109}
  \end{subfigure}
\qquad
  \begin{subfigure}[t]{.5\linewidth}
    \centering
    \includegraphics[width=0.7\columnwidth]{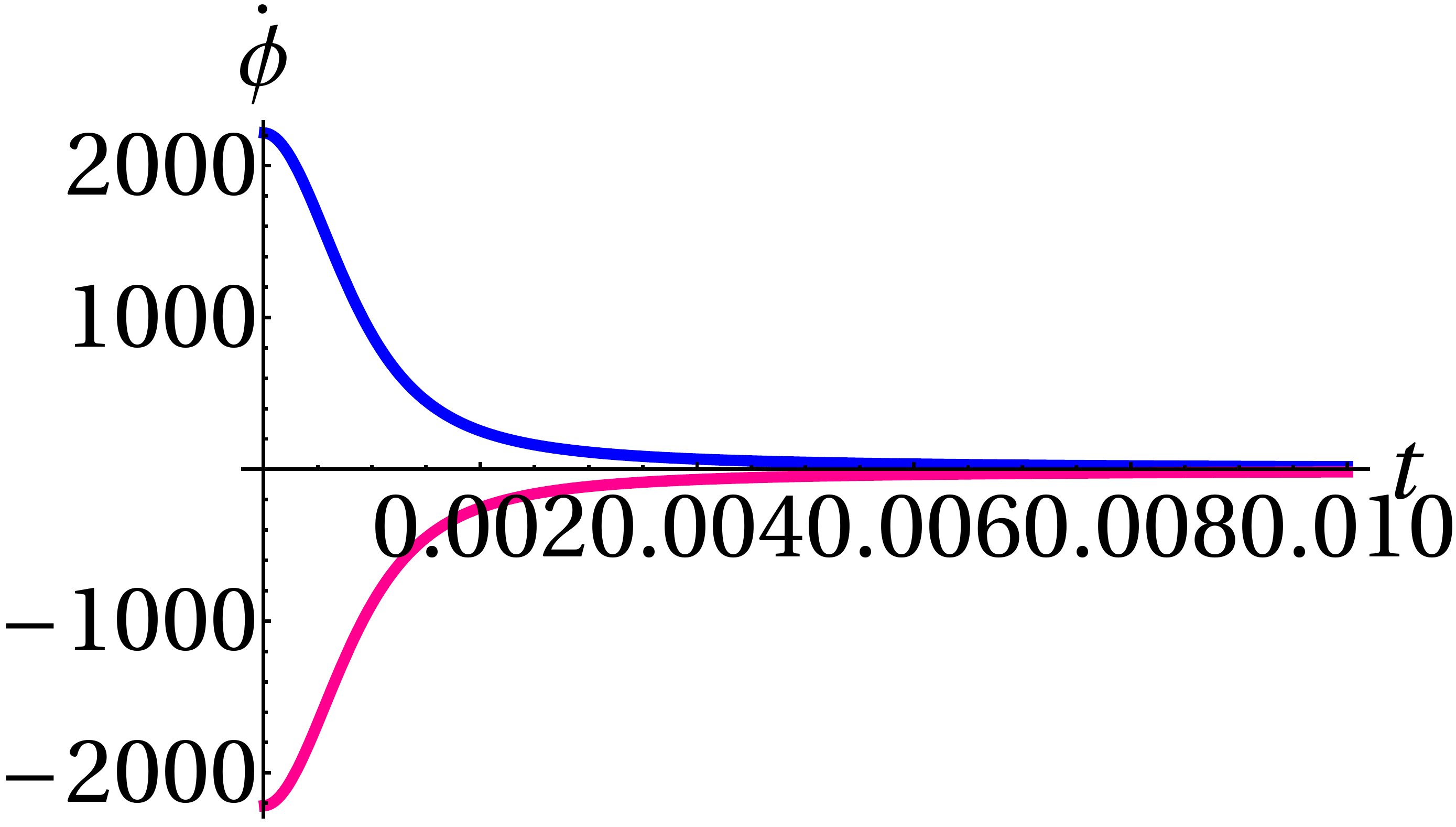}
    \caption{The axionic field strength $\dot\phi$ for $\dot k\left(0\right) = 1$ (blue curve), and $\dot k\left(0\right) = -1$ (red curve).}
    \label{110}
  \end{subfigure}
\\[4em]
 \begin{subfigure}[t]{.5\linewidth}
    \centering
    \includegraphics[width=0.7\columnwidth]{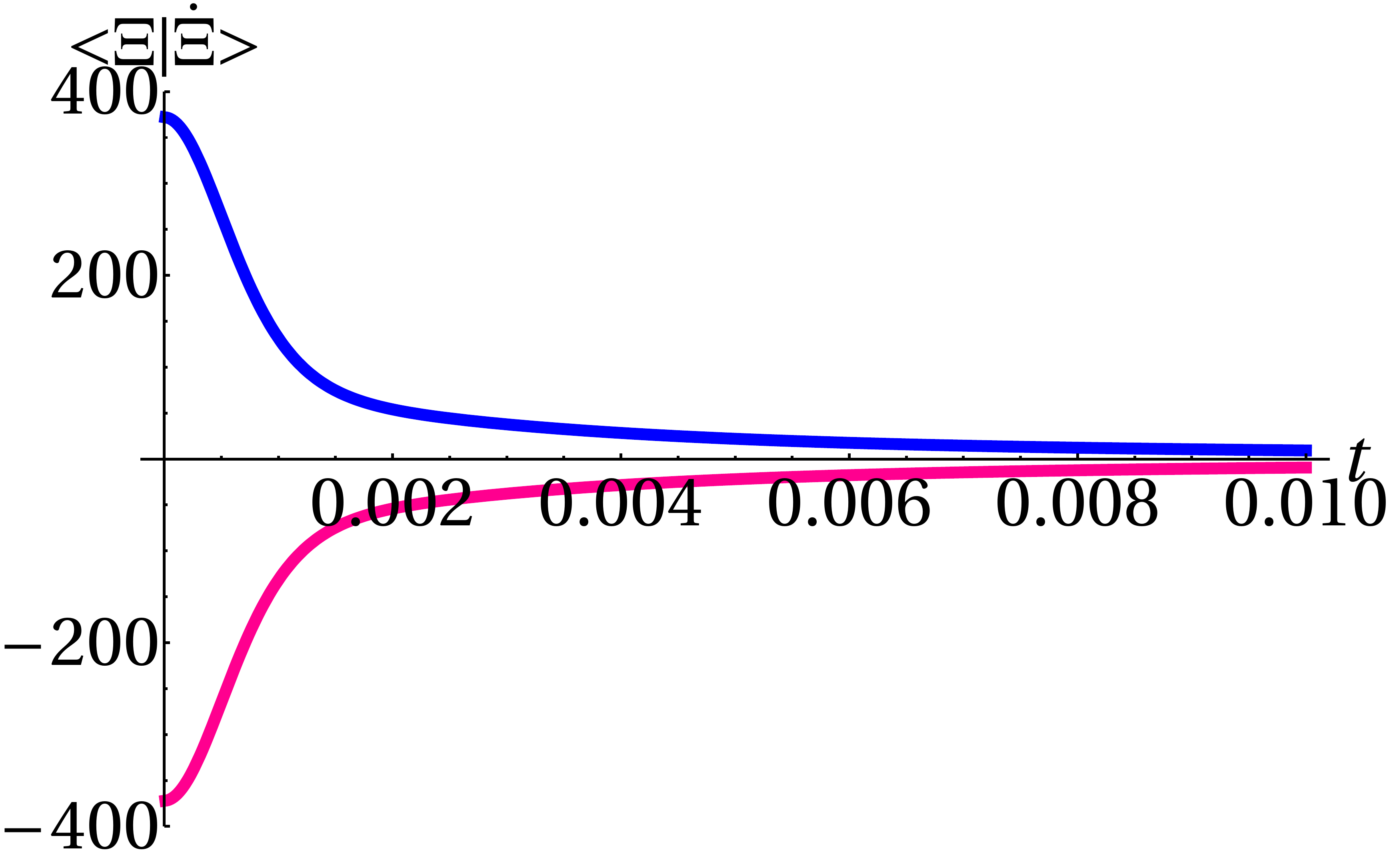}
    \caption{$ \langle \Xi | \dot{\Xi} \rangle$ for $\dot{k}(0)= 1$  (blue), and $\dot{k}(0)  = -1$ (red).}
    \label{111}
  \end{subfigure}
\qquad
  \begin{subfigure}[t]{.5\linewidth}
    \centering
    \includegraphics[width=0.7\columnwidth]{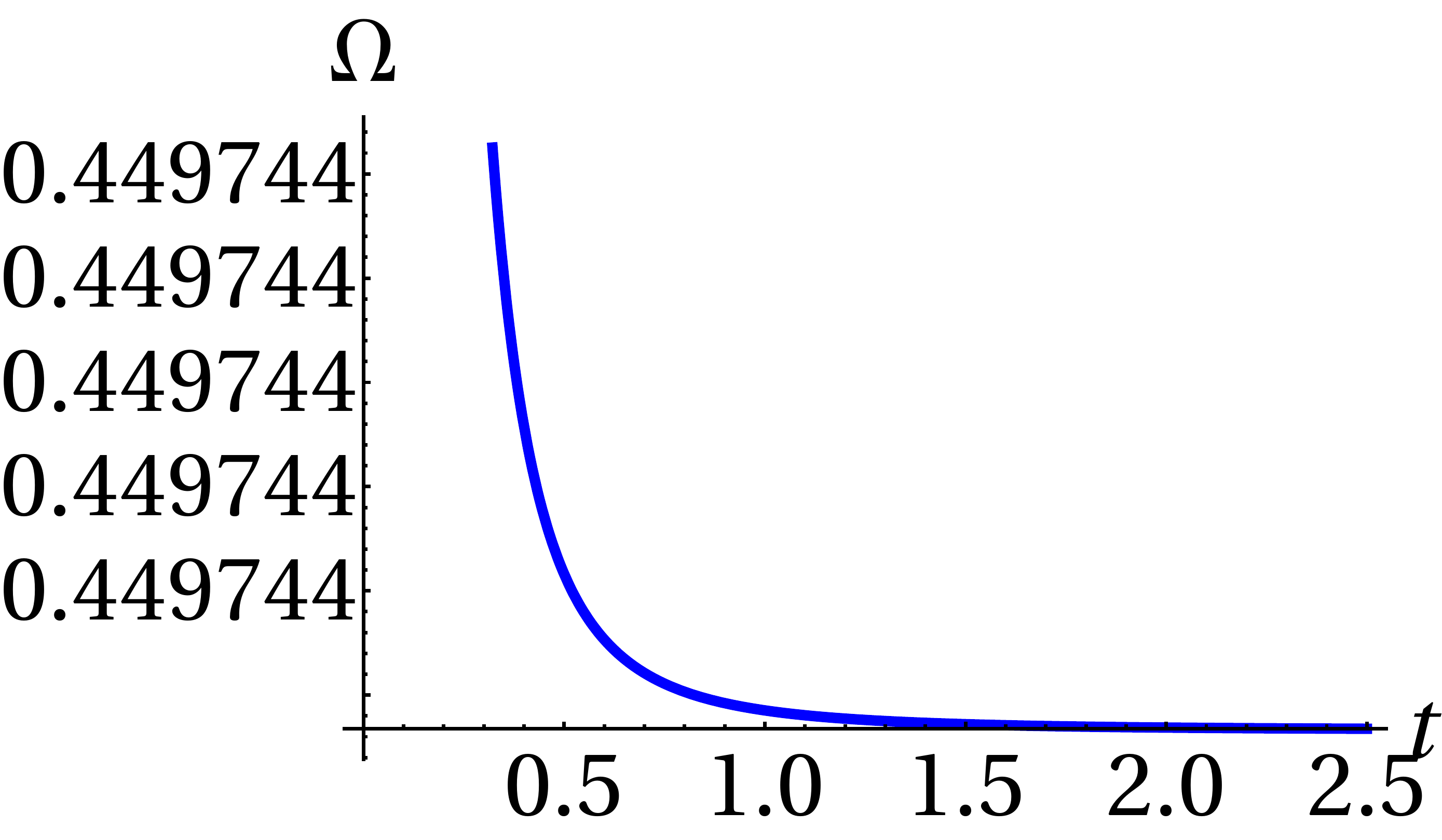}
    \caption{$\Omega$ at $\dot k\left(0\right) = 1$ and  $ \dot\sigma \left(0\right) =0 $.}
    \label{112}
  \end{subfigure}
  \caption{Dust-filled brane world with initial conditions set number 6 (continued).}
  \label{Fig21}
\end{figure}
Comparing the graphs for the various initial conditions we conclude the following: It is clear that the behavior of the scaling factors $a$ and $b$ is directly related to the moduli. The norm $G_{i\bar j} \dot z^i \dot z^{\bar j}$ always starts at a large value then decays. Even when it increases a bit as in IC5 Fig. (\ref{43}) it then peaks and decays. As a consequence the brane inevitably expands; \emph{even} when given contracting initial conditions (IC5). The larger the norm's initial value the smaller the starting size of the brane. The norm asymptotes to zero as the brane settles in its expected negative acceleration after an initial short phase of rapid accelerated expansion in most cases (except IC4). This last is particularly suggestive if one considers it in cosmological terms; as it implies an early and short inflationary epoch that coincides with the initial dynamics of the hypermultiplet fields. The next layer of initial conditions belongs to the harmonic function $k$; for which we have chosen positive, negative and zero initial velocities. The vanishing initial velocity case leads to a vanishing $k$ for all times, which is problematic for $\phi$, $\dot\phi$ and $\langle \Xi | \dot{\Xi} \rangle$ and thus we take as trivial, while the positive and negative values lead to a well-defined behavior for the $\phi$, $\dot\phi$ and $\langle \Xi | \dot{\Xi} \rangle$ fields. It is clear that for most initial conditions (the exception is IC5) $k$ expands rapidly then asymptotes to a constant value which leads to a similar behavior for $\phi$, $\dot\phi$ and $\langle \Xi | \dot{\Xi} \rangle$. The dilaton $\sigma$ (with initial conditions $\sigma=\dot \sigma=0$ at $t=0$) and its field strength $\dot\sigma$ do not seem to be sensitive to the $k$ initial conditions and neither does the function $\Omega$. In short we conclude that an initial phase of large moduli decay coincides with a rapid change in the rest of the hypermultiplet fields; leading to inflation. While the large $t$ behavior for all fields is a Friedmann-like negatively accelerated expansion, and vanishing hypermultiplet field strengths. If we consider our own universe as such a brane, then all of this certainly makes perfect sense.

\section{A radiation-filled brane}
\label{rr}
For a radiation- filled brane 
\be
    T_{tt}^{{\emph Brane}}  = \rho\left(t\right) = \frac{1}{{a^4 }}, ~~~~~~~~~~ p(t) =\frac{1}{{3a^4 }},
\label{Rad}
\ee
where $p\left( t \right)$ is the pressure of the fluid related to the density via the equation of state $p = \frac{1}{3}\rho$.
This leads to the Friedmann-like equations:
\bea
 3\left[ {\left( {\frac{{\dot a}}{a}} \right)^2  + \left( {\frac{{\dot a}}{a}} \right)\left( {\frac{{\dot b}}{b}} \right)} \right] &=& G_{i\bar j} \dot z^i \dot z^{\bar j}  + \frac{1}{{a^4 }} \nonumber\\
 2\frac{{\ddot a}}{a} + \left( {\frac{{\dot a}}{a}} \right)^2  + \frac{{\ddot b}}{b} + 2\left( {\frac{{\dot a}}{a}} \right)\left( {\frac{{\dot b}}{b}} \right) &=&  - G_{i\bar j} \dot z^i \dot z^{\bar j}  - \frac{1}{{3a^4 }}\nonumber\\
 3\left[ {\frac{{\ddot a}}{a} + \left( {\frac{{\dot a}}{a}} \right)^2 } \right] &=&  - G_{i\bar j} \dot z^i \dot z^{\bar j}.
\eea
The majority of the solutions in this section the norm $G_{i\bar j} \dot z^i \dot z^{\bar j}$ is negative for all times and we plot its absolute value, \emph{except} for IC4 where it is positive from the start.


\begin{figure}[H]
  \begin{subfigure}[t]{.5\linewidth}
    \centering
    \includegraphics[width=0.7\columnwidth]{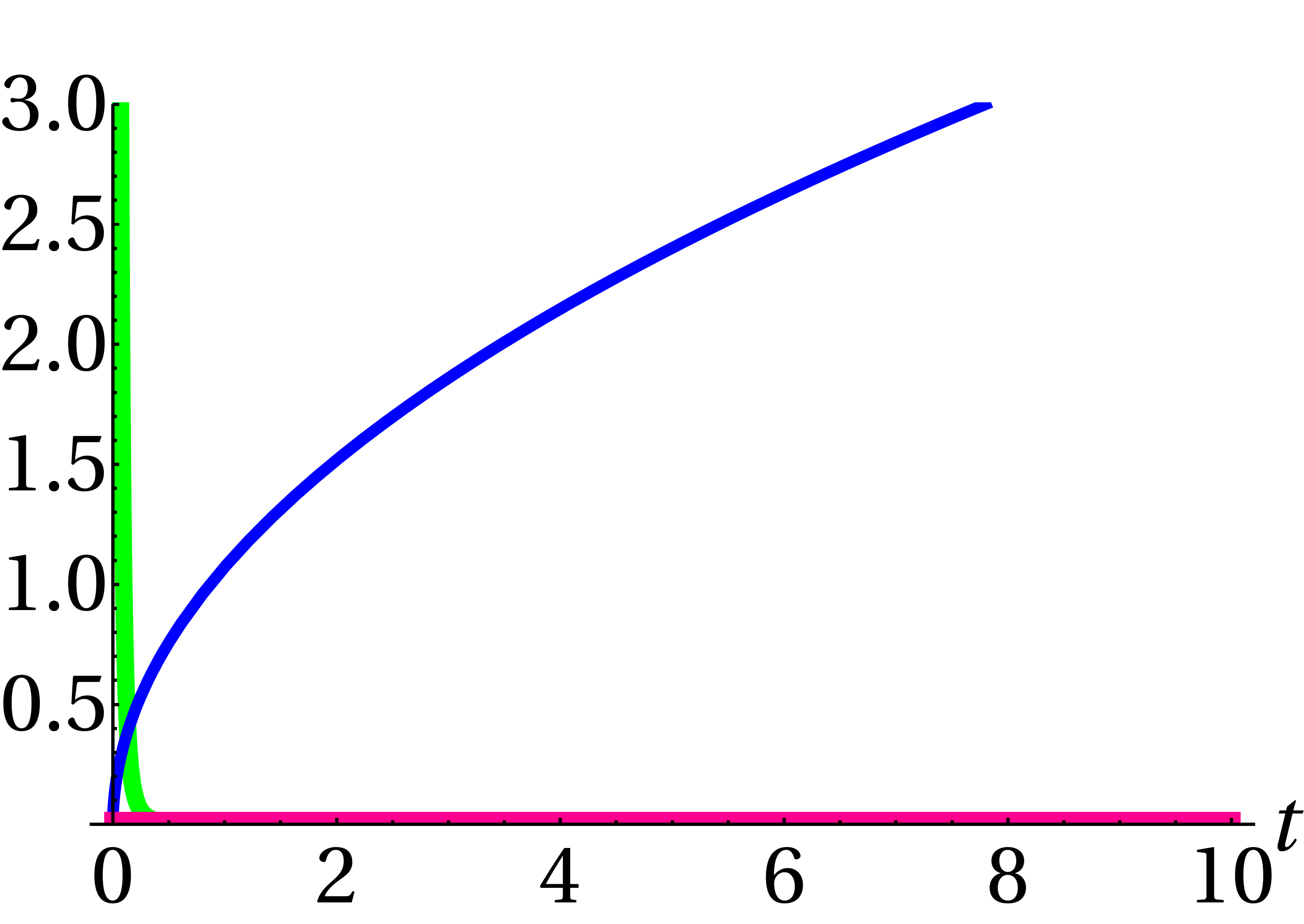}
    \caption{The scale factor $a$ is represented by the blue curve, $b$ by the red curve (flat on the $t$ axis), while $\left| {G_{i\bar j} \dot z^i \dot z^{\bar j}} \right|$ by the green curve.}
    \label{53}
  \end{subfigure}
\qquad
  \begin{subfigure}[t]{.5\linewidth}
    \centering
    \includegraphics[width=0.7\columnwidth]{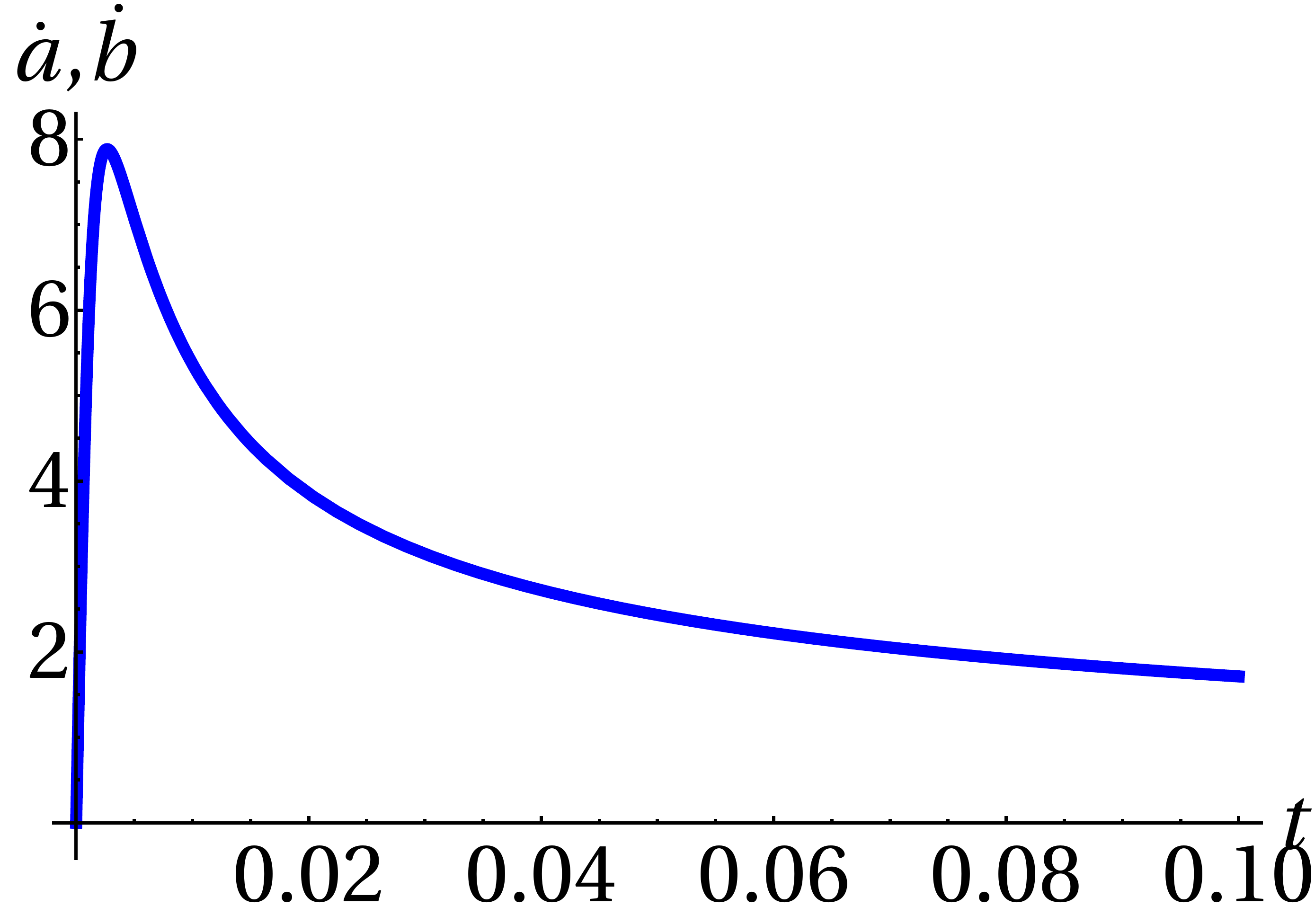}
    \caption{The expansion rates of the scale factors: $\dot a$ is represented by the blue curve, and $\dot b$ by the red curve (flat on the $t$ axis).}
    \label{54}
  \end{subfigure}
\\[9em]
  \begin{subfigure}[t]{.5\linewidth}
    \centering
    \includegraphics[width=0.7\columnwidth]{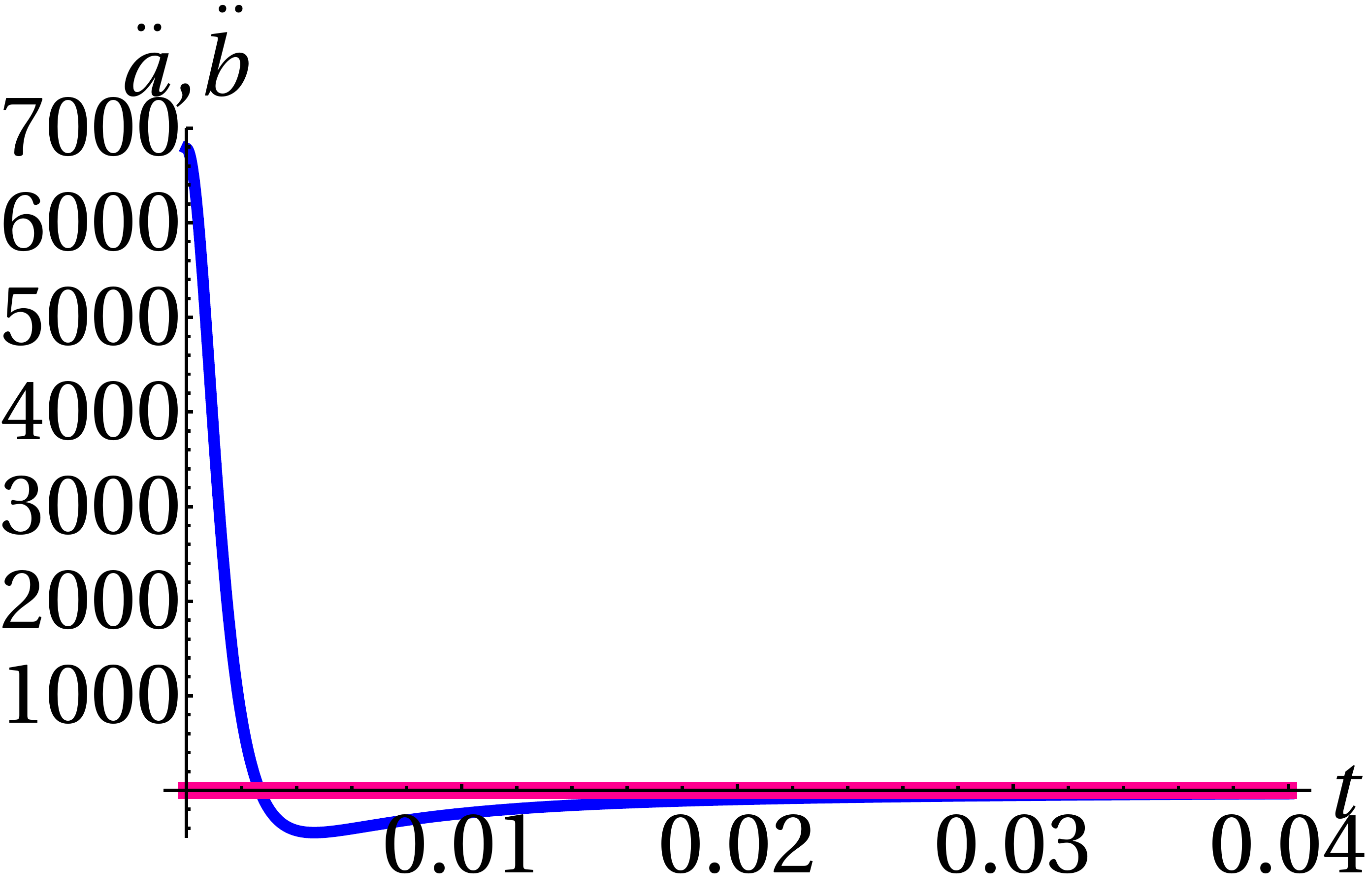}
    \caption{The accelerations of the scale factors: $\ddot a$ is represented by the blue curve, and $\ddot b$ by the red curve (flat on the $t$ axis).}
    \label{55}
  \end{subfigure}
\qquad
  \begin{subfigure}[t]{.5\linewidth}
    \centering
    \includegraphics[width=0.7\columnwidth]{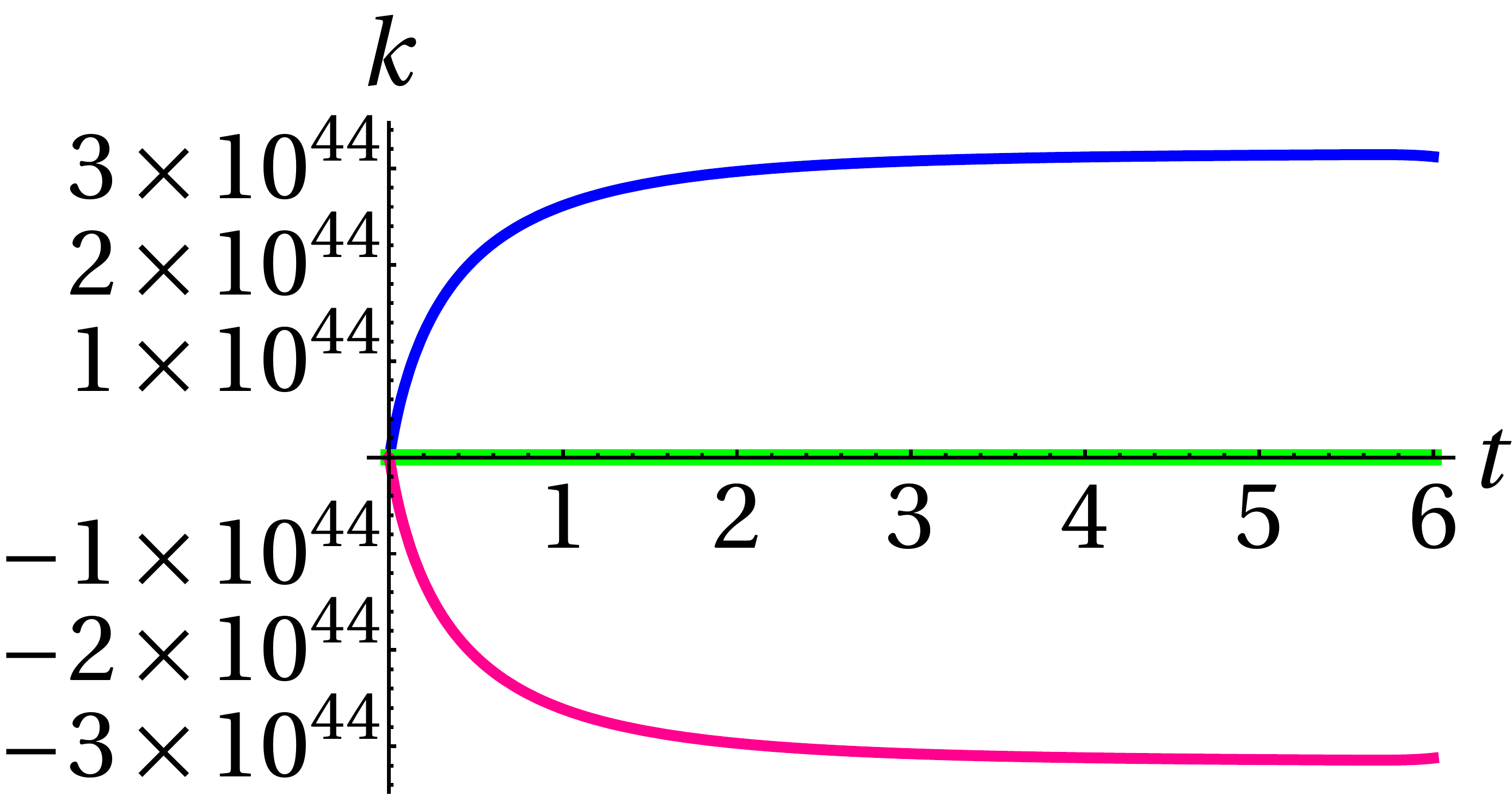}
    \caption{The harmonic function $k$ using: $\dot k\left(0\right)=1$ (blue curve), $\dot k\left(0\right)=0$ (green line), and $\dot k\left(0\right)=-1$ (red curve).}
    \label{56}
  \end{subfigure}
\caption{Radiation-filled brane world with initial conditions set number 1.}
  \label{Fig11}
\end{figure}
\begin{figure}[H]
\begin{subfigure}[t]{.5\linewidth}
    \centering
    \includegraphics[width=0.7\columnwidth]{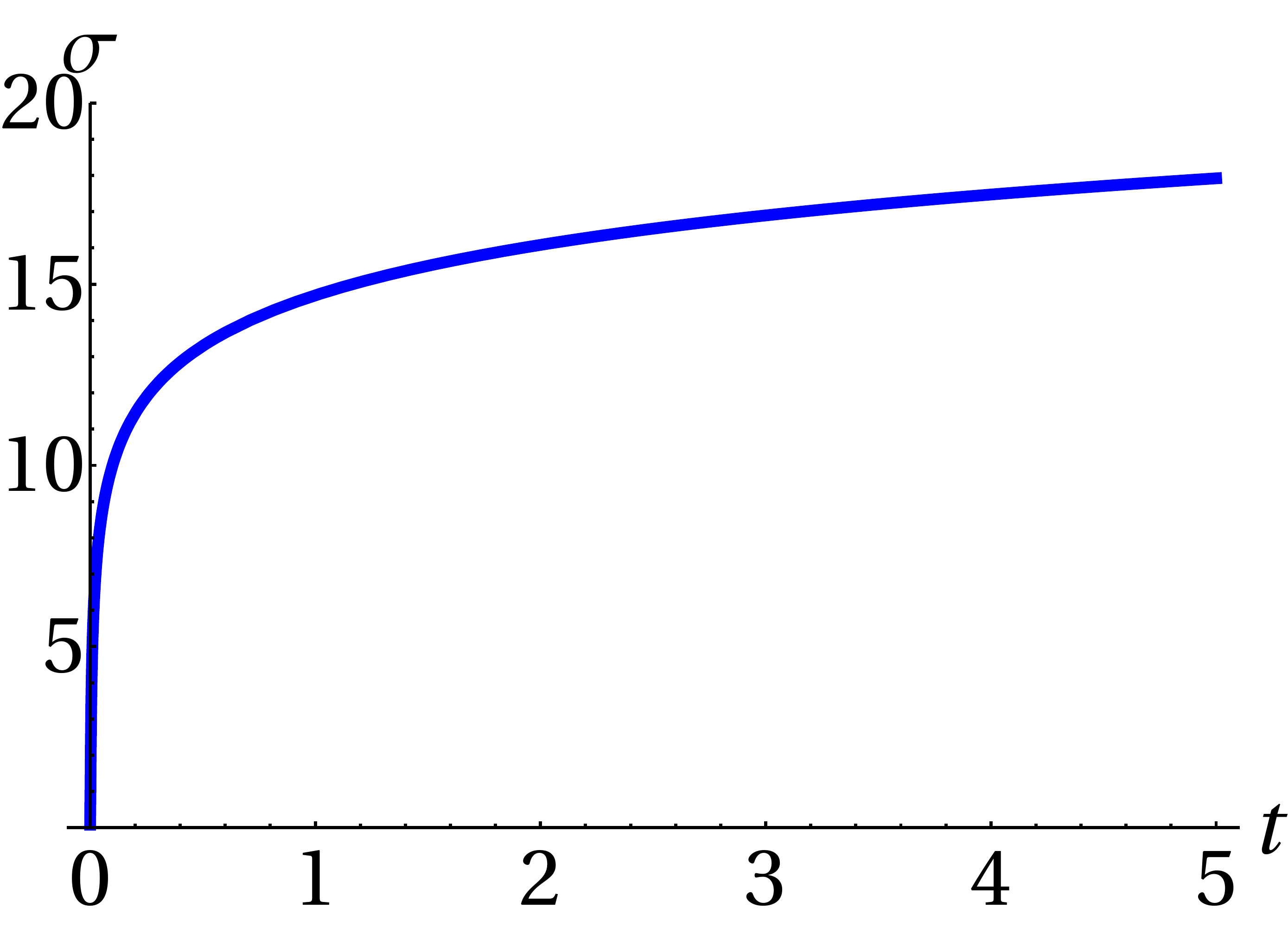}
    \caption{The dilaton $\sigma$; same for all three $\dot k\left(0\right)$.}
    \label{57}
  \end{subfigure}
\qquad
 \begin{subfigure}[t]{.5\linewidth}
    \centering
    \includegraphics[width=0.7\columnwidth]{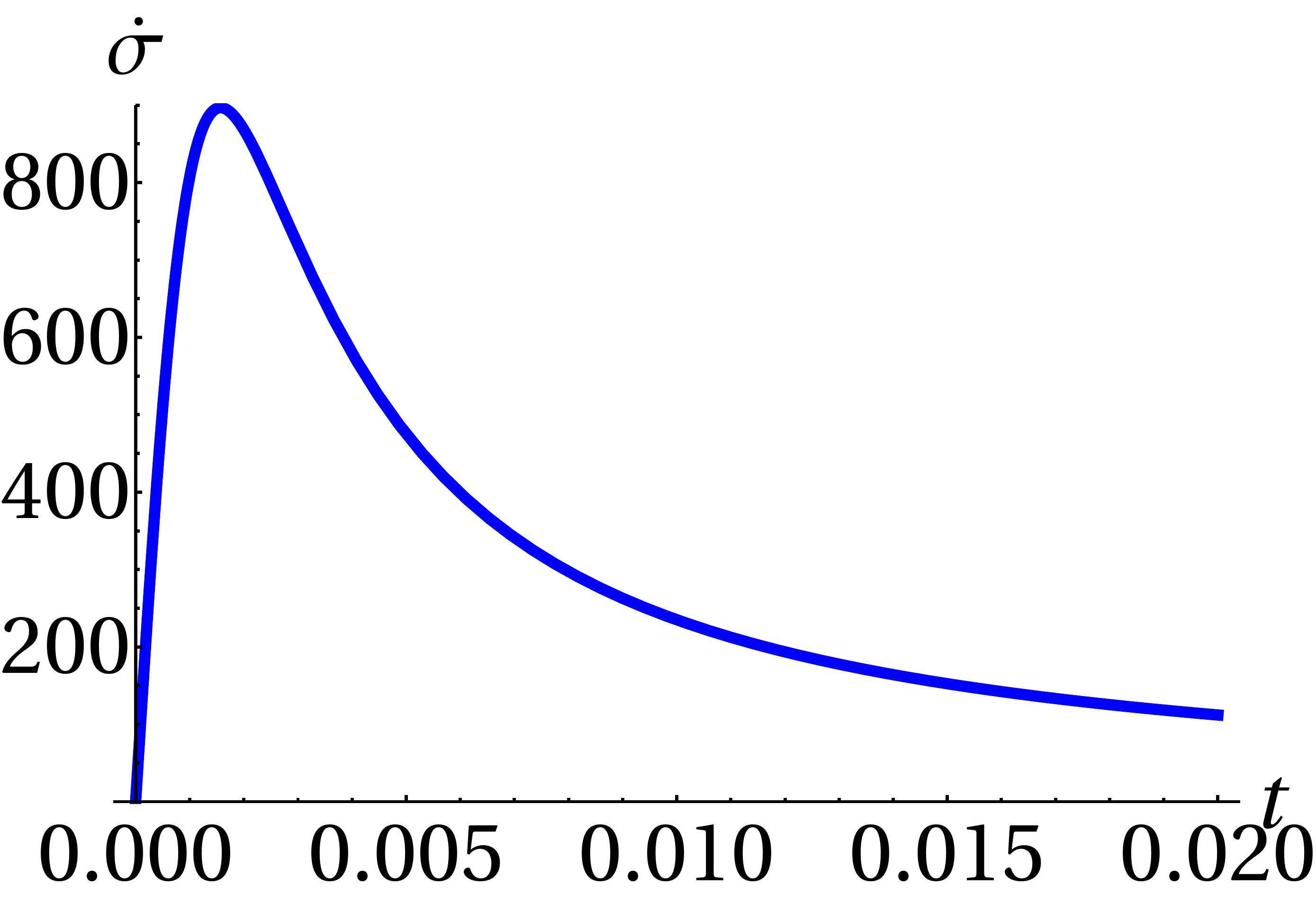}
    \caption{The dilatonic field strength $\dot\sigma$.}
    \label{58}
  \end{subfigure}
\\[4em]
 \begin{subfigure}[t]{.5\linewidth}
    \centering
    \includegraphics[width=0.7\columnwidth]{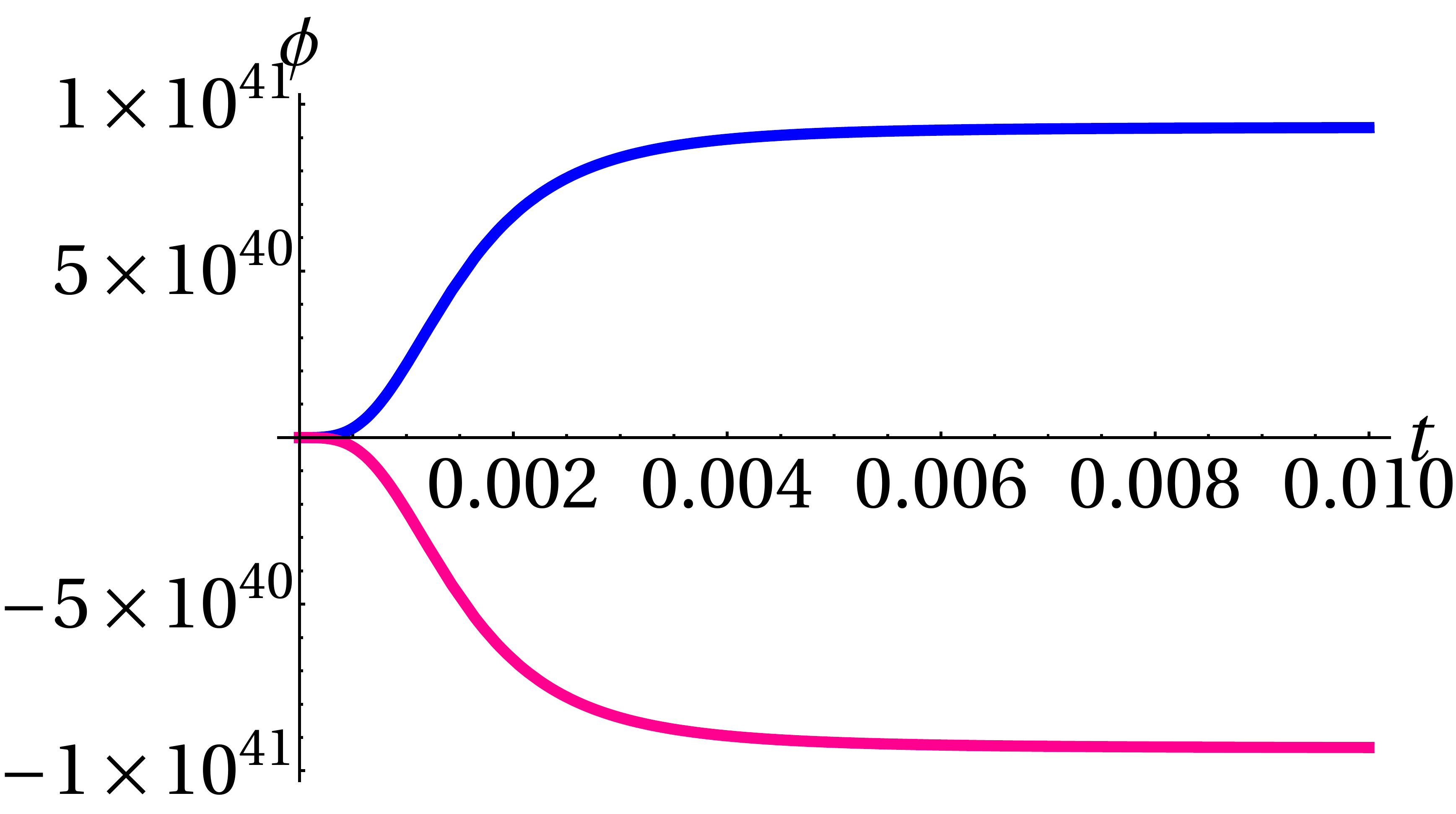}
    \caption{The universal axion $\phi$ for $\dot k\left(0\right) = 1$ (blue curve), and $\dot k\left(0\right) = -1$ (red curve). The solution diverges for $\dot k\left(0\right)=0$.}
    \label{59}
  \end{subfigure}
\qquad
  \begin{subfigure}[t]{.5\linewidth}
    \centering
    \includegraphics[width=0.7\columnwidth]{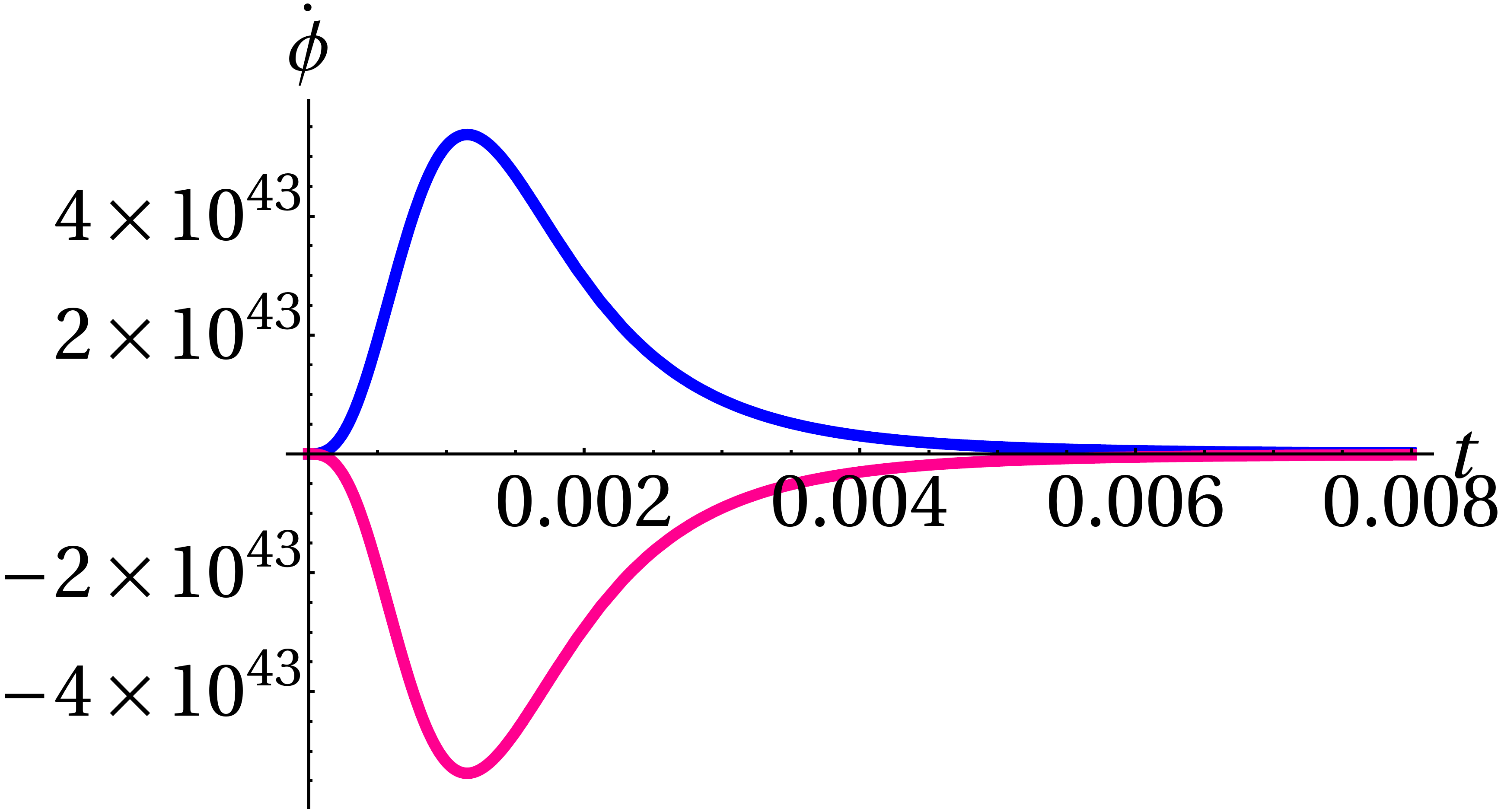}
    \caption{The axionic field strength $\dot\phi$ for $\dot k\left(0\right) = 1$ (blue curve), and $\dot k\left(0\right) = -1$ (red curve).}
    \label{60}
  \end{subfigure}
\\[4em]
\begin{subfigure}[t]{.5\linewidth}
    \centering
    \includegraphics[width=0.7\columnwidth]{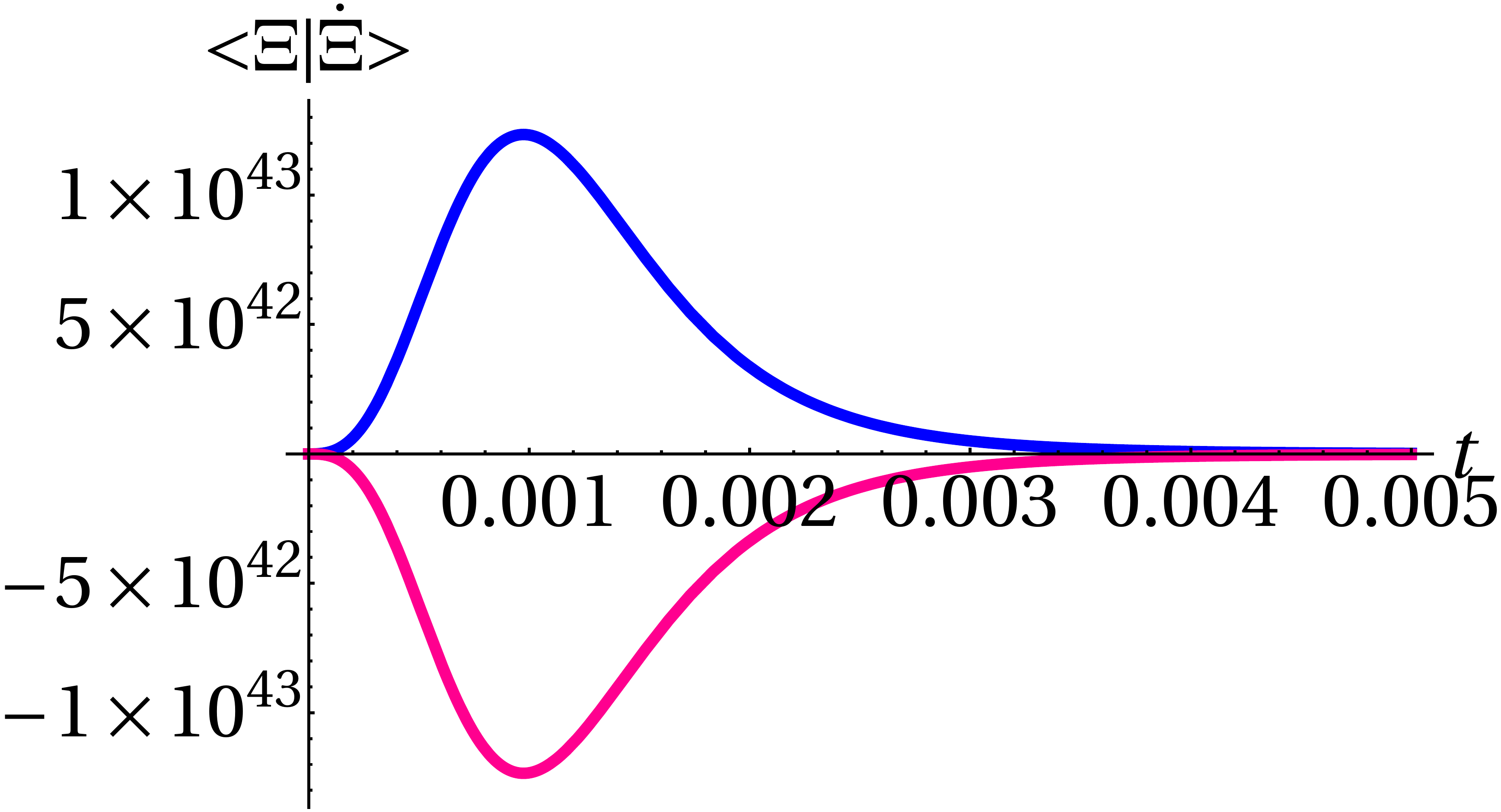}
    \caption{$ \langle \Xi | \dot{\Xi} \rangle$ for $\dot{k}(0)= 1$  (blue), and $\dot{k}(0)  = -1$ (red).}
    \label{61}
  \end{subfigure}
\qquad
  \begin{subfigure}[t]{.5\linewidth}
    \centering
    \includegraphics[width=0.7\columnwidth]{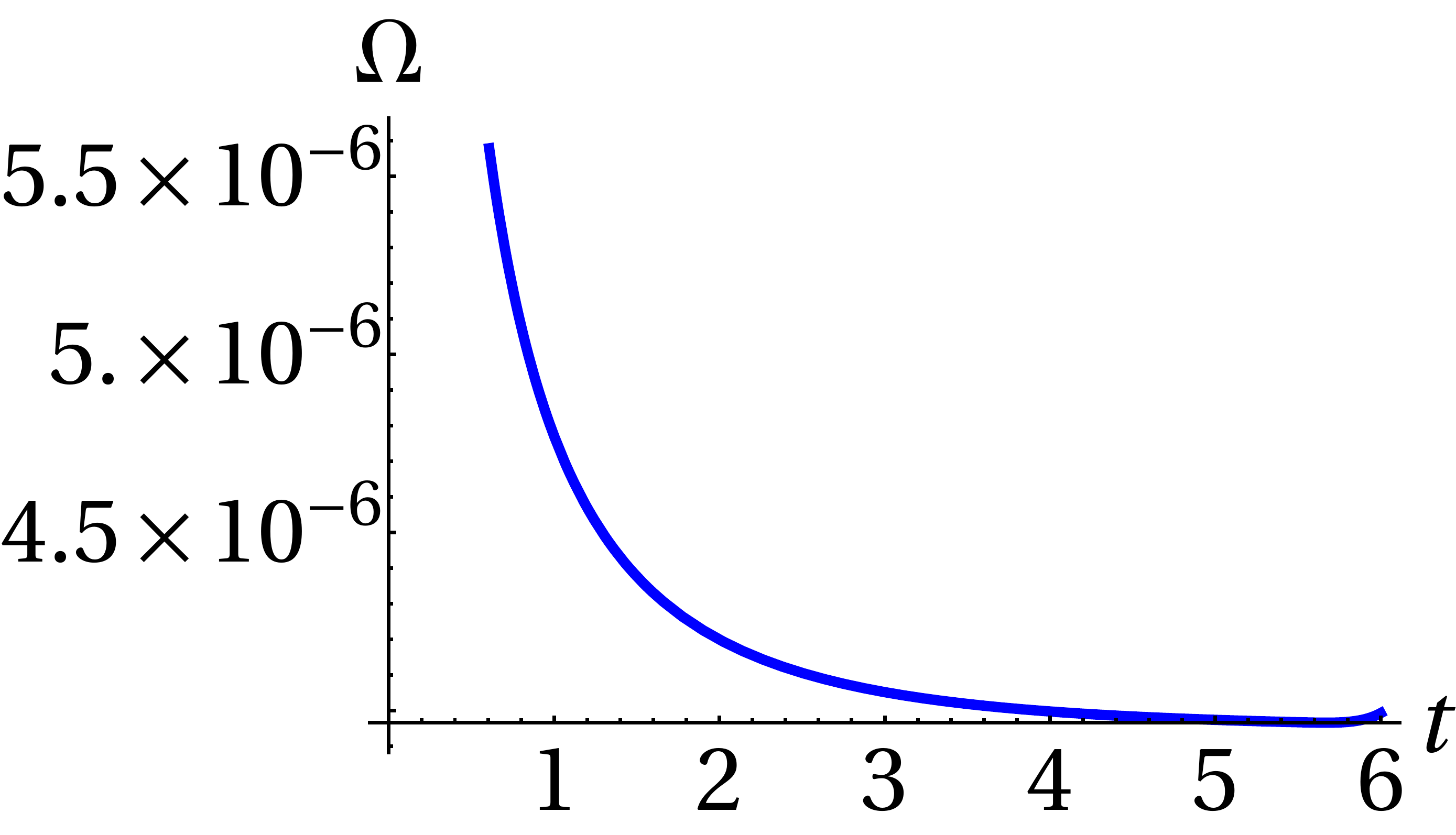}
    \caption{$\Omega$ at $\dot k\left(0\right) = 1$ and  $ \dot\sigma \left(0\right) =0 $.}
    \label{62}
  \end{subfigure}
\caption{Radiation-filled brane world with initial conditions set number 1 (continued).}
  \label{Fig12}
\end{figure}


\begin{figure}[H]
  \begin{subfigure}[t]{.5\linewidth}
    \centering
    \includegraphics[width=0.7\columnwidth]{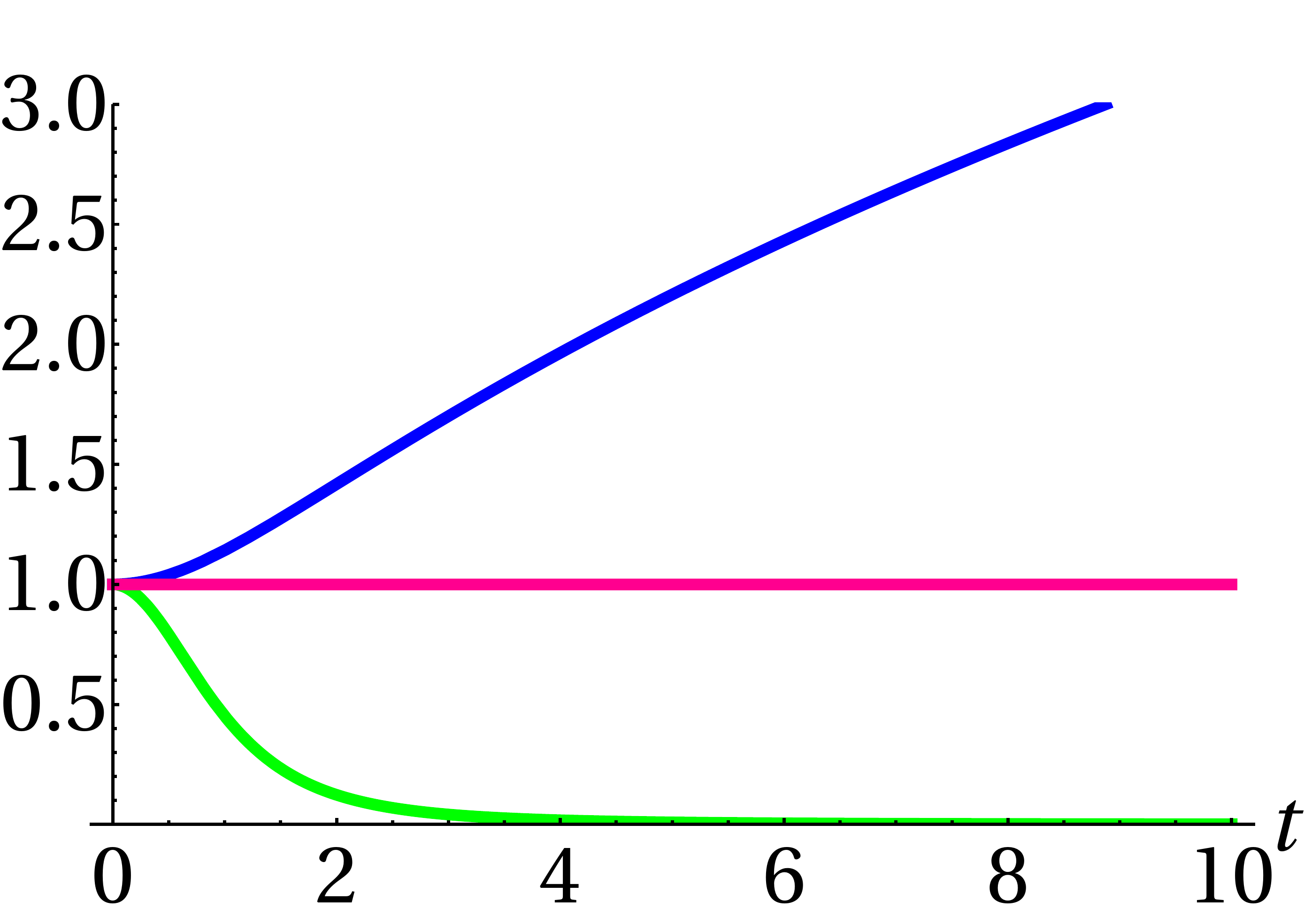}
    \caption{The scale factor $a$ is represented by the blue curve, $b$ by the red curve, while $\left| {G_{i\bar j} \dot z^i \dot z^{\bar j}} \right|$ by the green curve.}
    \label{63}
  \end{subfigure}
\qquad
  \begin{subfigure}[t]{.5\linewidth}
    \centering
    \includegraphics[width=0.7\columnwidth]{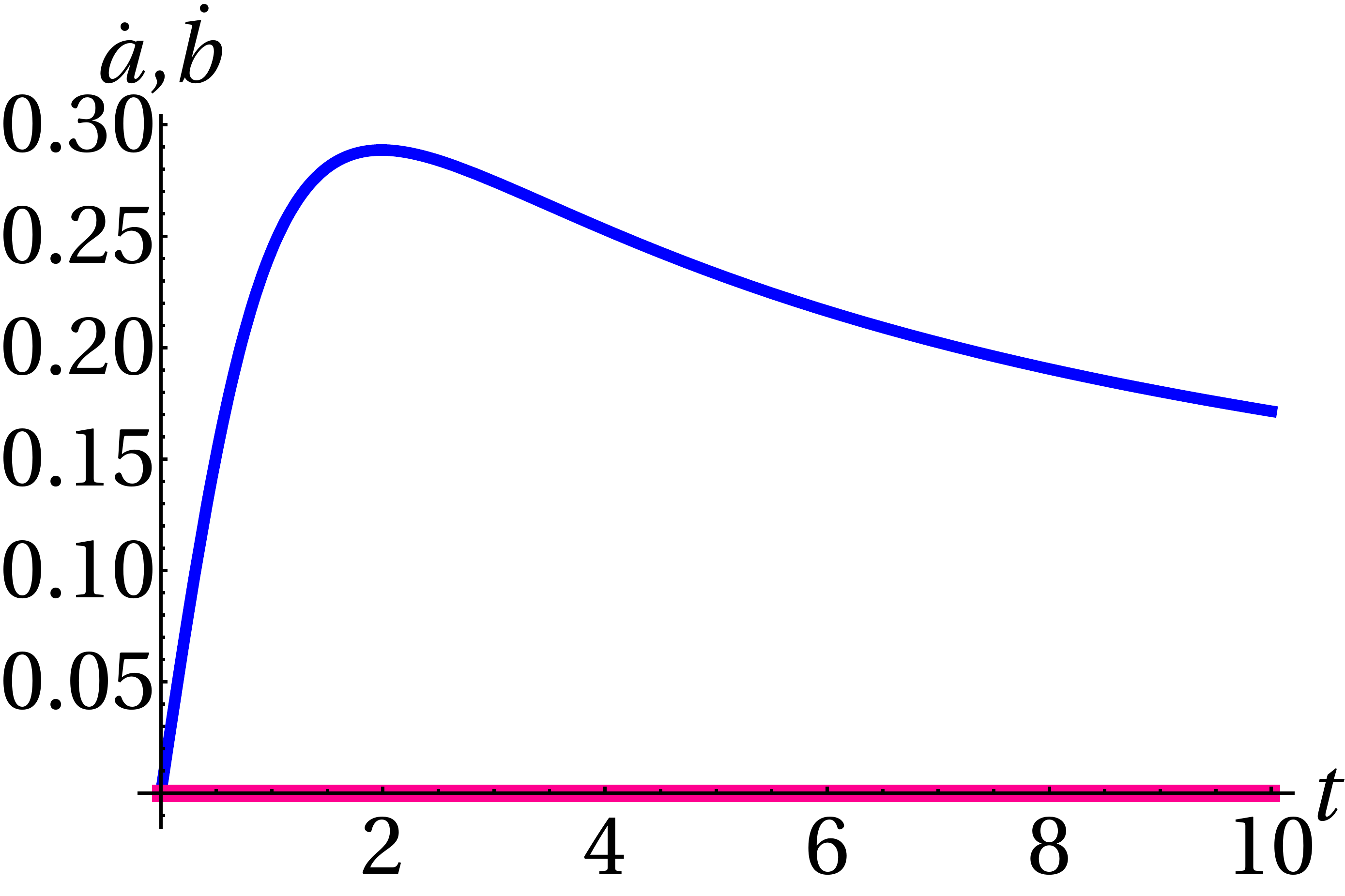}
    \caption{The expansion rates of the scale factors: $\dot a$ is represented by the blue curve, and $\dot b$ by the red curve (flat on the $t$ axis).}
    \label{64}
  \end{subfigure}
\\[9em]
  \begin{subfigure}[t]{.5\linewidth}
    \centering
    \includegraphics[width=0.7\columnwidth]{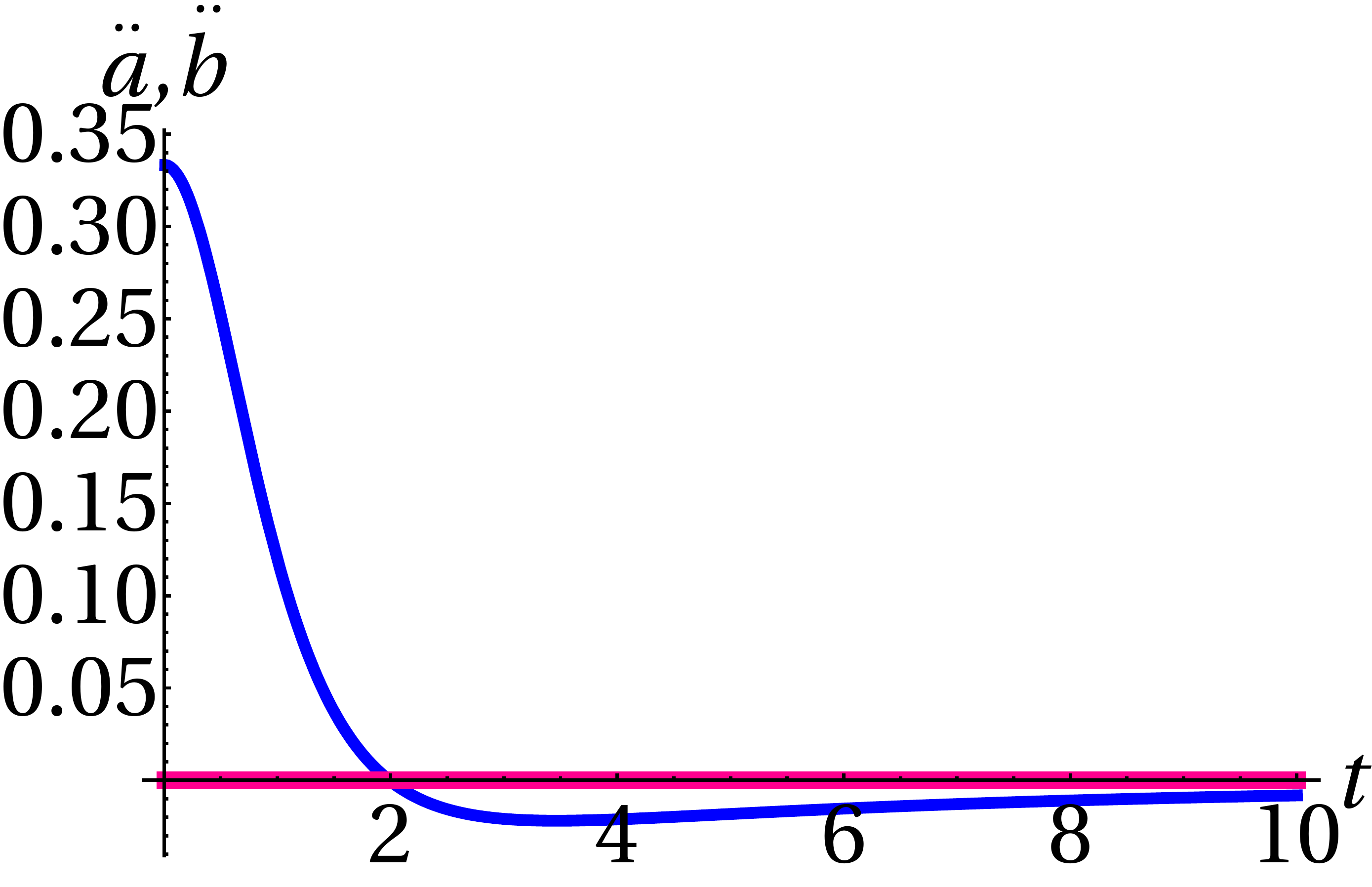}
    \caption{The accelerations of the scale factors: $\ddot a$ is represented by the blue curve, and $\ddot b$ by the red curve (flat on the $t$ axis).}
    \label{65}
  \end{subfigure}
\qquad
  \begin{subfigure}[t]{.5\linewidth}
    \centering
    \includegraphics[width=0.7\columnwidth]{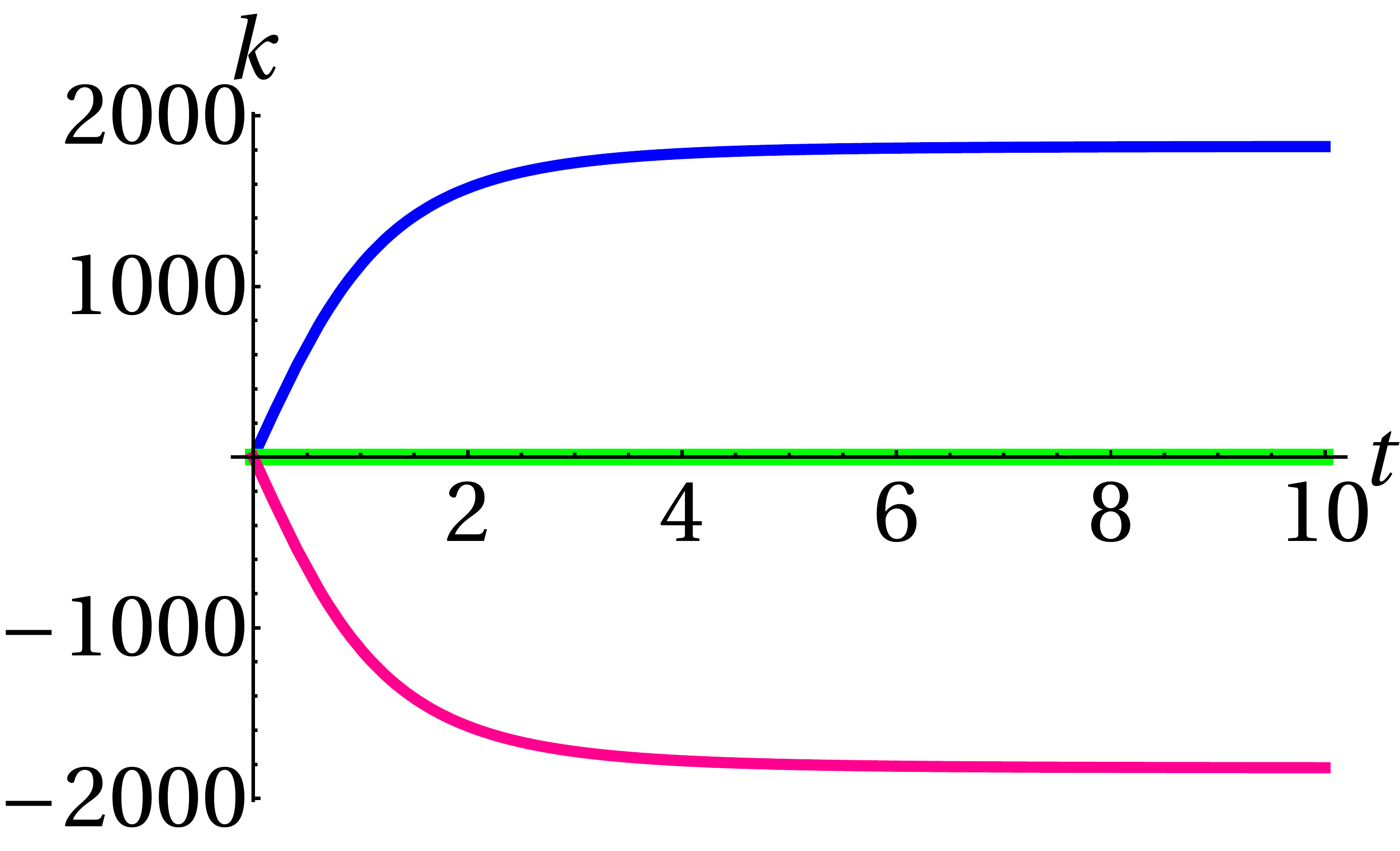}
    \caption{The harmonic function $k$ using: $\dot k\left(0\right)=1$ (blue curve), $\dot k\left(0\right)=0$ (green line), and $\dot k\left(0\right)=-1$ (red curve).}
    \label{66}
  \end{subfigure}
  \caption{Radiation-filled brane world with initial conditions set number 2.}
  \label{Fig13}
\end{figure}
\begin{figure}[H]
\begin{subfigure}[t]{.5\linewidth}
    \centering
    \includegraphics[width=0.7\columnwidth]{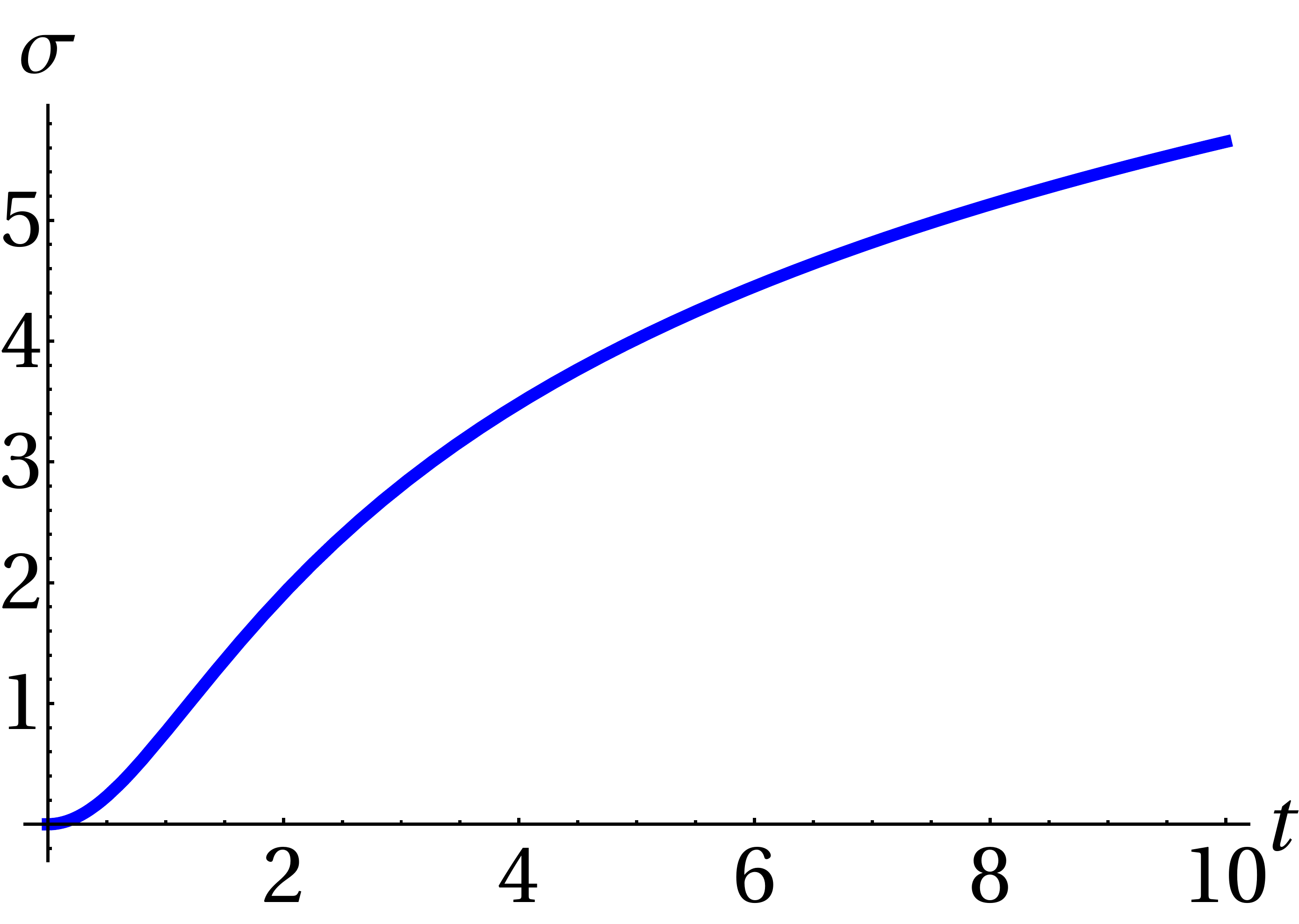}
    \caption{The dilaton $\sigma$; same for all three $\dot k\left(0\right)$.}
    \label{67}
  \end{subfigure}
\qquad
  \begin{subfigure}[t]{.5\linewidth}
    \centering
    \includegraphics[width=0.7\columnwidth]{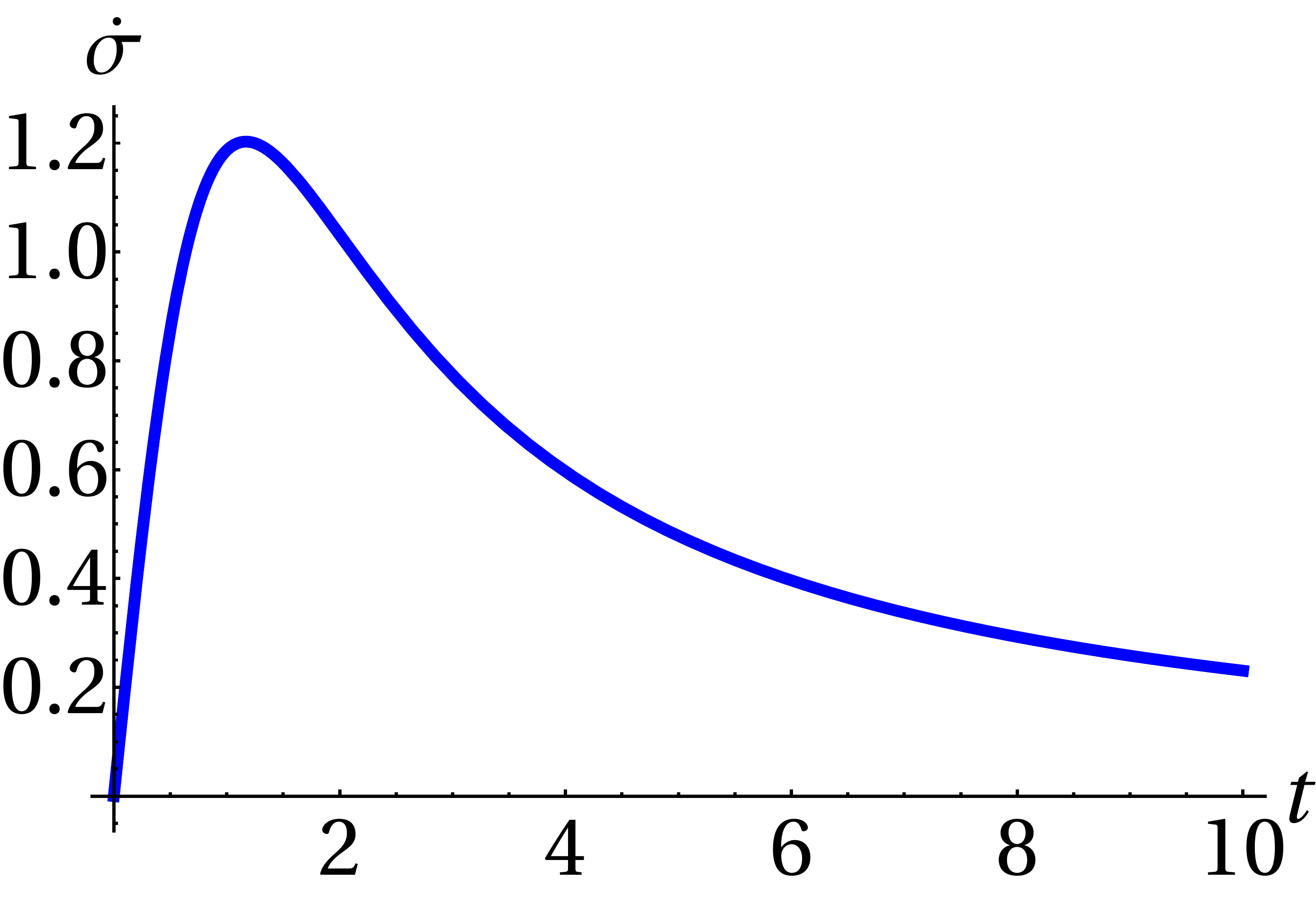}
    \caption{The dilatonic field strength $\dot\sigma$.}
    \label{68}
  \end{subfigure}
  \\[4em]
  \begin{subfigure}[t]{.5\linewidth}
    \centering
    \includegraphics[width=0.7\columnwidth]{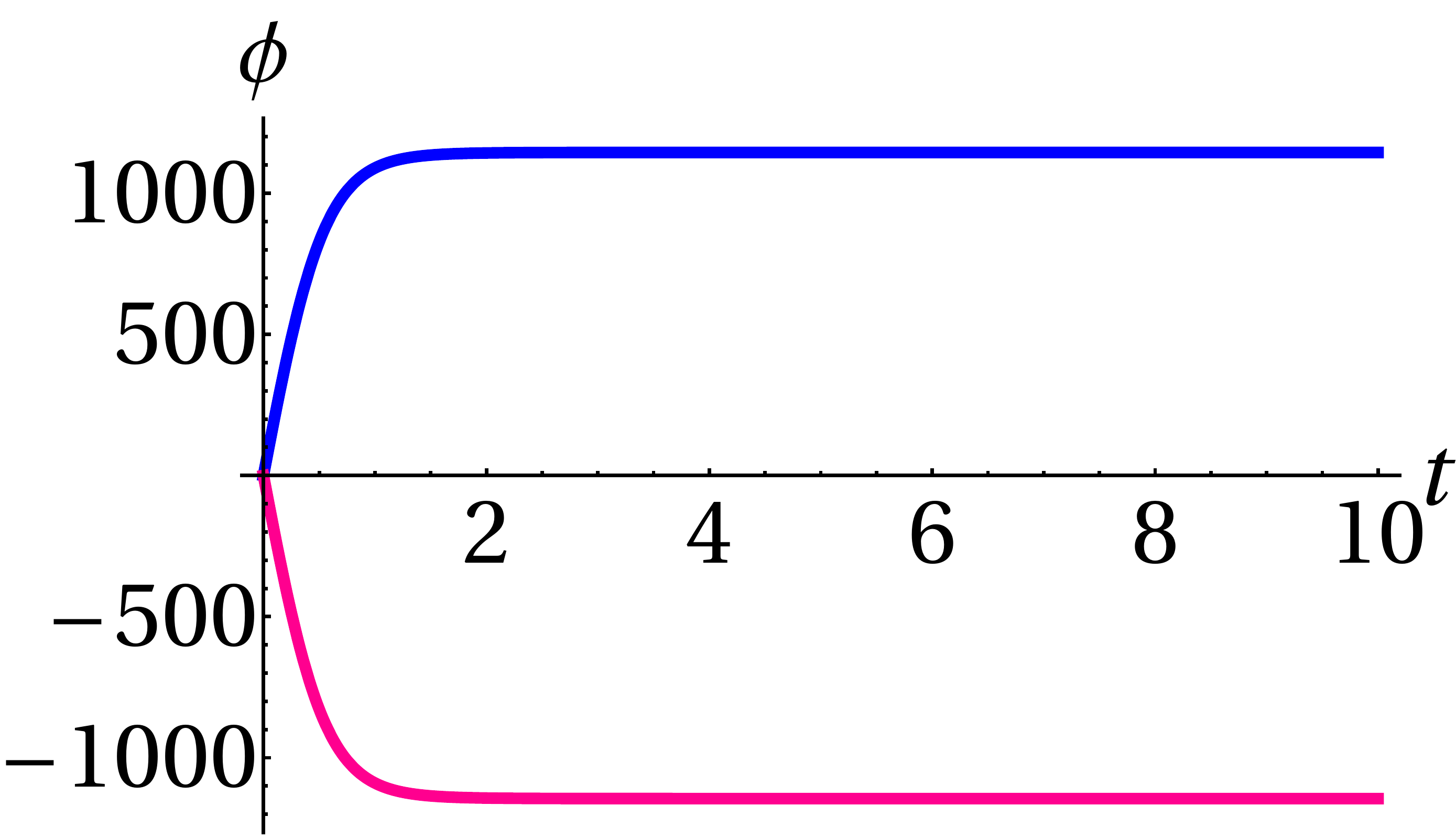}
    \caption{The universal axion $\phi$ for $\dot k\left(0\right) = 1$ (blue curve), and $\dot k\left(0\right) = -1$ (red curve). The solution diverges for $\dot k\left(0\right)=0$.}
    \label{69}
  \end{subfigure}
\qquad
  \begin{subfigure}[t]{.5\linewidth}
    \centering
    \includegraphics[width=0.7\columnwidth]{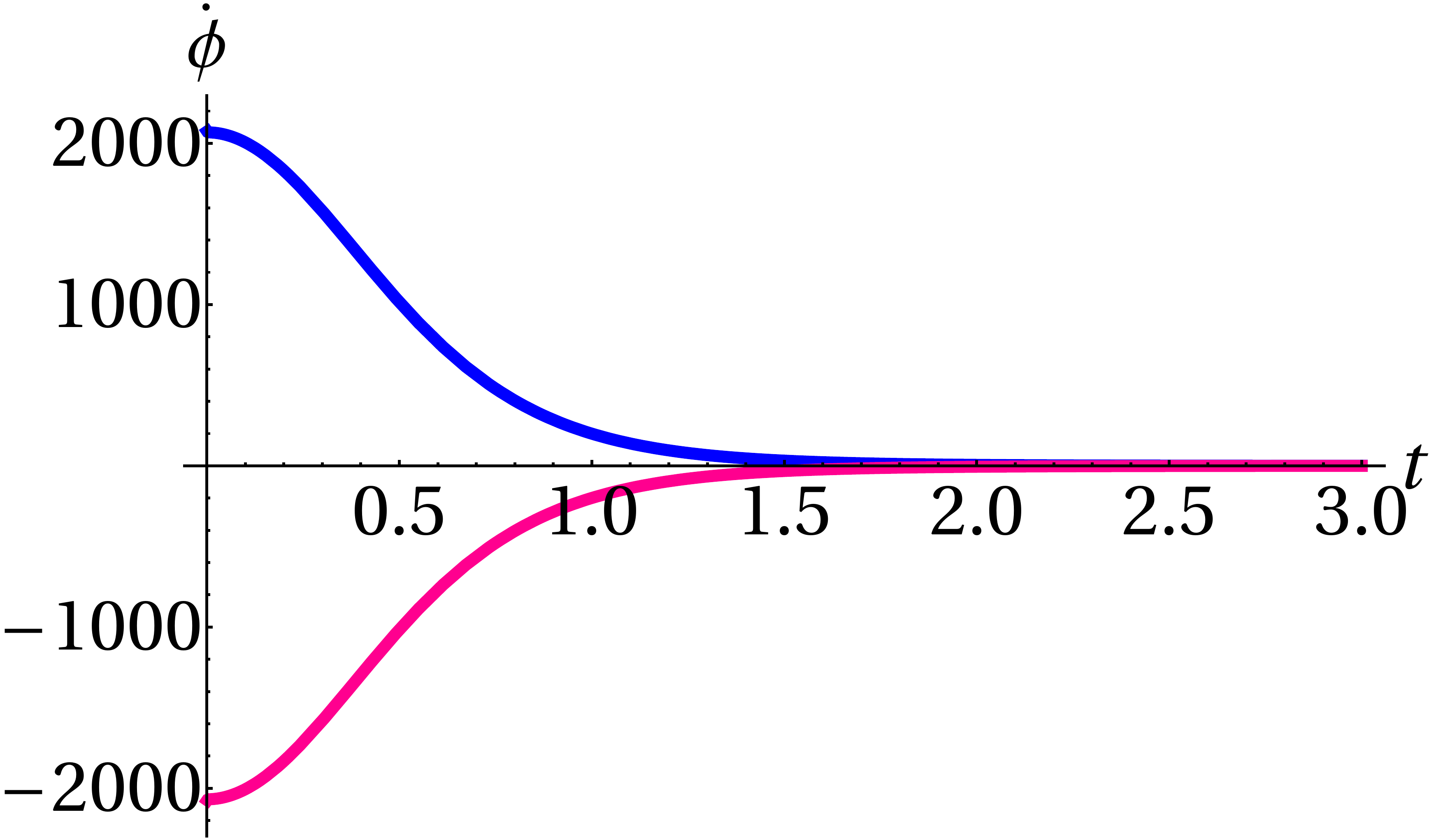}
    \caption{The axionic field strength $\dot\phi$ for $\dot k\left(0\right) = 1$ (blue curve), and $\dot k\left(0\right) = -1$ (red curve).}
    \label{70}
  \end{subfigure}
\\[4em]
\begin{subfigure}[t]{.5\linewidth}
    \centering
    \includegraphics[width=0.7\columnwidth]{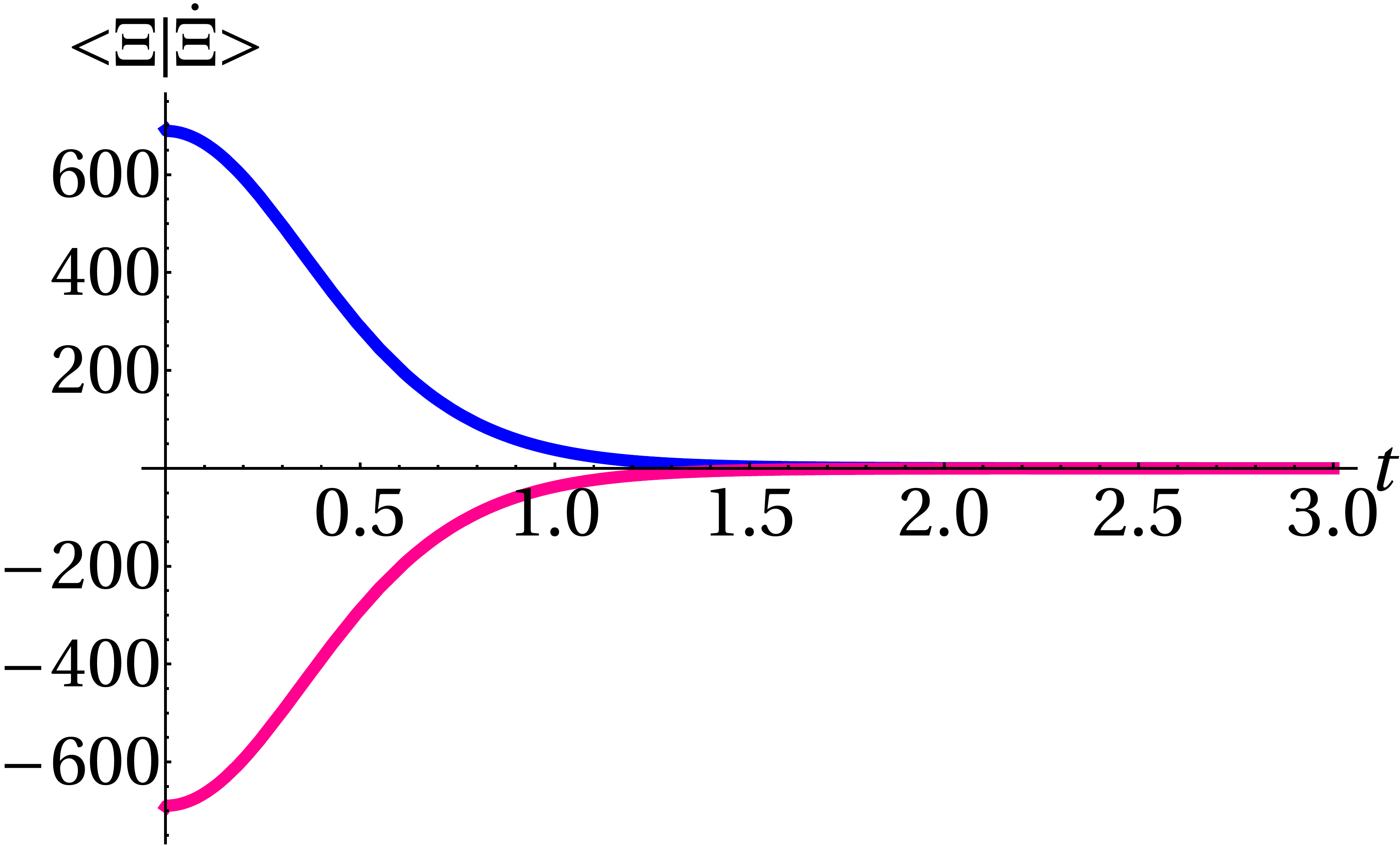}
    \caption{$ \langle \Xi | \dot{\Xi} \rangle$ for $\dot{k}(0)= 1$  (blue), and $\dot{k}(0)  = -1$ (red).}
    \label{71}
  \end{subfigure}
\qquad
  \begin{subfigure}[t]{.5\linewidth}
    \centering
    \includegraphics[width=0.7\columnwidth]{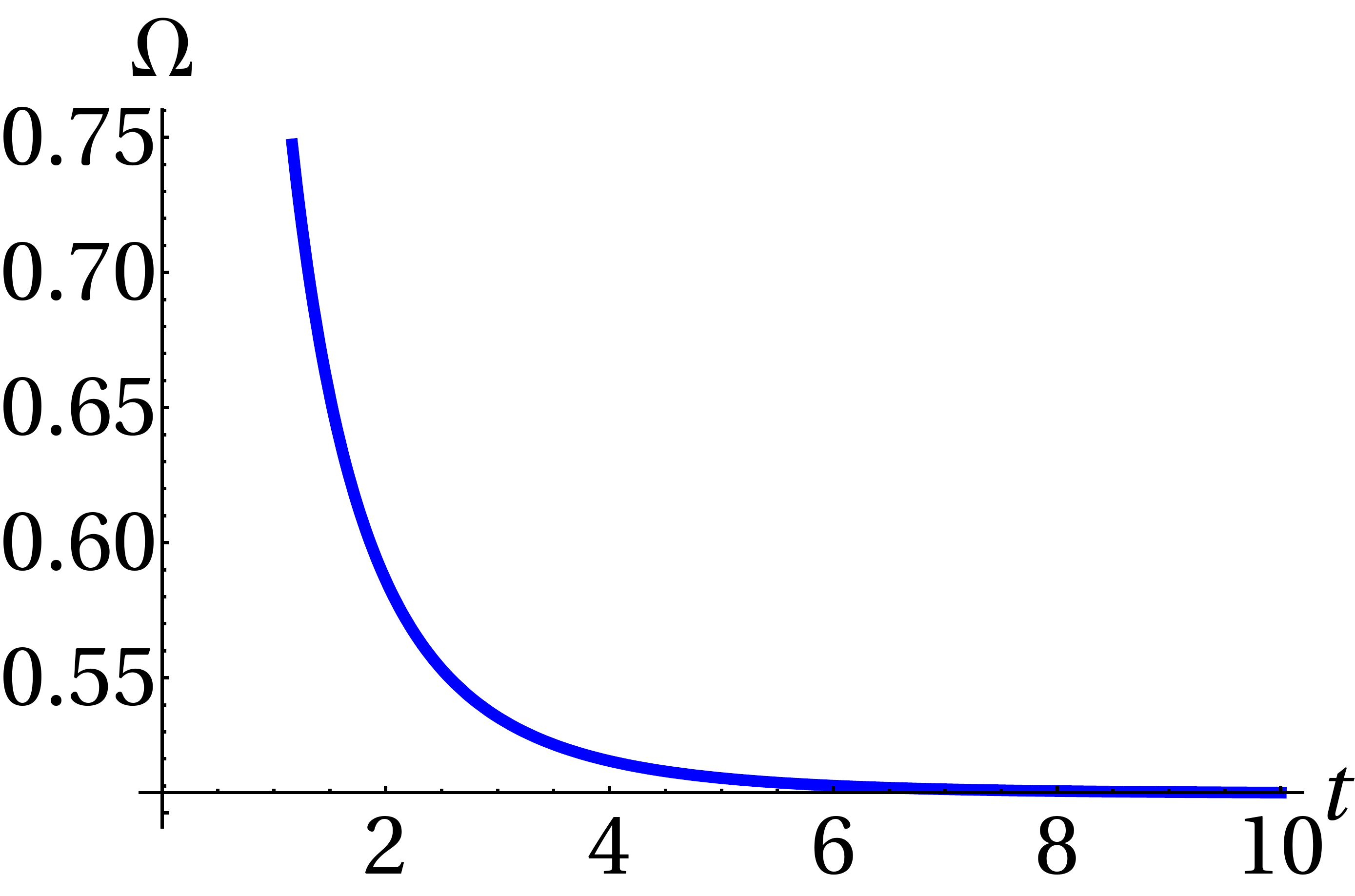}
    \caption{$\Omega$ at $\dot k\left(0\right) = 1$ and  $ \dot\sigma \left(0\right) =0 $.}
    \label{72}
  \end{subfigure}
  \caption{Radiation-filled brane world with initial conditions set number 2 (continued).}
  \label{Fig14}
\end{figure}


\begin{figure}[H]
  \begin{subfigure}[t]{.5\linewidth}
    \centering
    \includegraphics[width=0.7\columnwidth]{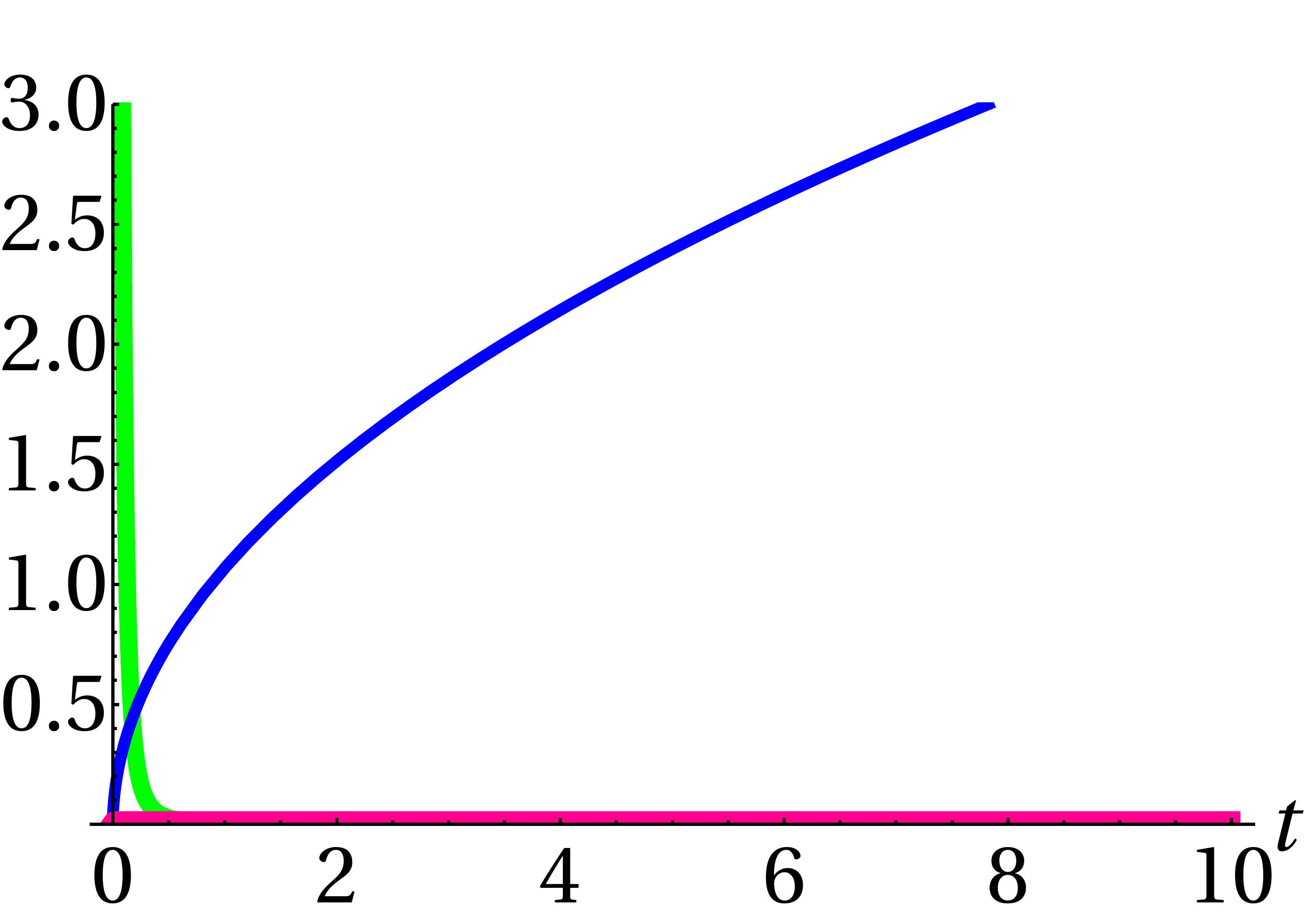}
    \caption{The scale factor $a$ is represented by the blue curve, $b$ by the red curve (almost flat on the $t$ axis), while $\left| {G_{i\bar j} \dot z^i \dot z^{\bar j}} \right|$ by the green curve.}
    \label{73}
  \end{subfigure}
\qquad
  \begin{subfigure}[t]{.5\linewidth}
    \centering
    \includegraphics[width=0.7\columnwidth]{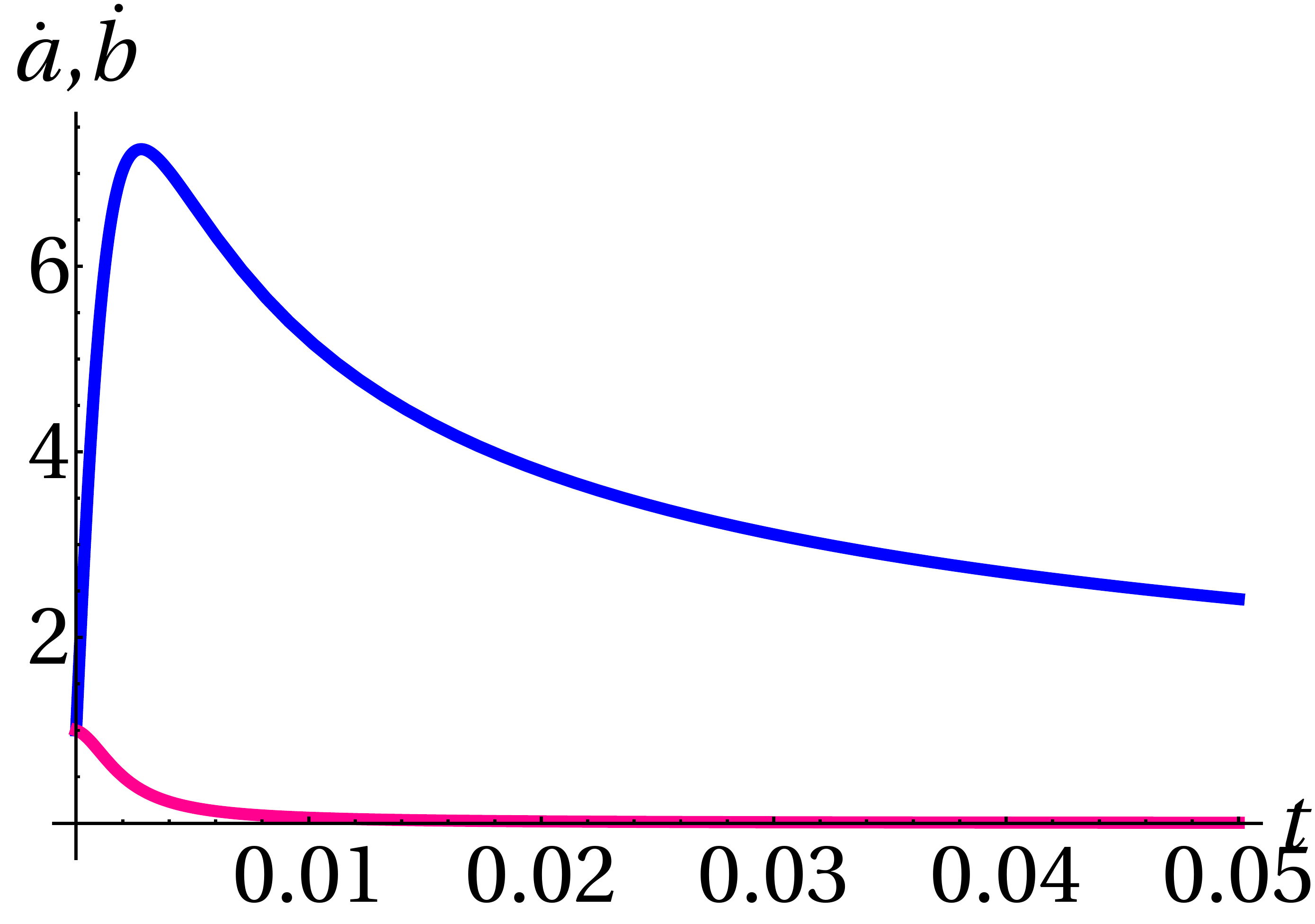}
    \caption{The expansion rates of the scale factors: $\dot a$ is represented by the blue curve, and $\dot b$ by the red curve.}
    \label{74}
  \end{subfigure}
\\[9em]  
 \begin{subfigure}[t]{.5\linewidth}
    \centering
    \includegraphics[width=0.7\columnwidth]{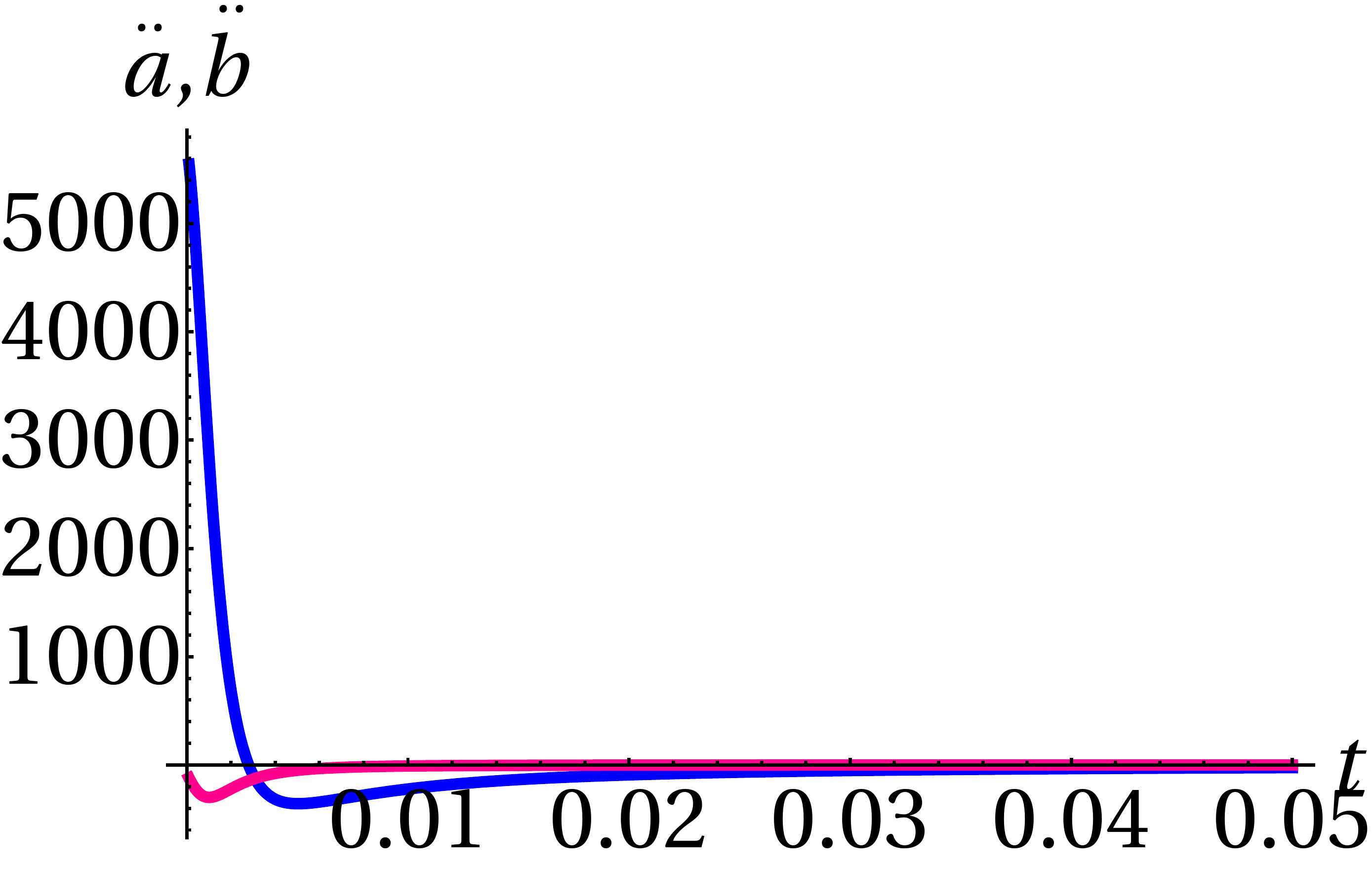}
    \caption{The accelerations of the scale factors: $\ddot a$ is represented by the blue curve, and $\ddot b$ by the red curve.}
    \label{75}
  \end{subfigure}
\qquad
  \begin{subfigure}[t]{.5\linewidth}
    \centering
    \includegraphics[width=0.7\columnwidth]{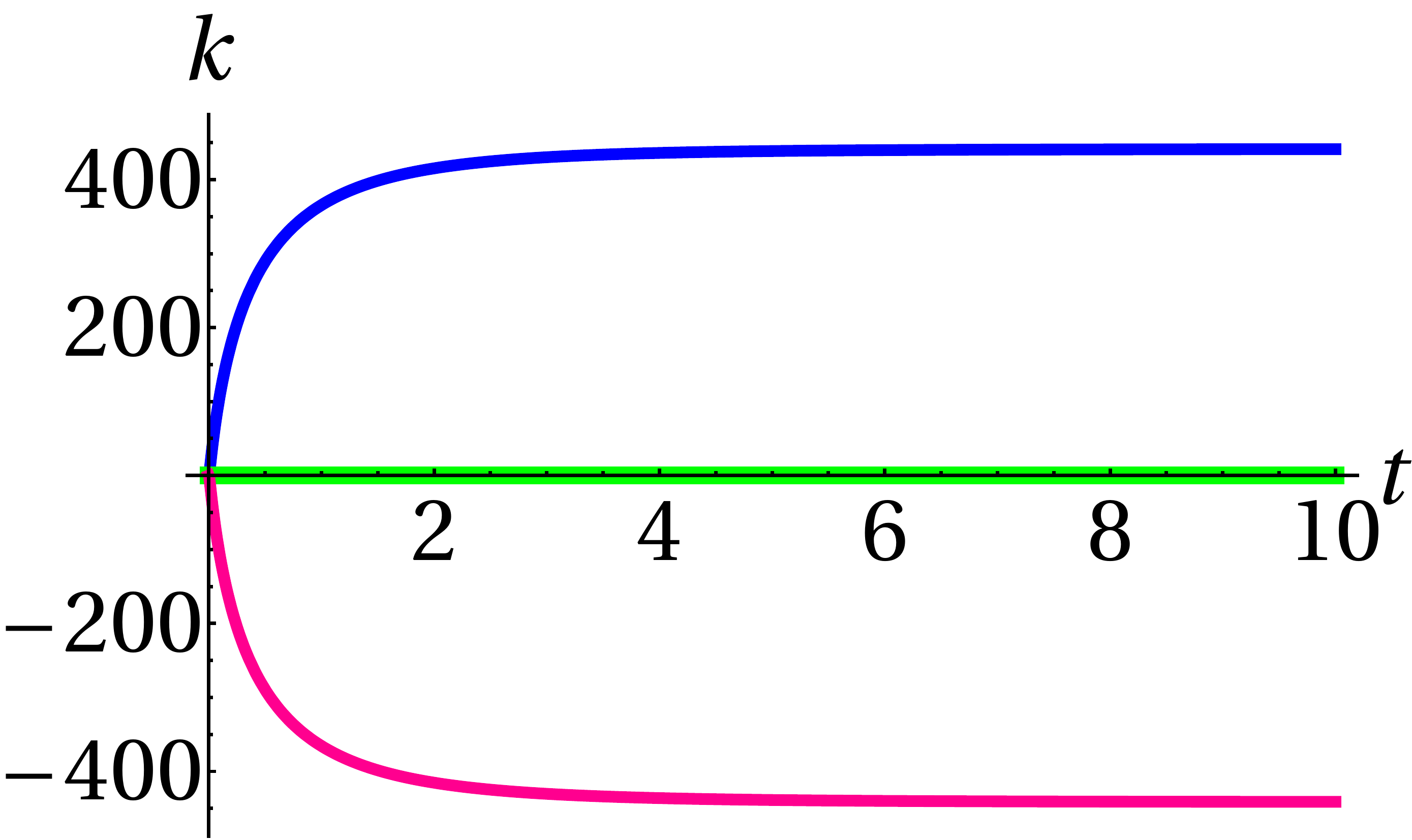}
    \caption{The harmonic function $k$ using: $\dot k\left(0\right)=1$ (blue curve), $\dot k\left(0\right)=0$ (green line), and $\dot k\left(0\right)=-1$ (red curve).}
    \label{76}
  \end{subfigure}
  \caption{Radiation-filled brane world with initial conditions set number 3.}
  \label{Fig15}
\end{figure}
\begin{figure}[H]
  \begin{subfigure}[t]{.5\linewidth}
    \centering
    \includegraphics[width=0.7\columnwidth]{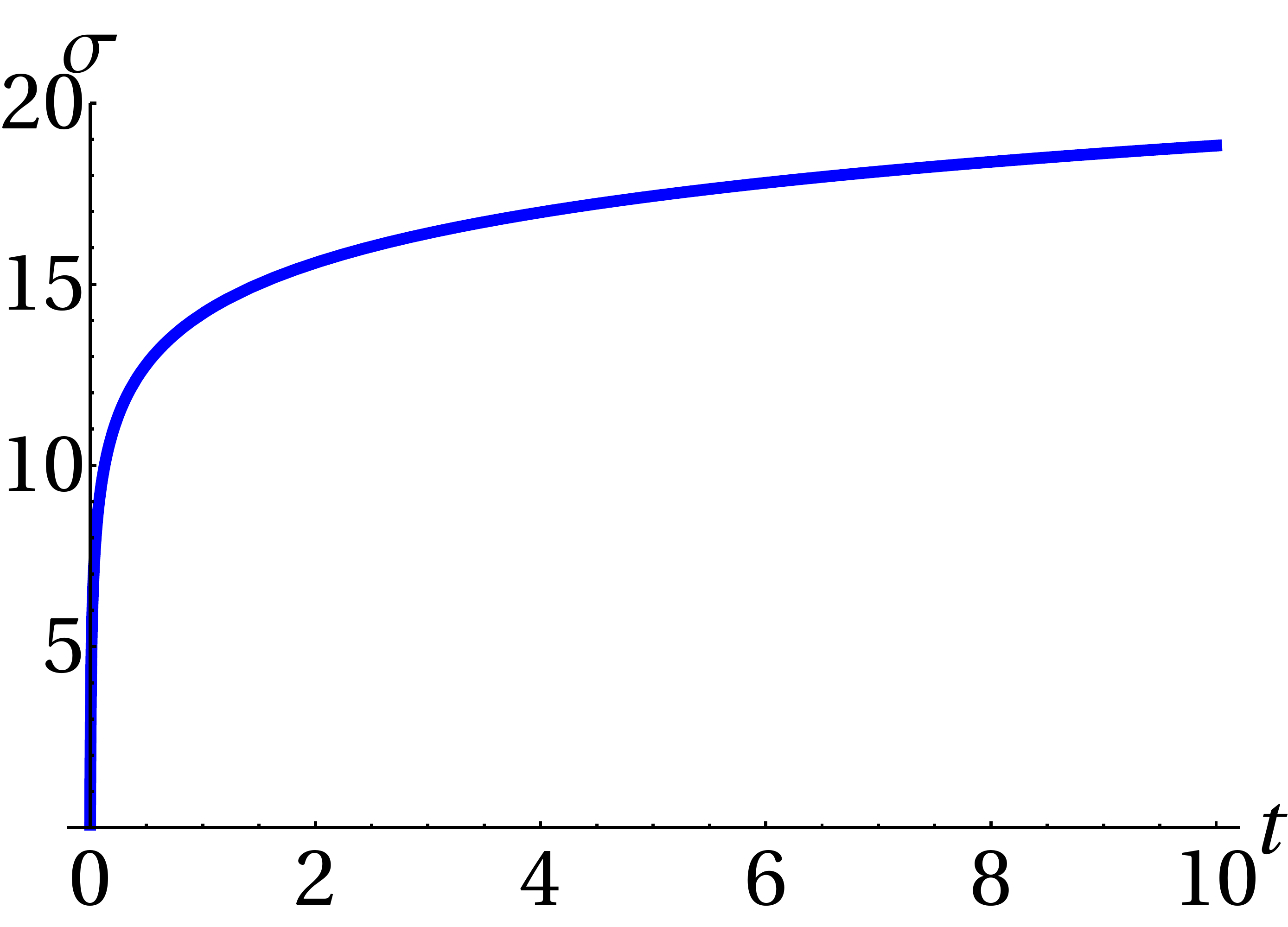}
    \caption{The dilaton $\sigma$; same for all three $\dot k\left(0\right)$.}
    \label{77}
  \end{subfigure}
\qquad
  \begin{subfigure}[t]{.5\linewidth}
    \centering
    \includegraphics[width=0.7\columnwidth]{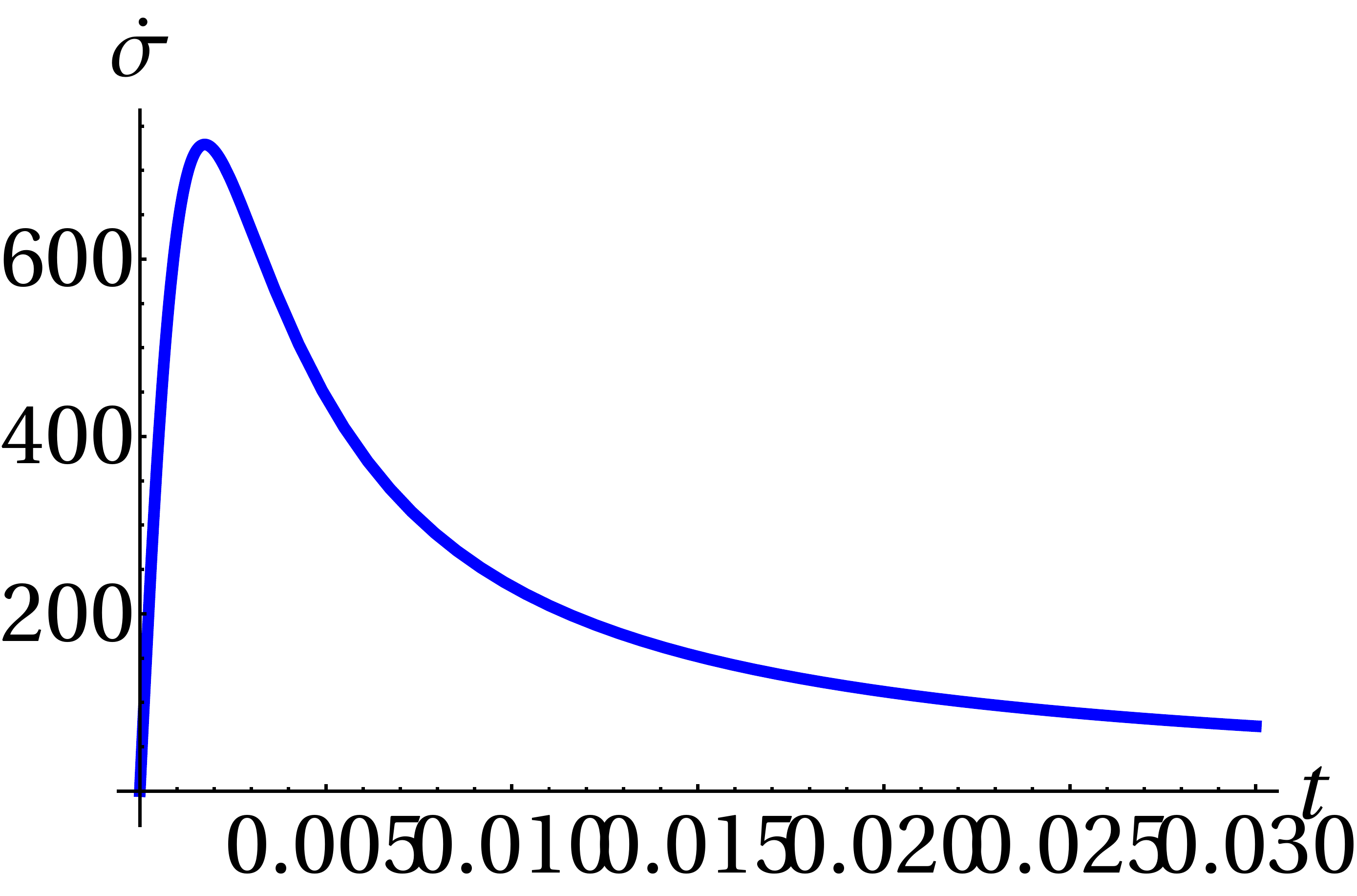}
    \caption{The dilatonic field strength $\dot\sigma$.}
    \label{78}
  \end{subfigure}
\\[4em]
  \begin{subfigure}[t]{.5\linewidth}
    \centering
    \includegraphics[width=0.7\columnwidth]{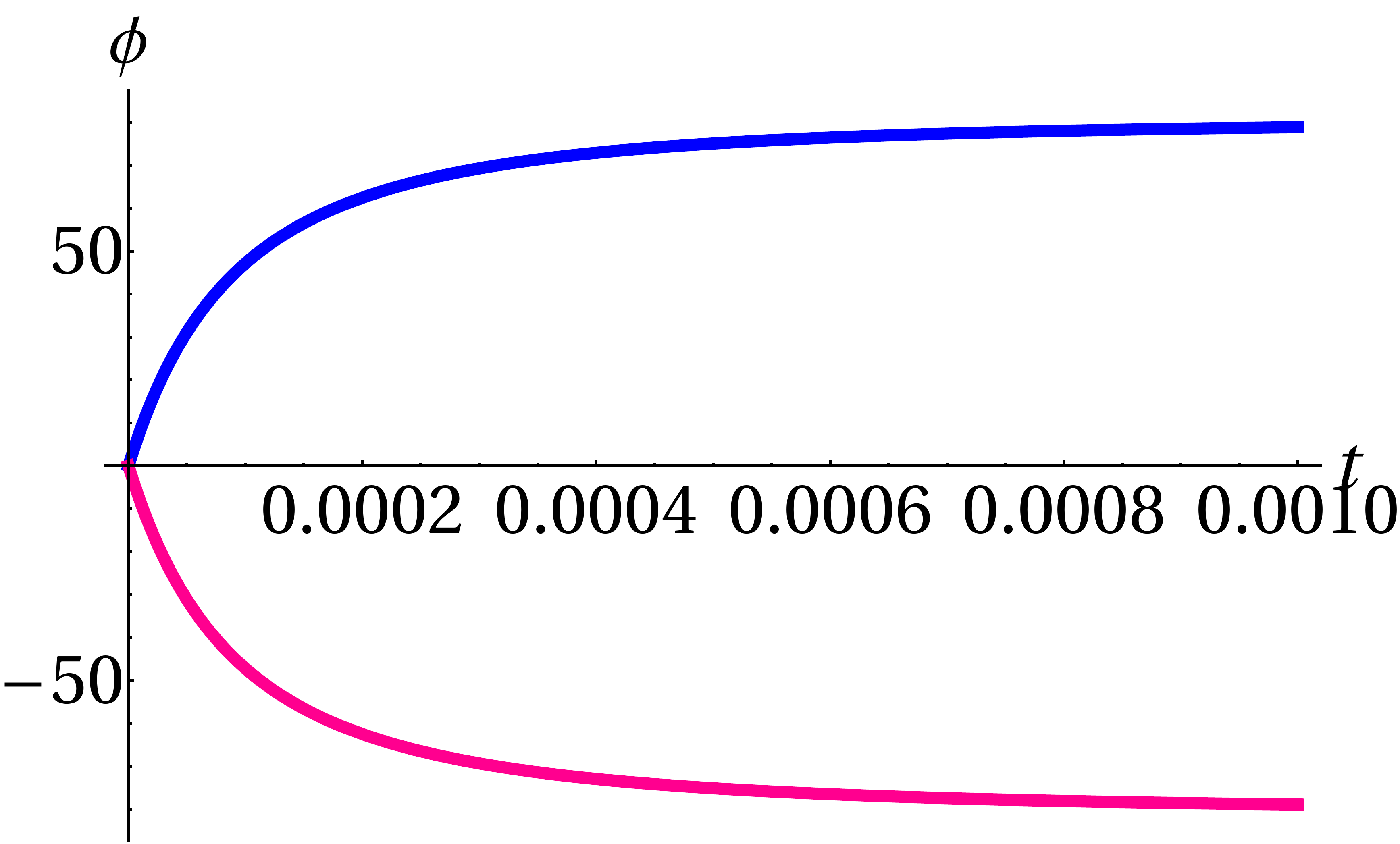}
    \caption{The universal axion $\phi$ for $\dot k\left(0\right) = 1$ (blue curve), and $\dot k\left(0\right) = -1$ (red curve). The solution diverges for $\dot k\left(0\right)=0$.}
    \label{79}
  \end{subfigure}
\qquad
 \begin{subfigure}[t]{.5\linewidth}
    \centering
    \includegraphics[width=0.7\columnwidth]{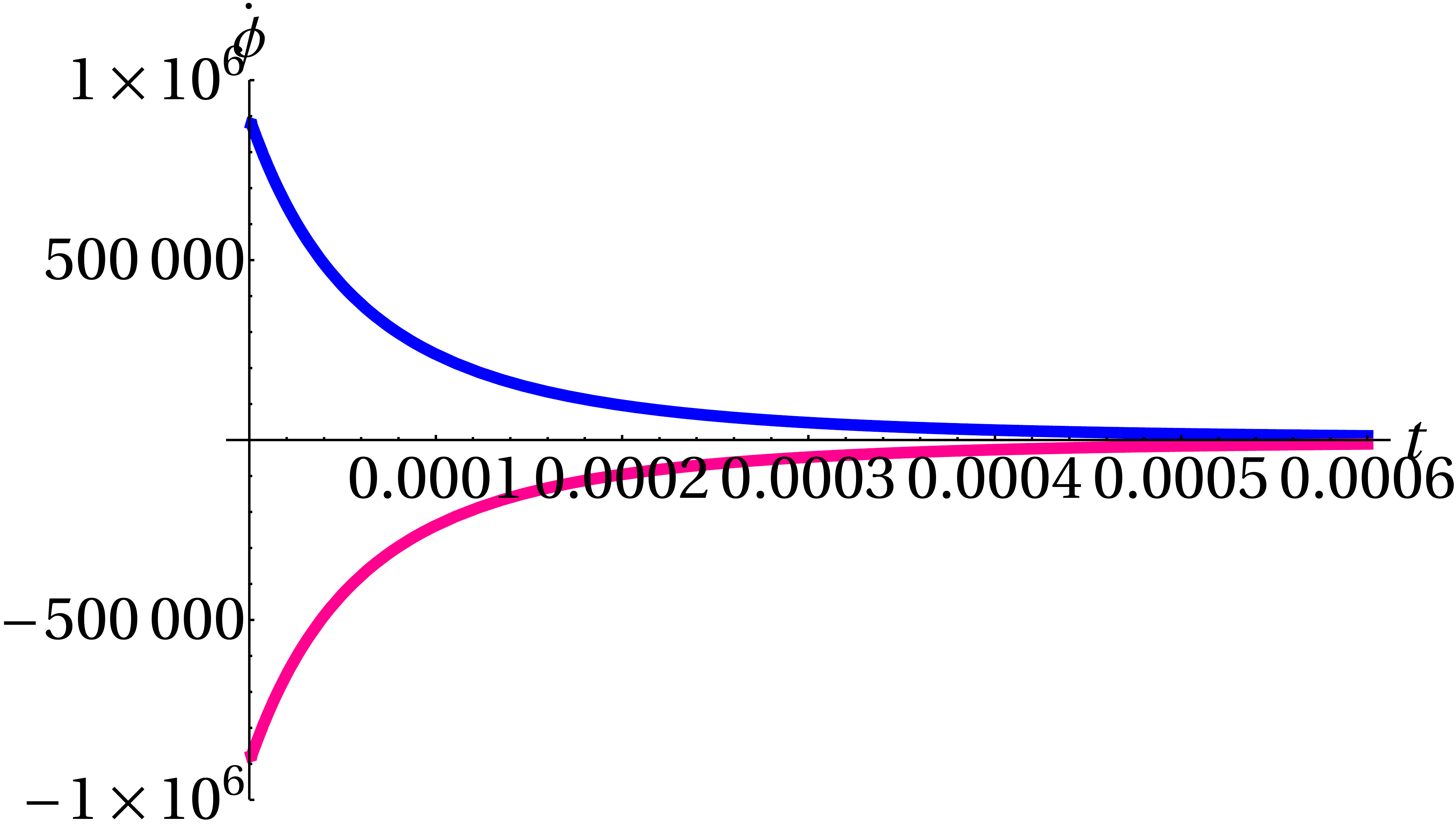}
    \caption{The axionic field strength $\dot\phi$ for $\dot k\left(0\right) = 1$ (blue curve), and $\dot k\left(0\right) = -1$ (red curve).}
    \label{80}
  \end{subfigure}
 \\[4em]
 \begin{subfigure}[t]{.5\linewidth}
    \centering
    \includegraphics[width=0.7\columnwidth]{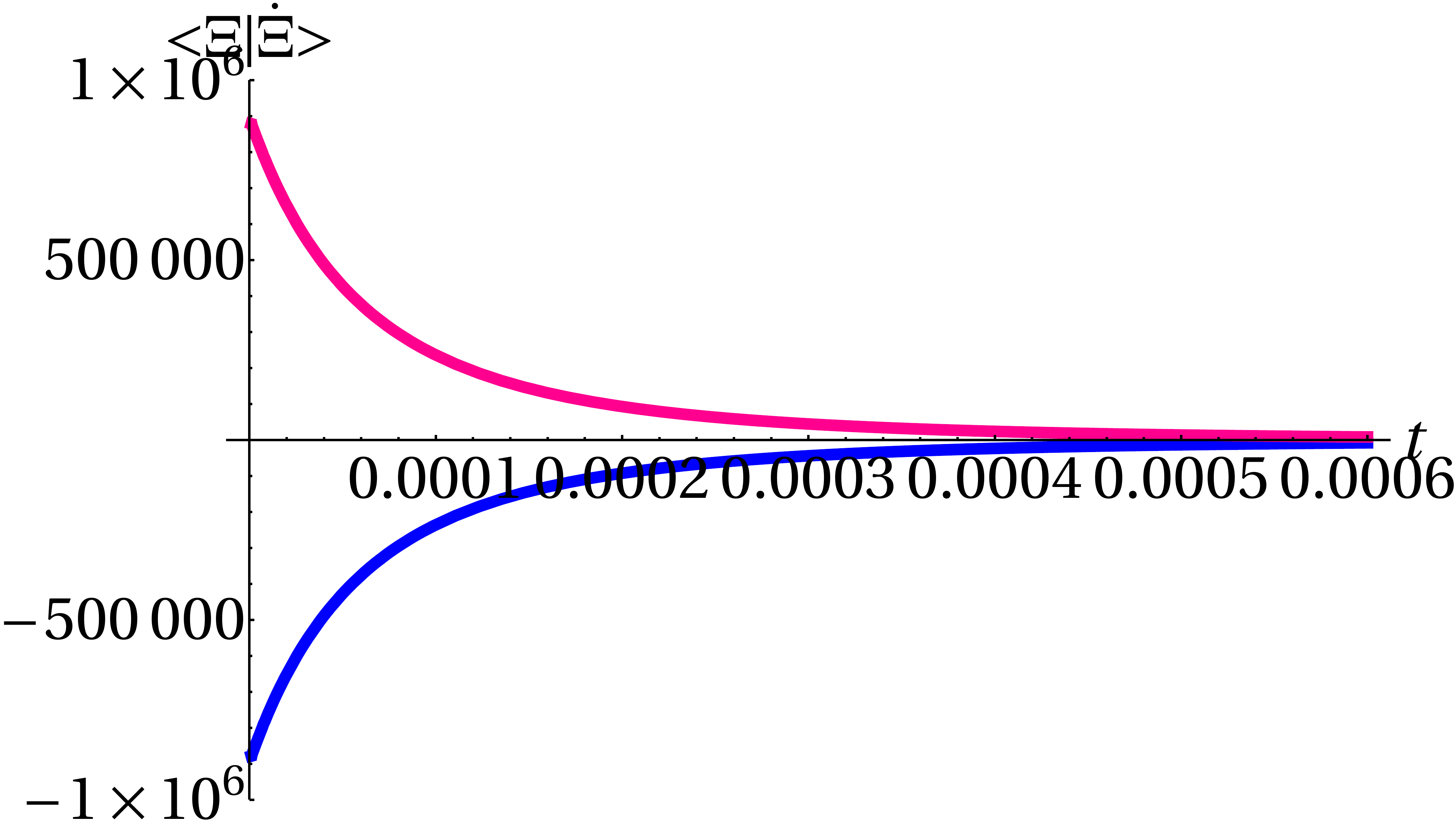}
    \caption{ $ \langle \Xi | \dot{\Xi} \rangle$ for $\dot{k}(0)= 1$  (blue), and $\dot{k}(0)  = -1$ (red).}
    \label{81}
  \end{subfigure}
\qquad
  \begin{subfigure}[t]{.5\linewidth}
    \centering
    \includegraphics[width=0.7\columnwidth]{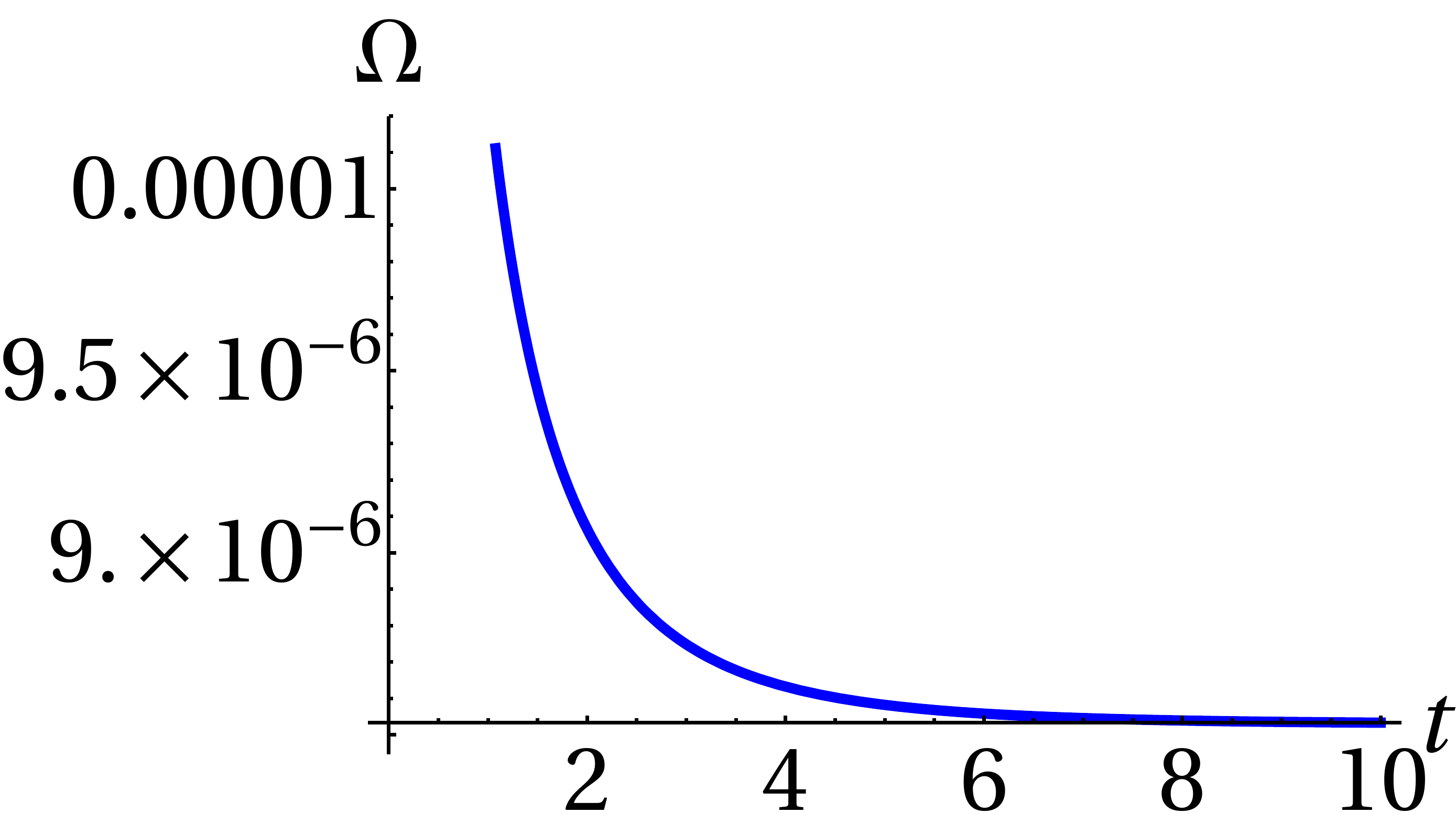}
    \caption{$\Omega$ at $\dot k\left(0\right) = 1$ and  $ \dot\sigma \left(0\right) =0 $.}
    \label{82}
  \end{subfigure}
  \caption{Radiation-filled brane world with initial conditions set number 3 (continued).}
  \label{Fig16}
\end{figure}


\begin{figure}[H]
  \begin{subfigure}[t]{.5\linewidth}
    \centering
    \includegraphics[width=0.7\columnwidth]{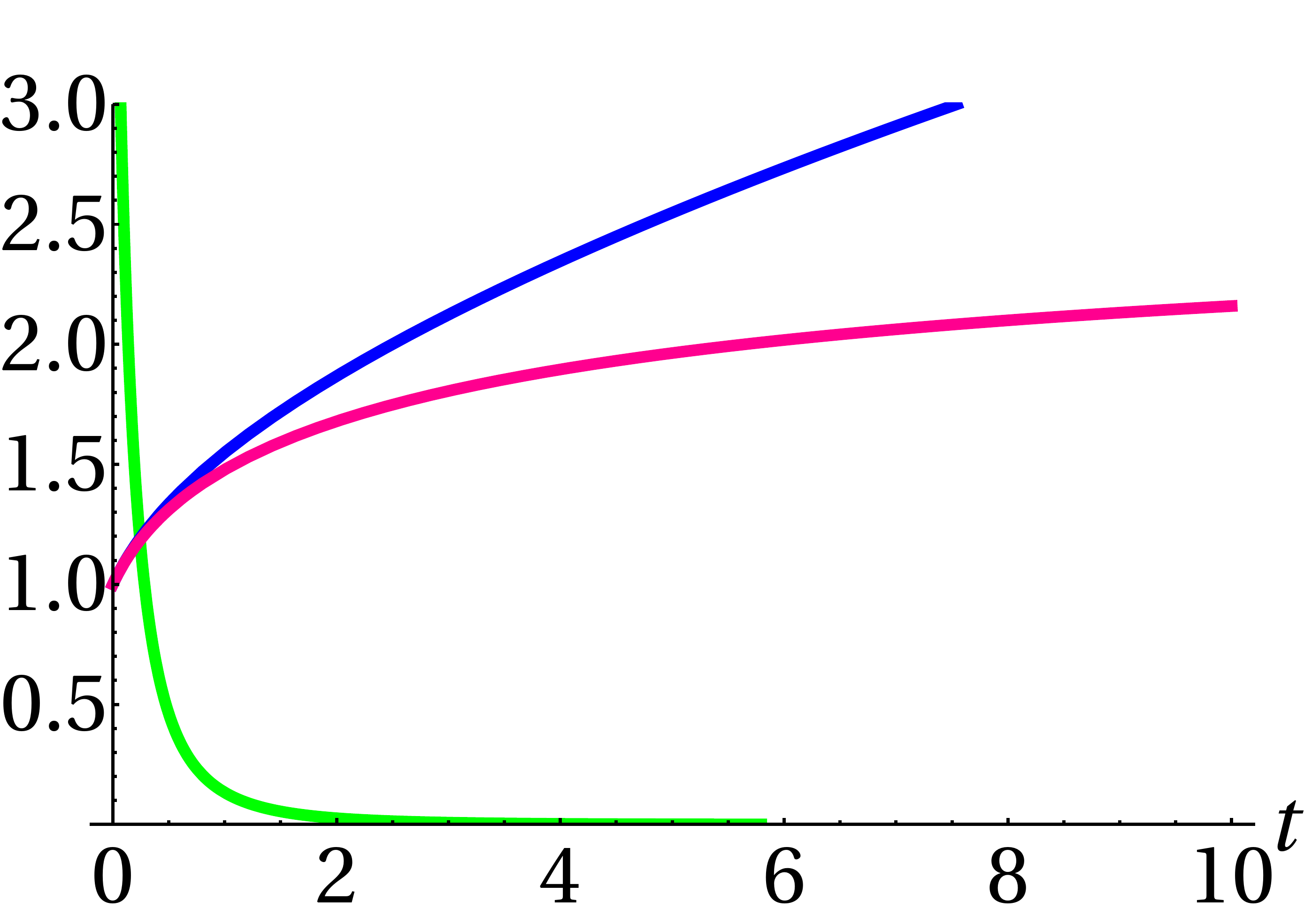}
    \caption{The scale factor $a$ is represented by the blue curve, $b$ by the red curve, while $ {G_{i\bar j} \dot z^i \dot z^{\bar j}} $ (already positive) by the green curve.}
    \label{83}
  \end{subfigure}
\qquad
  \begin{subfigure}[t]{.5\linewidth}
    \centering
    \includegraphics[width=0.7\columnwidth]{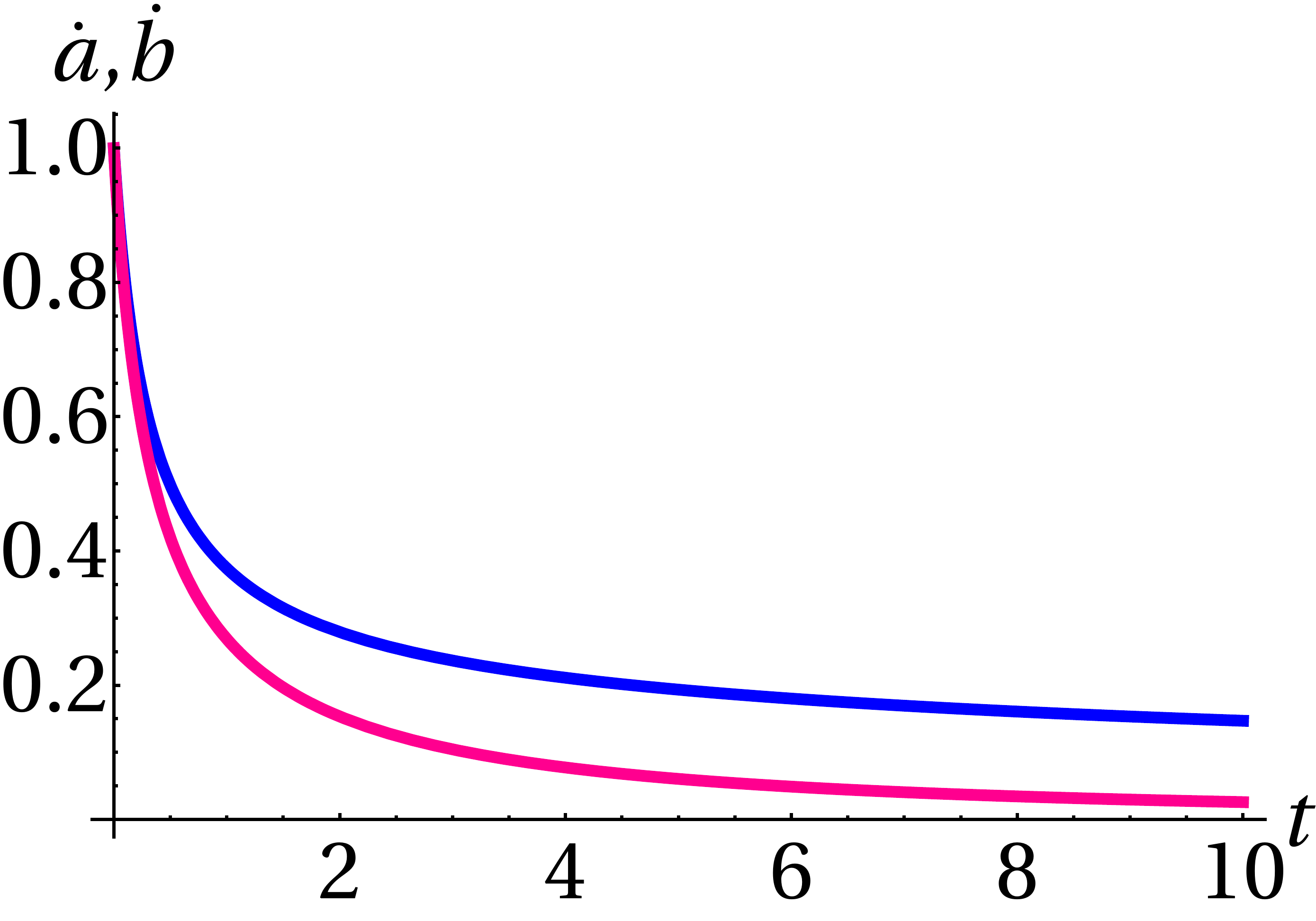}
    \caption{The expansion rates of the scale factors: $\dot a$ is represented by the blue curve, and $\dot b$ by the red curve.}
    \label{84}
  \end{subfigure}
\\[9em]
  \begin{subfigure}[t]{.5\linewidth}
    \centering
    \includegraphics[width=0.7\columnwidth]{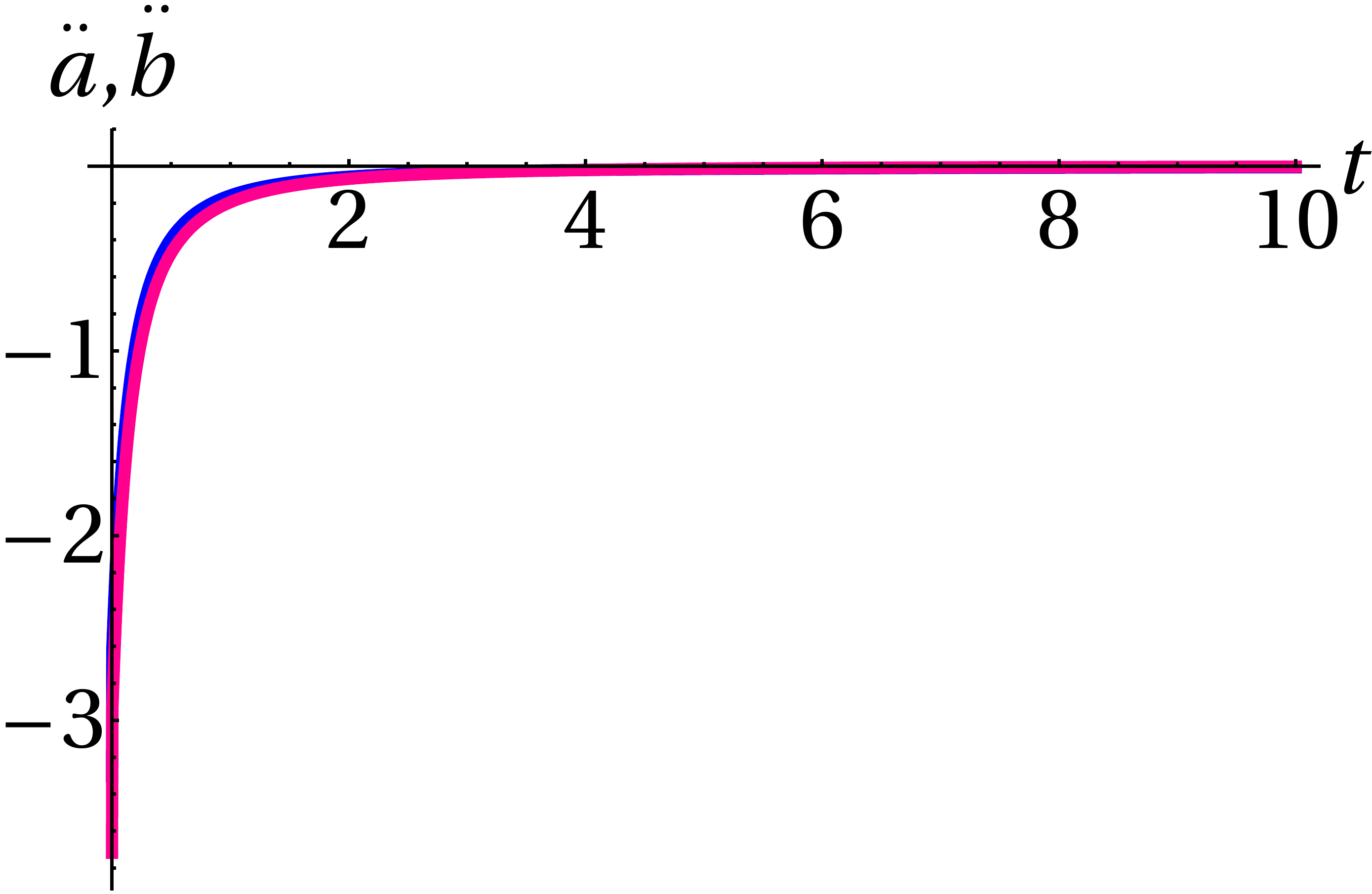}
    \caption{The accelerations of the scale factors: $\ddot a$ is represented by the blue curve, and $\ddot b$ by the red curve.}
    \label{85}
  \end{subfigure}
\qquad
  \begin{subfigure}[t]{.5\linewidth}
    \centering
    \includegraphics[width=0.7\columnwidth]{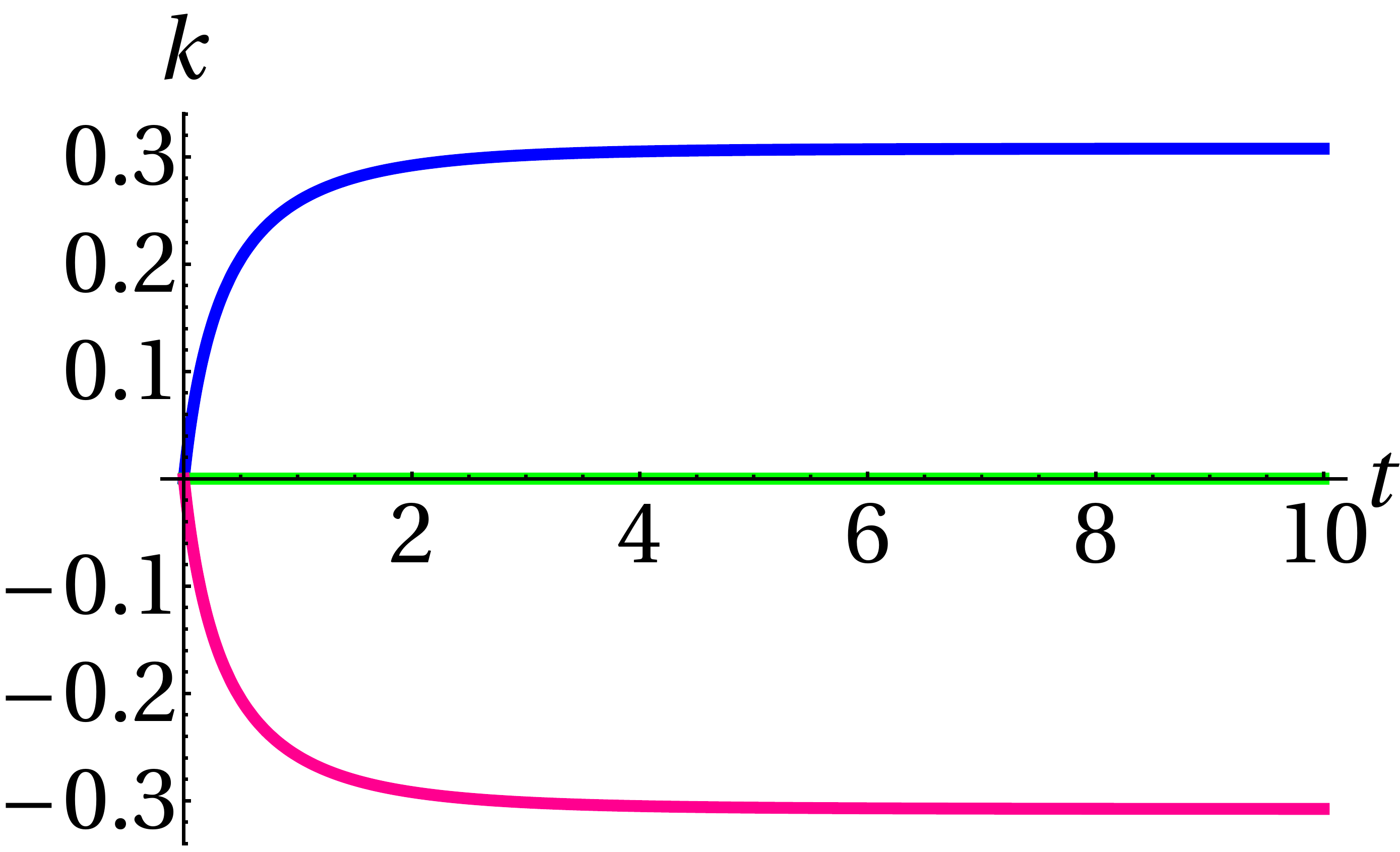}
    \caption{The harmonic function $k$ using: $\dot k\left(0\right)=1$ (blue curve), $\dot k\left(0\right)=0$ (green line), and $\dot k\left(0\right)=-1$ (red curve).}
    \label{86}
  \end{subfigure}
    \caption{Radiation-filled brane world with initial conditions set number 4.}
  \label{Fig17}
\end{figure}
\begin{figure}[H]
\begin{subfigure}[t]{.5\linewidth}
    \centering
    \includegraphics[width=0.7\columnwidth]{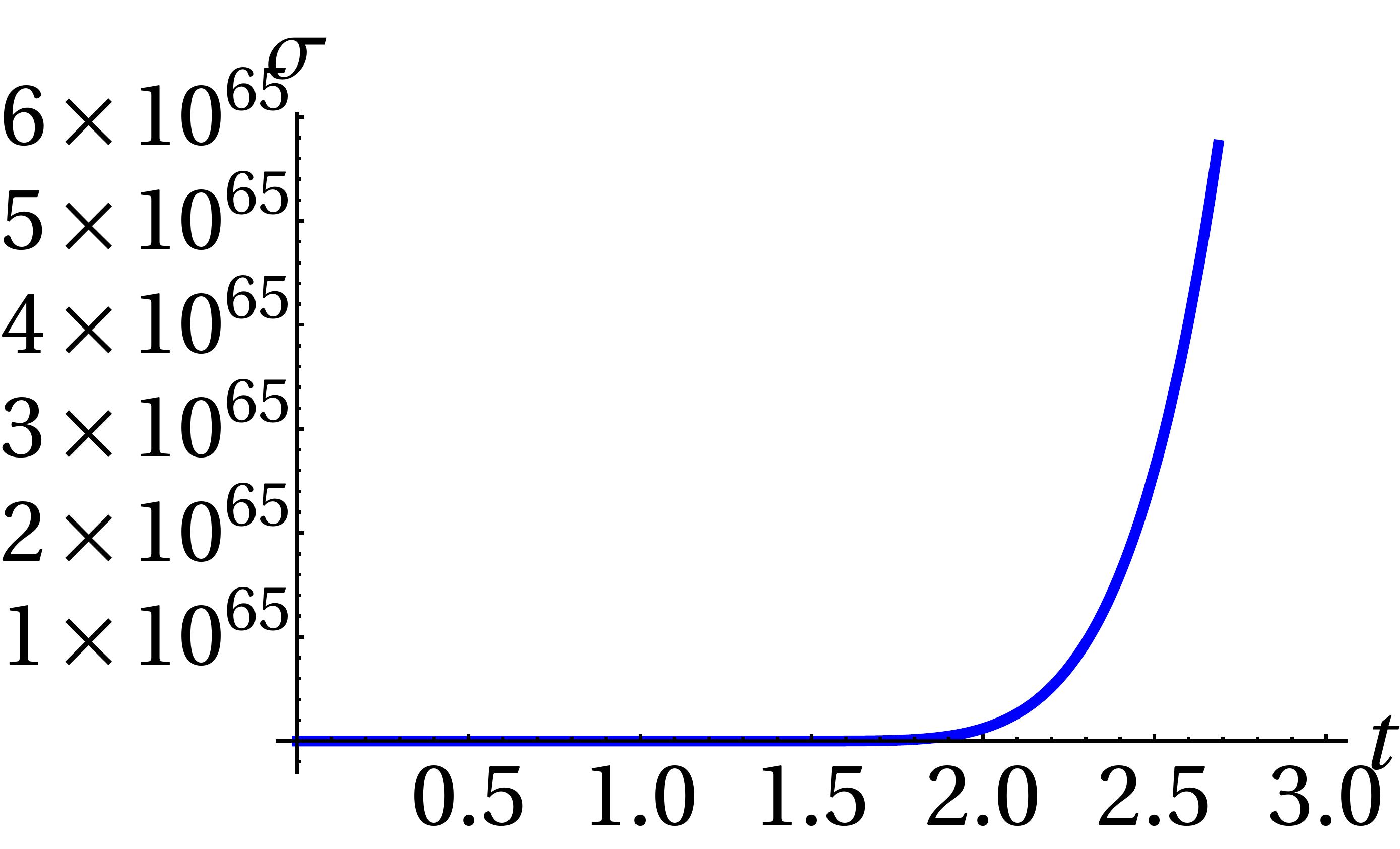}
    \caption{The dilaton $\sigma$; same for all three $\dot k\left(0\right)$.}
    \label{87}
  \end{subfigure}
\qquad
  \begin{subfigure}[t]{.5\linewidth}
    \centering
    \includegraphics[width=0.7\columnwidth]{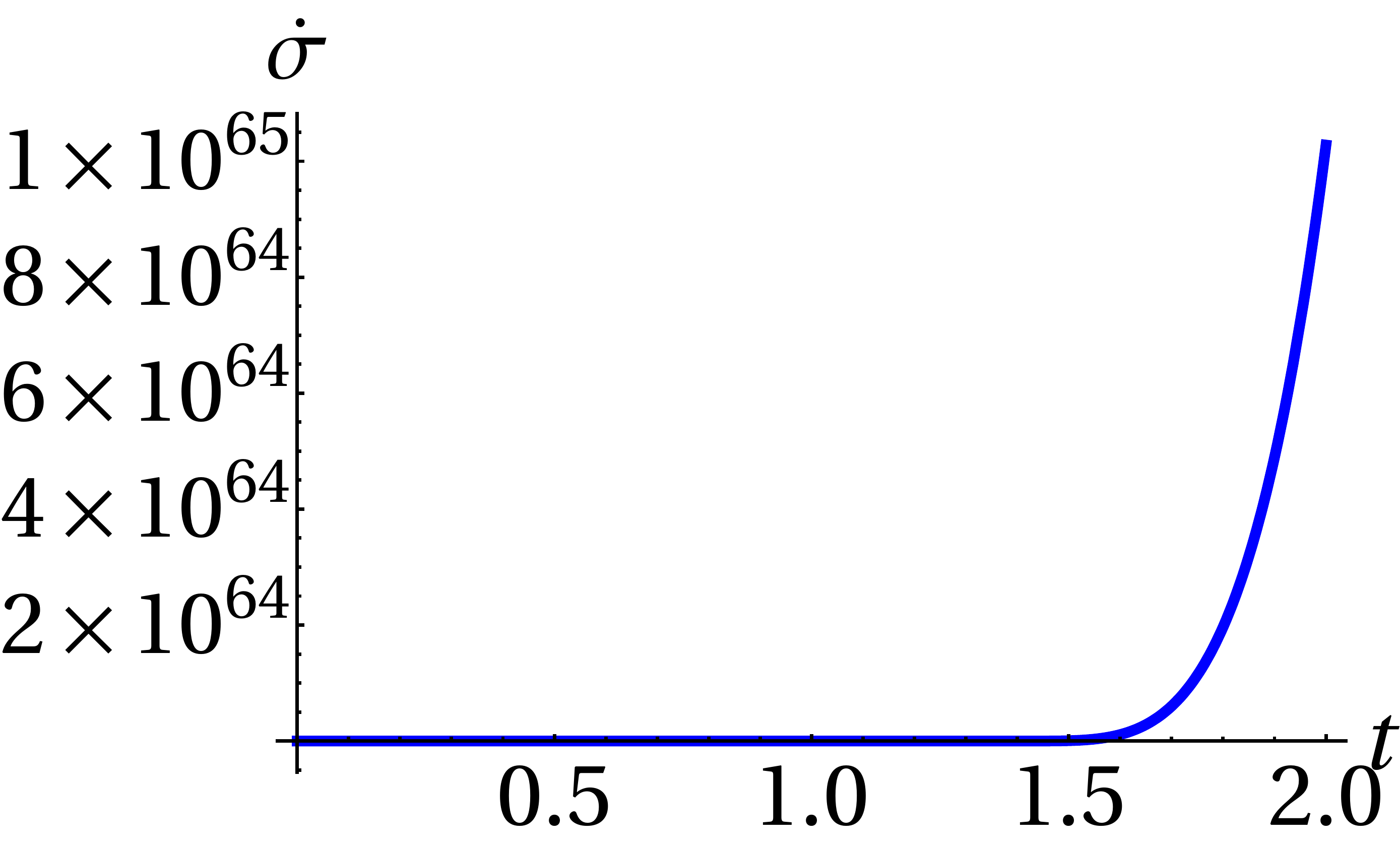}
    \caption{The dilatonic field strength $\dot\sigma$.}
   \label{88}
  \end{subfigure}
\\[4em]
  \begin{subfigure}[t]{.5\linewidth}
    \centering
    \includegraphics[width=0.7\columnwidth]{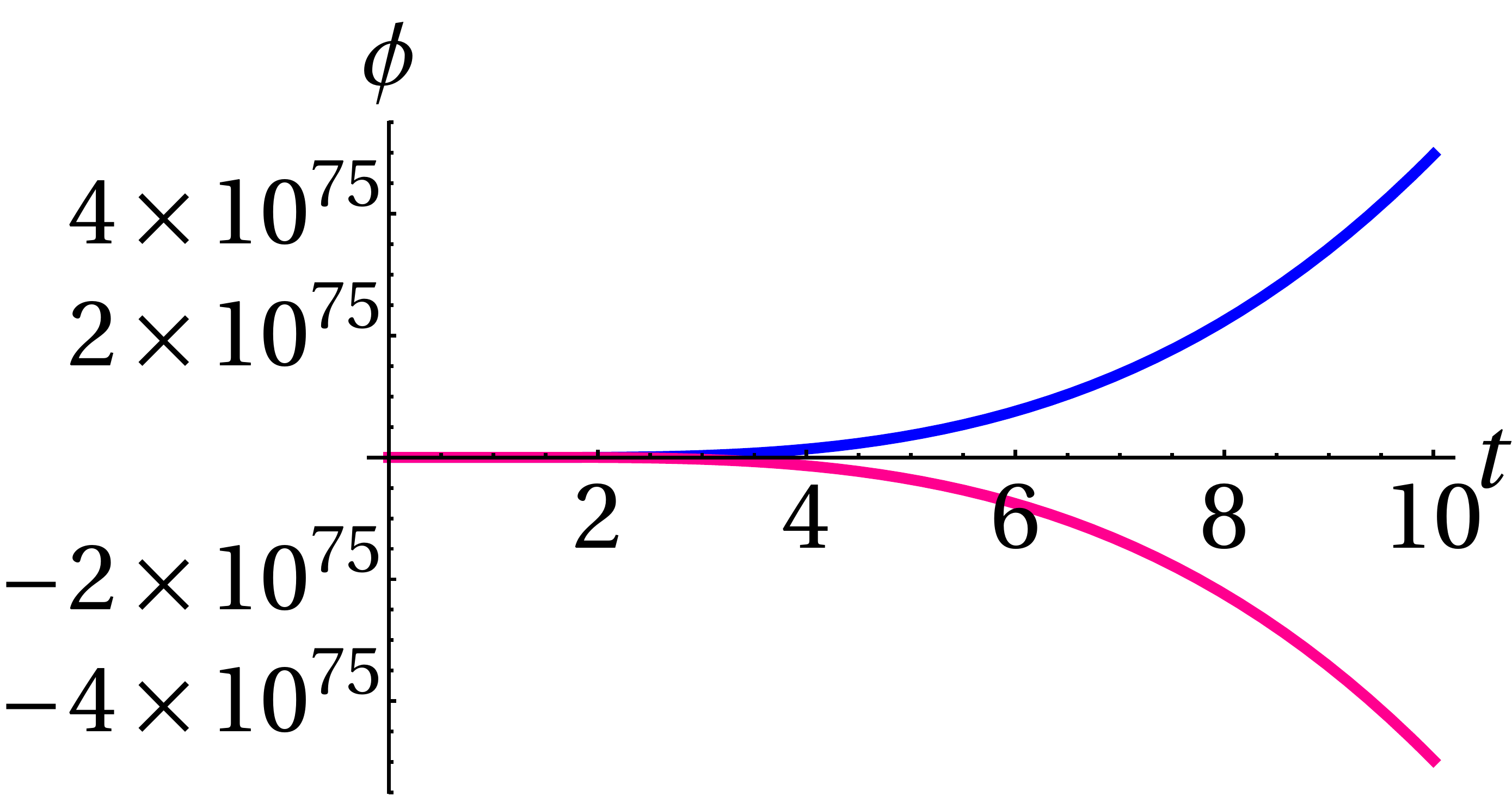}
    \caption{The universal axion $\phi$ for $\dot k\left(0\right) = 1$ (blue curve), and $\dot k\left(0\right) = -1$ (red curve). The solution diverges for $\dot k\left(0\right)=0$.}
    \label{89}
  \end{subfigure}
\qquad
  \begin{subfigure}[t]{.5\linewidth}
    \centering
    \includegraphics[width=0.7\columnwidth]{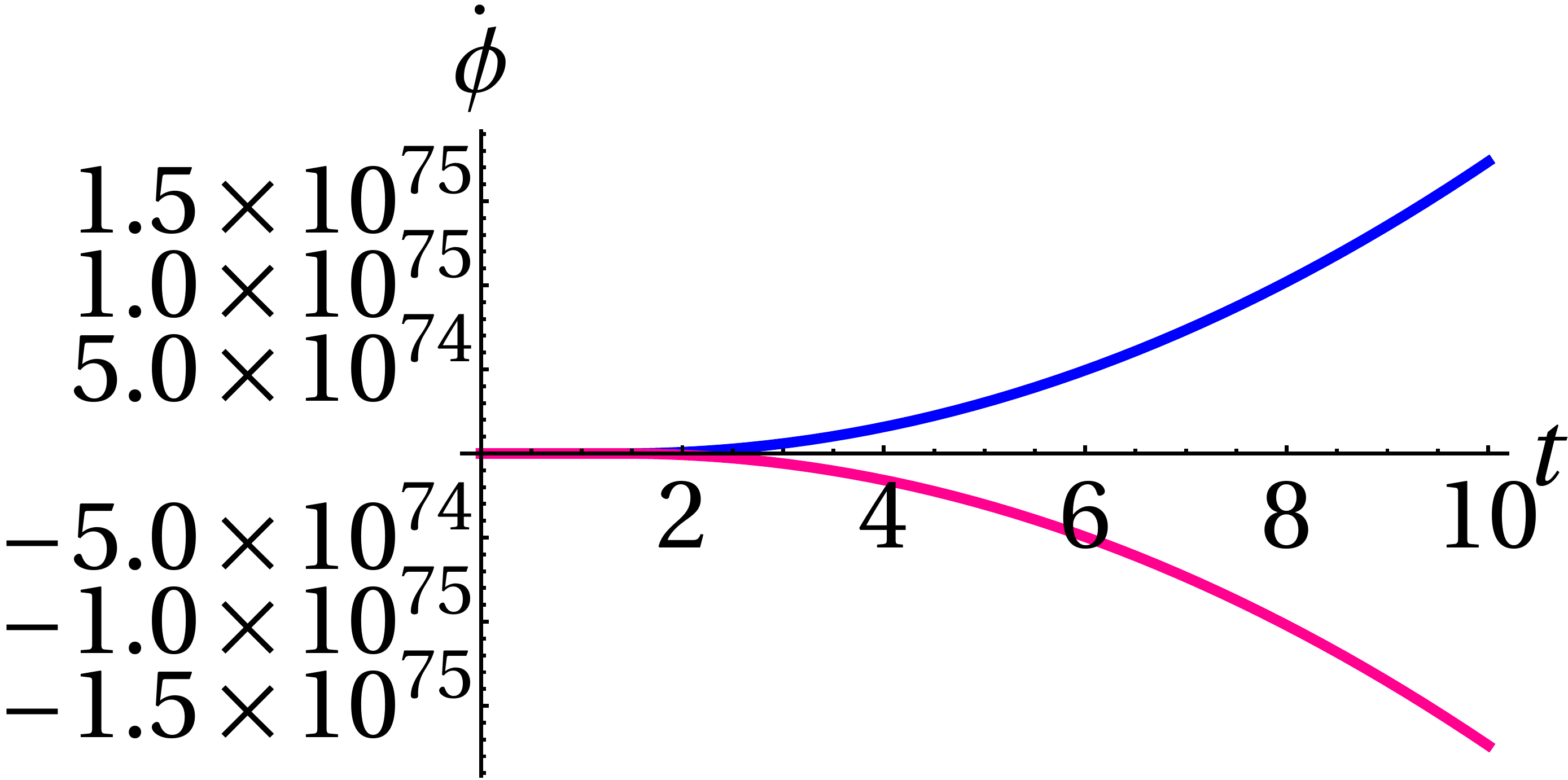}
    \caption{The axionic field strength $\dot\phi$ for $\dot k\left(0\right) = 1$ (blue curve), and $\dot k\left(0\right) = -1$ (red curve).}
    \label{90}
  \end{subfigure}
\\[4em]
    \begin{subfigure}[t]{.5\linewidth}
    \centering
    \includegraphics[width=0.7\columnwidth]{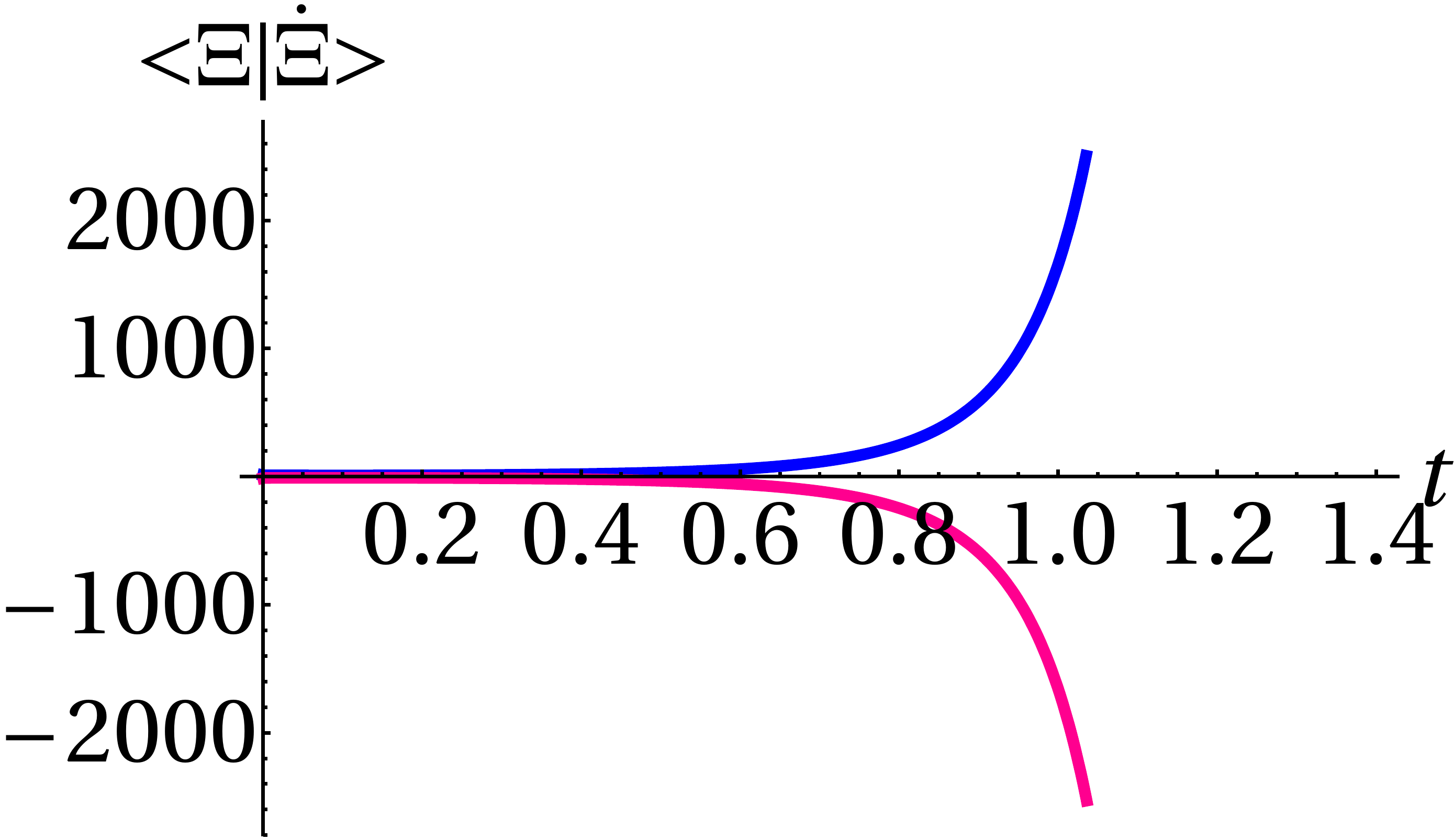}
    \caption{$ \langle \Xi | \dot{\Xi} \rangle$ for $\dot{k}(0)= 1$  (blue), and $\dot{k}(0)  = -1$ (red).}
    \label{91}
  \end{subfigure}
\qquad
  \begin{subfigure}[t]{.5\linewidth}
    \centering
    \includegraphics[width=0.7\columnwidth]{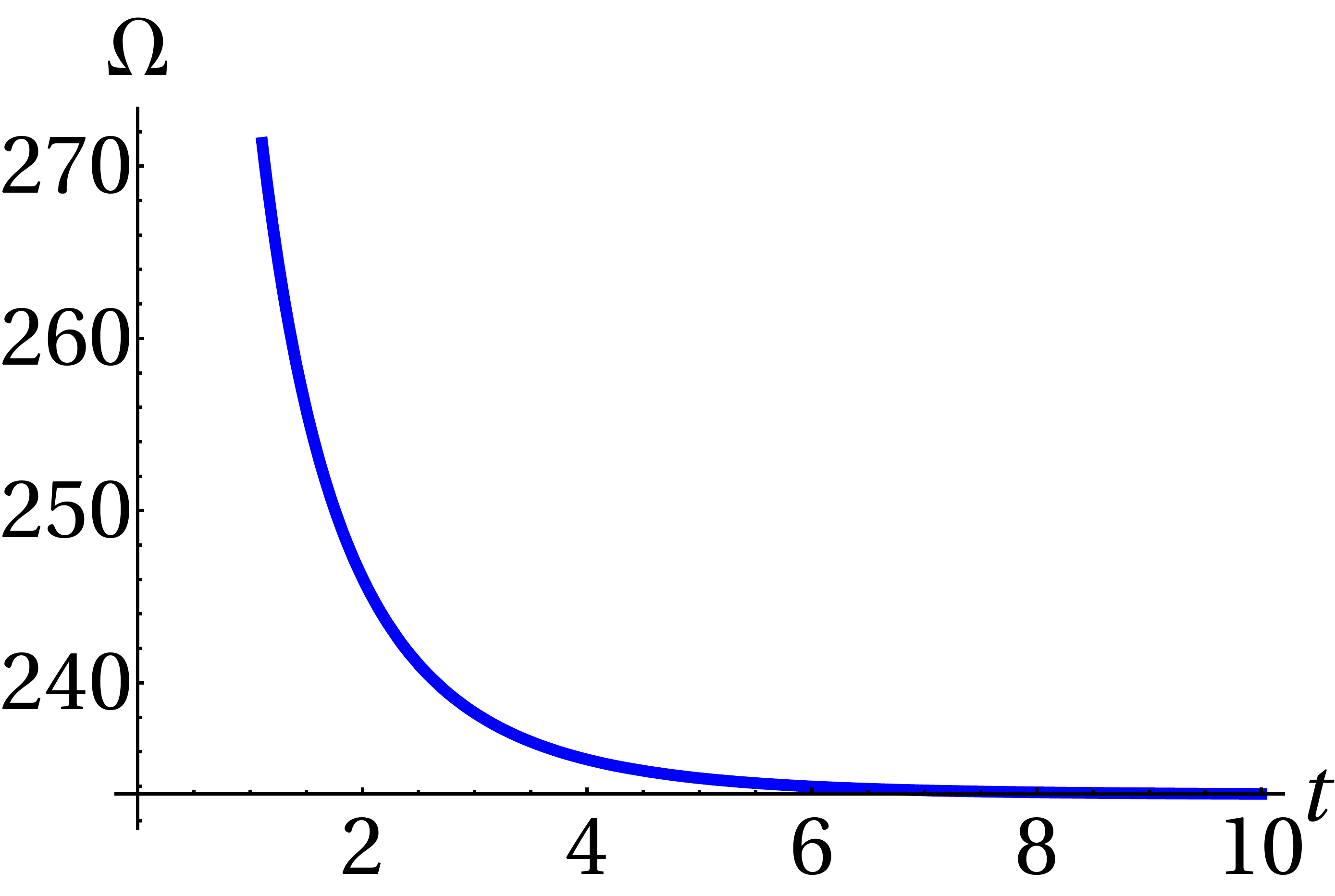}
    \caption{$\Omega$ at $\dot k\left(0\right) = 1$ and  $ \dot\sigma \left(0\right) =0 $.}
    \label{92}
  \end{subfigure}
  \caption{Radiation-filled brane world with initial conditions set number 4 (continued).}
  \label{Fig18}
\end{figure}


\begin{figure}[H]
  \begin{subfigure}[t]{.5\linewidth}
    \centering
    \includegraphics[width=0.7\columnwidth]{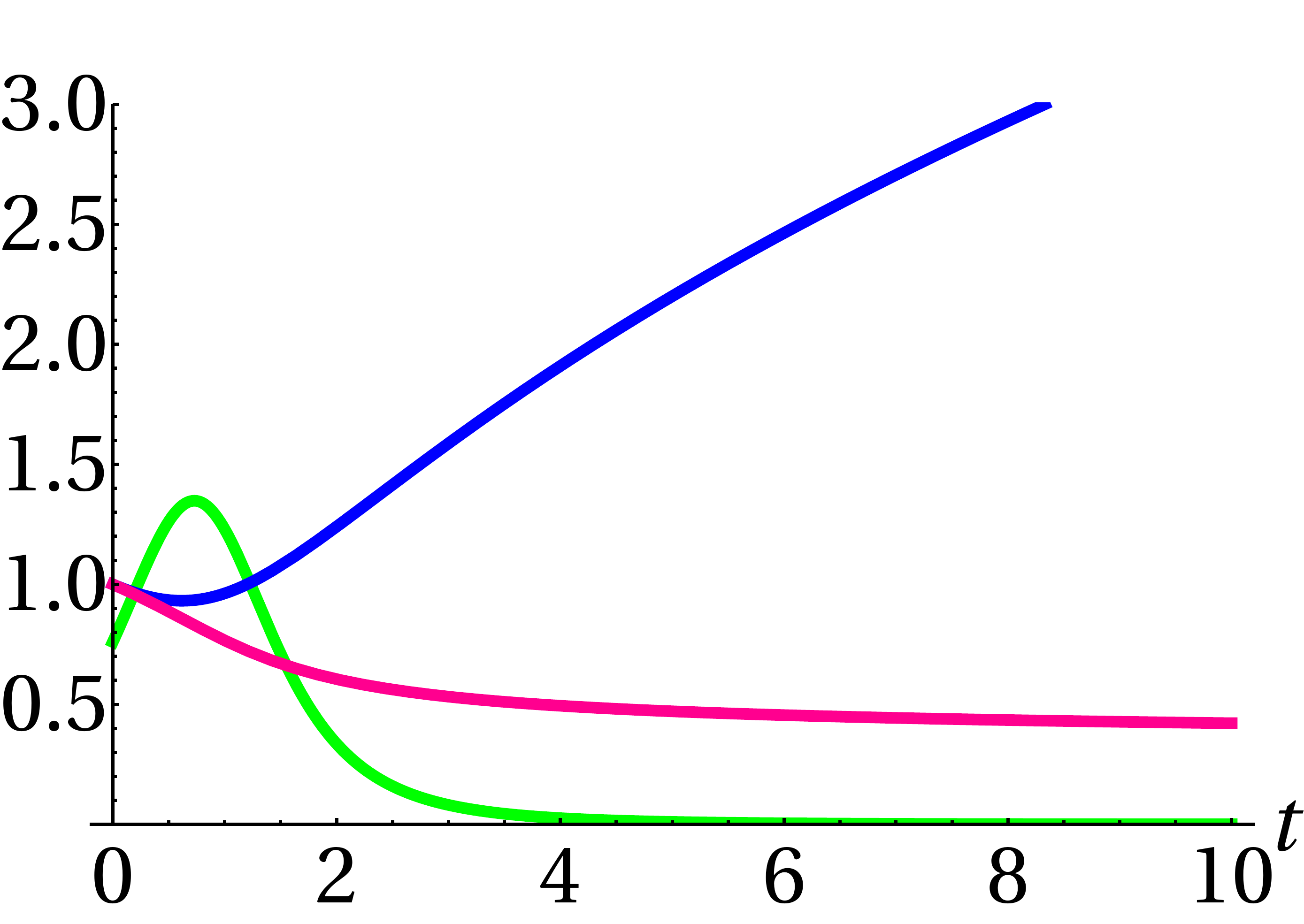}
    \caption{The scale factor $a$ is represented by the blue curve, $b$ by the red curve, while $\left| {G_{i\bar j} \dot z^i \dot z^{\bar j}} \right|$ by the green curve.}
    \label{93}
  \end{subfigure}
\qquad
  \begin{subfigure}[t]{.5\linewidth}
    \centering
    \includegraphics[width=0.7\columnwidth]{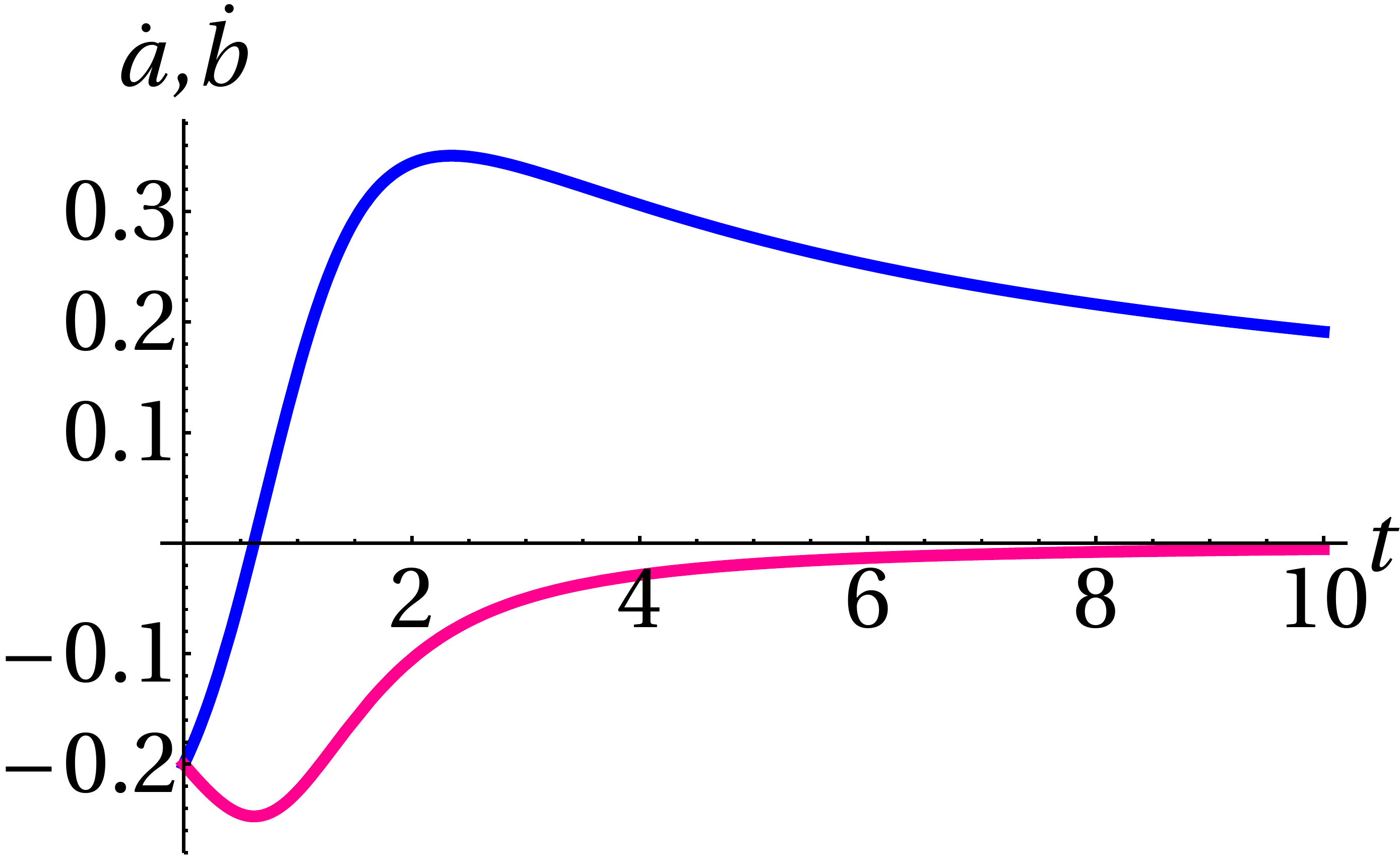}
    \caption{The expansion rates of the scale factors: $\dot a$ is represented by the blue curve, and $\dot b$ by the red curve.}
    \label{94}
  \end{subfigure}
\\[9em]
  \begin{subfigure}[t]{.5\linewidth}
    \centering
    \includegraphics[width=0.7\columnwidth]{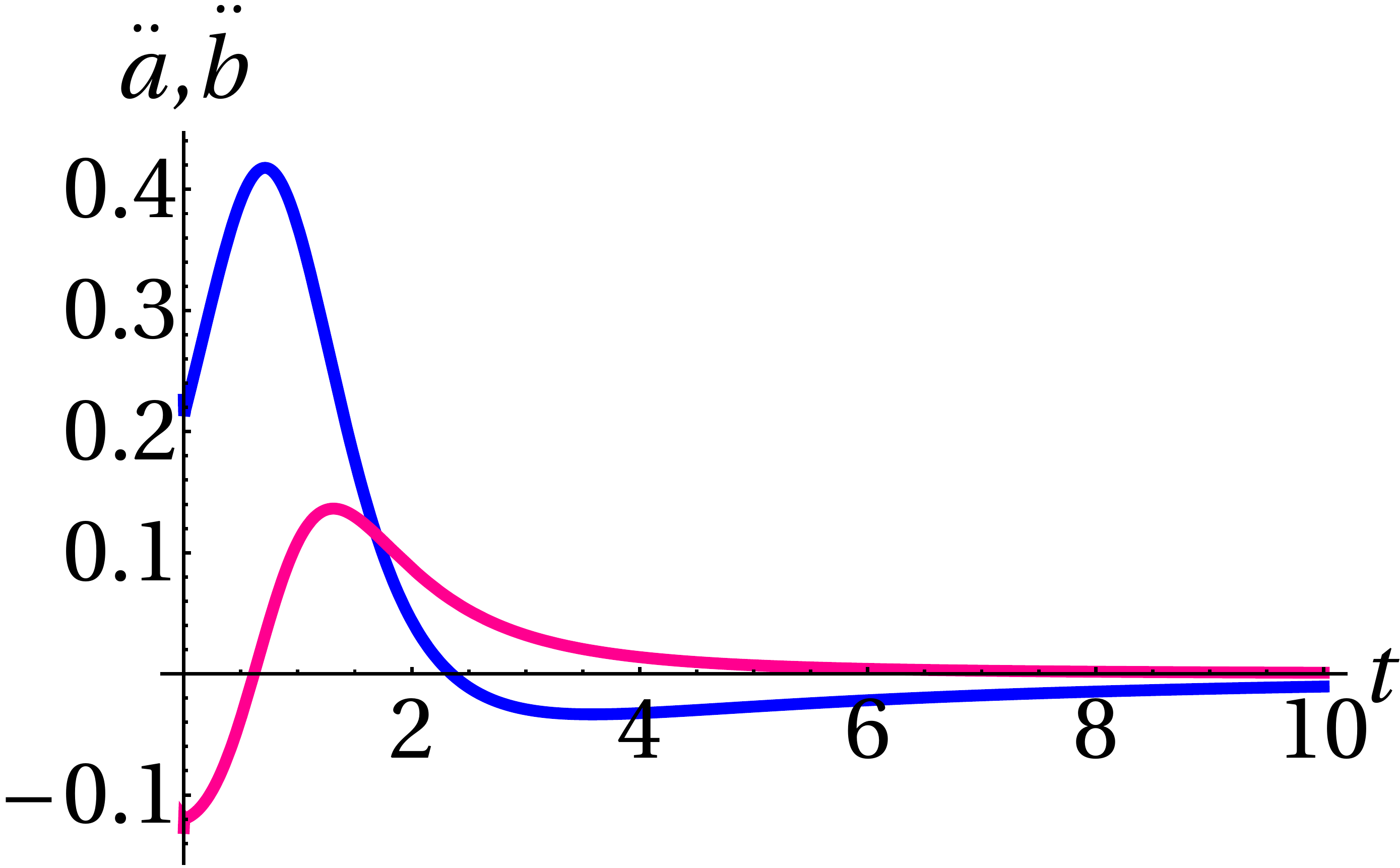}
    \caption{The accelerations of the scale factors: $\ddot a$ is represented by the blue curve, and $\ddot b$ by the red curve.}
    \label{95}
  \end{subfigure}
\qquad
  \begin{subfigure}[t]{.5\linewidth}
    \centering
    \includegraphics[width=0.7\columnwidth]{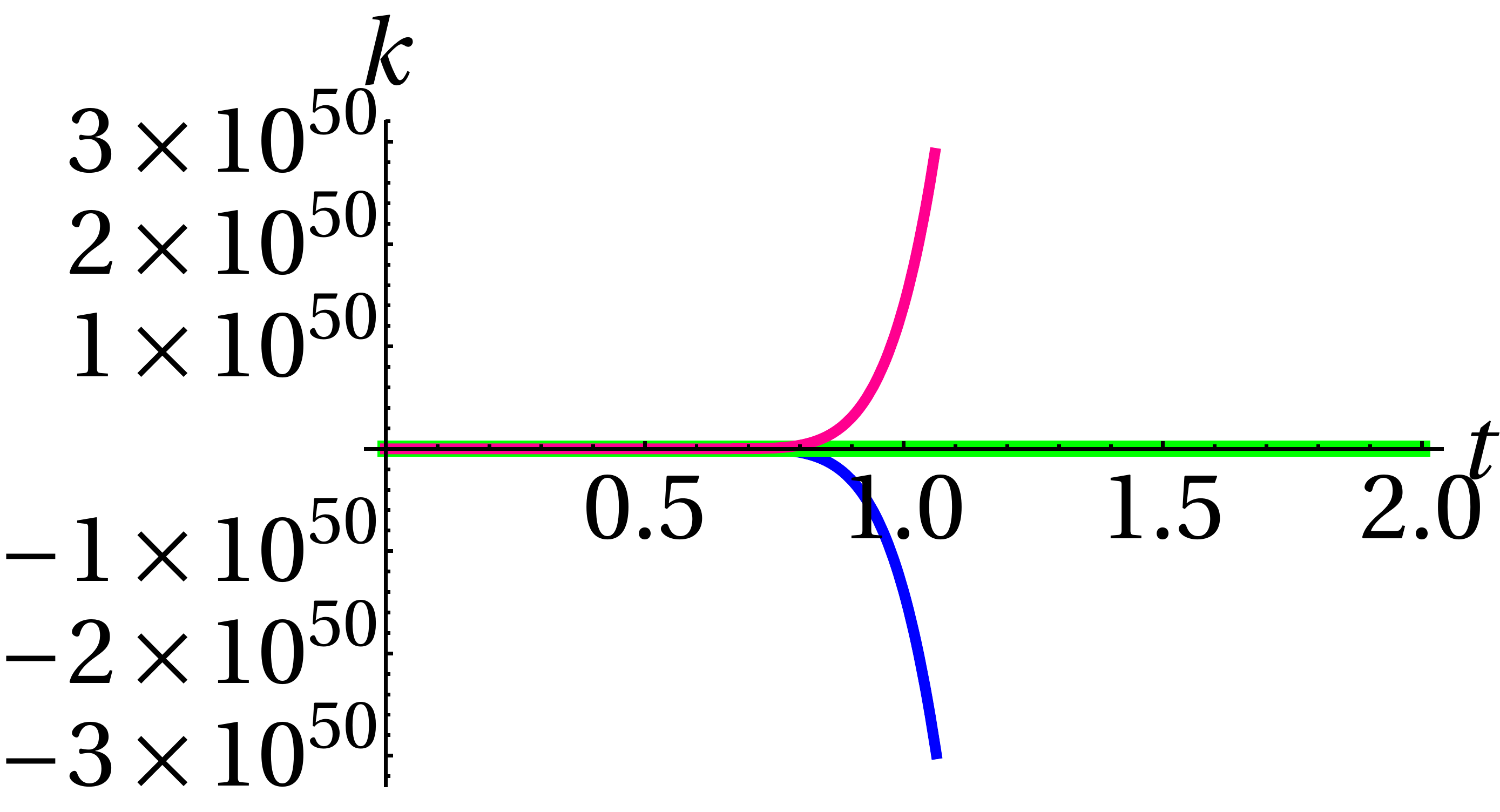}
    \caption{The harmonic function $k$ using: $\dot k\left(0\right)=1$ (blue curve), $\dot k\left(0\right)=0$ (green line), and $\dot k\left(0\right)=-1$ (red curve).}
    \label{96}
  \end{subfigure}
  \caption{Radiation-filled brane world with initial conditions set number 5.}
  \label{Fig19}
\end{figure}
\begin{figure}[H]
\begin{subfigure}[t]{.5\linewidth}
    \centering
    \includegraphics[width=0.7\columnwidth]{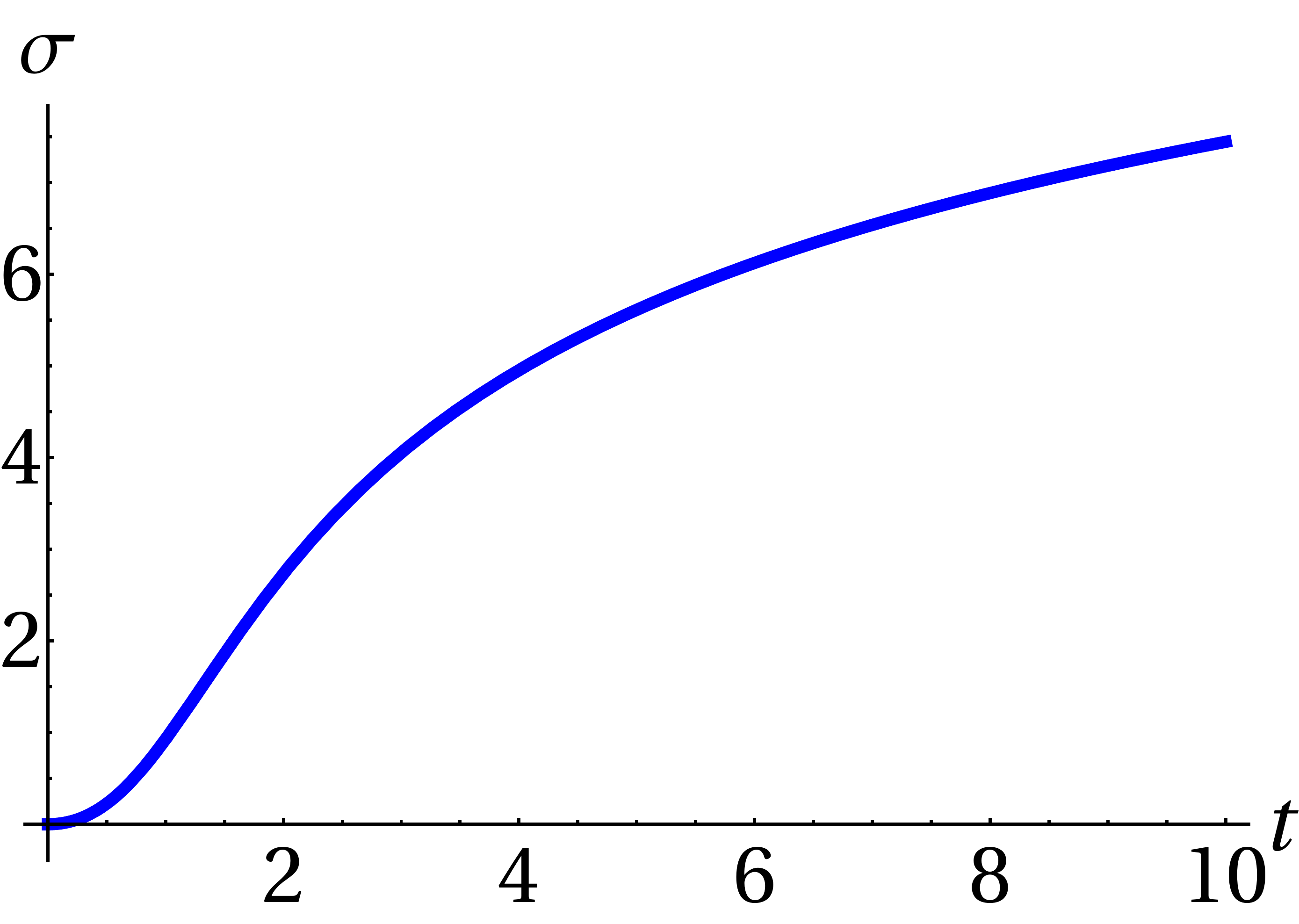}
    \caption{The dilaton $\sigma$; same for all three $\dot k\left(0\right)$.}
    \label{97}
  \end{subfigure}
\qquad
  \begin{subfigure}[t]{.5\linewidth}
    \centering
    \includegraphics[width=0.7\columnwidth]{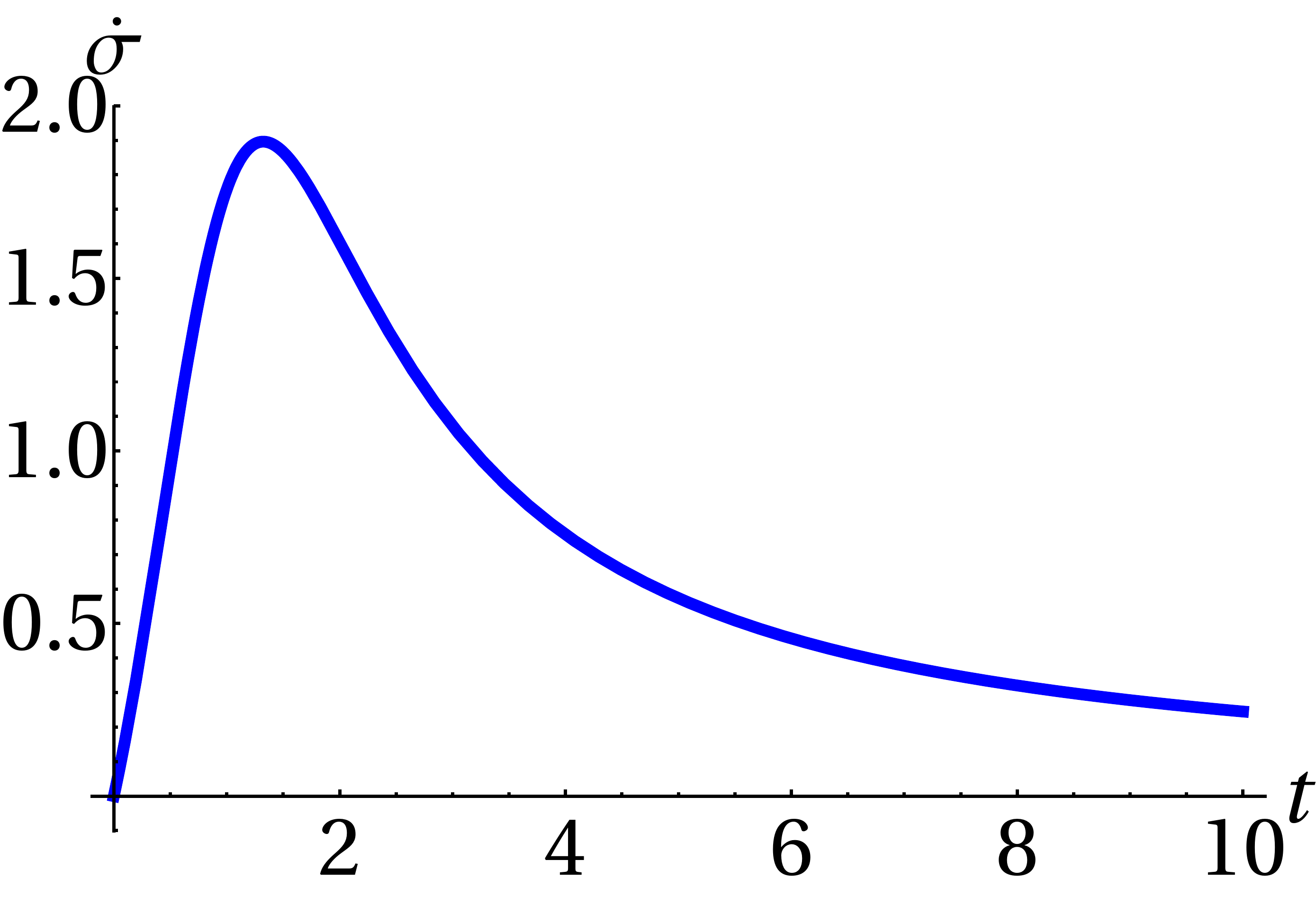}
    \caption{The dilatonic field strength $\dot\sigma$.}
    \label{98}
  \end{subfigure}    
\\[4em]
  \begin{subfigure}[t]{.5\linewidth}
    \centering
    \includegraphics[width=0.7\columnwidth]{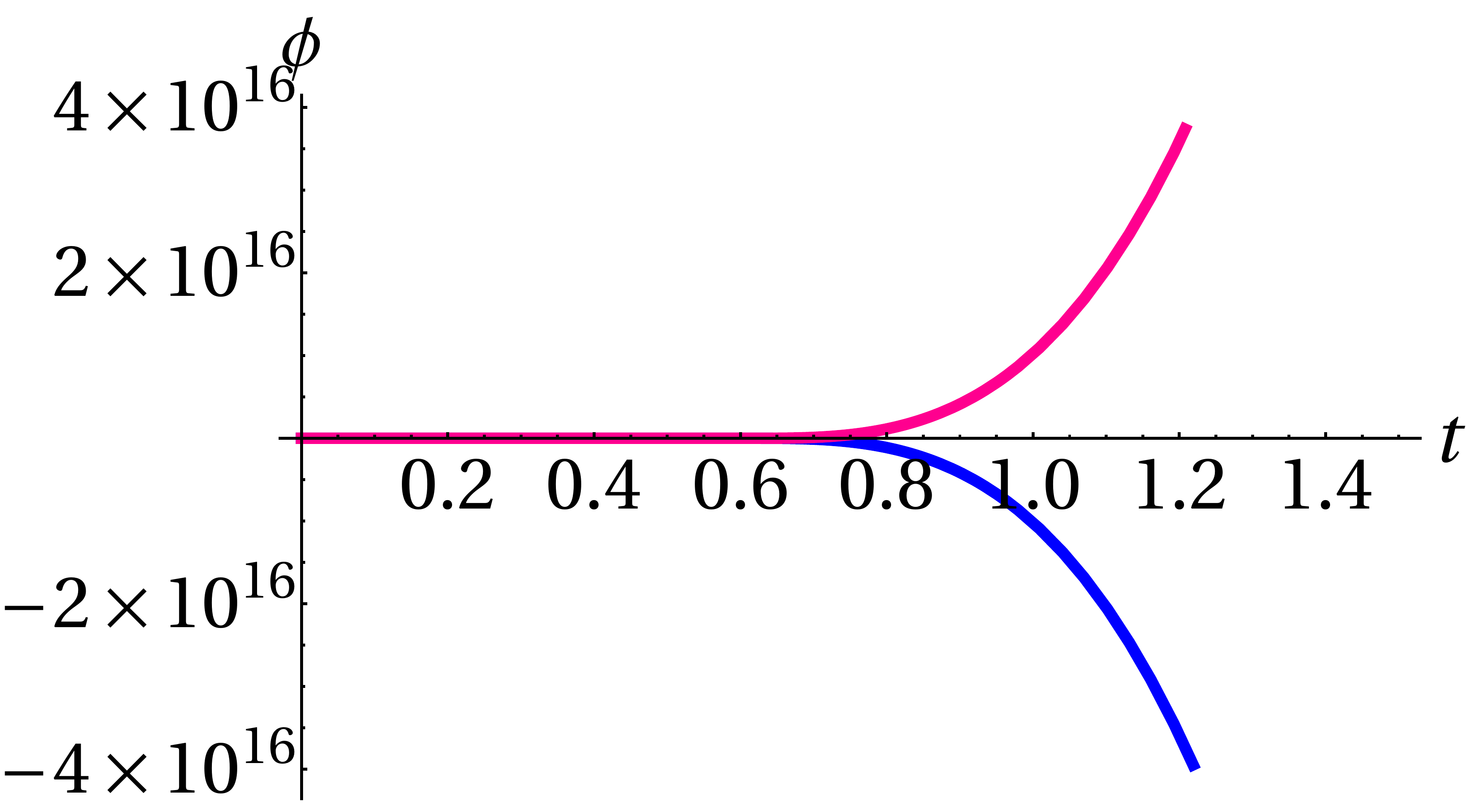}
    \caption{The universal axion $\phi$ for $\dot k\left(0\right) = 1$ (blue curve), and $\dot k\left(0\right) = -1$ (red curve). The solution diverges for $\dot k\left(0\right)=0$.}
    \label{99}
  \end{subfigure}
\qquad
  \begin{subfigure}[t]{.5\linewidth}
    \centering
    \includegraphics[width=0.7\columnwidth]{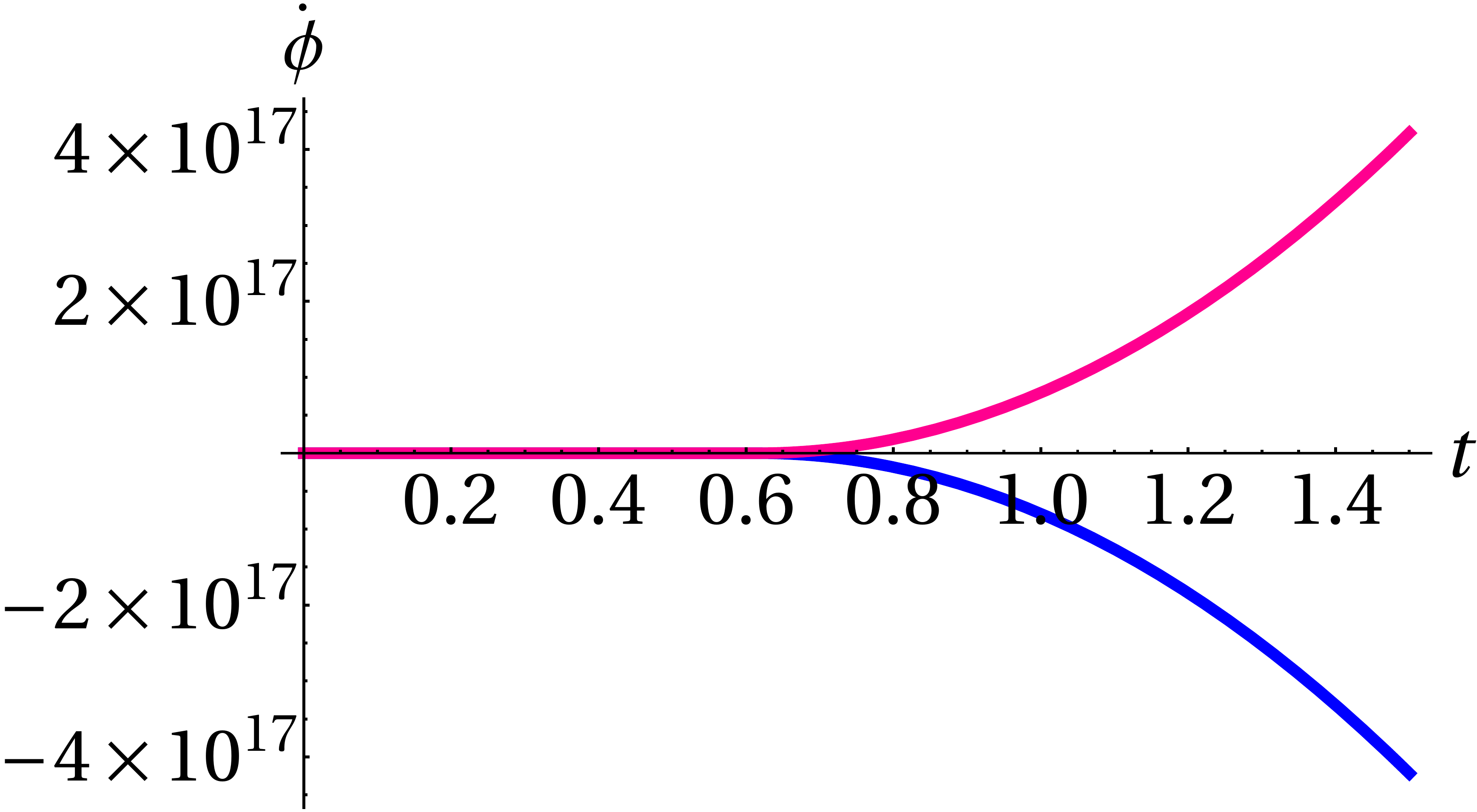}
    \caption{The axionic field strength $\dot\phi$ for $\dot k\left(0\right) = 1$ (blue curve), and $\dot k\left(0\right) = -1$ (red curve).}
    \label{100}
  \end{subfigure}
\\[4em]
  \begin{subfigure}[t]{.5\linewidth}
    \centering
    \includegraphics[width=0.7\columnwidth]{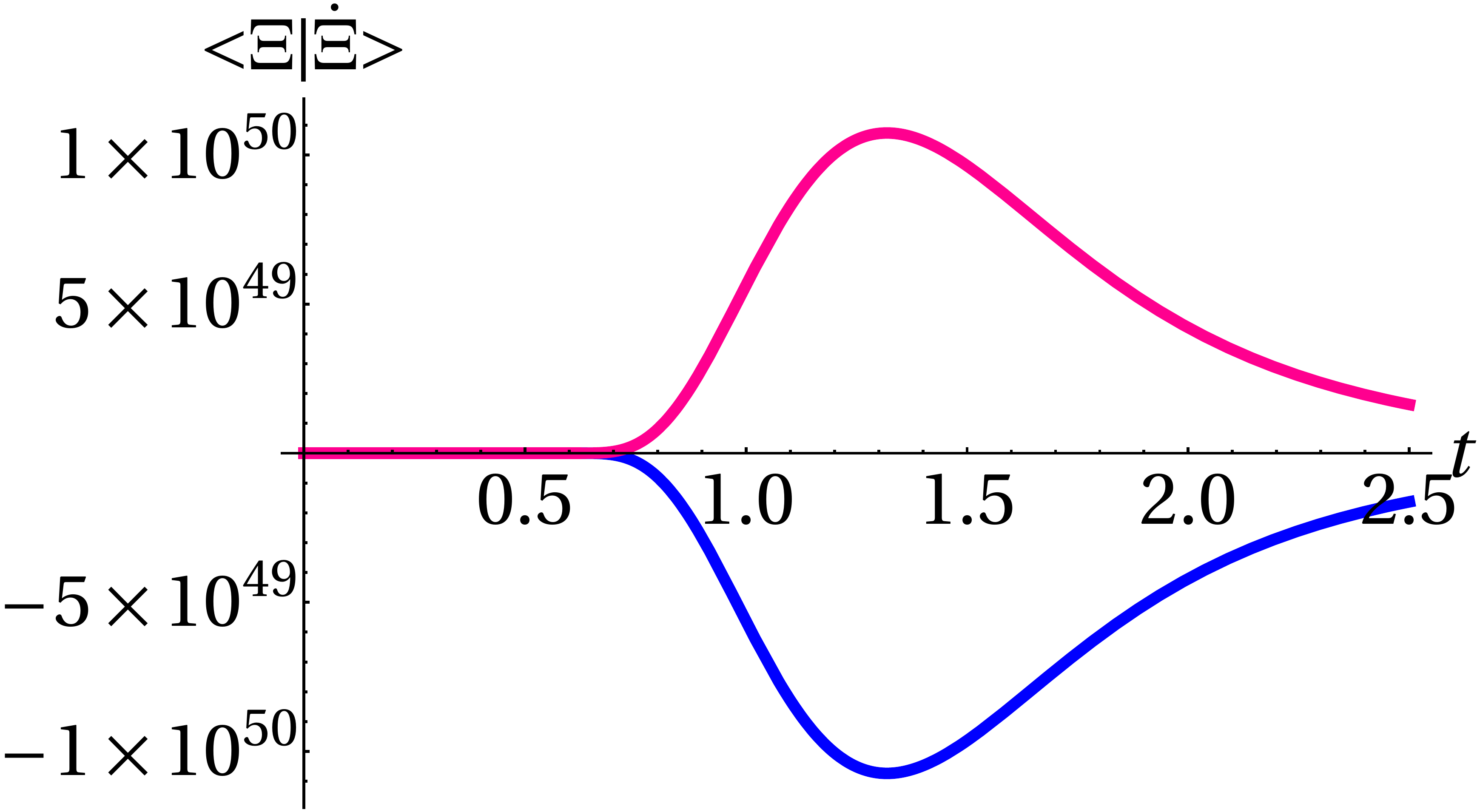}
    \caption{$ \langle \Xi | \dot{\Xi} \rangle$ for $\dot{k}(0)= 1$  (blue), and $\dot{k}(0)  = -1$ (red).}
    \label{101}
  \end{subfigure}
\qquad
  \begin{subfigure}[t]{.5\linewidth}
    \centering
    \includegraphics[width=0.7\columnwidth]{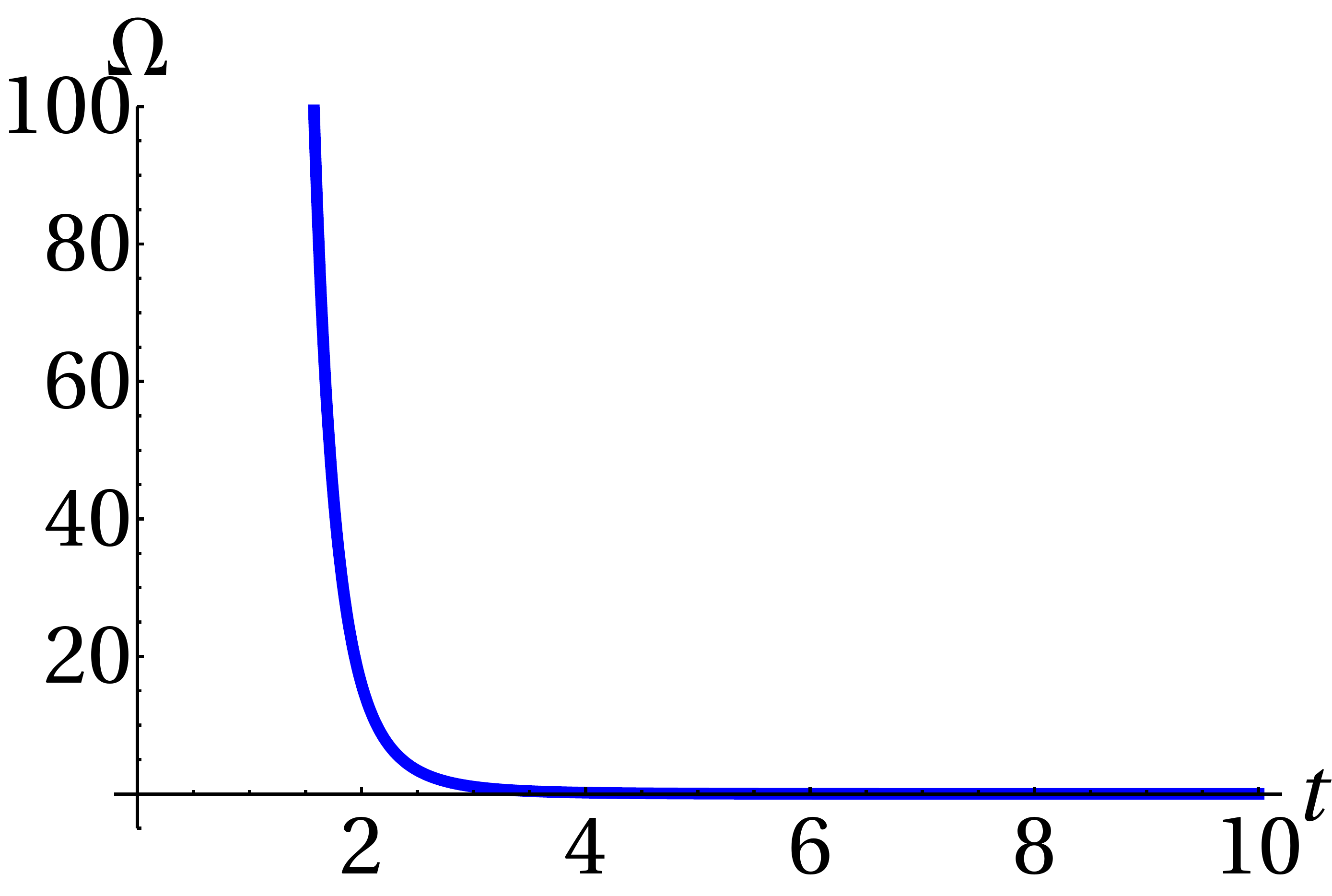}
    \caption{$\Omega$ at $\dot k\left(0\right) = 1$ and  $ \dot\sigma \left(0\right) =0 $.}
    \label{102}
  \end{subfigure}
  \caption{Radiation-filled brane world with initial conditions set number 5 (continued).}
  \label{Fig20}
\end{figure}


\begin{figure}[H]
  \begin{subfigure}[t]{.5\linewidth}
    \centering
    \includegraphics[width=0.7\columnwidth]{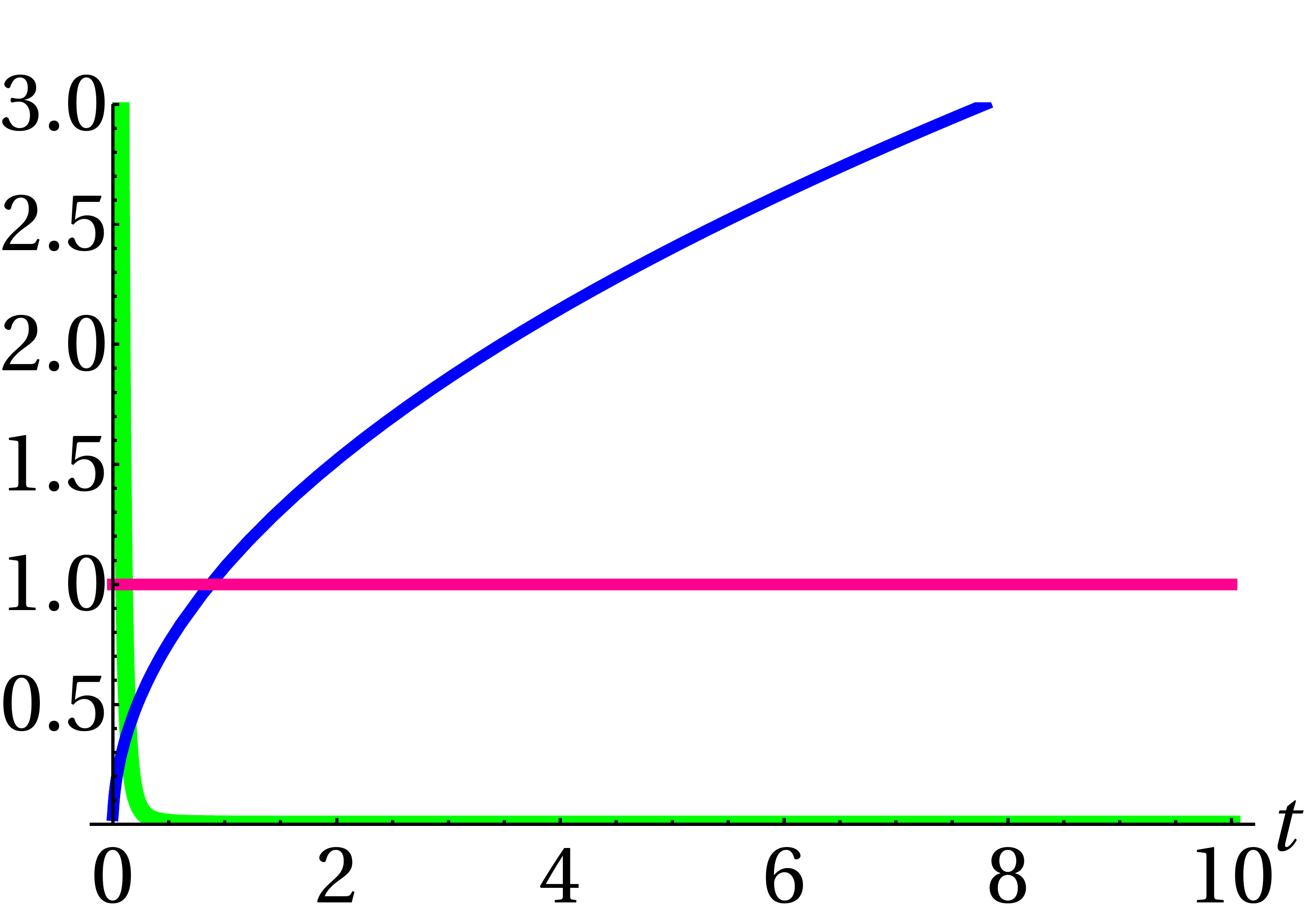}
    \caption{The scale factor $a$ is represented by the blue curve, $b$ by the red curve, while $\left| {G_{i\bar j} \dot z^i \dot z^{\bar j}} \right|$ by the green curve.}
    \label{113}
  \end{subfigure}
\qquad
  \begin{subfigure}[t]{.5\linewidth}
    \centering
    \includegraphics[width=0.7\columnwidth]{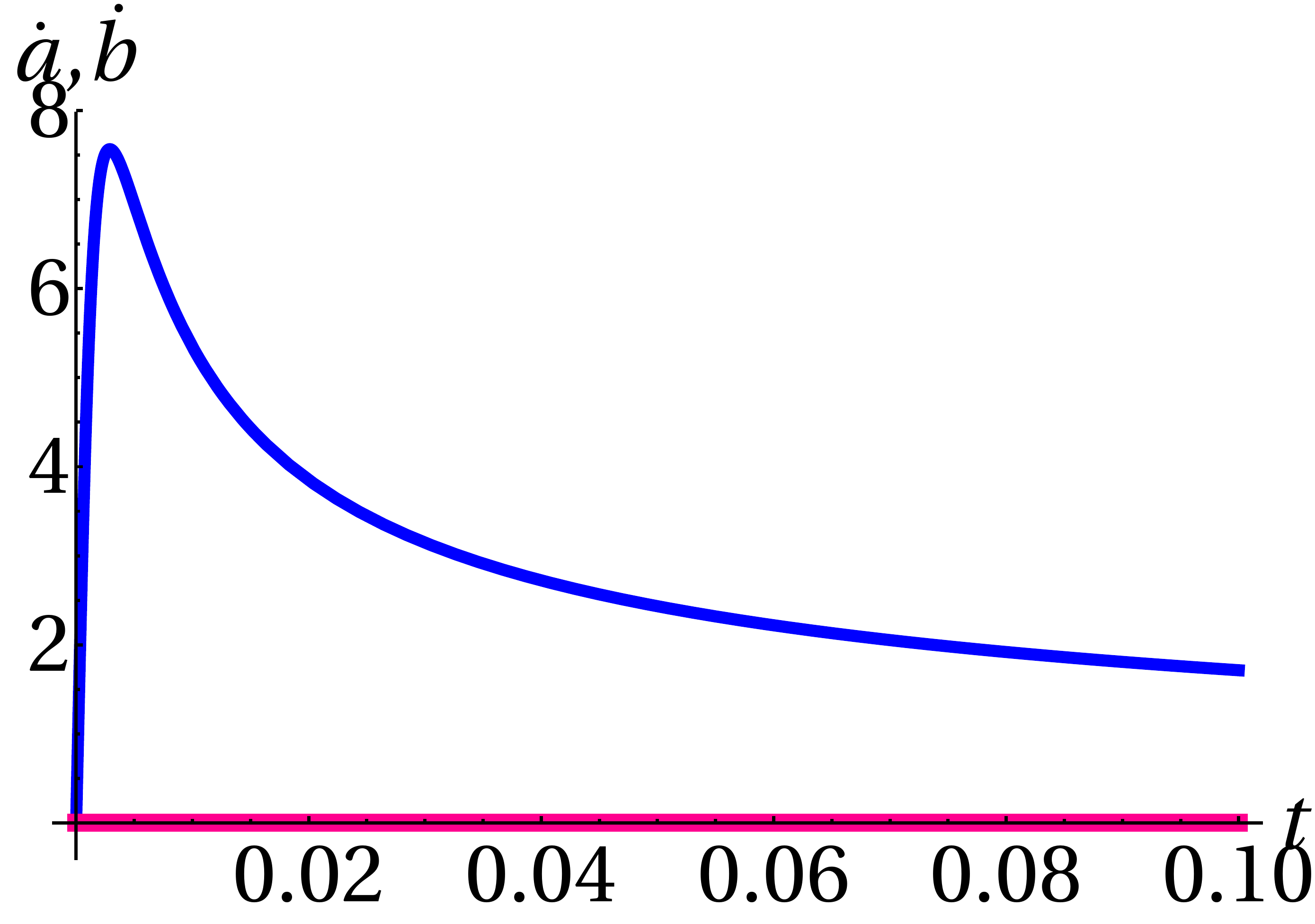}
    \caption{The expansion rates of the scale factors: $\dot a$ is represented by the blue curve, and $\dot b$ by the red curve (flat on the $t$ axis).}
    \label{114}
  \end{subfigure}
  \\[9em]
  \begin{subfigure}[t]{.5\linewidth}
    \centering
    \includegraphics[width=0.7\columnwidth]{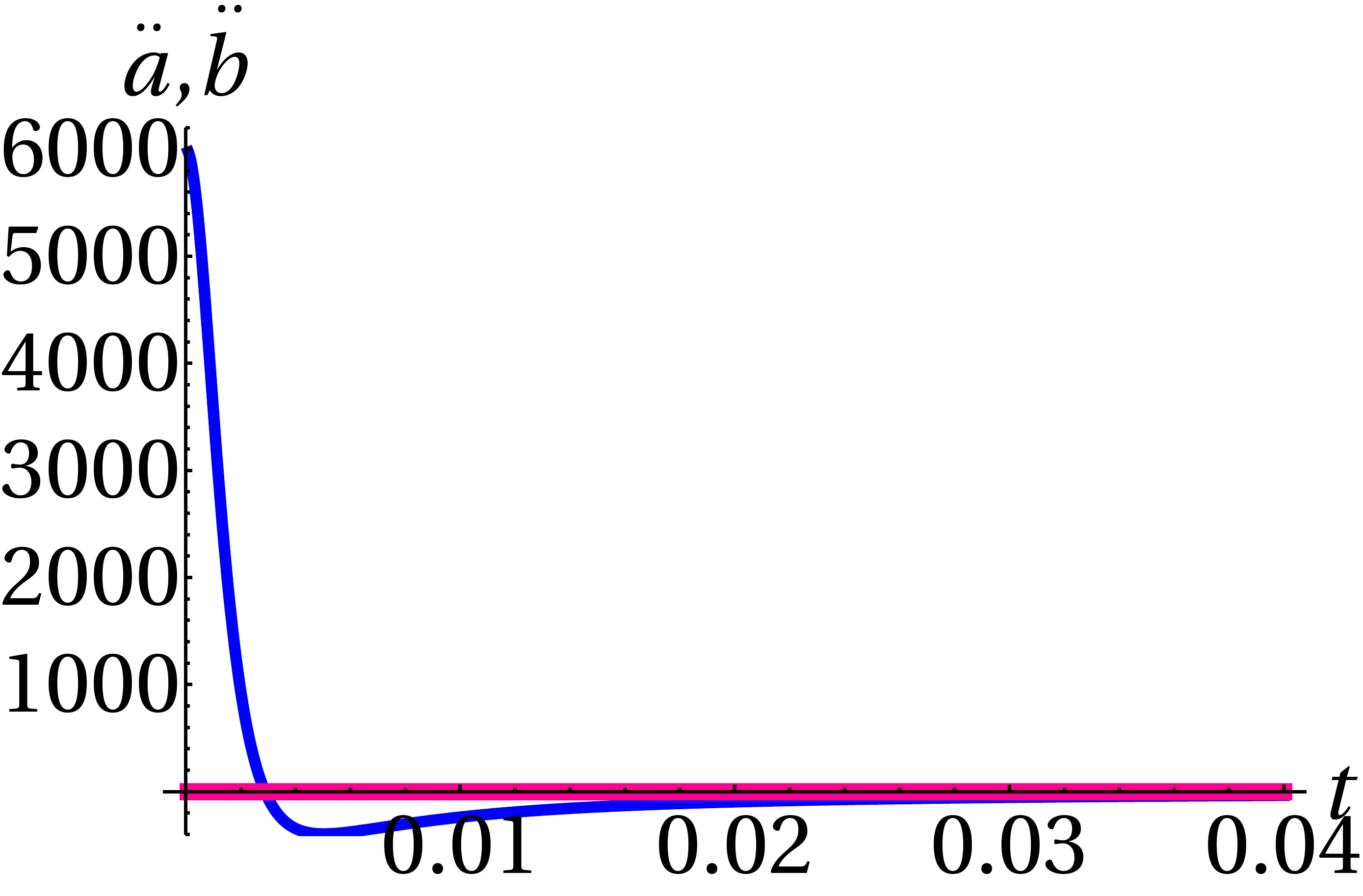}
    \caption{The accelerations of the scale factors: $\ddot a$ is represented by the blue curve, and $\ddot b$ by the red curve (flat on the $t$ axis).}
    \label{115}
  \end{subfigure}
\qquad
  \begin{subfigure}[t]{.5\linewidth}
    \centering
    \includegraphics[width=0.7\columnwidth]{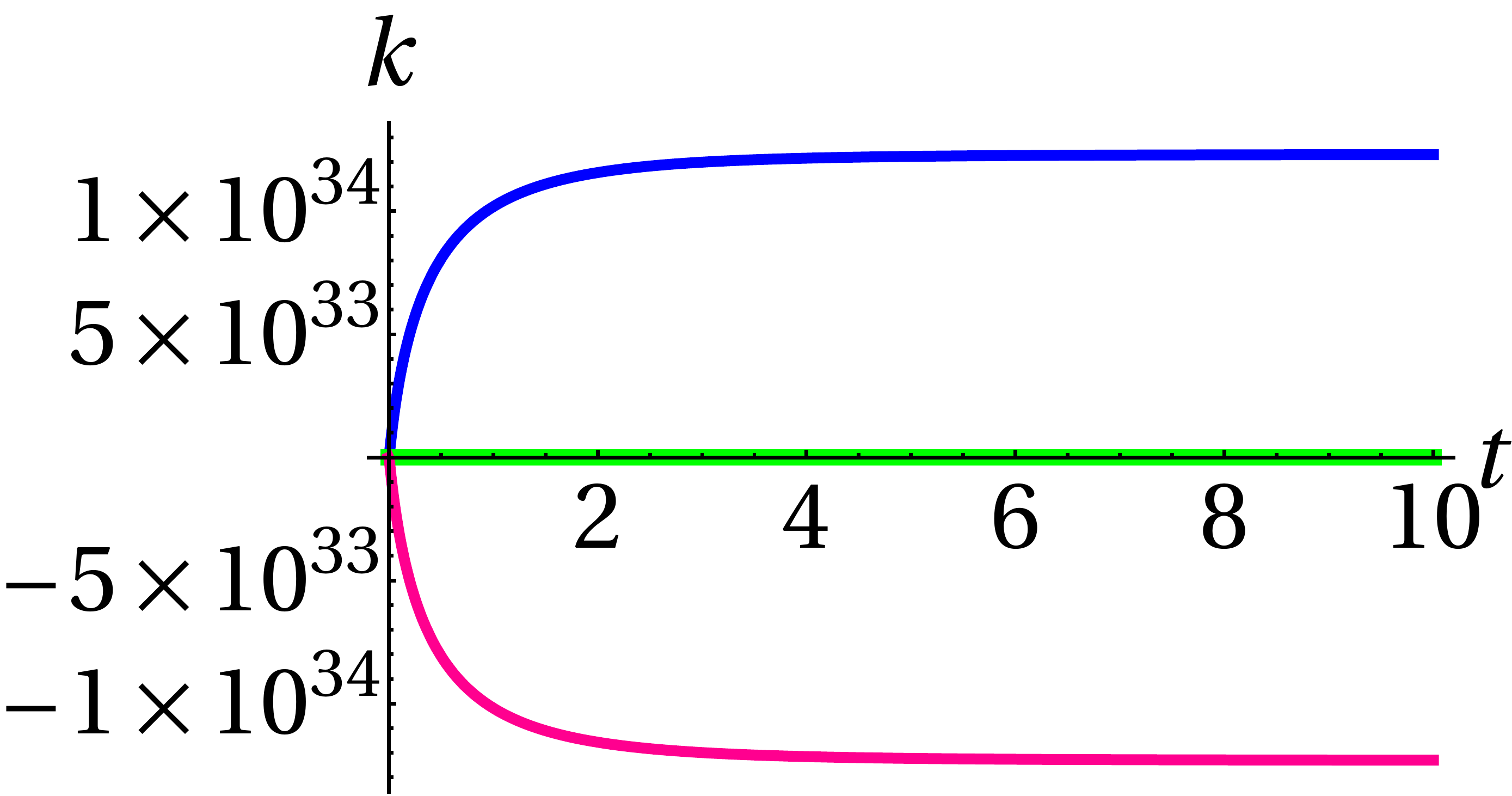}
    \caption{The harmonic function $k$ using: $\dot k\left(0\right)=1$ (blue curve), $\dot k\left(0\right)=0$ (green line), and $\dot k\left(0\right)=-1$ (red curve).}
    \label{116}
  \end{subfigure}
\caption{Radiation-filled brane world with initial conditions set number 6.}
  \label{Fig23}
\end{figure}
\begin{figure}[H]
  \begin{subfigure}[t]{.5\linewidth}
    \centering
    \includegraphics[width=0.7\columnwidth]{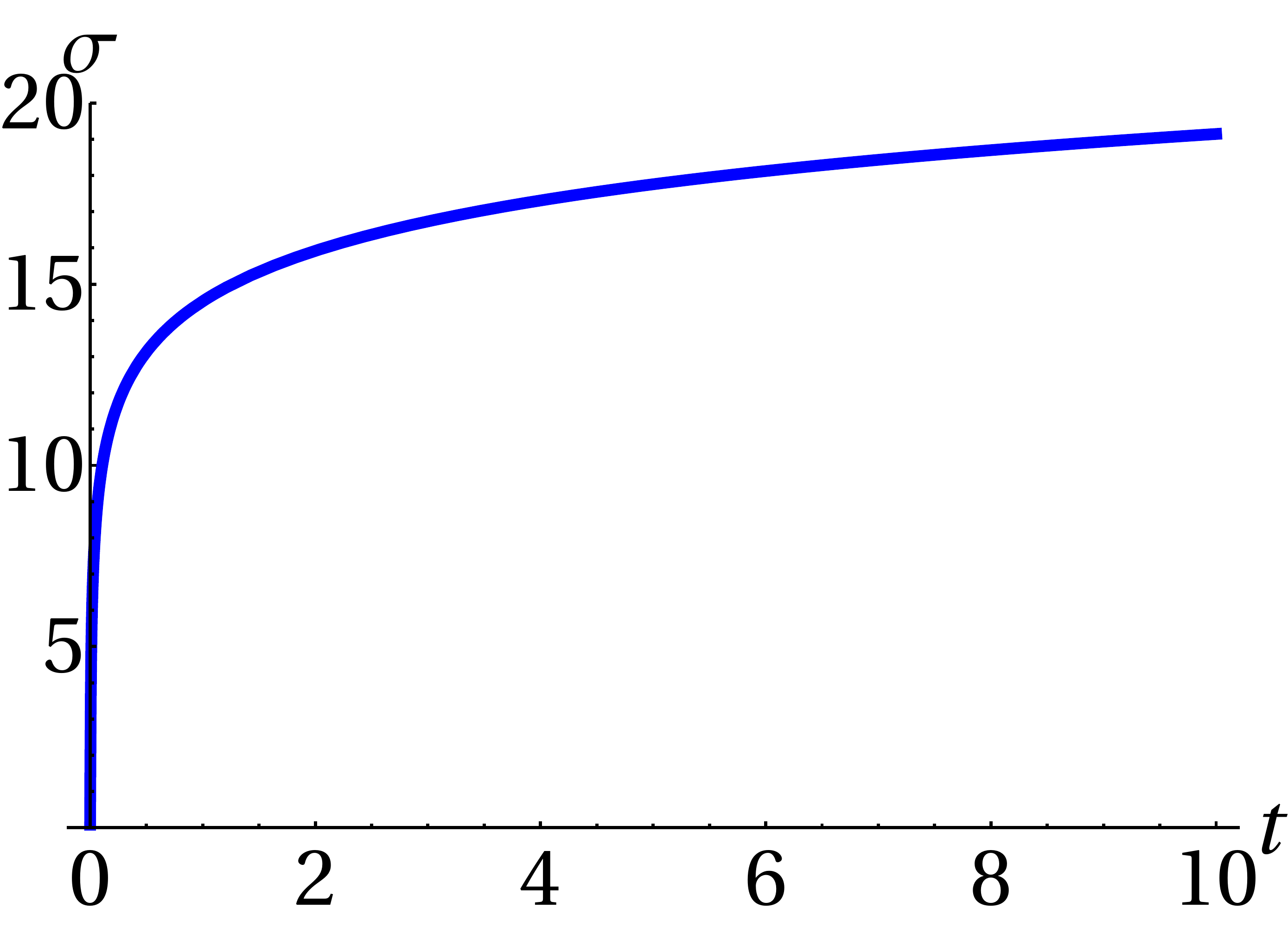}
    \caption{The dilaton $\sigma$; same for all three $\dot k\left(0\right)$.}
    \label{117}
  \end{subfigure}
\qquad
  \begin{subfigure}[t]{.5\linewidth}
    \centering
    \includegraphics[width=0.7\columnwidth]{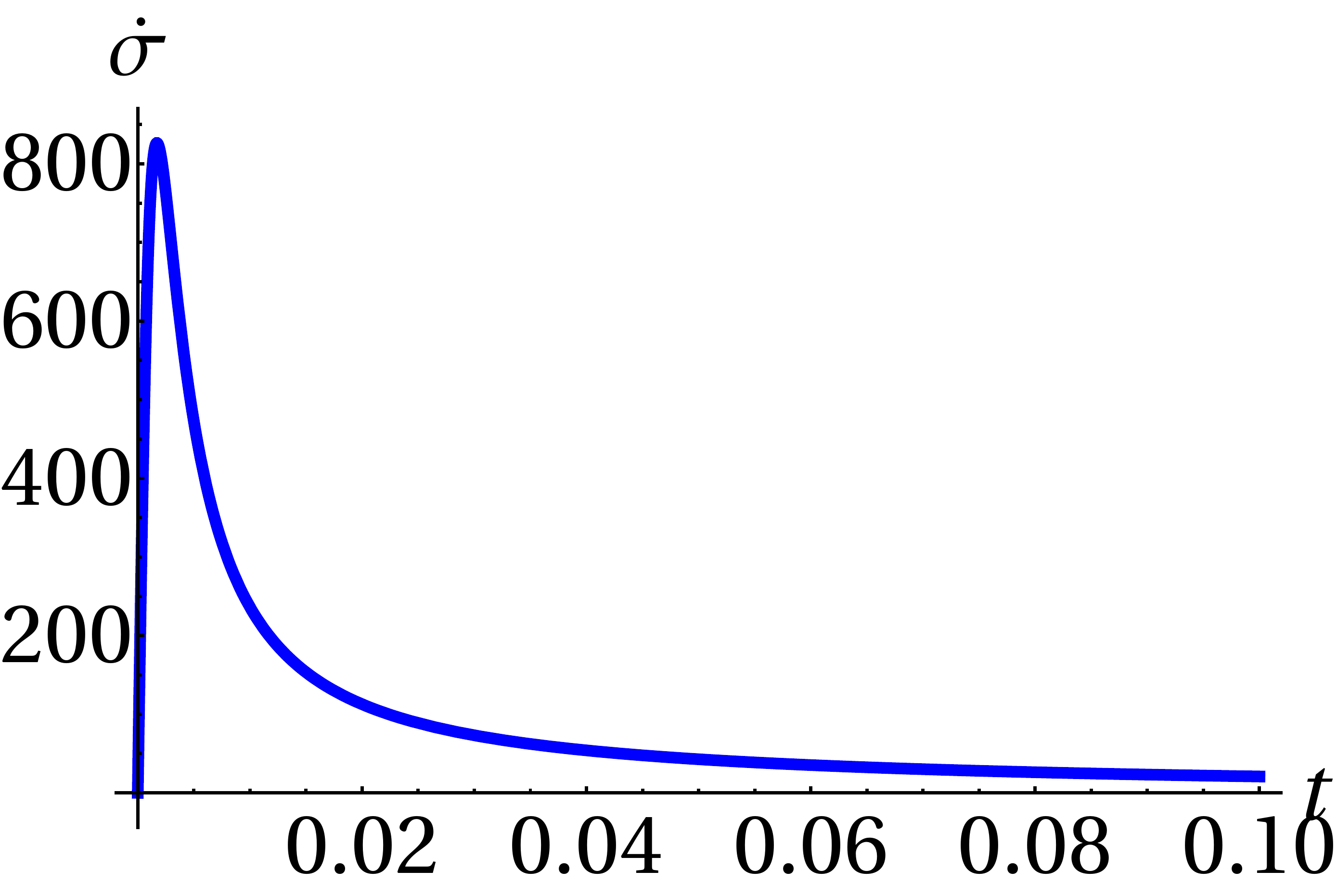}
    \caption{The dilatonic field strength $\dot\sigma$.}
    \label{118}
  \end{subfigure}
\\[4em]  
  \begin{subfigure}[t]{.5\linewidth}
    \centering
    \includegraphics[width=0.7\columnwidth]{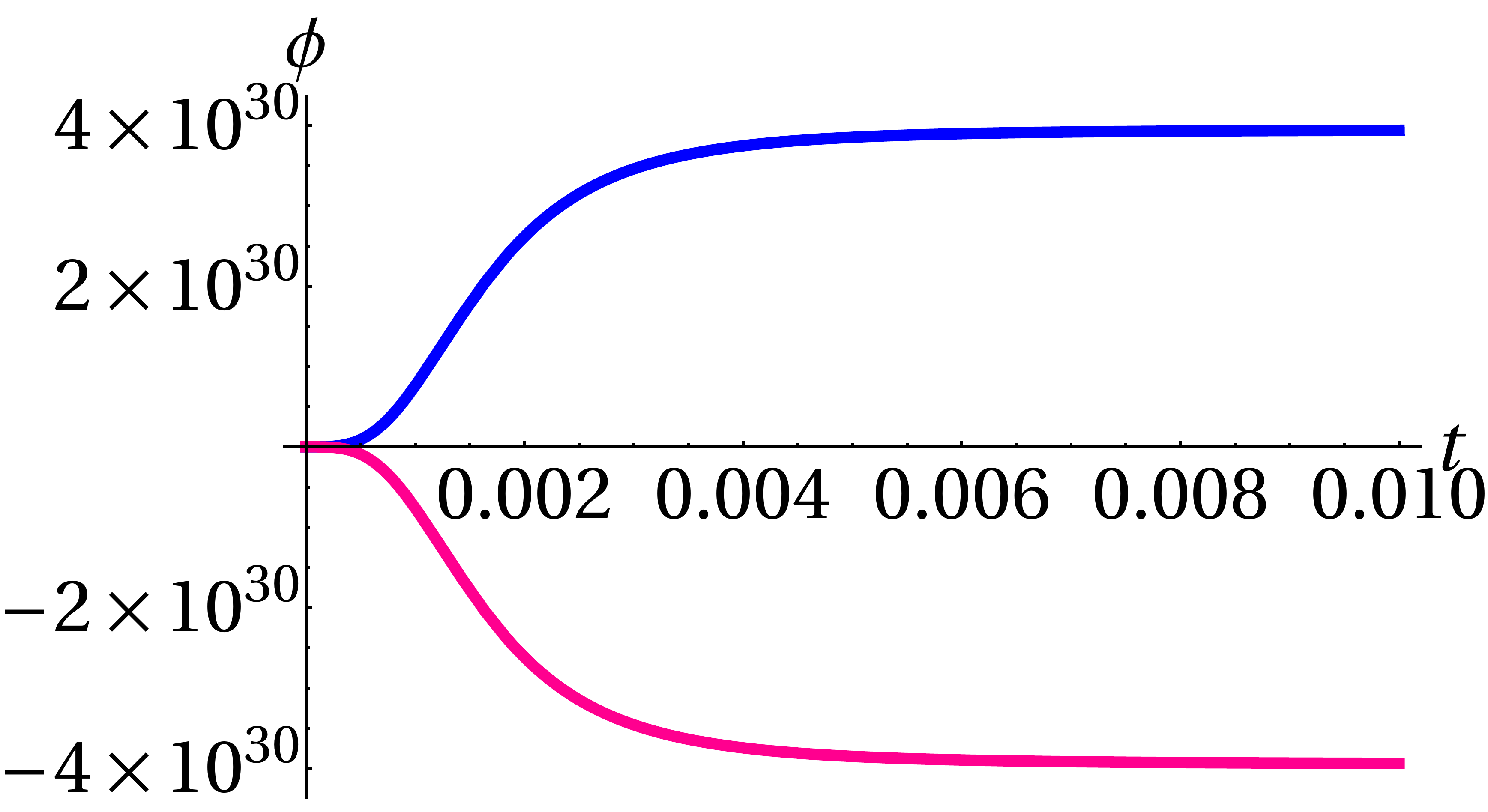}
    \caption{The universal axion $\phi$ for $\dot k\left(0\right) = 1$ (blue curve), and $\dot k\left(0\right) = -1$ (red curve). The solution diverges for $\dot k\left(0\right)=0$.}
    \label{119}
  \end{subfigure}
\qquad
  \begin{subfigure}[t]{.5\linewidth}
    \centering
    \includegraphics[width=0.7\columnwidth]{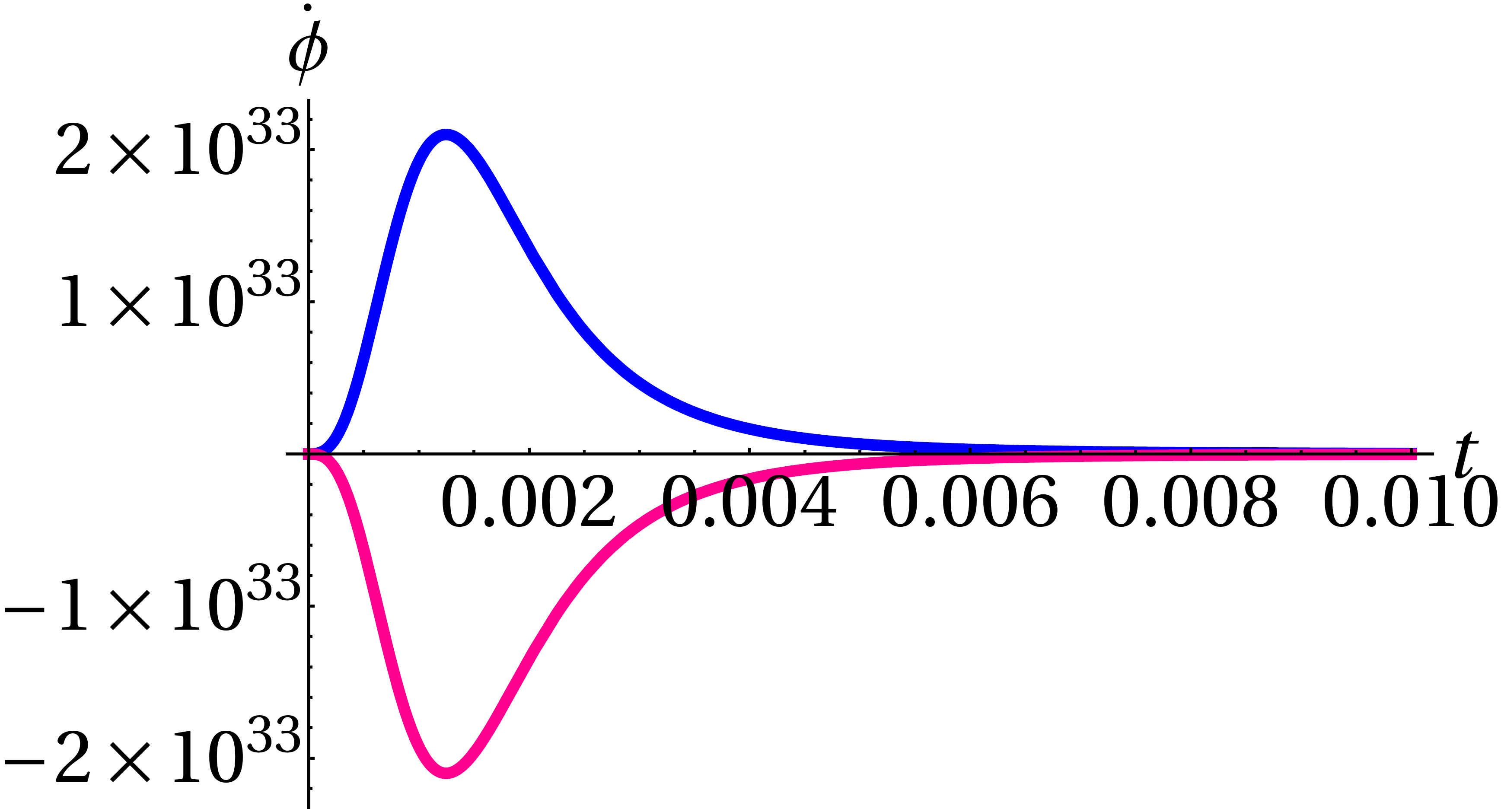}
    \caption{The axionic field strength $\dot\phi$ for $\dot k\left(0\right) = 1$ (blue curve), and $\dot k\left(0\right) = -1$ (red curve).}
    \label{120}
  \end{subfigure}
\\[4em]
  \begin{subfigure}[t]{.5\linewidth}
    \centering
    \includegraphics[width=0.7\columnwidth]{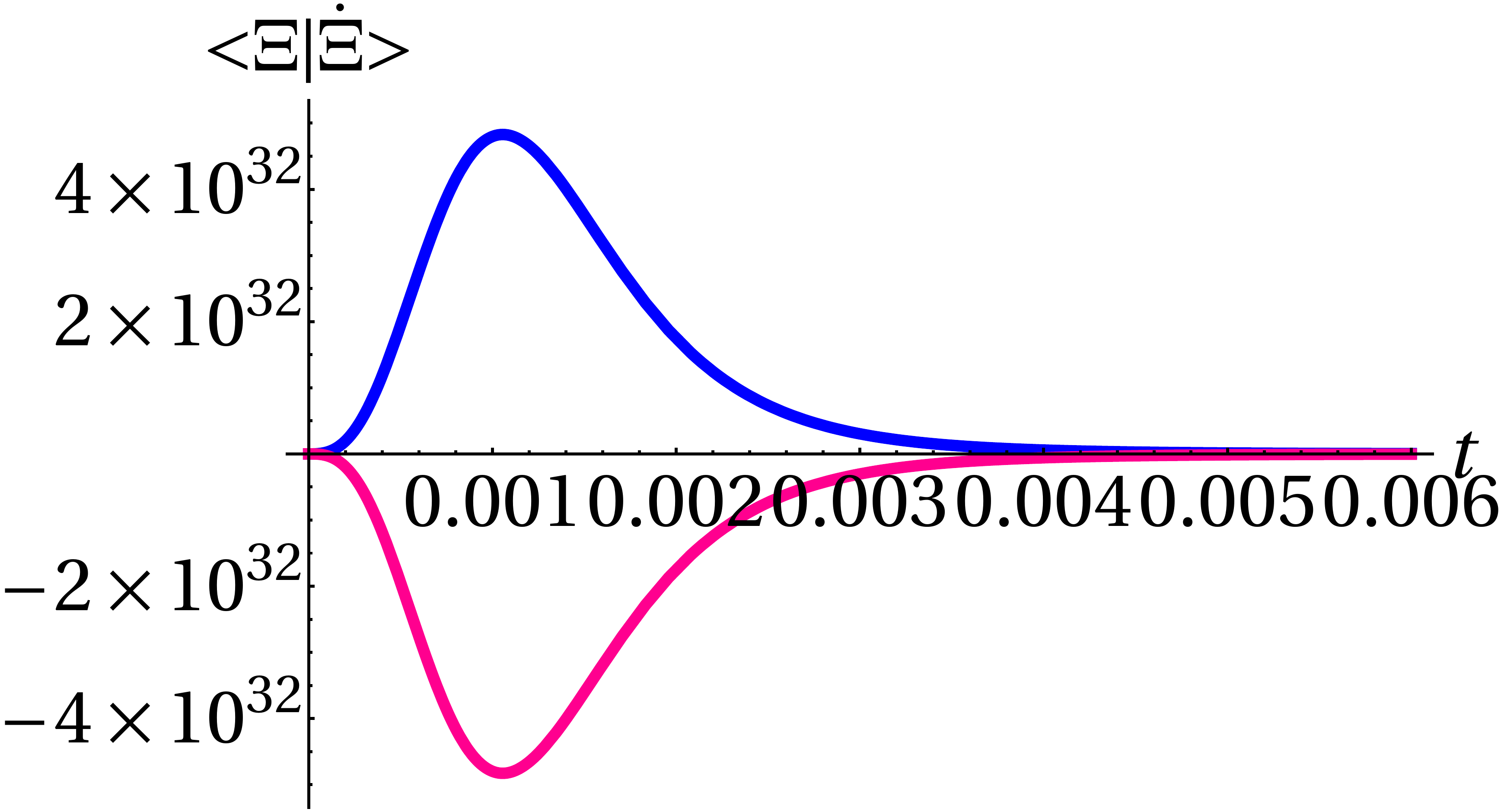}
    \caption{$ \langle \Xi | \dot{\Xi} \rangle$ for $\dot{k}(0)= 1$  (blue), and $\dot{k}(0)  = -1$ (red).}
    \label{121}
  \end{subfigure}
\qquad
  \begin{subfigure}[t]{.5\linewidth}
    \centering
    \includegraphics[width=0.7\columnwidth]{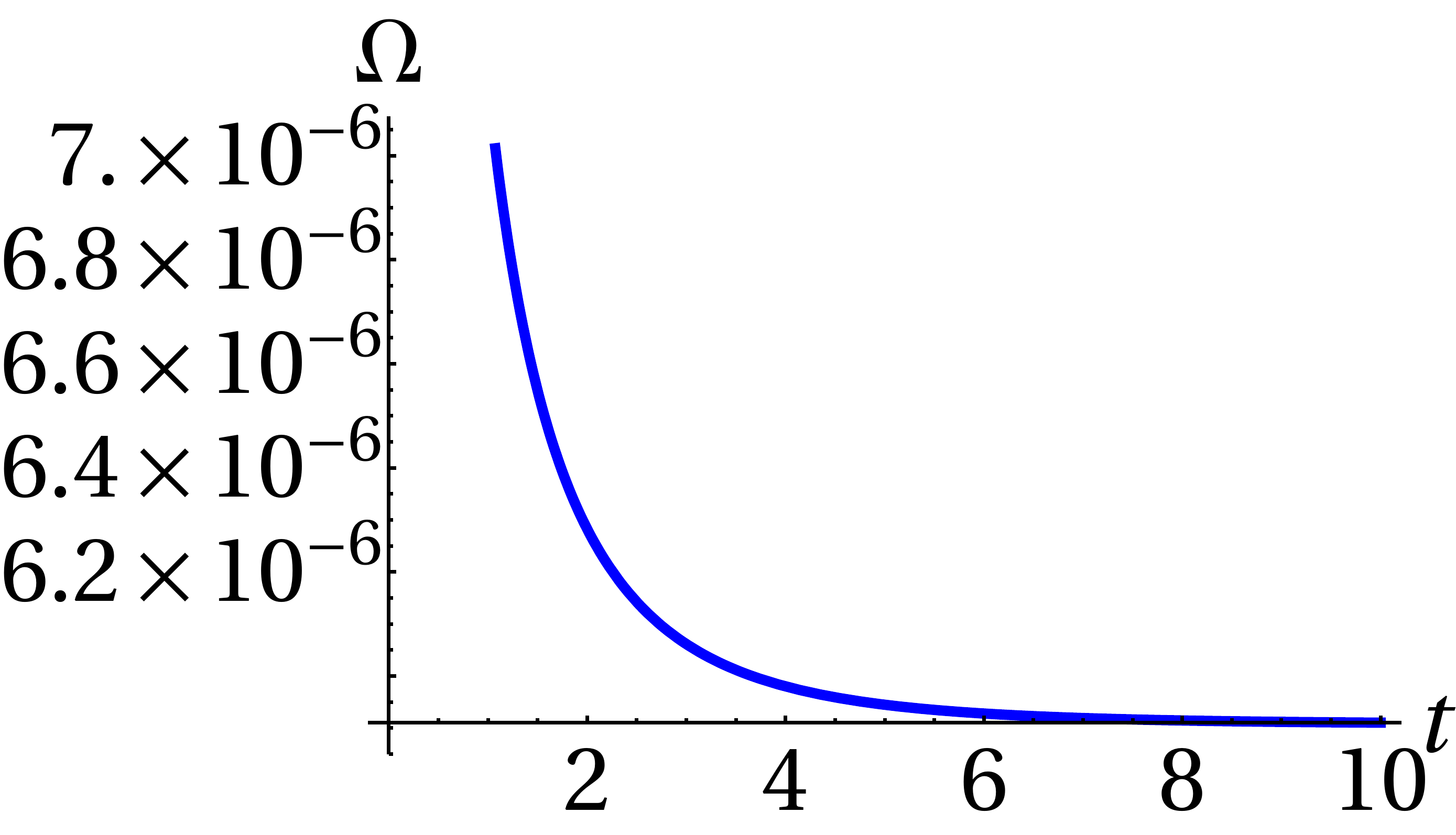}
    \caption{$\Omega$ at $\dot k\left(0\right) = 1$ and  $ \dot\sigma \left(0\right) =0 $.}
    \label{122}
  \end{subfigure}
  \caption{Radiation-filled brane world with initial conditions set number 6 (continued).}
  \label{Fig24}
\end{figure}
\newpage 

The first thing we notice in this solution is that the bulk's scale factor $b$ does not match the brane's scale factor $a$ as it generally did for the dust case, even when they have the same initial conditions. It either stays constant at its initial value, or quickly asymptotes to some specific number. Comparing with the dust case, it is also clear that $a$ expands faster for the radiation filled brane, which is expected. It is also obvious here that the norm ${G_{i\bar j} \dot z^i \dot z^{\bar j}} $ is more \emph{strongly} coupled to $a$ than it is to $b$. The behavior of the remaining fields is very similar to the dust case, albeit with different rates of change.

\setcounter{equation}{0}

\chapter{Possible Universes Coupled to the Hypermultiplets of $\N = 2$ Supergravity}
\label{RDE}

We continue our study started in the previous chapter. To study the current era of the universe, we have to add a cosmological 
constant term ($\Lambda$) that represents the dark energy responsible for the accelerating expansion of the universe. So here we study a brane filled with all radiation, dust and energy that represents our present 4- dimensional universe.
When studying the time evolution of such brane, we have found that the moduli $G_{i\bar{j}} \dot z^{i} \dot z^{\bar{j}}$ 
with $\Lambda$ (the 3-brane cosmological constant) can induce a 3-brane cosmic evolution same as our world, and even in many
cases (initial conditions) when removing $\Lambda$ and let only the bulk cosmological constant ($\tilde{\Lambda}$)
the late time acceleration of the brane- universe is kept.

Means that the whole cosmological history of our own universe,
whether- early time inflationary epoch or late time accelerated expansion-  
may be interpreted only by the effects of a bulk (extra dimension). Moreover studying different initial conditions (ICs) enables to 
find out other possible universes that may be exist with our known universe. Finally our model expects 
possible scenarios of the future behaviour of the brane- world depending on the ICs, $\Lambda$ and $\tilde{\Lambda}$.         

\section{Brane embedding and the fields equations}
Consider adding the brane cosmological constant term $\Lambda(t)$ in addition to the bulk cosmological constant term
$\tilde{\Lambda}(t)$ to the Einstein equtions $G_{\mu\nu} + \Lambda g_{\mu\nu} = T^{Brane}_{\mu\nu}+T^{Bulk}_{yy} $,
and $G_{yy} + \tilde{\Lambda} g_{yy} = T^{Bulk}_{yy} $. The matter desnsity and pressure of 
a brane filled with radition, dust and pure energy are: 
\be 
\rho(t)= \frac{1}{a^3}+ \frac{1}{a^4}, ~~~~~~ p(t)=  \frac{1}{3a^4}.
\label{all}
\ee
Then the Friedmann-like equations become:
\bea
 3\left[ {\left( {\frac{{\dot a}}{a}} \right)^2  + \left( {\frac{{\dot a}}{a}} \right)\left( {\frac{{\dot b}}{b}} \right)} \right] &=& G_{i\bar j} \dot z^i \dot z^{\bar j}  + \frac{{1 }}{{a^3 }} + \frac{1}{{a^4 }} + \Lambda\left(t\right) \nonumber\\
 2\frac{{\ddot a}}{a} + \left( {\frac{{\dot a}}{a}} \right)^2  + \frac{{\ddot b}}{b} + 2\left( {\frac{{\dot a}}{a}} \right)\left( {\frac{{\dot b}}{b}} \right) &=& \Lambda\left(t\right) - \frac{1}{3a^4} - G_{i\bar j} \dot z^i \dot z^{\bar j}  \nonumber\\
 3\left[ {\frac{{\ddot a}}{a} + \left( {\frac{{\dot a}}{a}} \right)^2 } \right] &=&  \tilde\Lambda \left(t\right) - G_{i\bar j} \dot z^i \dot z^{\bar j} ,
\label{EE3}
\eea
Analysis of the fields equations are performed numerically following the initial conditions schemes outlined in table (\ref{tab2})
and (\ref{tabb}) . This is broken down into two major categories: branes with vanishing initial conditions on the scale factors, and those with non-vanishing values but initially decreasing in velocity. 
The studies are performed in two cases: first for constant $\Lambda$ and $\tilde \Lambda$ while 
solving for the brane scale factor $a$, the bulk scale factor $b$ and the moduli norm $G_{i\bar j} \dot z^i \dot z^{\bar j}$, 
second for constant $b$ and $\tilde \Lambda$ while solving for the brane scale factor $a$, $\Lambda$ 
and $G_{i\bar j} \dot z^i \dot z^{\bar j}$.
For all cases for IC set \#1  I demonstrate all the results for $a$, $b$, $G_{i\bar j} \dot z^i \dot z^{\bar j}$ 
and the hypermultiplets fields, while for the last of ICs, we demonstrate the different universes (ICs) only according to
$a$, $b$, the moduli and $k$.
Anyways, the behavior of the scalar fields are similar,  albeit with different rates of change.

\begin{table}[H]
\centering
\caption{The eighteen sets of initial conditions (IC) used in the constant $\Lambda$ computations.}
\label{tab2}
\vspace{0.5cm}
\begin{adjustbox}{max width=\textwidth}
\begin{tabular}{|c|c|c|c|c|c|c|l|}
\hline
\textbf{IC Set \#} & \multicolumn{1}{l|}{$a\left(0\right)$}                  & $b\left(0\right)$                     & $\dot a\left(0\right)$               & $\dot b\left(0\right)$                  & $\Lambda$  & $\tilde \Lambda$ & \textbf{Description} \\ \hline
\textbf{1}  & \multirow{9}{*}{0}                      & \multirow{9}{*}{0}    & \multirow{9}{*}{0} & \multirow{9}{*}{0}    & 0  & 0  & \multirow{9}{*}{Big bang-like IC with vanishing initial velocities}   \\ \cline{1-1} \cline{6-7}
\textbf{2}  &                                         &                       &                    &                      & 1  & 0  &    \\ \cline{1-1} \cline{6-7}
\textbf{3}  &                                         &                       &                    &                       & 0  & 1  &    \\ \cline{1-1} \cline{6-7}
\textbf{4}  &                                         &                       &                    &                       & 1  & 1  &    \\ \cline{1-1} \cline{6-7}
\textbf{5}  &                                         &                       &                    &                       & -1 & 0  &    \\ \cline{1-1} \cline{6-7}
\textbf{6}  &                                         &                       &                    &                       & 0  & -1 &    \\ \cline{1-1} \cline{6-7}
\textbf{7}  &                                         &                       &                    &                       & -1 & -1 &    \\ \cline{1-1} \cline{6-7}
\textbf{8}  &                                         &                       &                    &                       & 1  & -1 &    \\ \cline{1-1} \cline{6-7}
\textbf{9}  &                                         &                       &                    &                       & -1 & 1  &    \\ \hline\hline
\textbf{10} & \multirow{9}{*}{1}                      & \multirow{9}{*}{-0.2} & \multirow{9}{*}{1} & \multirow{9}{*}{-0.2} & 0  & 0  & 
\multirow{9}{*}{Non-singular IC with initial negative velocities}   \\ \cline{1-1} \cline{6-7}
\textbf{11} &                                         &                       &                    &                       & 1  & 0  &    \\ \cline{1-1} \cline{6-7}
\textbf{12} &                                         &                       &                    &                       & 0  & 1  &    \\ \cline{1-1} \cline{6-7}
\textbf{13} &                                         &                       &                    &                       & 1  & 1  &    \\ \cline{1-1} \cline{6-7}
\textbf{14} &                                         &                       &                    &                       & -1 & 0  &    \\ \cline{1-1} \cline{6-7}
\textbf{15} &                                         &                       &                    &                       & 0 & -1  &    \\ \cline{1-1} \cline{6-7}
\textbf{16} &                                         &                       &                    &                       & -1 & -1  &    \\ \cline{1-1} \cline{6-7}
\textbf{17} &                                         &                       &                    &                       & 1 & -1  &    \\ \cline{1-1} \cline{6-7}
\textbf{18} &                                         &                       &                    &                       & -1 & 1  &    \\ \hline
\end{tabular}
\end{adjustbox}
\end{table}
\subsection{Constant $\Lambda$}
The cosmological constants $\Lambda$ and $\tilde \Lambda$ are both taken to be either vanishing, positive, or negative. The calculation is performed over all possible permutations of these values.


\begin{figure}[H]
  \begin{subfigure}[t]{.5\linewidth}
    \centering
    \includegraphics[width=0.7\columnwidth]{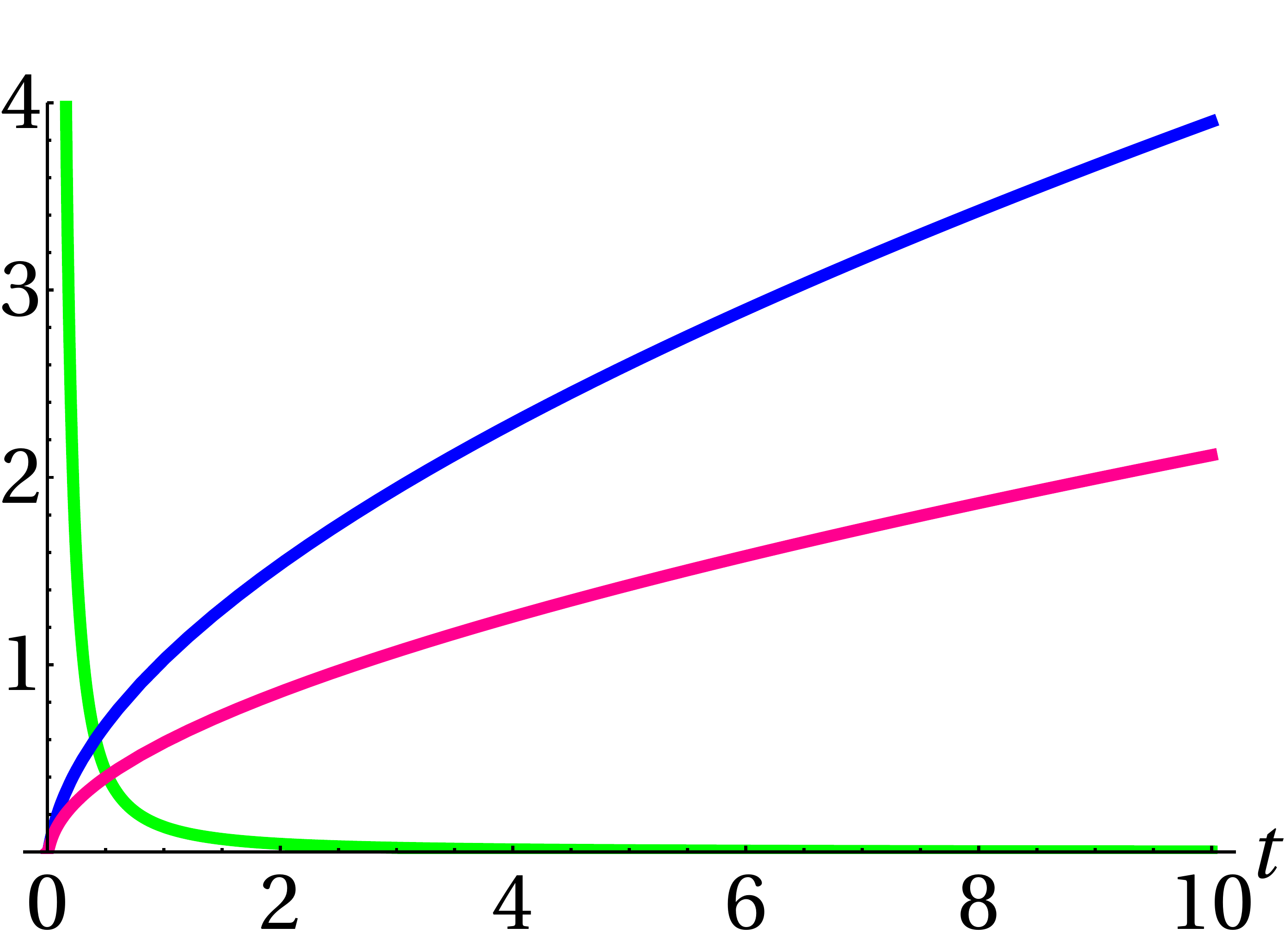}
    \caption{The scale factor $a$ is represented by the blue curve, $b$ by the red curve, while $\left| {G_{i\bar j} \dot z^i \dot z^{\bar j}} \right|$ by the green curve. The curve for $b$ is scaled up by a factor of 10.}
    \label{abzz00001constantL}
  \end{subfigure}
\qquad
  \begin{subfigure}[t]{.5\linewidth}
    \centering
    \includegraphics[width=0.7\columnwidth]{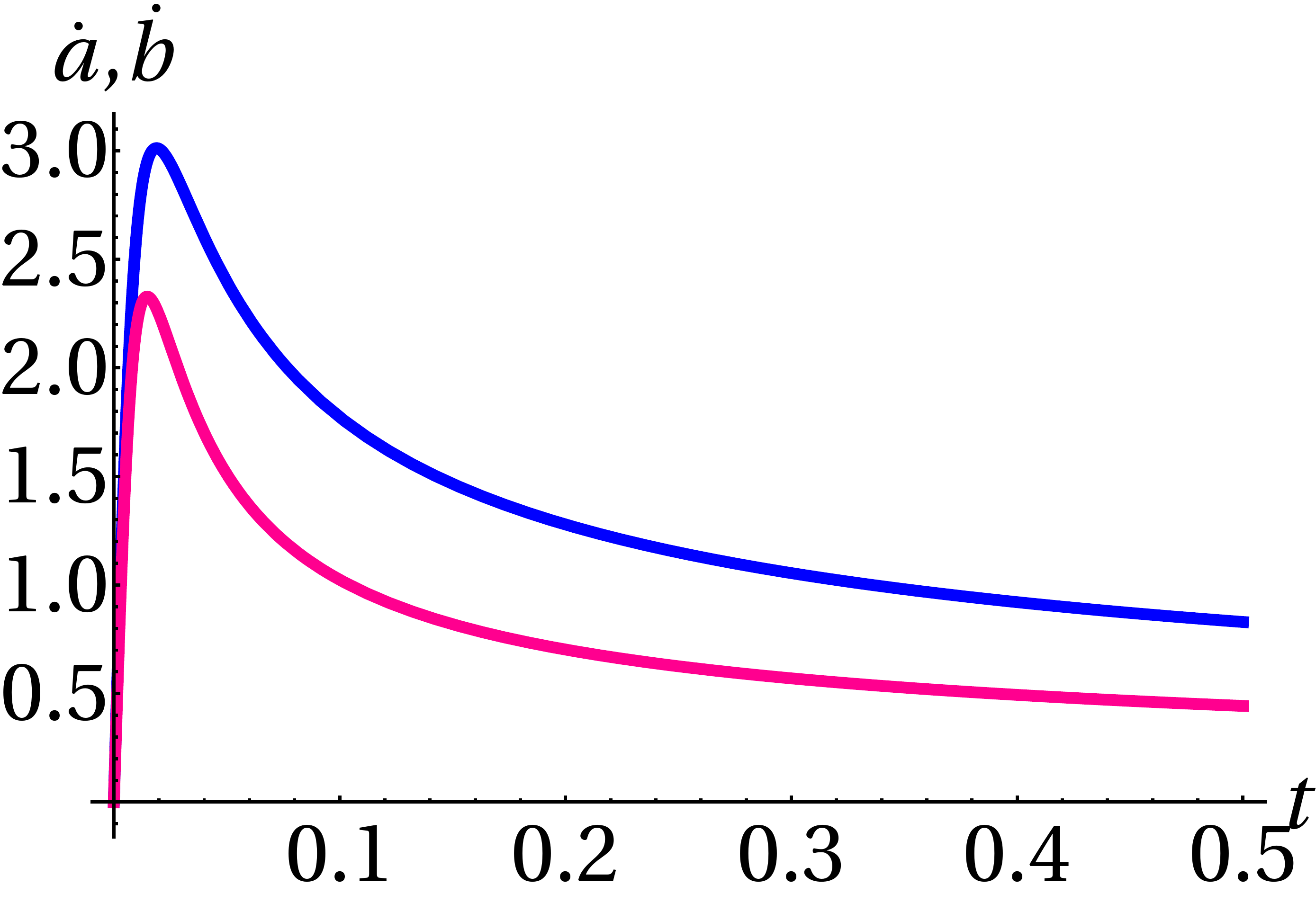}
    \caption{The expansion rates of the scale factors: $\dot a$ is represented by the blue curve, and $\dot b$ by the red curve. The curve for $\dot b$ is scaled up by a factor of 10.}
    \label{adotbdot00001constantL}
  \end{subfigure}
\\[9em]
  \begin{subfigure}[t]{.5\linewidth}
    \centering
    \includegraphics[width=0.7\columnwidth]{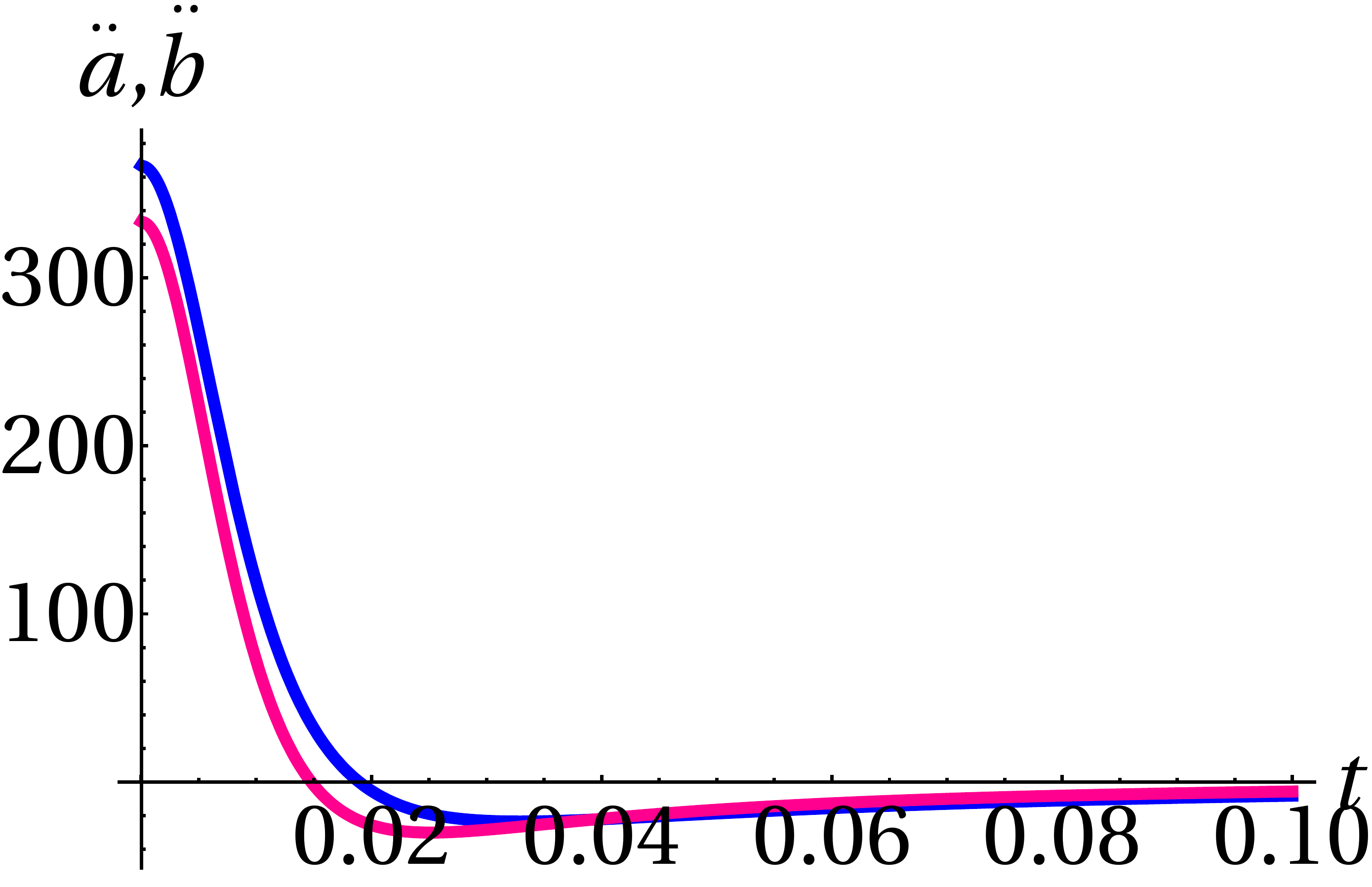}
    \caption{The accelerations of the scale factors: $\ddot a$ is represented by the blue curve, and $\ddot b$ by the red curve. The curve for $\ddot b$ is scaled up by a factor of 10.}
    \label{addotbddot00001constantL}
  \end{subfigure}
\qquad
  \begin{subfigure}[t]{.5\linewidth}
    \centering
    \includegraphics[width=0.7\columnwidth]{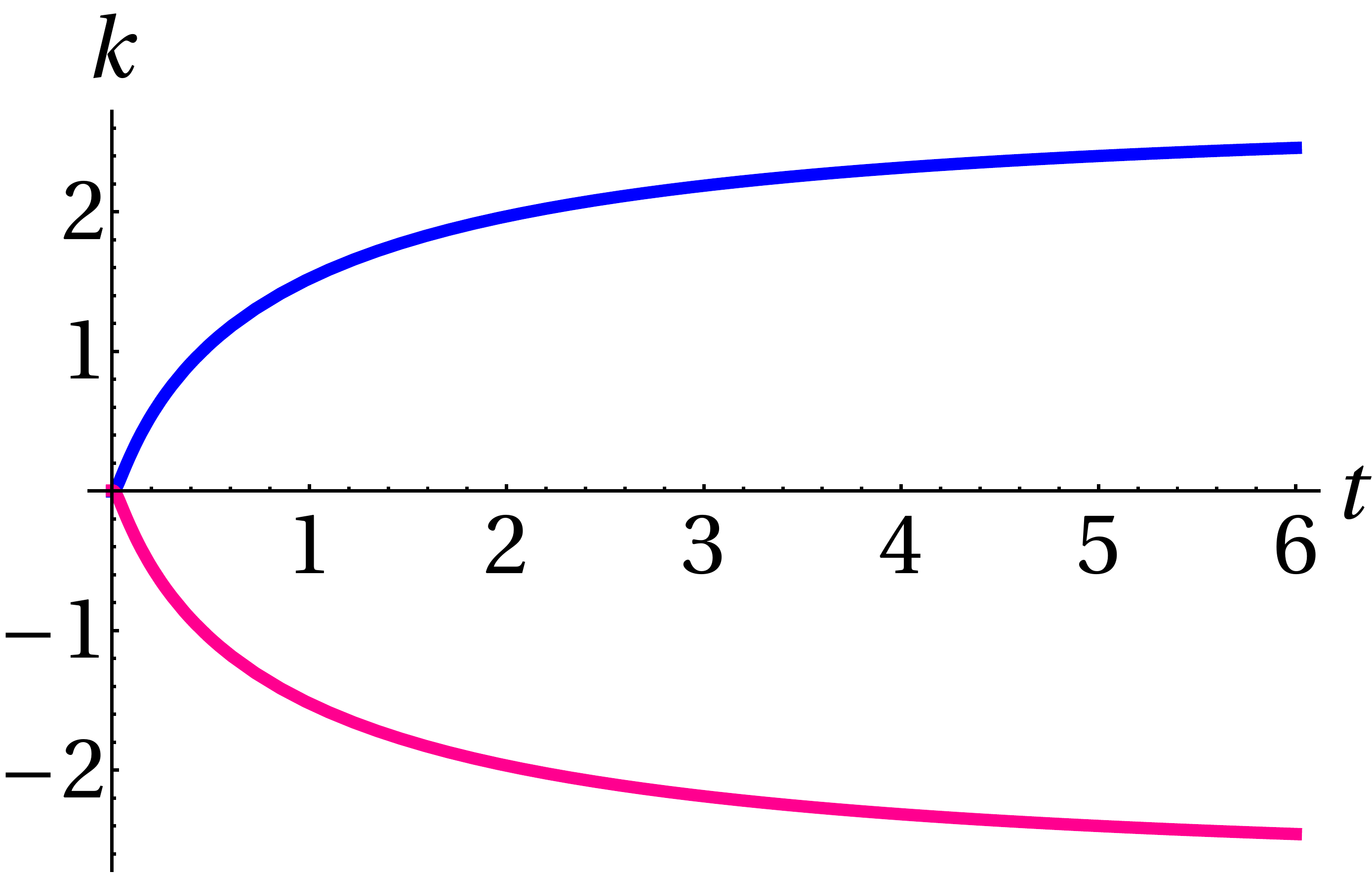}
    \caption{The harmonic function $k$ using: $\dot k\left(0\right)=1$ (blue curve), and $\dot k\left(0\right)=-1$ (red curve). While $\dot k\left(0\right)=0$ diverges.}
    \label{k00001constantL}
  \end{subfigure}
    \caption{Initial conditions set number 1 for constant $\Lambda$.}
  \label{Fig1}
  \end{figure}
\begin{figure}[H]
\begin{subfigure}[t]{.5\linewidth}
    \centering
    \includegraphics[width=0.7\columnwidth]{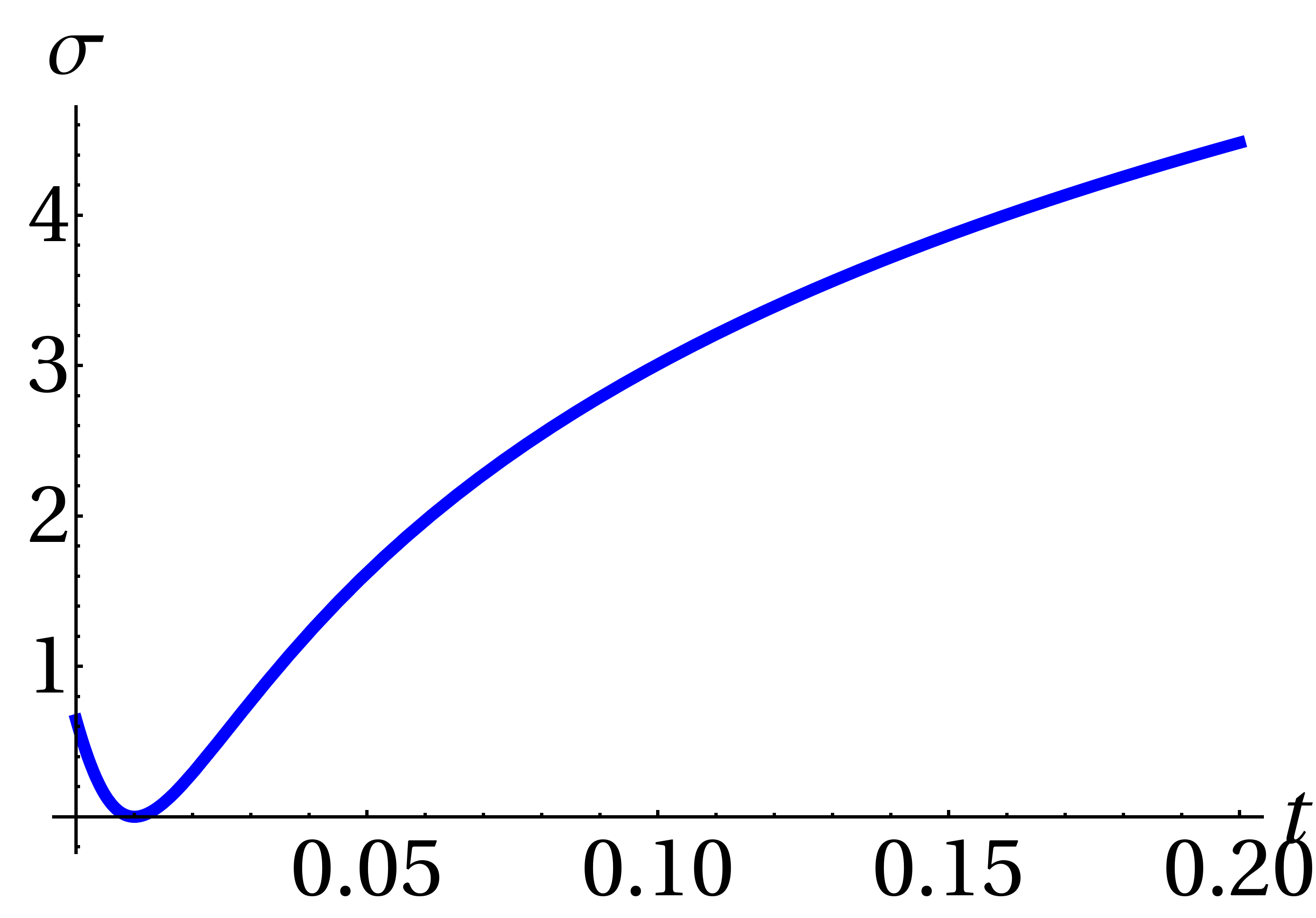}
    \caption{The dilaton $\sigma$; same for all three $\dot k\left(0\right)$.}
    \label{sigma00001constantL}
  \end{subfigure}
  \qquad
  \begin{subfigure}[t]{.5\linewidth}
    \centering
    \includegraphics[width=0.7\columnwidth]{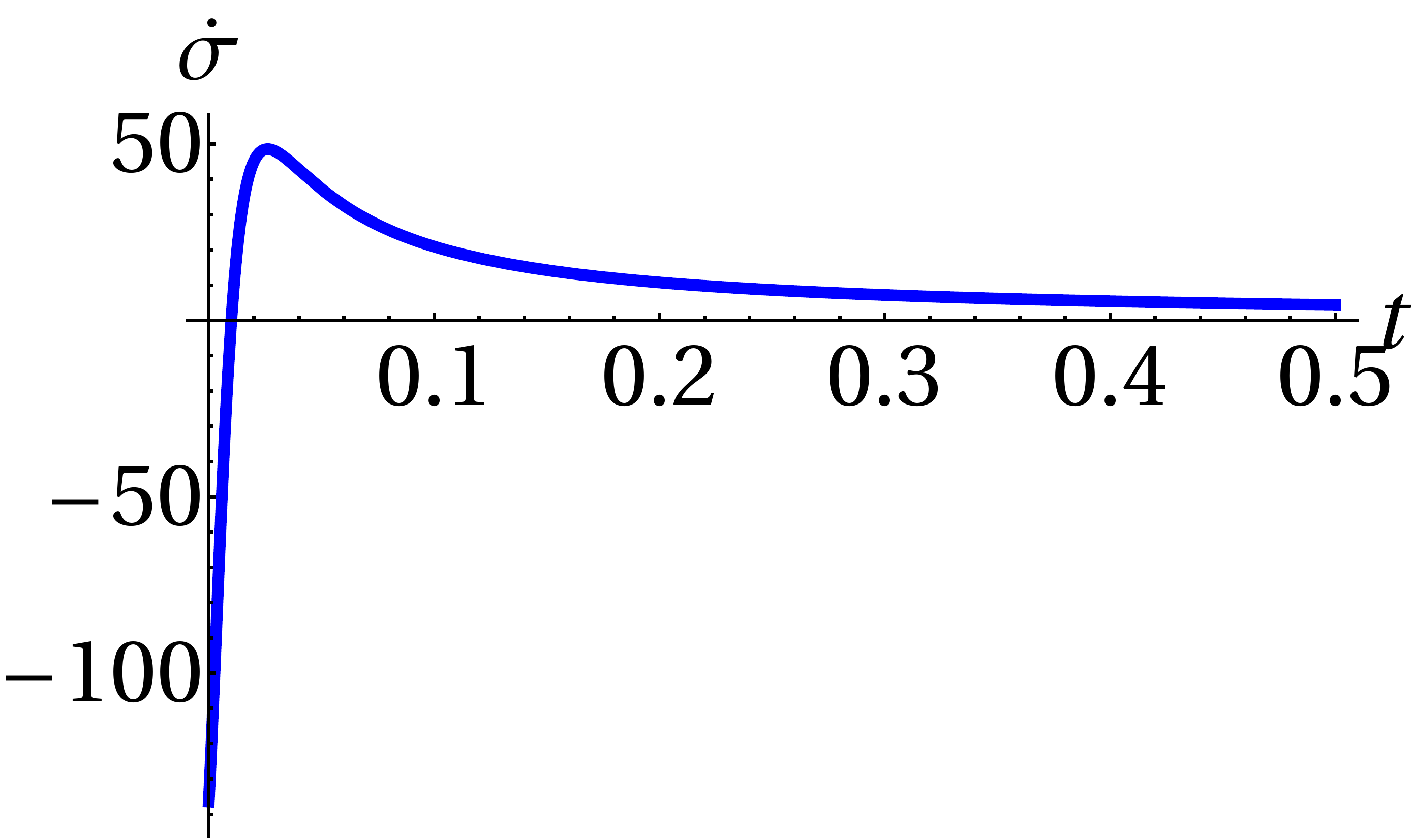}
    \caption{The dilatonic field strength $\dot\sigma$.}
    \label{sigmadot00001constantL}
  \end{subfigure}
\\[4em]  
\begin{subfigure}[t]{.5\linewidth}
    \centering
    \includegraphics[width=0.7\columnwidth]{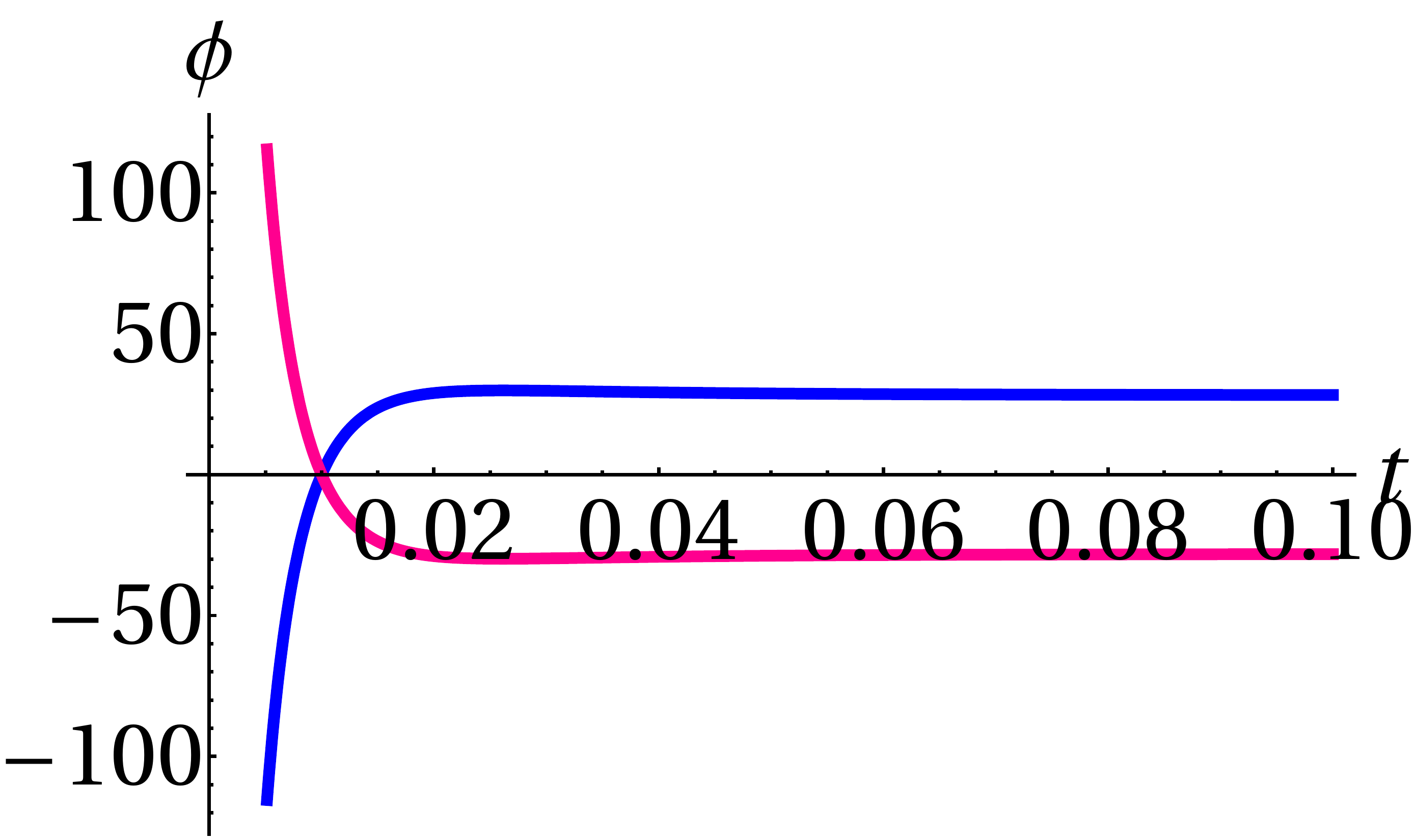}
    \caption{The universal axion $\phi$ for $\dot k\left(0\right) = 1$ (blue curve), and $\dot k\left(0\right) = -1$ (red curve). The solution diverges for $\dot k\left(0\right)=0$.}
    \label{phi00001constantL}
  \end{subfigure}
\qquad
  \begin{subfigure}[t]{.5\linewidth}
    \centering
    \includegraphics[width=0.7\columnwidth]{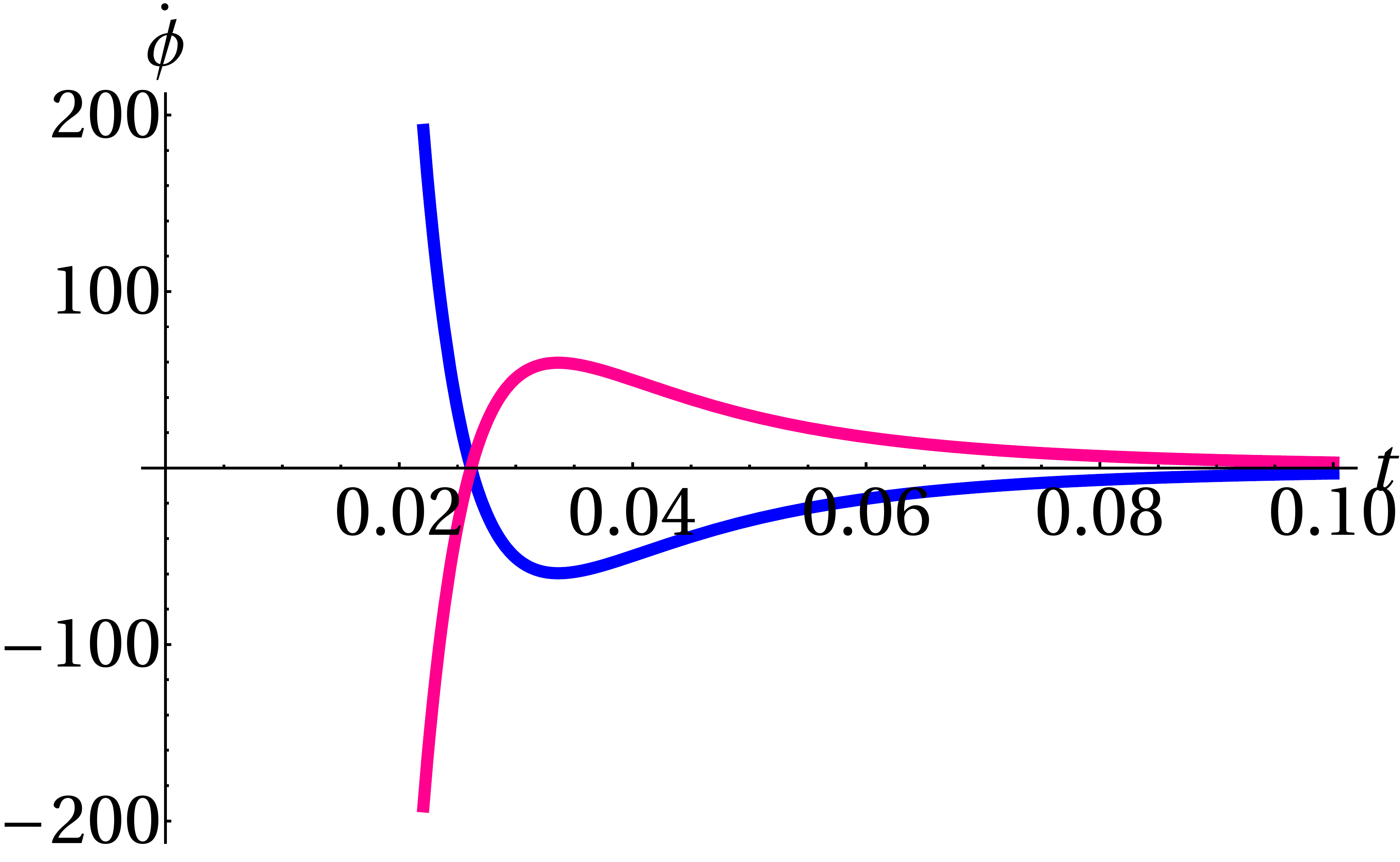}
    \caption{The axionic field strength $\dot\phi$ for $\dot k\left(0\right) = 1$ (blue curve), and $\dot k\left(0\right) = -1$ (red curve).}
    \label{phidot00001constantL}
  \end{subfigure}
\\[4em]    
\begin{subfigure}[t]{.5\linewidth}
    \centering
    \includegraphics[width=0.7\columnwidth]{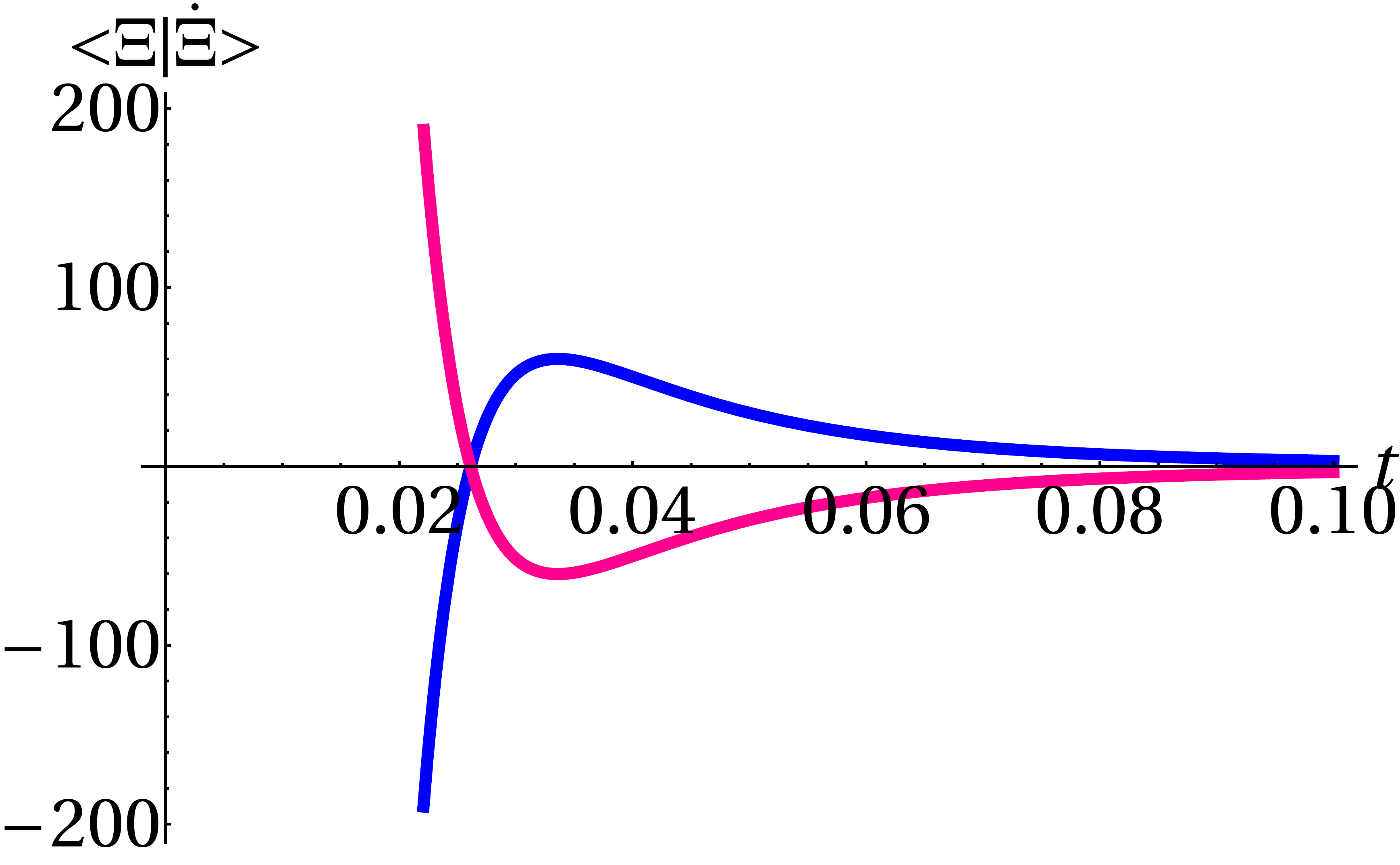}
    \caption{$ \langle \Xi | \dot{\Xi} \rangle$ for $\dot{k}(0)= 1$  (blue), and $\dot{k}(0)  = -1$ (red).}
    \label{xx00001constantL}
  \end{subfigure}
\qquad
 \begin{subfigure}[t]{.5\linewidth}
    \centering
    \includegraphics[width=0.7\columnwidth]{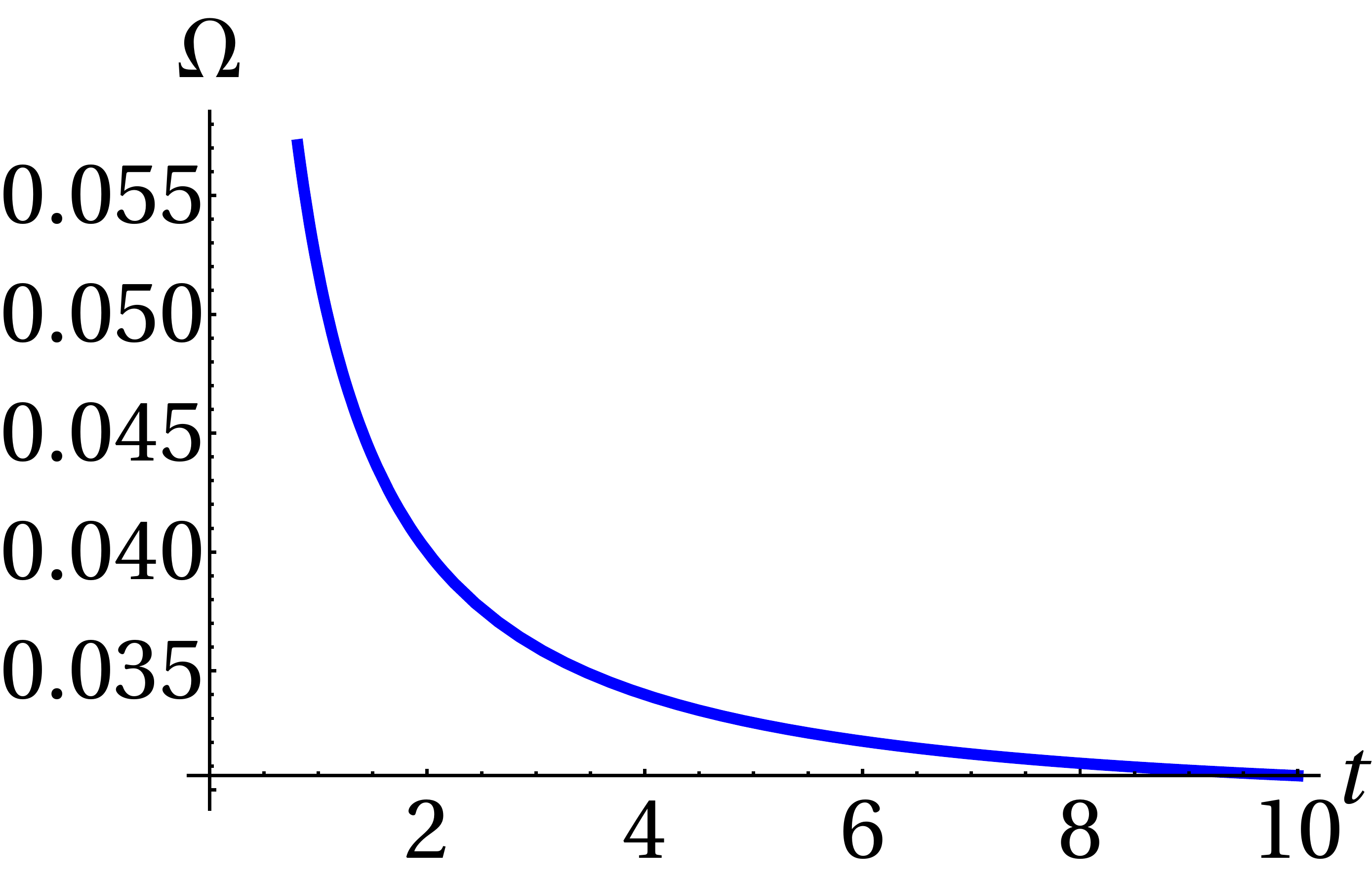}
    \caption{$\Omega$ at $\dot k\left(0\right) = 1$ and  $ \dot\sigma \left(0\right) =0 $.}
    \label{omega00001constantL}
  \end{subfigure}
\vspace{0.3cm}
  \caption{Initial conditions set number 1 for constant $\Lambda$ (continued) .}
  \label{Fig2}
\end{figure}


\begin{figure}[H]
  \begin{subfigure}[t]{.5\linewidth}
    \centering
    \includegraphics[width=0.7\columnwidth]{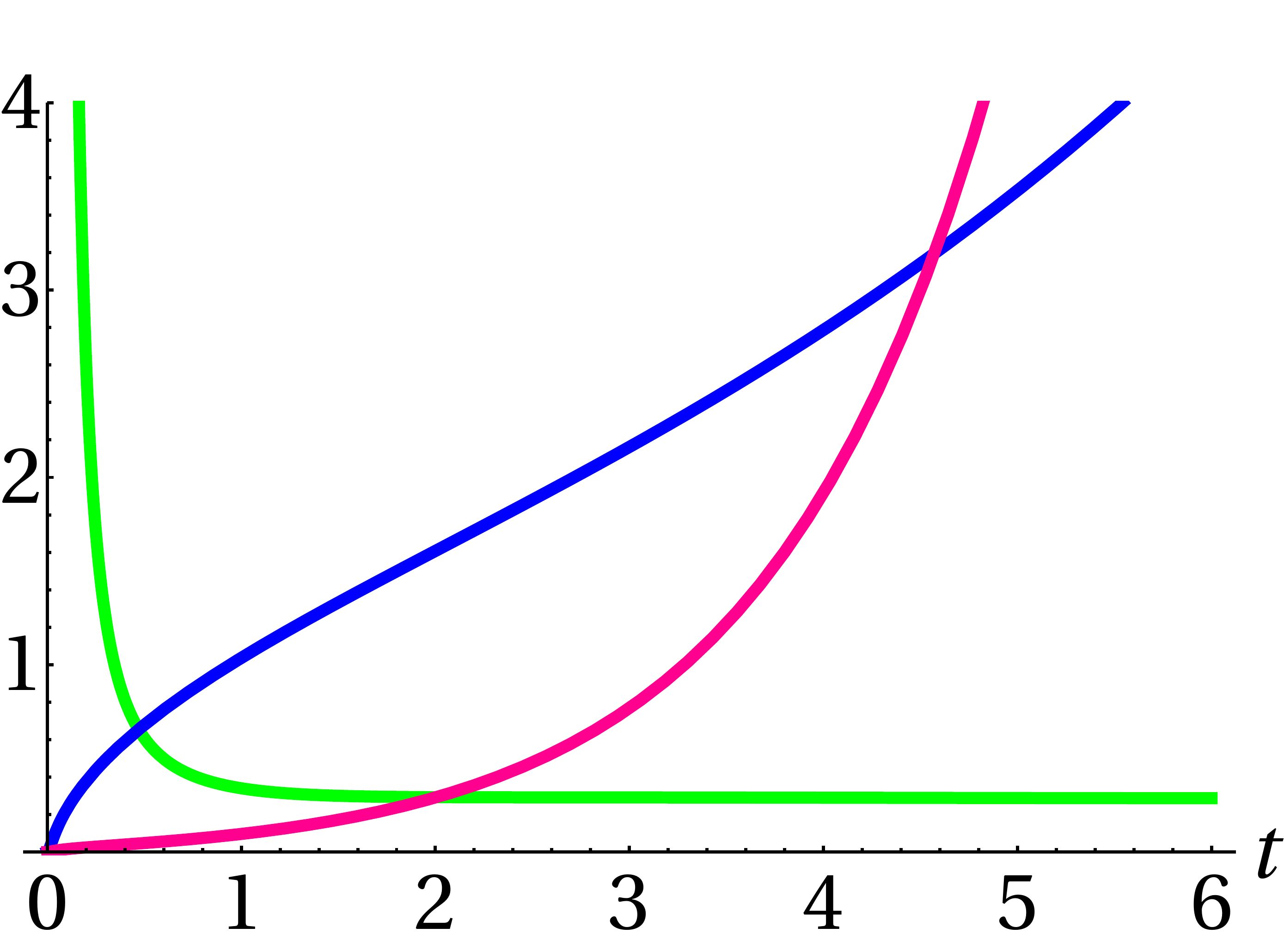}
    \caption{The scale factor $a$ is represented by the blue curve, $b$ by the red curve, while $\left| {G_{i\bar j} \dot z^i \dot z^{\bar j}} \right|$ by the green curve.}
    \label{abzz00002constantL}
  \end{subfigure}
\qquad
  \begin{subfigure}[t]{.5\linewidth}
    \centering
    \includegraphics[width=0.7\columnwidth]{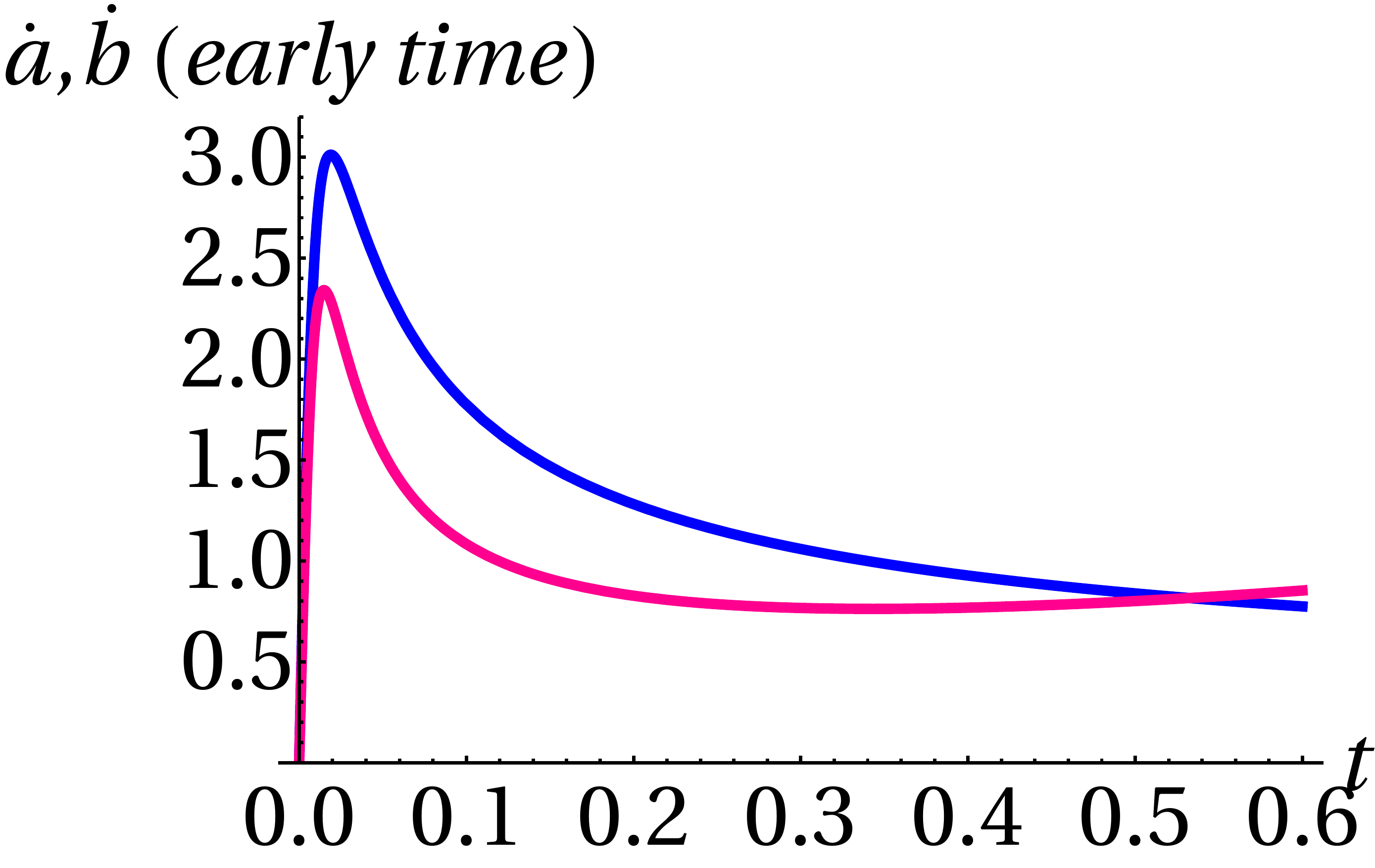}
    \caption{The early time expansion rates of the scale factors: $\dot a$ is represented by the blue curve, and $\dot b$ by the red curve. The curve for $\dot b$ is scaled up by a factor of 10.}
    \label{adotbdotEARLY00002constantL}
  \end{subfigure}
\\[9em]
  \begin{subfigure}[t]{.5\linewidth}
    \centering
    \includegraphics[width=0.7\columnwidth]{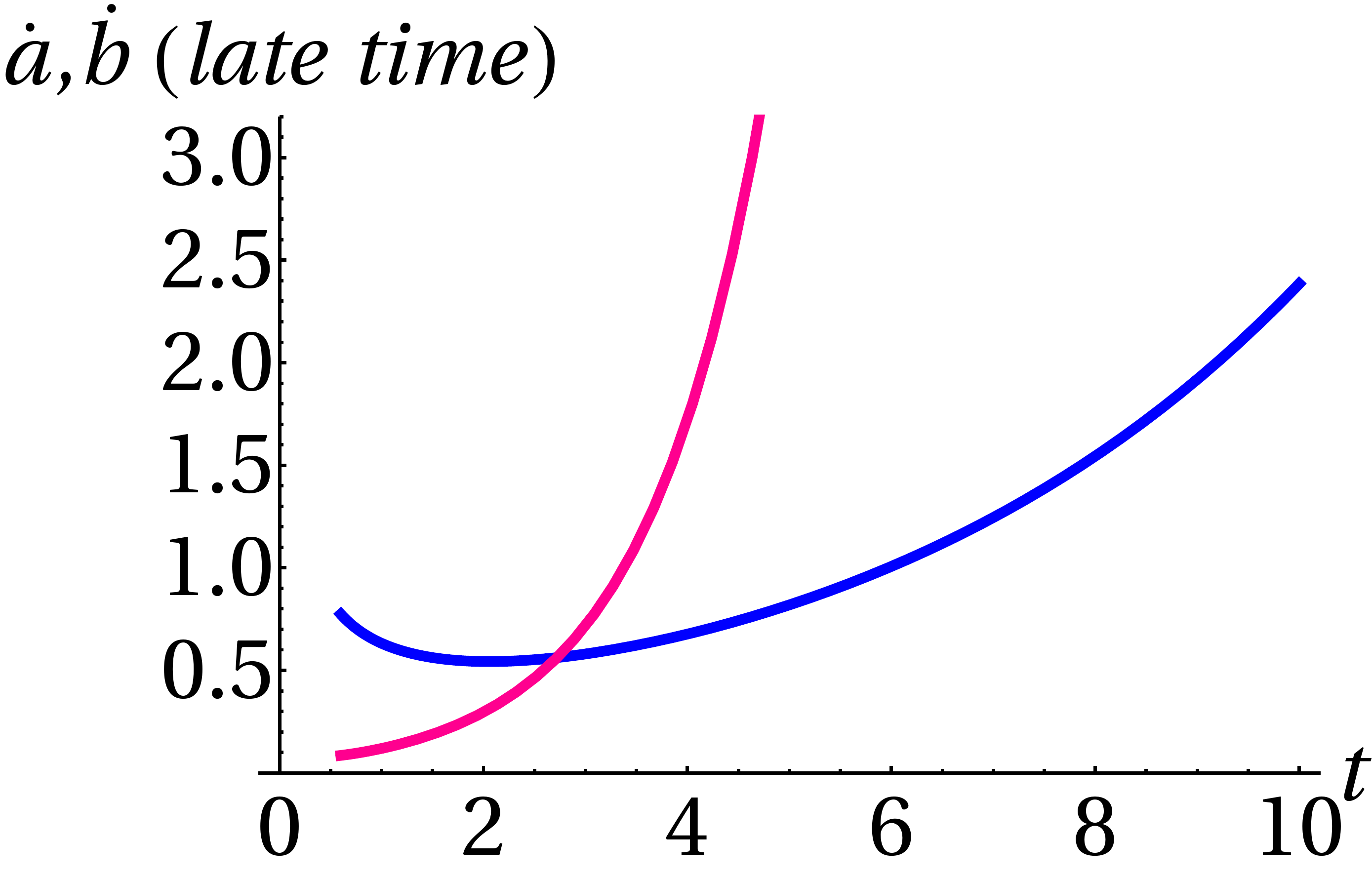}
    \caption{The late time expansion rates of the scale factors: $\dot a$ is represented by the blue curve, and $\dot b$ by the red curve.}
    \label{adotbdotLATE00002constantL}
  \end{subfigure}
\qquad
  \begin{subfigure}[t]{.5\linewidth}
    \centering
    \includegraphics[width=0.7\columnwidth]{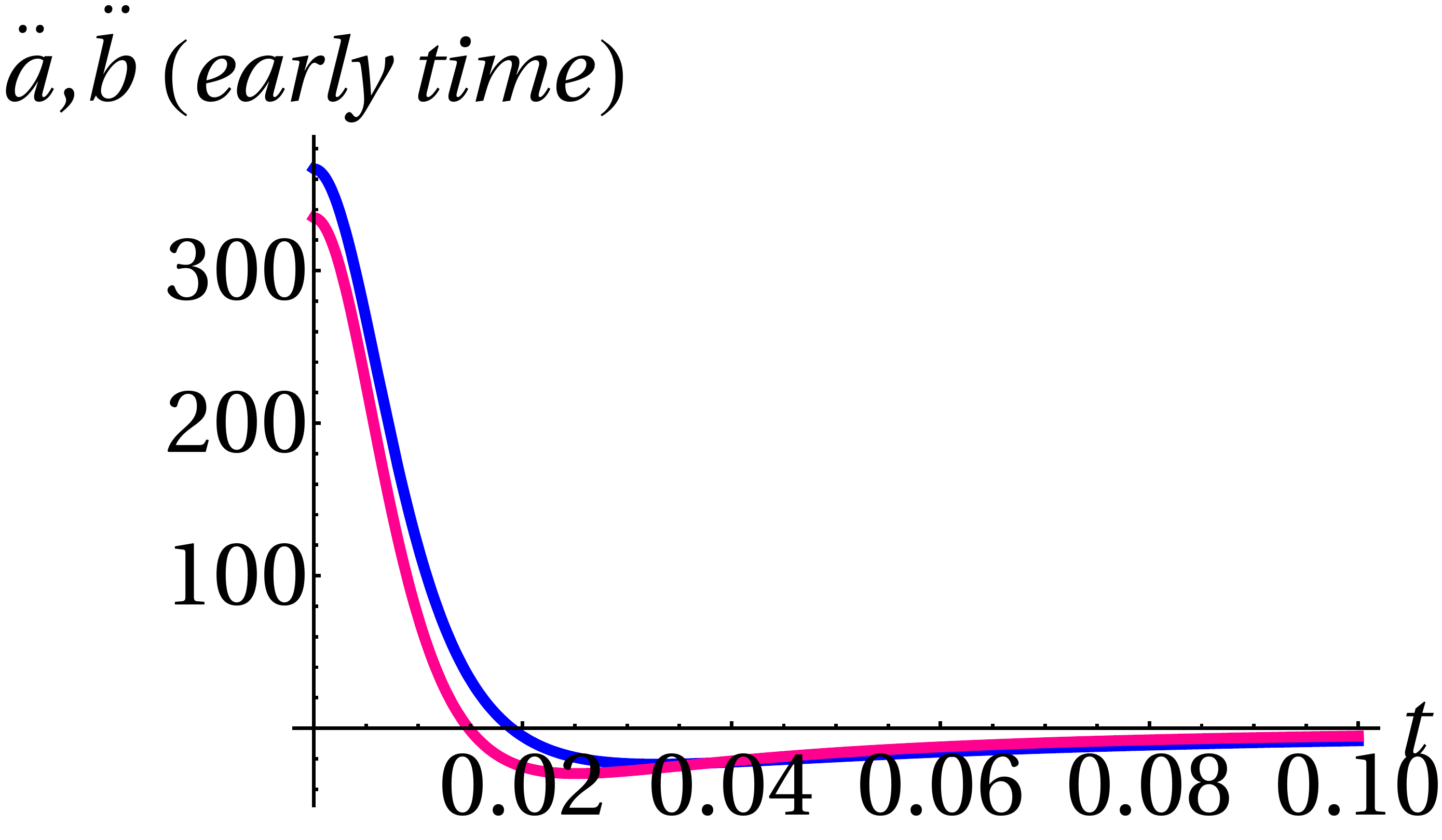}
    \caption{The early time accelerations of the scale factors: $\ddot a$ is represented by the blue curve, and $\ddot b$ by the red curve. The curve for $\ddot b$ is scaled up by a factor of 10.}
    \label{addotbddotEARLY00002constantL}
  \end{subfigure}
    \caption{Initial conditions set number 2 for constant $\Lambda$.}
  \label{Fig3}
\end{figure}
\begin{figure}[H]
\begin{subfigure}[t]{.5\linewidth}
    \centering
    \includegraphics[width=0.7\columnwidth]{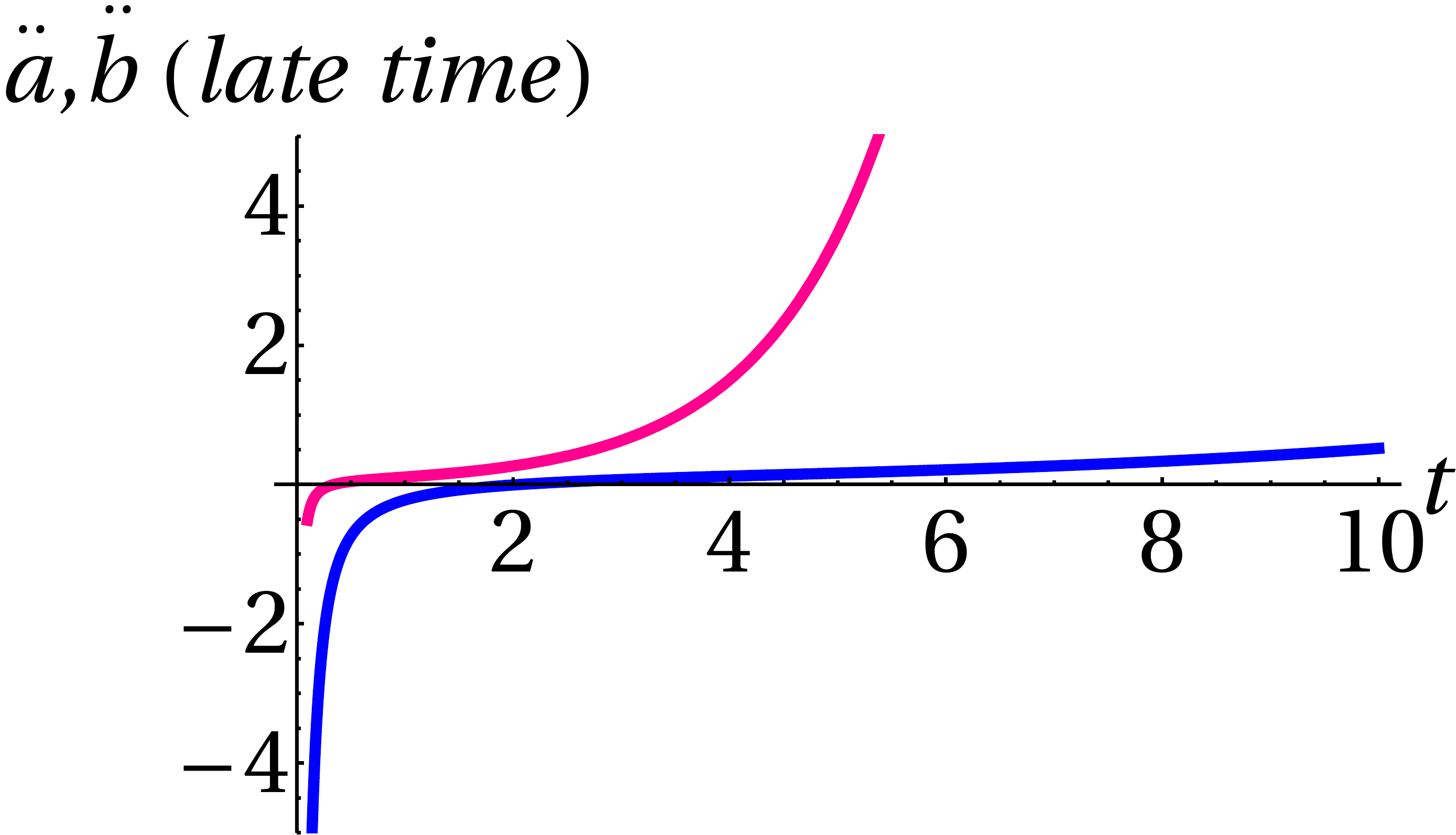}
    \caption{The late time accelerations of the scale factors: $\ddot a$ is represented by the blue curve, and $\ddot b$ by the red curve.}
    \label{addotbddotLATE00002constantL}
  \end{subfigure}
\qquad
  \begin{subfigure}[t]{.5\linewidth}
    \centering
    \includegraphics[width=0.7\columnwidth]{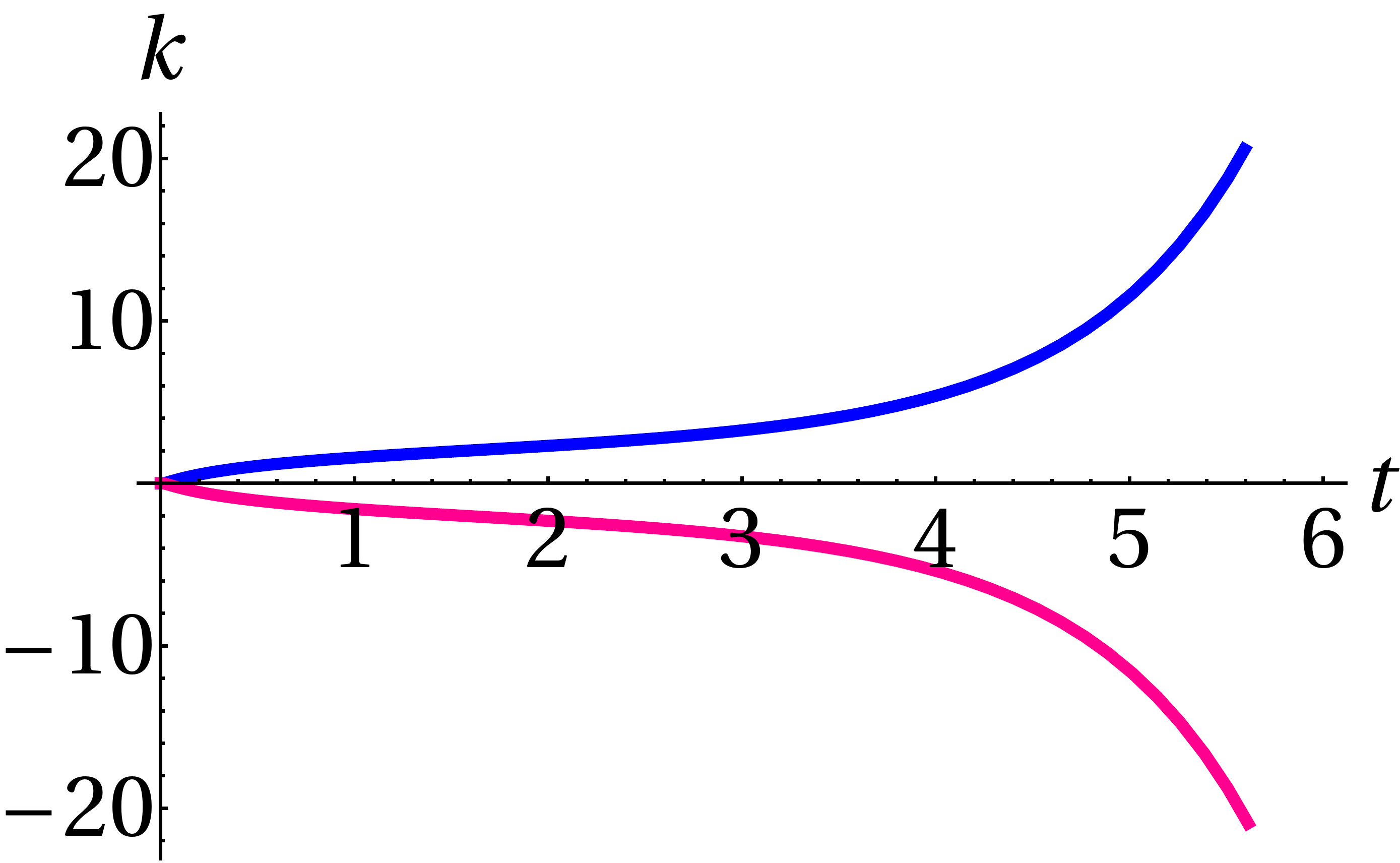}
    \caption{The harmonic function $k$ using: $\dot k\left(0\right)=1$ (blue curve), and $\dot k\left(0\right)=-1$ (red curve). While $\dot k\left(0\right)=0$ diverges.}
    \label{k00002constantL}
  \end{subfigure}
  \caption{Initial conditions set number 2 for constant $\Lambda$ (continued).}
  \label{Fig33}
\end{figure}

\vspace{3cm}


\begin{figure}[H]
\begin{subfigure}[t]{.5\linewidth}
    \centering
    \includegraphics[width=0.7\columnwidth]{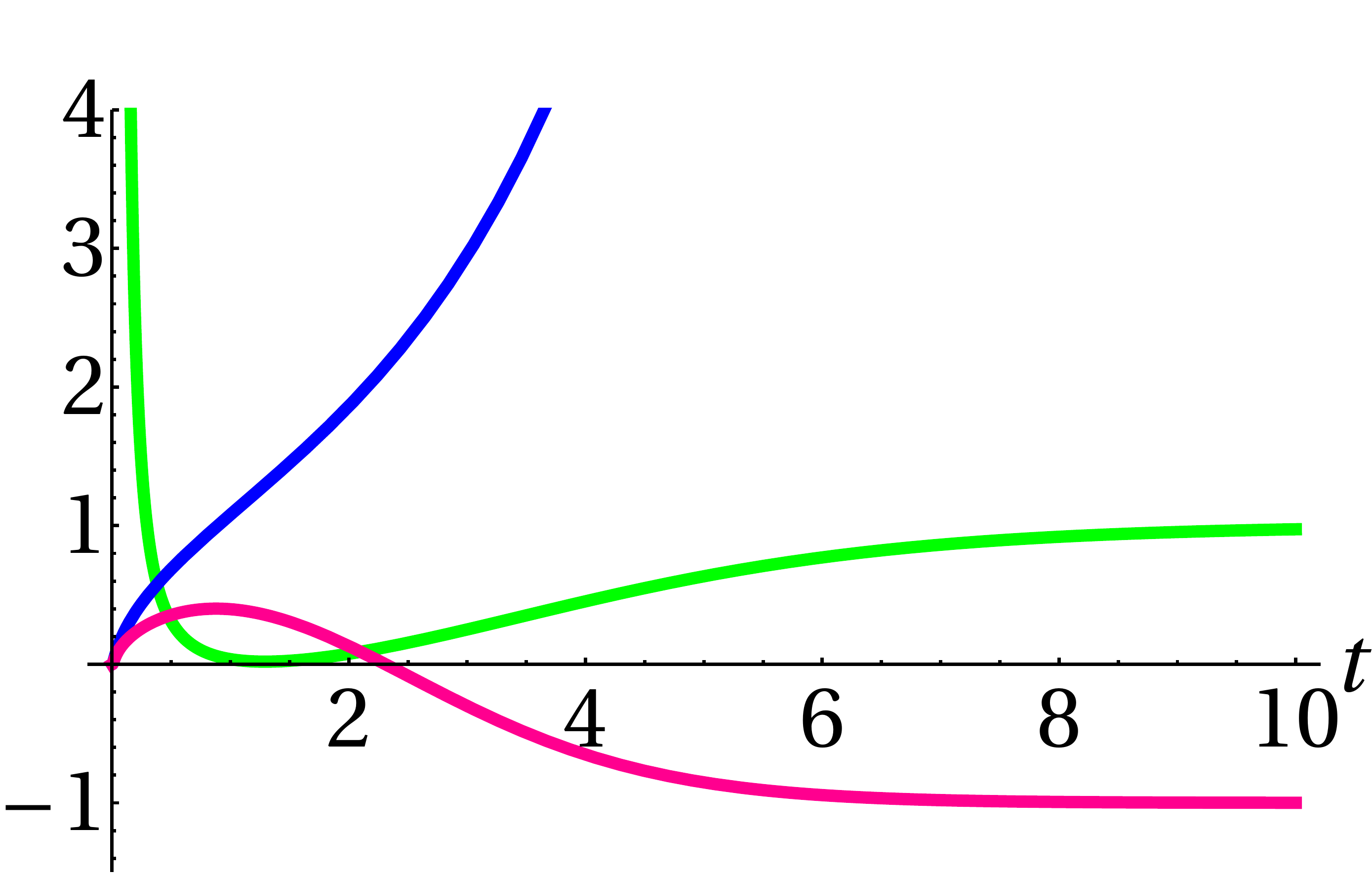}
    \caption{The scale factor $a$ is represented by the blue curve, $b$ by the red  curve, while $\left| {G_{i\bar j} \dot z^i \dot z^{\bar j}} \right|$ by the green  curve. The curve for $b$ is scaled up by a factor of 10.}
    \label{abzz00003constantL}
  \end{subfigure}
\qquad
  \begin{subfigure}[t]{.5\linewidth}
    \centering
    \includegraphics[width=0.7\columnwidth]{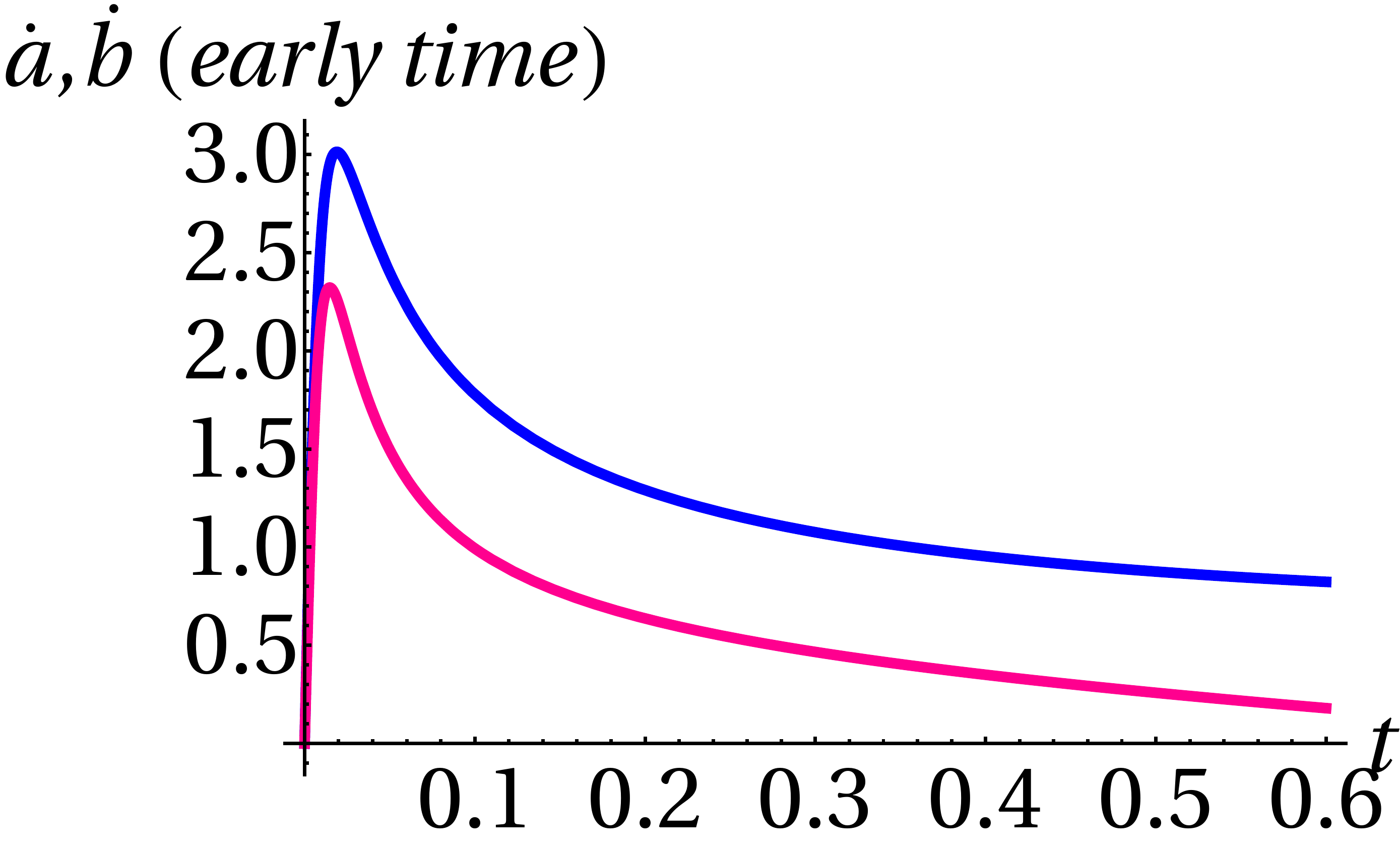}
    \caption{The early time expansion rates of the scale factors: $\dot a$ is represented by the blue curve, and $\dot b$ by the red curve. The curve for $\dot b$ is scaled up by a factor of 10.}
    \label{adotbdotEARLY00003constantL}
  \end{subfigure}
 \caption{Initial conditions set number 3 for constant $\Lambda$.}
  \label{Fig5dash}
  \end{figure}
\begin{figure}[H]
\begin{subfigure}[t]{.5\linewidth}
    \centering
    \includegraphics[width=0.7\columnwidth]{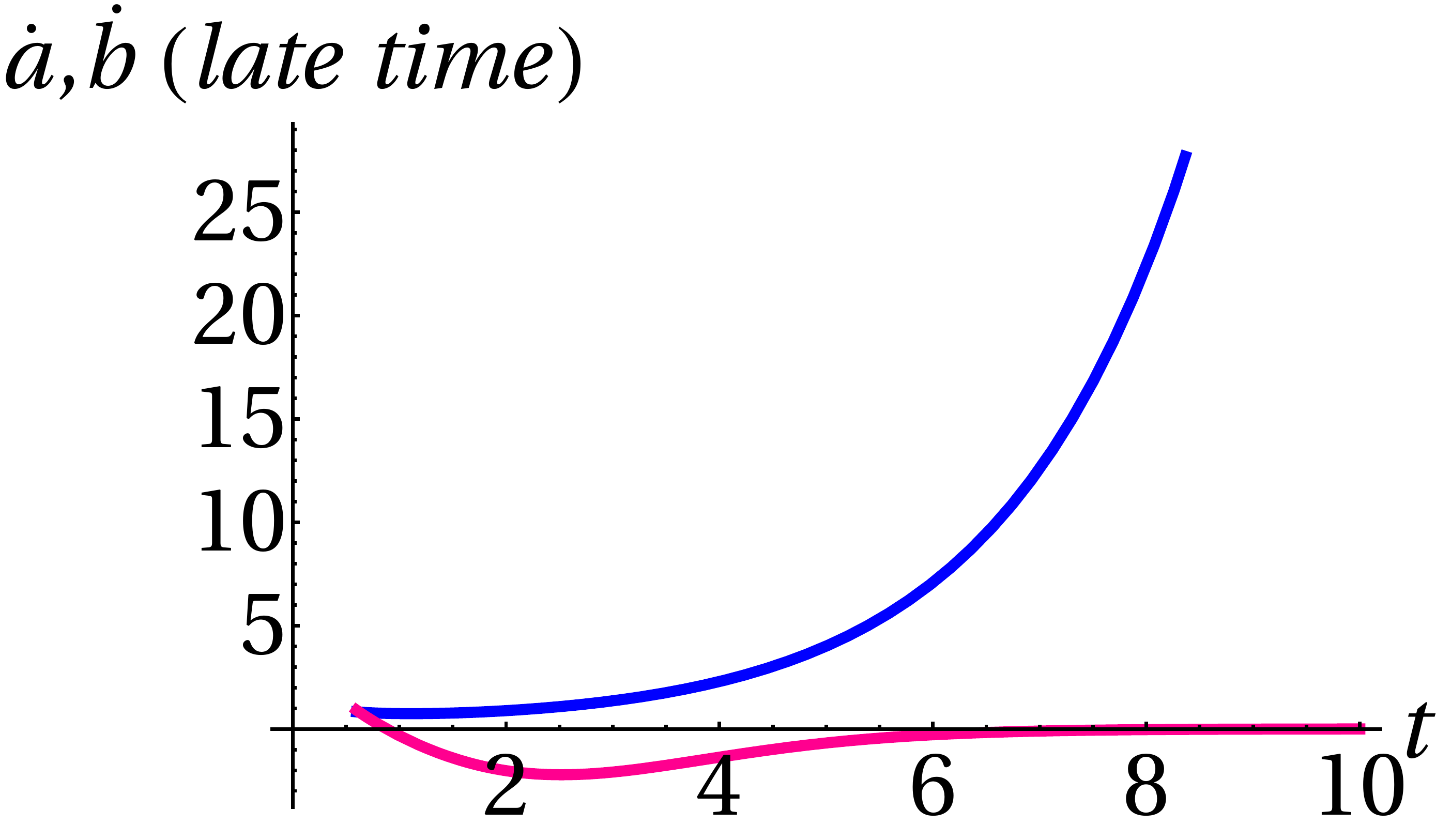}
    \caption{The late time expansion rates of the scale factors: $\dot a$ is represented by the blue curve, and $\dot b$ by the red curve. The curve for $\dot b$ is scaled up by a factor of 50.}
    \label{adotbdotLATE00003constantL}
  \end{subfigure}
\qquad
  \begin{subfigure}[t]{.5\linewidth}
    \centering
    \includegraphics[width=0.7\columnwidth]{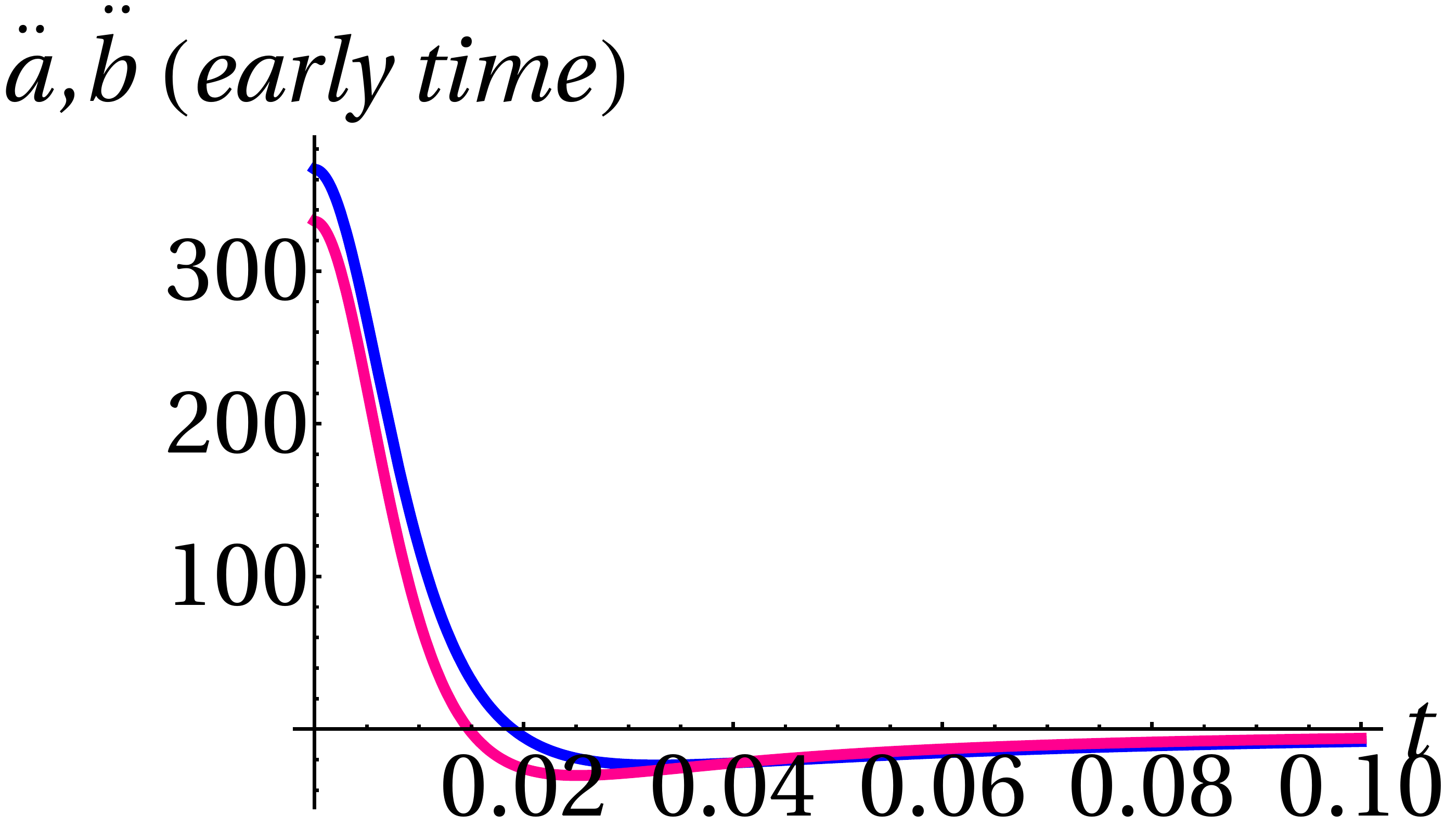}
    \caption{The early time accelerations of the scale factors: $\ddot a$ is represented by the blue curve, and $\ddot b$ by the red curve. The curve for $\dot b$ is scaled up by a factor of 10.}
    \label{addotbddotEARLY00003constantL}
  \end{subfigure}
\\[9em]
  \begin{subfigure}[t]{.5\linewidth}
    \centering
    \includegraphics[width=0.7\columnwidth]{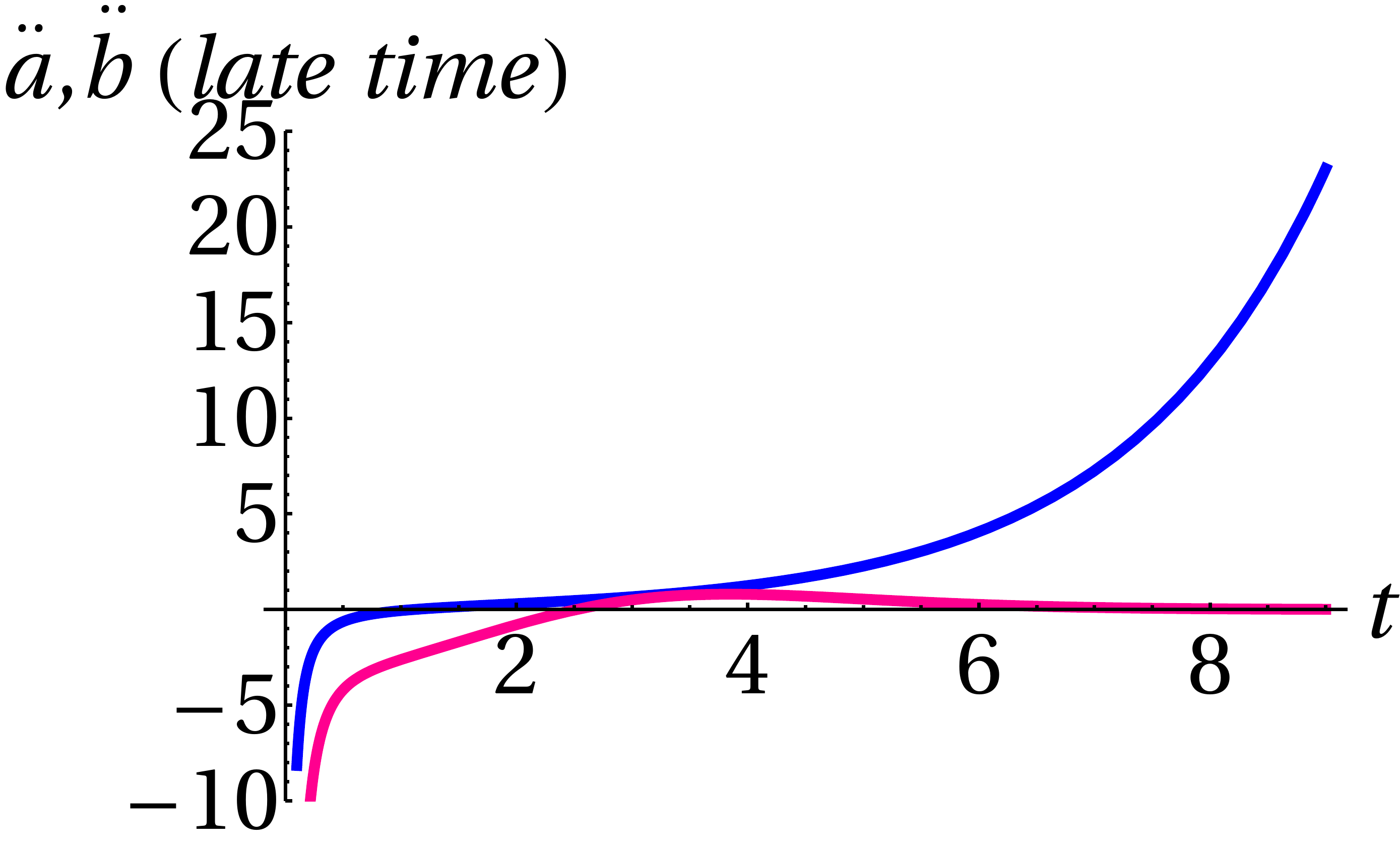}
    \caption{The late time accelerations of the scale factors: $\ddot a$ is represented by the blue curve, and $\ddot b$ by the red curve. The curve for $\ddot b$ is scaled up by a factor of 50.}
    \label{addotbddotLATE00003constantL}
  \end{subfigure}
\qquad
  \begin{subfigure}[t]{.5\linewidth}
    \centering
    \includegraphics[width=0.7\columnwidth]{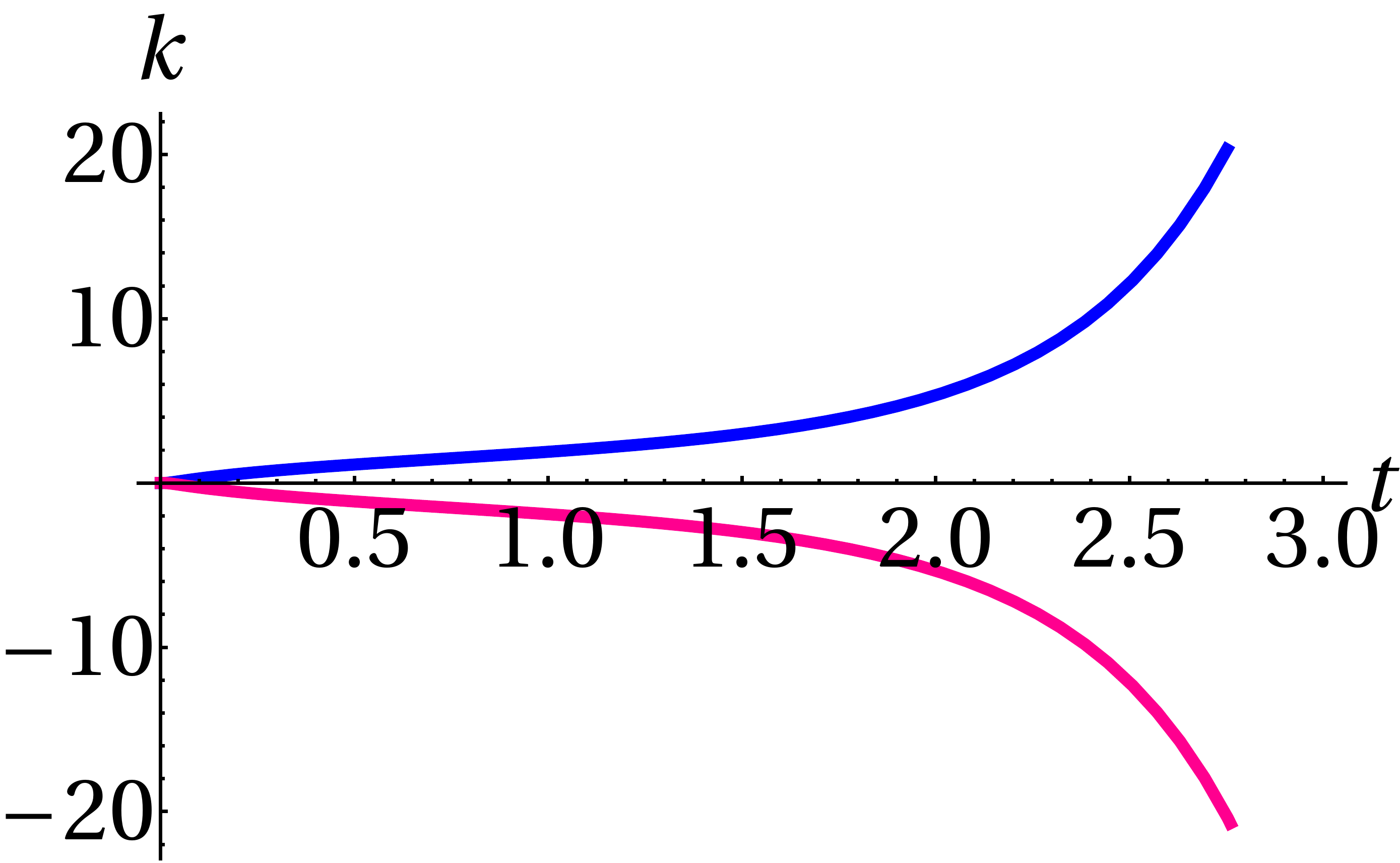}
    \caption{The harmonic function $k$ using: $\dot k\left(0\right)=1$ (blue curve), and $\dot k\left(0\right)=-1$ (red curve). While $\dot k\left(0\right)=0$ diverges.}
    \label{k00003constantL}
  \end{subfigure}
 \caption{Initial conditions set number 3 for constant $\Lambda$ (continued).}
  \label{Fig5dash}
  \end{figure}

\begin{figure}[H]
  \begin{subfigure}[t]{.5\linewidth}
    \centering
    \includegraphics[width=0.7\columnwidth]{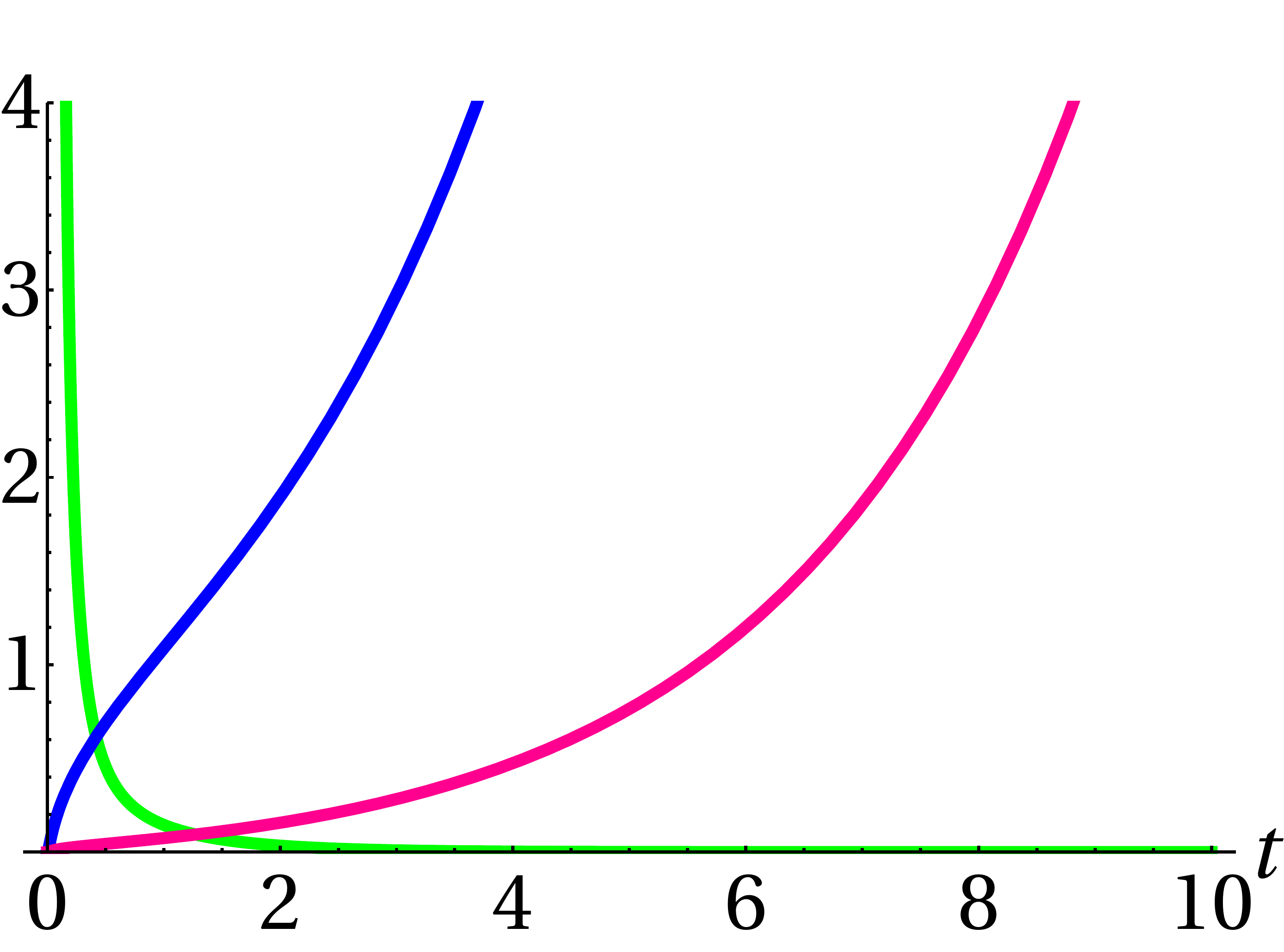}
    \caption{The scale factor $a$ is represented by the blue curve, $b$ by the red curve, while $\left| {G_{i\bar j} \dot z^i \dot z^{\bar j}} \right|$ by the green curve.}
    \label{abzz00004constantL}
  \end{subfigure}
\qquad
  \begin{subfigure}[t]{.5\linewidth}
    \centering
    \includegraphics[width=0.7\columnwidth]{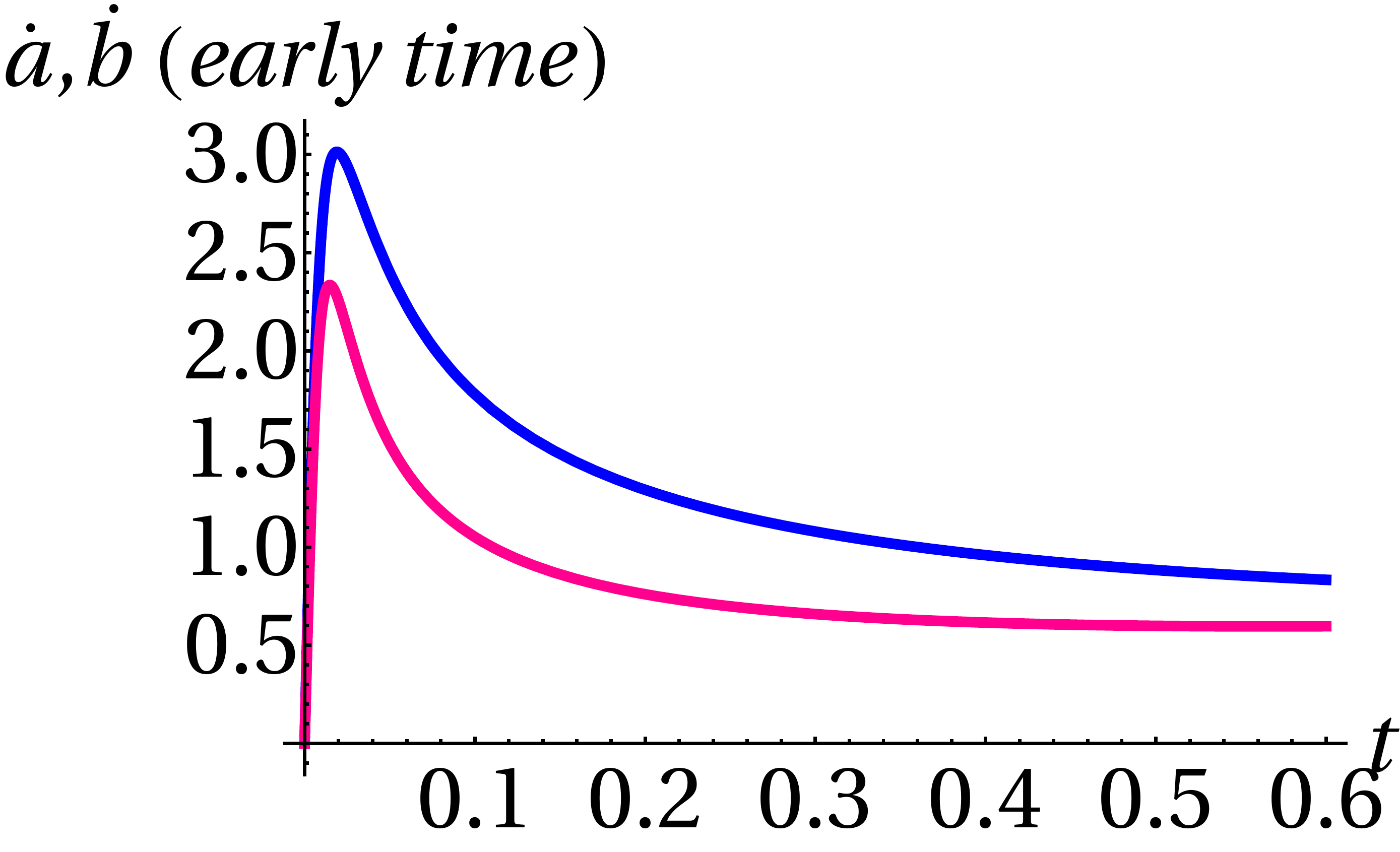}
    \caption{The early time expansion rates of the scale factors: $\dot a$ is represented by the blue curve, and $\dot b$ by the red curve. The curve for $\dot b$ is scaled up by a factor of 10.}
    \label{adotbdotEARLY00004constantL}
  \end{subfigure}
\\[9em]
  \begin{subfigure}[t]{.5\linewidth}
    \centering
    \includegraphics[width=0.7\columnwidth]{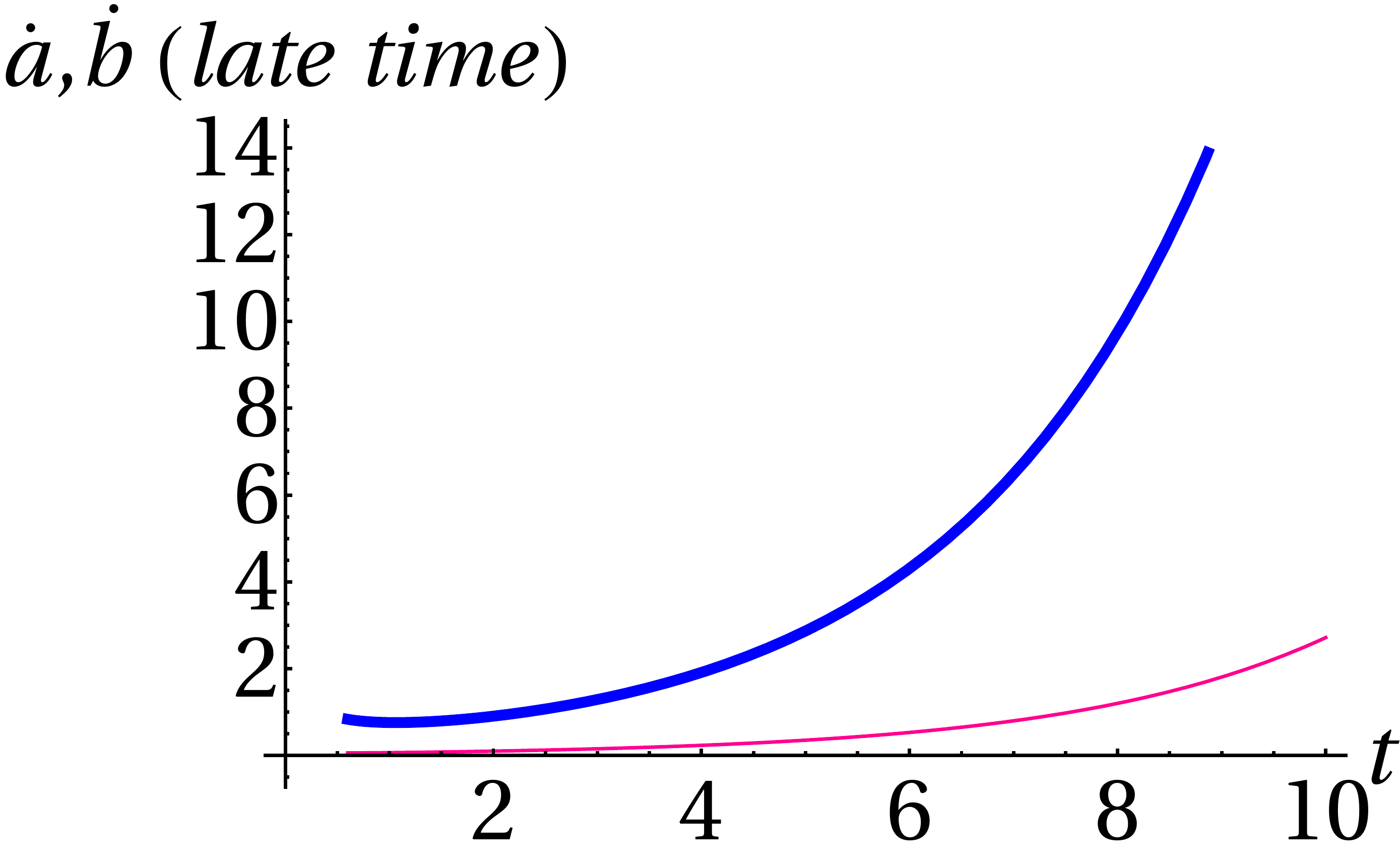}
    \caption{The late time expansion rates of the scale factors: $\dot a$ is represented by the blue curve, and $\dot b$ by the red curve.}
    \label{adotbdotLATE00004constantL}
  \end{subfigure}
\qquad
  \begin{subfigure}[t]{.5\linewidth}
    \centering
    \includegraphics[width=0.7\columnwidth]{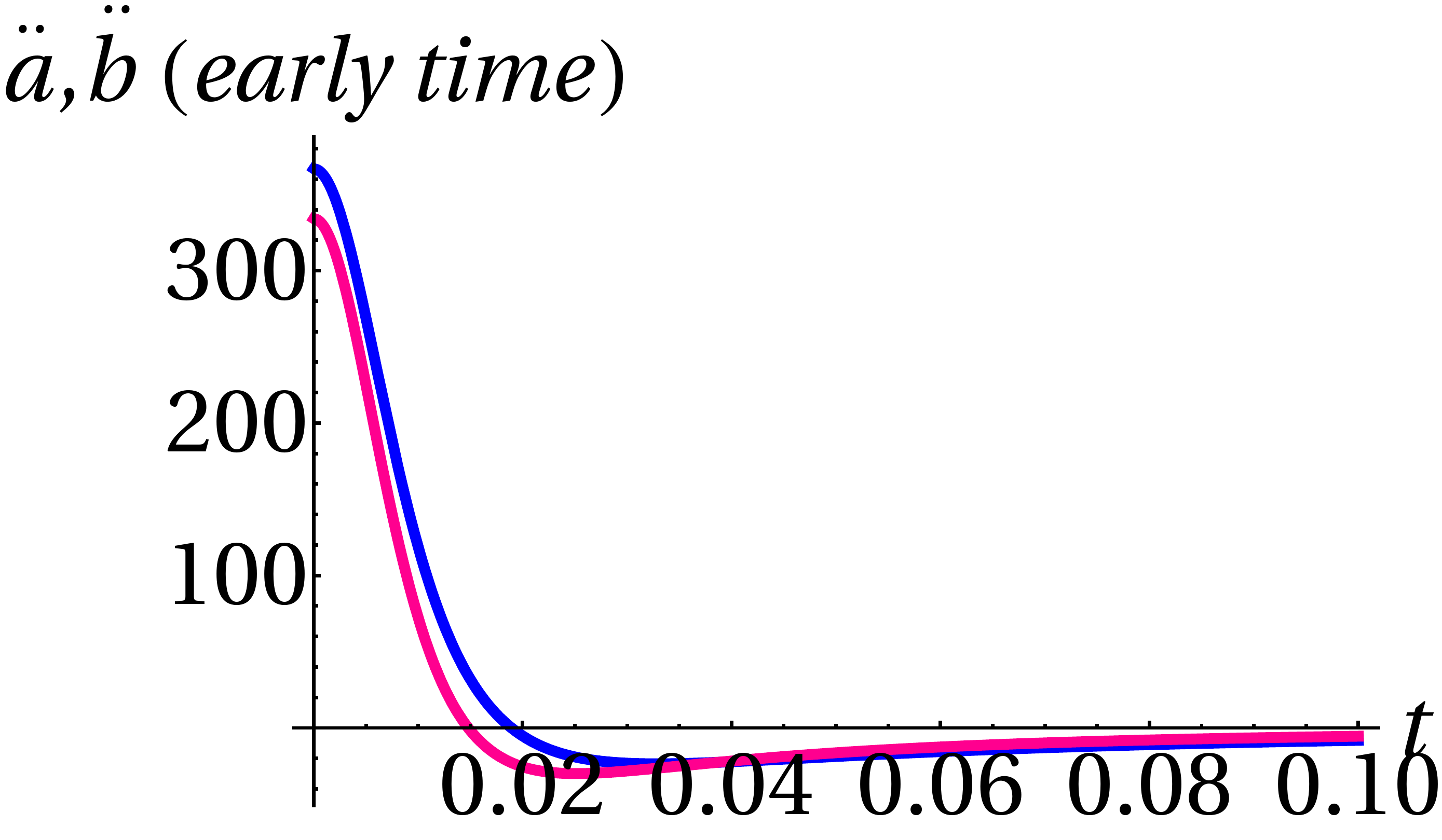}
    \caption{The early time accelerations of the scale factors: $\ddot a$ is represented by the blue curve, and $\ddot b$ by the red curve. The curve for $\ddot b$ is scaled up by a factor of 10.}
    \label{addotbddotEARLY00004constantL}
  \end{subfigure}
    \caption{Initial conditions set number 4 for constant $\Lambda$.}
  \label{Fig7}
  \end{figure}
\begin{figure}[H]
\begin{subfigure}[t]{.5\linewidth}
    \centering
    \includegraphics[width=0.7\columnwidth]{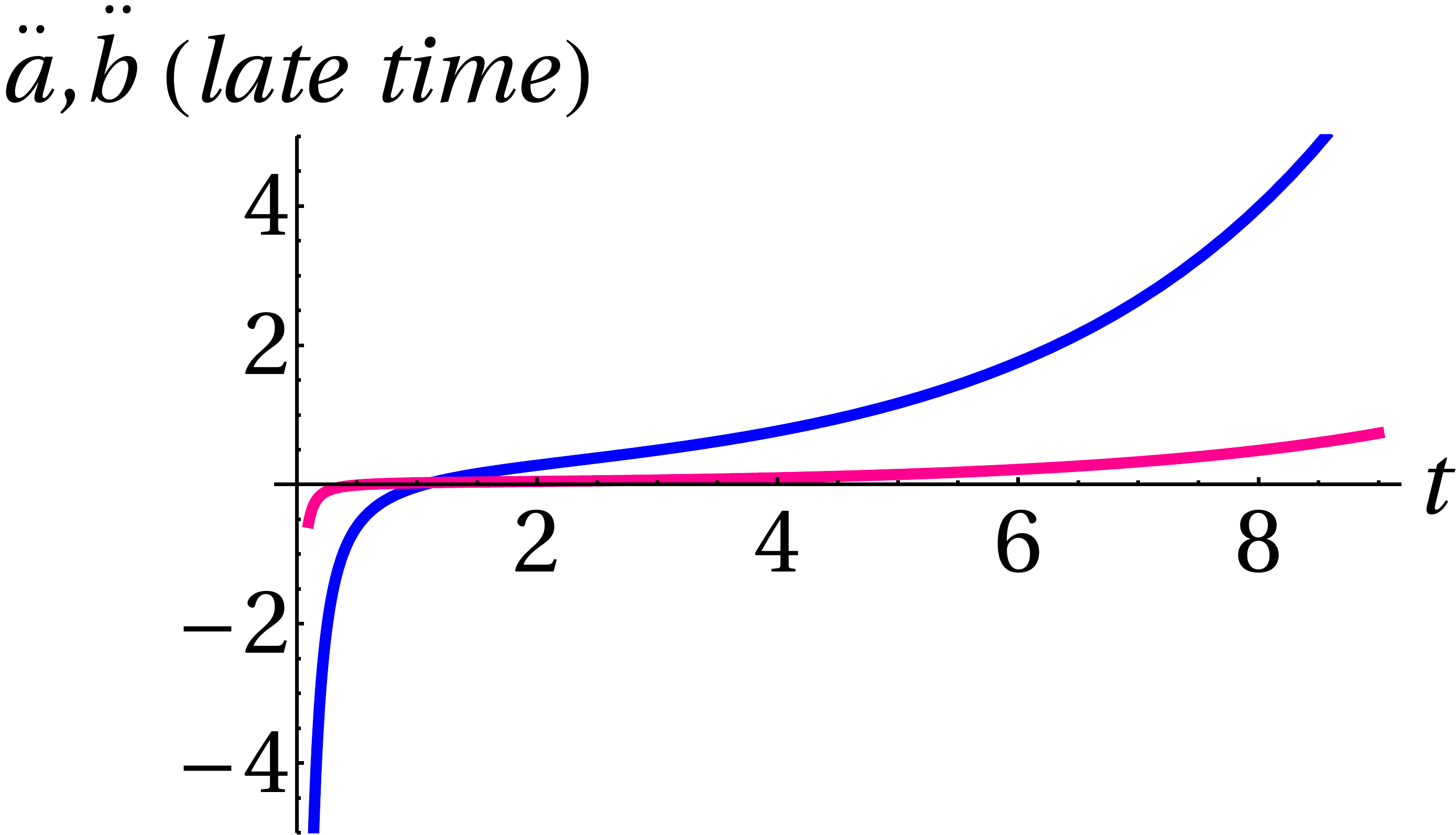}
    \caption{The late time accelerations of the scale factors: $\ddot a$ is represented by the blue curve, and $\ddot b$ by the red curve.}
    \label{addotbddotLATE00004constantL}
  \end{subfigure}
\qquad
  \begin{subfigure}[t]{.5\linewidth}
    \centering
    \includegraphics[width=0.7\columnwidth]{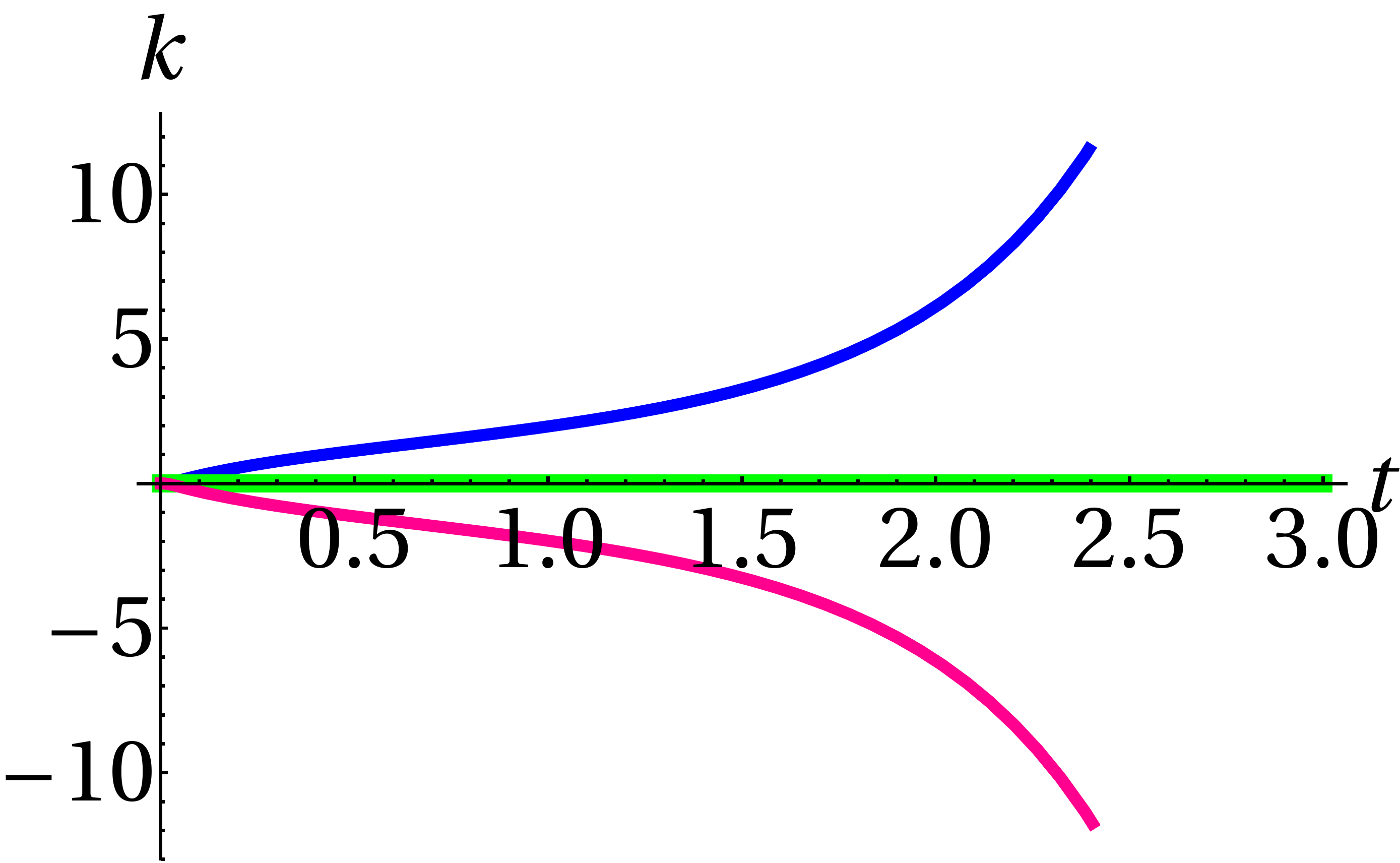}
    \caption{The harmonic function $k$ using: $\dot k\left(0\right)=1$ (blue curve), $\dot k\left(0\right)=0$ (green line), and $\dot k\left(0\right)=-1$ (red curve).}
    \label{k00004constantL}
  \end{subfigure}
    \caption{Initial conditions set number 4 for constant $\Lambda$ (continued).}
  \label{Fig77}
  \end{figure}
  
\vspace{3cm}


\begin{figure}[H]
  \begin{subfigure}[t]{.5\linewidth}
    \centering
    \includegraphics[width=0.7\columnwidth]{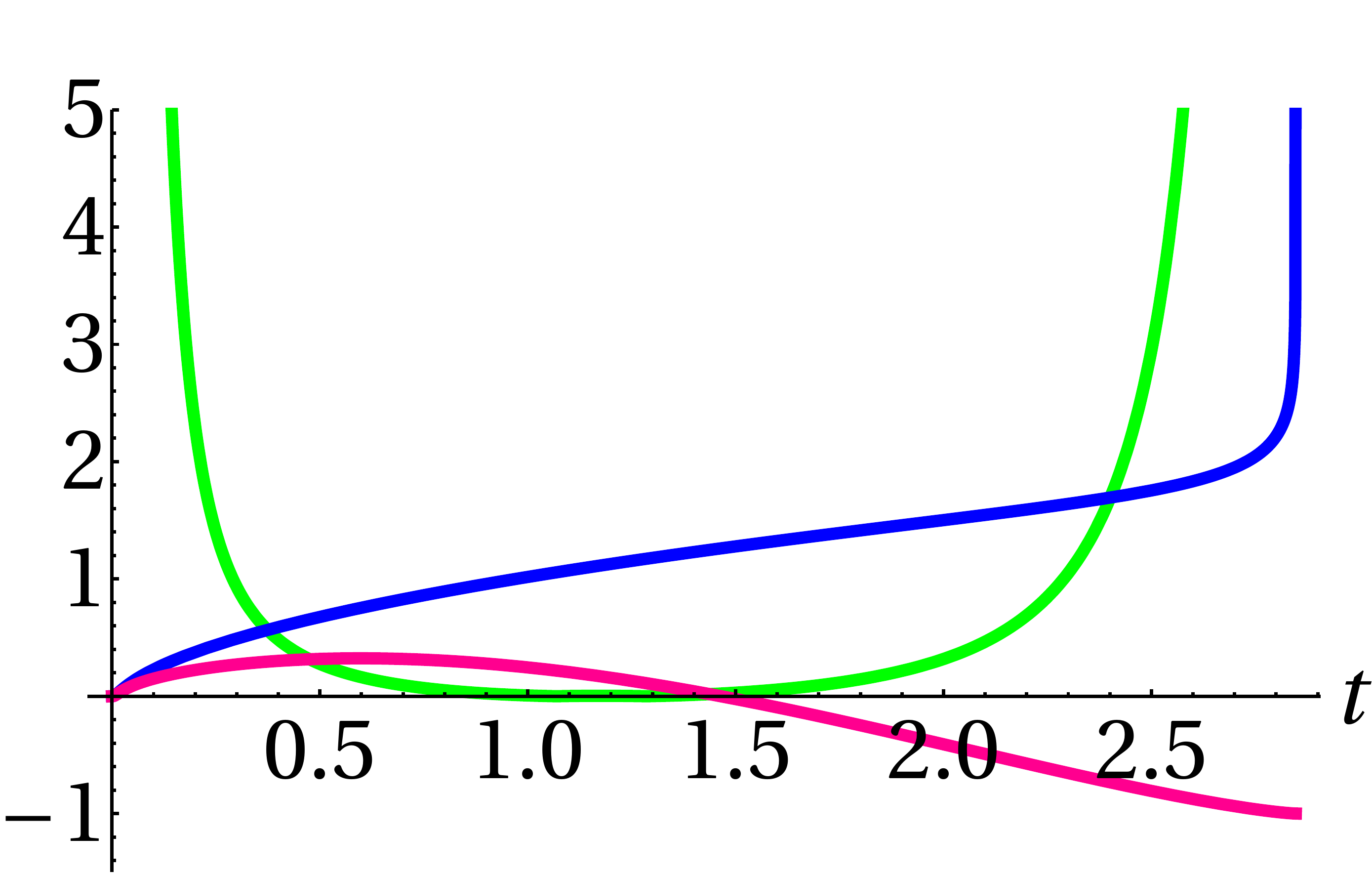}
    \caption{The scale factor $a$ is represented by the blue curve, $b$ by the red curve, while $\left| {G_{i\bar j} \dot z^i \dot z^{\bar j}} \right|$ by the green curve. The curve for $b$ is scaled up by a factor of 10.}
    \label{abzz00005constantL}
  \end{subfigure}
\qquad
  \begin{subfigure}[t]{.5\linewidth}
    \centering
    \includegraphics[width=0.7\columnwidth]{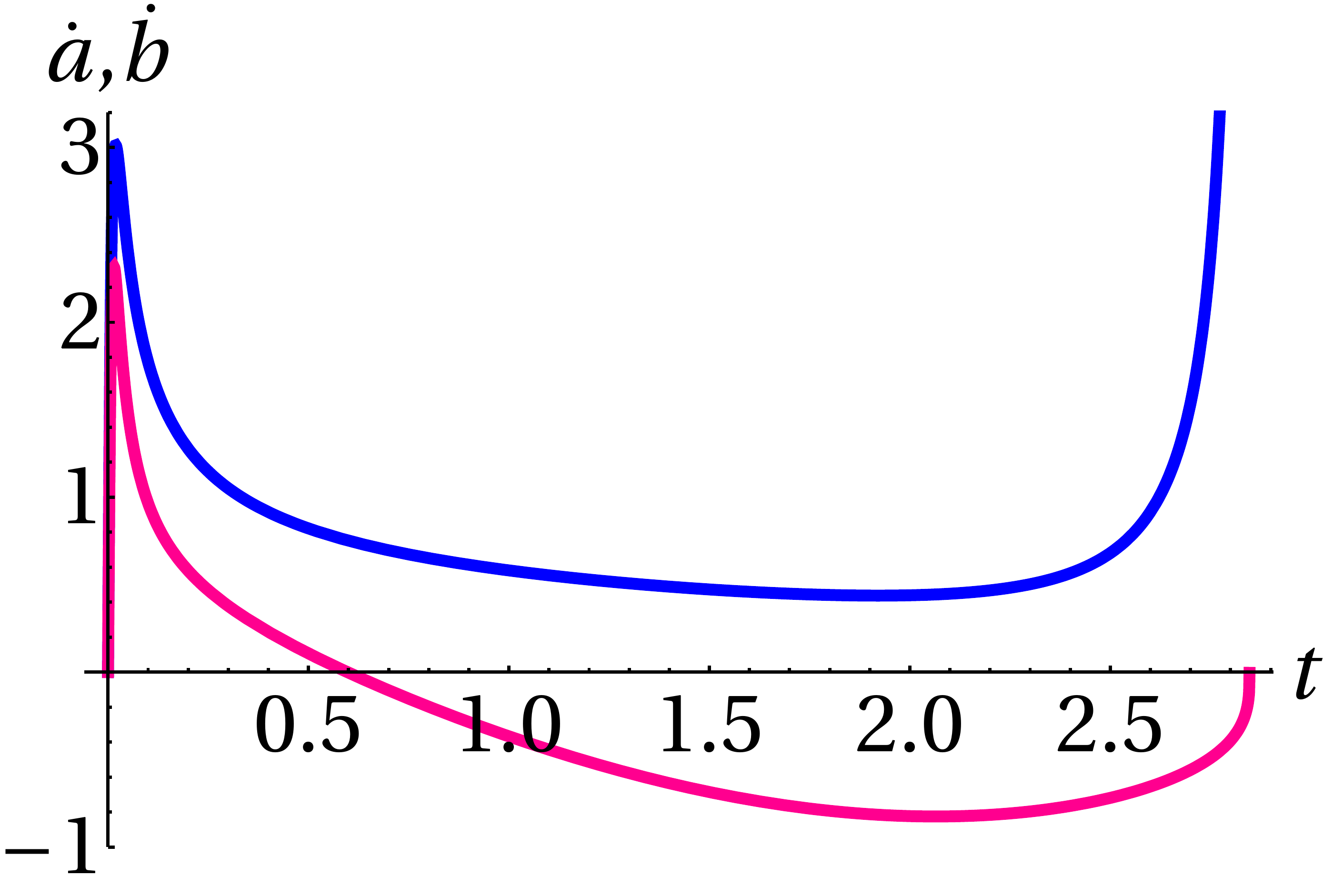}
    \caption{The expansion rates of the scale factors: $\dot a$ is represented by the blue curve, and $\dot b$ by the red curve. The curve for $\dot b$ is scaled up by a factor of 10.}
    \label{adotbdotEARLY00005constantL}
  \end{subfigure}
      \caption{Initial conditions set number 5 for constant $\Lambda$ .}
  \label{Fig9}
  \end{figure}
\begin{figure}[H]
  \begin{subfigure}[t]{.5\linewidth}
    \centering
    \includegraphics[width=0.7\columnwidth]{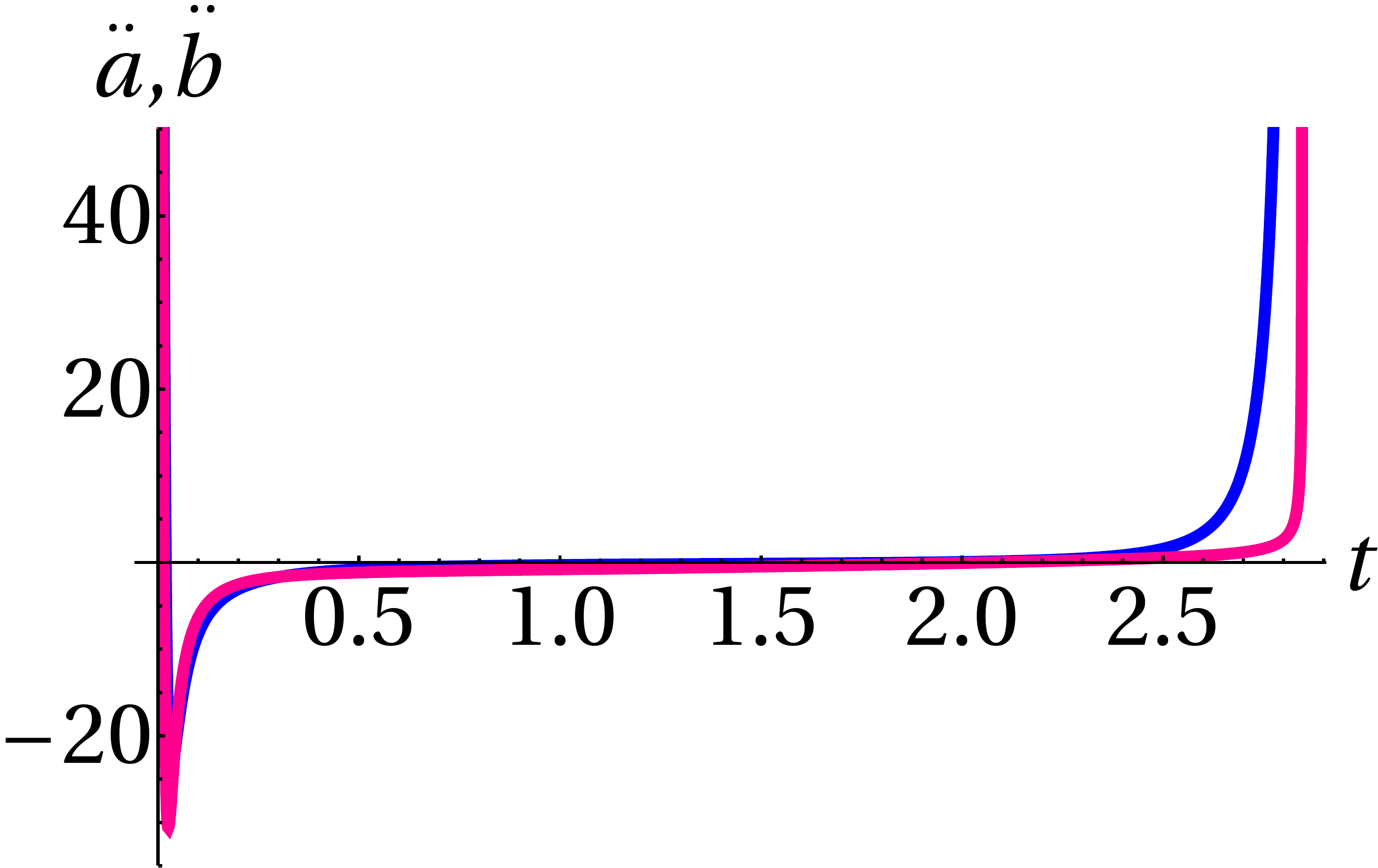}
    \caption{The accelerations of the scale factors: $\ddot a$ is represented by the blue curve, and $\ddot b$ by the red curve.}
    \label{addotbddotEARLY00005constantL}
  \end{subfigure}
\qquad
  \begin{subfigure}[t]{.5\linewidth}
    \centering
    \includegraphics[width=0.7\columnwidth]{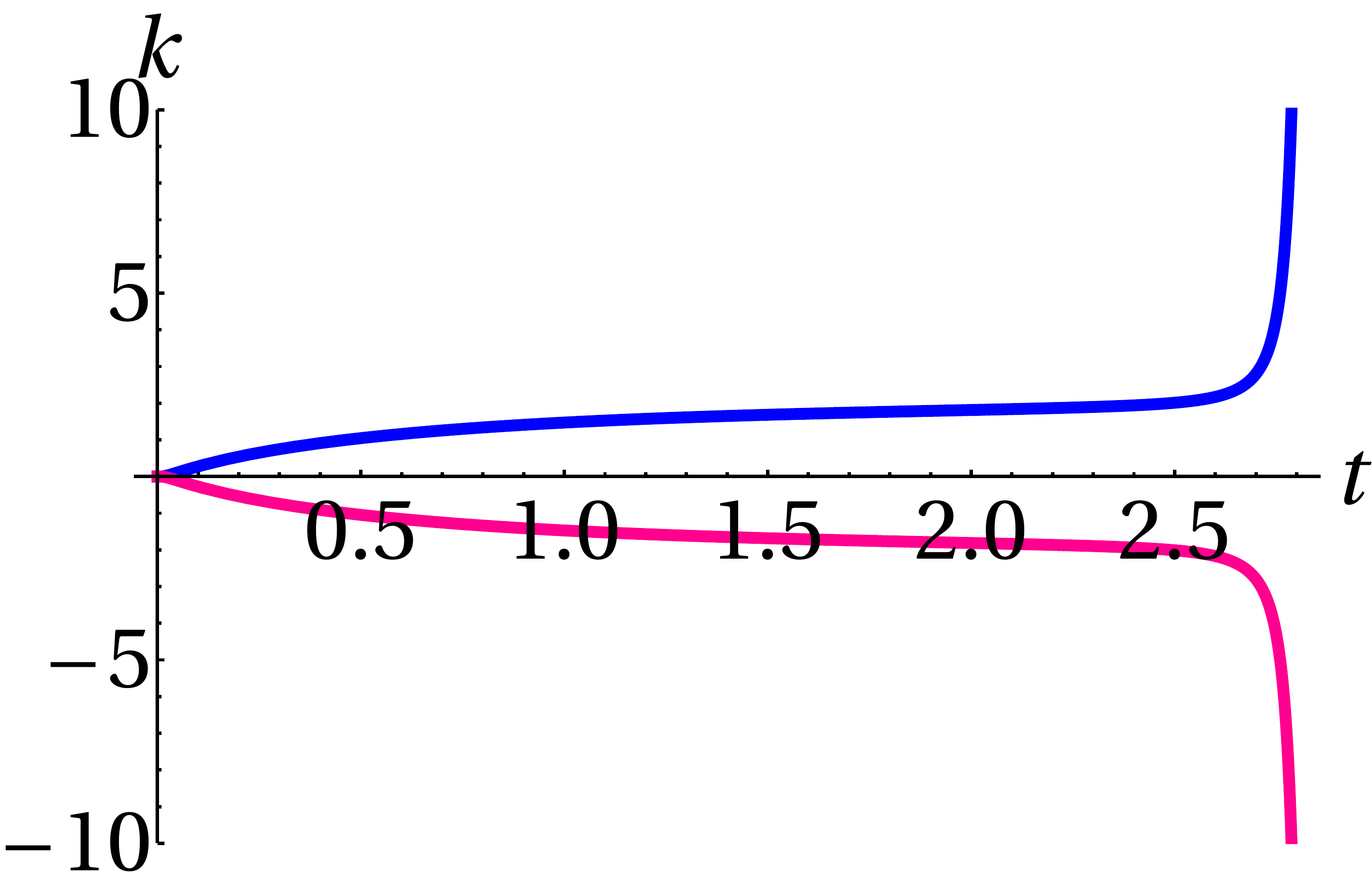}
    \caption{The harmonic function $k$ using: $\dot k\left(0\right)=1$ (blue curve), and $\dot k\left(0\right)=-1$ (red curve). While $\dot k\left(0\right)=0$ diverges.}
    \label{k00005constantL}
  \end{subfigure}
 \caption{Initial conditions set number 5 for constant $\Lambda$ (continued).}
  \label{Fig10}
\end{figure}

\vspace{3cm}

\vspace{-0.7cm}
\begin{figure}[H]
\begin{subfigure}[t]{.5\linewidth}
    \centering
    \includegraphics[width=0.7\columnwidth]{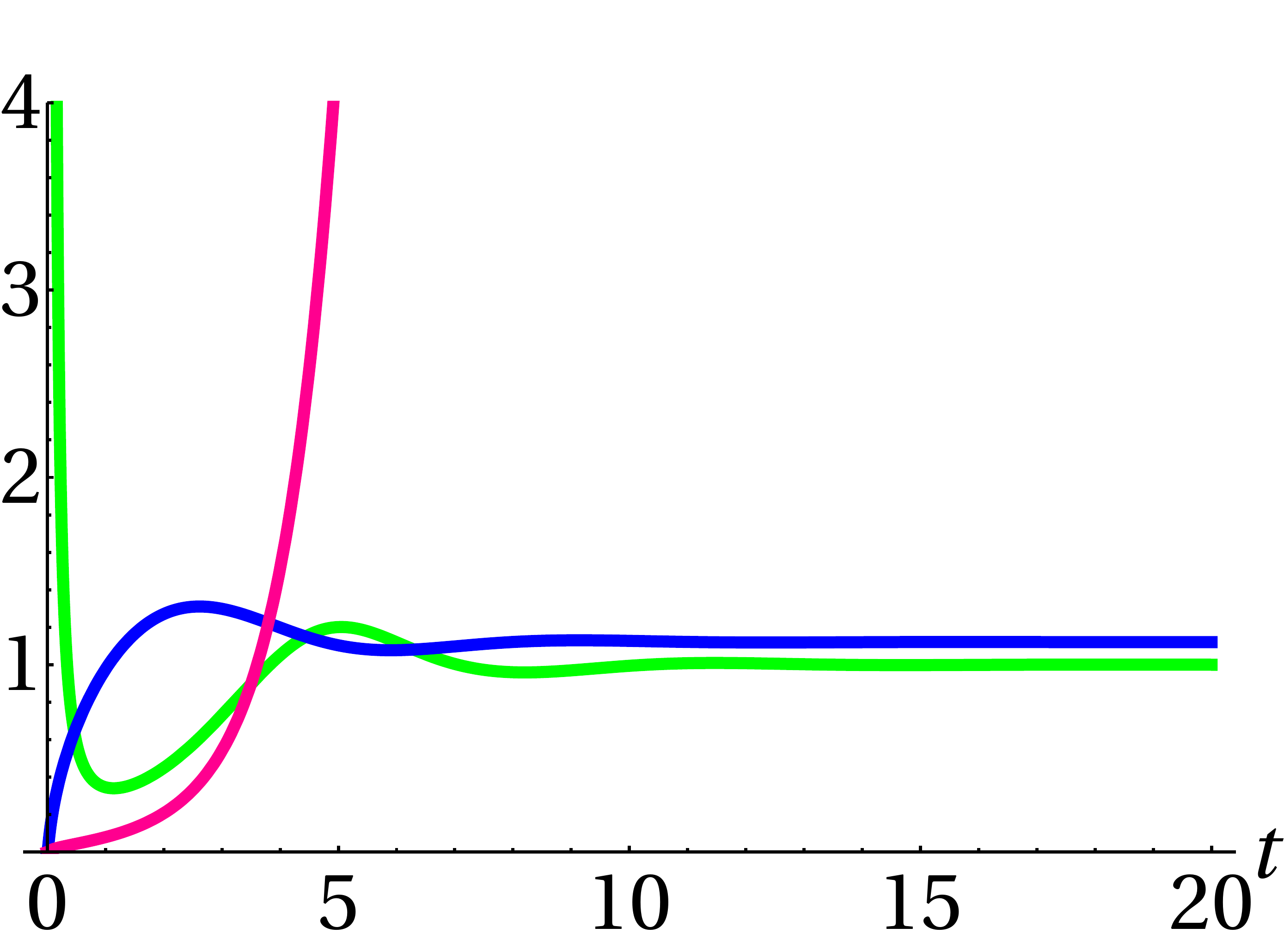}
    \caption{The scale factor $a$ is represented by the blue curve, $b$ by the red curve, while $\left| {G_{i\bar j} \dot z^i \dot z^{\bar j}} \right|$ by the green curve.}
    \label{abzz00006constantL}
  \end{subfigure}
\qquad
  \begin{subfigure}[t]{.5\linewidth}
    \centering
    \includegraphics[width=0.7\columnwidth]{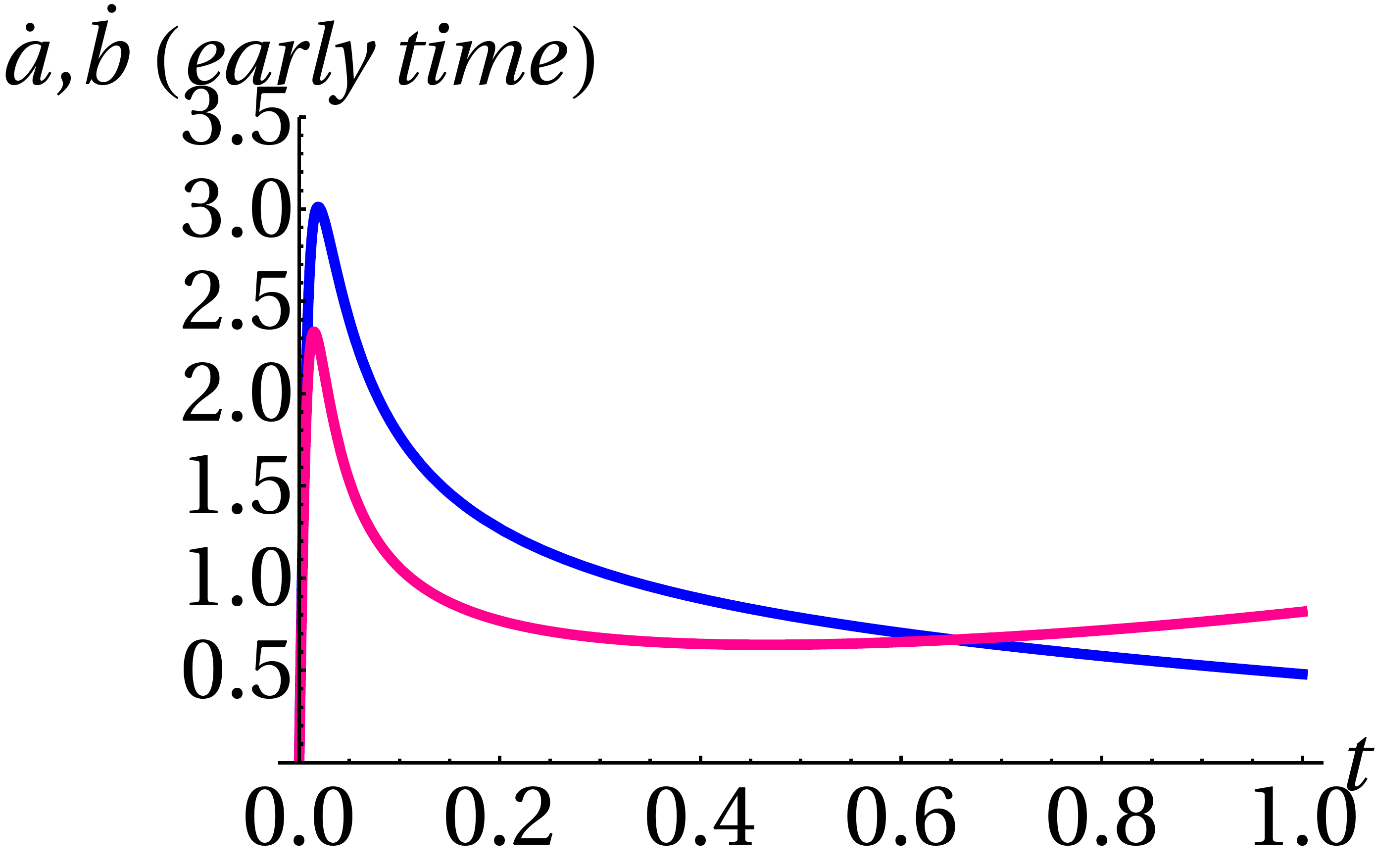}
    \caption{The early time expansion rates of the scale factors: $\dot a$ is represented by the blue curve, and $\dot b$ by the red curve. The curve for $\dot b$ is scaled up by a factor of 10.}
    \label{adotbdotEARLY00006constantL}
  \end{subfigure}
\caption{Initial conditions set number 6 for constant $\Lambda$.}
  \label{Fig112}
\end{figure}
\begin{figure}[H]
  \begin{subfigure}[t]{.5\linewidth}
    \centering
    \includegraphics[width=0.7\columnwidth]{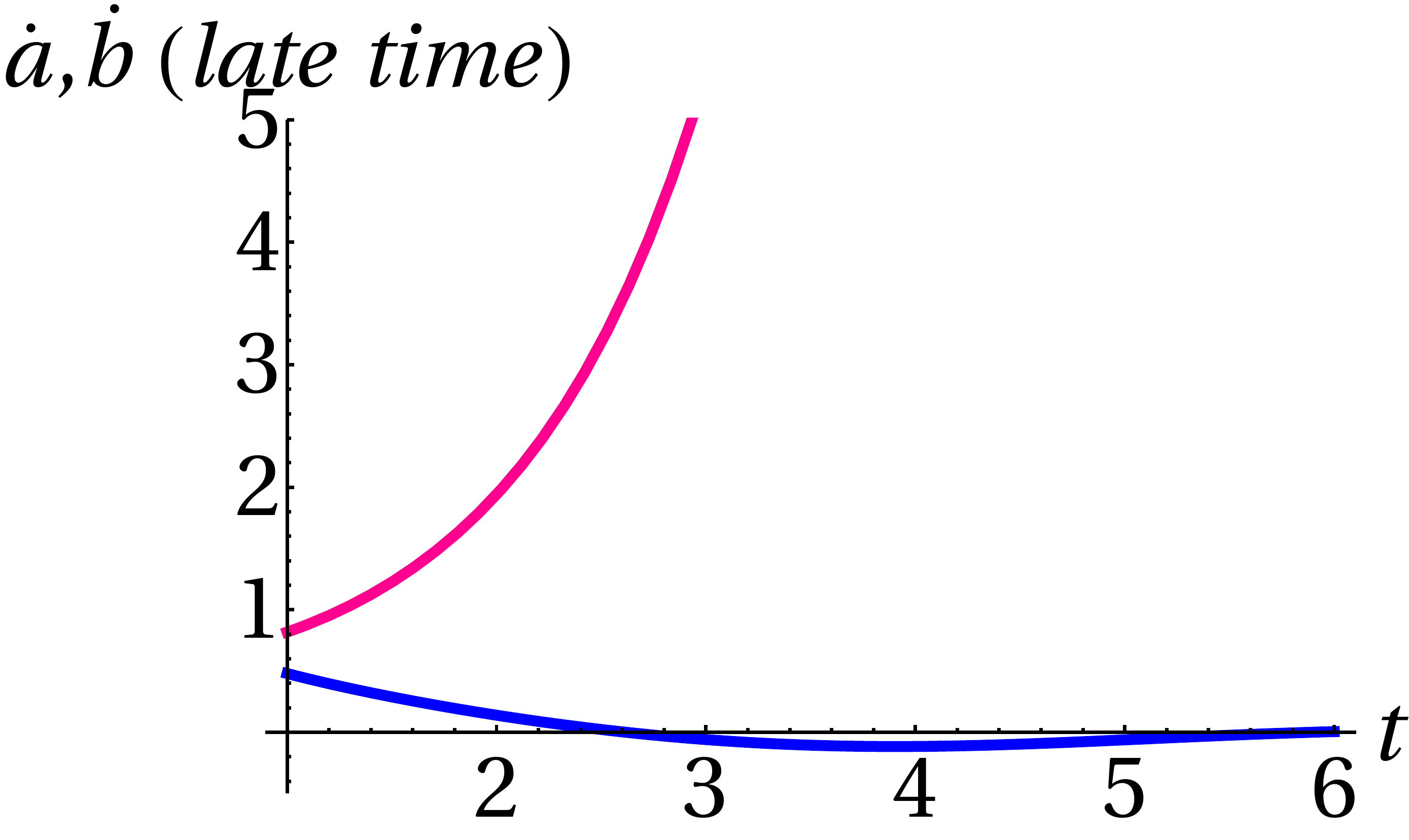}
    \caption{The late time expansion rates of the scale factors: $\dot a$ is represented by the blue curve, and $\dot b$ by the red curve. The curve for $\dot b$ is scaled up by a factor of 10.}
    \label{adotbdotLATE00006constantL}
  \end{subfigure}
\qquad
  \begin{subfigure}[t]{.5\linewidth}
    \centering
    \includegraphics[width=0.7\columnwidth]{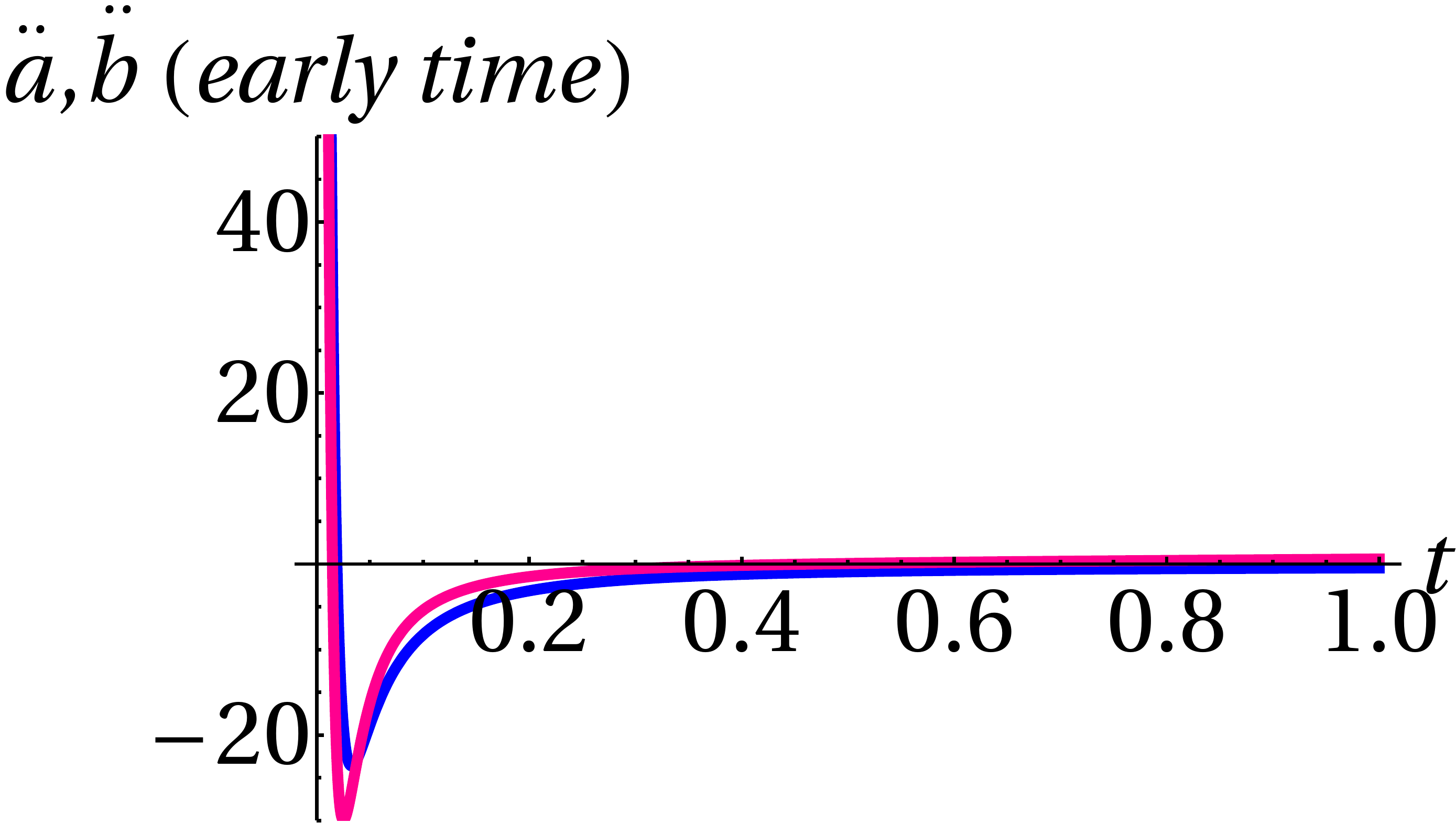}
    \caption{The early time accelerations of the scale factors: $\ddot a$ is represented by the blue curve, and $\ddot b$ by the red curve. The curve for $\ddot b$ is scaled up by a factor of 10.}
    \label{addotbddotEARLY00006constantL}
  \end{subfigure}
\\[9em]
  \begin{subfigure}[t]{.5\linewidth}
    \centering
    \includegraphics[width=0.7\columnwidth]{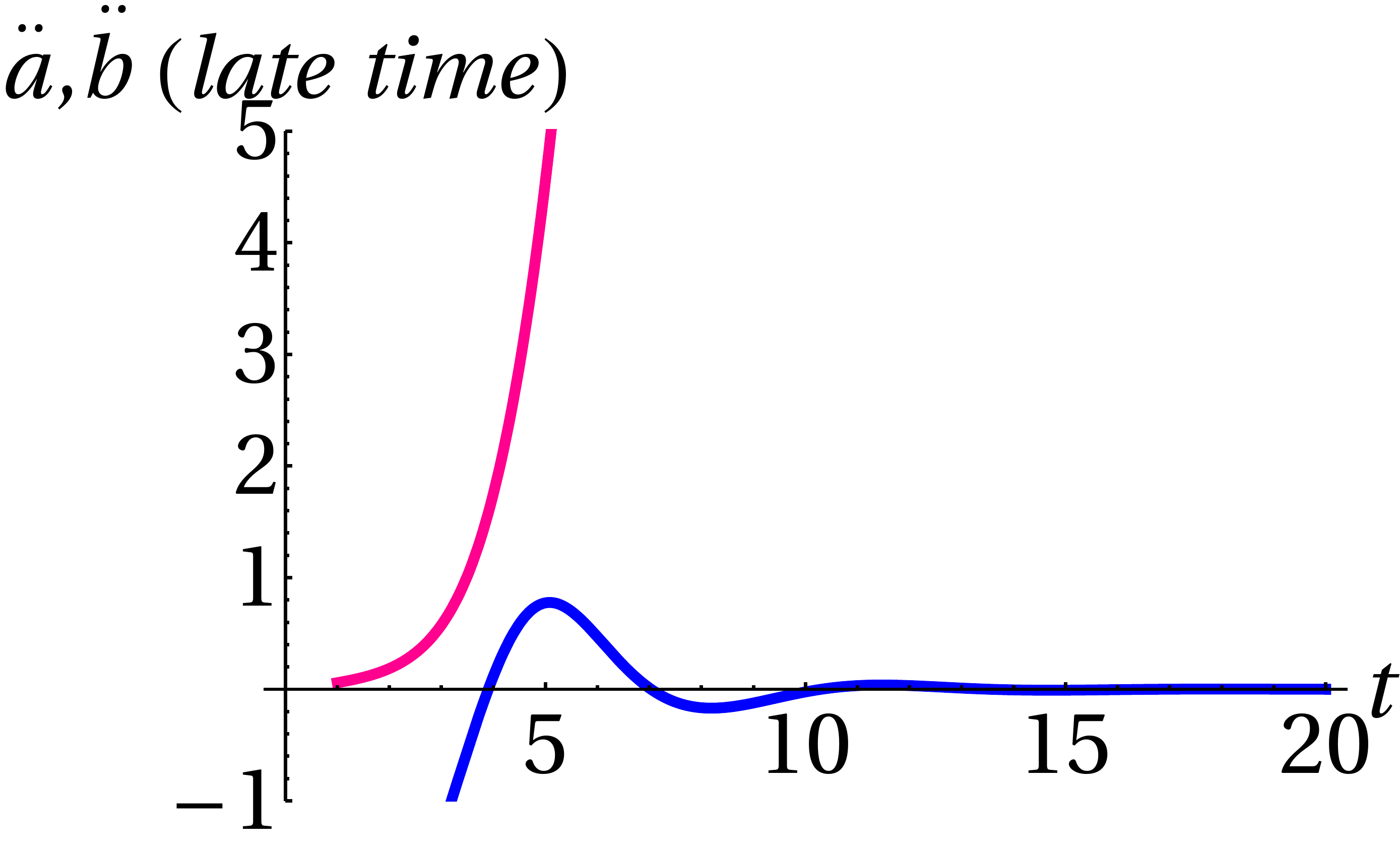}
    \caption{The late time accelerations of the scale factors: $\ddot a$ is represented by the blue curve, and $\ddot b$ by the red curve. The curve for $\ddot a$ is scaled up by a factor of 10.}
    \label{addotbddotEARLY00006constantL}
  \end{subfigure}
\qquad
  \begin{subfigure}[t]{.5\linewidth}
    \centering
    \includegraphics[width=0.7\columnwidth]{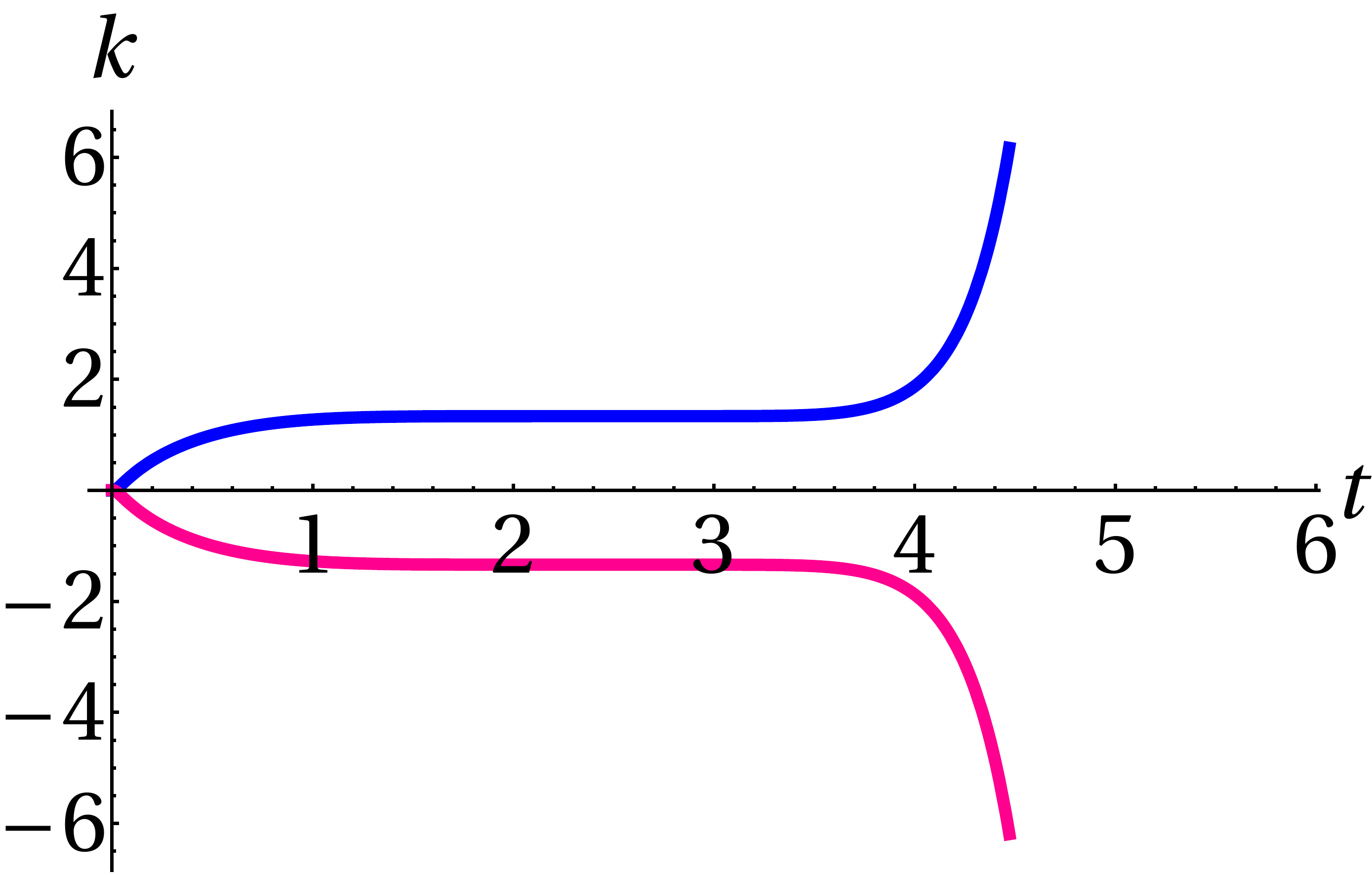}
    \caption{The harmonic function $k$ using: $\dot k\left(0\right)=1$ (blue curve), and $\dot k\left(0\right)=-1$ (red curve). While $\dot k\left(0\right)=0$ diverges.}
    \label{k00006constantL}
  \end{subfigure}
\caption{Initial conditions set number 6 for constant $\Lambda$ (continued).}
  \label{Fig12}
\end{figure}
%


\begin{figure}[H]
  \begin{subfigure}[t]{.5\linewidth}
    \centering
    \includegraphics[width=0.7\columnwidth]{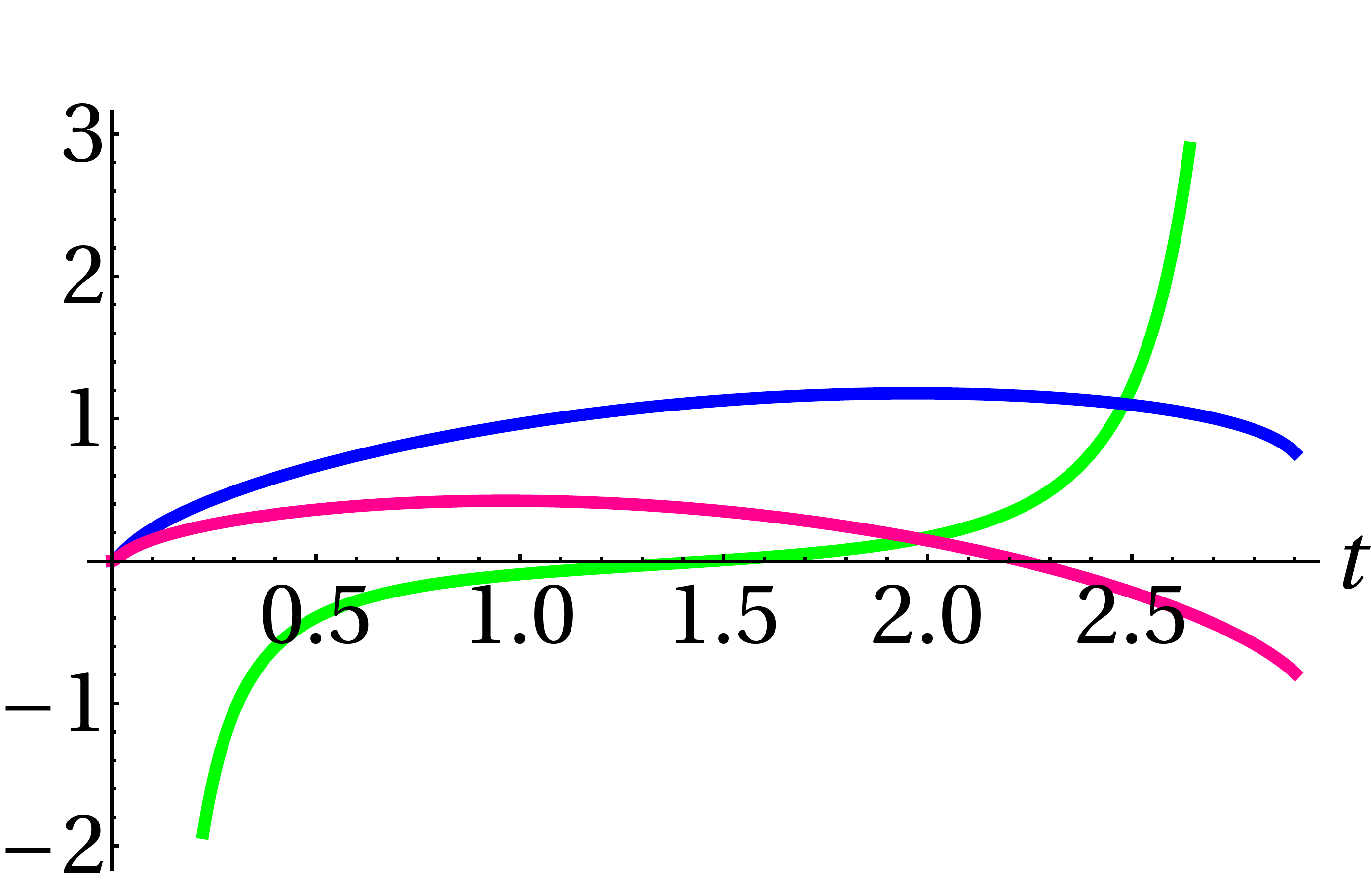}
    \caption{The scale factor $a$ is represented by the blue curve, $b$ by the red curve, while ${G_{i\bar j} \dot z^i \dot z^{\bar j}}$ (not the absolute value) by the green curve. The curve for $b$ is scaled up by a factor of 10.}
    \label{abzz00007constantL}
  \end{subfigure}
\qquad
  \begin{subfigure}[t]{.5\linewidth}
    \centering
    \includegraphics[width=0.7\columnwidth]{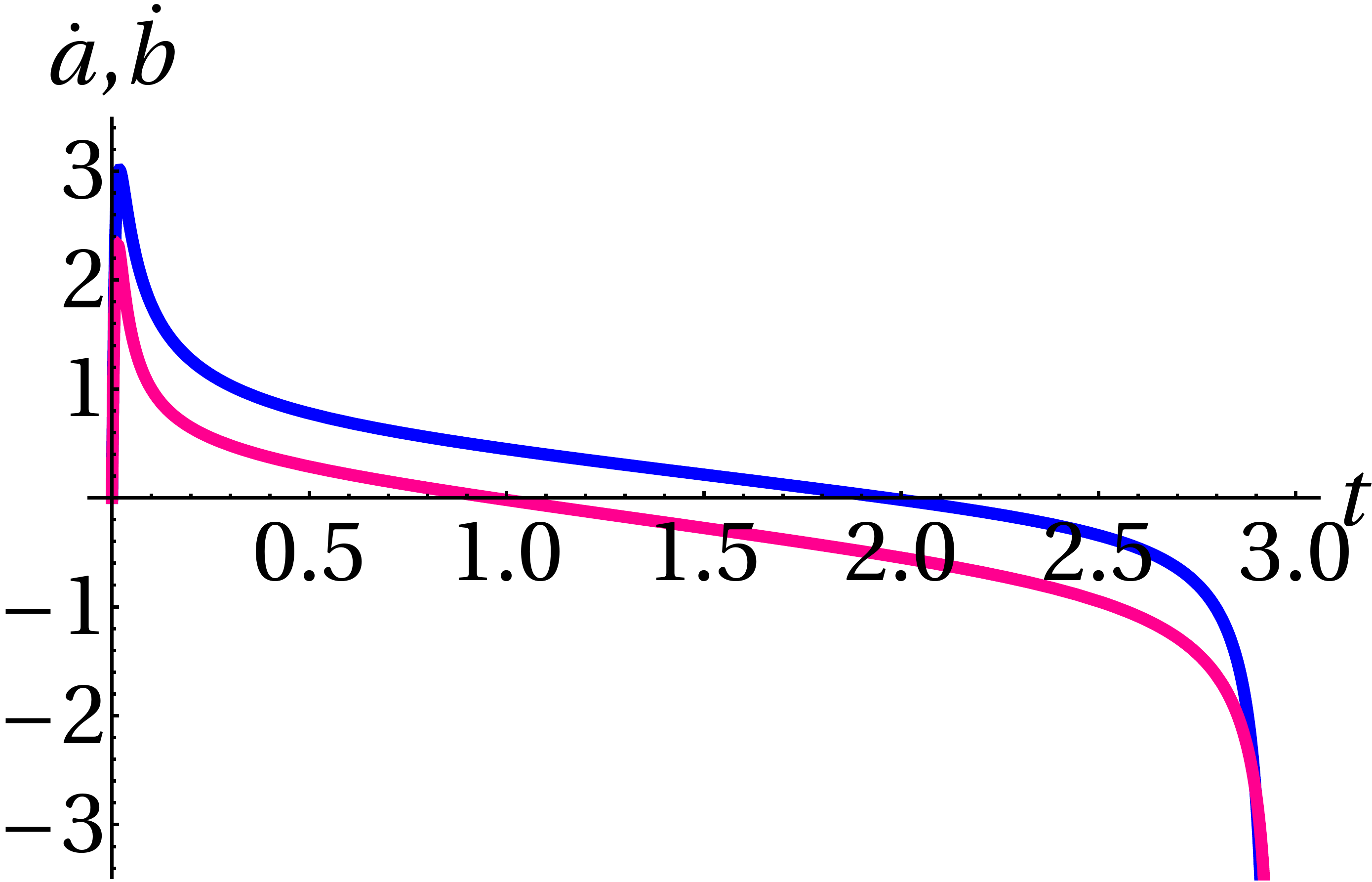}
    \caption{The expansion rates of the scale factors: $\dot a$ is represented by the blue curve, and $\dot b$ by the red curve. The curve for $\dot b$ is scaled up by a factor of 10.}
    \label{adotbdot00007constantL}
  \end{subfigure}
\\[9em]
  \begin{subfigure}[t]{.5\linewidth}
    \centering
    \includegraphics[width=0.7\columnwidth]{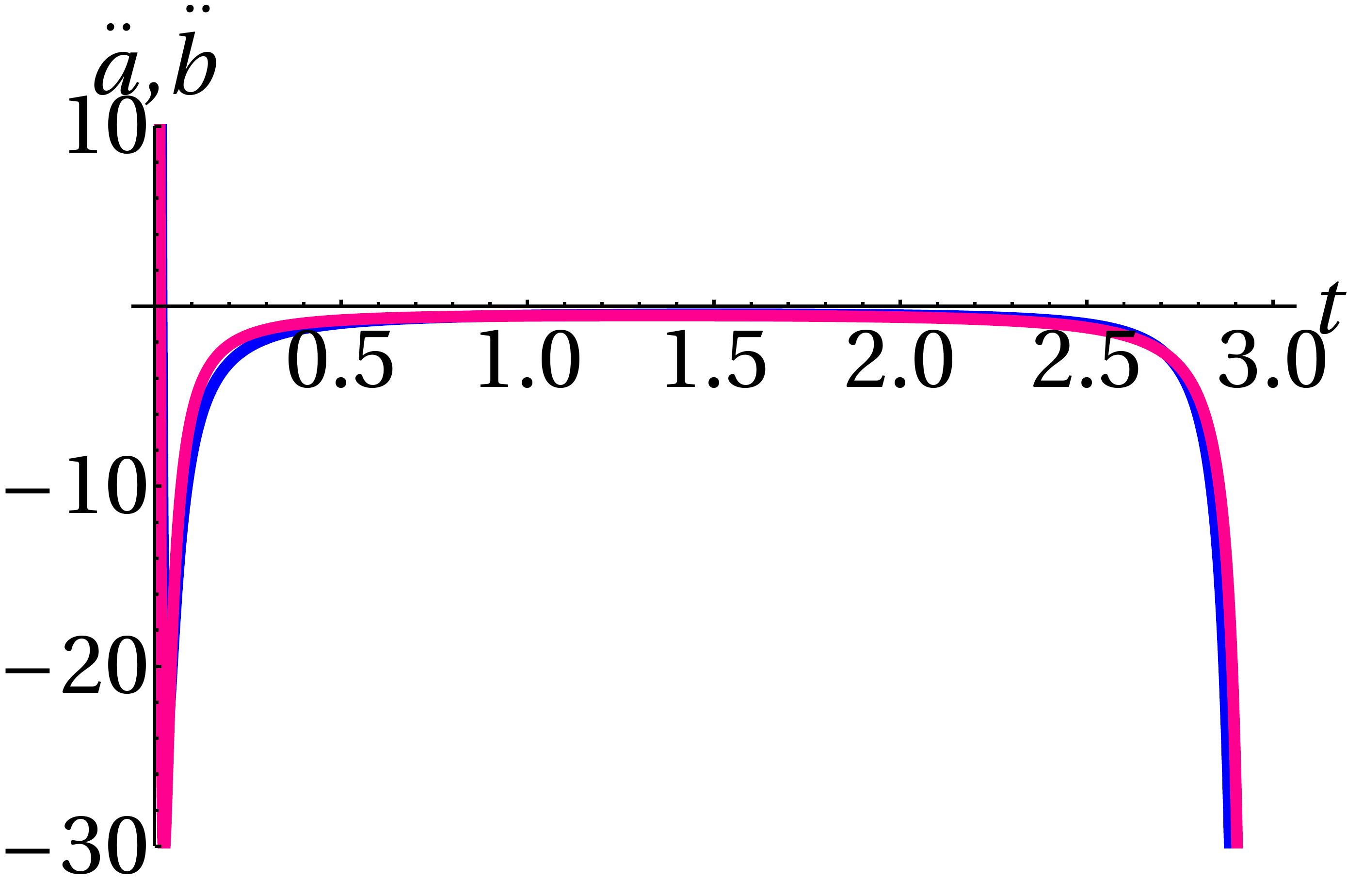}
    \caption{The accelerations of the scale factors: $\ddot a$ is represented by the blue curve, and $\ddot b$ by the red curve. The curve for $\ddot b$ is scaled up by a factor of 10.}
    \label{addotbddot00007constantL}
  \end{subfigure}
\qquad
  \begin{subfigure}[t]{.5\linewidth}
    \centering
    \includegraphics[width=0.7\columnwidth]{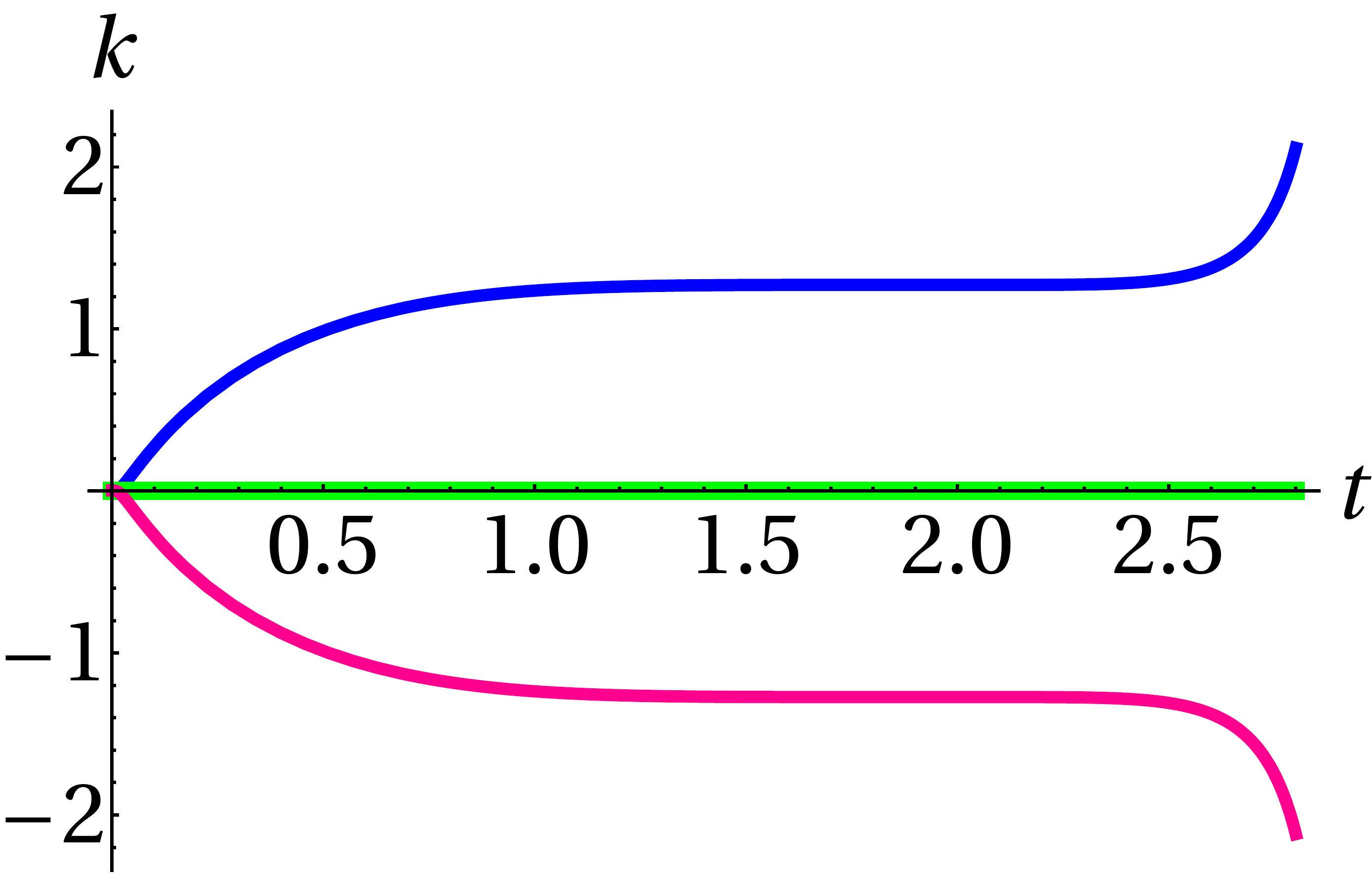}
    \caption{The harmonic function $k$ using: $\dot k\left(0\right)=1$ (blue curve), $\dot k\left(0\right)=0$ (green line), and $\dot k\left(0\right)=-1$ (red curve).}
    \label{k00007constantL}
  \end{subfigure}
      \caption{Initial conditions set number 7 for constant $\Lambda$.}
  \label{Fig13}
  \end{figure}


\begin{figure}[H]
  \begin{subfigure}[t]{.5\linewidth}
    \centering
    \includegraphics[width=0.7\columnwidth]{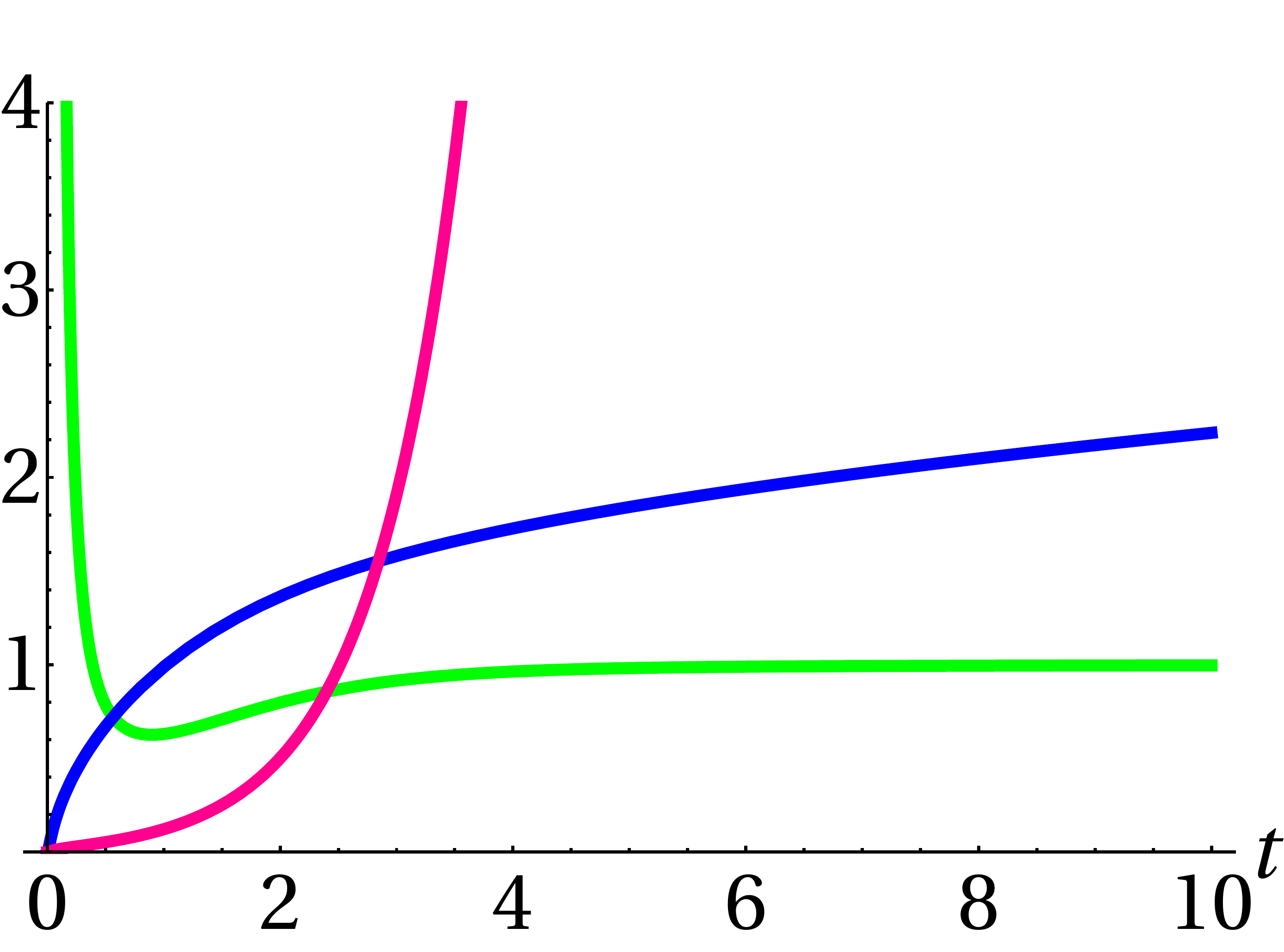}
    \caption{The scale factor $a$ is represented by the blue curve, $b$ by the red curve, while $\left| {G_{i\bar j} \dot z^i \dot z^{\bar j}} \right|$ by the green curve.}
    \label{abzz00008constantL}
  \end{subfigure}
\qquad
  \begin{subfigure}[t]{.5\linewidth}
    \centering
    \includegraphics[width=0.7\columnwidth]{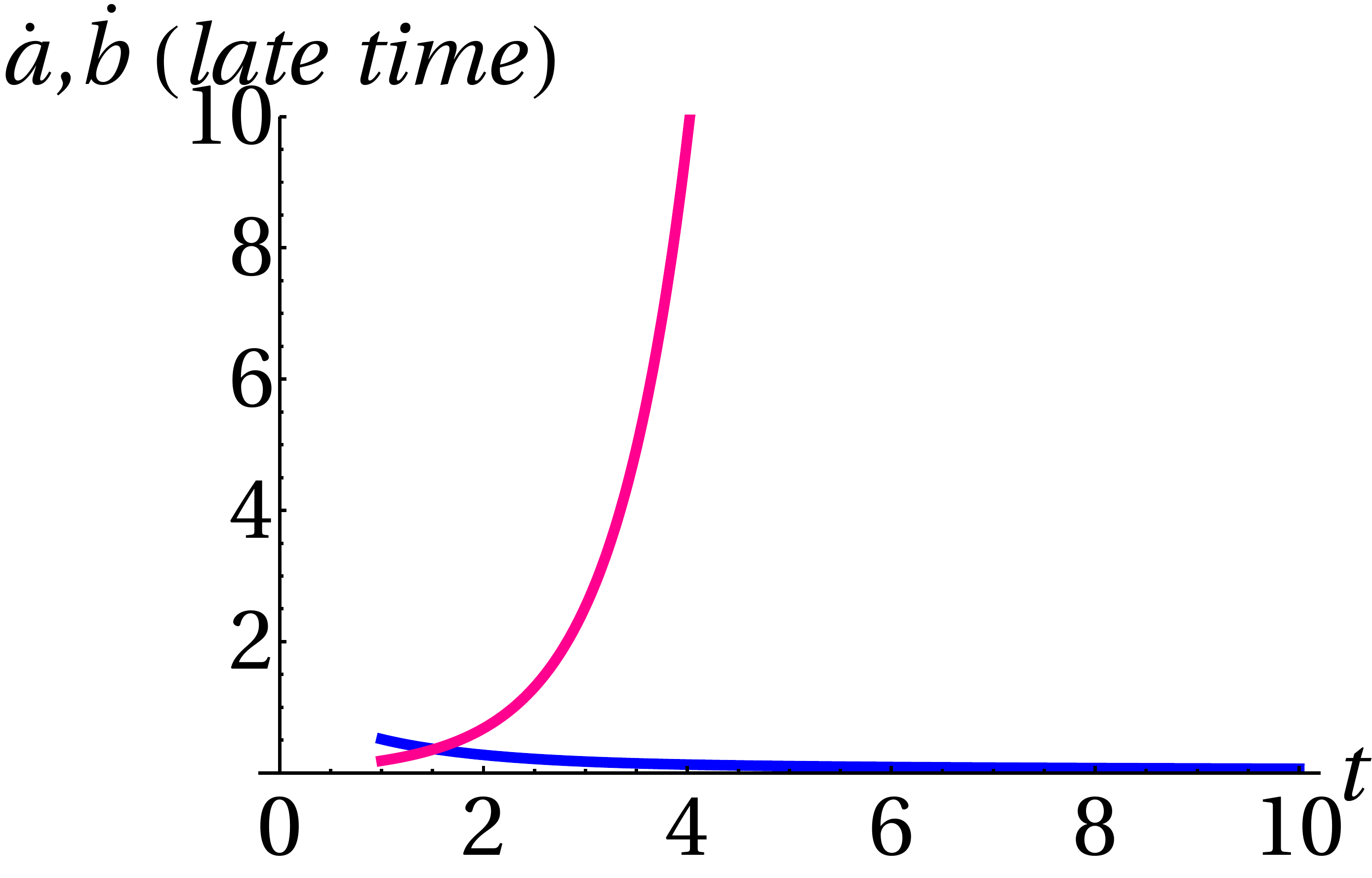}
    \caption{The early time expansion rates of the scale factors: $\dot a$ is represented by the blue curve, and $\dot b$ by the red curve. The curve for $\dot b$ is scaled up by a factor of 10.}
    \label{adotbdotEARLY00008constantL}
  \end{subfigure}
\\[9em]
  \begin{subfigure}[t]{.5\linewidth}
    \centering
    \includegraphics[width=0.7\columnwidth]{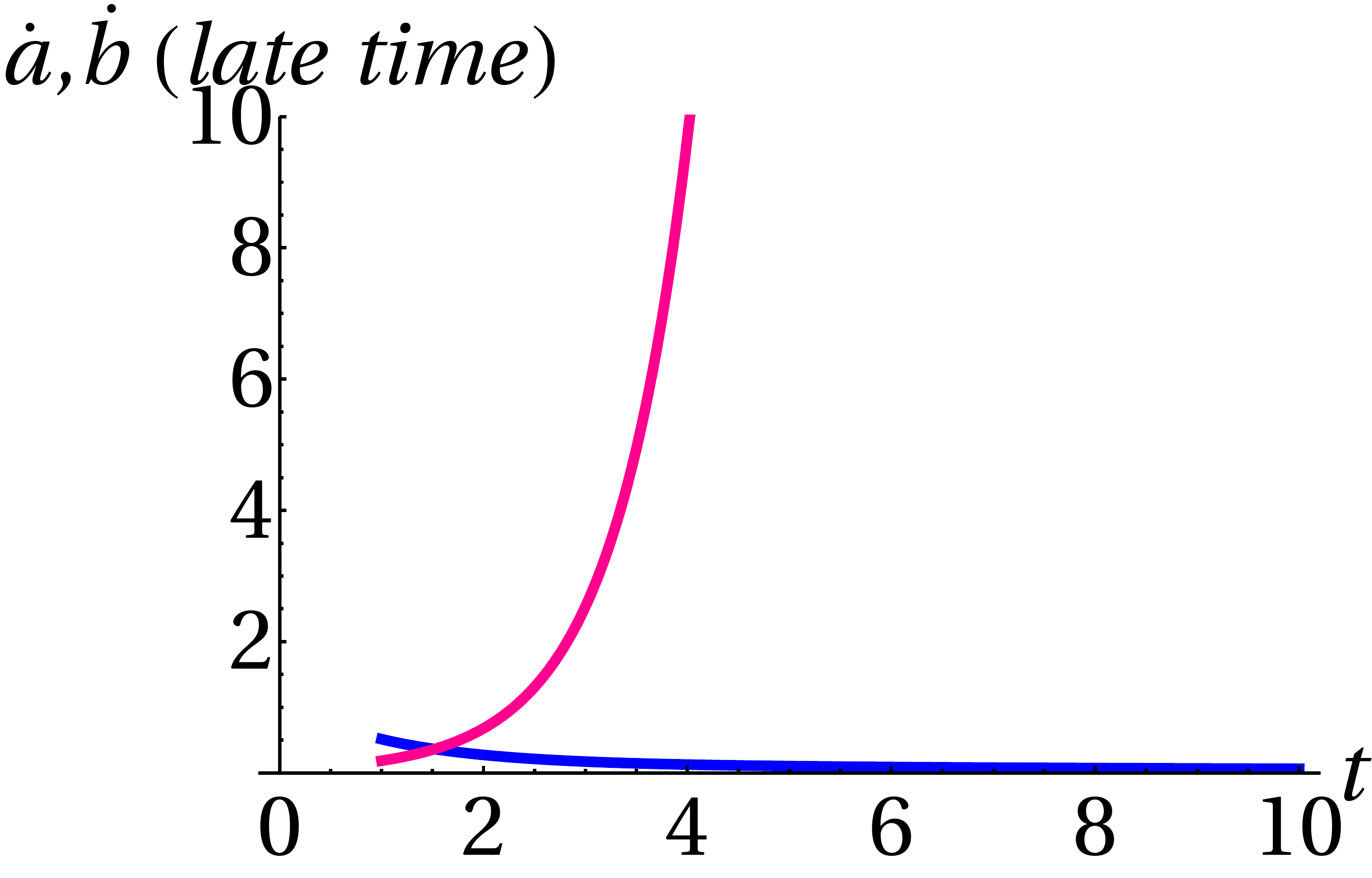}
    \caption{The late time expansion rates of the scale factors: $\dot a$ is represented by the blue curve, and $\dot b$ by the red curve.}
    \label{adotbdotLATE00008constantL}
  \end{subfigure}
\qquad
  \begin{subfigure}[t]{.5\linewidth}
    \centering
    \includegraphics[width=0.7\columnwidth]{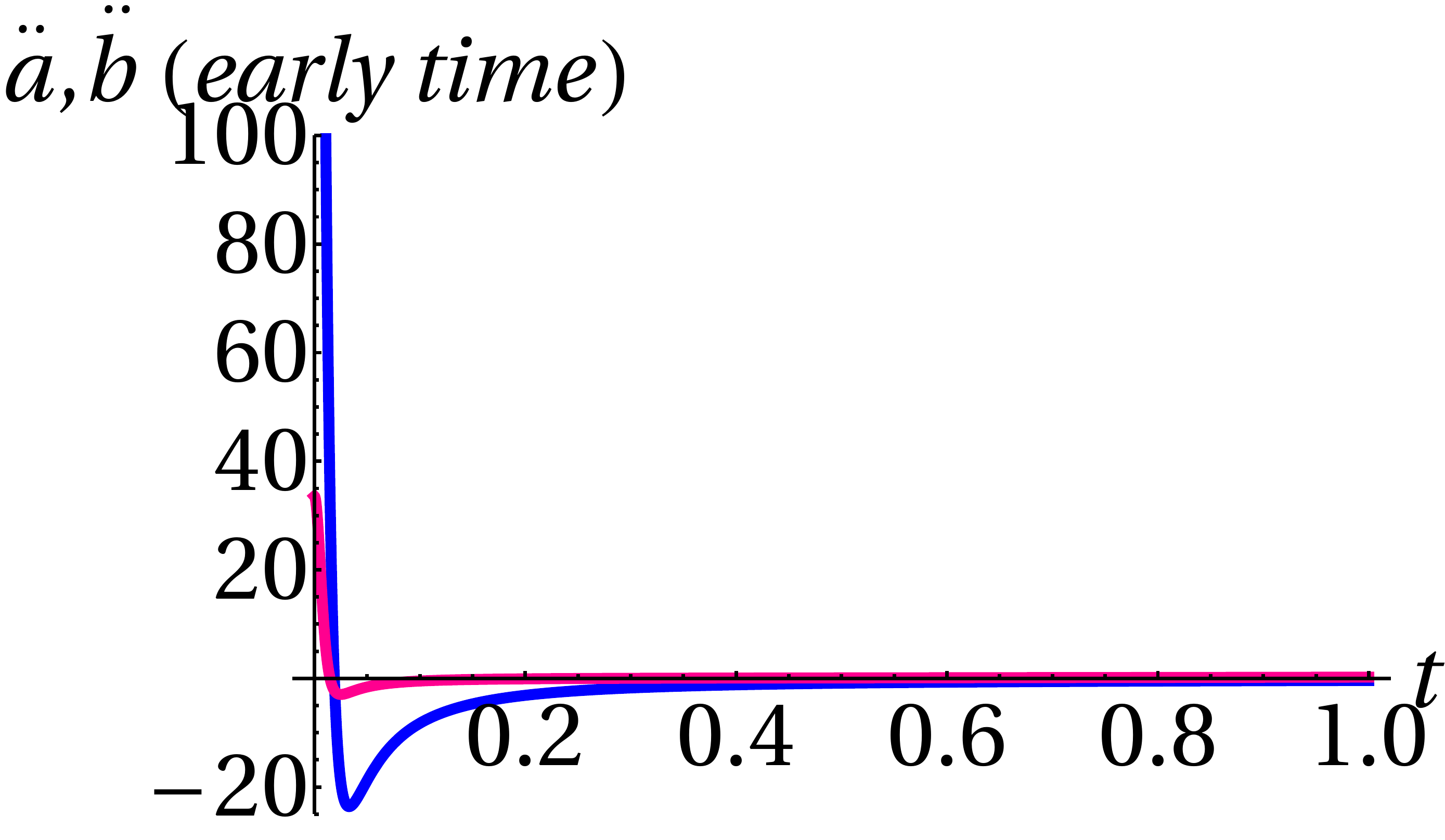}
    \caption{The early time accelerations of the scale factors: $\ddot a$ is represented by the blue curve, and $\ddot b$ by the red curve.}
    \label{addotbddotEARLY00008constantL}
  \end{subfigure}
\caption{Initial conditions set number 8 for constant $\Lambda$.}
  \label{Fig161}
\end{figure}
\begin{figure}[H]
  \begin{subfigure}[t]{.5\linewidth}
    \centering
    \includegraphics[width=0.7\columnwidth]{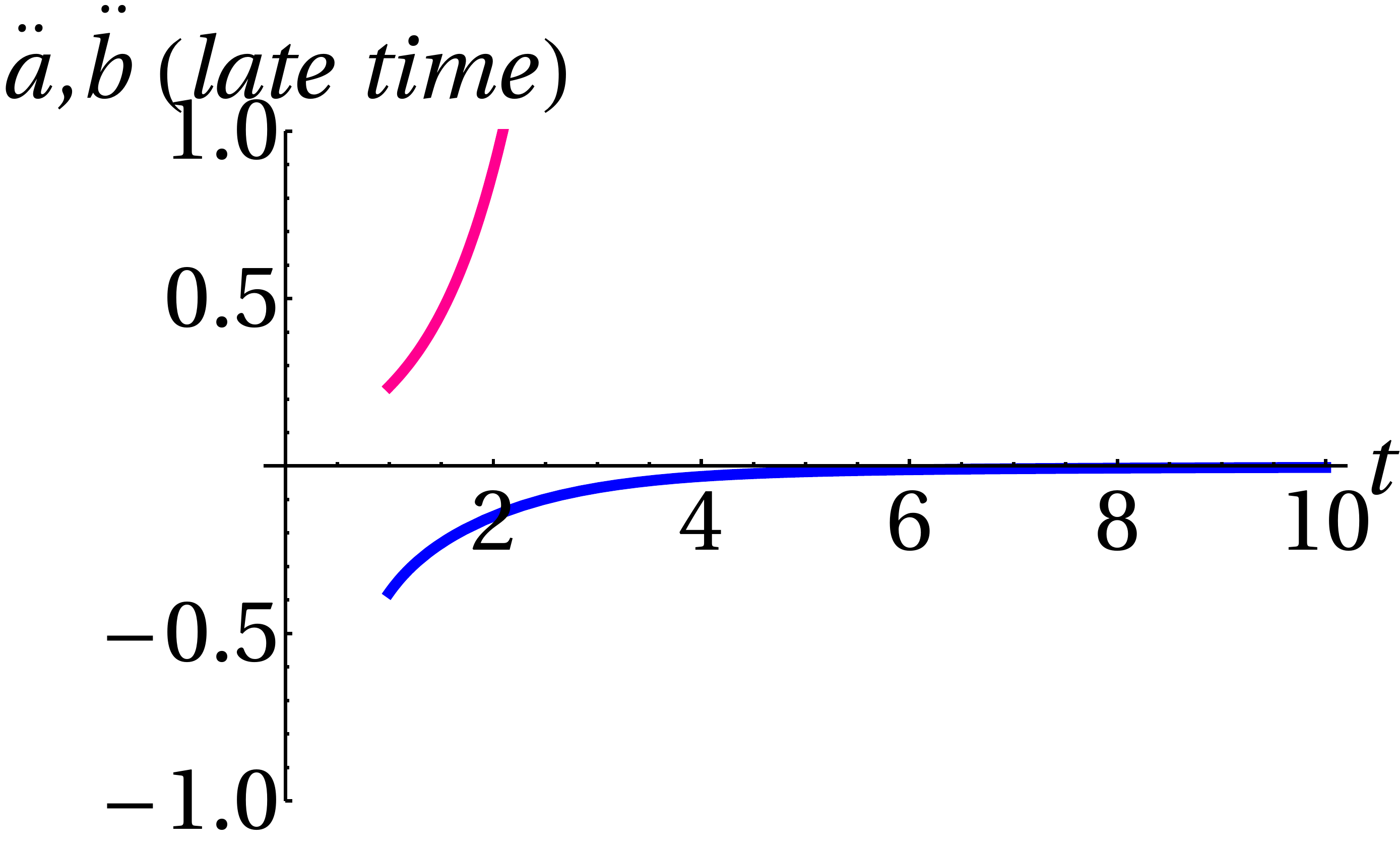}
    \caption{The late time accelerations of the scale factors: $\ddot a$ is represented by the blue curve, and $\ddot b$ by the red curve.}
    \label{addotbddotLATE00008constantL}
  \end{subfigure}
\qquad
  \begin{subfigure}[t]{.5\linewidth}
    \centering
    \includegraphics[width=0.7\columnwidth]{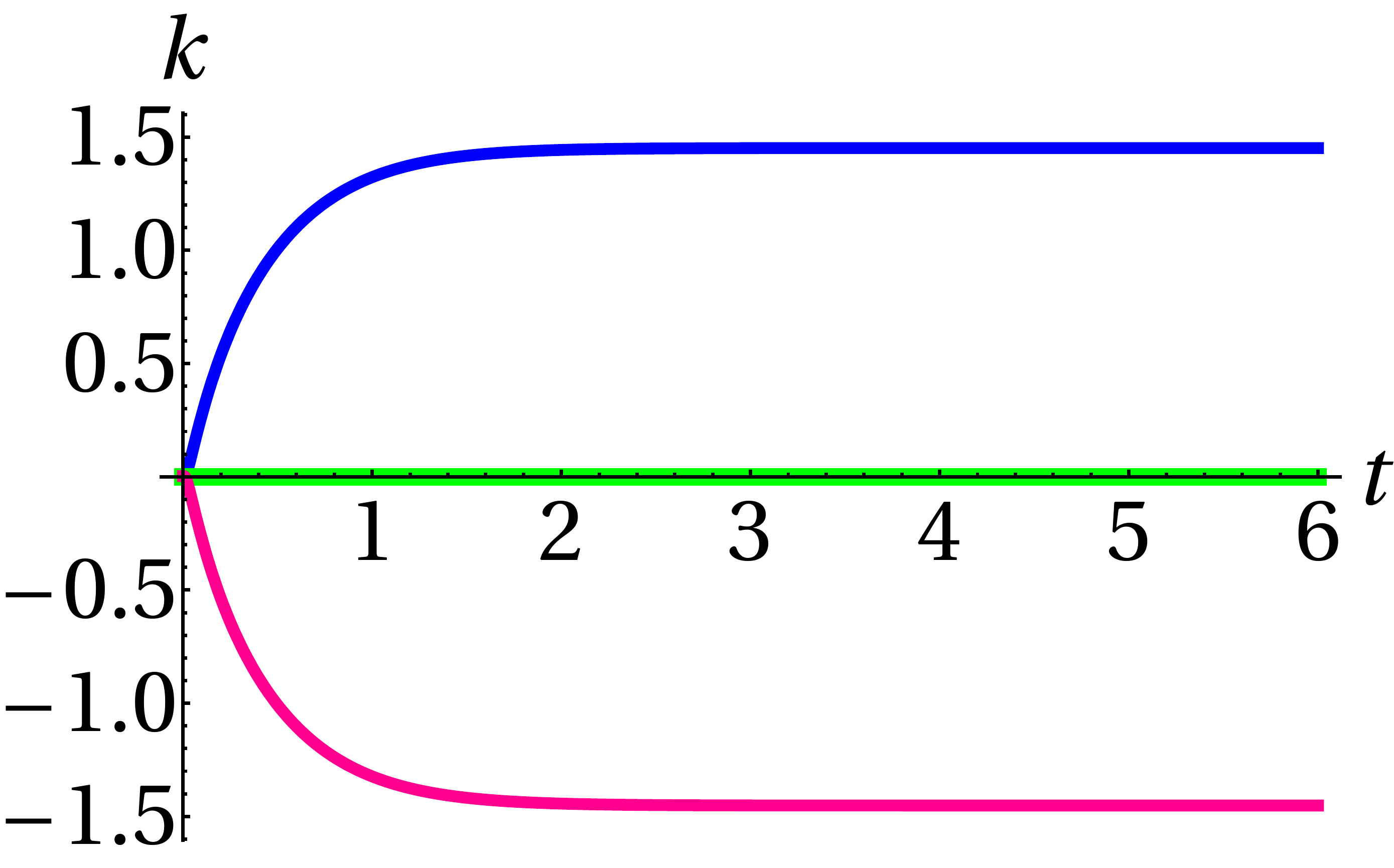}
    \caption{The harmonic function $k$ using: $\dot k\left(0\right)=1$ (blue curve), $\dot k\left(0\right)=0$ (green line), and $\dot k\left(0\right)=-1$ (red curve).}
    \label{k00008constantL}
  \end{subfigure}
\caption{Initial conditions set number 8 for constant $\Lambda$ (continued).}
  \label{Fig16}
\end{figure}
\vspace{-0.1cm}
\begin{figure}[H]
  \begin{subfigure}[t]{.5\linewidth}
    \centering
    \includegraphics[width=0.7\columnwidth]{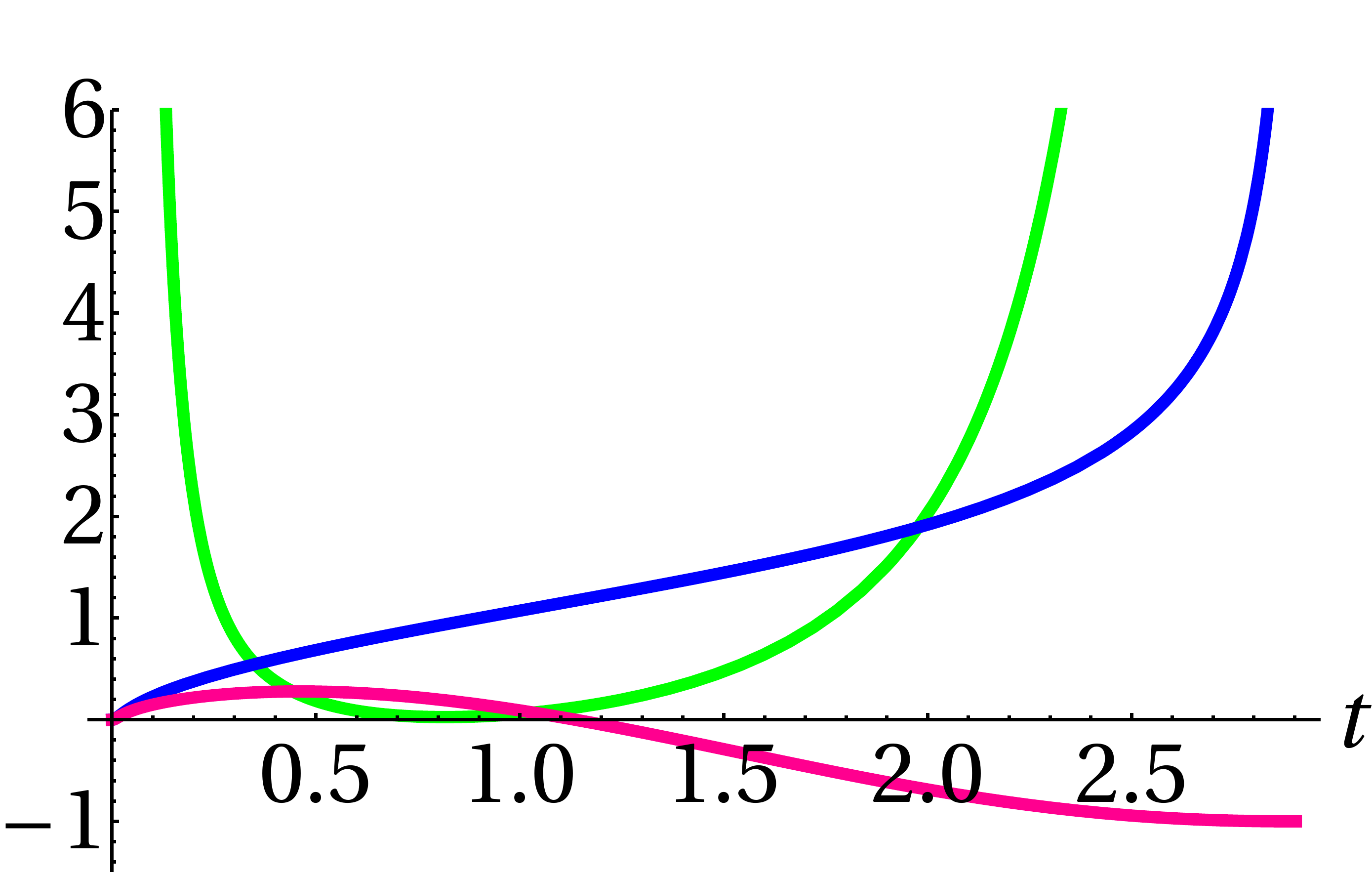}
    \caption{The scale factor $a$ is represented by the blue curve, $b$ by the red curve, while $\left| {G_{i\bar j} \dot z^i \dot z^{\bar j}} \right|$ by the green curve.}
    \label{abzz00009constantL}
  \end{subfigure}
\qquad
  \begin{subfigure}[t]{.5\linewidth}
    \centering
    \includegraphics[width=0.7\columnwidth]{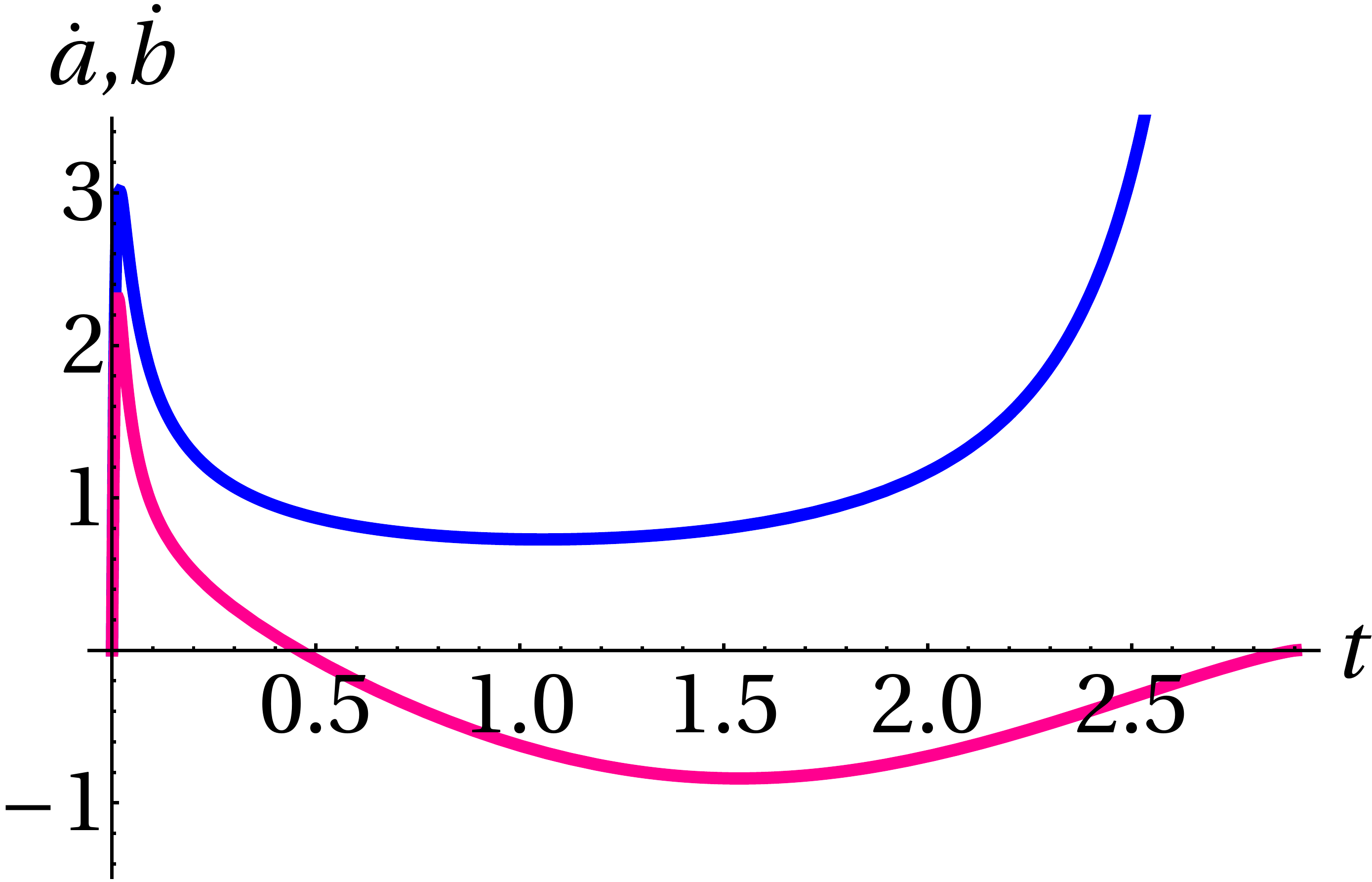}
    \caption{The expansion rates of the scale factors: $\dot a$ is represented by the blue curve, and $\dot b$ by the red curve. The curve for $\dot b$ is scaled up by a factor of 10.}
    \label{adotbdot00009constantL}
  \end{subfigure}
\\[4em]
  \begin{subfigure}[t]{.5\linewidth}
    \centering
    \includegraphics[width=0.7\columnwidth]{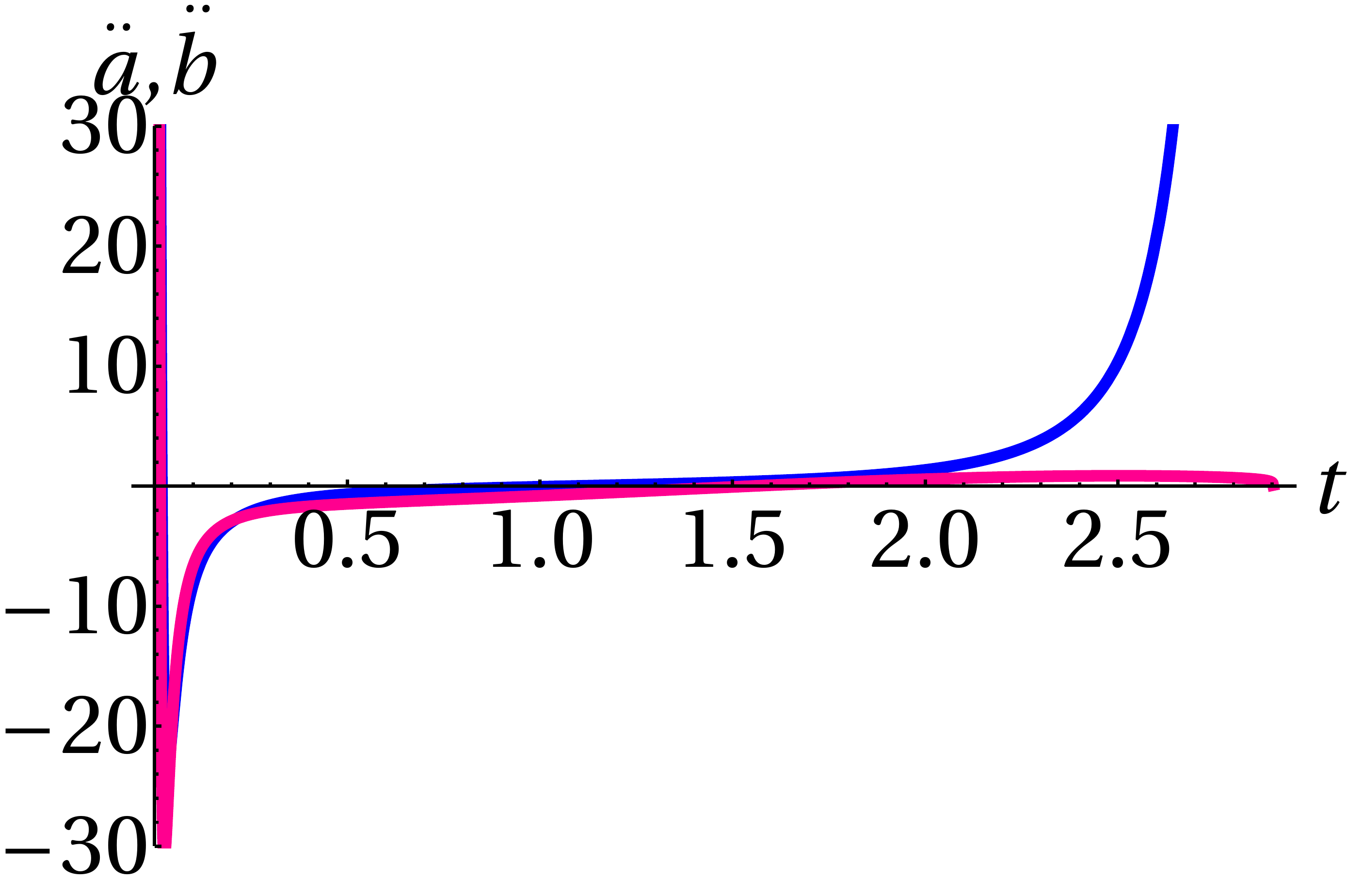}
    \caption{The accelerations of the scale factors: $\ddot a$ is represented by the blue curve, and $\ddot b$ by the red curve. The curve for $\ddot b$ is scaled up by a factor of 10.}
    \label{addotbddot00009constantL}
  \end{subfigure}
\qquad
  \begin{subfigure}[t]{.5\linewidth}
    \centering
    \includegraphics[width=0.7\columnwidth]{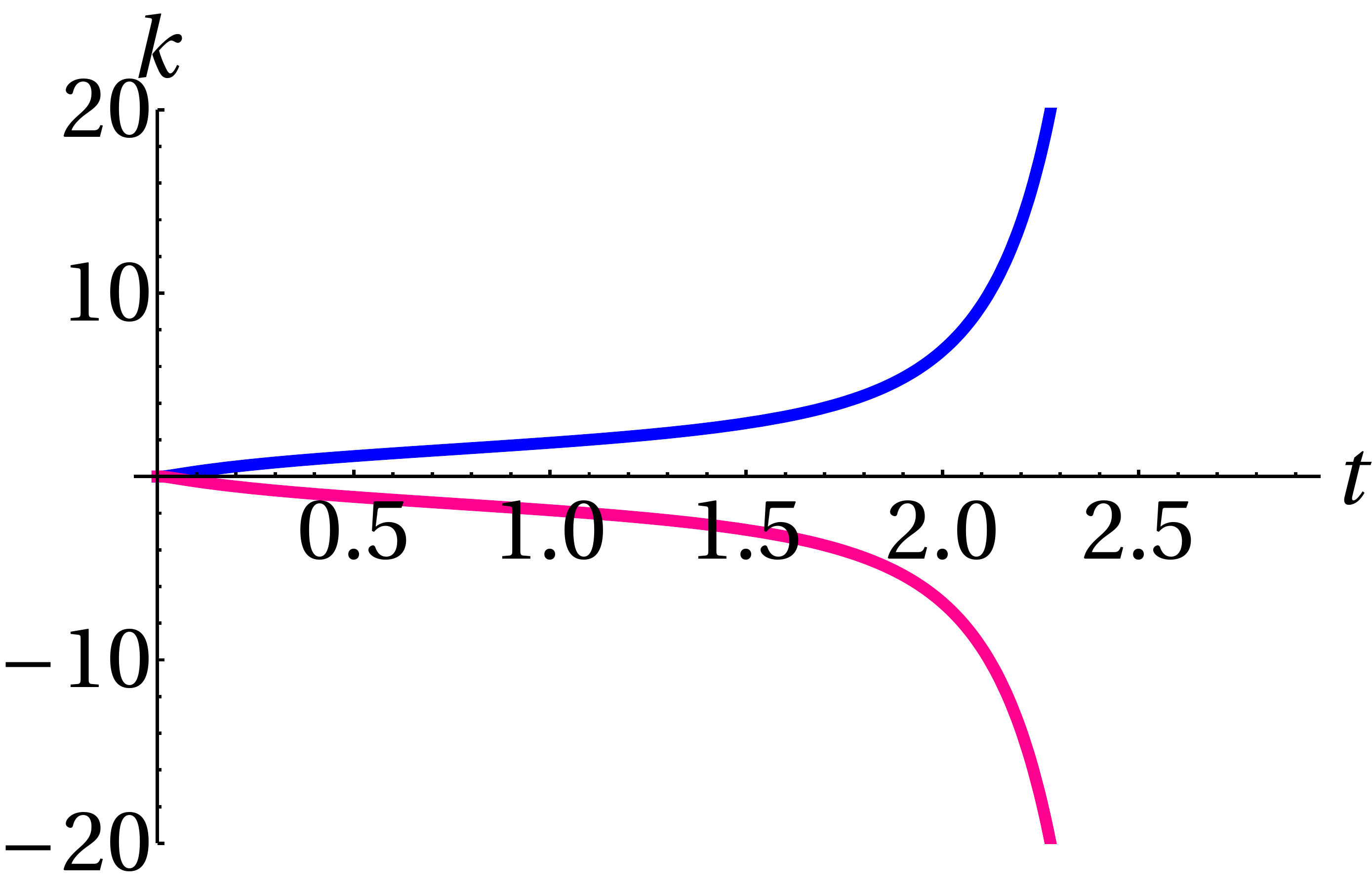}
    \caption{The harmonic function $k$ using: $\dot k\left(0\right)=1$ (blue curve), and $\dot k\left(0\right)=-1$ (red curve). While $\dot k\left(0\right)=0$ diverges.}
    \label{k00009constantL}
  \end{subfigure}
\caption{Initial conditions set number 9 for constant $\Lambda$.}
  \label{Fig17}
  \end{figure}


\begin{figure}[H]
  \begin{subfigure}[t]{.5\linewidth}
    \centering
    \includegraphics[width=0.7\columnwidth]{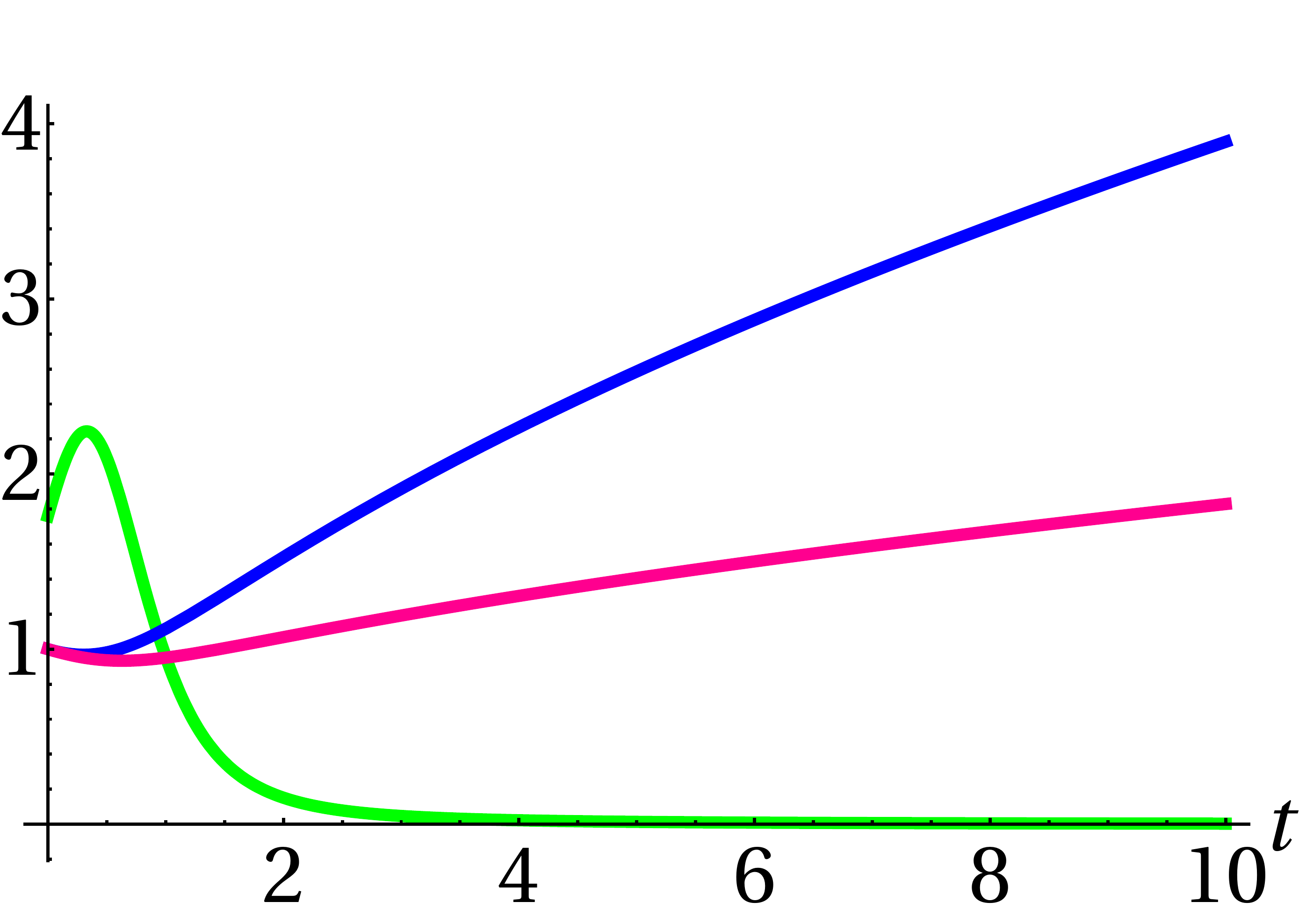}
    \caption{The scale factor $a$ is represented by the blue curve, $b$ by the red curve, while $\left| {G_{i\bar j} \dot z^i \dot z^{\bar j}} \right|$ by the green curve.}
    \label{abzz1-02-1-021constantL}
  \end{subfigure}
\qquad
  \begin{subfigure}[t]{.5\linewidth}
    \centering
    \includegraphics[width=0.7\columnwidth]{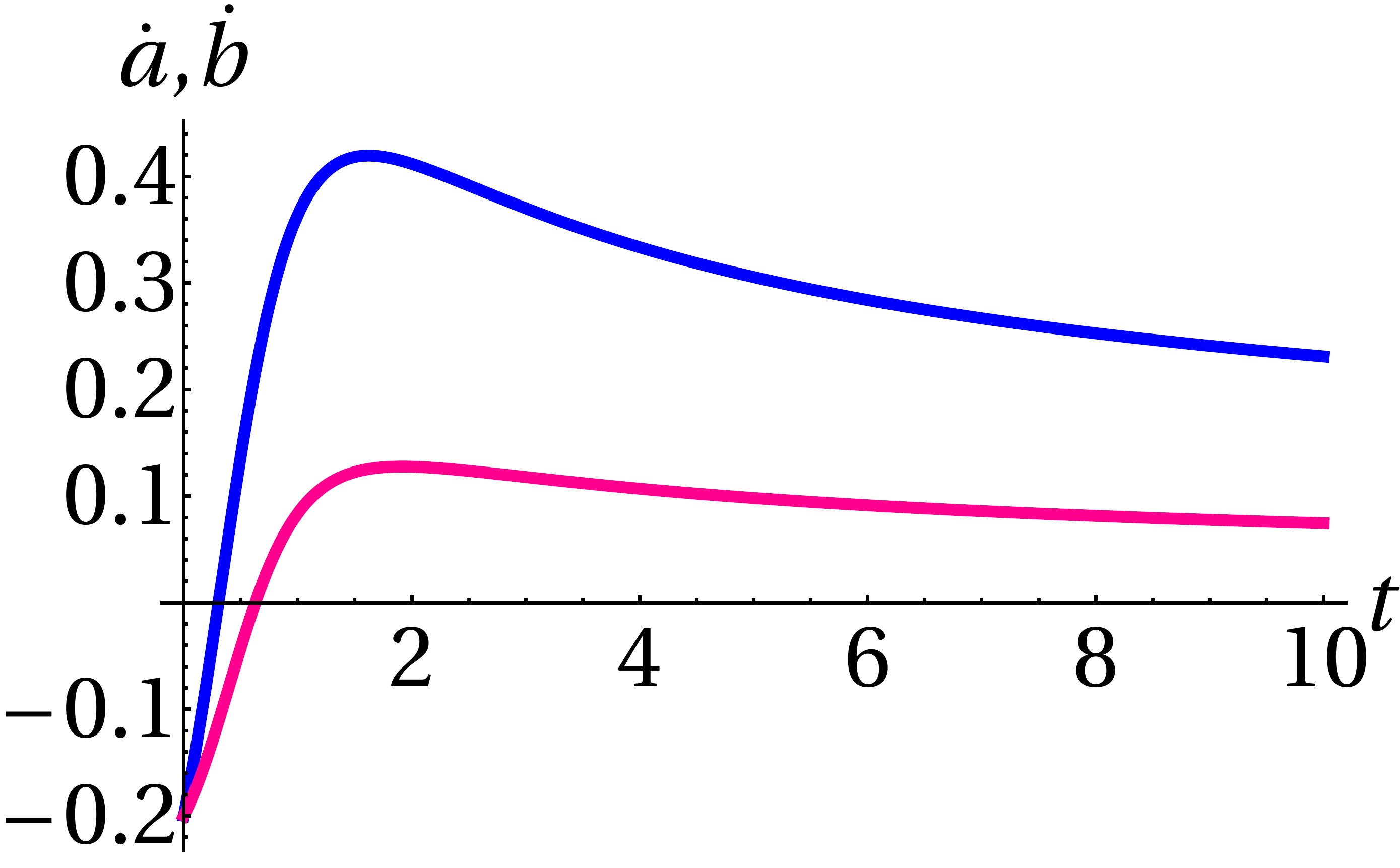}
    \caption{The expansion rates of the scale factors: $\dot a$ is represented by the blue curve, and $\dot b$ by the red curve.}
    \label{adotbdot1-02-1-021constantL}
  \end{subfigure}
\\[9em]
  \begin{subfigure}[t]{.5\linewidth}
    \centering
    \includegraphics[width=0.7\columnwidth]{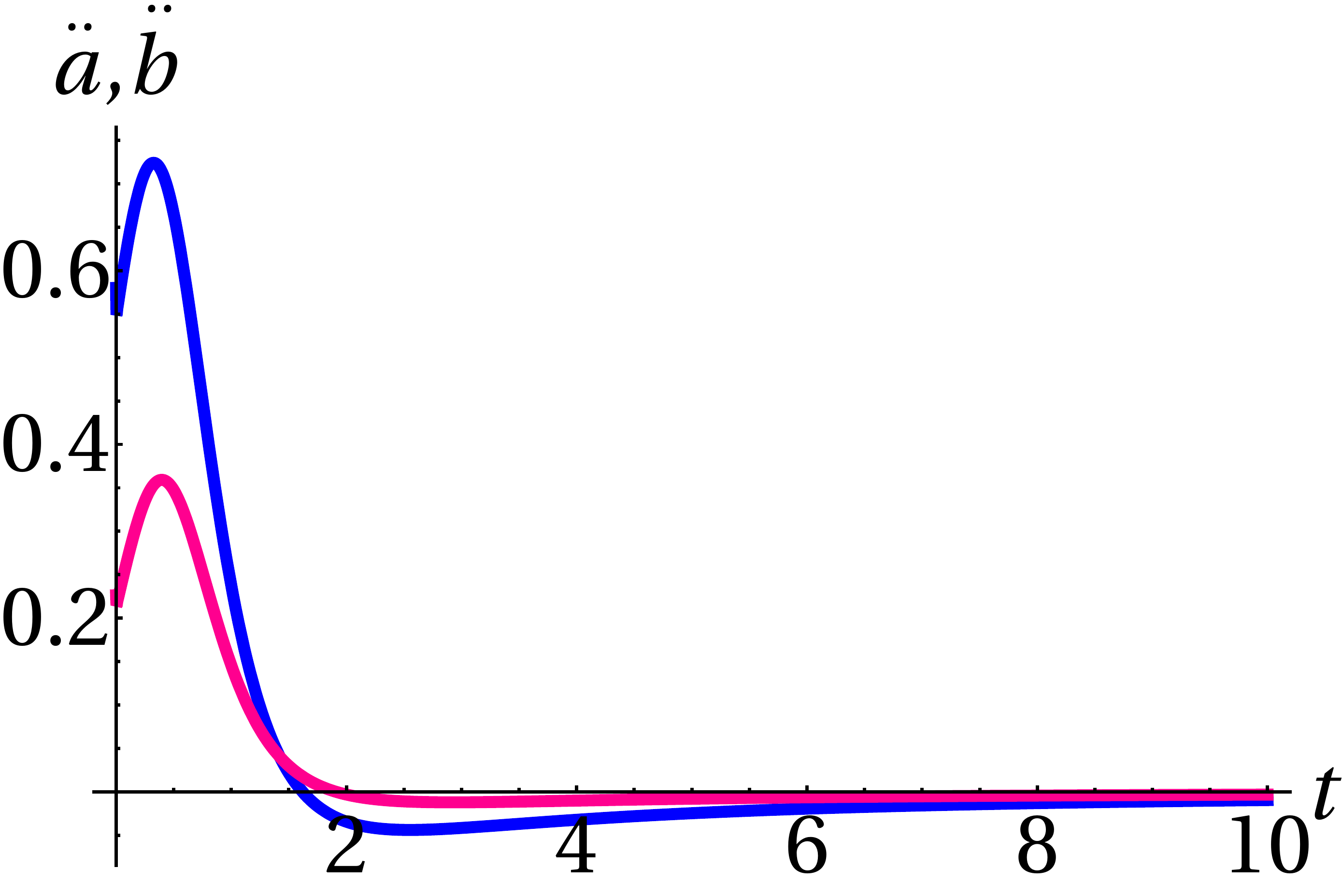}
    \caption{The accelerations of the scale factors: $\ddot a$ is represented by the blue curve, and $\ddot b$ by the red curve.}
    \label{addotbddot1-02-1-021constantL}
  \end{subfigure}
\qquad
  \begin{subfigure}[t]{.5\linewidth}
    \centering
    \includegraphics[width=0.7\columnwidth]{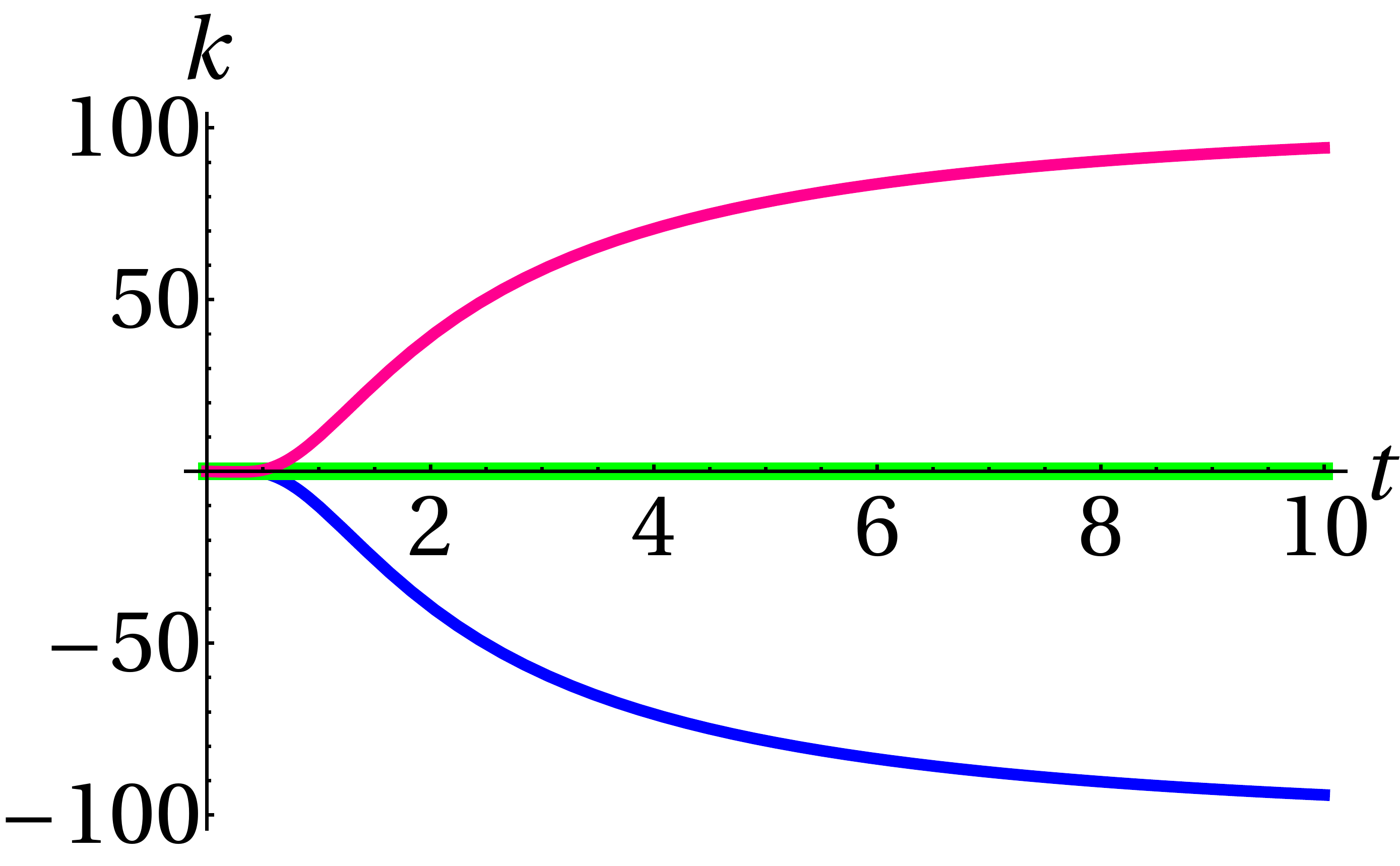}
    \caption{The harmonic function $k$ using: $\dot k\left(0\right)=1$ (blue curve), $\dot k\left(0\right)=0$ (green line), and $\dot k\left(0\right)=-1$ (red curve).}
    \label{k1-02-1-021constantL}
  \end{subfigure}
    \caption{Initial conditions set number 10 for constant $\Lambda$.}
  \label{Fig19}
  \end{figure}
%


\begin{figure}[H]
  \begin{subfigure}[t]{.5\linewidth}
    \centering
    \includegraphics[width=0.7\columnwidth]{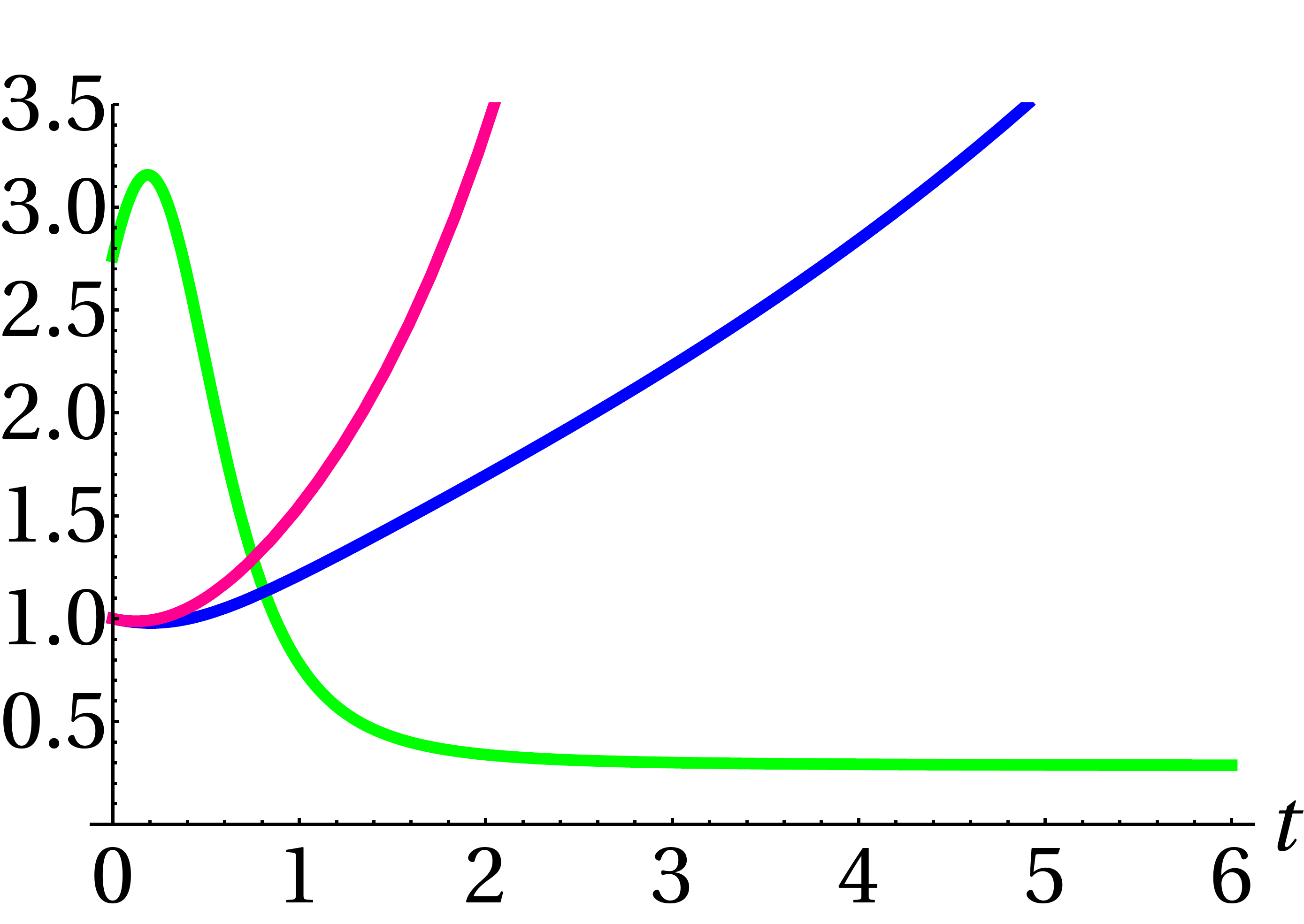}
    \caption{The scale factor $a$ is represented by the blue curve, $b$ by the red curve, while $\left| {G_{i\bar j} \dot z^i \dot z^{\bar j}} \right|$ by the green curve.}
    \label{abzz1-02-1-022constantL}
  \end{subfigure}
\qquad
  \begin{subfigure}[t]{.5\linewidth}
    \centering
    \includegraphics[width=0.7\columnwidth]{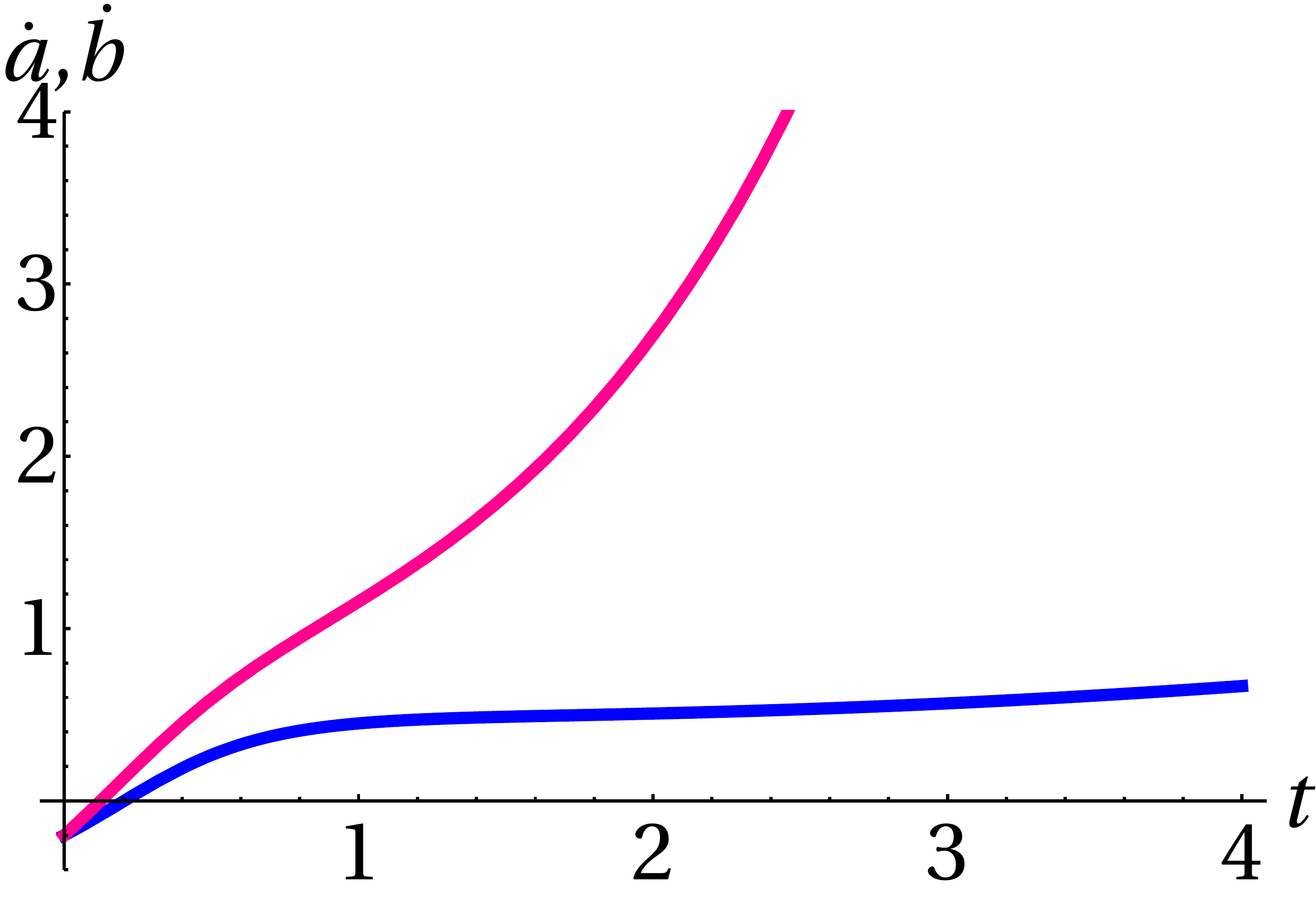}
    \caption{The expansion rates of the scale factors: $\dot a$ is represented by the blue curve, and $\dot b$ by the red curve.}
    \label{adotbdot1-02-1-022constantL}
  \end{subfigure}
\\[9em]
  \begin{subfigure}[t]{.5\linewidth}
    \centering
    \includegraphics[width=0.7\columnwidth]{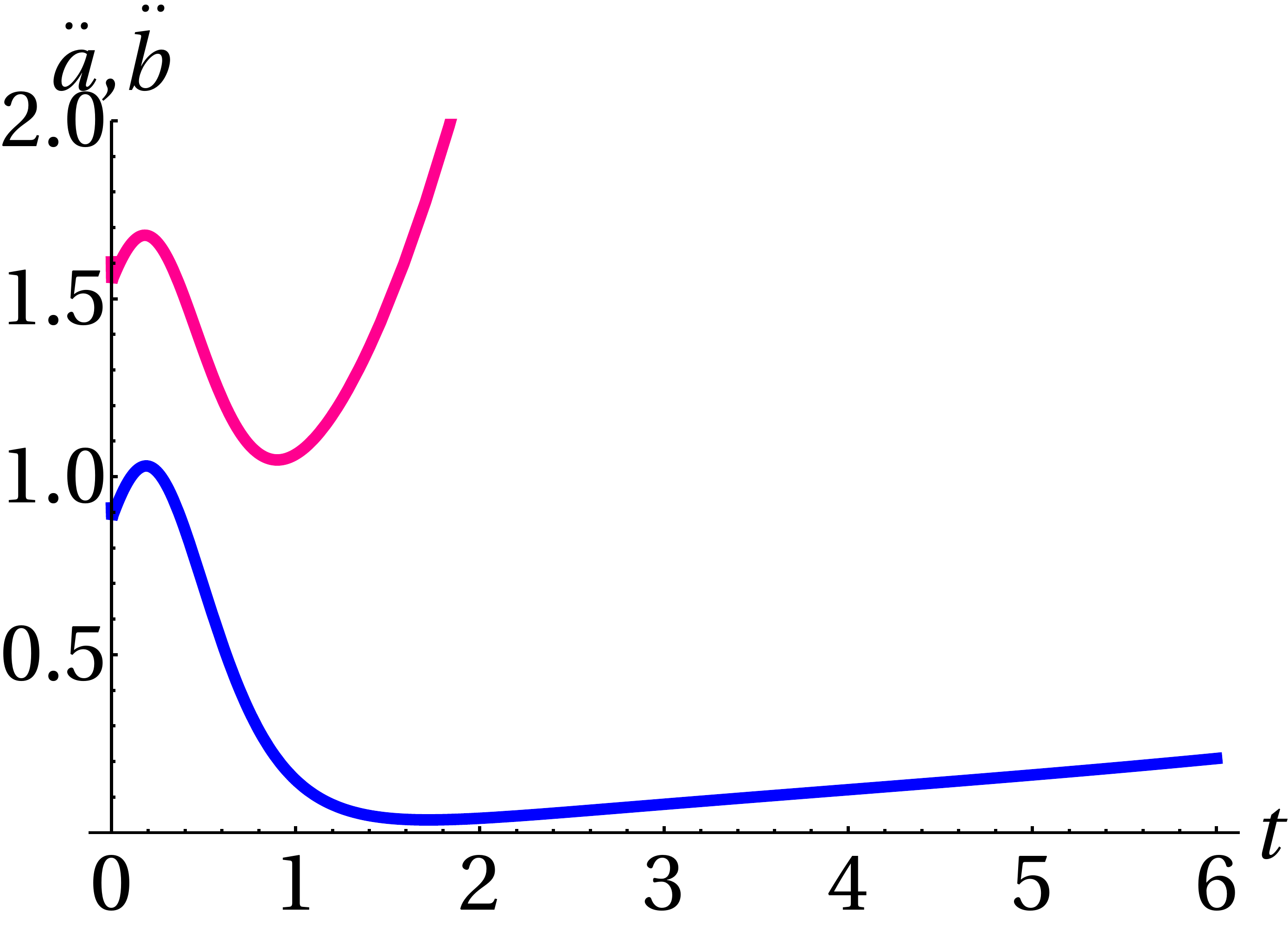}
    \caption{The accelerations of the scale factors: $\ddot a$ is represented by the blue curve, and $\ddot b$ by the red curve.}
    \label{addotbddot1-02-1-022constantL}
  \end{subfigure}
\qquad
  \begin{subfigure}[t]{.5\linewidth}
    \centering
    \includegraphics[width=0.7\columnwidth]{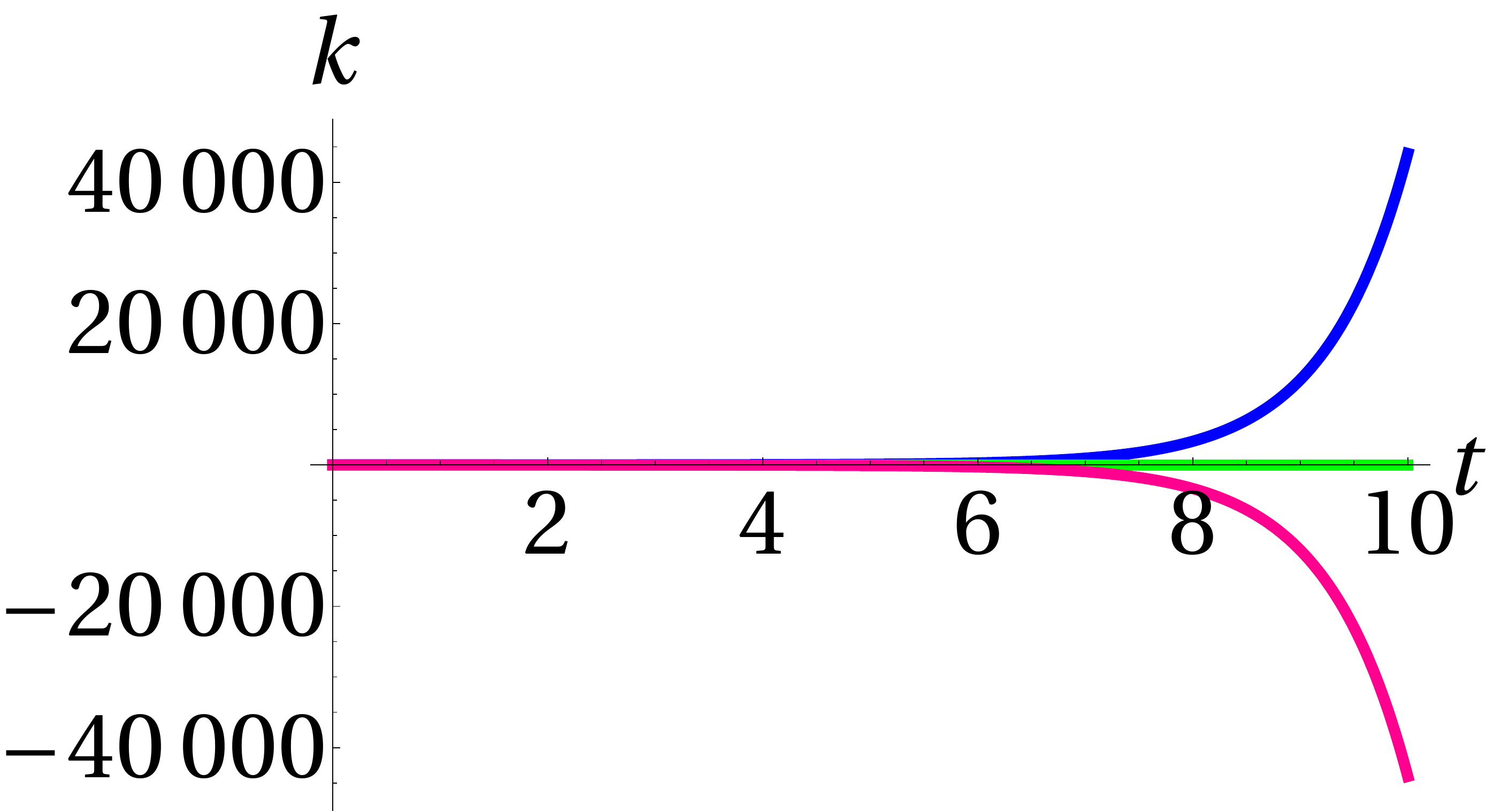}
    \caption{The harmonic function $k$ using: $\dot k\left(0\right)=1$ (blue curve), $\dot k\left(0\right)=0$ (green line), and $\dot k\left(0\right)=-1$ (red curve).}
    \label{k1-02-1-022constantL}
  \end{subfigure}
  \caption{Initial conditions set number 11 for constant $\Lambda$.}
  \label{Fig22}
\end{figure}

\begin{figure}[H]
  \begin{subfigure}[t]{.5\linewidth}
    \centering
    \includegraphics[width=0.7\columnwidth]{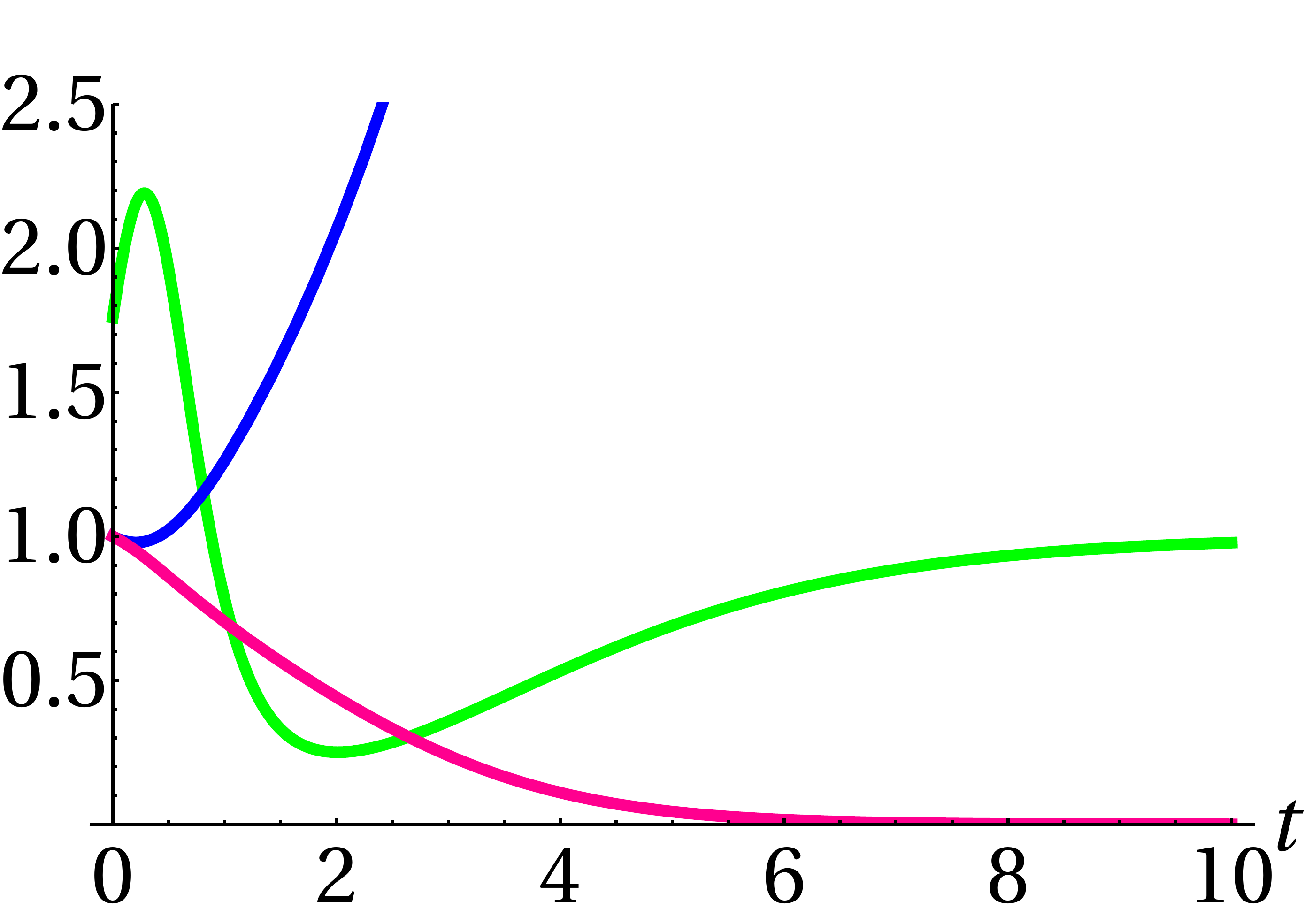}
    \caption{The scale factor $a$ is represented by the blue curve, $b$ by the red curve, while $\left| {G_{i\bar j} \dot z^i \dot z^{\bar j}} \right|$ by the green curve.}
    \label{abzz1-02-1-023constantL}
  \end{subfigure}
\qquad
  \begin{subfigure}[t]{.5\linewidth}
    \centering
    \includegraphics[width=0.7\columnwidth]{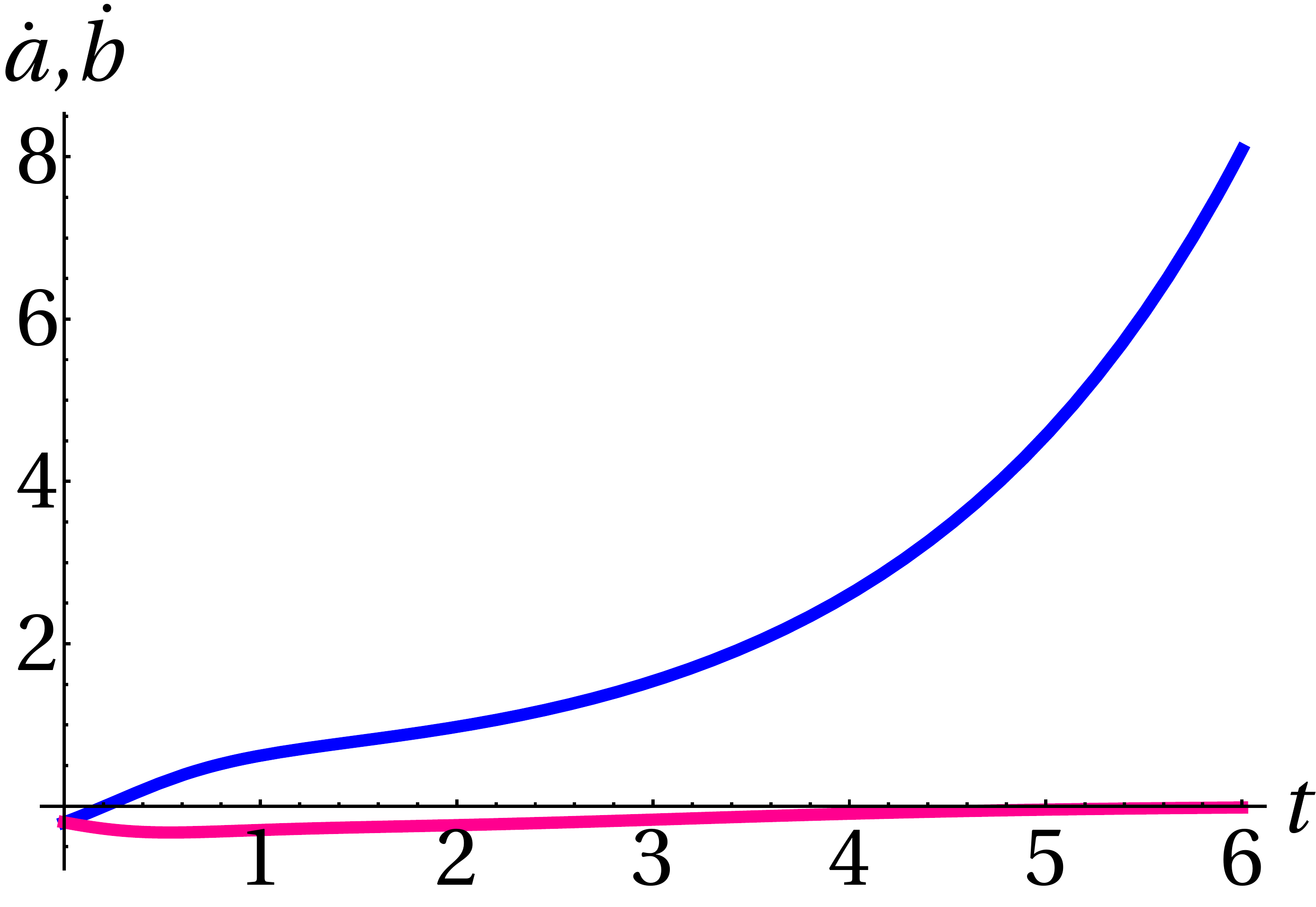}
    \caption{The expansion rates of the scale factors: $\dot a$ is represented by the blue curve, and $\dot b$ by the red curve.}
    \label{adotbdot1-02-1-023constantL}
  \end{subfigure}
\\[9em]
  \begin{subfigure}[t]{.5\linewidth}
    \centering
    \includegraphics[width=0.7\columnwidth]{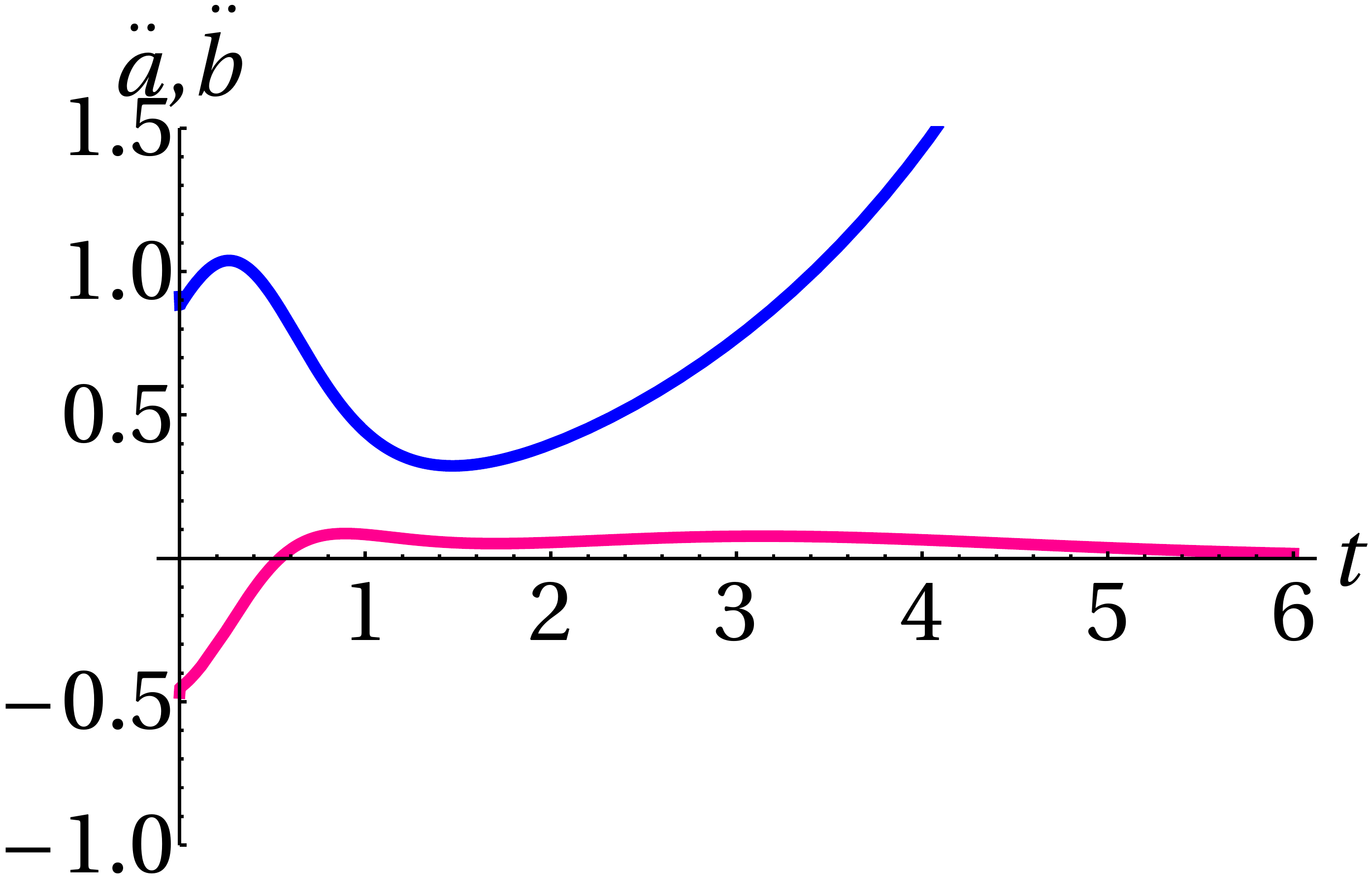}
    \caption{The accelerations of the scale factors: $\ddot a$ is represented by the blue curve, and $\ddot b$ by the red curve.}
    \label{addotbddot1-02-1-023constantL}
  \end{subfigure}
\qquad
  \begin{subfigure}[t]{.5\linewidth}
    \centering
    \includegraphics[width=0.7\columnwidth]{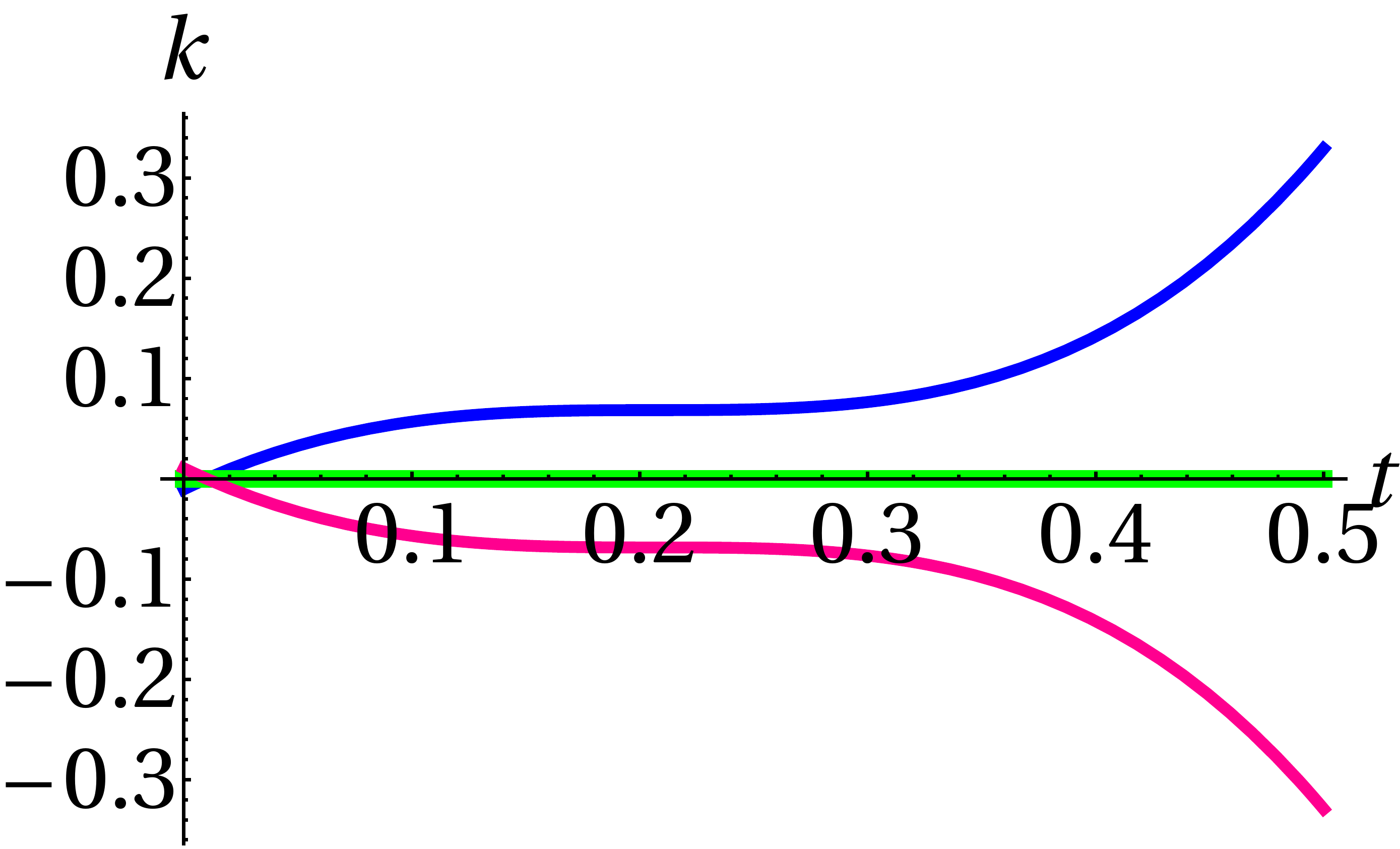}
    \caption{The harmonic function $k$ using: $\dot k\left(0\right)=1$ (blue curve), $\dot k\left(0\right)=0$ (green line), and $\dot k\left(0\right)=-1$ (red curve).}
    \label{k1-02-1-023constantL}
  \end{subfigure}
 \caption{Initial conditions set number 12 for constant $\Lambda$.}
  \label{Fig24}
\end{figure}


\begin{figure}[H]
  \begin{subfigure}[t]{.5\linewidth}
    \centering
    \includegraphics[width=0.7\columnwidth]{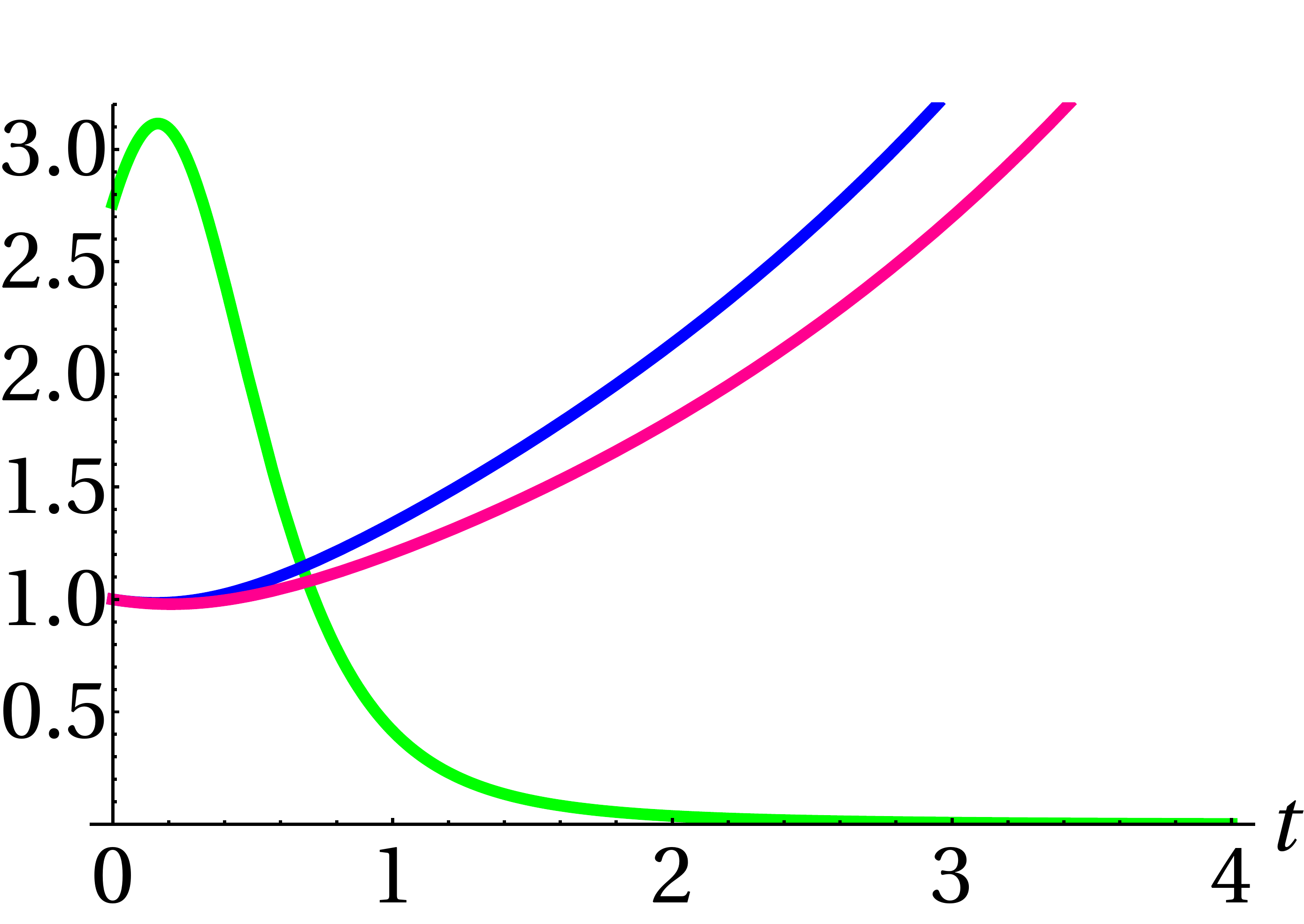}
    \caption{The scale factor $a$ is represented by the blue curve, $b$ by the red curve, while $\left| {G_{i\bar j} \dot z^i \dot z^{\bar j}} \right|$ by the green curve.}
    \label{abzz1-02-1-024constantL}
  \end{subfigure}
\qquad
  \begin{subfigure}[t]{.5\linewidth}
    \centering
    \includegraphics[width=0.7\columnwidth]{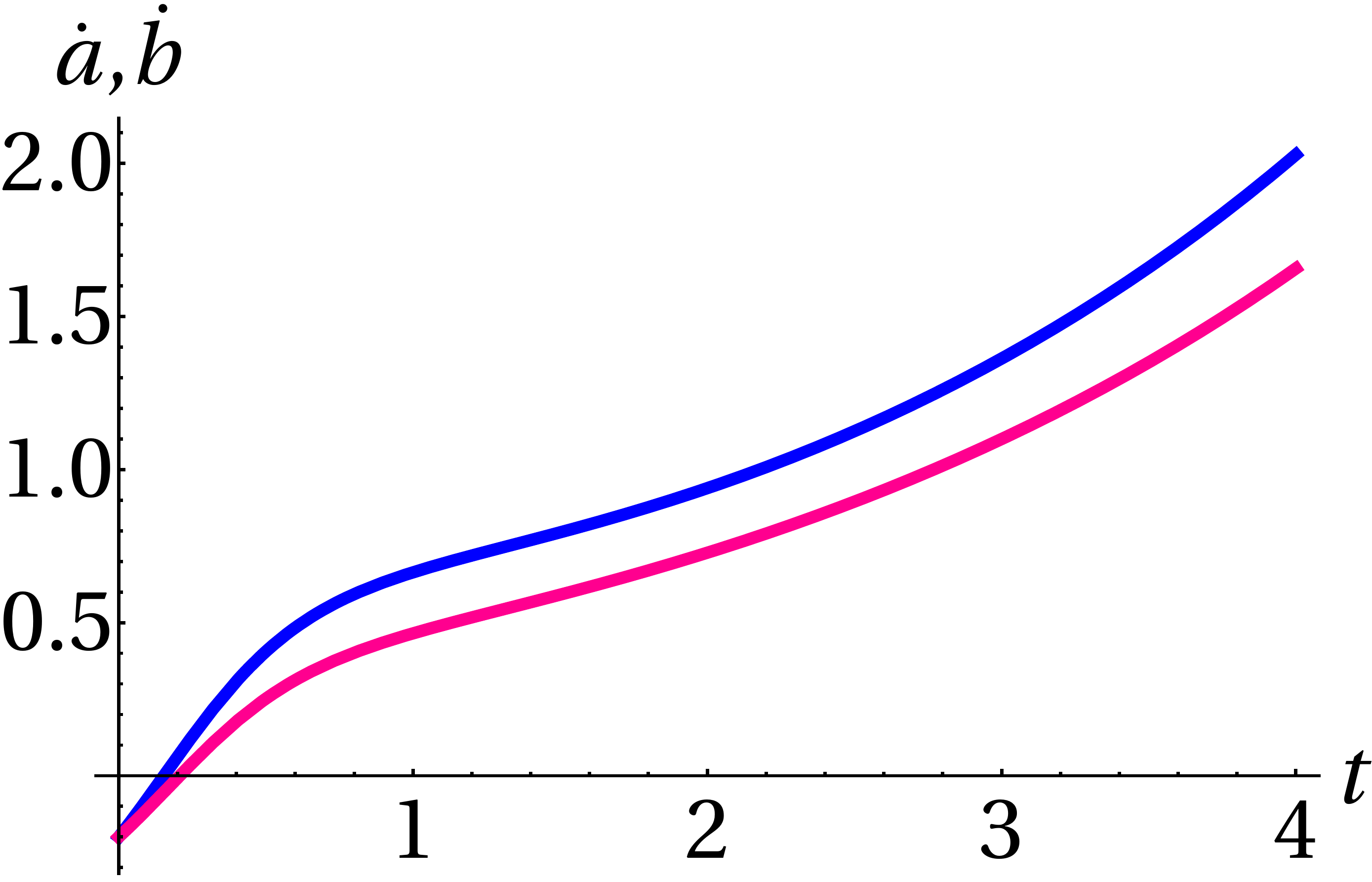}
    \caption{The expansion rates of the scale factors: $\dot a$ is represented by the blue curve, and $\dot b$ by the red curve.}
    \label{adotbdot1-02-1-024constantL}
  \end{subfigure}
\\[9em]
  \begin{subfigure}[t]{.5\linewidth}
    \centering
    \includegraphics[width=0.7\columnwidth]{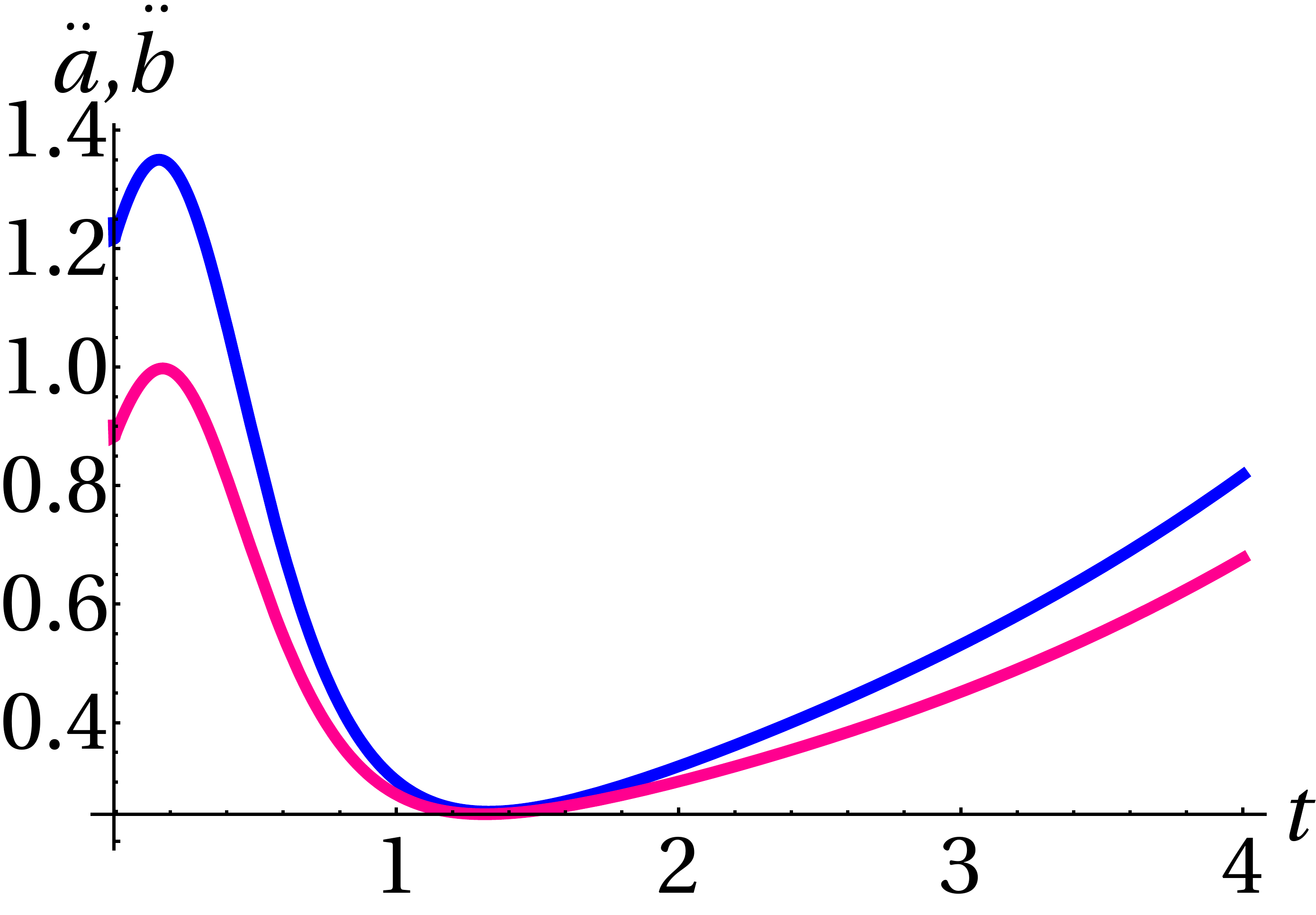}
    \caption{The accelerations of the scale factors: $\ddot a$ is represented by the blue curve, and $\ddot b$ by the red curve.}
    \label{addotbddot1-02-1-024constantL}
  \end{subfigure}
\qquad
  \begin{subfigure}[t]{.5\linewidth}
    \centering
    \includegraphics[width=0.7\columnwidth]{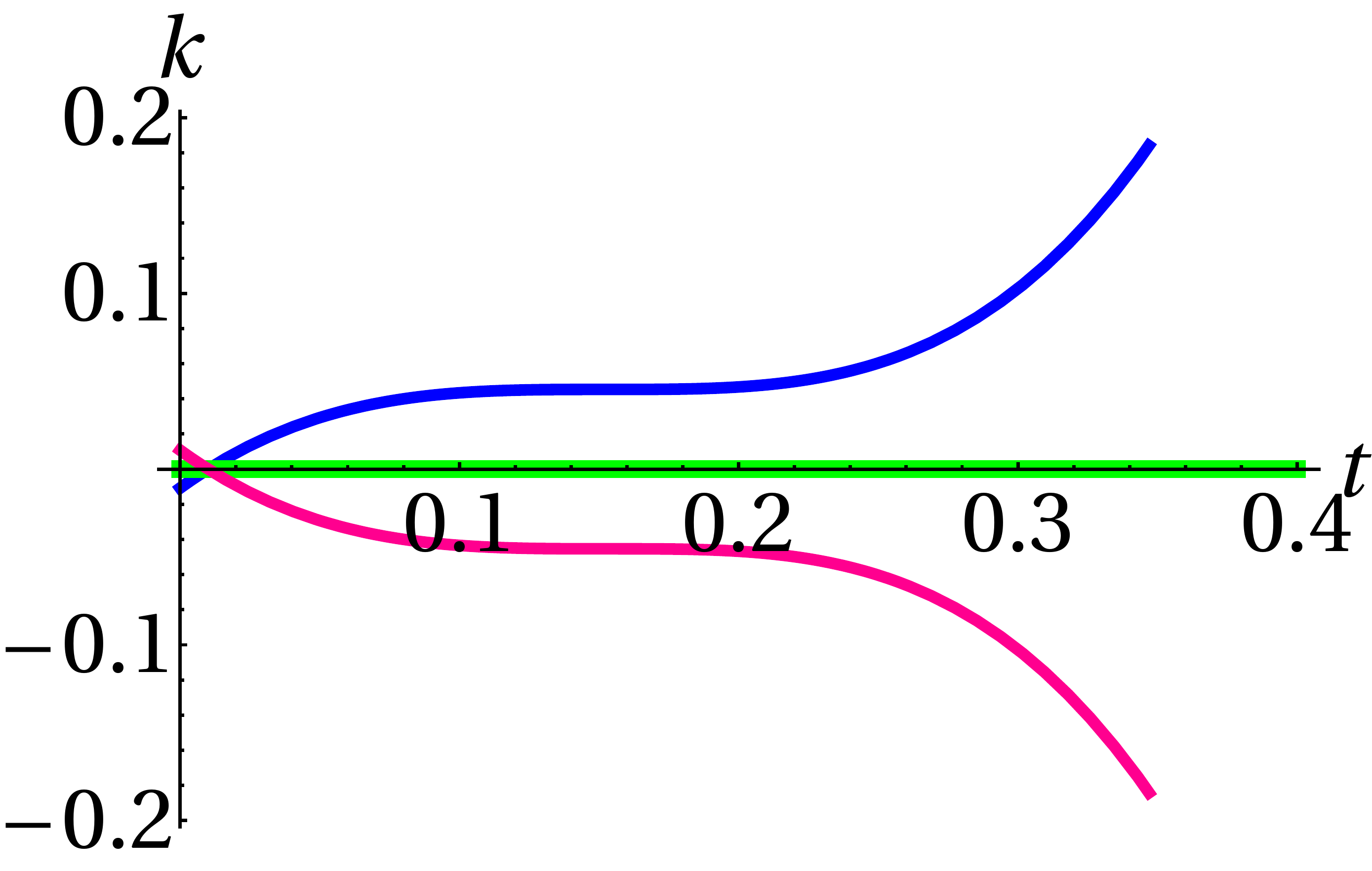}
    \caption{The harmonic function $k$ using: $\dot k\left(0\right)=1$ (blue curve), $\dot k\left(0\right)=0$ (green line), and $\dot k\left(0\right)=-1$ (red curve).}
    \label{k1-02-1-024constantL}
  \end{subfigure}
 \caption{Initial conditions set number 13 for constant $\Lambda$.}
  \label{Fig26}
\end{figure}


\begin{figure}[H]
  \begin{subfigure}[t]{.5\linewidth}
    \centering
    \includegraphics[width=0.7\columnwidth]{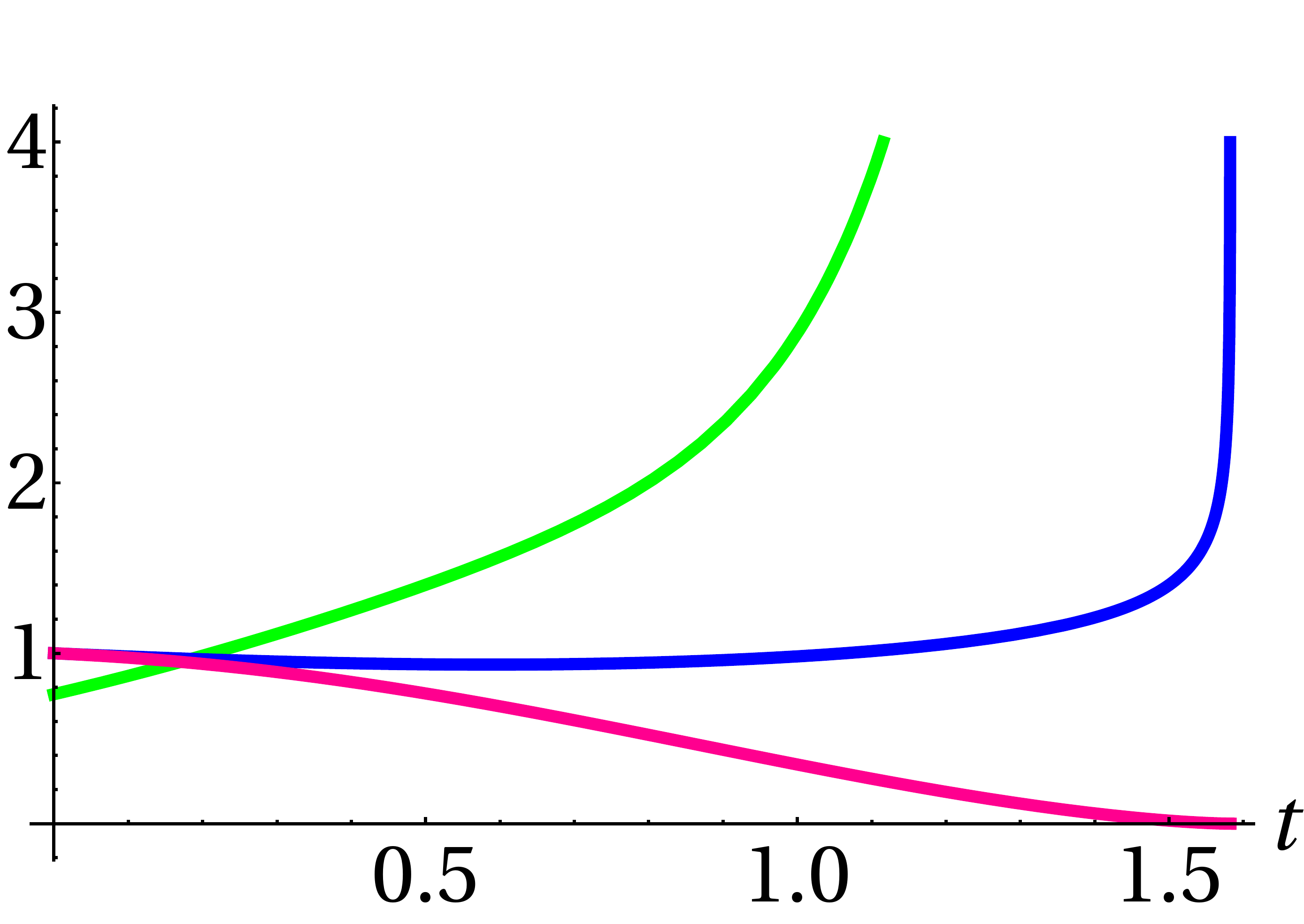}
    \caption{The scale factor $a$ is represented by the blue curve, $b$ by the red curve, while $\left| {G_{i\bar j} \dot z^i \dot z^{\bar j}} \right|$ by the green curve.}
    \label{abzz1-02-1-025constantL}
  \end{subfigure}
\qquad
  \begin{subfigure}[t]{.5\linewidth}
    \centering
    \includegraphics[width=0.7\columnwidth]{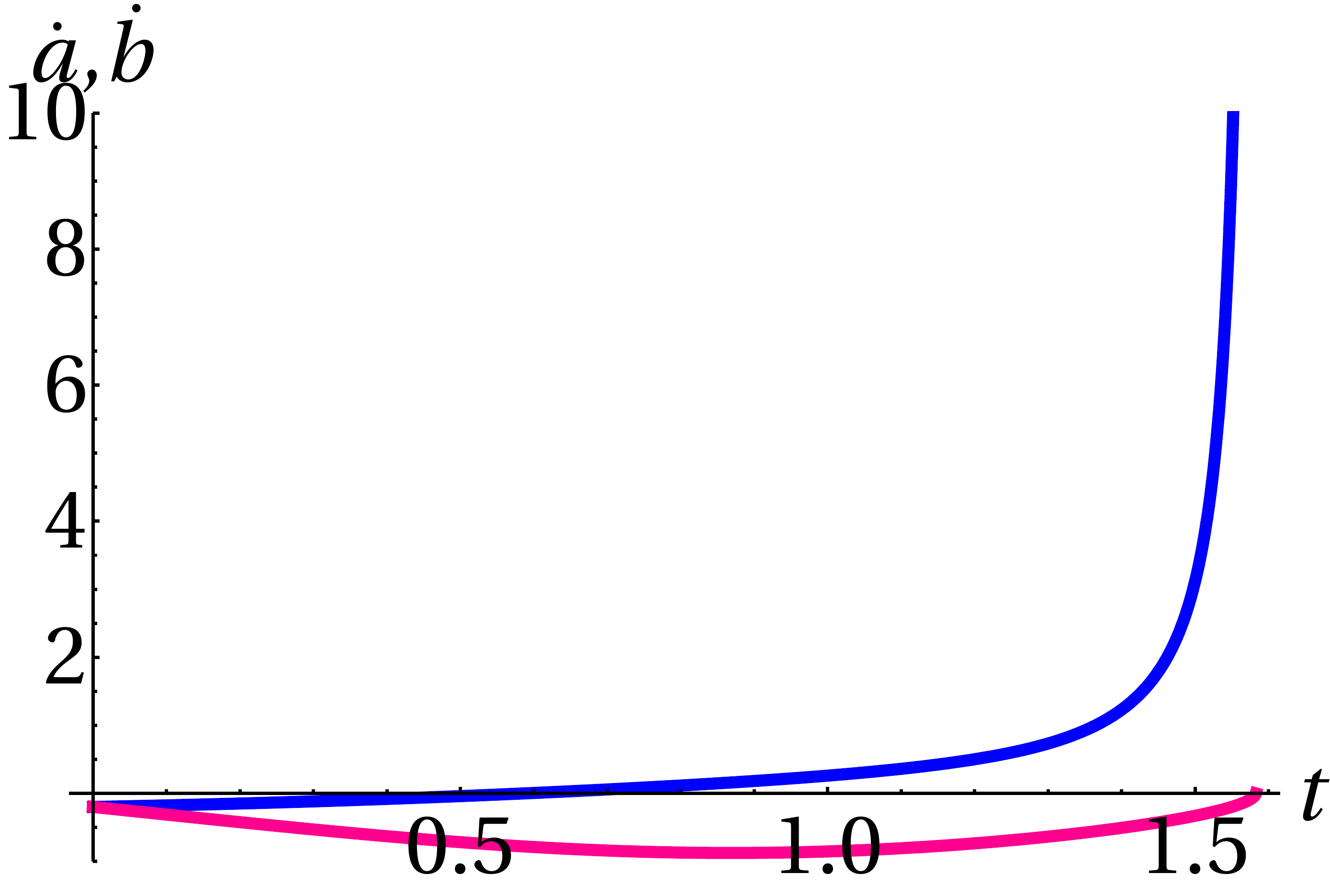}
    \caption{The expansion rates of the scale factors: $\dot a$ is represented by the blue curve, and $\dot b$ by the red curve.}
    \label{adotbdotEARLY1-02-1-025constantL}
  \end{subfigure}
\\[9em]
  \begin{subfigure}[t]{.5\linewidth}
    \centering
    \includegraphics[width=0.7\columnwidth]{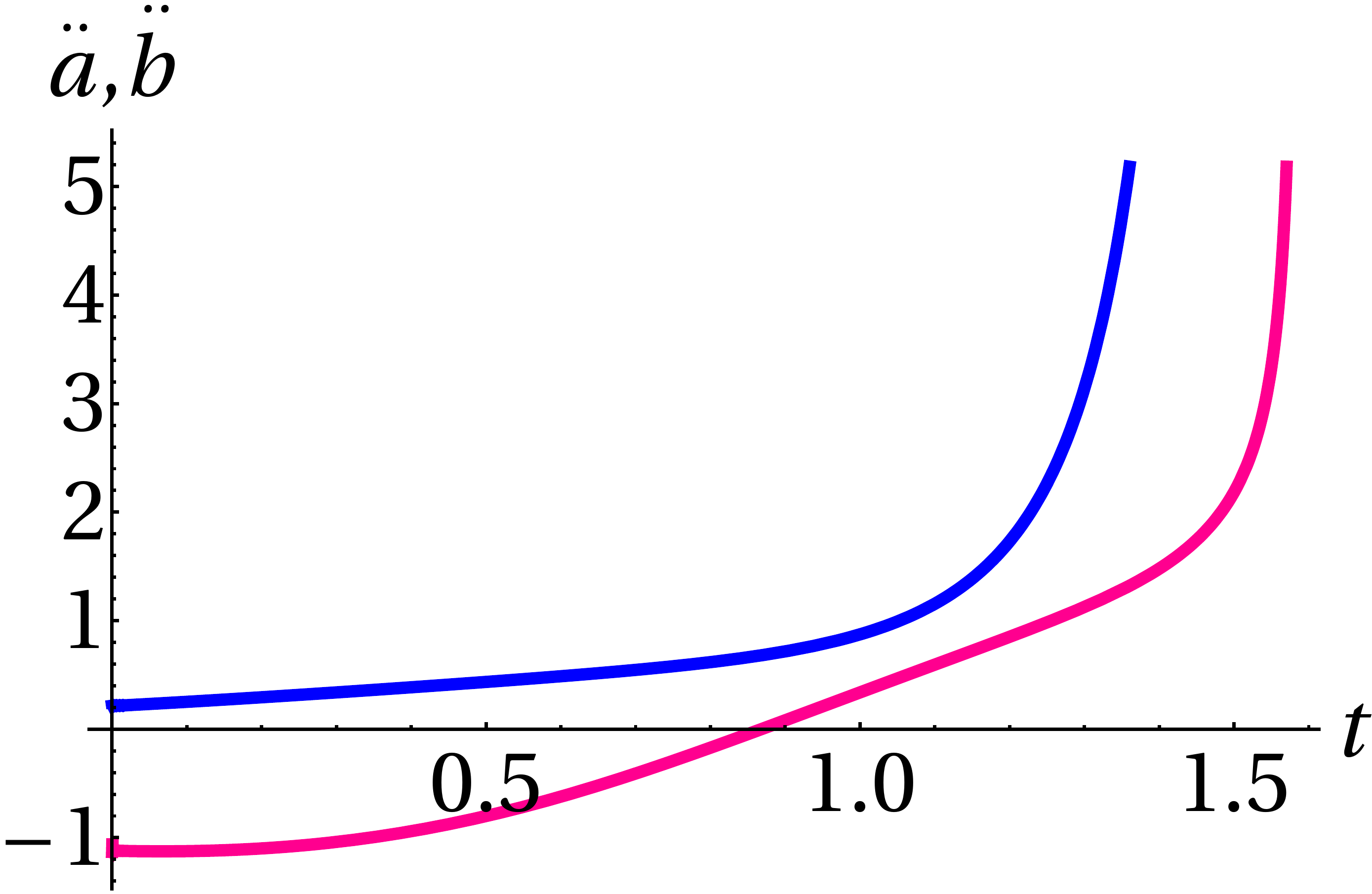}
    \caption{The accelerations of the scale factors: $\ddot a$ is represented by the blue curve, and $\ddot b$ by the red curve.}
    \label{addotbddotEARLY1-02-1-025constantL}
  \end{subfigure}
\qquad
  \begin{subfigure}[t]{.5\linewidth}
    \centering
    \includegraphics[width=0.7\columnwidth]{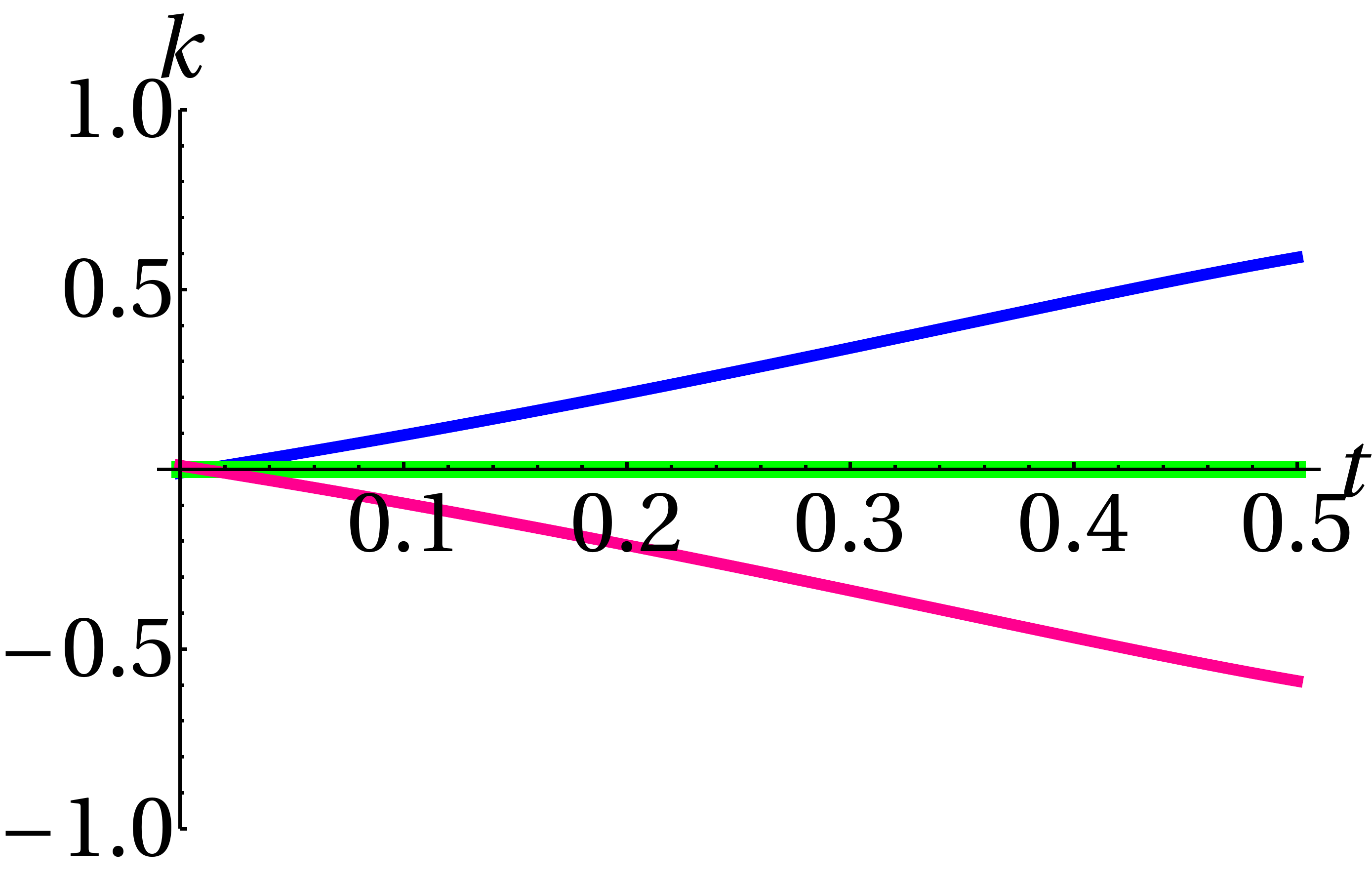}
    \caption{The harmonic function $k$ using: $\dot k\left(0\right)=1$ (blue curve), $\dot k\left(0\right)=0$ (green line), and $\dot k\left(0\right)=-1$ (red curve).}
    \label{k1-02-1-025constantL}
  \end{subfigure}
      \caption{Initial conditions set number 14 for constant $\Lambda$.}
  \label{Fig27}
  \end{figure}


\begin{figure}[H]
  \begin{subfigure}[t]{.5\linewidth}
    \centering
    \includegraphics[width=0.7\columnwidth]{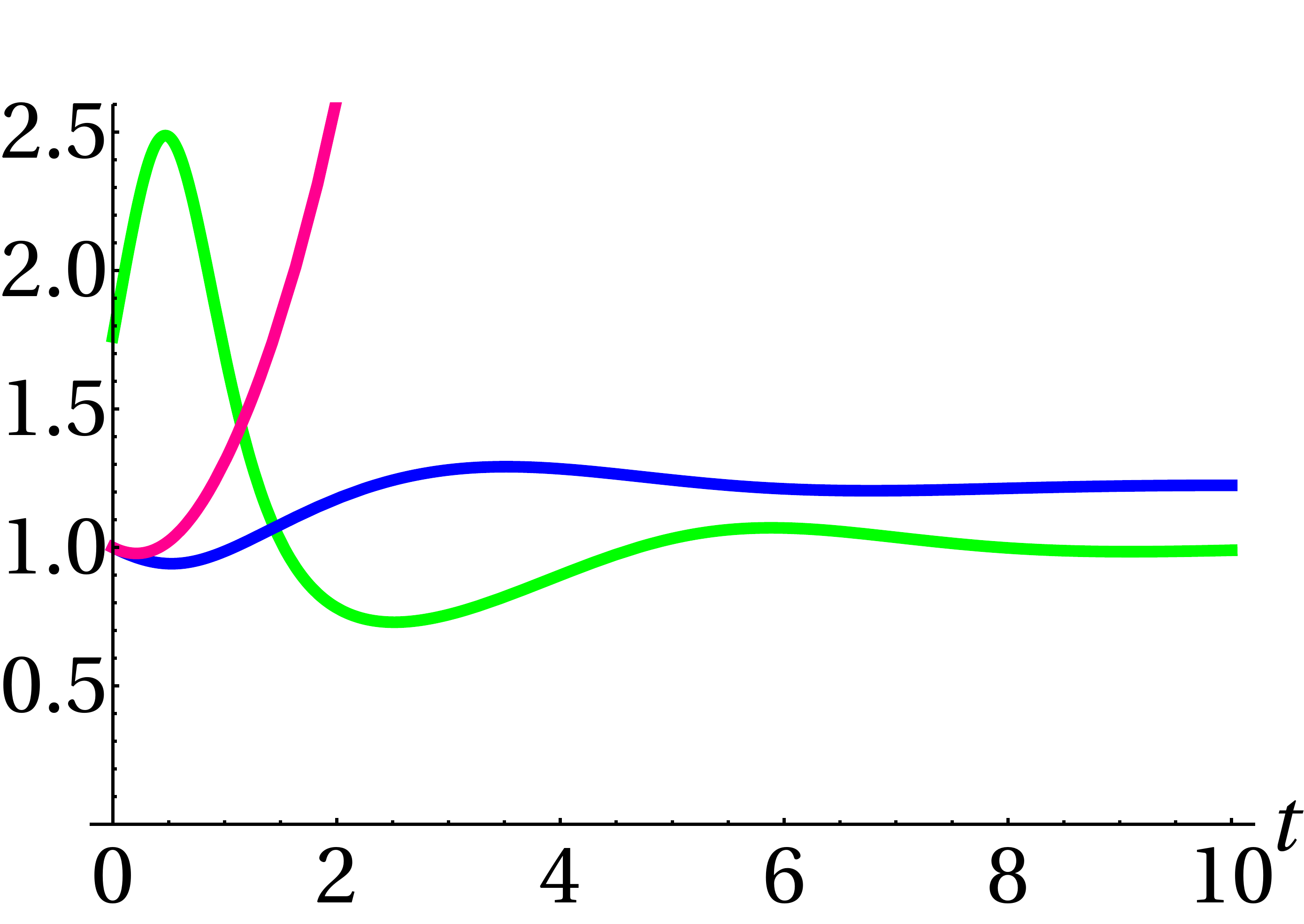}
    \caption{The scale factor $a$ is represented by the blue curve, $b$ by the red curve, while $\left| {G_{i\bar j} \dot z^i \dot z^{\bar j}} \right|$ by the green curve.}
    \label{abzz1-02-1-026constantL}
  \end{subfigure}
\qquad
  \begin{subfigure}[t]{.5\linewidth}
    \centering
    \includegraphics[width=0.7\columnwidth]{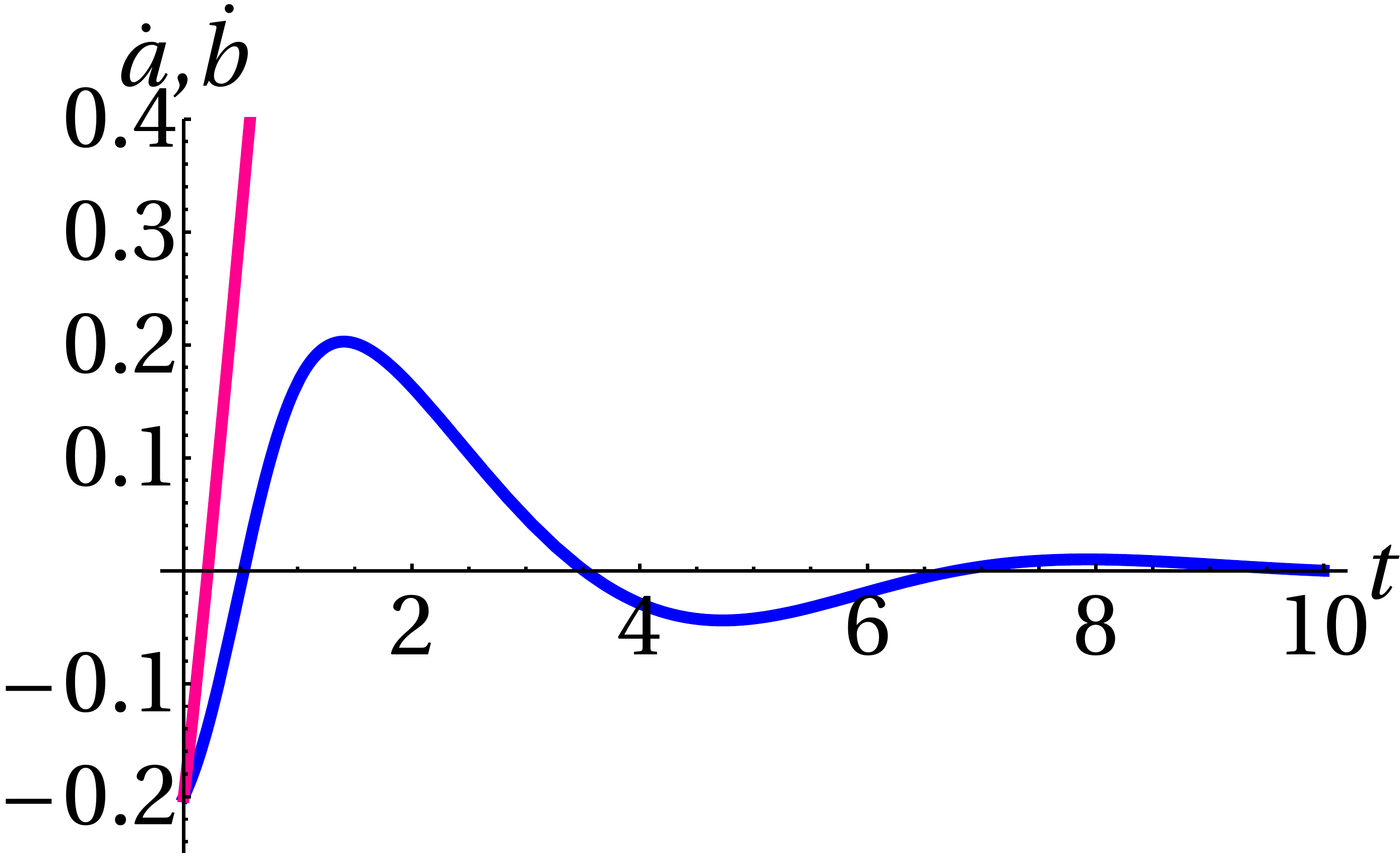}
    \caption{The expansion rates of the scale factors: $\dot a$ is represented by the blue curve, and $\dot b$ by the red curve.}
    \label{adotbdot1-02-1-026constantL}
  \end{subfigure}
\\[9em]
  \begin{subfigure}[t]{.5\linewidth}
    \centering
    \includegraphics[width=0.7\columnwidth]{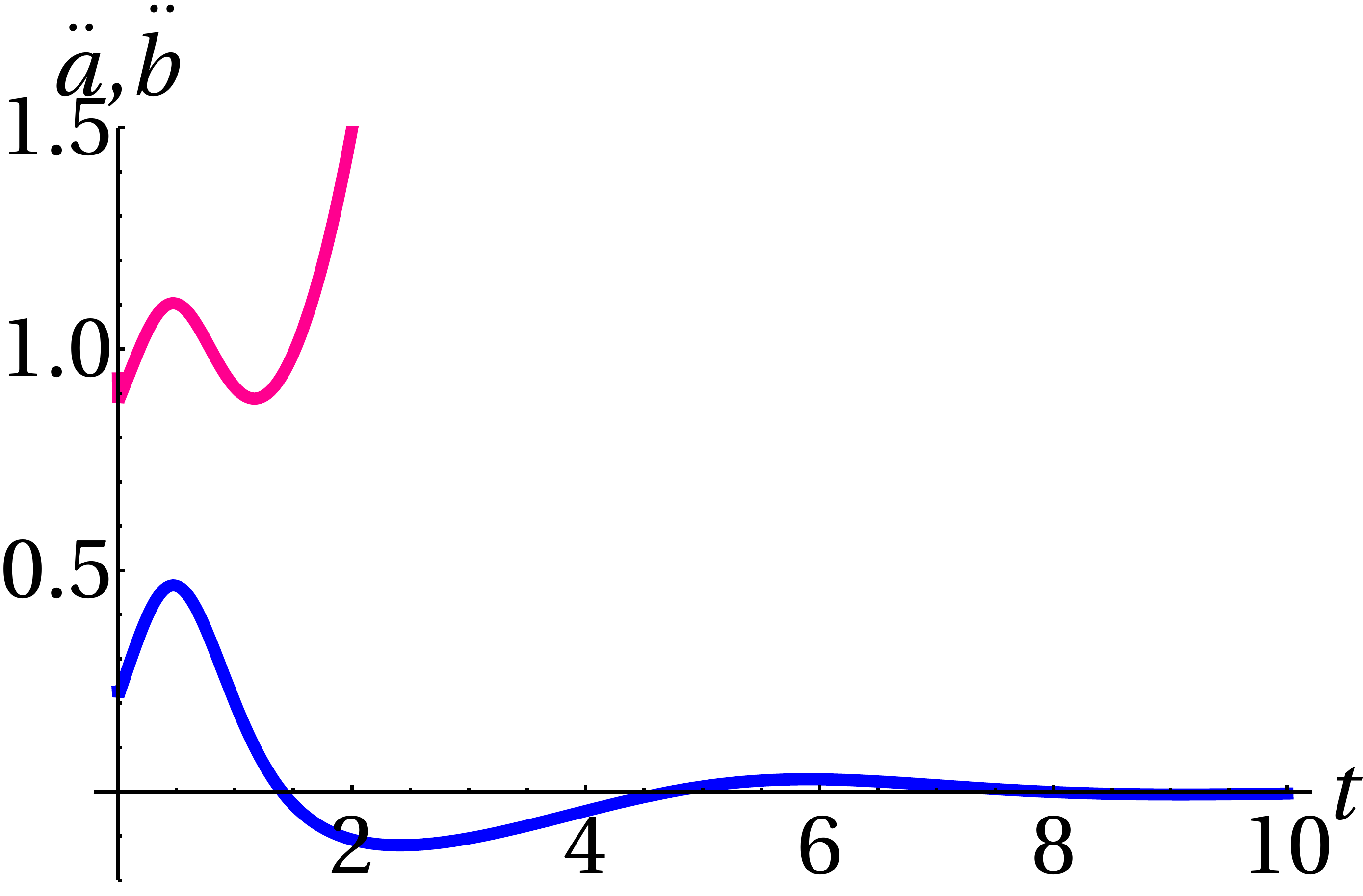}
    \caption{The accelerations of the scale factors: $\ddot a$ is represented by the blue curve, and $\ddot b$ by the red curve.}
    \label{addotbddot1-02-1-026constantL}
  \end{subfigure}
\qquad
  \begin{subfigure}[t]{.5\linewidth}
    \centering
    \includegraphics[width=0.7\columnwidth]{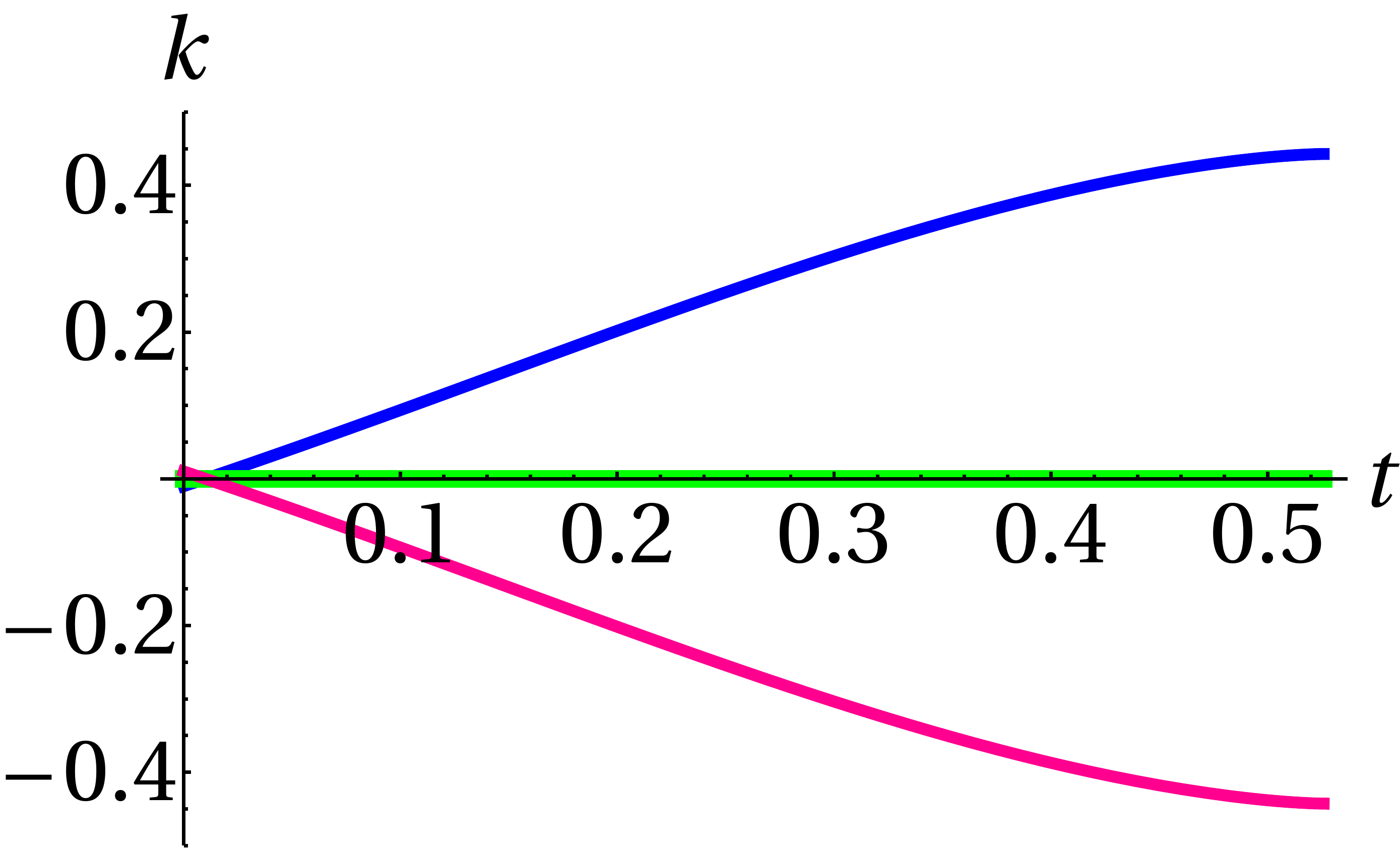}
    \caption{The harmonic function $k$ using: $\dot k\left(0\right)=1$ (blue curve), $\dot k\left(0\right)=0$ (green line), and $\dot k\left(0\right)=-1$ (red curve).}
    \label{k1-02-1-026constantL}
  \end{subfigure}
\caption{Initial conditions set number 15 for constant $\Lambda$.}
  \label{Fig30}
\end{figure}

\begin{figure}[H]
  \begin{subfigure}[t]{.5\linewidth}
    \centering
    \includegraphics[width=0.7\columnwidth]{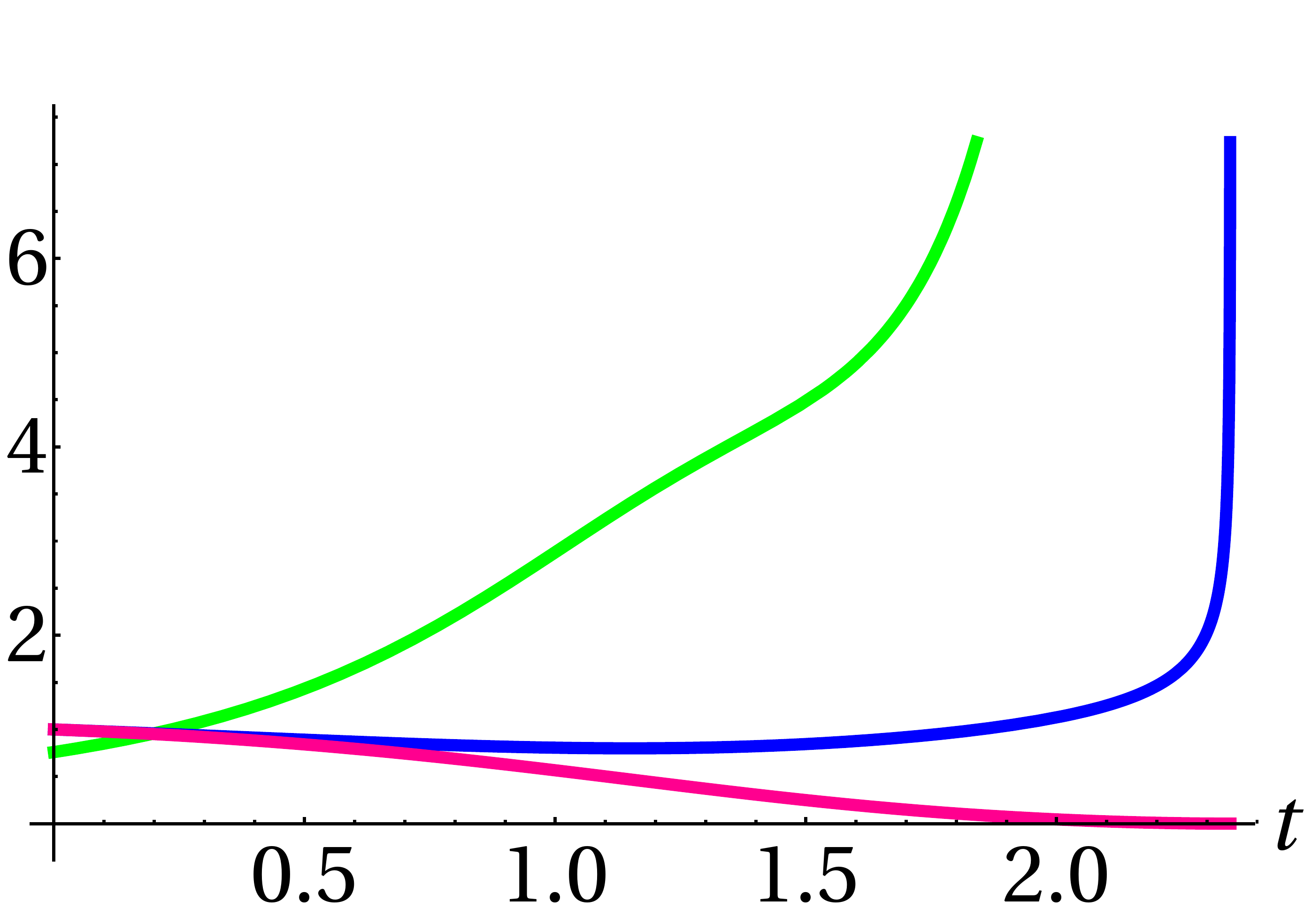}
    \caption{The scale factor $a$ is represented by the blue curve, $b$ by the red curve, while ${G_{i\bar j} \dot z^i \dot z^{\bar j}}$ (not the absolute value) by the green curve.}
    \label{abzz1-02-1-027constantL}
  \end{subfigure}
\qquad
  \begin{subfigure}[t]{.5\linewidth}
    \centering
    \includegraphics[width=0.7\columnwidth]{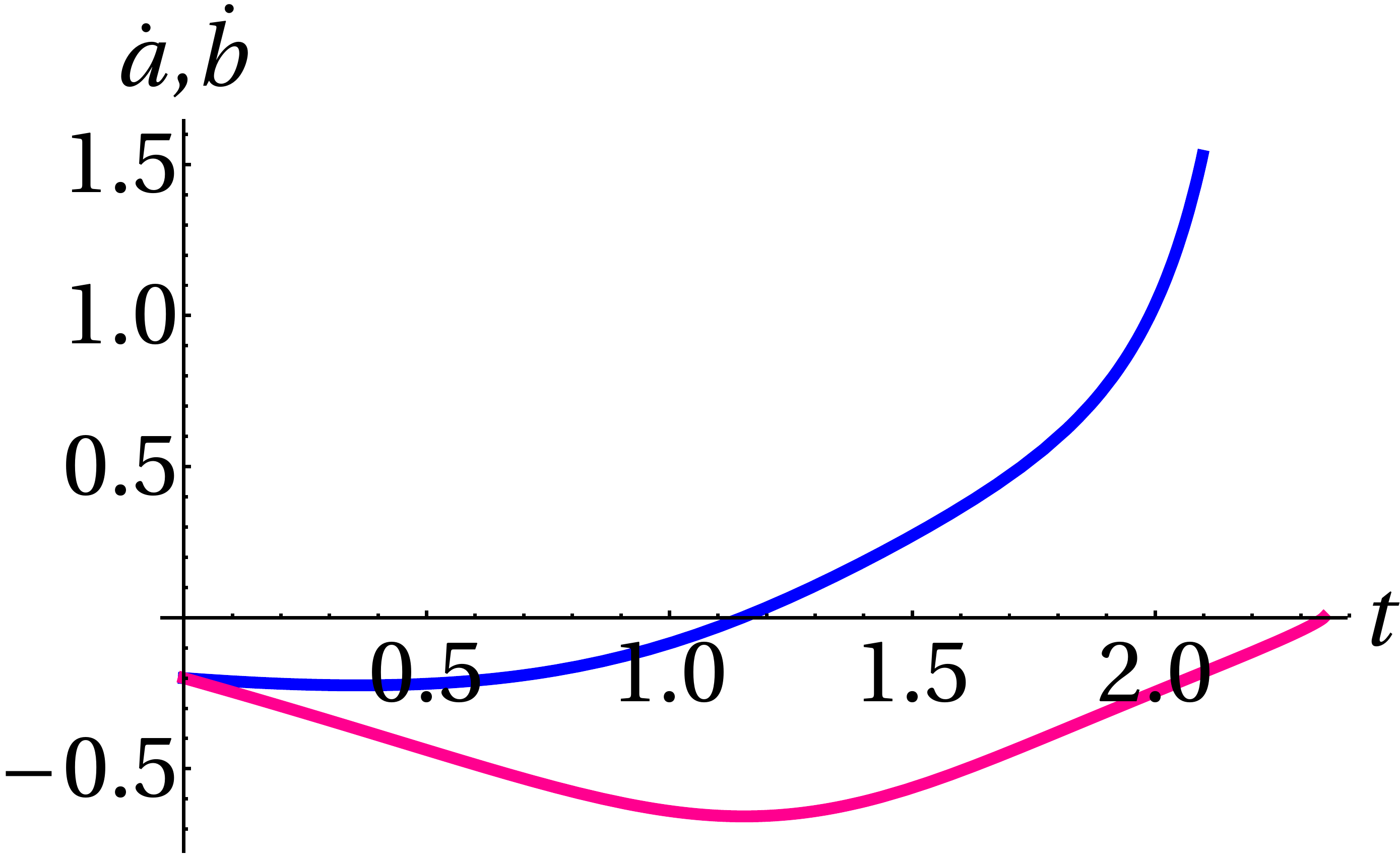}
    \caption{The expansion rates of the scale factors: $\dot a$ is represented by the blue curve, and $\dot b$ by the red curve.}
    \label{adotbdot1-02-1-027constantL}
  \end{subfigure}
\\[9em]
  \begin{subfigure}[t]{.5\linewidth}
    \centering
    \includegraphics[width=0.7\columnwidth]{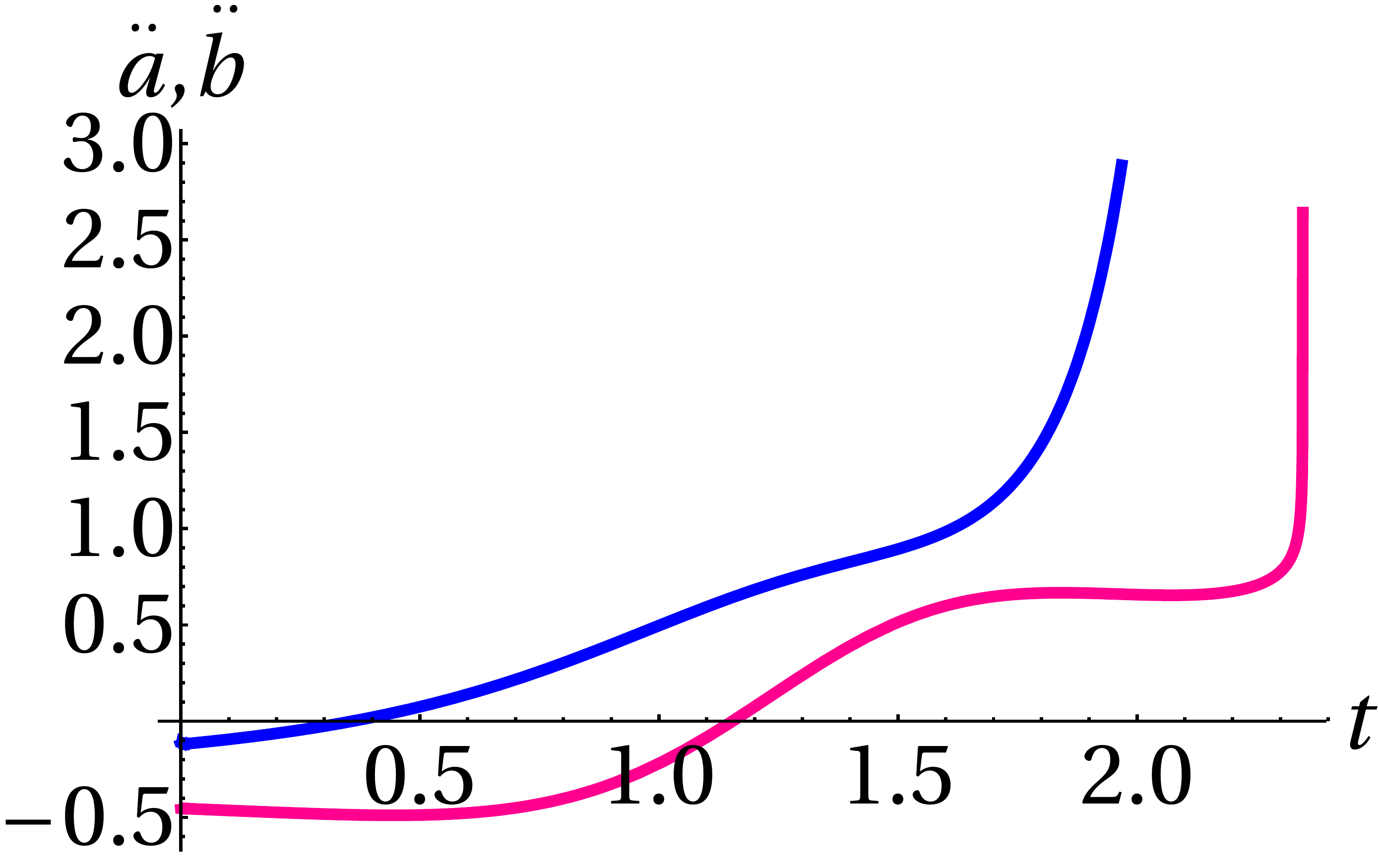}
    \caption{The accelerations of the scale factors: $\ddot a$ is represented by the blue curve, and $\ddot b$ by the red curve.}
    \label{addotbddot1-02-1-027constantL}
  \end{subfigure}
\qquad
  \begin{subfigure}[t]{.5\linewidth}
    \centering
    \includegraphics[width=0.7\columnwidth]{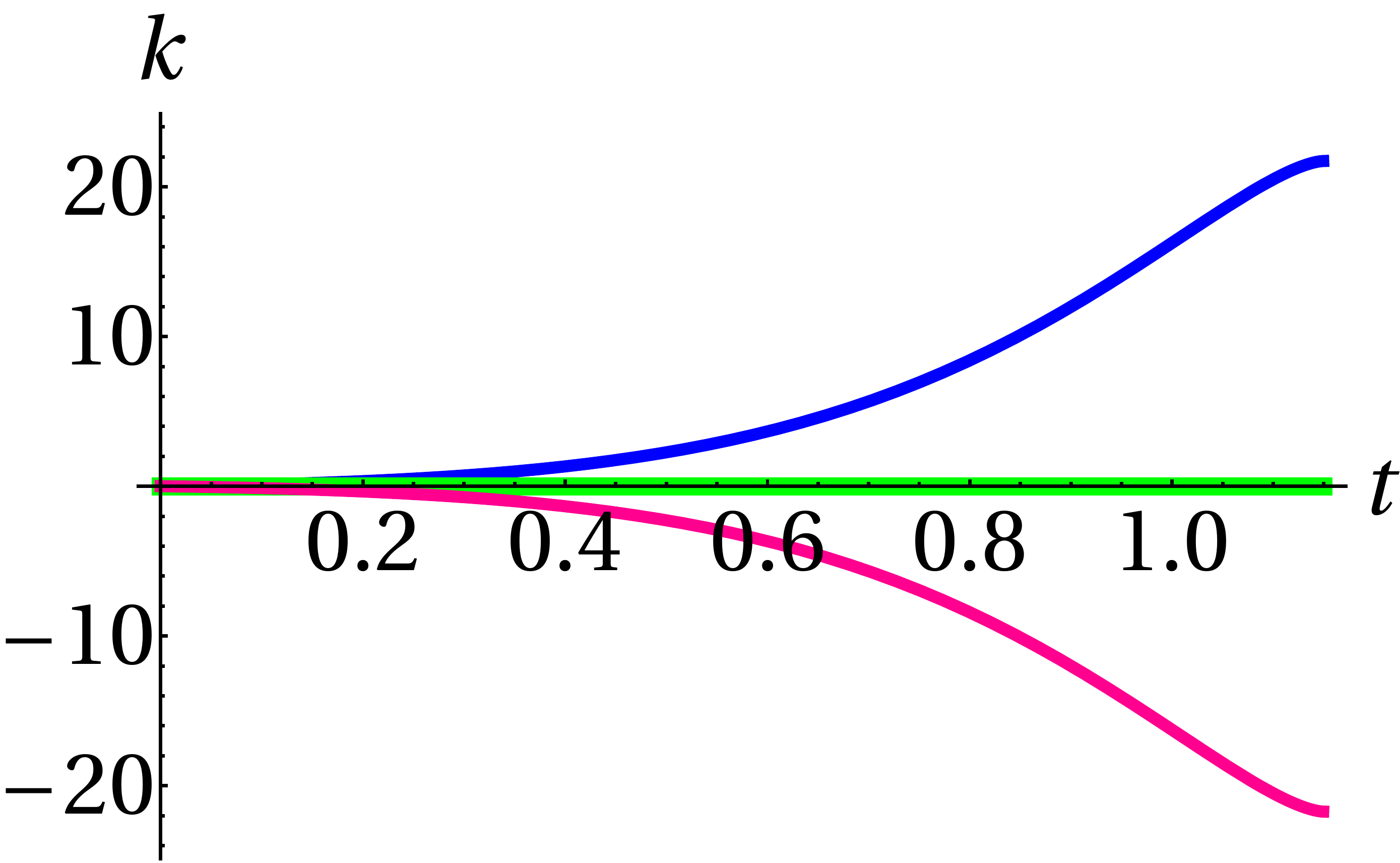}
    \caption{The harmonic function $k$ using: $\dot k\left(0\right)=1$ (blue curve), $\dot k\left(0\right)=0$ (green line), and $\dot k\left(0\right)=-1$ (red curve).}
    \label{k1-02-1-027constantL}
  \end{subfigure}
 \caption{Initial conditions set number 16 for constant $\Lambda$.}
  \label{Fig32}
\end{figure}


\begin{figure}[H]
  \begin{subfigure}[t]{.5\linewidth}
    \centering
    \includegraphics[width=0.7\columnwidth]{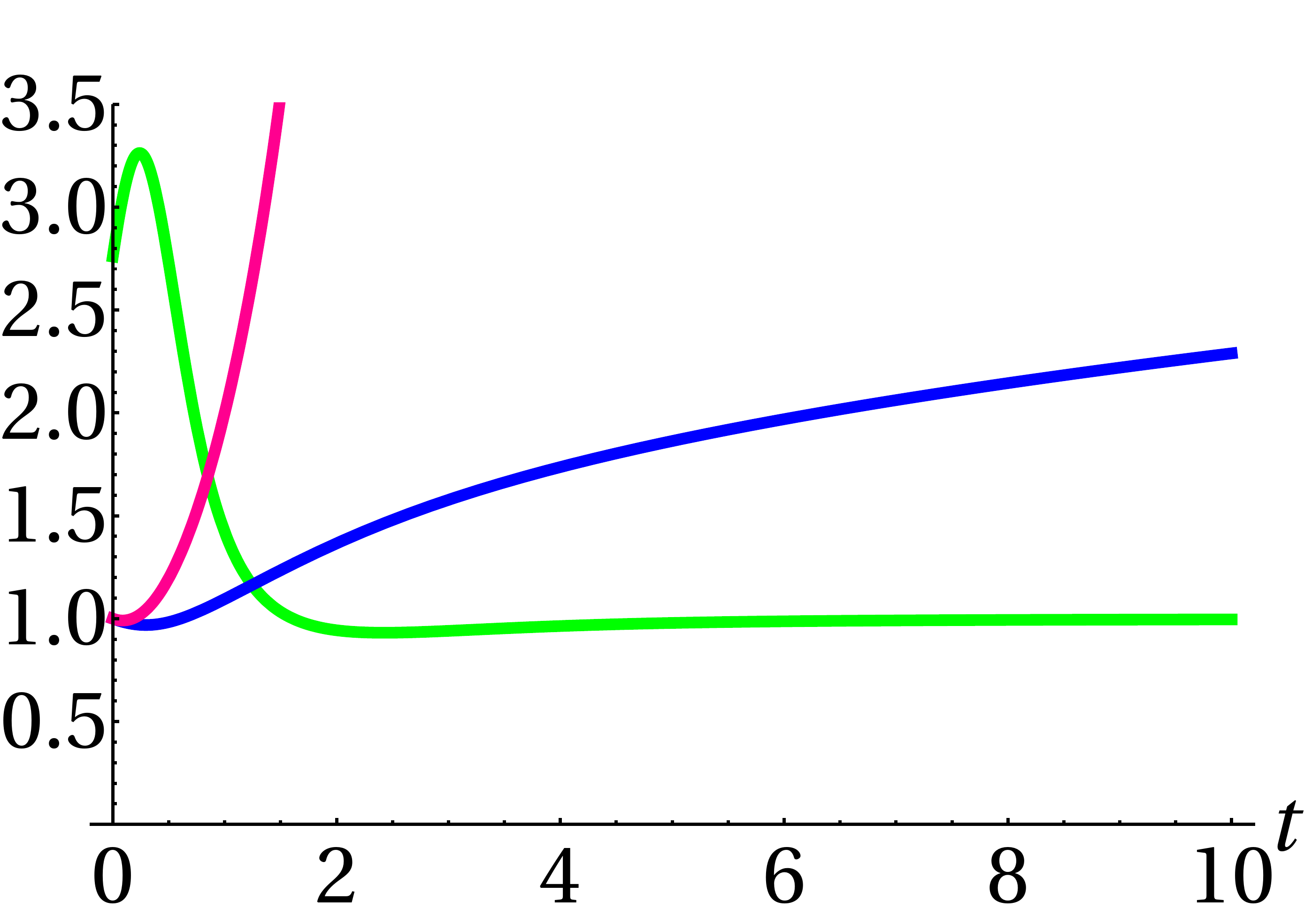}
    \caption{The scale factor $a$ is represented by the blue curve, $b$ by the red curve, while $\left| {G_{i\bar j} \dot z^i \dot z^{\bar j}} \right|$ by the green curve.}
    \label{abzz1-02-1-028constantL}
  \end{subfigure}
\qquad
  \begin{subfigure}[t]{.5\linewidth}
    \centering
    \includegraphics[width=0.7\columnwidth]{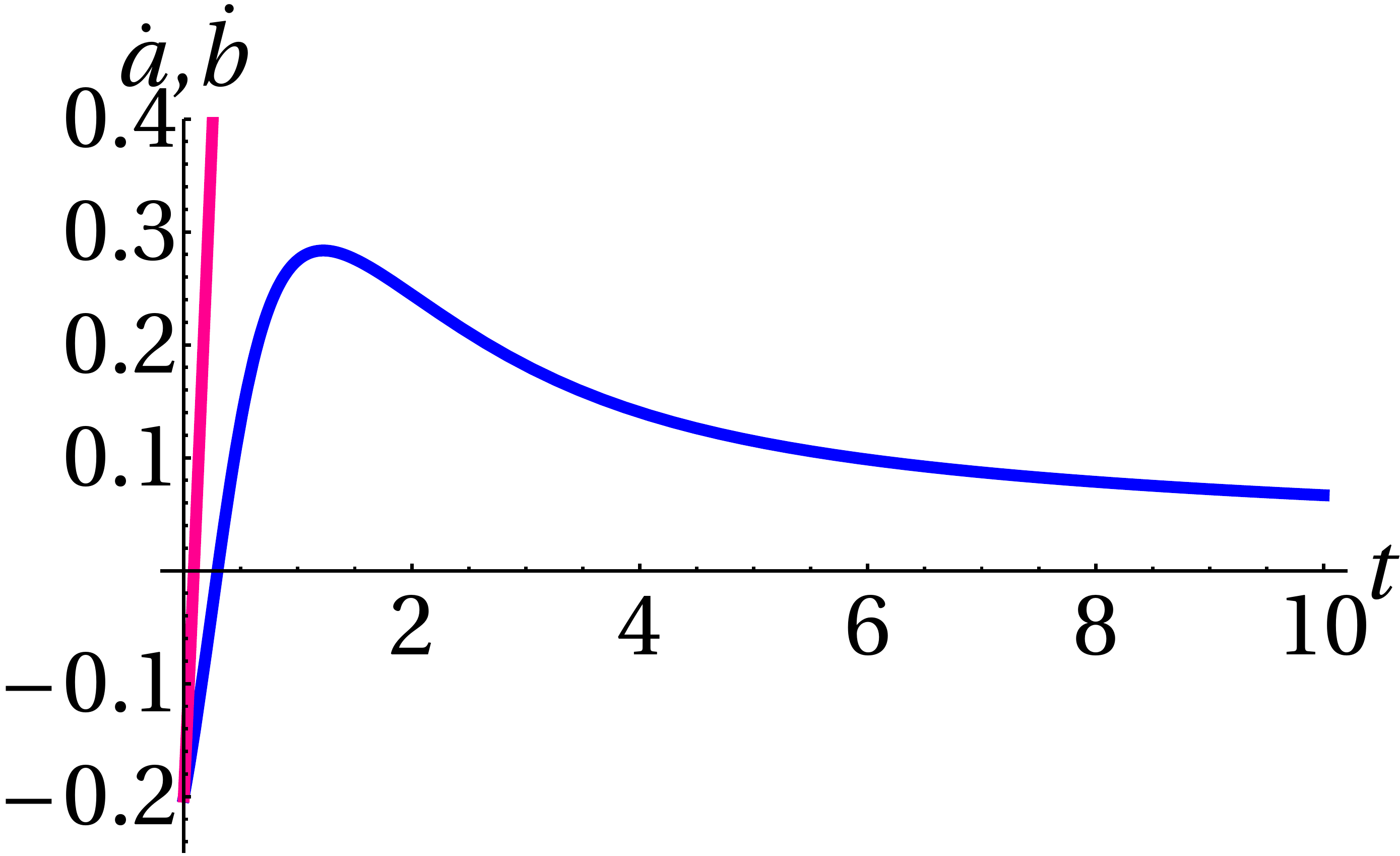}
    \caption{The expansion rates of the scale factors: $\dot a$ is represented by the blue curve, and $\dot b$ by the red curve.}
    \label{adotbdot1-02-1-028constantL}
  \end{subfigure}
\\[9em]
  \begin{subfigure}[t]{.5\linewidth}
    \centering
    \includegraphics[width=0.7\columnwidth]{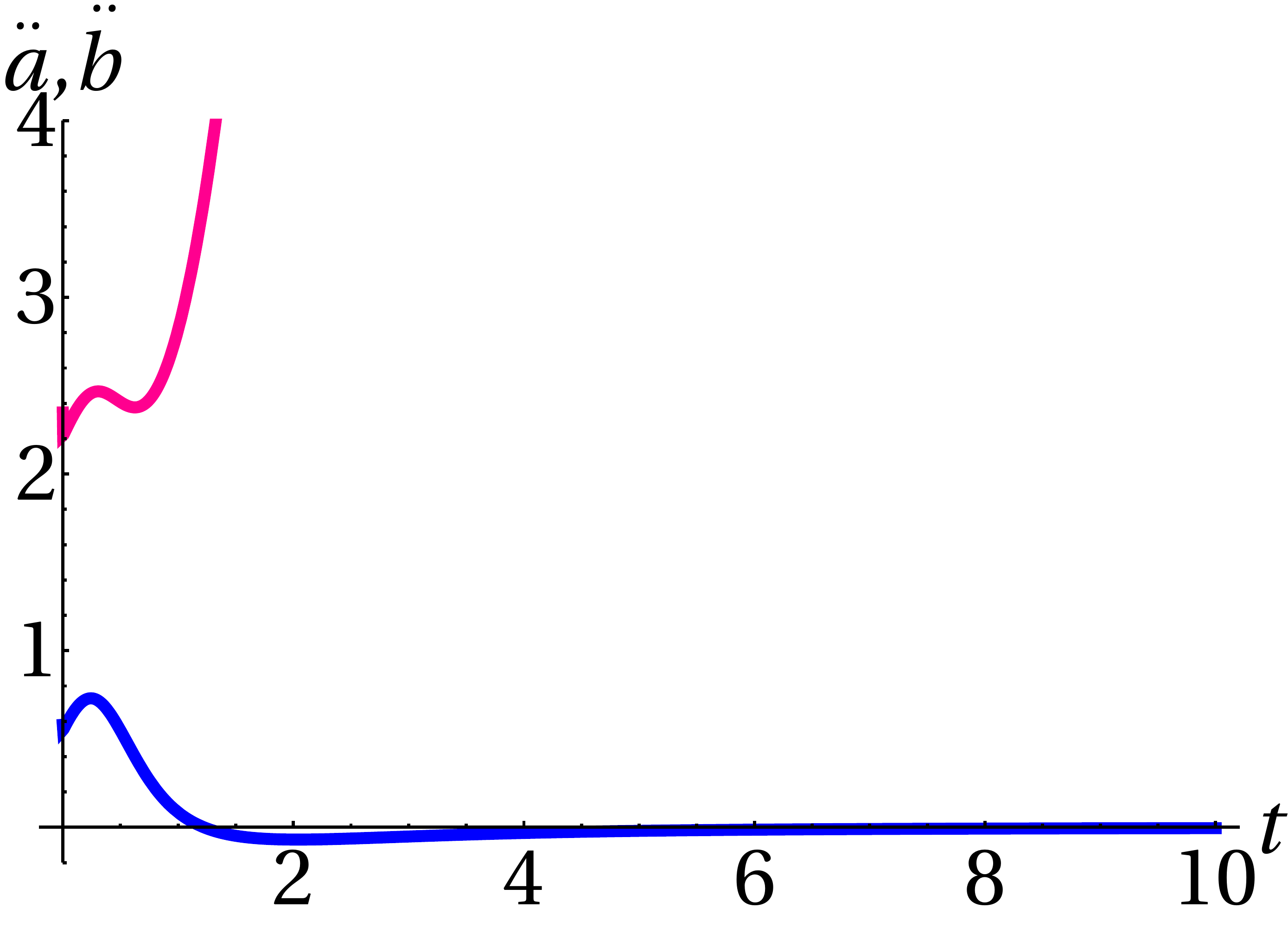}
    \caption{The accelerations of the scale factors: $\ddot a$ is represented by the blue curve, and $\ddot b$ by the red curve.}
    \label{addotbddot1-02-1-028constantL}
  \end{subfigure}
\qquad
  \begin{subfigure}[t]{.5\linewidth}
    \centering
    \includegraphics[width=0.7\columnwidth]{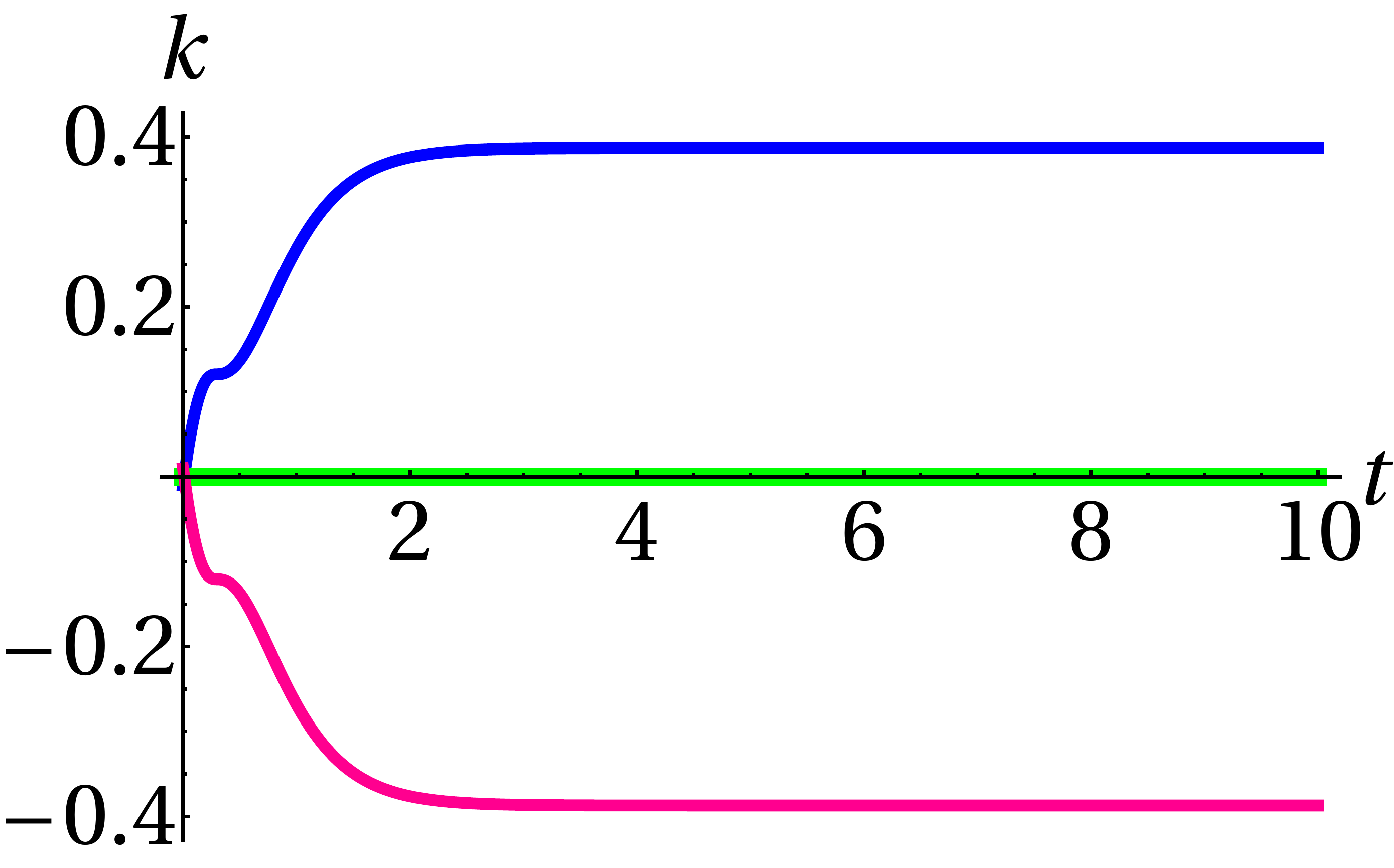}
    \caption{The harmonic function $k$ using: $\dot k\left(0\right)=1$ (blue curve), $\dot k\left(0\right)=0$ (green line), and $\dot k\left(0\right)=-1$ (red curve).}
    \label{k1-02-1-028constantL}
  \end{subfigure}
\caption{Initial conditions set number 17 for constant $\Lambda$.}
  \label{Fig34}
\end{figure}

\begin{figure}[H]
  \begin{subfigure}[t]{.5\linewidth}
    \centering
    \includegraphics[width=0.7\columnwidth]{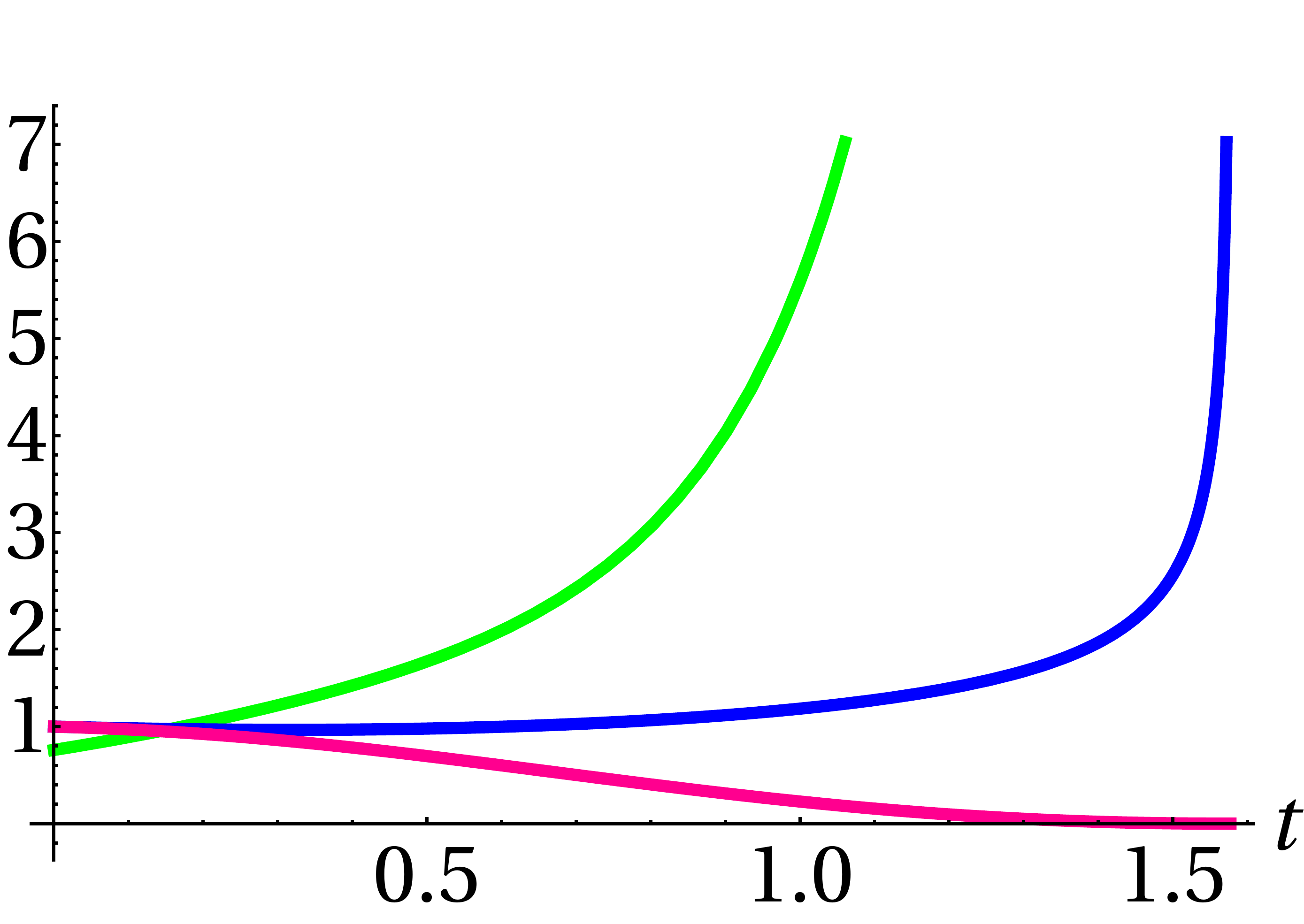}
    \caption{The scale factor $a$ is represented by the blue curve, $b$ by the red curve, while $\left| {G_{i\bar j} \dot z^i \dot z^{\bar j}} \right|$ by the green curve.}
    \label{abzz1-02-1-029constantL}
  \end{subfigure}
\qquad
  \begin{subfigure}[t]{.5\linewidth}
    \centering
    \includegraphics[width=0.7\columnwidth]{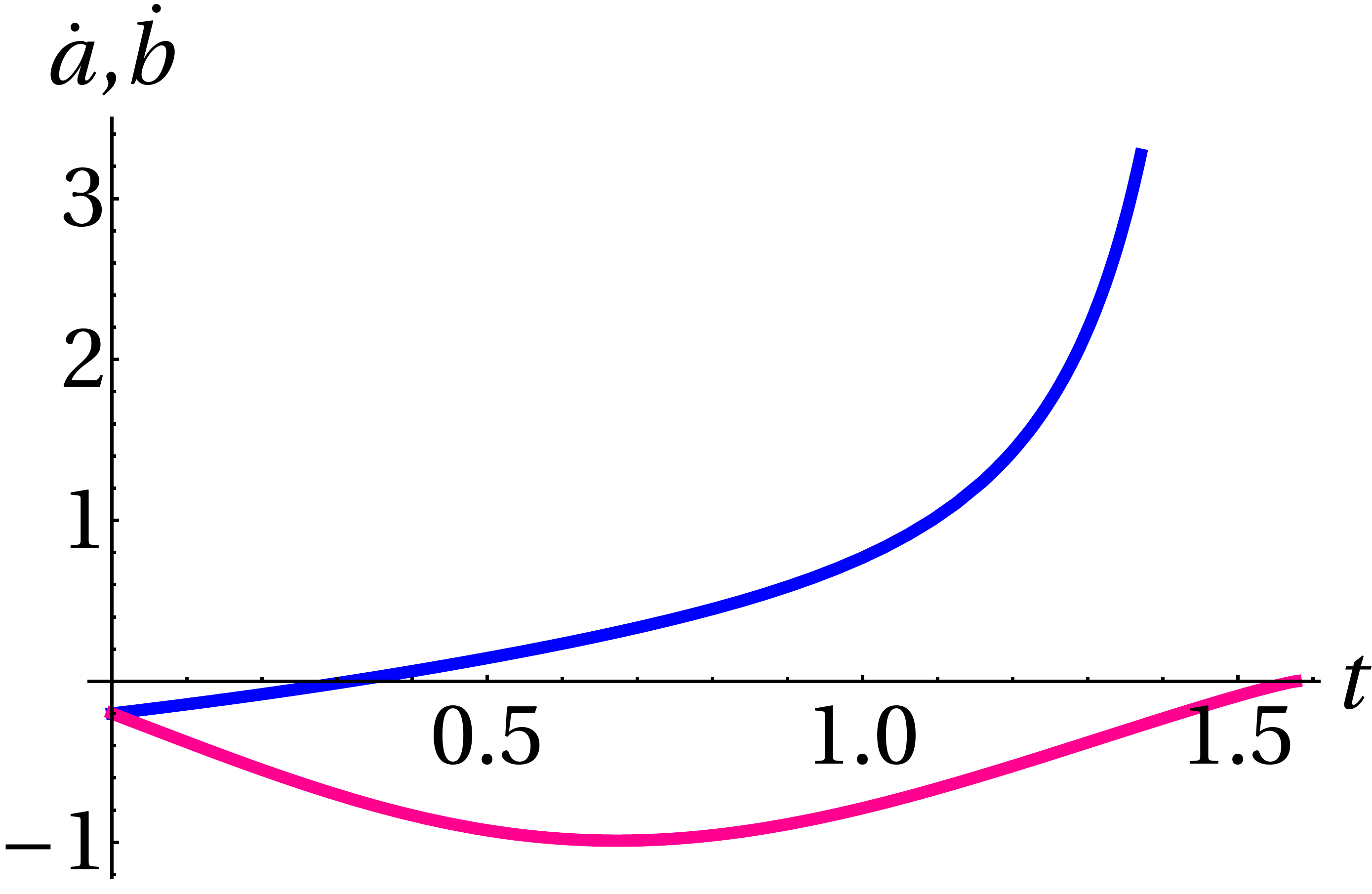}
    \caption{The expansion rates of the scale factors: $\dot a$ is represented by the blue curve, and $\dot b$ by the red curve.}
    \label{adotbdot1-02-1-029constantL}
  \end{subfigure}
\\[9em]
  \begin{subfigure}[t]{.5\linewidth}
    \centering
    \includegraphics[width=0.7\columnwidth]{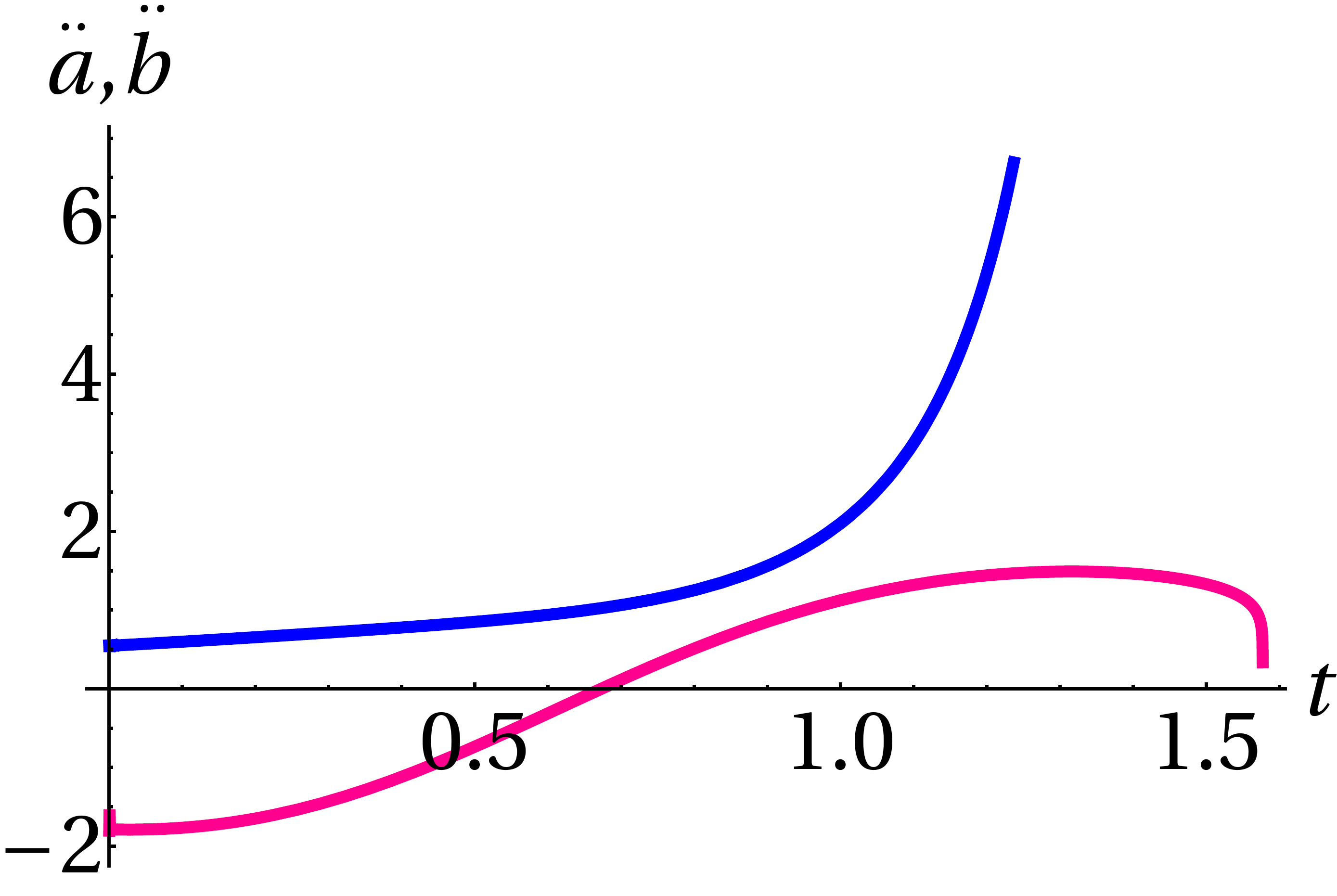}
    \caption{The accelerations of the scale factors: $\ddot a$ is represented by the blue curve, and $\ddot b$ by the red curve.}
    \label{addotbddot1-02-1-029constantL}
  \end{subfigure}
\qquad
  \begin{subfigure}[t]{.5\linewidth}
    \centering
    \includegraphics[width=0.7\columnwidth]{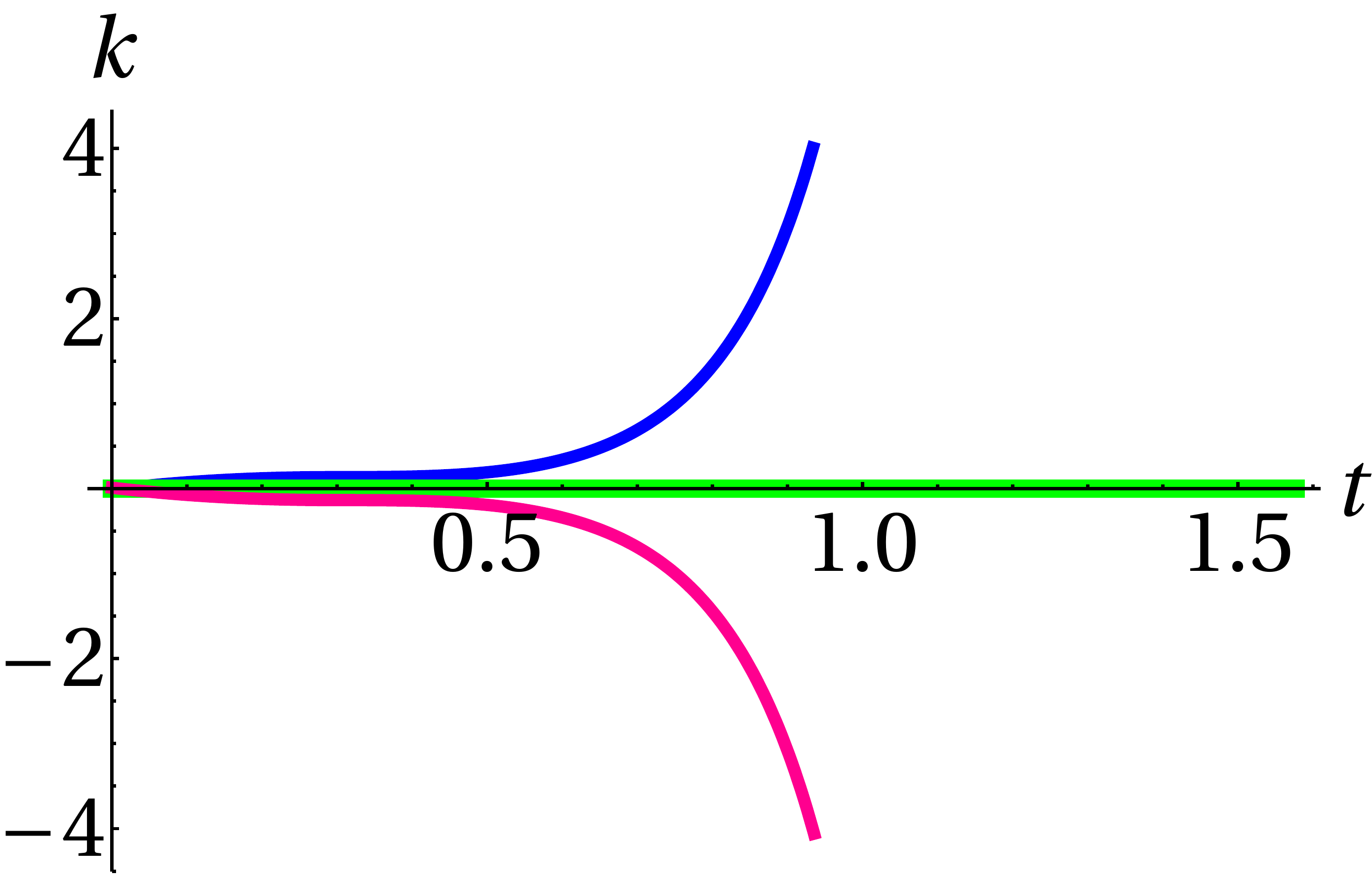}
    \caption{The harmonic function $k$ using: $\dot k\left(0\right)=1$ (blue curve), $\dot k\left(0\right)=0$ (green line), and $\dot k\left(0\right)=-1$ (red curve).}
    \label{k1-02-1-029constantL}
  \end{subfigure}
      \caption{Initial conditions set number 18 for constant $\Lambda$.}
  \label{Fig35}
  \end{figure}
The conclusions one may draw from all of these cases are straightforwardly obtained by inspection. The scale factors $a$ of these `universes' are strongly correlated to the values of the moduli flow norm $ G_{i\bar j} \dot z^i \dot z^{\bar j} $; large values of the later correspond to small values of the former, maximum values of the later correspond to minimum values of the former, and so on. Particularly important is that the flow norm always starts at large, almost infinite, values and decays rapidly, inducing an inflationary behavior for $a\left(t\right)$ \emph{without} the unnecessary introduction of an inflaton \cite{Emam:2020oyb}. It is also interesting that later time accelerations do not necessarily arise because of the presence of a cosmological constant $\Lambda$. For instance in IC set \#3 there exists a late time accelerating behavior of $a$ that is \emph{not} induced by the presence of $\Lambda$, but rather seems to be related to $\tilde \Lambda$ mediated by the moduli flow, and similarly for IC set \#6. This last is particularly rich in that it exhibits inflation, deceleration, acceleration, and then continues to an almost constant value for $a$. If taken seriously from a cosmological perspective it seems to predict a universe that at later times oscillates and then decays to a constant `size.' If this represents our universe then our current time may, perhaps, be between $t=4$ and $t=5$ on the plot's scale. Another interesting behavior is exhibited by the bulk scale factor $b$ in IC sets \#3, 5, 7, and 9. It starts at positive values then smoothly switches to negative! This `spacelike to timelike' behavior of metric components indicates that there is a causal discontinuity (a horizon if you will) in the bulk at a certain critical time $t_c$; where $b$ vanishes. In general, our own universe may be described by the solutions following IC sets \#2, 3, 4, 5 (which has a future big rip despite a negative $\Lambda$), 6, 8, and 9 (also with a big rip).

If we look at IC set \#1 particularly, we can describe the behaviour of the remaining fields. For example the dilaton is known to be proportional to the volume of the underlying Calabi-Yau metric. In most solutions this initially drops, in correlation with the inflationary epoch, then expands mostly to constant values and in some cases (correlated with the big rip) to infinity. This is consistent with the behavior of the scale factors in that all eleven dimensions behave similarly. The axionic values are mostly related to the initial inflation as one might expect.

\subsection{Constant $b$}

Analysis here follows the initial conditions schemes outlined in table (\ref{tabb}). Once again this is broken down into the same two major categories with varying values for $\tilde \Lambda$. The first six sets are for constant $\tilde \Lambda$ allowing for varying $\Lambda$, while the second set of six are for constant $\Lambda$ allowing for varying $\tilde\Lambda$.

\begin{table}[H]
\centering
\caption{The twelve sets of initial conditions (IC) used in the constant $b$ computations.}
\label{tabb}
\vspace{0.5cm}
\begin{adjustbox}{max width=\textwidth}
\begin{tabular}{|c|c|c|c|l|}
\hline
\textbf{IC Set \#} & $a\left(0\right)$  & $\dot a\left(0\right)$         & $\tilde \Lambda$ & \textbf{Description} \\ \hline
\textbf{1}         & \multirow{3}{*}{0} & \multirow{3}{*}{0}             & 0                & \multirow{3}{*}{Big bang-like IC}           \\ \cline{1-1} \cline{4-4}
\textbf{2}         &                    &                                & 1                &           \\ \cline{1-1} \cline{4-4}
\textbf{3}      &                    &                                & -1               &           \\ \hline
\textbf{4}         & \multirow{3}{*}{1} & \multirow{3}{*}{-0.2}          & 0                & \multirow{3}{*}{Non- singular IC with negative initial velocity}           \\ \cline{1-1} \cline{4-4}
\textbf{5}        &                    &                                & 1                &           \\ \cline{1-1} \cline{4-4}
\textbf{6}         &                    &                                & -1               &           \\ \hline\hline
\textbf{IC Set \#} & $a\left(0\right)$  & $\dot a\left(0\right)$         & $\Lambda$        & \textbf{Description} \\ \hline
\textbf{7}         & \multirow{3}{*}{0} & \multirow{3}{*}{0}             & 0                & \multirow{3}{*}{Big bang-like IC}        \\ \cline{1-1} \cline{4-4}
\textbf{8}         &                    &                                & 1                &           \\ \cline{1-1} \cline{4-4}
\textbf{9}         &                    &                                & -1               &           \\ \hline
\textbf{10}        & \multirow{3}{*}{1} & \multirow{3}{*}{-0.2}          & 0                & \multirow{3}{*}{Non- singular IC with negative initial velocity}                 \\ \cline{1-1} \cline{4-4}
\textbf{11}        &                    &                                & 1                &           \\ \cline{1-1} \cline{4-4}
\textbf{12}        &                    &                                & -1               &           \\ \hline
\end{tabular}
\end{adjustbox}
\end{table}
\clearpage
\newpage


\begin{figure}[H]
  \begin{subfigure}[t]{.5\linewidth}
    \centering
    \includegraphics[width=0.7\columnwidth]{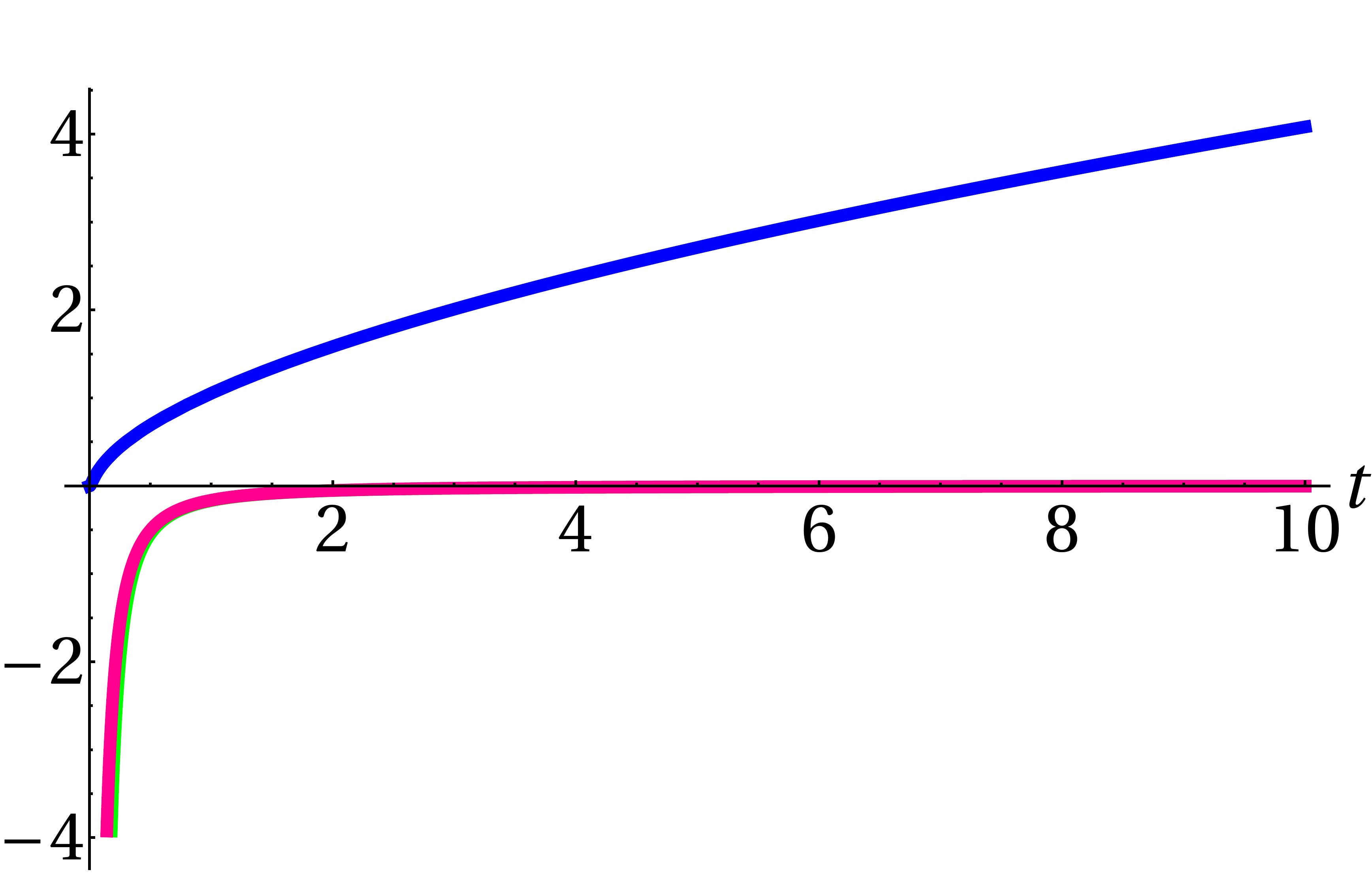}
    \caption{The scale factor $a$ is represented by the blue curve, $\Lambda$ by the red curve, while $ {G_{i\bar j} \dot z^i \dot z^{\bar j}} $ by the green curve.}
    \label{aLzz00001constantb}
  \end{subfigure}
\qquad
  \begin{subfigure}[t]{.5\linewidth}
    \centering
    \includegraphics[width=0.7\columnwidth]{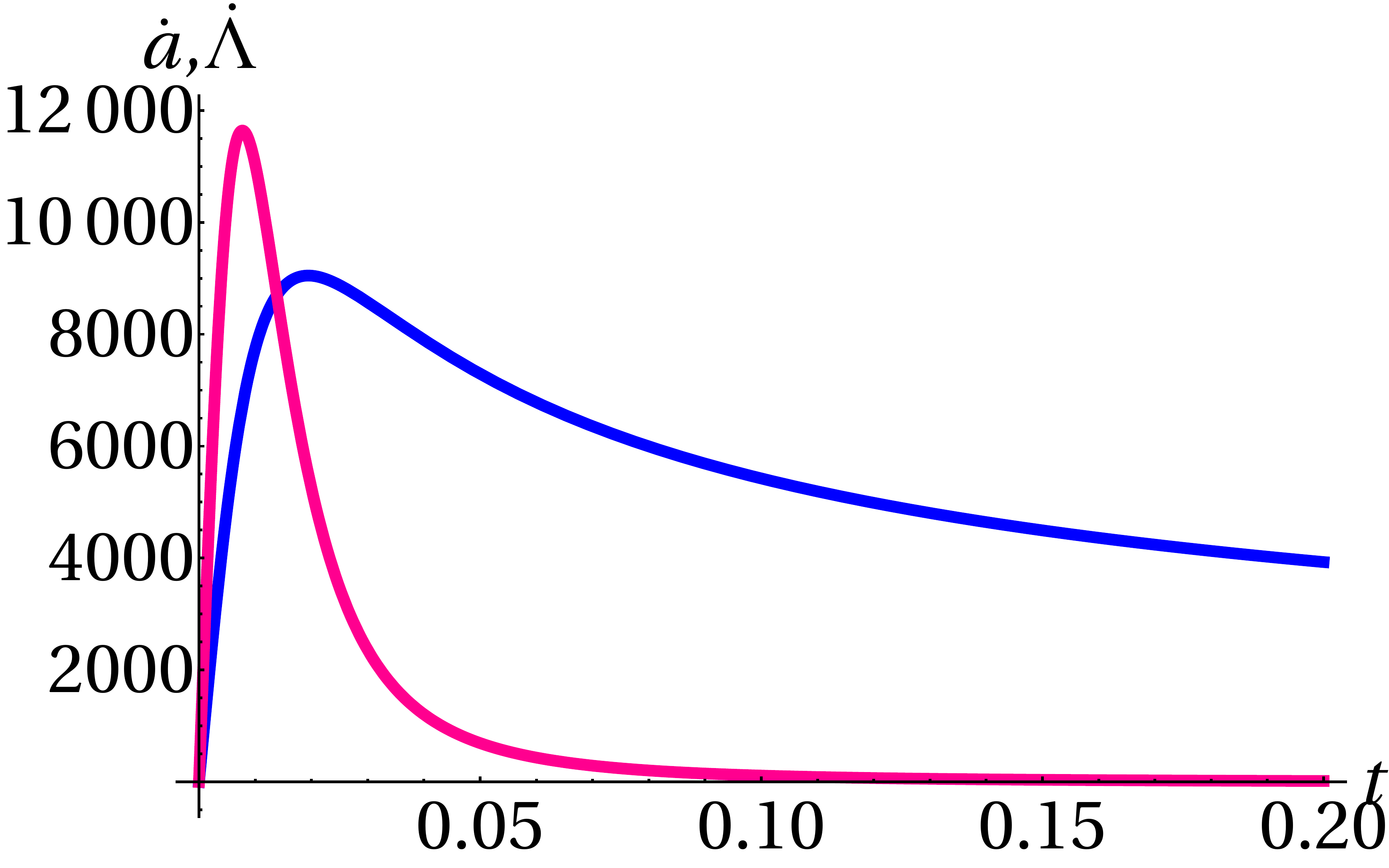}
    \caption{The expansion rates: $\dot a$ is represented by the blue curve, and $\dot \Lambda$ by the red curve. The curve for $\dot a$ is scaled up by a factor of 3000.}
    \label{adotLdot00001constantb}
  \end{subfigure}
\\[9em]
  \begin{subfigure}[t]{.5\linewidth}
    \centering
    \includegraphics[width=0.7\columnwidth]{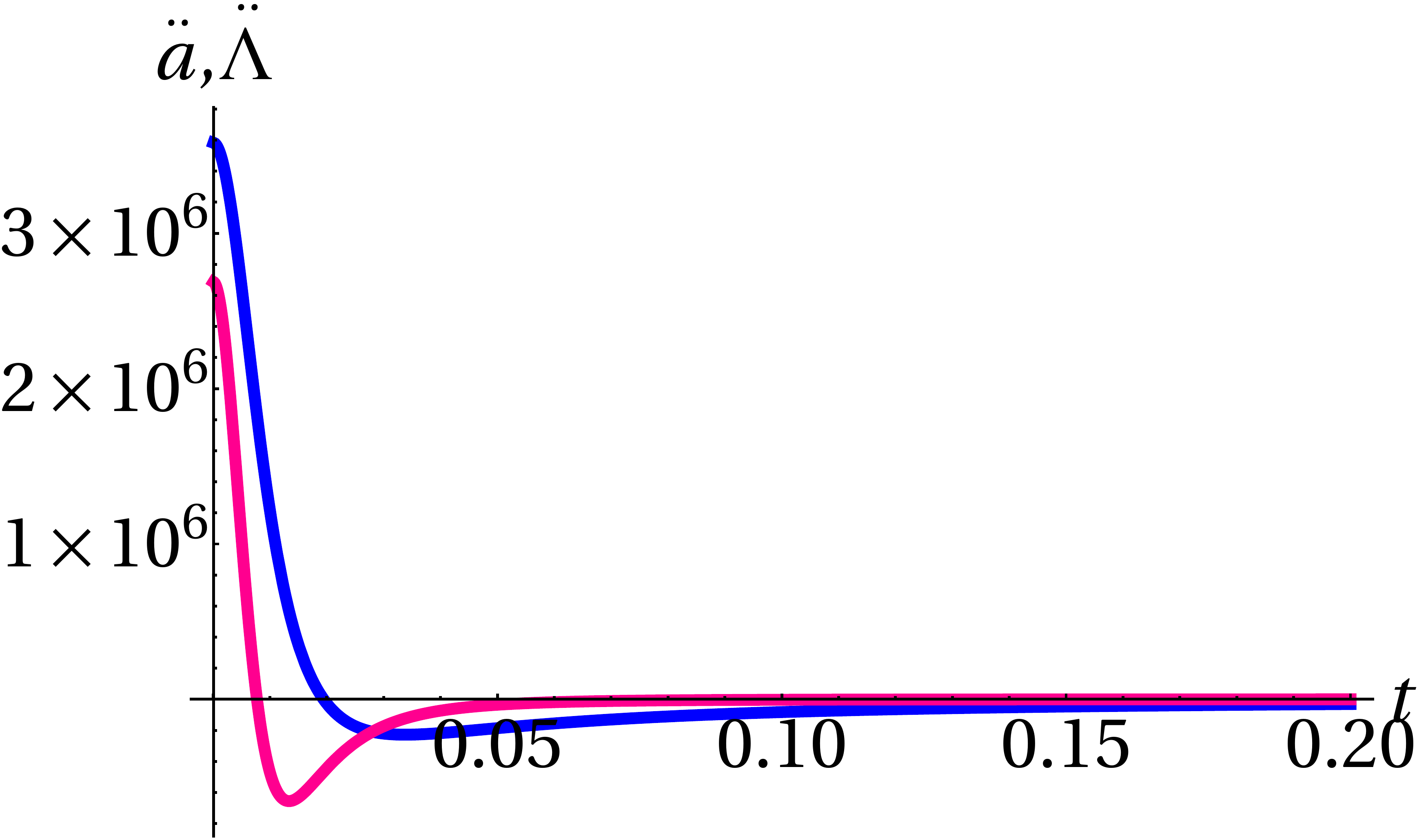}
    \caption{The accelerations: $\ddot a$ is represented by the blue curve, and $\ddot \Lambda$ by the red curve. The curve for $\dot a$ is scaled up by a factor of 10000.}
    \label{addotLddot00001constantb}
  \end{subfigure}
\qquad
  \begin{subfigure}[t]{.5\linewidth}
    \centering
    \includegraphics[width=0.7\columnwidth]{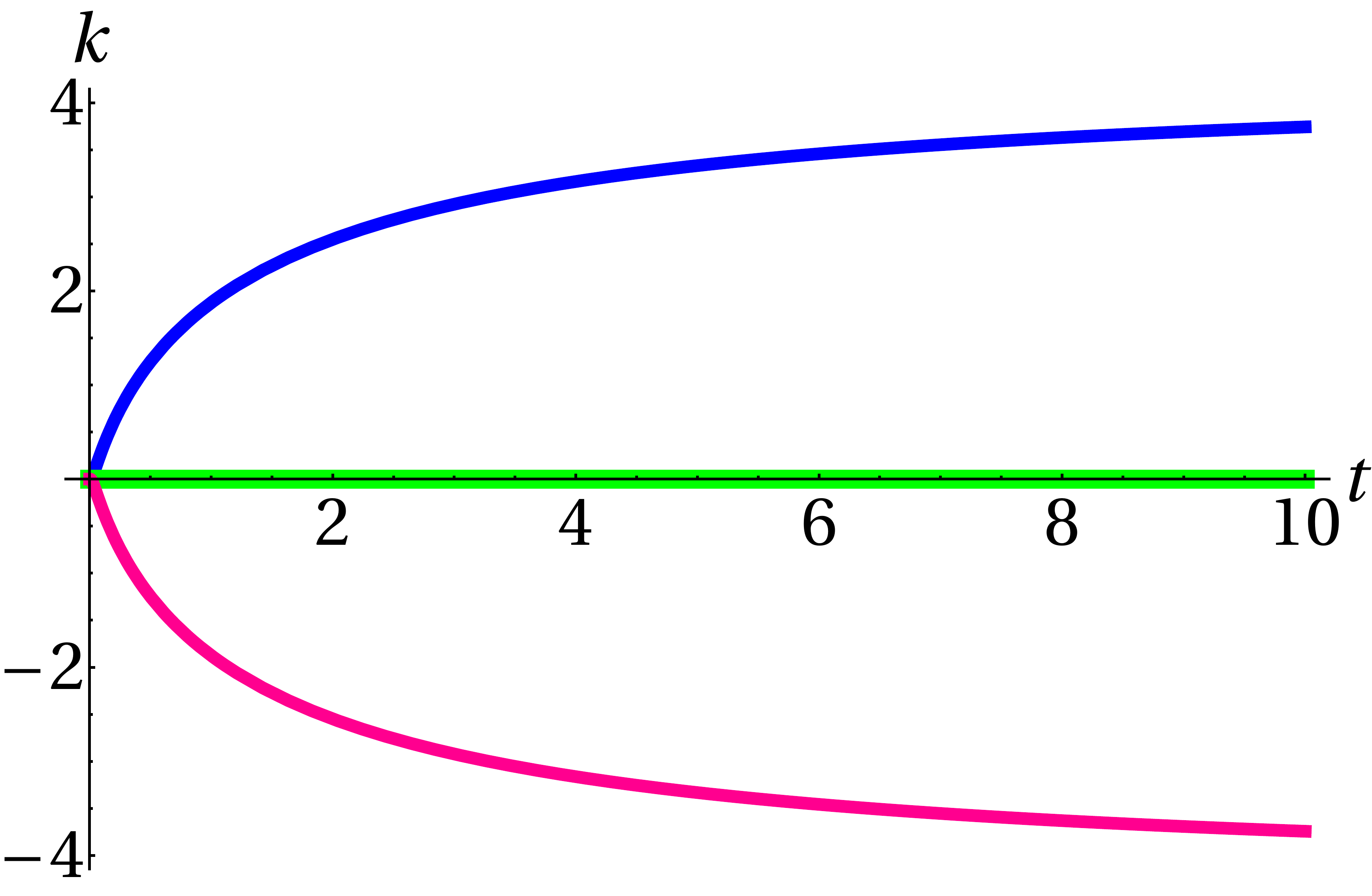}
    \caption{The harmonic function $k$ using: $\dot k\left(0\right)=1$ (blue curve), $\dot k\left(0\right)=0$ (green line), and $\dot k\left(0\right)=-1$ (red curve).}
    \label{k00001constantb}
  \end{subfigure}
  \caption{Initial conditions set number 1 for constant $b$.}
  \label{Fig38}
\end{figure}
\begin{figure}[H]
\begin{subfigure}[t]{.5\linewidth}
    \centering
    \includegraphics[width=0.7\columnwidth]{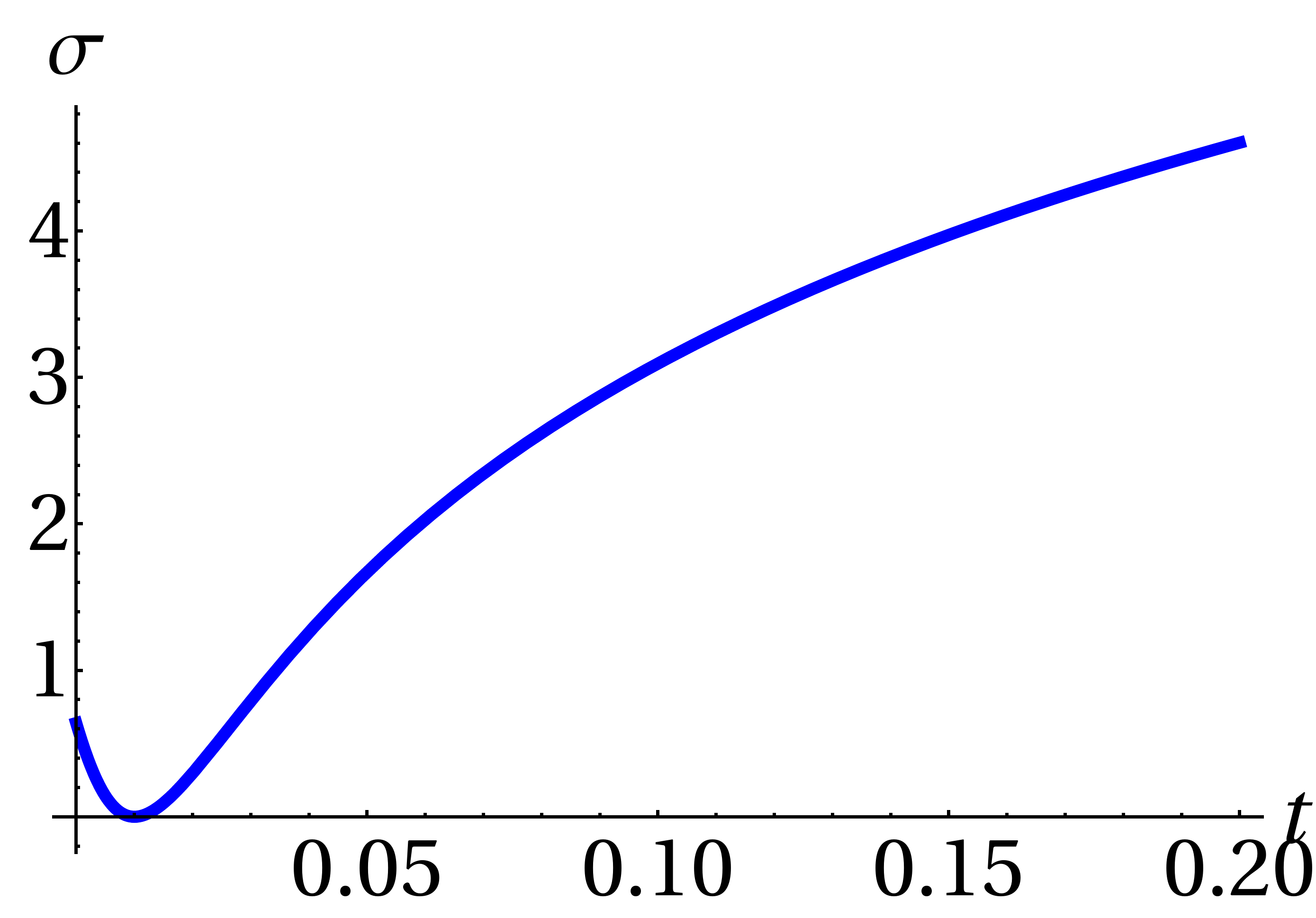}
    \caption{The dilaton $\sigma$; same for all three $\dot k\left(0\right)$.}
    \label{sigma00001constantb}
  \end{subfigure}
  \qquad
  \begin{subfigure}[t]{.5\linewidth}
    \centering
    \includegraphics[width=0.7\columnwidth]{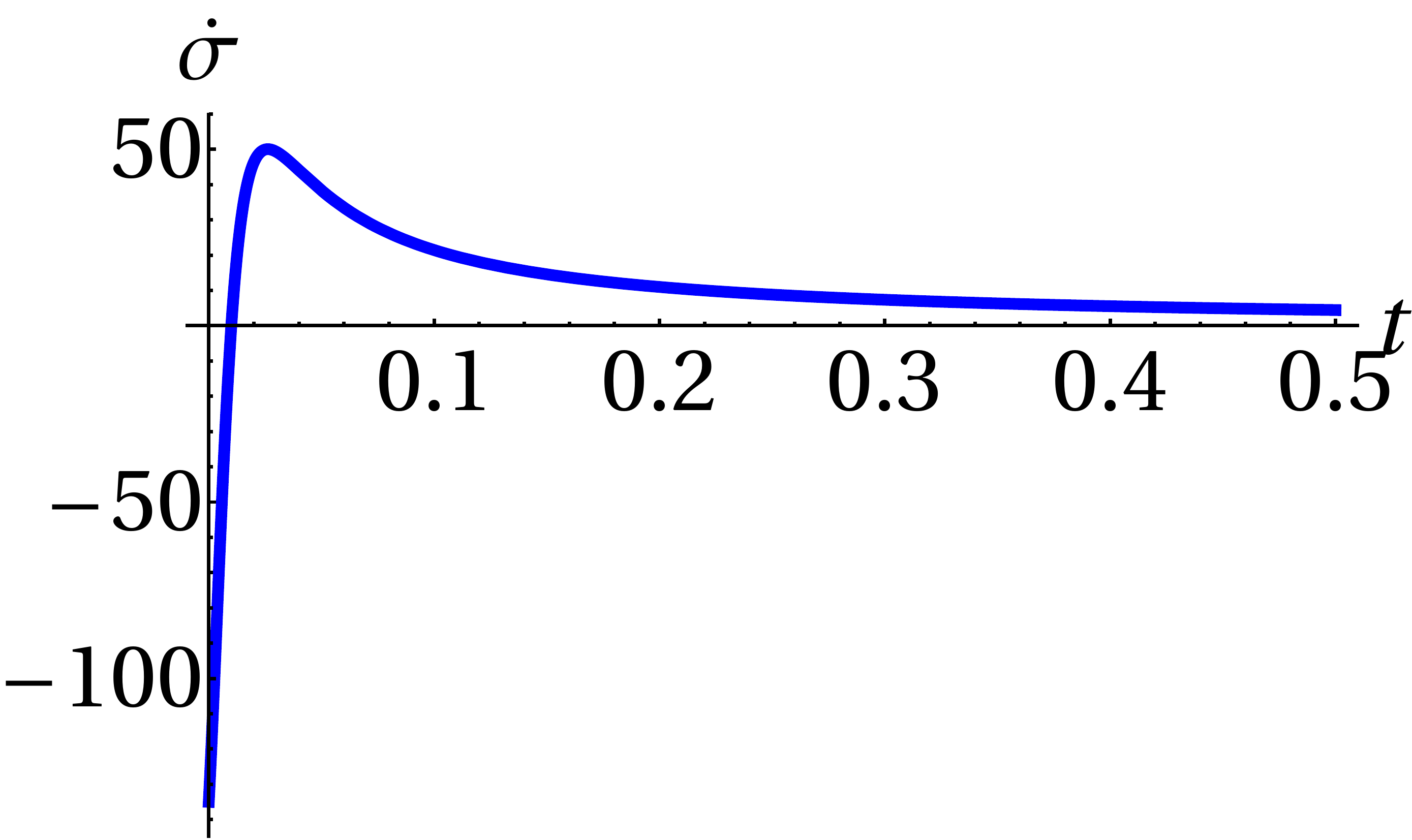}
    \caption{The dilatonic field strength $\dot\sigma$.}
    \label{sigmadot00001constantb}
  \end{subfigure}
\\[4em]
  \begin{subfigure}[t]{.5\linewidth}
    \centering
    \includegraphics[width=0.7\columnwidth]{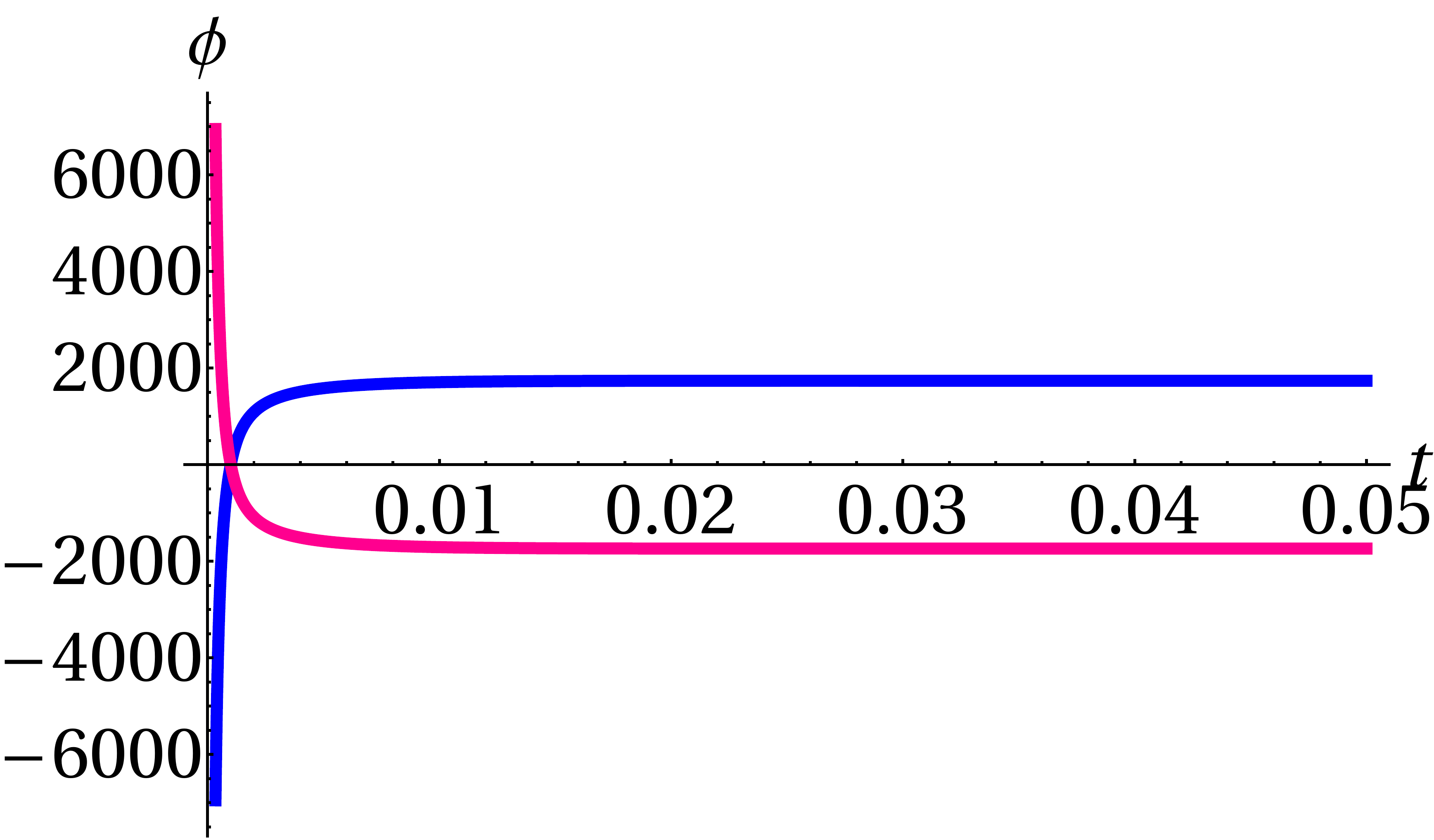}
    \caption{The universal axion $\phi$ for $\dot k\left(0\right) = 1$ (blue curve), and $\dot k\left(0\right) = -1$ (red curve). The solution diverges for $\dot k\left(0\right)=0$.}
    \label{phi00001constantb}
  \end{subfigure}
\qquad
  \begin{subfigure}[t]{.5\linewidth}
    \centering
    \includegraphics[width=0.7\columnwidth]{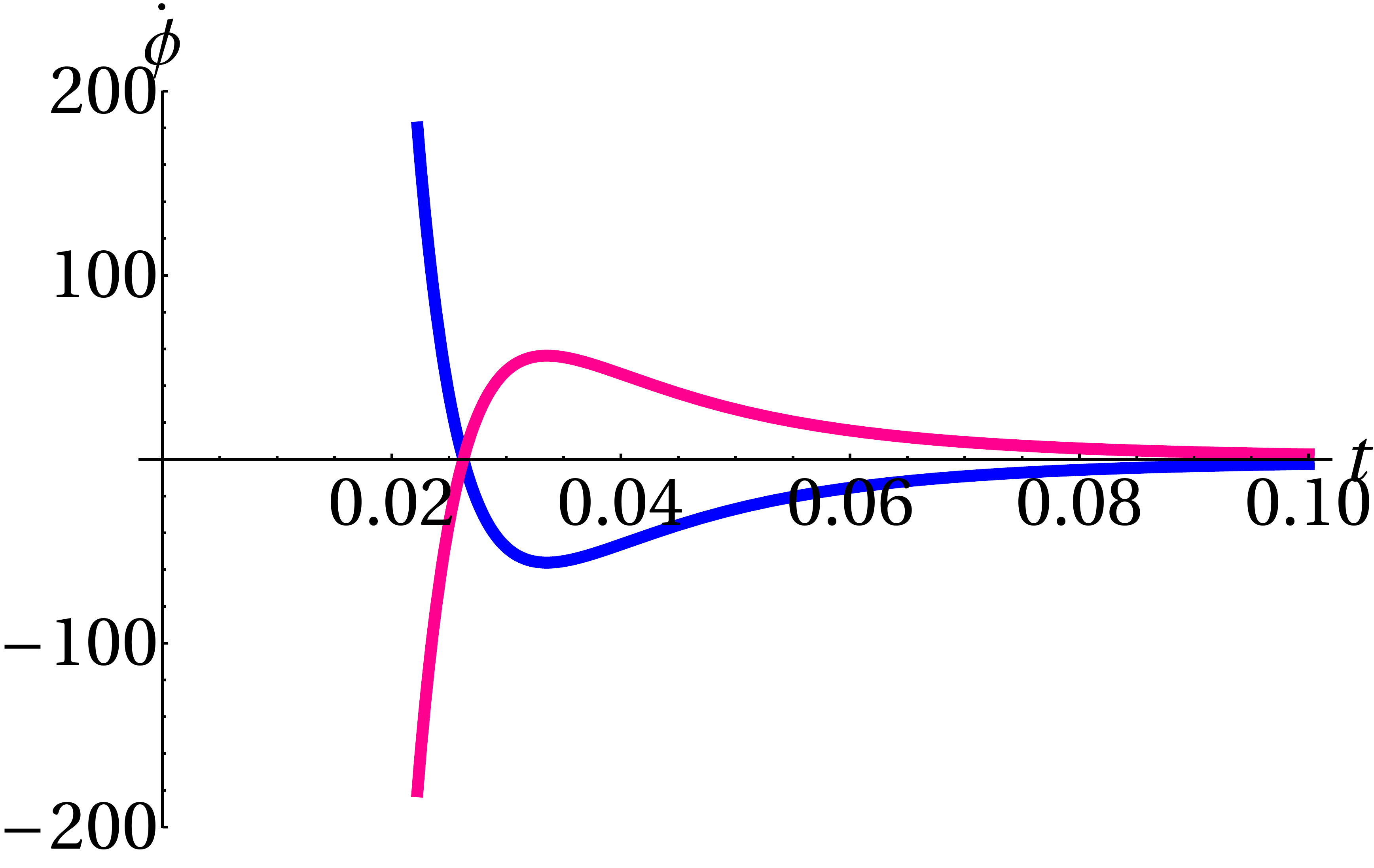}
    \caption{The axionic field strength $\dot\phi$ for $\dot k\left(0\right) = 1$ (blue curve), and $\dot k\left(0\right) = -1$ (red curve).}
    \label{phidot00001constantb}
  \end{subfigure}
\\[4em]
    \begin{subfigure}[t]{.5\linewidth}
    \centering
    \includegraphics[width=0.7\columnwidth]{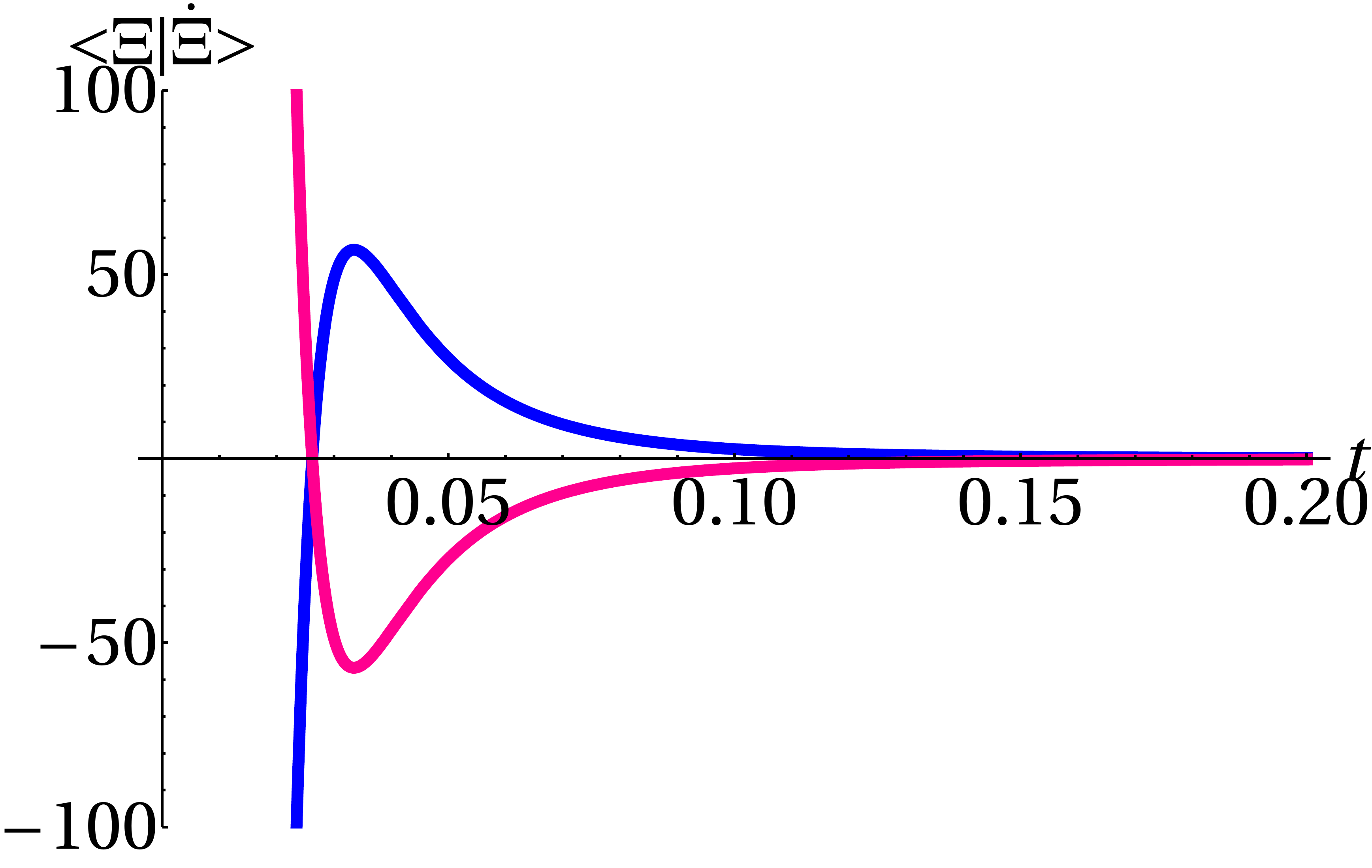}
    \caption{$ \langle \Xi | \dot \Xi \rangle$ for $\dot{k}(0)= 1$  (blue), and $\dot{k}(0)  = -1$ (red).}
    \label{xx00001constantb}
  \end{subfigure}
\qquad
 \begin{subfigure}[t]{.5\linewidth}
    \centering
    \includegraphics[width=0.7\columnwidth]{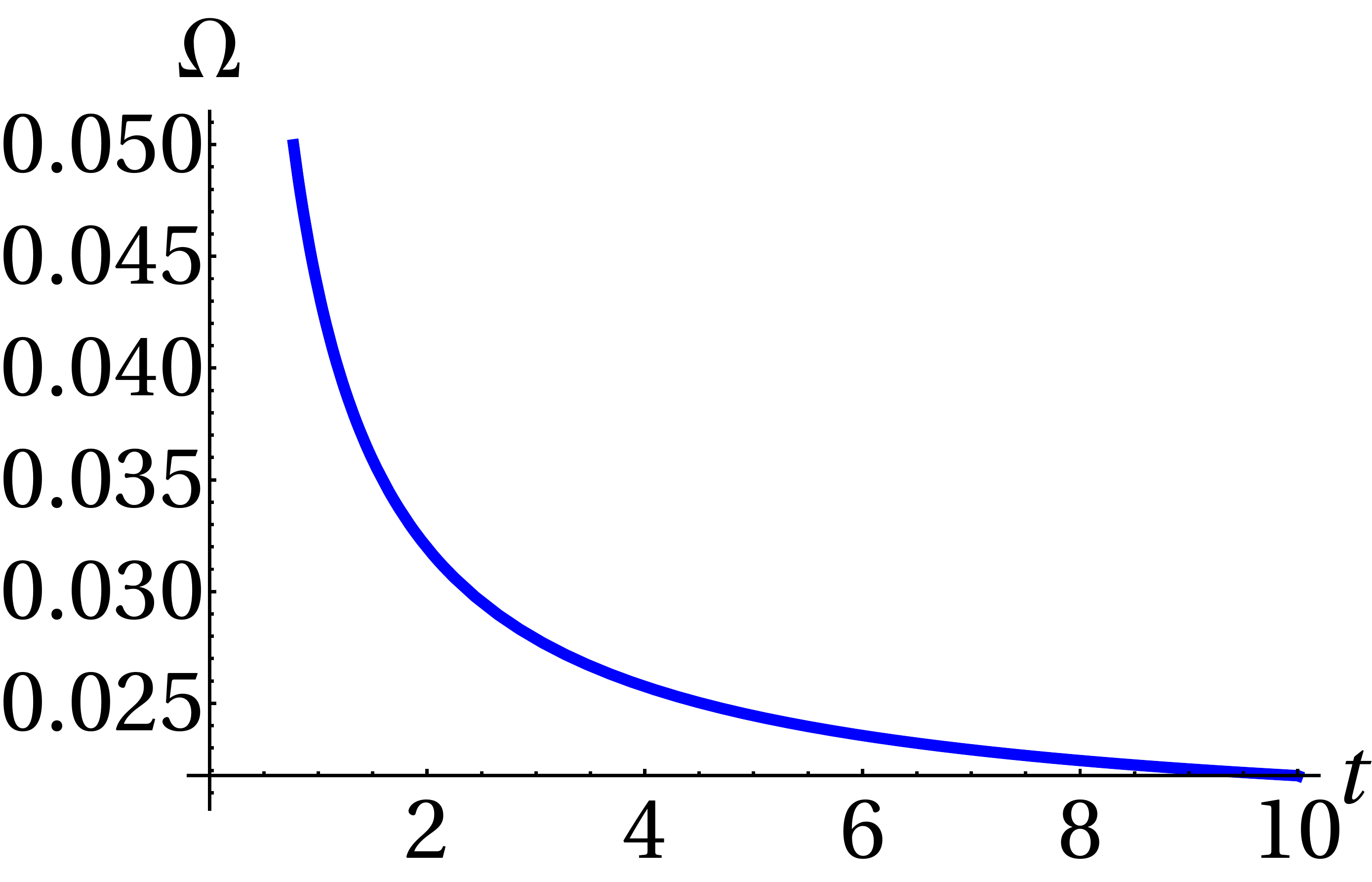}
    \caption{$\Omega$ at $\dot k\left(0\right) = 1$ and  $ \dot\sigma \left(0\right) =0 $.}
    \label{omega00001constantb}
  \end{subfigure}
\vspace{0.3cm}
  \caption{Initial conditions set number 1 for constant $b$ (continued).}
  \label{Fig37}
  \end{figure}


\begin{figure}[H]
  \begin{subfigure}[t]{.5\linewidth}
    \centering
    \includegraphics[width=0.7\columnwidth]{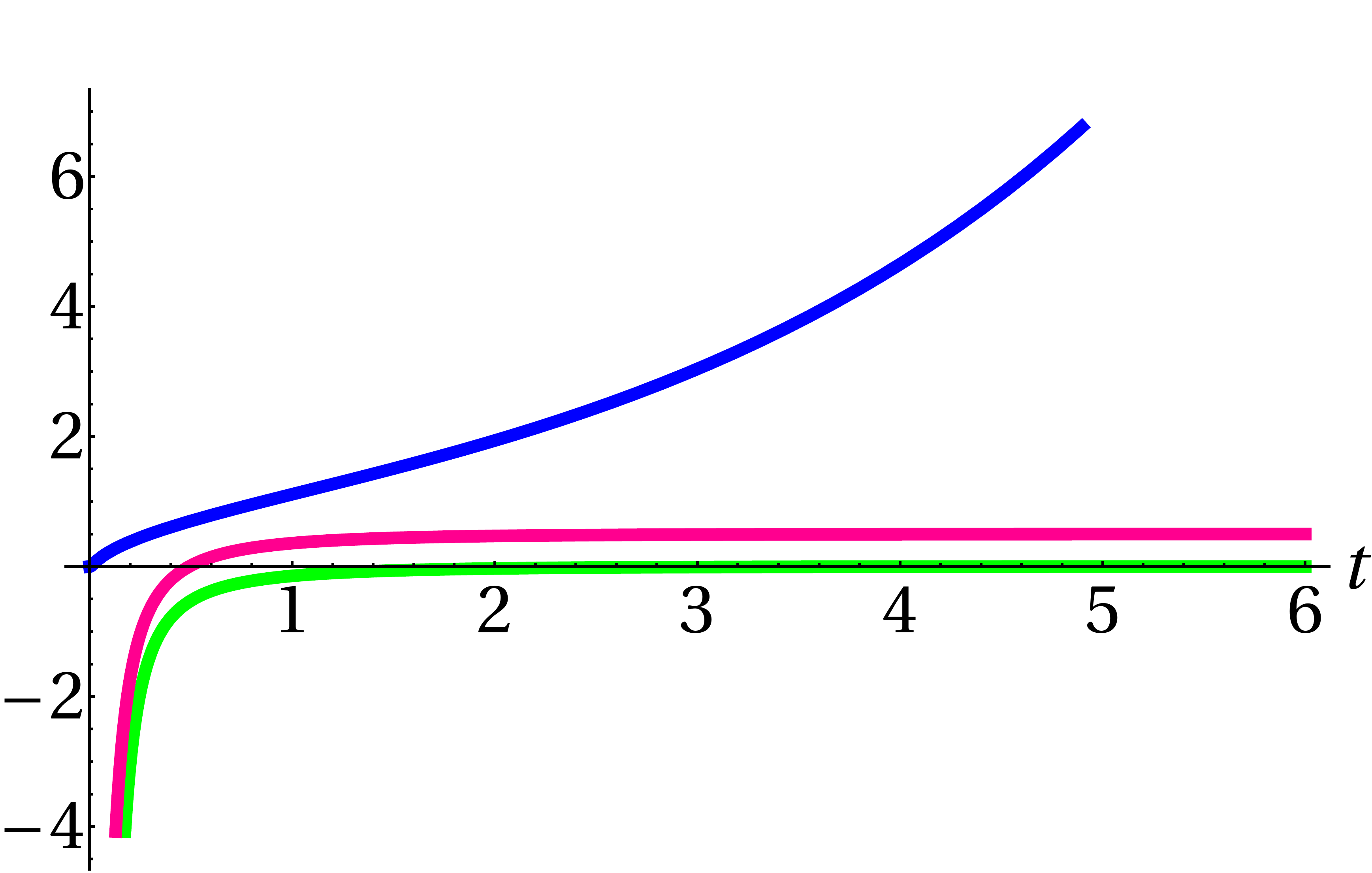}
    \caption{The scale factor $a$ is represented by the blue curve, $\Lambda$ by the red curve, while $ {G_{i\bar j} \dot z^i \dot z^{\bar j}} $ by the green curve.}
    \label{aLzz00002constantb}
  \end{subfigure}
\qquad
  \begin{subfigure}[t]{.5\linewidth}
    \centering
    \includegraphics[width=0.7\columnwidth]{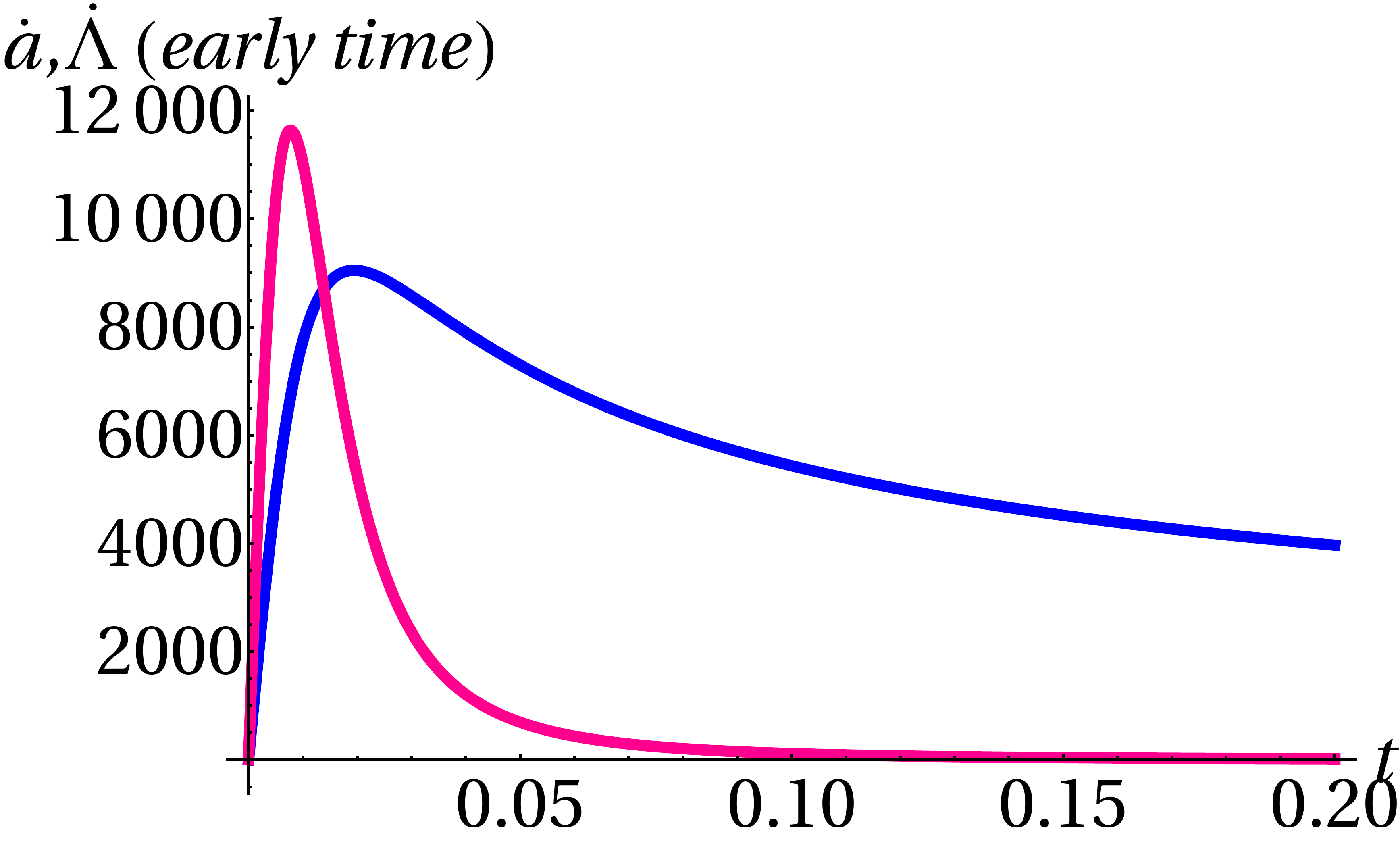}
    \caption{The early time expansion rates: $\dot a$ is represented by the blue curve, and $\dot \Lambda$ by the red curve. The curve for $\dot a$ is scaled up by a factor of 3000.}
    \label{adotLdotEARLY00002constantb}
  \end{subfigure}
\\[4em]
  \begin{subfigure}[t]{.5\linewidth}
    \centering
    \includegraphics[width=0.7\columnwidth]{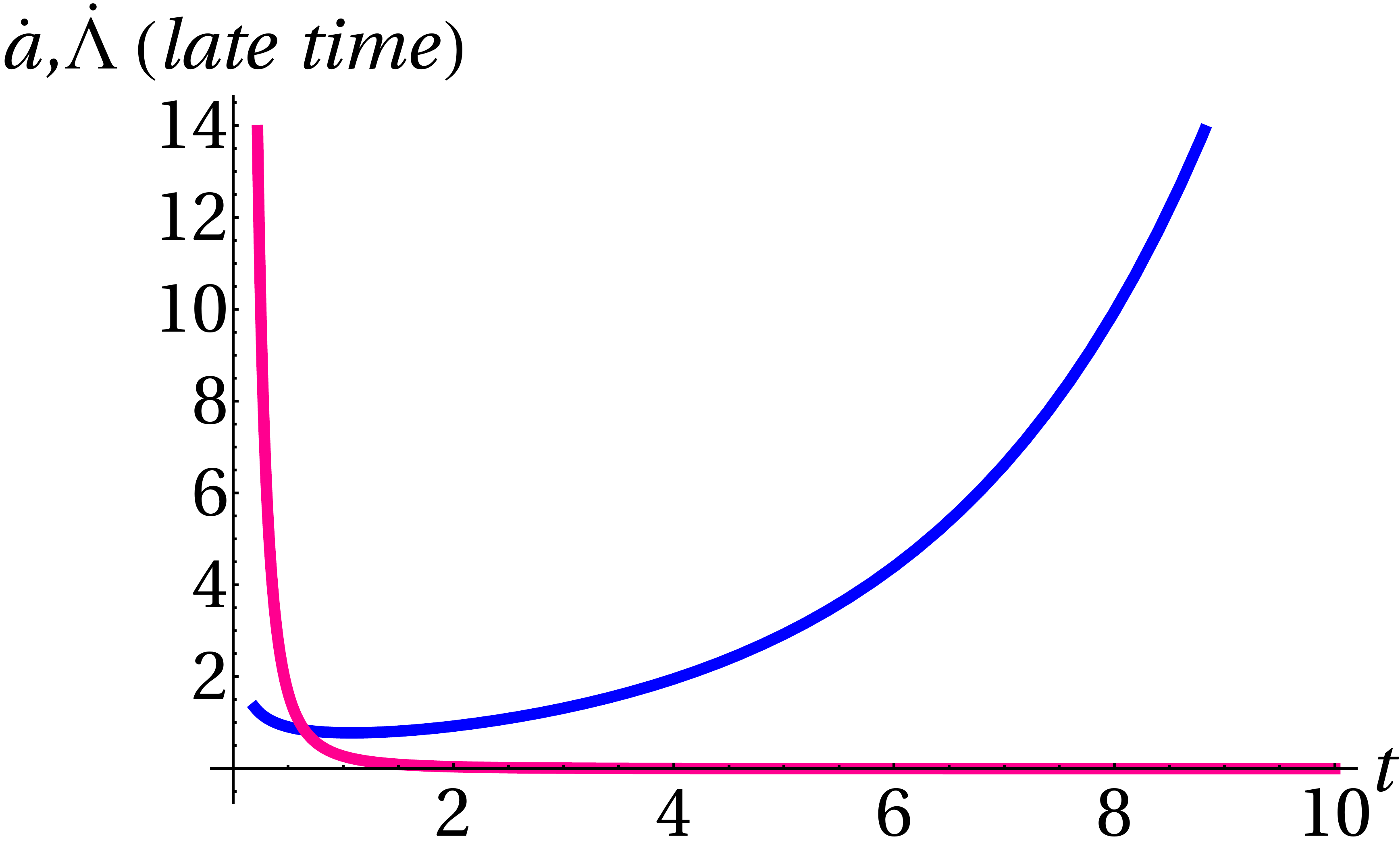}
    \caption{The late time expansion rates: $\dot a$ is represented by the blue curve, and $\dot \Lambda$ by the red curve.}
    \label{adotLdotLATE00002constantb}
  \end{subfigure}
\qquad
  \begin{subfigure}[t]{.5\linewidth}
    \centering
    \includegraphics[width=0.7\columnwidth]{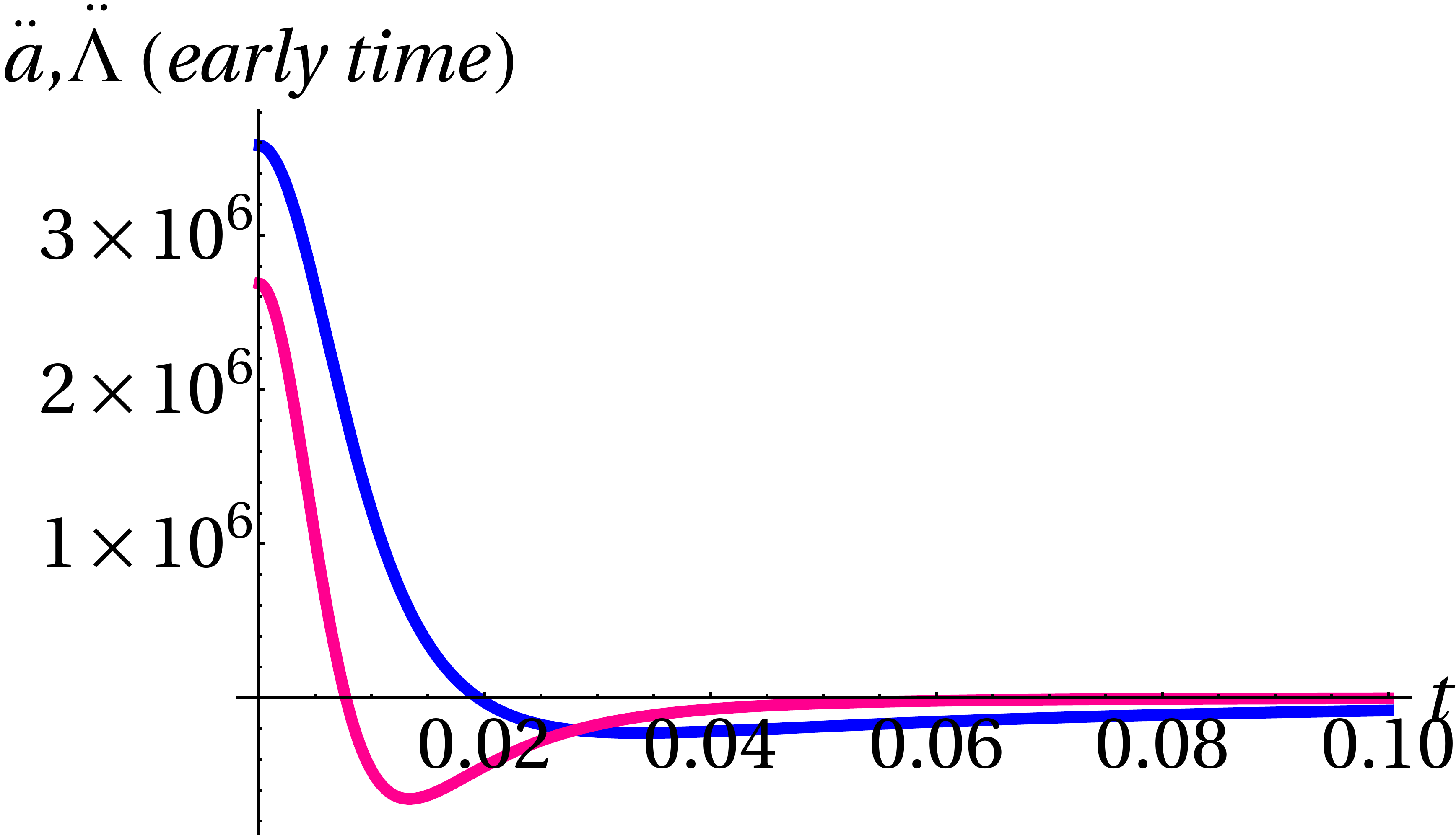}
    \caption{The early time accelerations: $\ddot a$ is represented by the blue curve, and $\ddot \Lambda$ by the rred curve. The curve for $\dot a$ is scaled up by a factor of 10000.}
    \label{addotLddotEARLY00002constantb}
  \end{subfigure}
\\[4em]
  \begin{subfigure}[t]{.5\linewidth}
    \centering
    \includegraphics[width=0.7\columnwidth]{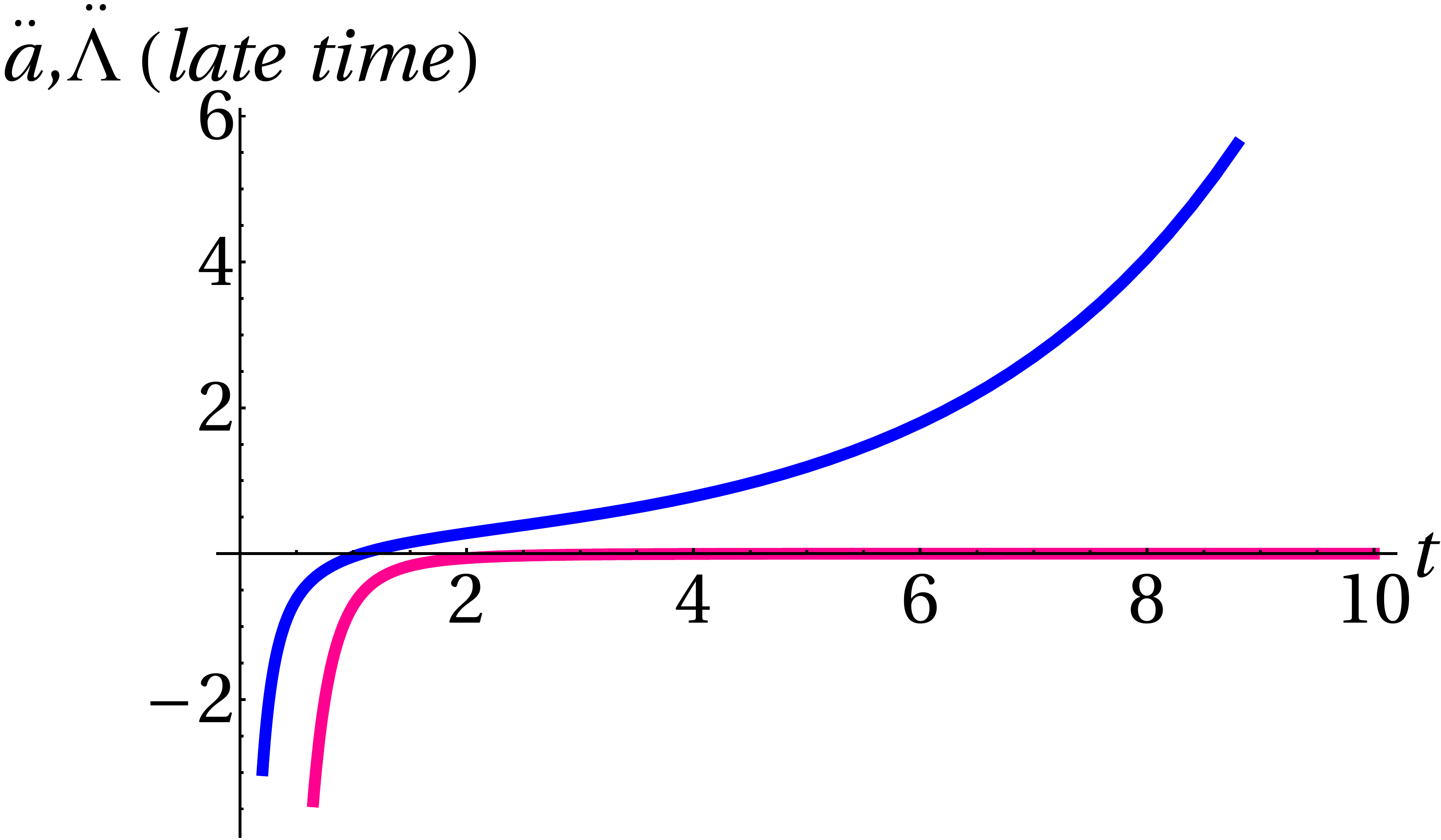}
    \caption{The late time accelerations: $\ddot a$ is represented by the blue curve, and $\ddot \Lambda$ by the red curve.}
    \label{addotLddotlate00002constantb}
  \end{subfigure}
\qquad
  \begin{subfigure}[t]{.5\linewidth}
    \centering
    \includegraphics[width=0.7\columnwidth]{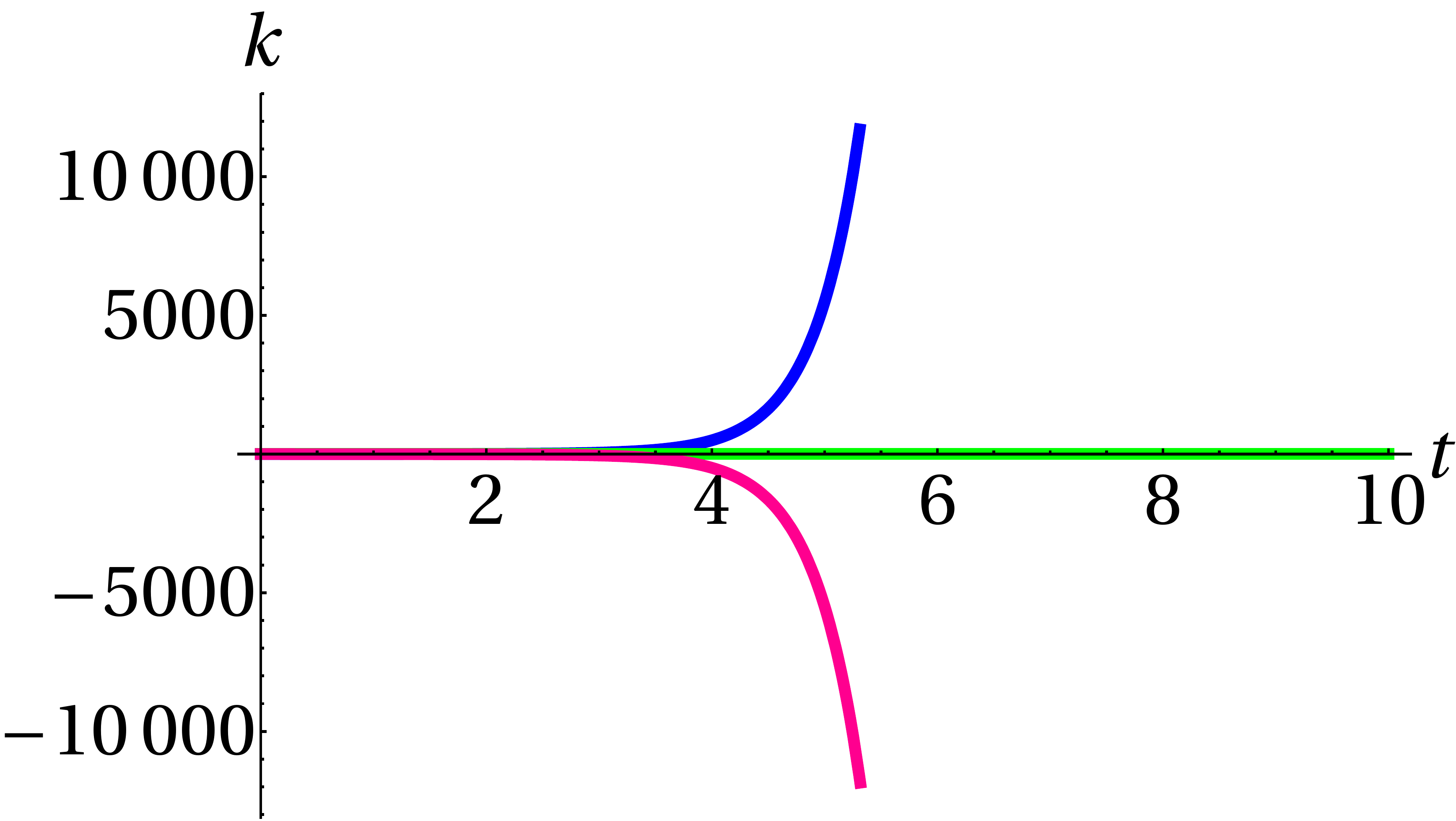}
    \caption{The harmonic function $k$ using: $\dot k\left(0\right)=1$ (blue curve), $\dot k\left(0\right)=0$ (green line), and $\dot k\left(0\right)=-1$ (red curve).}
    \label{k00002constantb}
  \end{subfigure}
    \caption{Initial conditions set number 2 for constant $b$.}
  \label{Fig391}
  \end{figure}


\begin{figure}[H]
  \begin{subfigure}[t]{.5\linewidth}
    \centering
    \includegraphics[width=0.7\columnwidth]{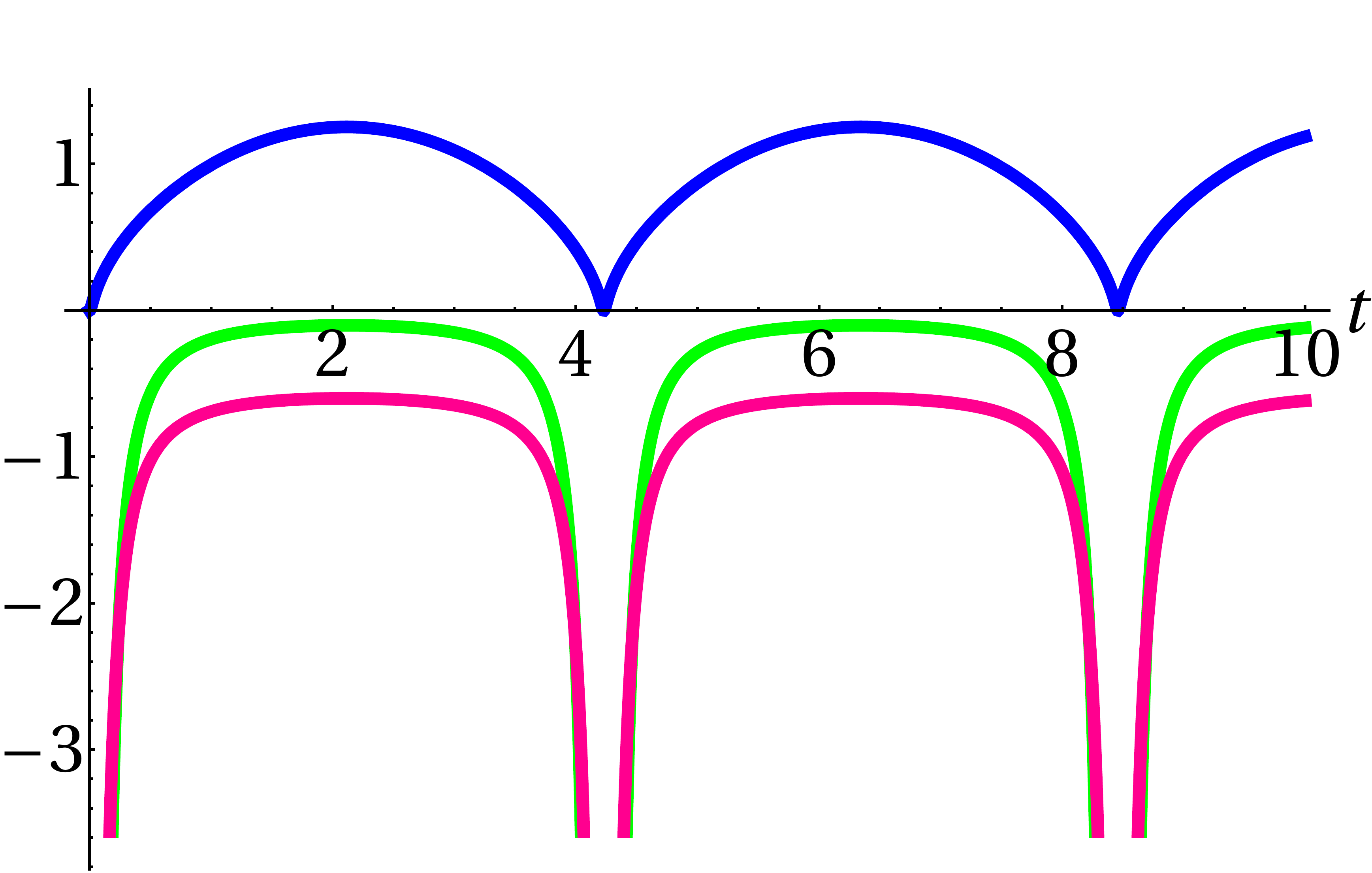}
    \caption{The scale factor $a$ is represented by the blue curve, $\Lambda$ by the red curve, while $ {G_{i\bar j} \dot z^i \dot z^{\bar j}} $ by the green curve.}
    \label{aLzz00003constantb}
  \end{subfigure}
\qquad
  \begin{subfigure}[t]{.5\linewidth}
    \centering
    \includegraphics[width=0.7\columnwidth]{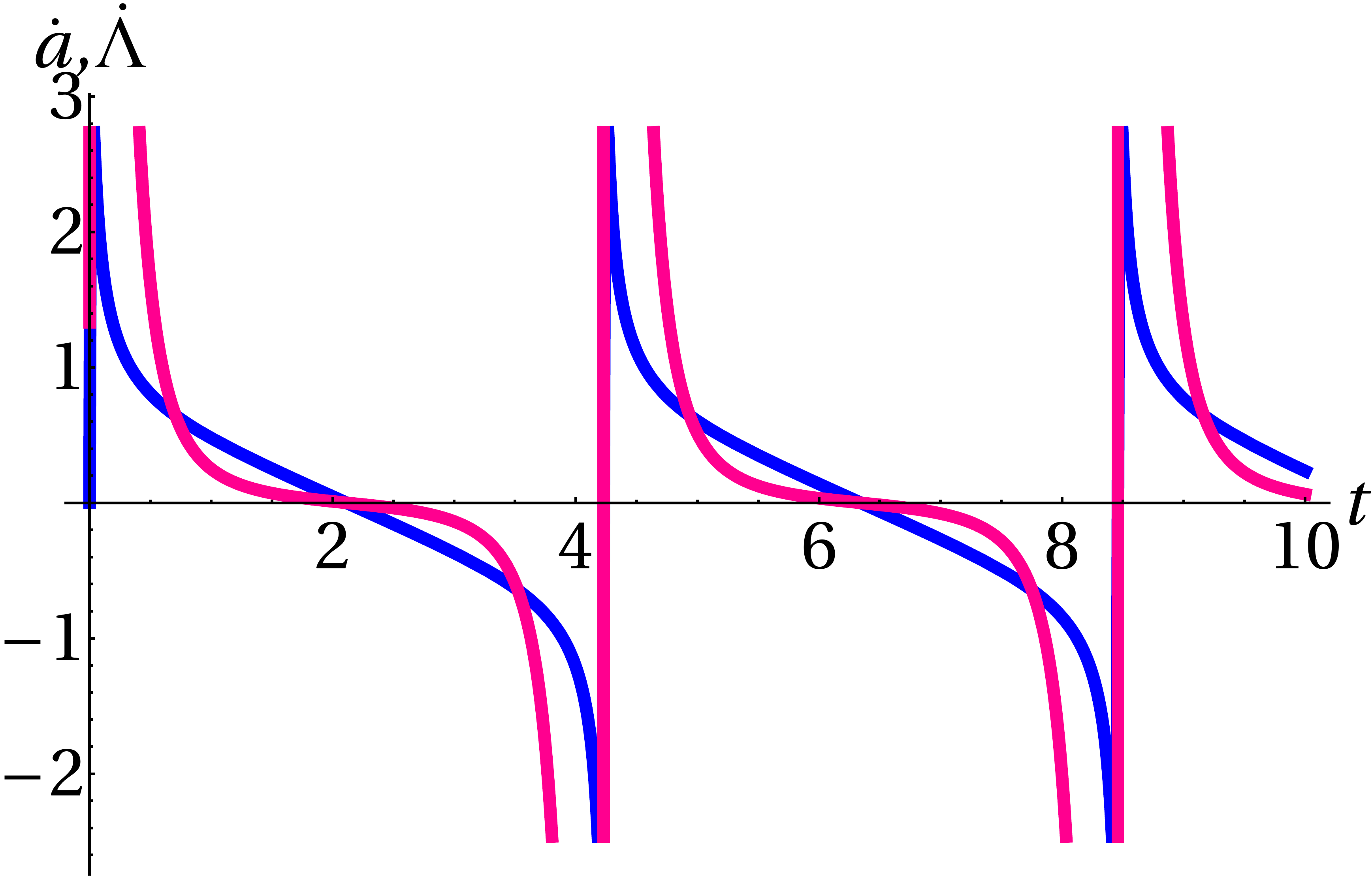}
    \caption{The expansion rates: $\dot a$ is represented by the blue curve, and $\dot \Lambda$ by the red curve.}
    \label{adotLdot00003constantb}
  \end{subfigure}
   \\[9em]
  \begin{subfigure}[t]{.5\linewidth}
    \centering
    \includegraphics[width=0.7\columnwidth]{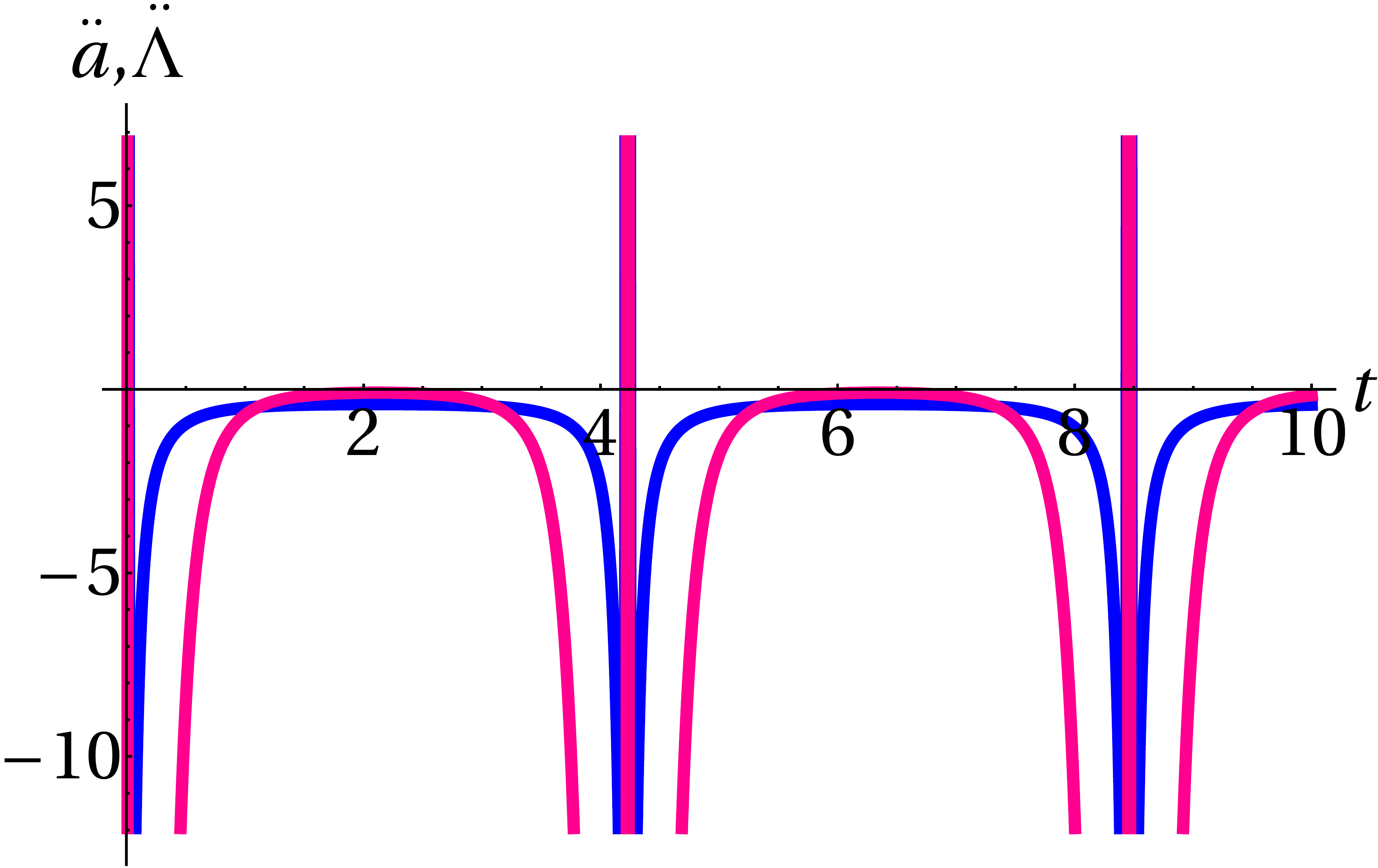}
    \caption{The accelerations: $\ddot a$ is represented by the blue curve, and $\ddot \Lambda$ by the red curve.}
    \label{addotLddot00003constantb}
  \end{subfigure}
\qquad
  \begin{subfigure}[t]{.5\linewidth}
    \centering
    \includegraphics[width=0.7\columnwidth]{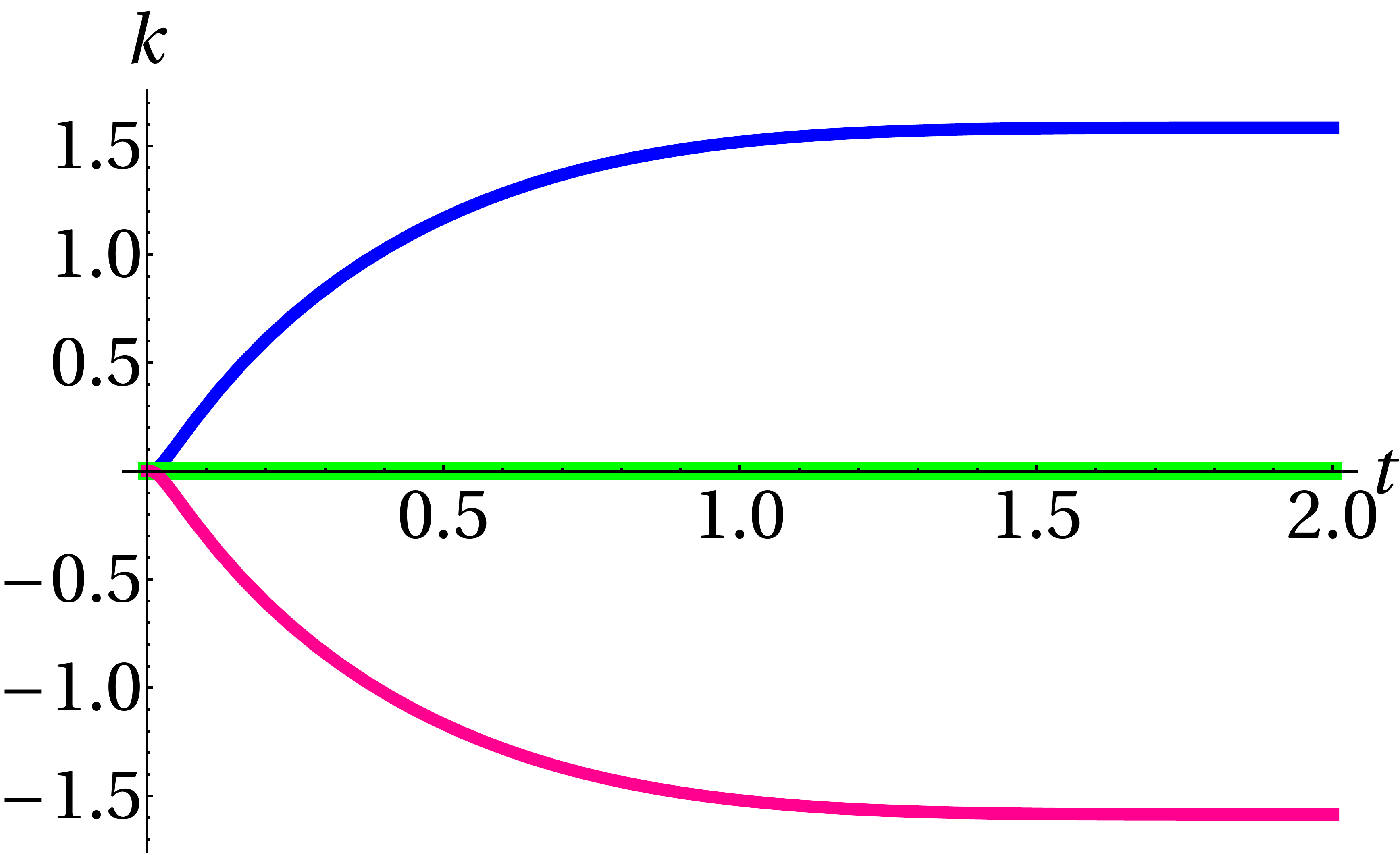}
    \caption{The harmonic function $k$ using: $\dot k\left(0\right)=1$ (blue curve), $\dot k\left(0\right)=0$ (green line), and $\dot k\left(0\right)=-1$ (red curve).}
    \label{k00003constantb}
  \end{subfigure}
\caption{Initial conditions set number 3 for constant $b$.}
  \label{Fig42}
\end{figure}

\begin{figure}[H]
  \begin{subfigure}[t]{.5\linewidth}
    \centering
    \includegraphics[width=0.7\columnwidth]{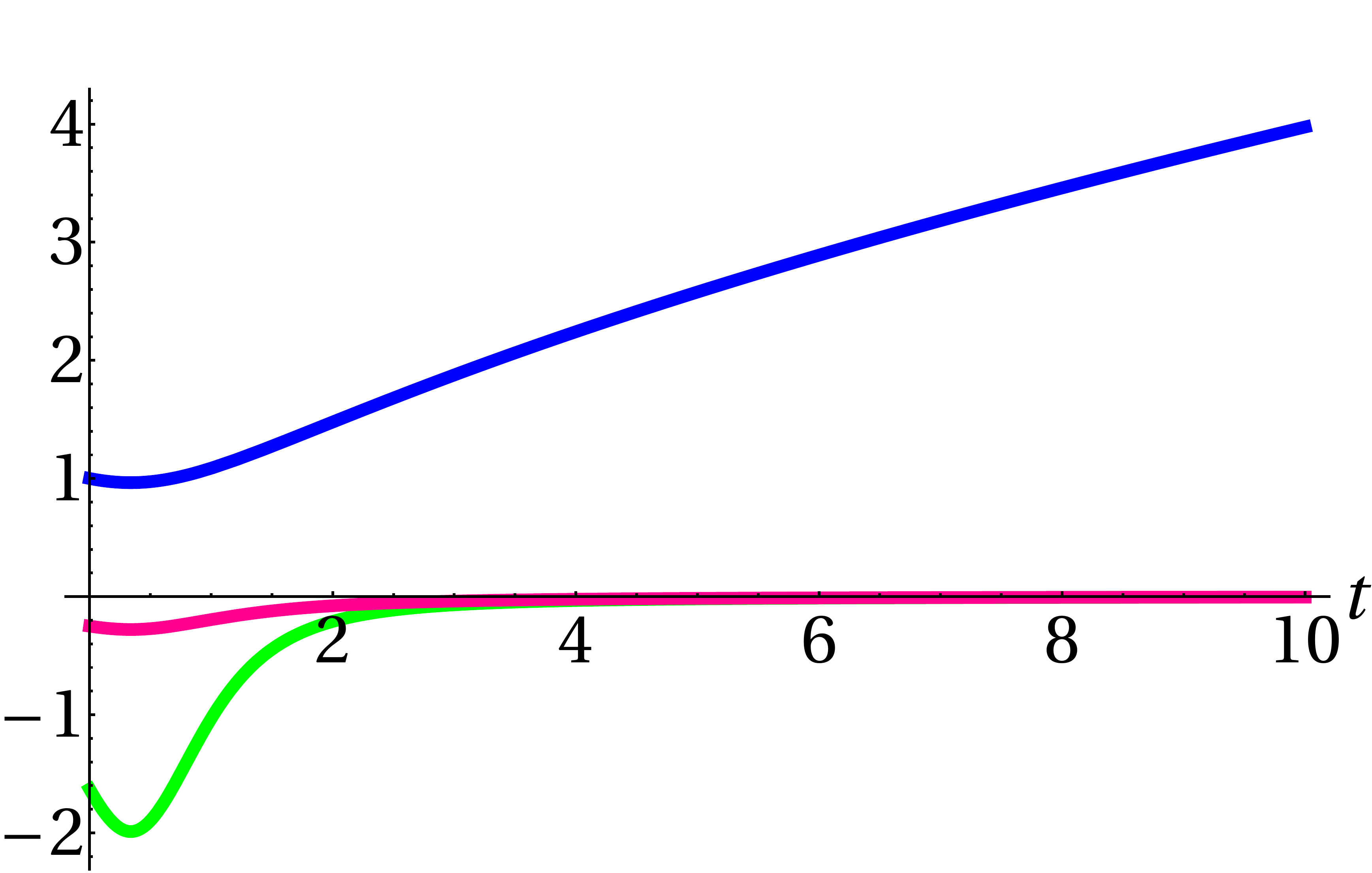}
    \caption{The scale factor $a$ is represented by the blue curve, $\Lambda$ by the red curve, while $ {G_{i\bar j} \dot z^i \dot z^{\bar j}} $ by the green curve.}
    \label{aLzz1-02-1-024constantb}
  \end{subfigure}
\qquad
  \begin{subfigure}[t]{.5\linewidth}
    \centering
    \includegraphics[width=0.7\columnwidth]{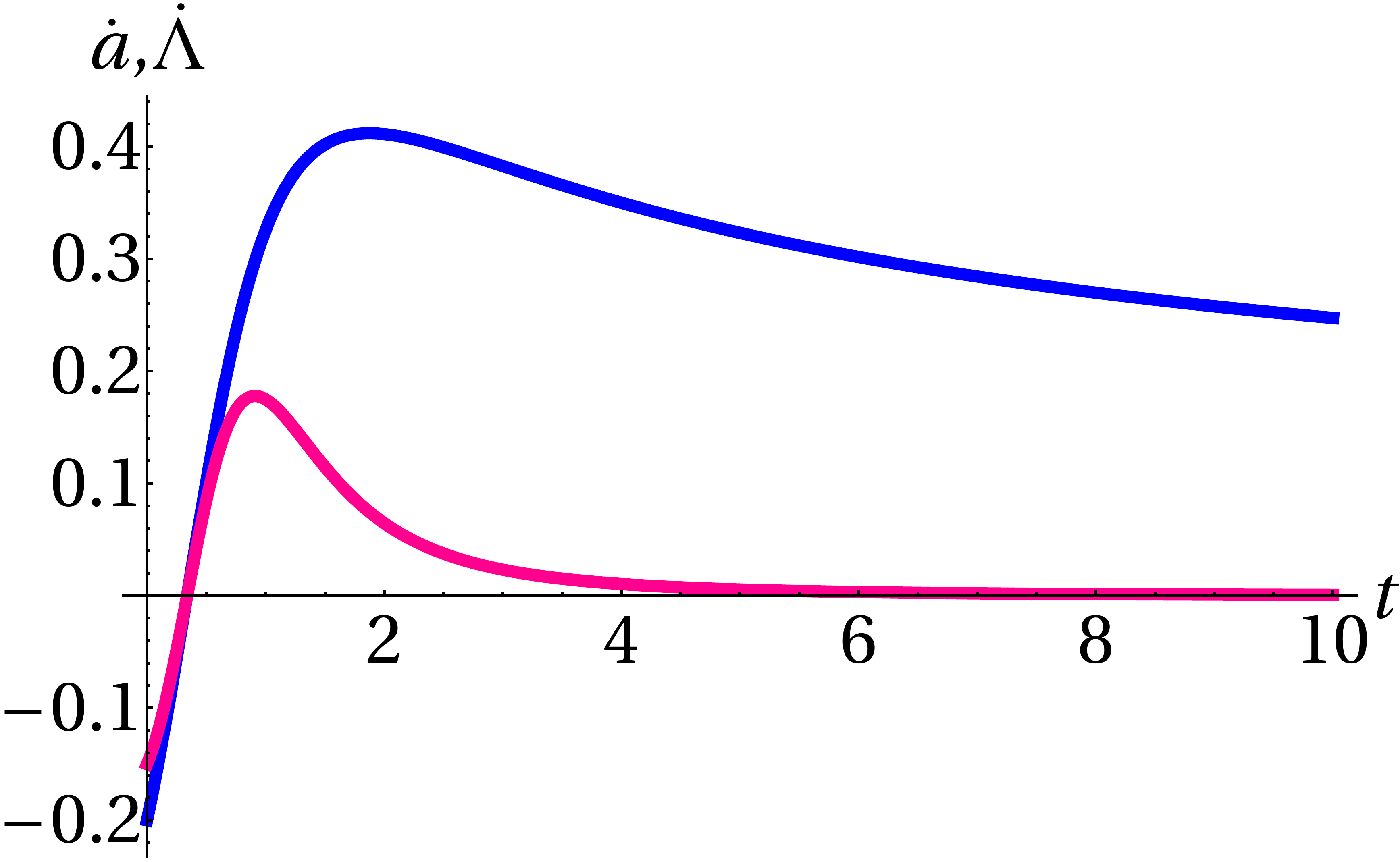}
    \caption{The expansion rates: $\dot a$ is represented by the blue curve, and $\dot \Lambda$ by the red curve.}
    \label{adotLdot1-02-1-024constantb}
  \end{subfigure}
\\[9em]
  \begin{subfigure}[t]{.5\linewidth}
    \centering
    \includegraphics[width=0.7\columnwidth]{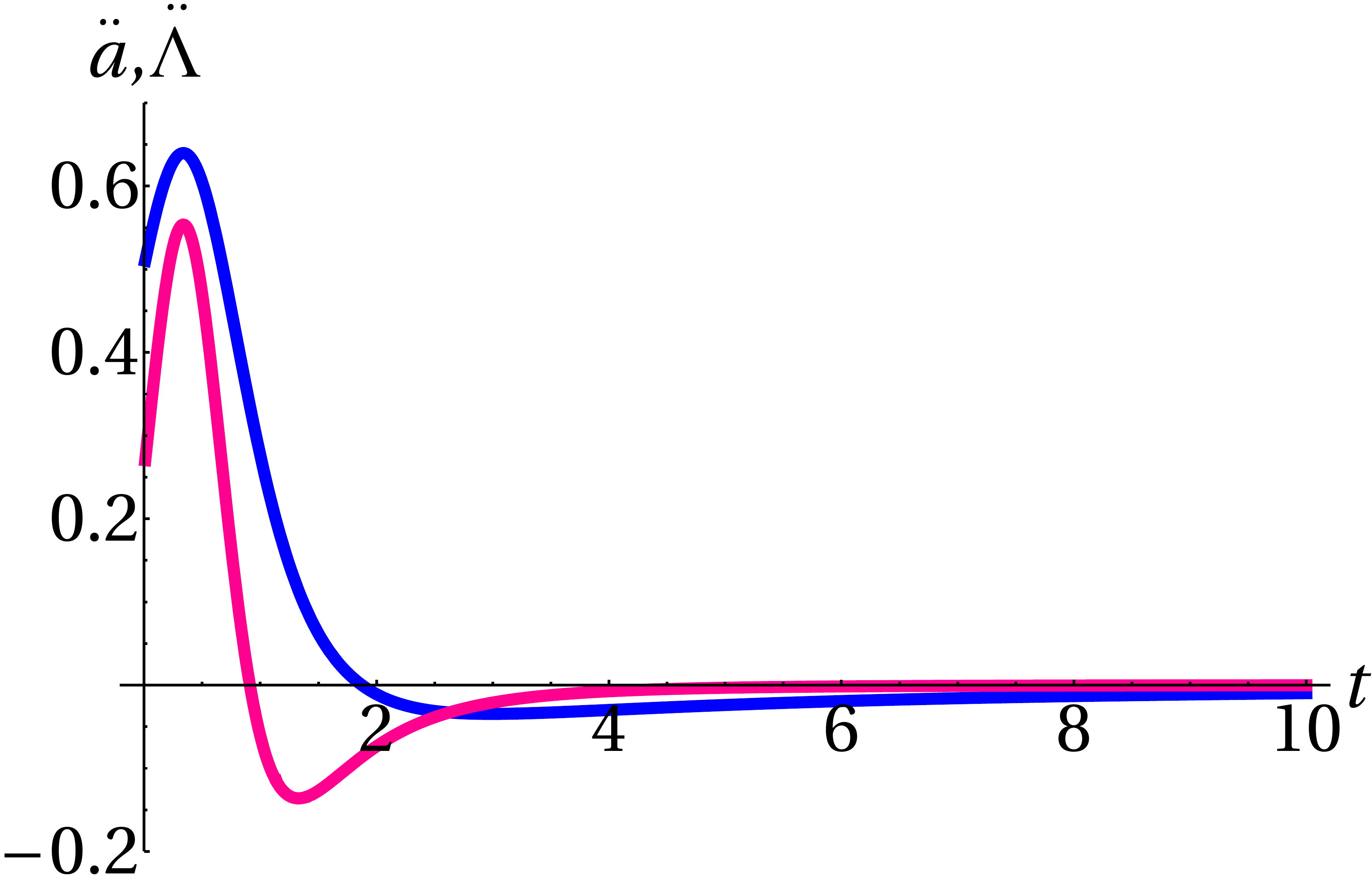}
    \caption{The accelerations: $\ddot a$ is represented by the blue curve, and $\ddot \Lambda$ by the red curve.}
    \label{addotLddot1-02-1-024constantb}
  \end{subfigure}
\qquad
  \begin{subfigure}[t]{.5\linewidth}
    \centering
    \includegraphics[width=0.7\columnwidth]{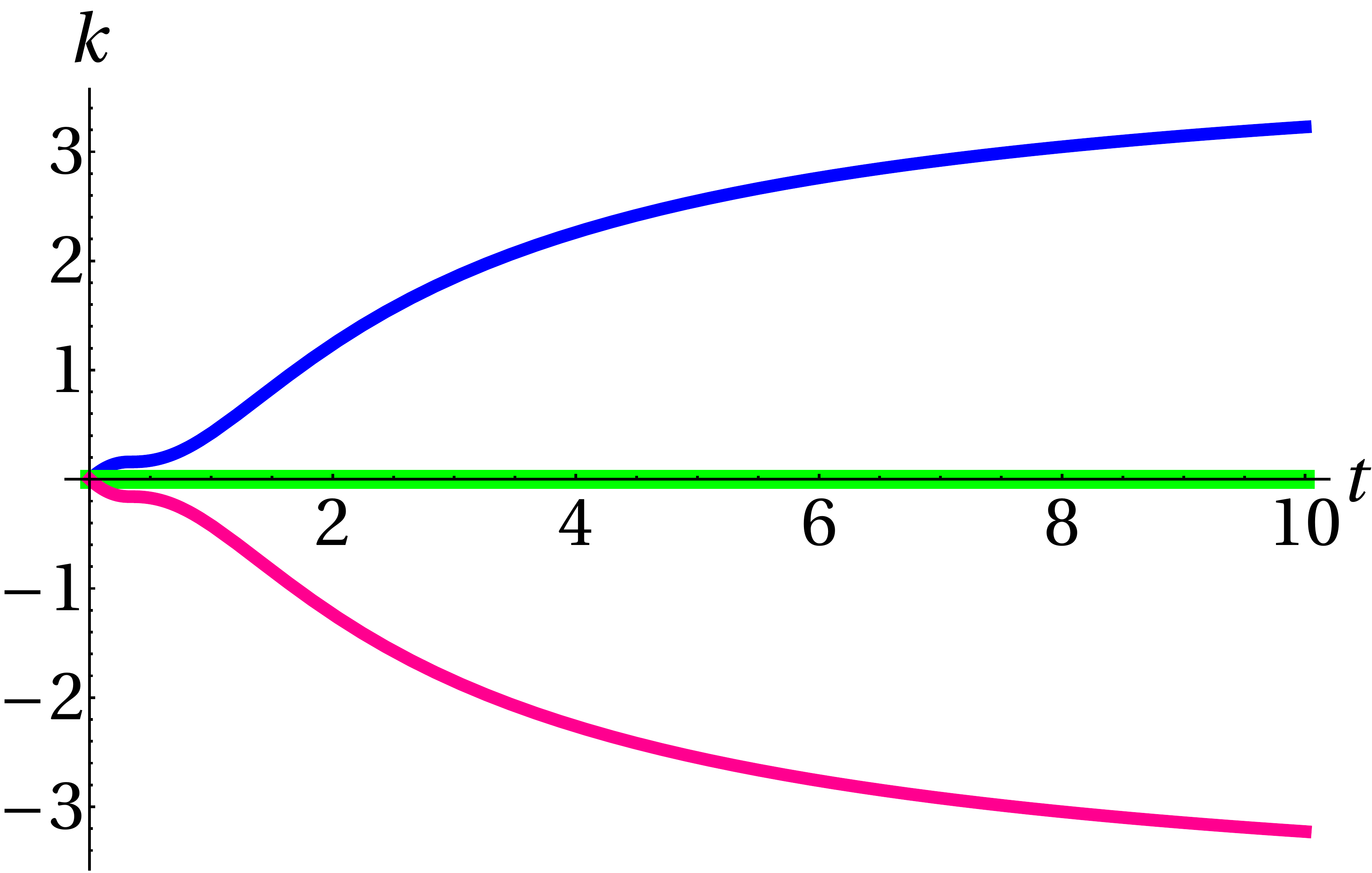}
    \caption{The harmonic function $k$ using: $\dot k\left(0\right)=1$ (blue curve), $\dot k\left(0\right)=0$ (green line), and $\dot k\left(0\right)=-1$ (red curve).}
    \label{k1-02-1-024constantb}
  \end{subfigure}
\caption{Initial conditions set number 4 for constant $b$.}
  \label{Fig44}
\end{figure}


\begin{figure}[H]
  \begin{subfigure}[t]{.5\linewidth}
    \centering
    \includegraphics[width=0.7\columnwidth]{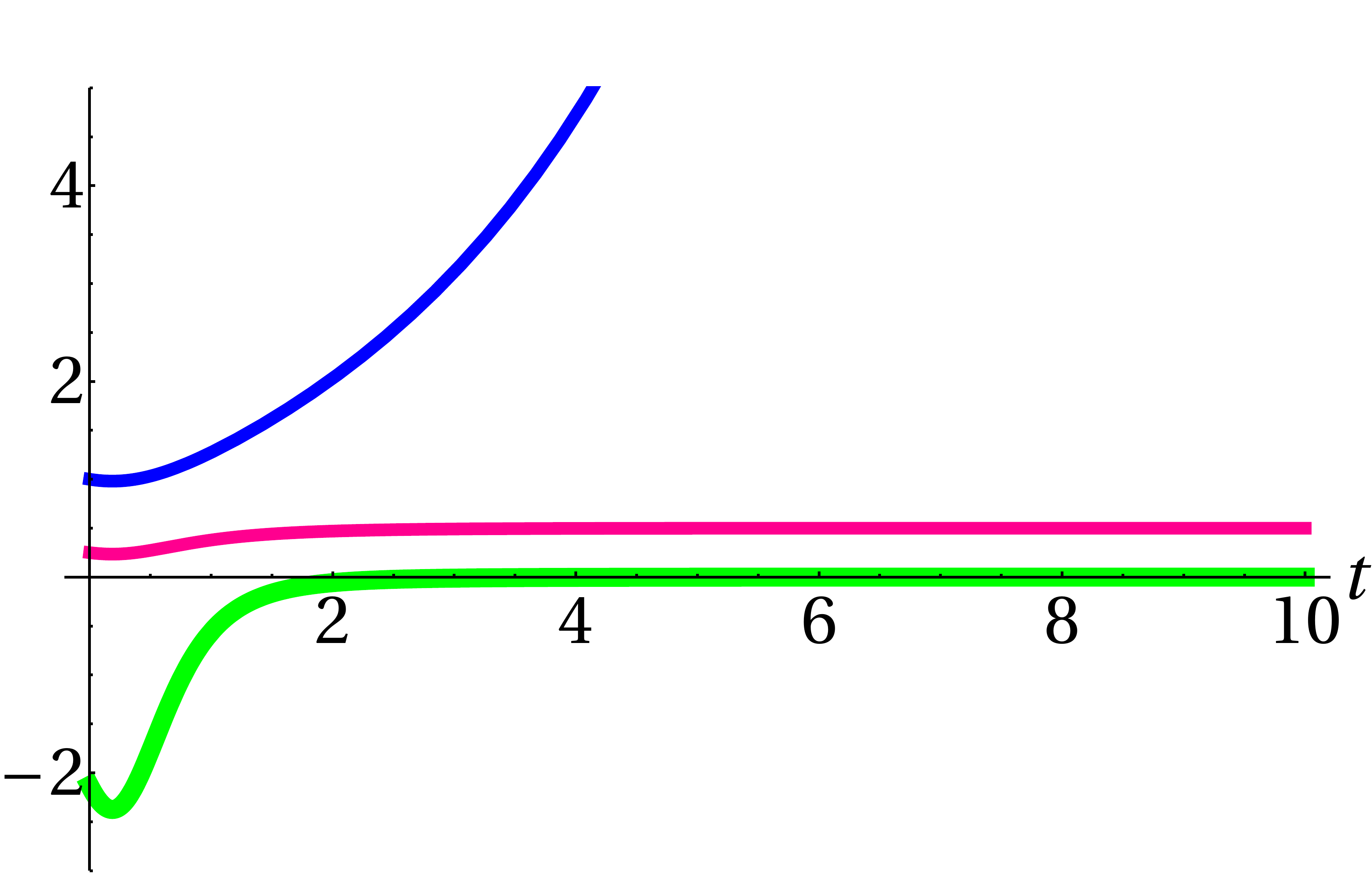}
    \caption{The scale factor $a$ is represented by the blue curve, $\Lambda$ by the red curve, while $ {G_{i\bar j} \dot z^i \dot z^{\bar j}} $ by the green curve.}
    \label{aLzz1-02-1-025constantb}
  \end{subfigure}
\qquad
  \begin{subfigure}[t]{.5\linewidth}
    \centering
    \includegraphics[width=0.7\columnwidth]{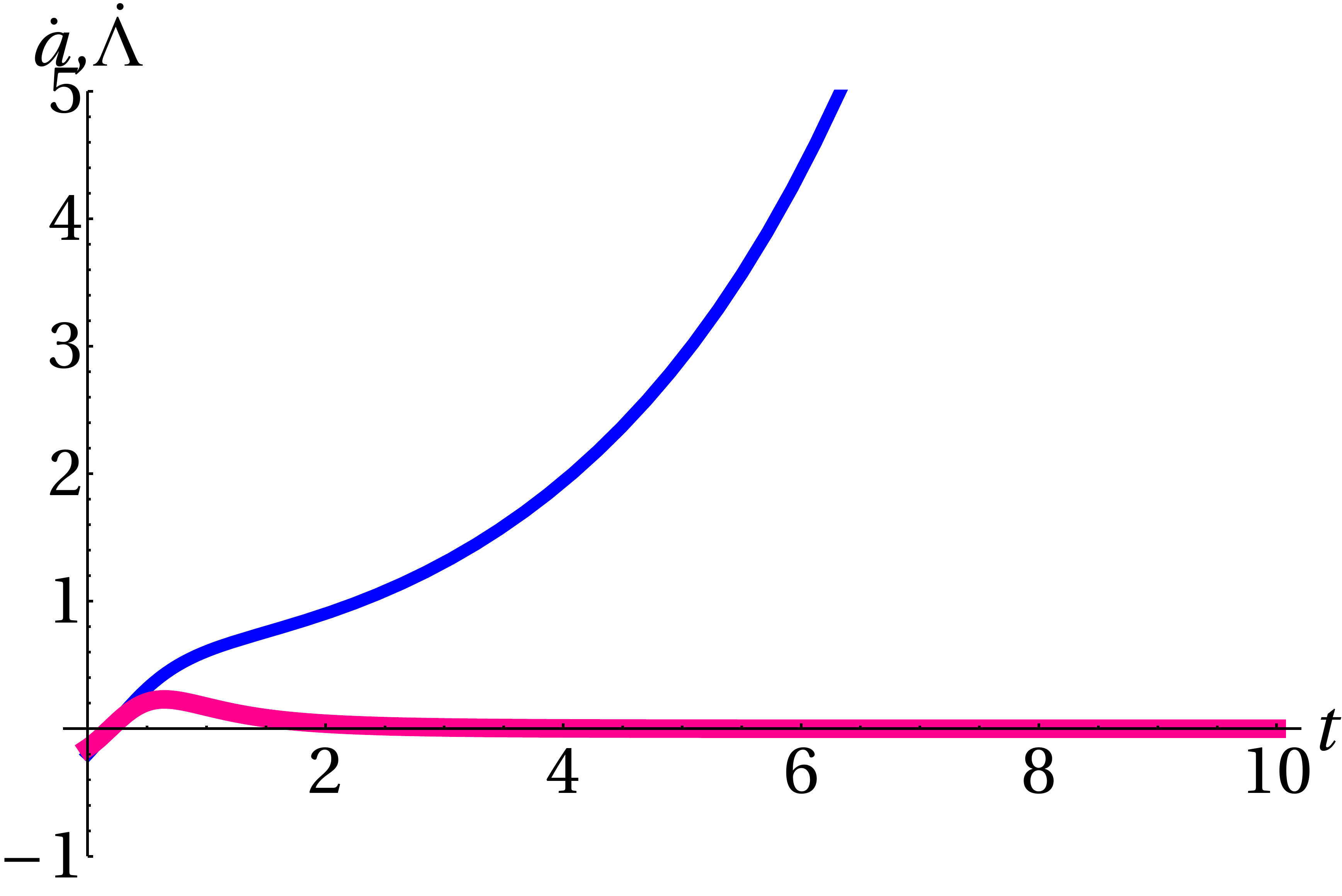}
    \caption{The expansion rates: $\dot a$ is represented by the blue curve, and $\dot \Lambda$ by the red curve.}
    \label{adotLdot1-02-1-025constantb}
  \end{subfigure}
\\[9em]
  \begin{subfigure}[t]{.5\linewidth}
    \centering
    \includegraphics[width=0.7\columnwidth]{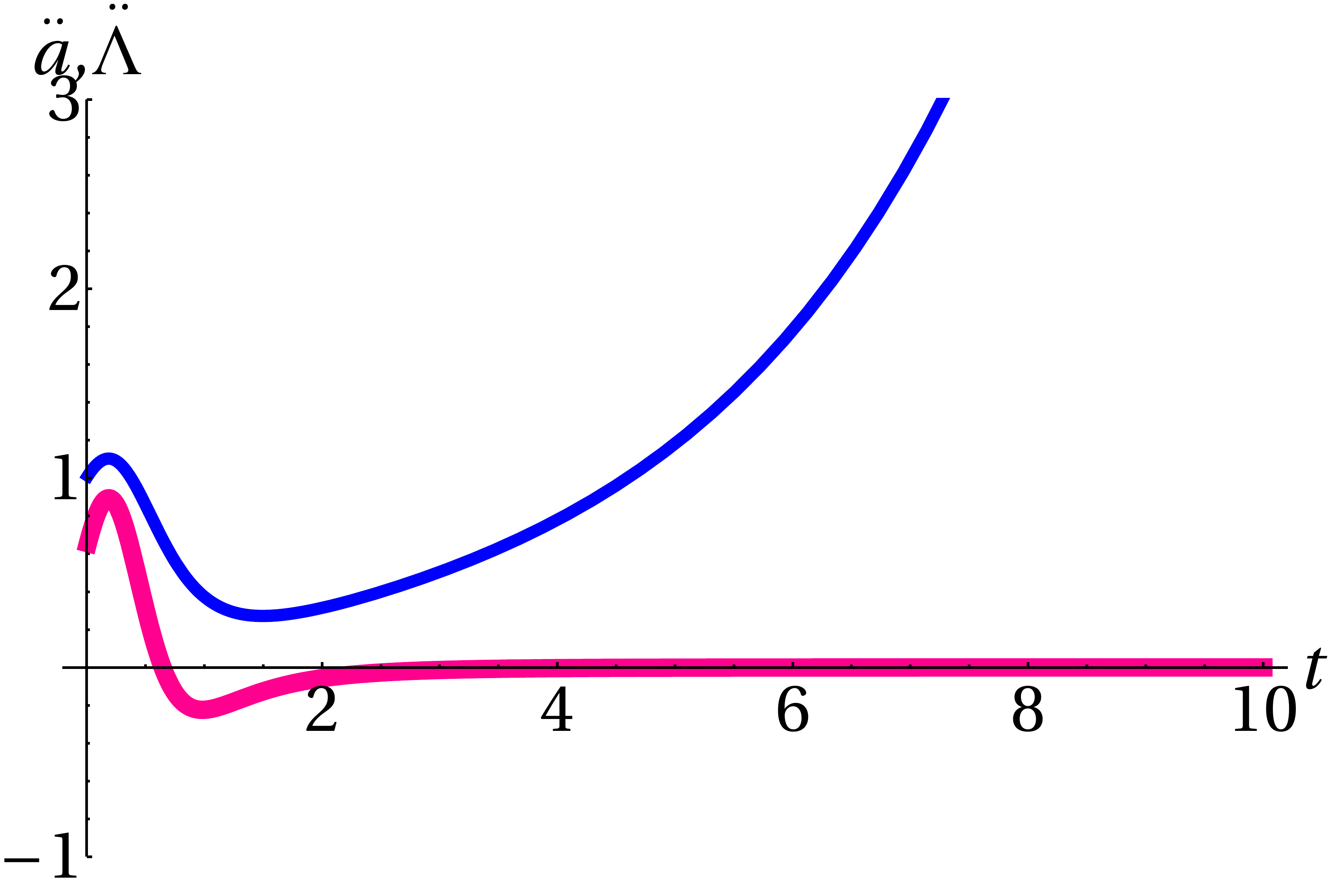}
    \caption{The accelerations: $\ddot a$ is represented by the blue curve, and $\ddot \Lambda$ by the red curve.}
    \label{addotLddot1-02-1-025constantb}
  \end{subfigure}
\qquad
  \begin{subfigure}[t]{.5\linewidth}
    \centering
    \includegraphics[width=0.7\columnwidth]{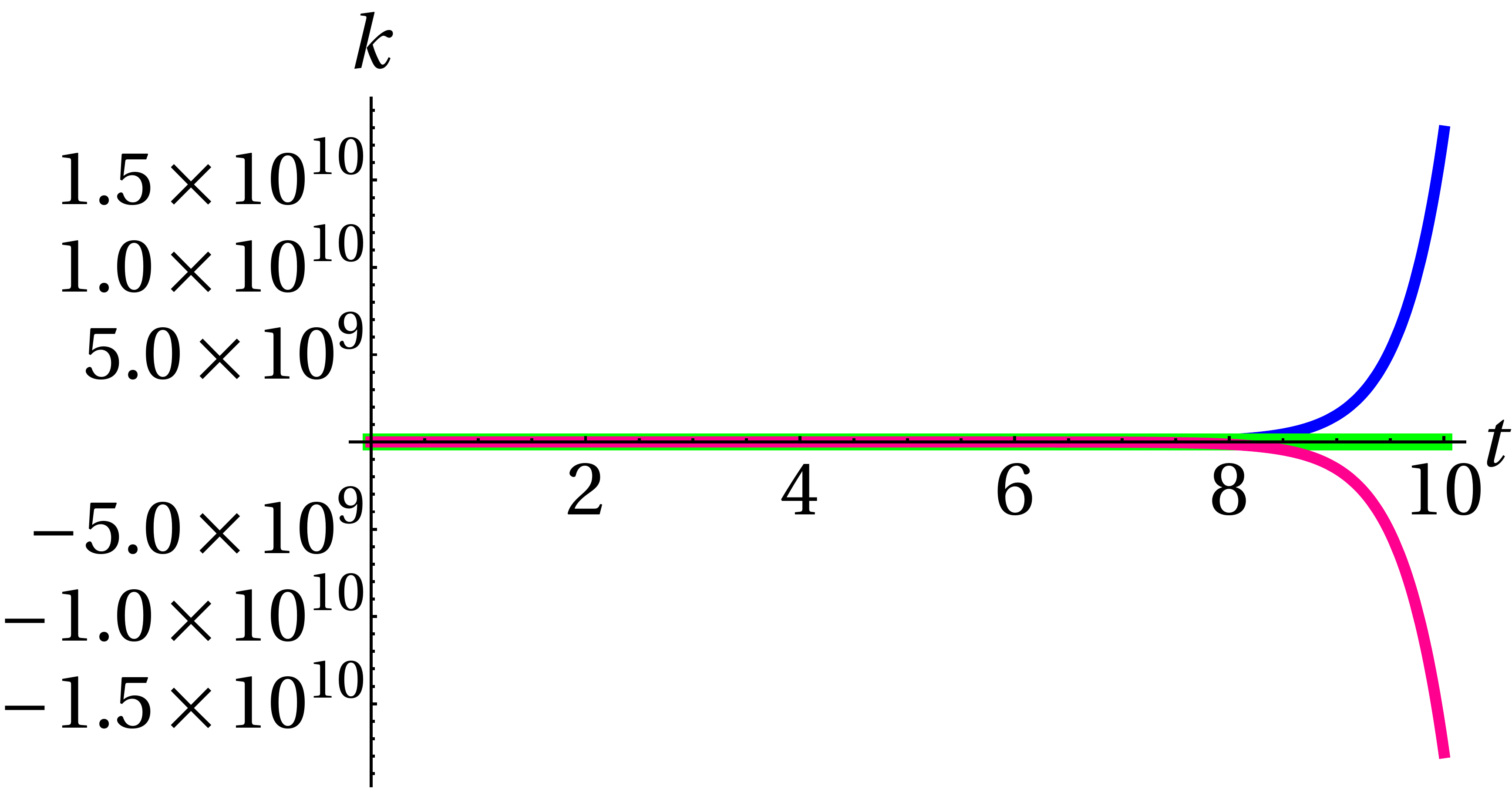}
    \caption{The harmonic function $k$ using: $\dot k\left(0\right)=1$ (blue curve), $\dot k\left(0\right)=0$ (green line), and $\dot k\left(0\right)=-1$ (red curve).}
    \label{k1-02-1-025constantb}
  \end{subfigure}
\caption{Initial conditions set number 5 for constant $b$.}
  \label{Fig46}
\end{figure}

\begin{figure}[H]
  \begin{subfigure}[t]{.5\linewidth}
    \centering
    \includegraphics[width=0.7\columnwidth]{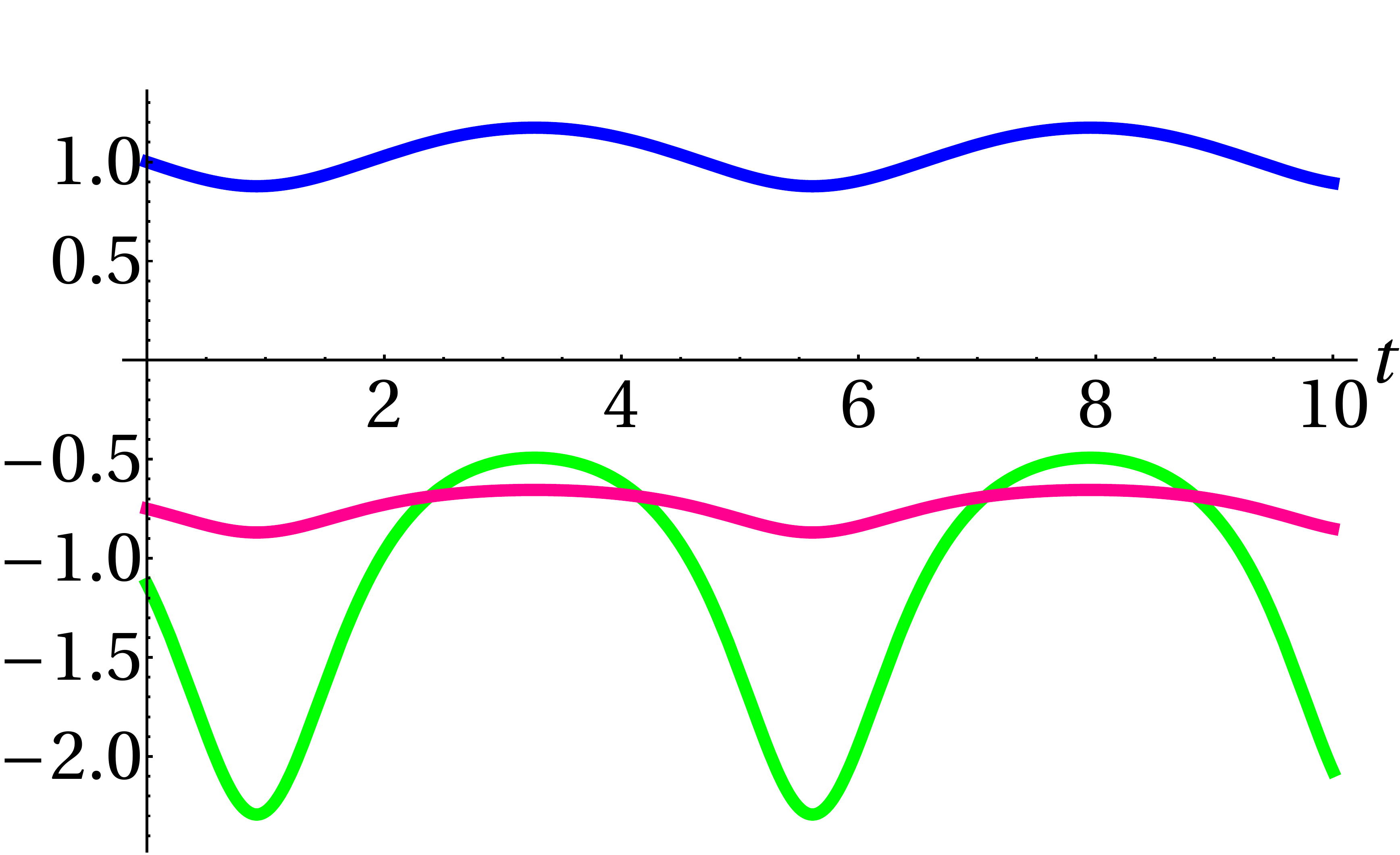}
    \caption{The scale factor $a$ is represented by the blue curve, $\Lambda$ by the red curve, while $ {G_{i\bar j} \dot z^i \dot z^{\bar j}} $ by the green curve.}
    \label{aLzz1-02-1-026constantb}
  \end{subfigure}
\qquad
  \begin{subfigure}[t]{.5\linewidth}
    \centering
    \includegraphics[width=0.7\columnwidth]{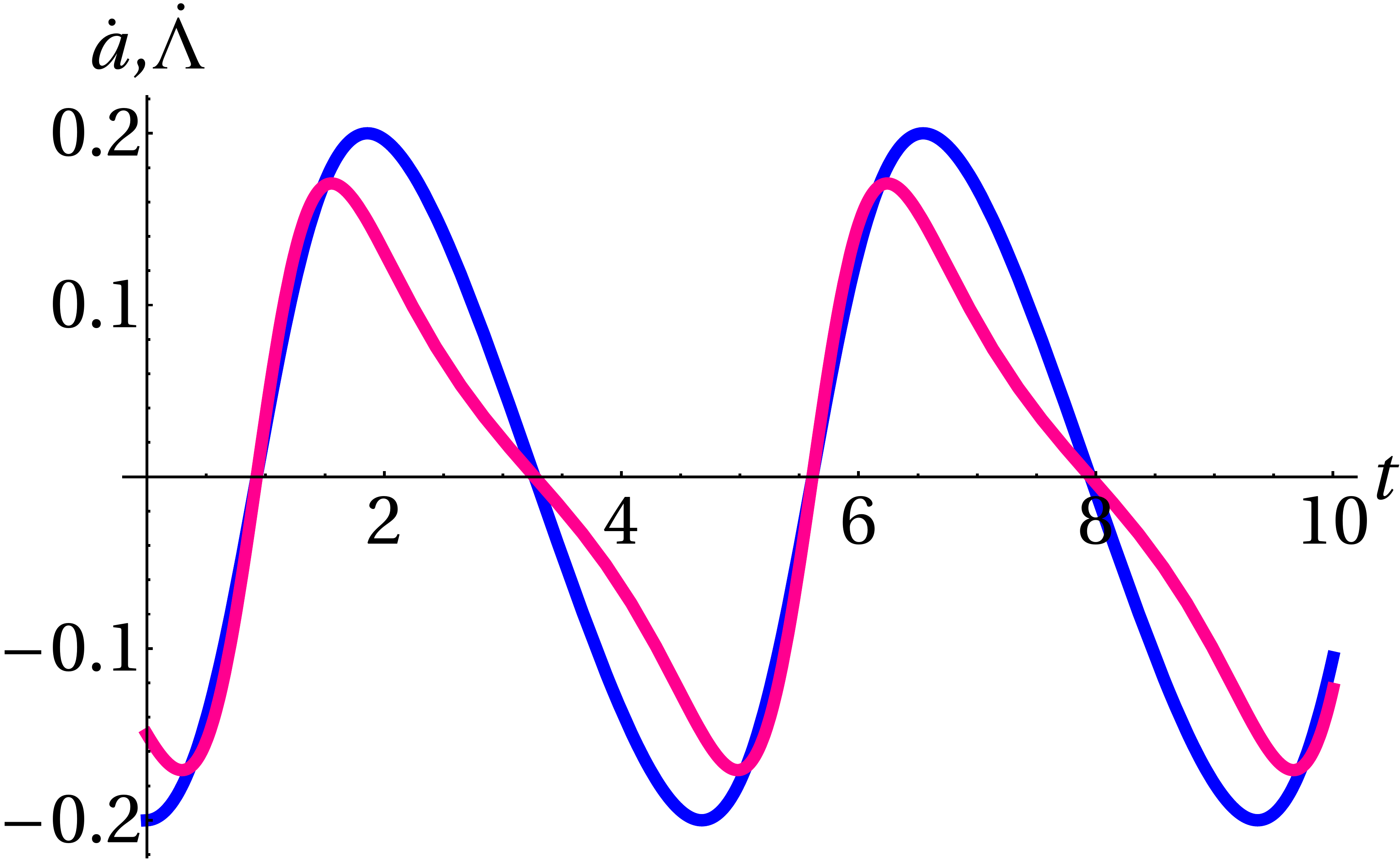}
    \caption{The expansion rates: $\dot a$ is represented by the blue curve, and $\dot \Lambda$ by the red curve.}
    \label{adotLdot1-02-1-026constantb}
  \end{subfigure}
\\[9em]
  \begin{subfigure}[t]{.5\linewidth}
    \centering
    \includegraphics[width=0.7\columnwidth]{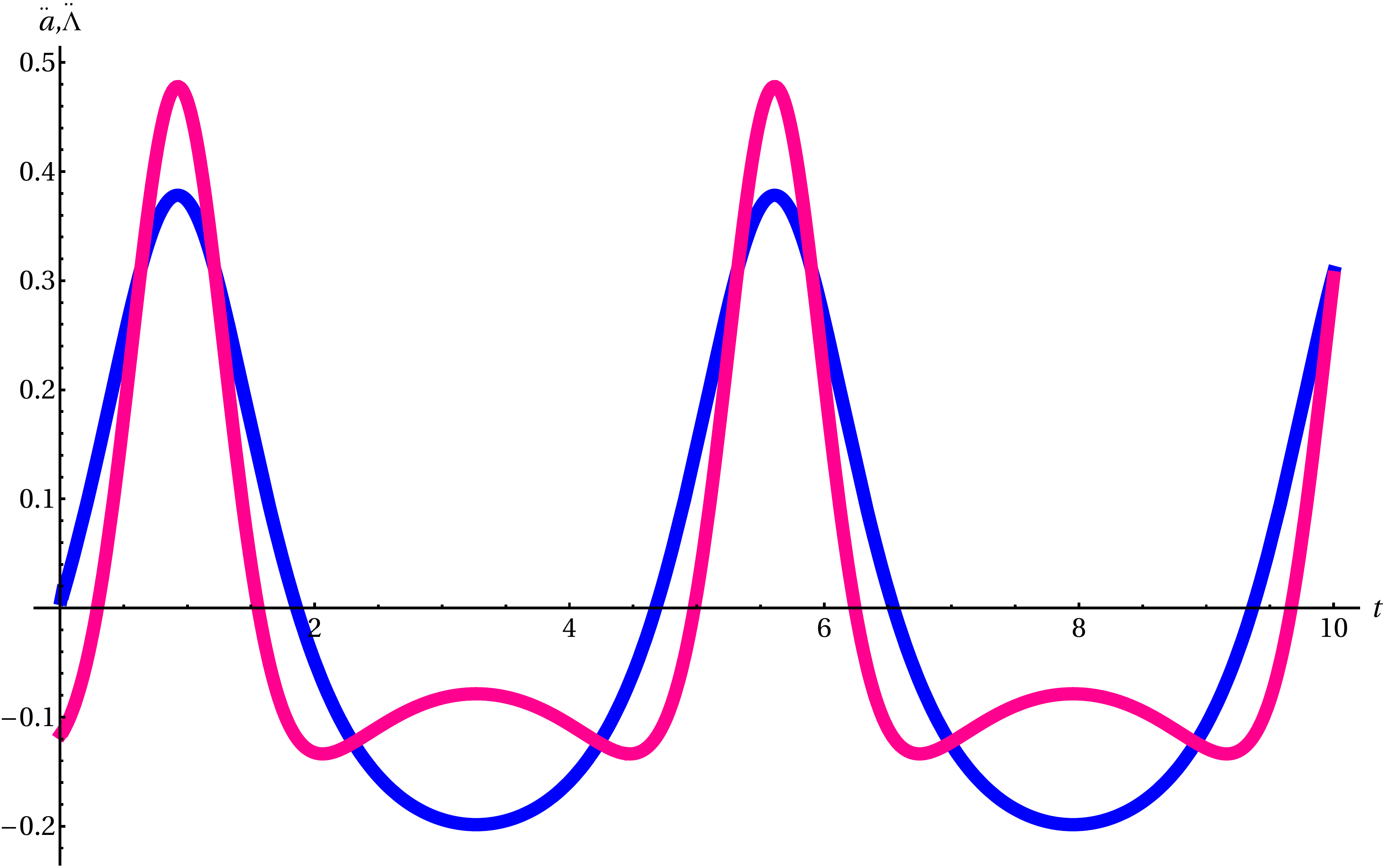}
    \caption{The accelerations: $\ddot a$ is represented by the blue curve, and $\ddot \Lambda$ by the red curve.}
    \label{addotLddot1-02-1-026constantb}
  \end{subfigure}
\qquad
  \begin{subfigure}[t]{.5\linewidth}
    \centering
    \includegraphics[width=0.7\columnwidth]{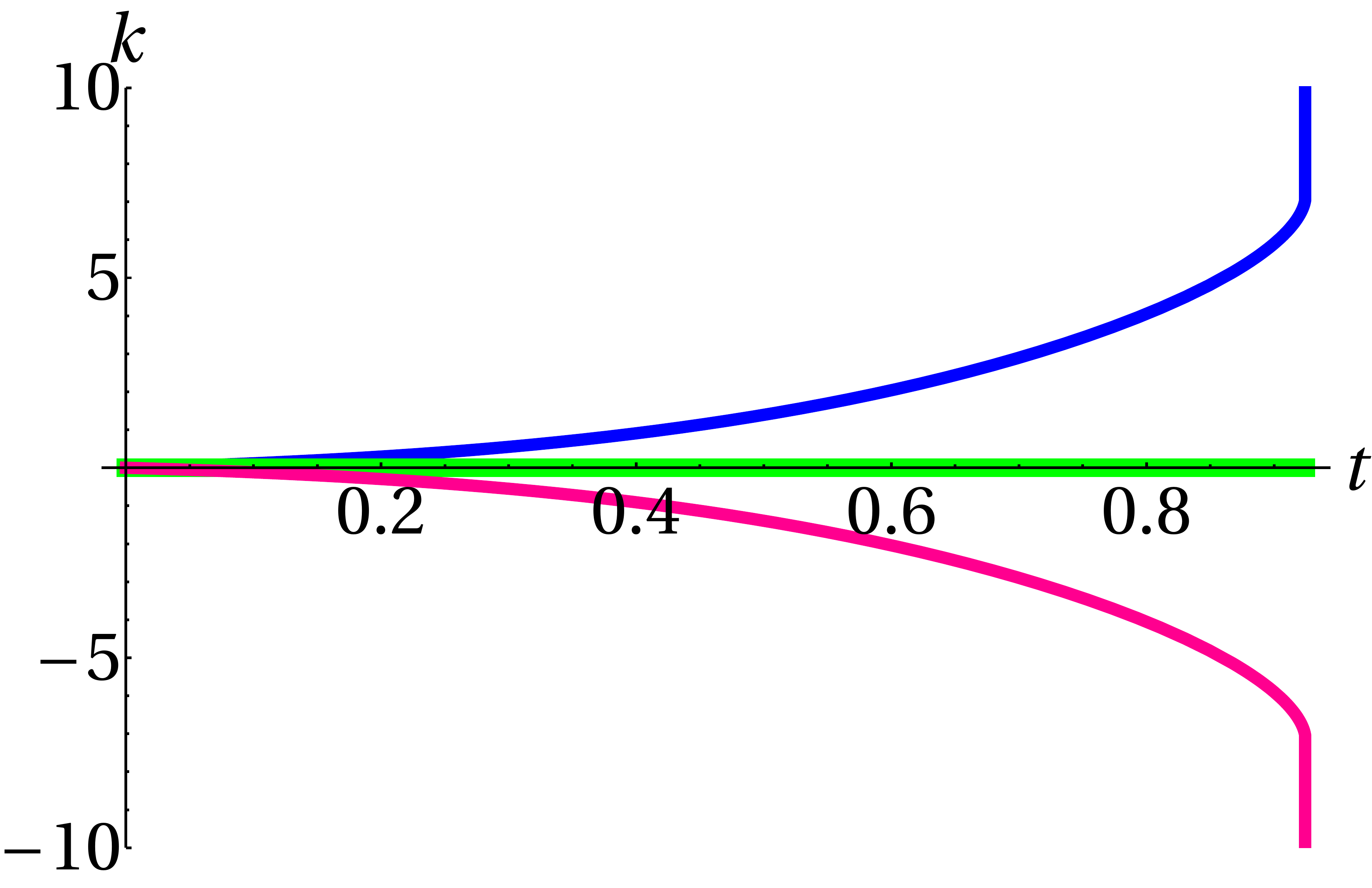}
    \caption{The harmonic function $k$ using: $\dot k\left(0\right)=1$ (blue curve), $\dot k\left(0\right)=0$ (green line), and $\dot k\left(0\right)=-1$ (red curve).}
    \label{k1-02-1-026constantb}
  \end{subfigure}
    \caption{Initial conditions set number 6 for constant $b$.}
  \label{Fig47}
  \end{figure}

\begin{figure}[H]
  \begin{subfigure}[t]{.5\linewidth}
    \centering
    \includegraphics[width=0.7\columnwidth]{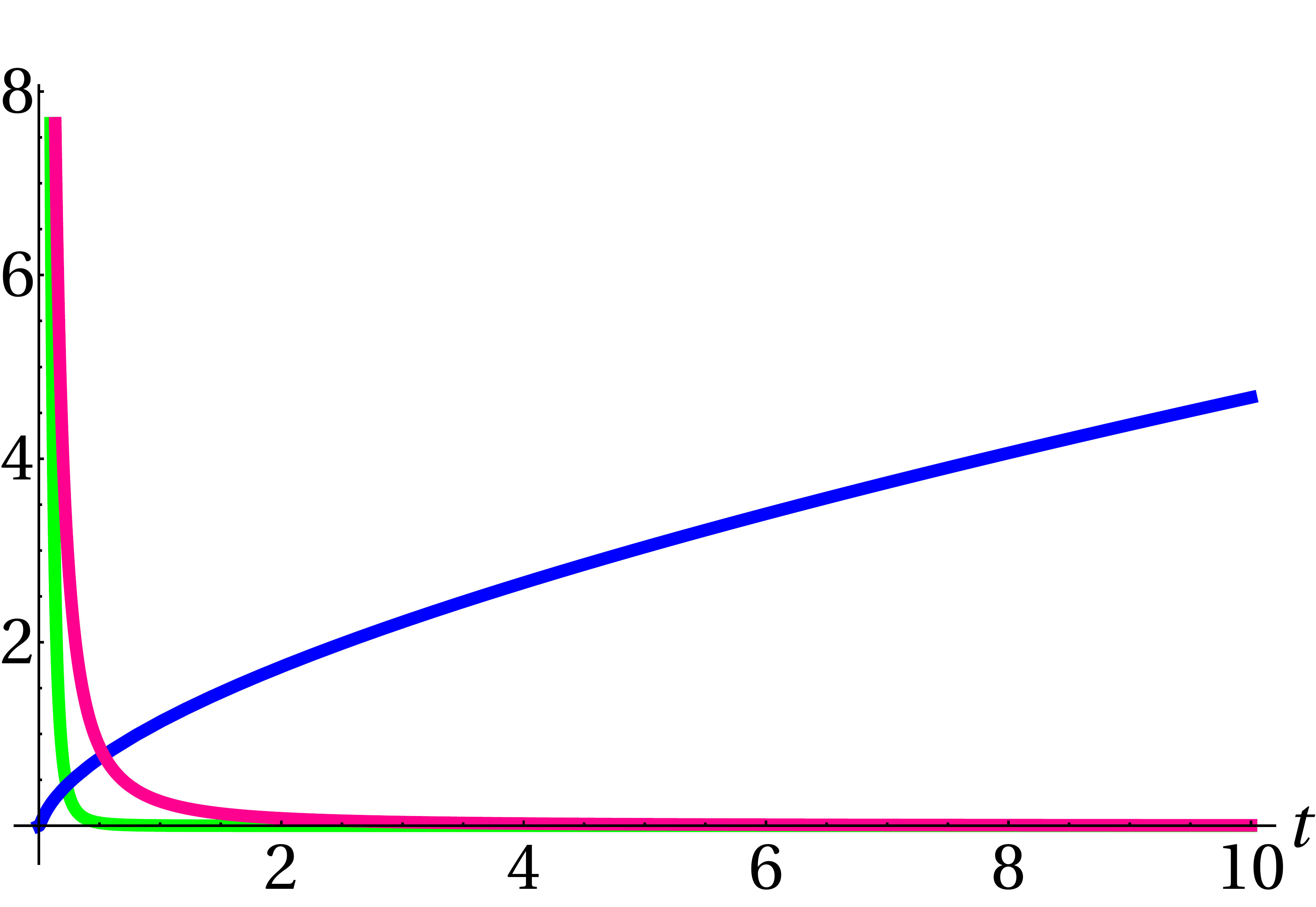}
    \caption{The scale factor $a$ is represented by the blue curve, $\tilde\Lambda$ by the red curve, while $ \left|{G_{i\bar j} \dot z^i \dot z^{\bar j}} \right|$ by the green curve.}
    \label{aLTzz00007constantb}
  \end{subfigure}
\qquad
  \begin{subfigure}[t]{.5\linewidth}
    \centering
    \includegraphics[width=0.7\columnwidth]{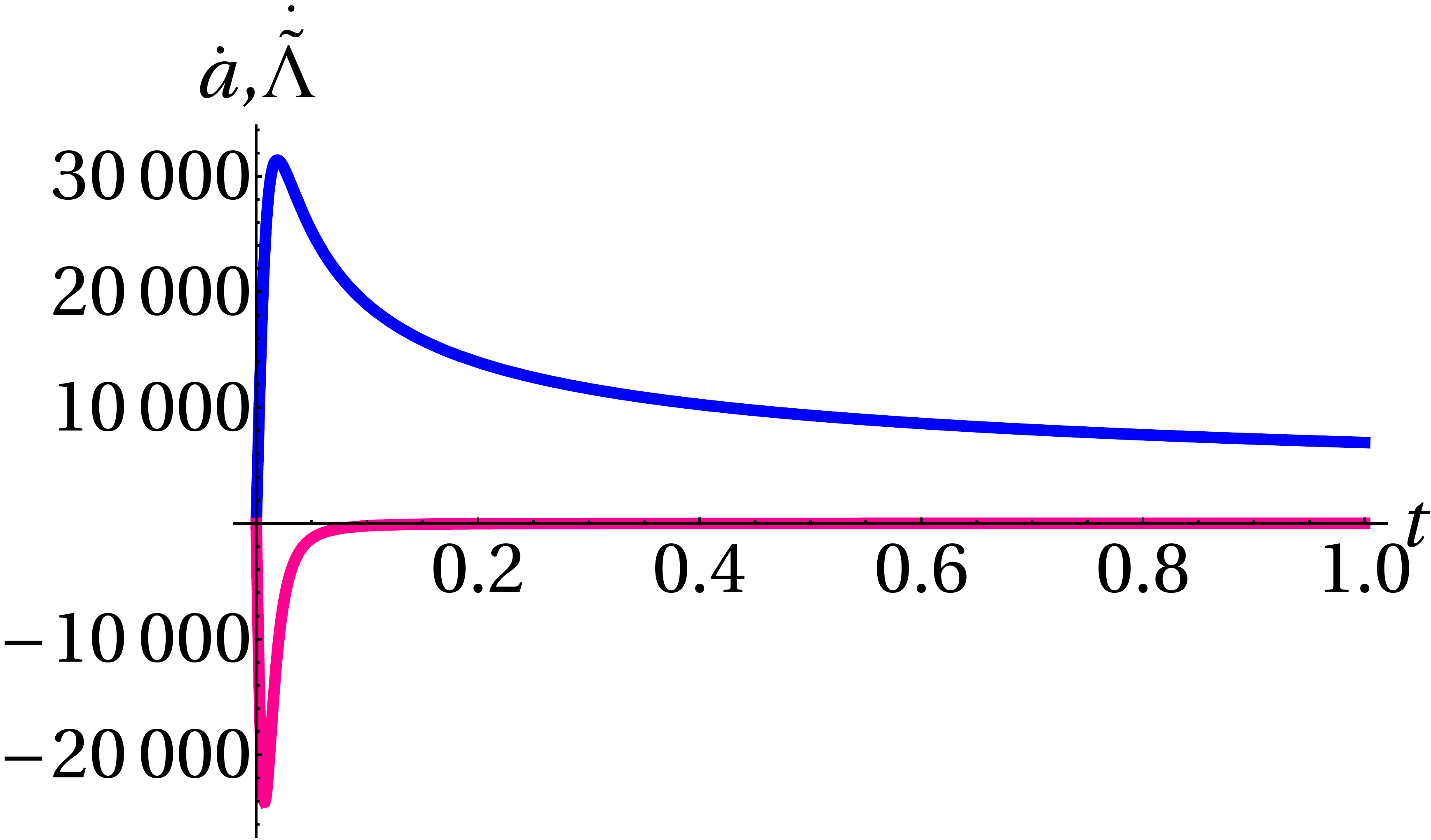}
    \caption{The expansion rates: $\dot a$ is represented by the blue curve, and $\dot \tilde\Lambda$ by the red curve. The curve for $\dot a$ is scaled up by a factor of 10000.}
    \label{adotLTdot00007constantb}
  \end{subfigure}
\\[9em]
  \begin{subfigure}[t]{.5\linewidth}
    \centering
    \includegraphics[width=0.7\columnwidth]{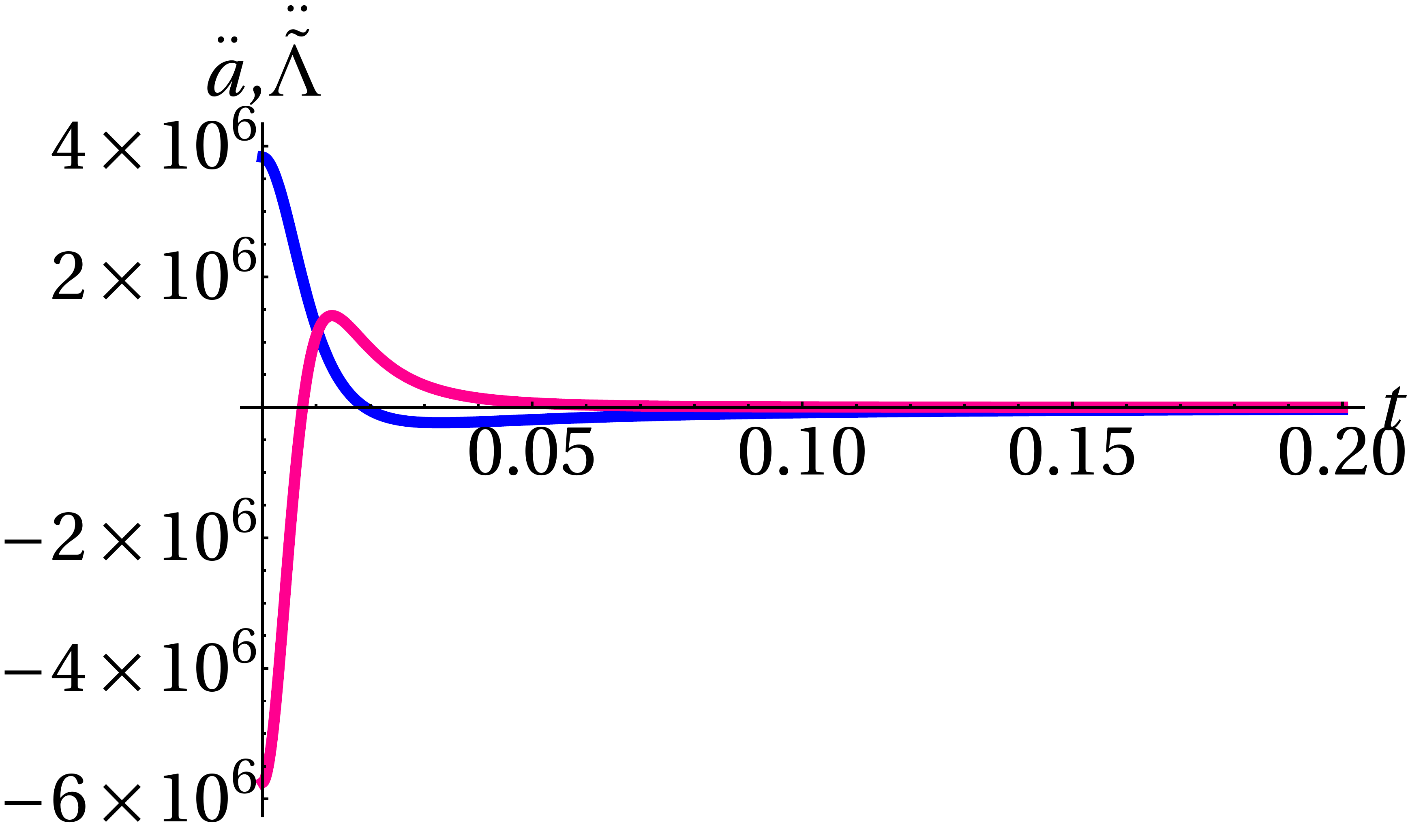}
    \caption{The accelerations: $\ddot a$ is represented by the blue curve, and $\ddot \tilde\Lambda$ by the red curve. The curve for $\ddot a$ is scaled up by a factor of 10000.}
    \label{addotLTddot00007constantb}
  \end{subfigure}
\qquad
  \begin{subfigure}[t]{.5\linewidth}
    \centering
    \includegraphics[width=0.7\columnwidth]{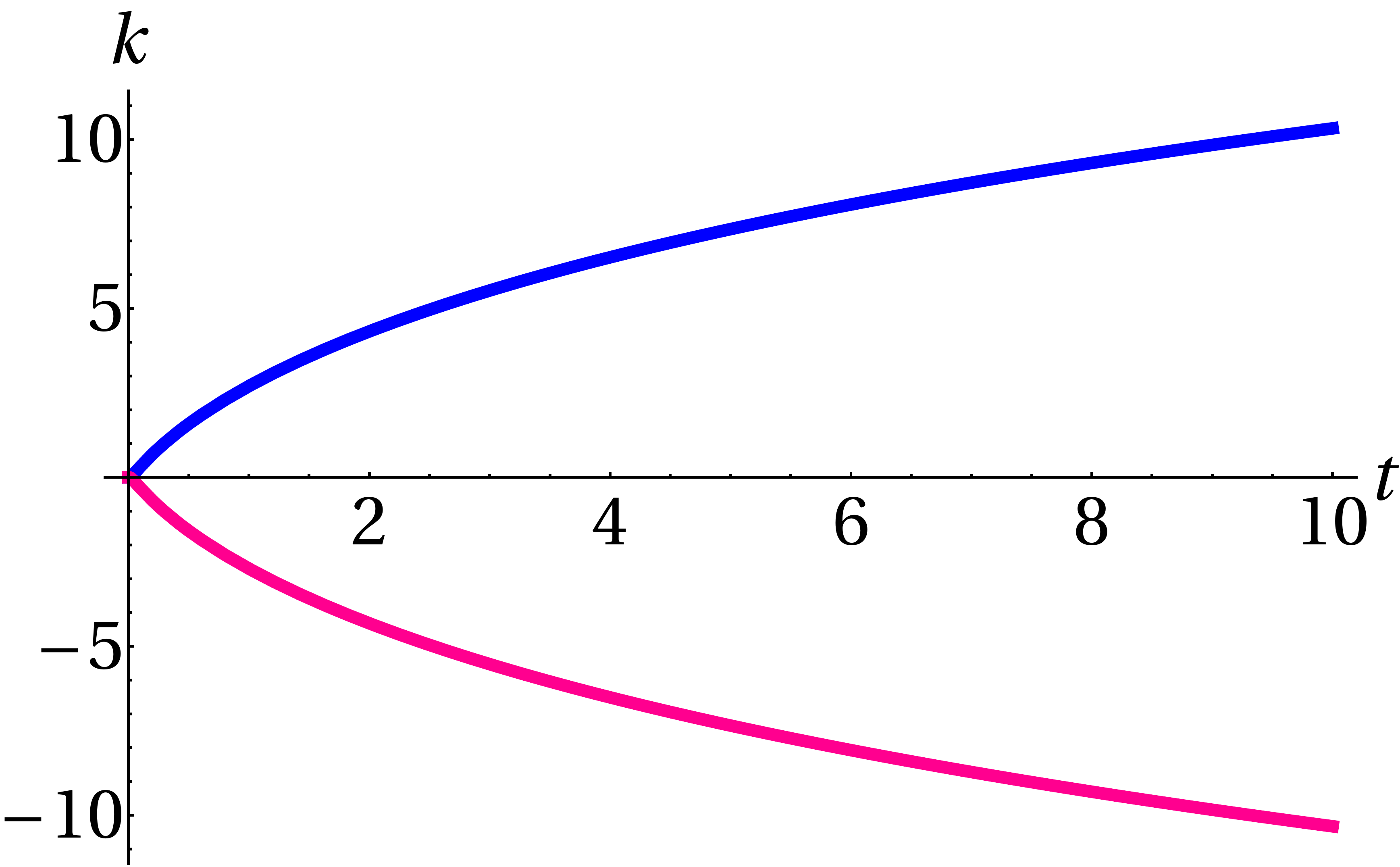}
    \caption{The harmonic function $k$ using: $\dot k\left(0\right)=1$ (blue curve), and $\dot k\left(0\right)=-1$ (red curve). While $\dot k\left(0\right)=0$ diverges.}
    \label{k00007constantb}
  \end{subfigure}
 \caption{Initial conditions set number 7 for constant $b$.}
  \label{Fig50}
\end{figure}


\begin{figure}[H]
  \begin{subfigure}[t]{.5\linewidth}
    \centering
    \includegraphics[width=0.7\columnwidth]{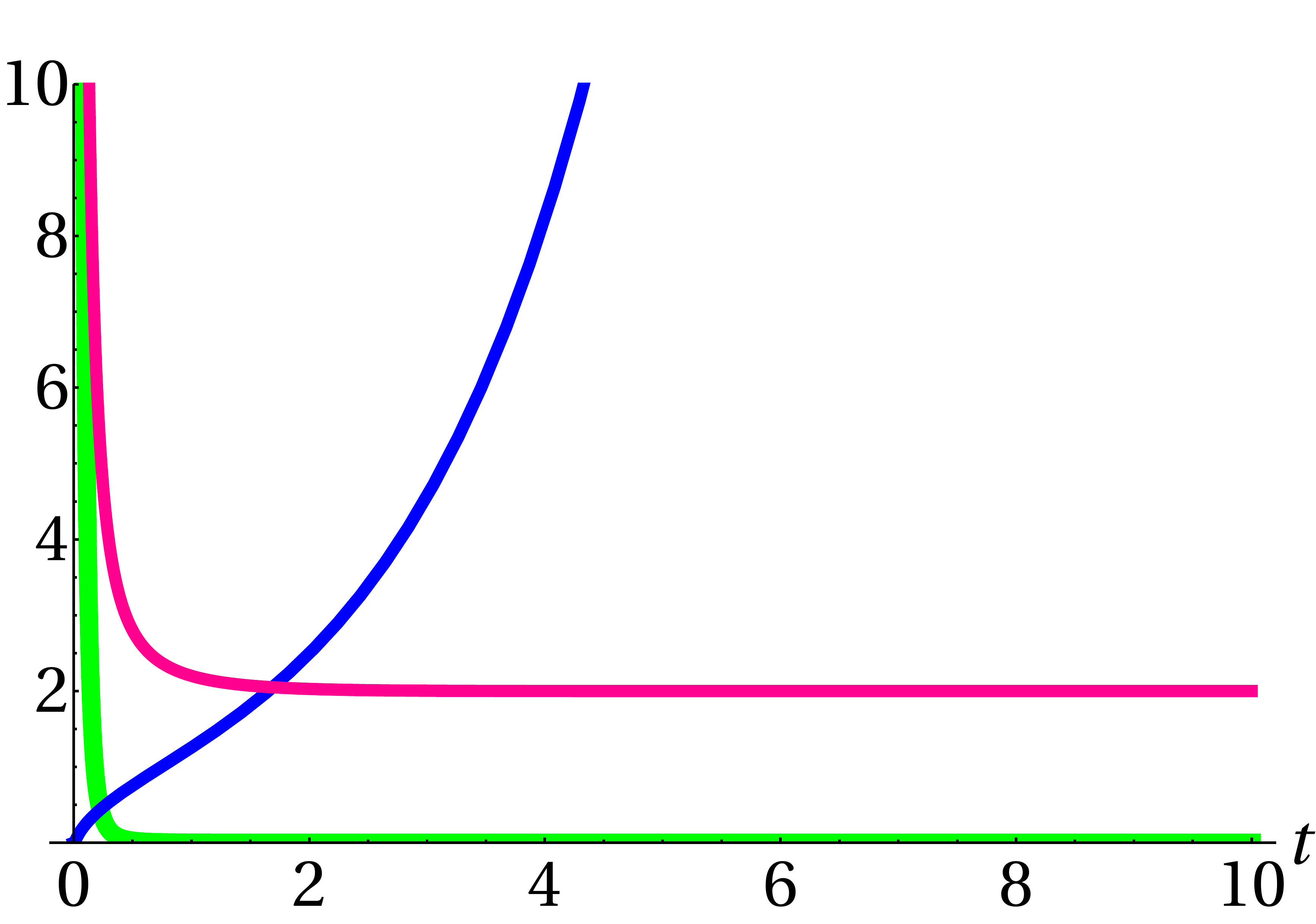}
    \caption{The scale factor $a$ is represented by the blue curve, $\tilde\Lambda$ by the red curve, while $ \left|{G_{i\bar j} \dot z^i \dot z^{\bar j}} \right|$ by the green curve.}
    \label{aLTzz00008constantb}
  \end{subfigure}
\qquad
  \begin{subfigure}[t]{.5\linewidth}
    \centering
    \includegraphics[width=0.7\columnwidth]{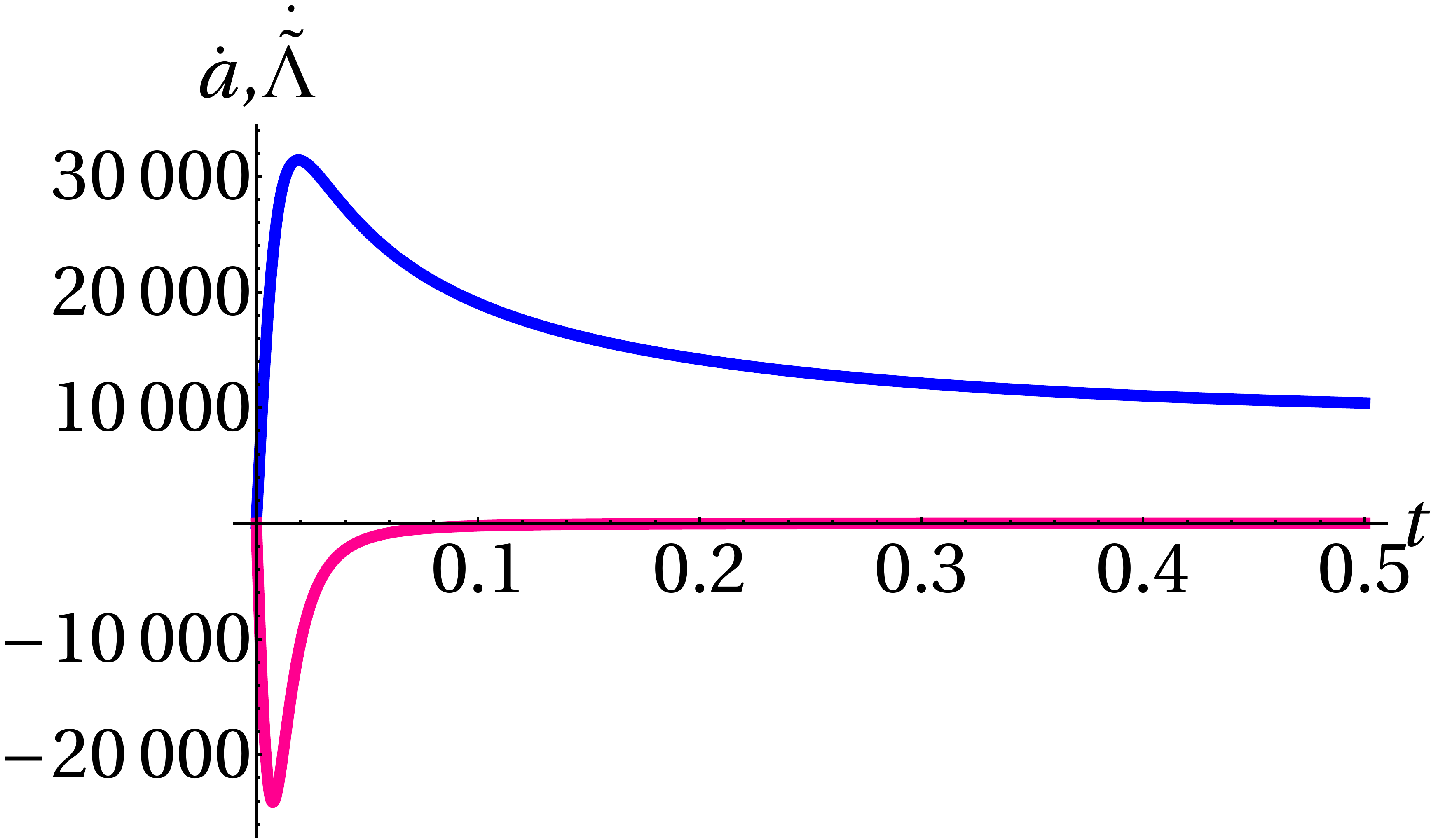}
    \caption{The expansion rates: $\dot a$ is represented by the blue curve, and $\dot \tilde\Lambda$ by the red curve. The curve for $\dot a$ is scaled up by a factor of 10000.}
    \label{adotLTdot00008constantb}
  \end{subfigure}
\\[9em]
  \begin{subfigure}[t]{.5\linewidth}
    \centering
    \includegraphics[width=0.7\columnwidth]{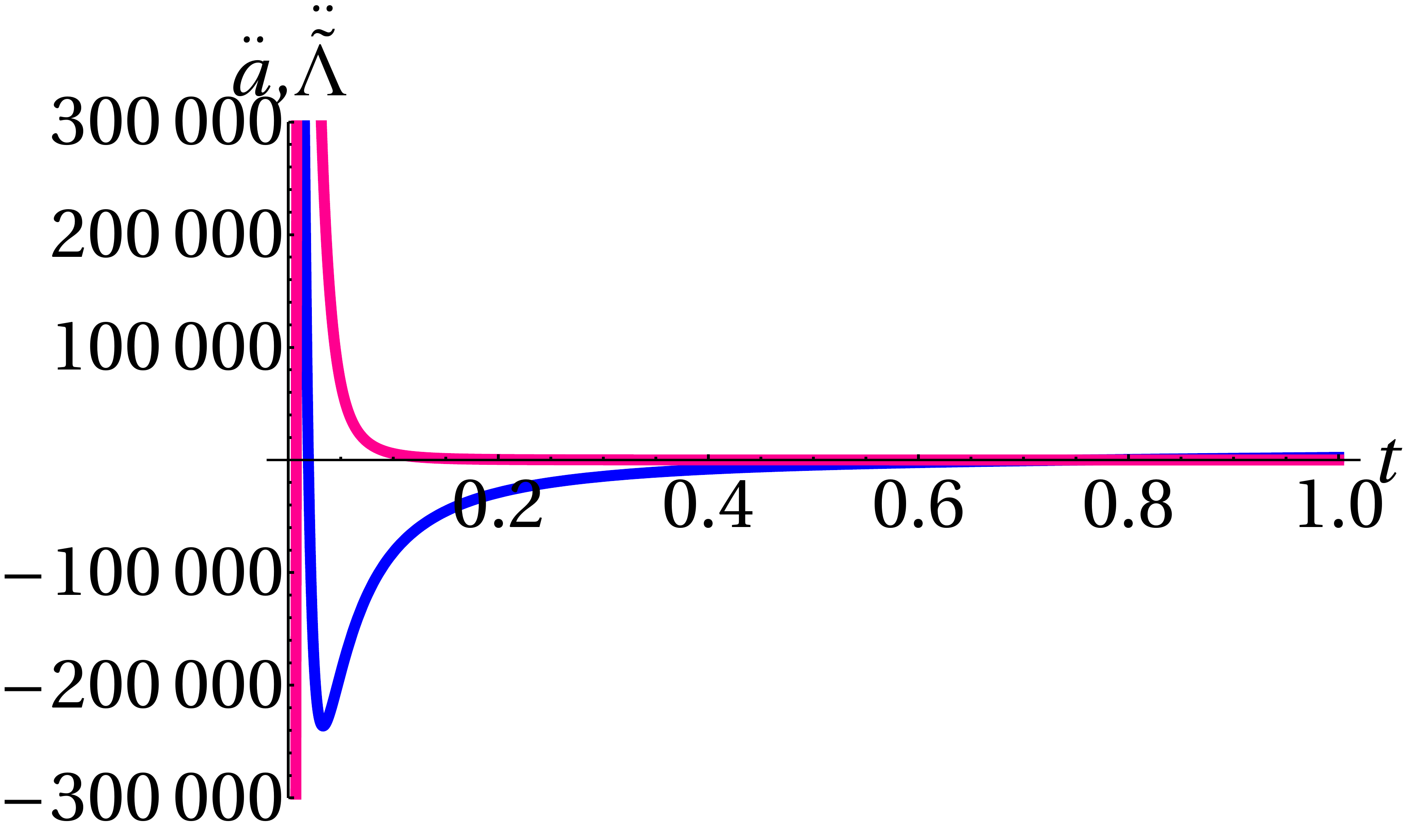}
    \caption{The accelerations: $\ddot a$ is represented by the blue curve, and $\ddot \tilde\Lambda$ by the red curve. The curve for $\ddot a$ is scaled up by a factor of 10000.}
    \label{addotLTddot00008constantb}
  \end{subfigure}
\qquad
  \begin{subfigure}[t]{.5\linewidth}
    \centering
    \includegraphics[width=0.7\columnwidth]{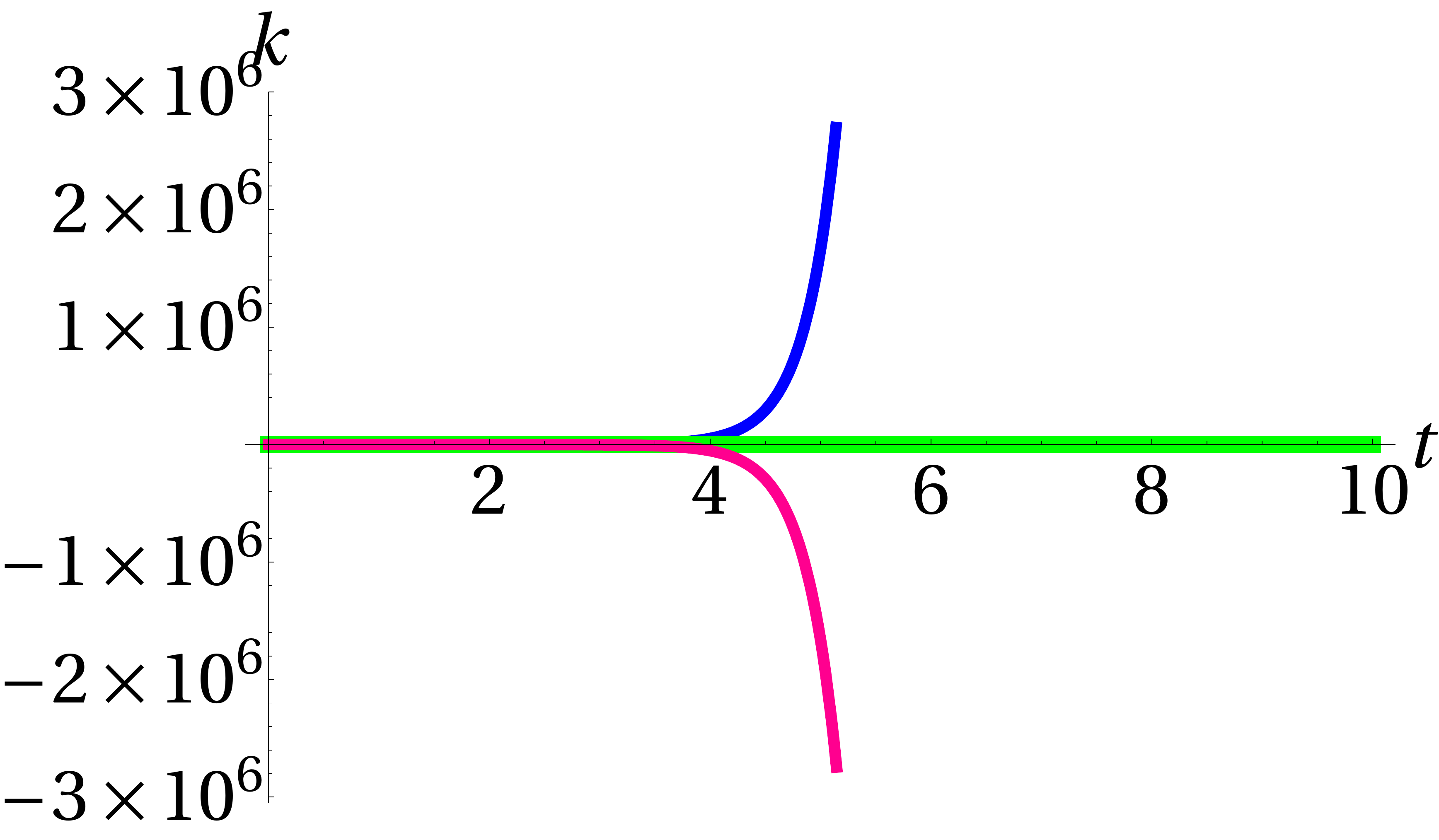}
    \caption{The harmonic function $k$ using: $\dot k\left(0\right)=1$ (blue curve), $\dot k\left(0\right)=0$ (green line), and $\dot k\left(0\right)=-1$ (red curve).}
    \label{k00008constantb}
  \end{subfigure}
  \caption{Initial conditions set number 8 for constant $b$.}
  \label{Fig51}
  \end{figure}

\begin{figure}[H]
  \begin{subfigure}[t]{.5\linewidth}
    \centering
    \includegraphics[width=0.7\columnwidth]{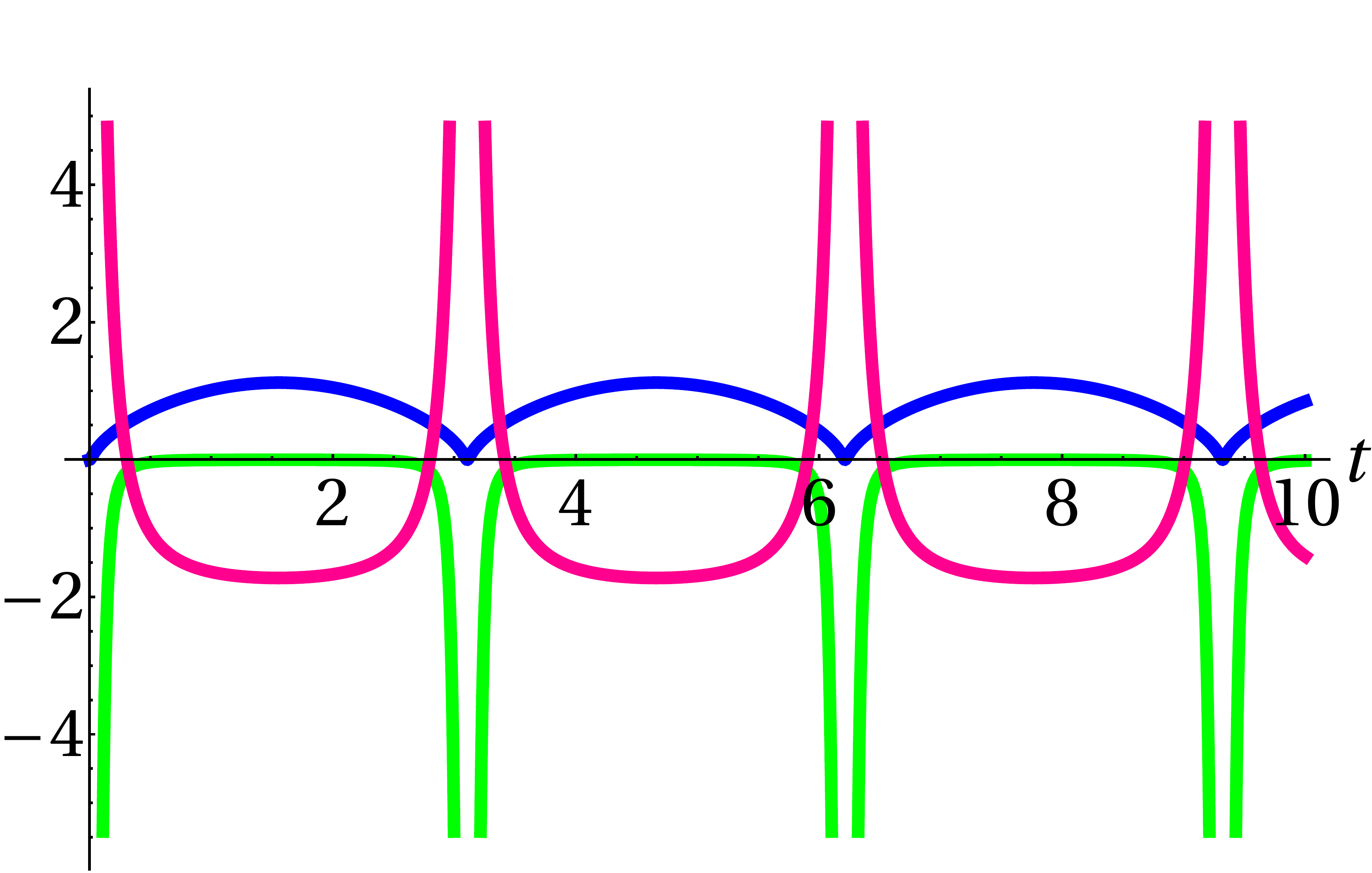}
    \caption{The scale factor $a$ is represented by the blue curve, $\tilde\Lambda$ by the red curve, while $ {G_{i\bar j} \dot z^i \dot z^{\bar j}} $ by the green curve.}
    \label{aLTzz00009constantb}
  \end{subfigure}
\qquad
  \begin{subfigure}[t]{.5\linewidth}
    \centering
    \includegraphics[width=0.7\columnwidth]{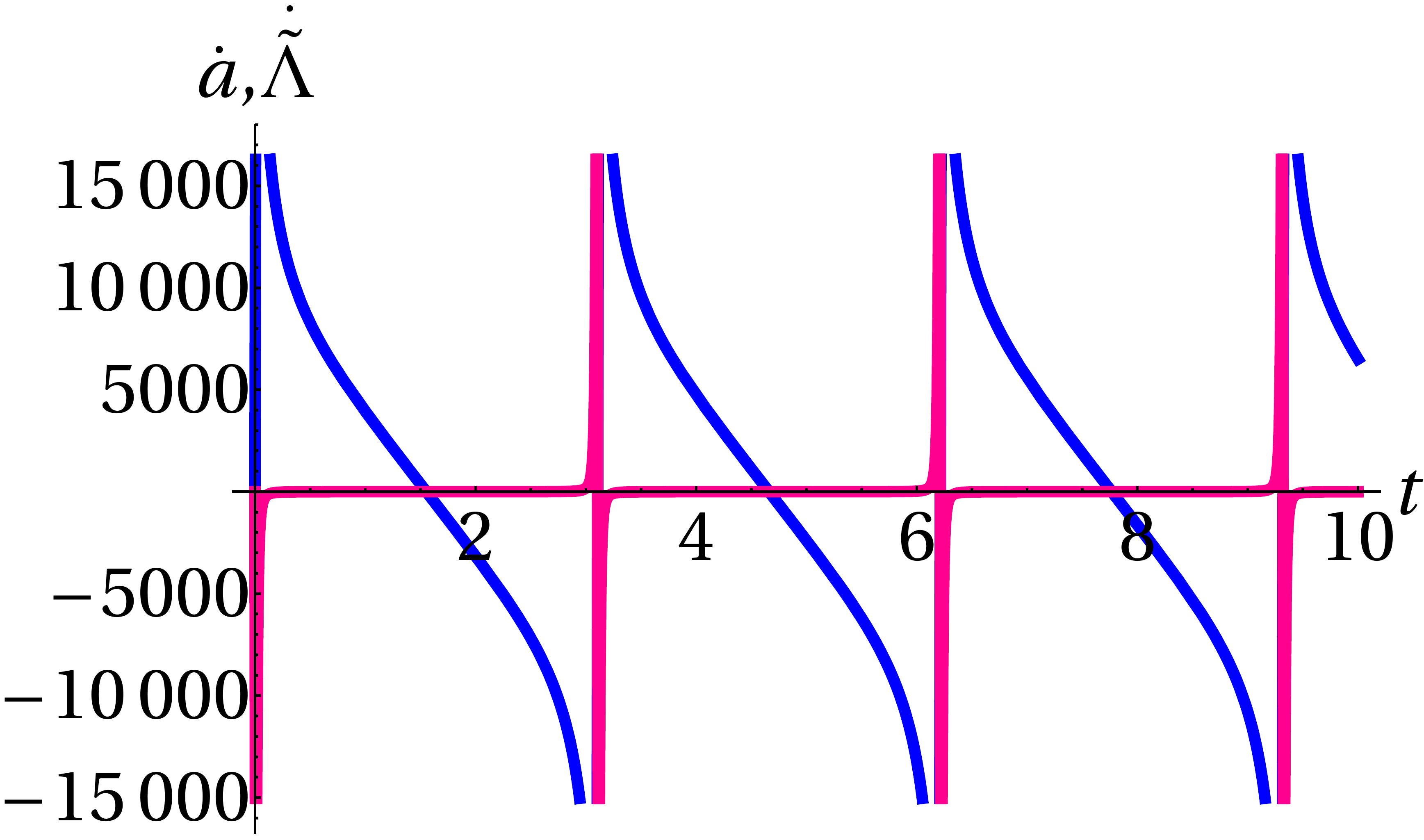}
    \caption{The expansion rates: $\dot a$ is represented by the blue curve, and $\dot \tilde\Lambda$ by the red curve. The curve for $\dot a$ is scaled up by a factor of 10000.}
    \label{adotLTdot00009constantb}
  \end{subfigure}
\\[9em]
  \begin{subfigure}[t]{.5\linewidth}
    \centering
    \includegraphics[width=0.7\columnwidth]{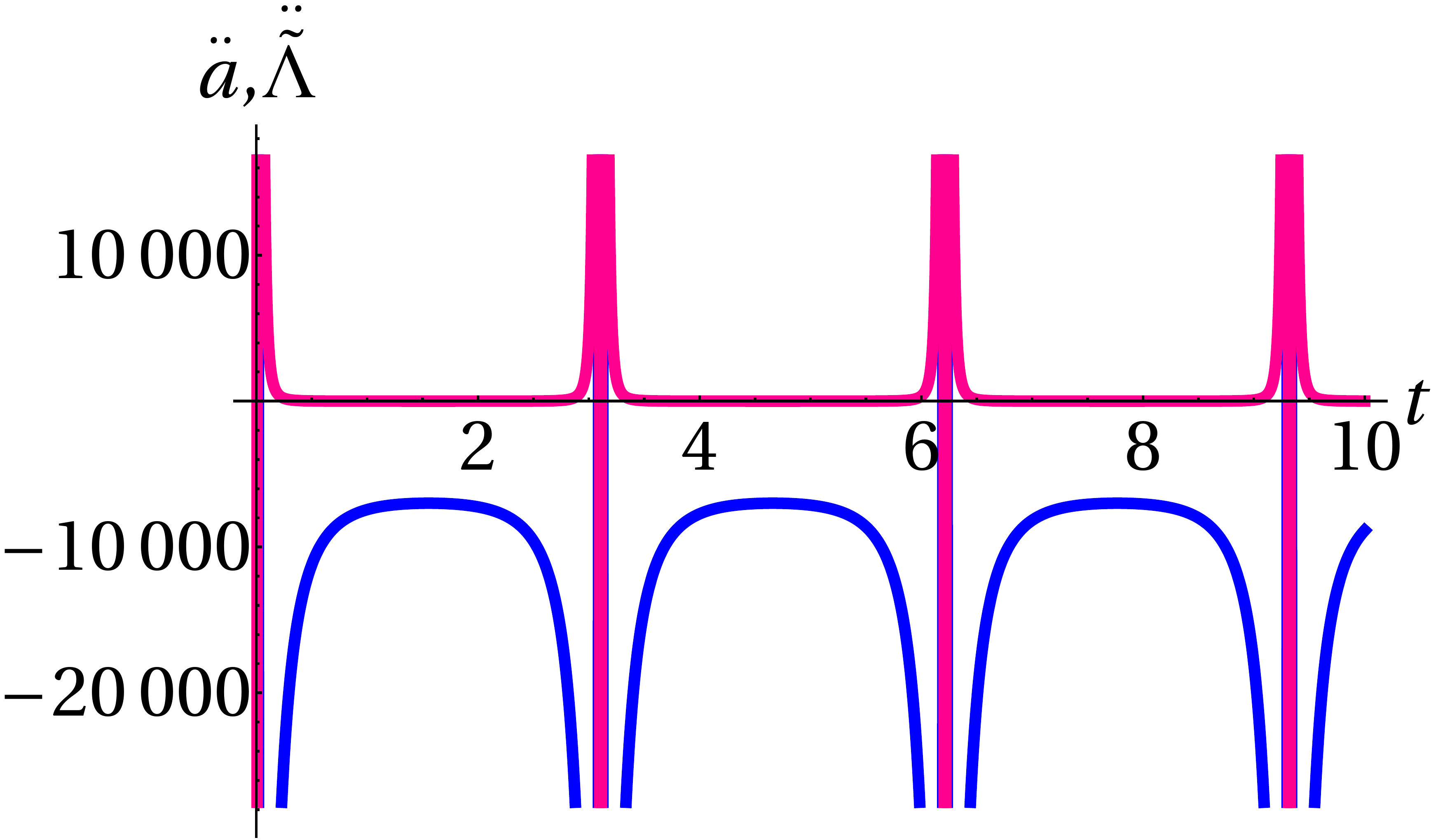}
    \caption{The accelerations: $\ddot a$ is represented by the blue curve, and $\ddot \tilde\Lambda$ by the red curve. The curve for $\ddot a$ is scaled up by a factor of 10000.}
    \label{addotLTddot00009constantb}
  \end{subfigure}
\qquad
  \begin{subfigure}[t]{.5\linewidth}
    \centering
    \includegraphics[width=0.7\columnwidth]{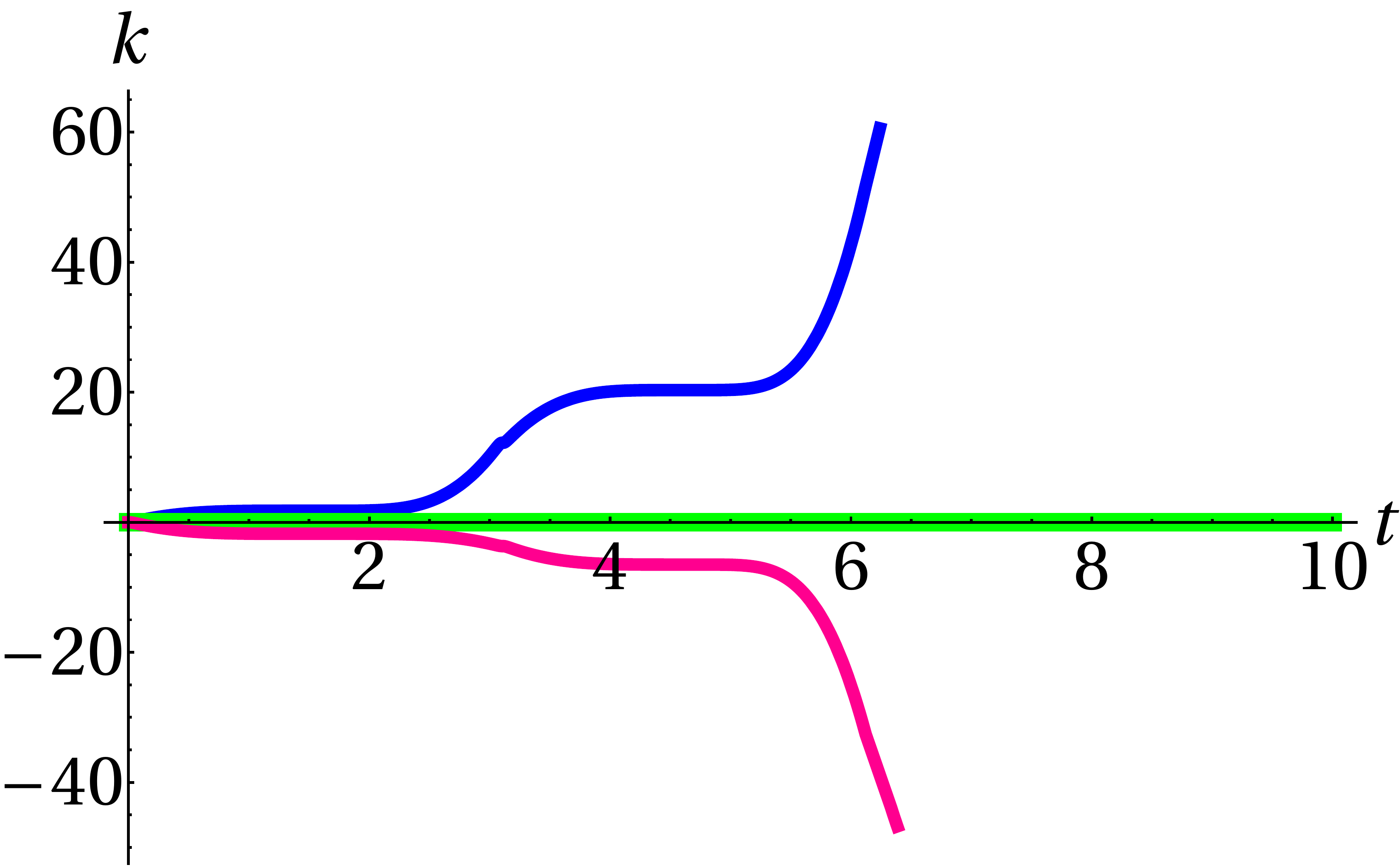}
    \caption{The harmonic function $k$ using: $\dot k\left(0\right)=1$ (blue curve), $\dot k\left(0\right)=0$ (green line), and $\dot k\left(0\right)=-1$ (red curve).}
    \label{k00009constantb}
  \end{subfigure}
\caption{Initial conditions set number 9 for constant $b$.}
  \label{Fig54}
\end{figure}


\begin{figure}[H]
  \begin{subfigure}[t]{.5\linewidth}
    \centering
    \includegraphics[width=0.7\columnwidth]{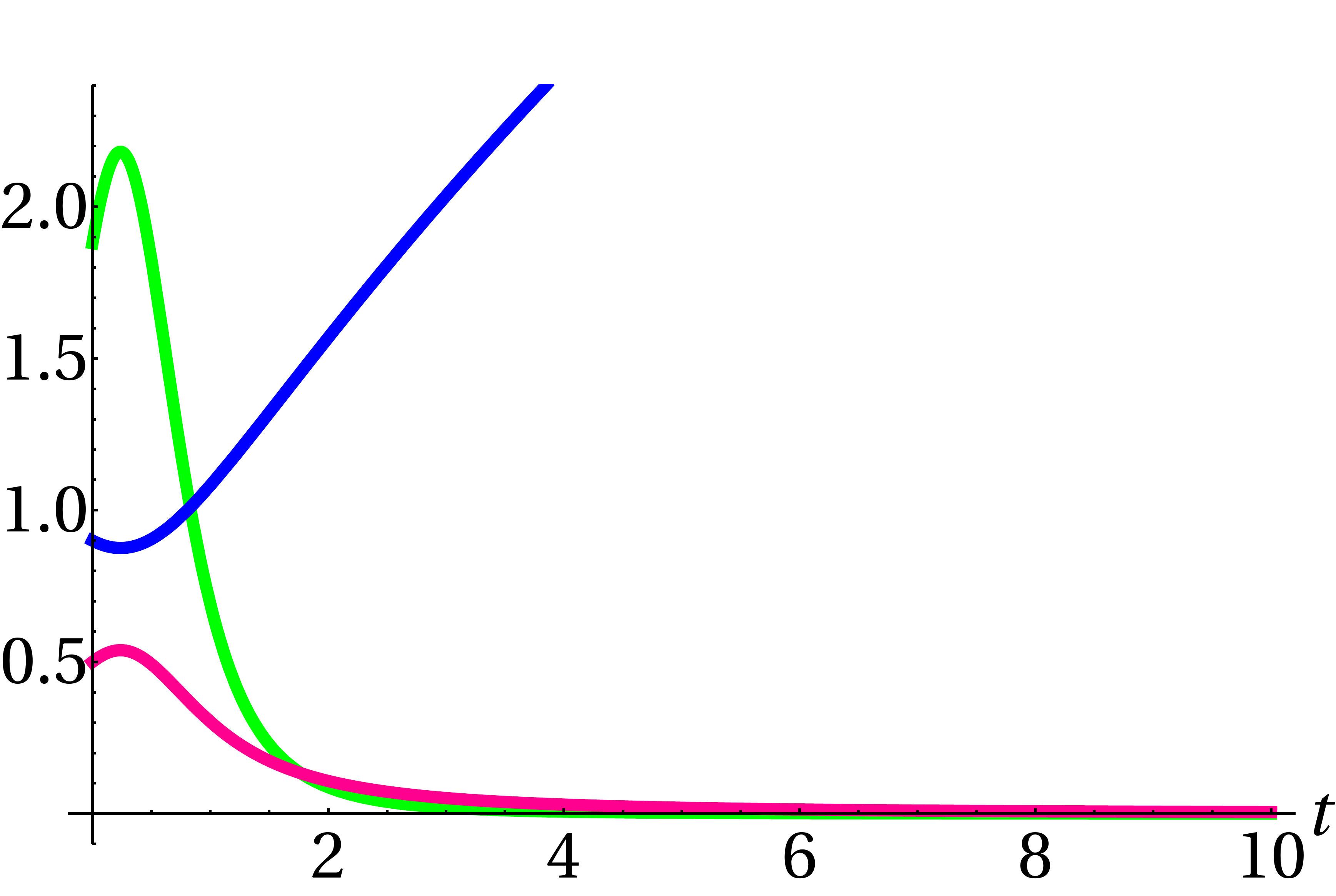}
    \caption{The scale factor $a$ is represented by the blue curve, $\tilde\Lambda$ by the red curve, while $ \left|{G_{i\bar j} \dot z^i \dot z^{\bar j}} \right|$ by the green curve.}
    \label{aLTzz000010constantb}
  \end{subfigure}
\qquad
  \begin{subfigure}[t]{.5\linewidth}
    \centering
    \includegraphics[width=0.7\columnwidth]{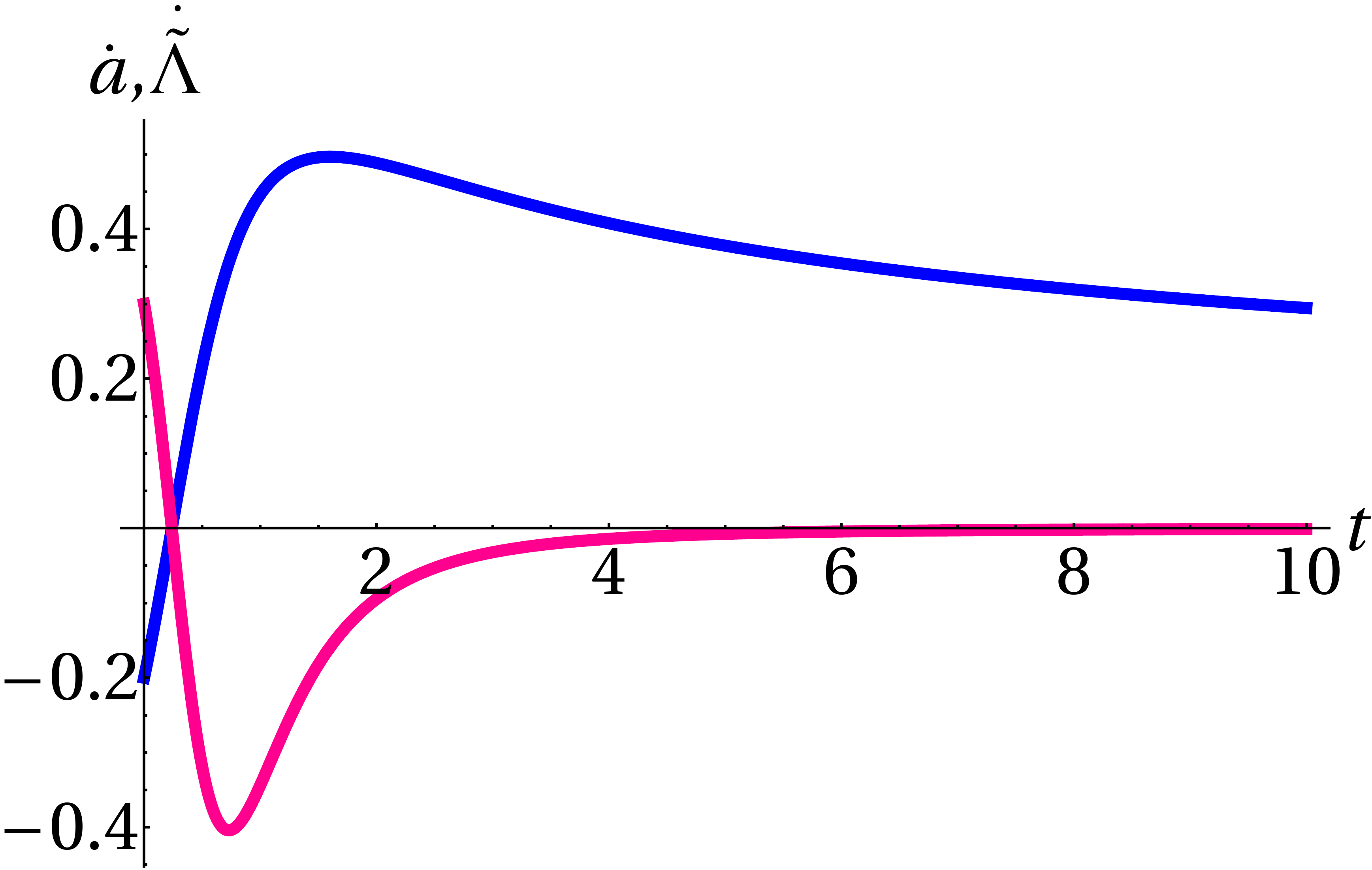}
    \caption{The expansion rates: $\dot a$ is represented by the blue curve, and $\dot \tilde\Lambda$ by the red curve.}
    \label{adotLTdot000010constantb}
  \end{subfigure}
\\[9em]
  \begin{subfigure}[t]{.5\linewidth}
    \centering
    \includegraphics[width=0.7\columnwidth]{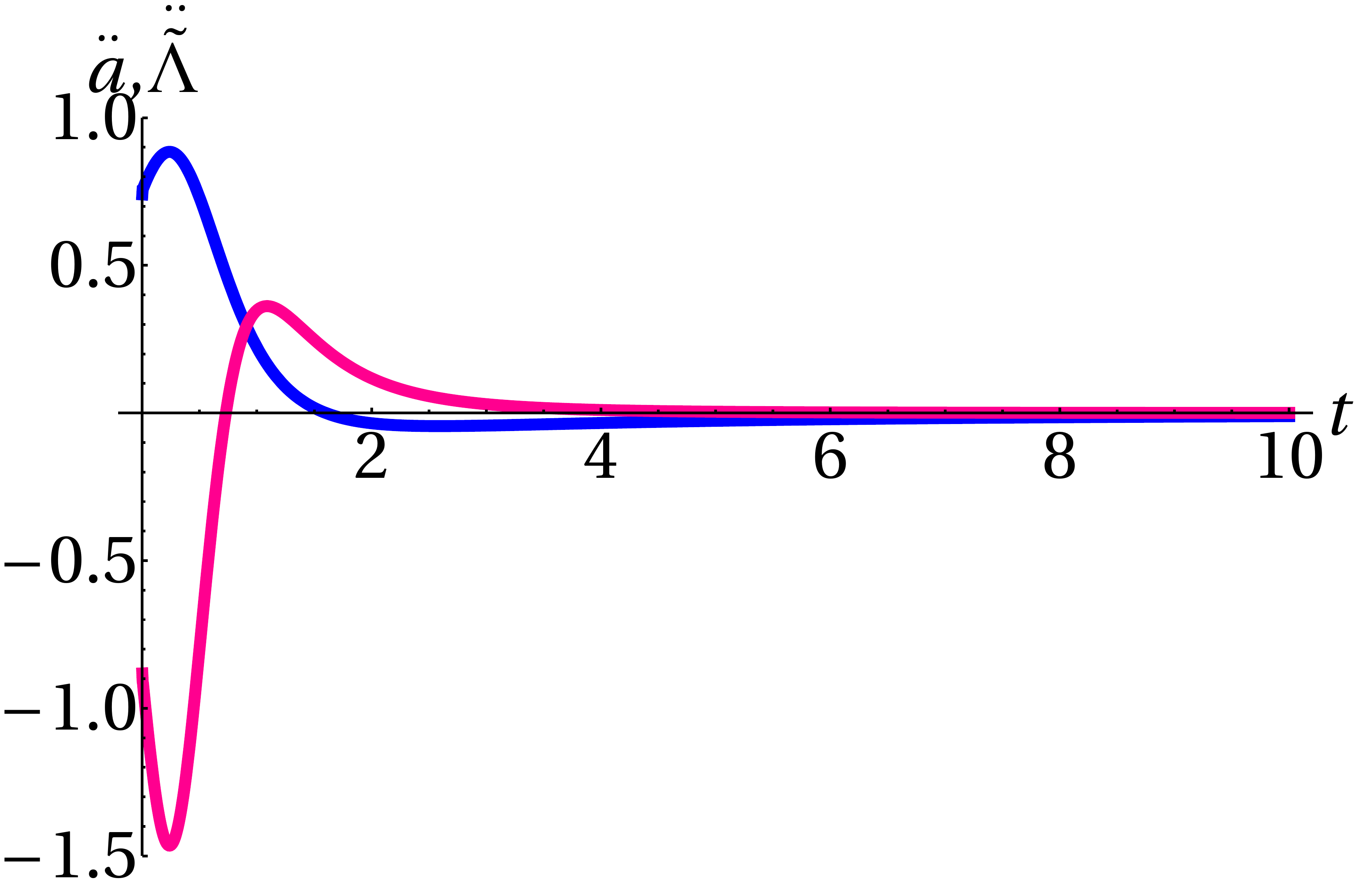}
    \caption{The accelerations: $\ddot a$ is represented by the blue curve, and $\ddot \tilde\Lambda$ by the red curve.}
    \label{addotLTddot000010constantb}
  \end{subfigure}
\qquad
  \begin{subfigure}[t]{.5\linewidth}
    \centering
    \includegraphics[width=0.7\columnwidth]{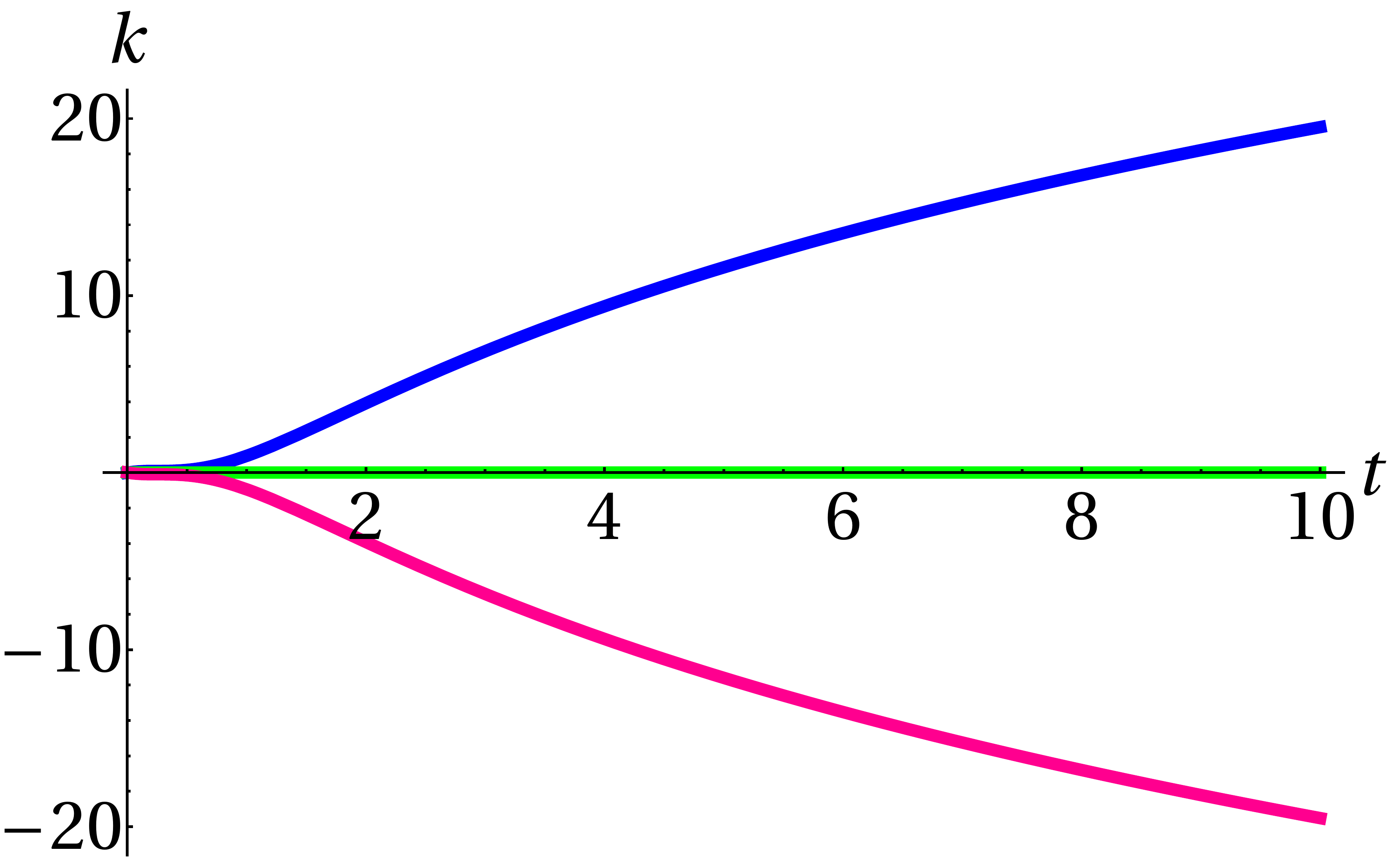}
    \caption{The harmonic function $k$ using: $\dot k\left(0\right)=1$ (blue curve), $\dot k\left(0\right)=0$ (green line), and $\dot k\left(0\right)=-1$ (red curve).}
    \label{k000010constantb}
  \end{subfigure}
    \caption{Initial conditions set number 10 for constant $b$.}
  \label{Fig55}
  \end{figure}

\begin{figure}[H]
  \begin{subfigure}[t]{.5\linewidth}
    \centering
    \includegraphics[width=0.7\columnwidth]{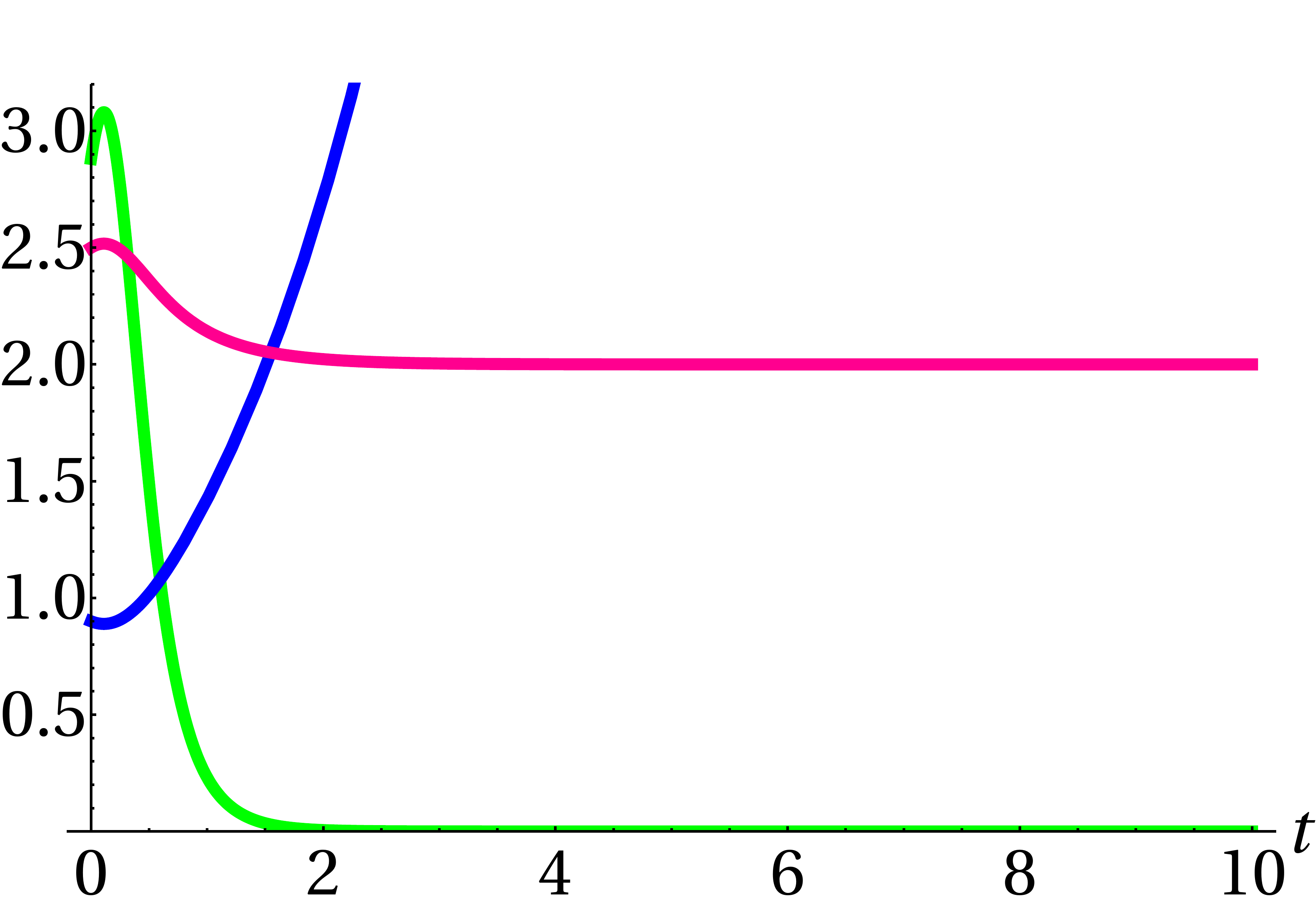}
    \caption{The scale factor $a$ is represented by the blue curve, $\tilde\Lambda$ by the red curve, while $ \left|{G_{i\bar j} \dot z^i \dot z^{\bar j}} \right|$ by the green curve.}
    \label{aLTzz000011constantb}
  \end{subfigure}
\qquad
  \begin{subfigure}[t]{.5\linewidth}
    \centering
    \includegraphics[width=0.7\columnwidth]{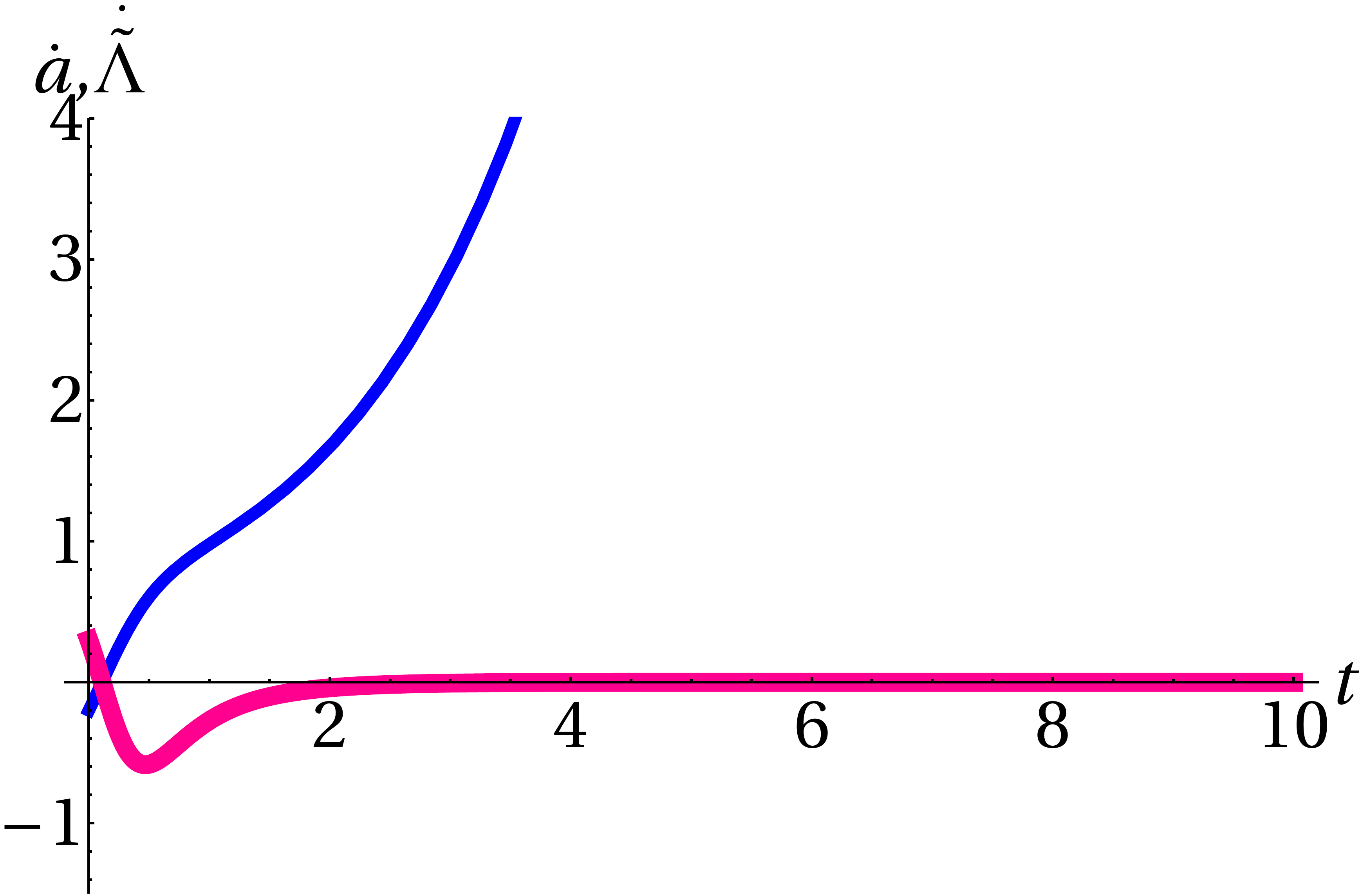}
    \caption{The expansion rates: $\dot a$ is represented by the blue curve, and $\dot \tilde\Lambda$ by the red curve.}
    \label{adotLTdot000011constantb}
  \end{subfigure}
\\[9em]
  \begin{subfigure}[t]{.5\linewidth}
    \centering
    \includegraphics[width=0.7\columnwidth]{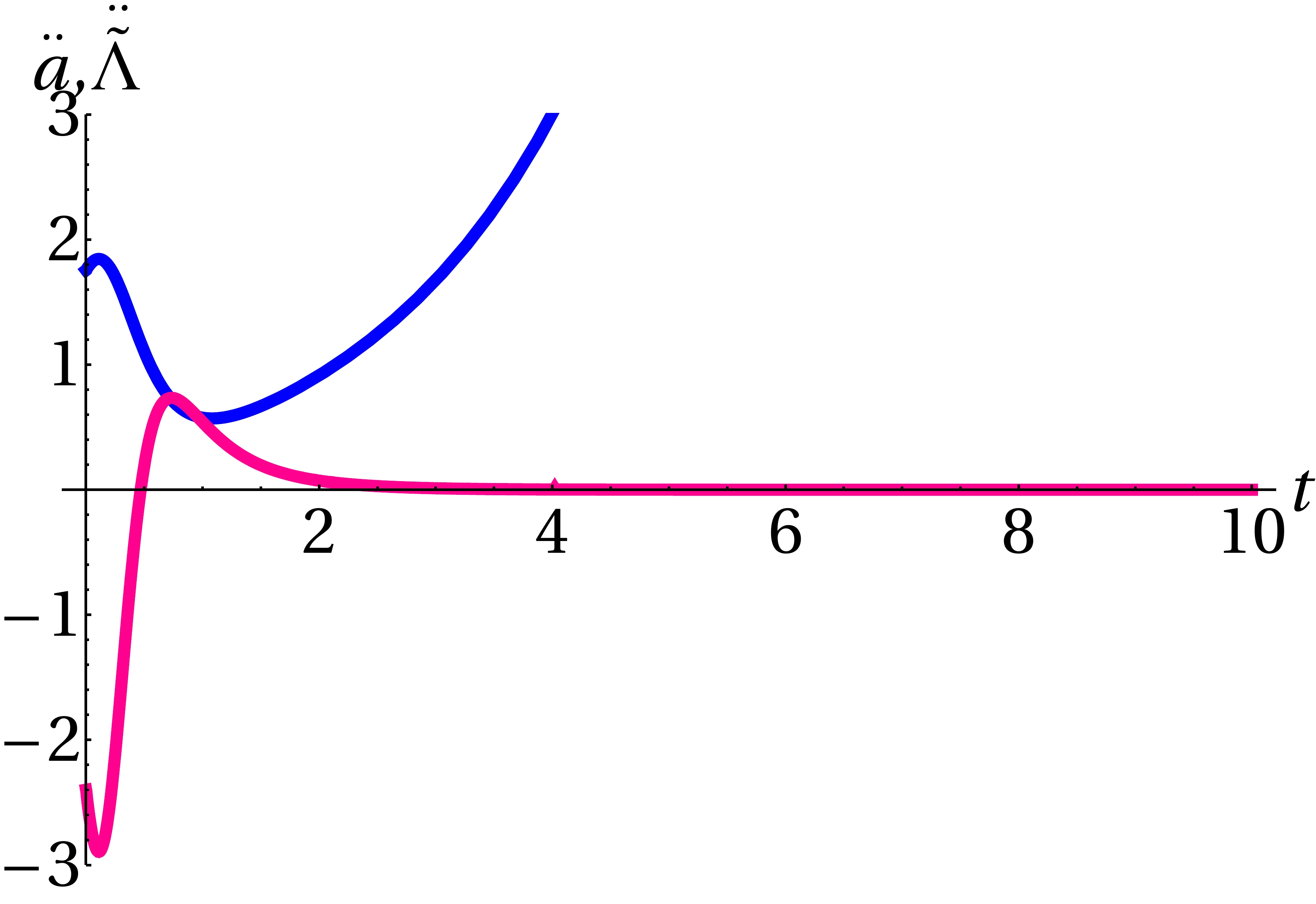}
    \caption{The accelerations: $\ddot a$ is represented by the blue curve, and $\ddot \tilde\Lambda$ by the red curve.}
    \label{addotLTddot000011constantb}
  \end{subfigure}
\qquad
  \begin{subfigure}[t]{.5\linewidth}
    \centering
    \includegraphics[width=0.7\columnwidth]{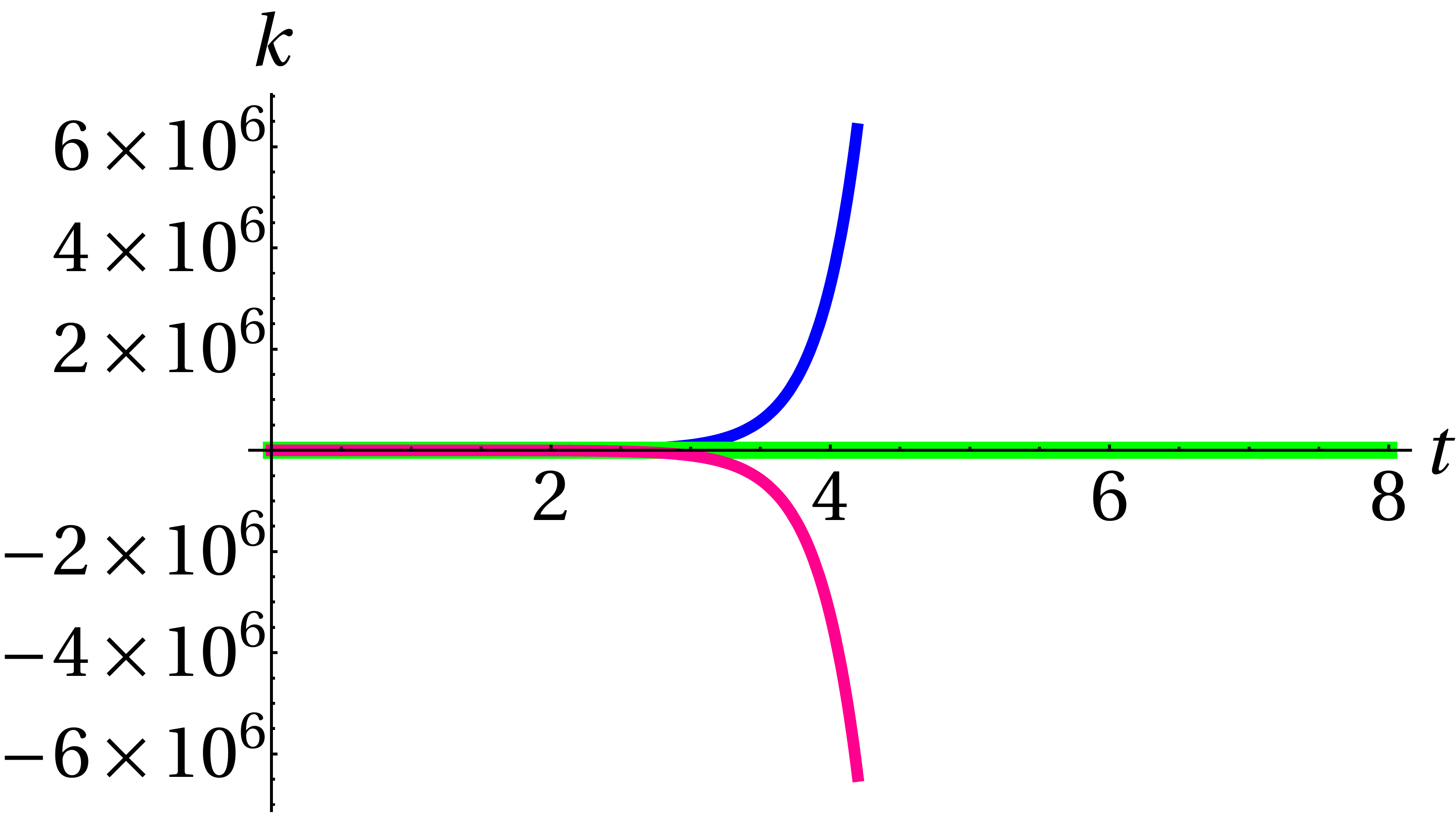}
    \caption{The harmonic function $k$ using: $\dot k\left(0\right)=1$ (blue curve), $\dot k\left(0\right)=0$ (green line), and $\dot k\left(0\right)=-1$ (red curve).}
    \label{k000011constantb}
  \end{subfigure}
\caption{Initial conditions set number 11 for constant $b$.}
  \label{Fig58}
\end{figure}


\begin{figure}[H]
  \begin{subfigure}[t]{.5\linewidth}
    \centering
    \includegraphics[width=0.7\columnwidth]{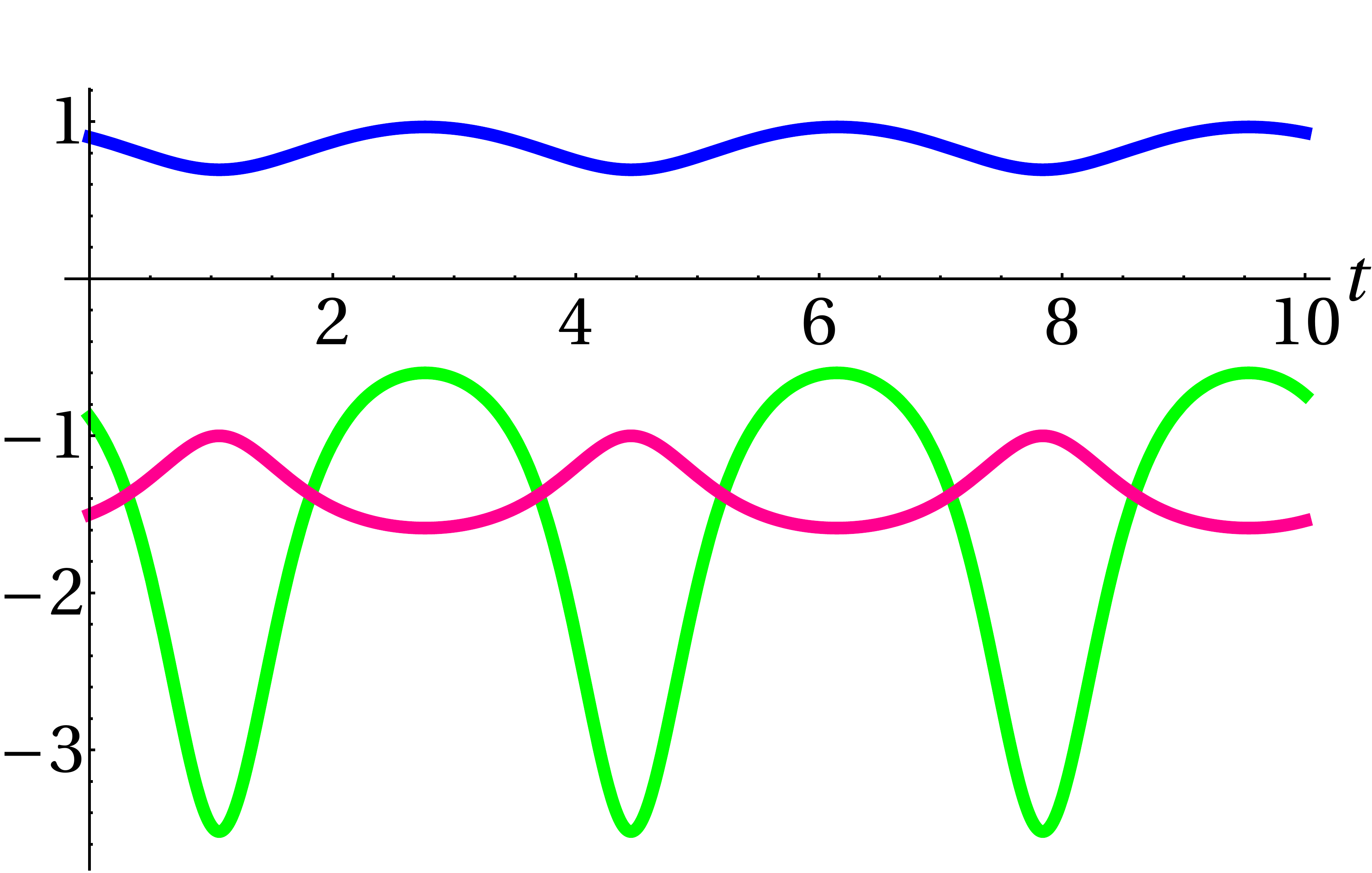}
    \caption{The scale factor $a$ is represented by the blue curve, $\tilde\Lambda$ by the red curve, while $ {G_{i\bar j} \dot z^i \dot z^{\bar j}} $ by the green curve.}
    \label{aLTzz000012constantb}
  \end{subfigure}
\qquad
  \begin{subfigure}[t]{.5\linewidth}
    \centering
    \includegraphics[width=0.7\columnwidth]{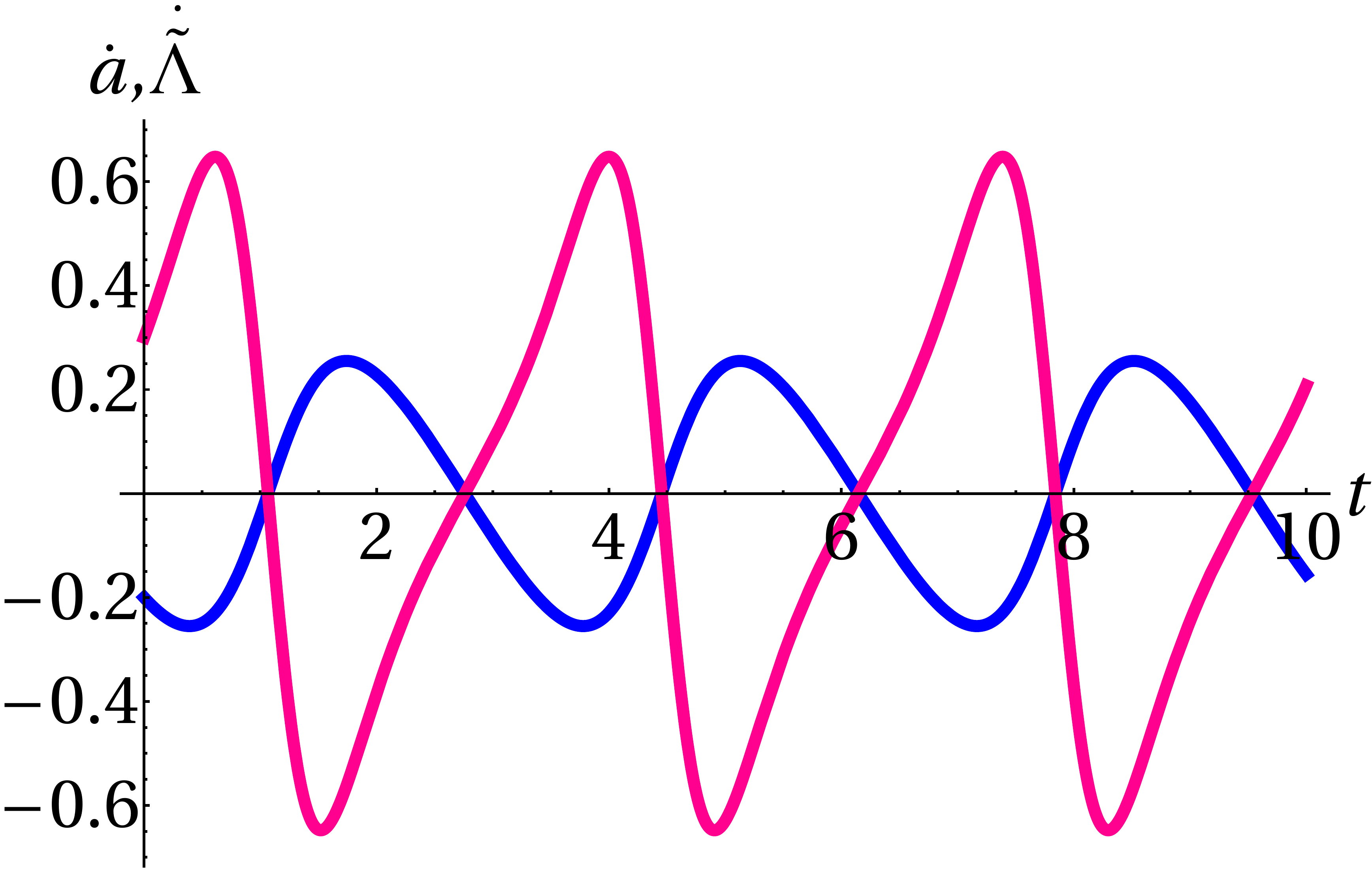}
    \caption{The expansion rates: $\dot a$ is represented by the blue curve, and $\dot \tilde\Lambda$ by the red curve.}
    \label{adotLTdot000012constantb}
  \end{subfigure}
\\[5em]
  \begin{subfigure}[t]{.5\linewidth}
    \centering
    \includegraphics[width=0.7\columnwidth]{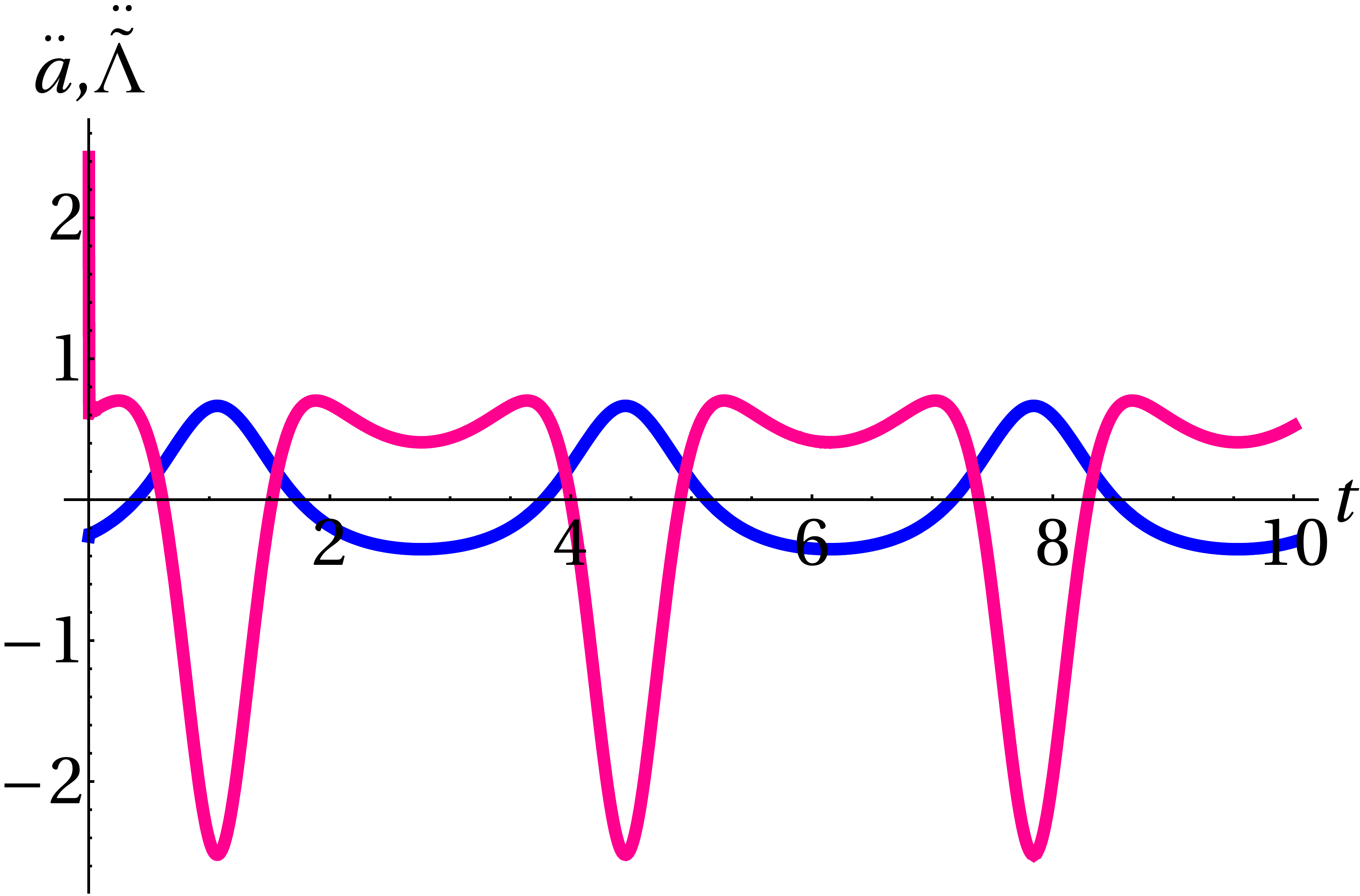}
    \caption{The accelerations: $\ddot a$ is represented by the blue curve, and $\ddot \tilde\Lambda$ by the red curve.}
    \label{addotLTddot000012constantb}
  \end{subfigure}
\qquad
  \begin{subfigure}[t]{.5\linewidth}
    \centering
    \includegraphics[width=0.7\columnwidth]{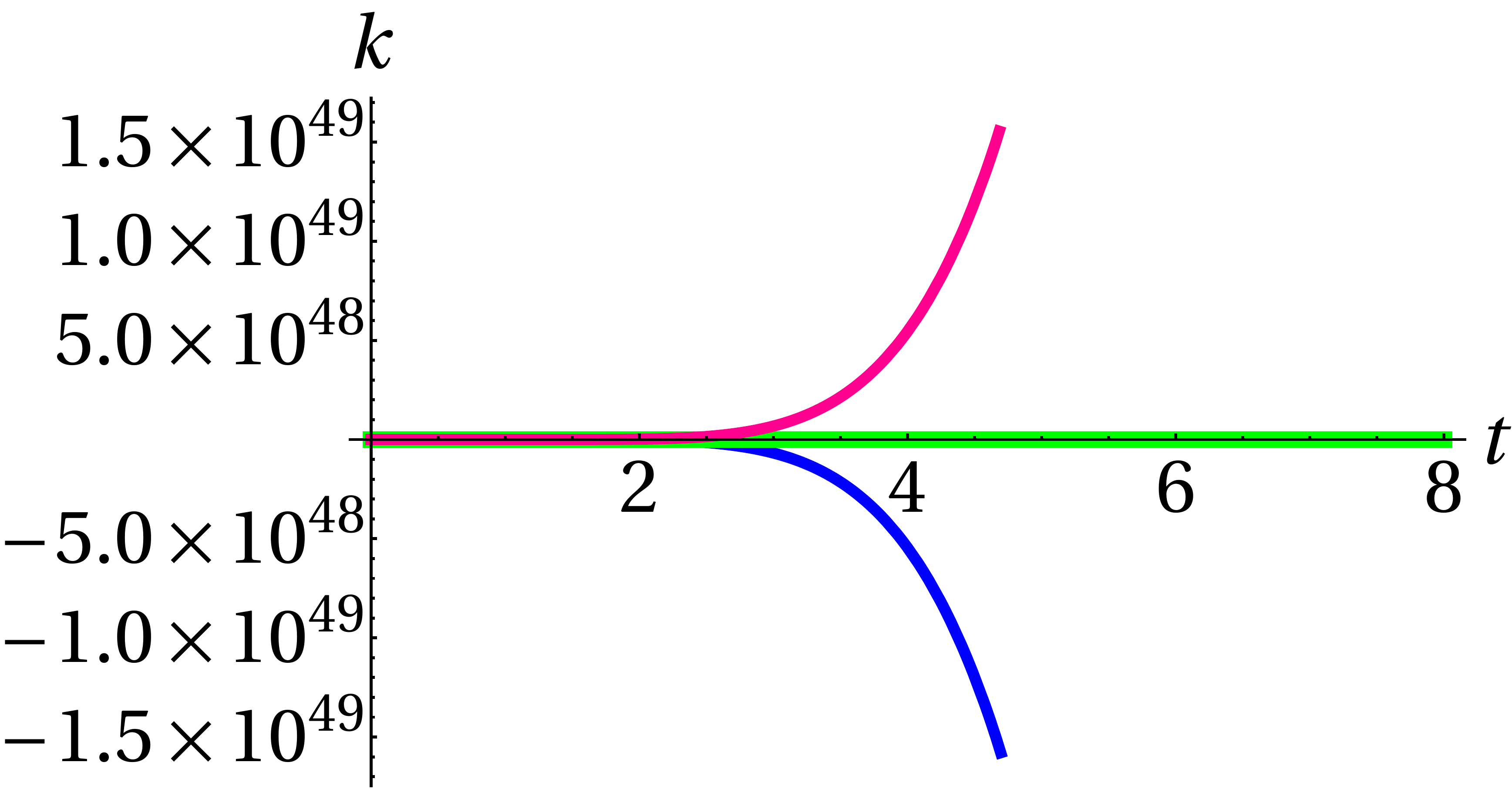}
    \caption{The harmonic function $k$ using: $\dot k\left(0\right)=1$ (blue curve), $\dot k\left(0\right)=0$ (green line), and $\dot k\left(0\right)=-1$ (red curve).}
    \label{k000012constantb}
  \end{subfigure}
    \caption{Initial conditions set number 12 for constant $b$.}
  \label{Fig57}
  \end{figure}
Once again the correlation between the scale factor $a$ and the moduli flow norm is clear. For the first set of six IC's, it is interesting to note that a constant value of the bulk scale $b$ seems to necessarily lead to mostly negative values $\Lambda\left(t\right)$. As such one cannot seriously consider those from a cosmological perspective. However, IC sets \#2, and 5 are clear exceptions. In fact these are the only two cases that may be thought of as possible histories of our universe. There are two interesting cases of periodical universes; these are IC sets \# 3 and 6, corresponding to negative $\tilde \Lambda$. For the second set of six initial conditions only set\#8 can be a model of our universe, although a trivial one since this is a set with a positive $\Lambda$. Oscillating universes with negative $\tilde \Lambda$ appear in sets 9 and 12. Finally the behavior of the remaining fields is similar to those studied in the constant $\Lambda$ cases.
\section{A brane- world depends on the bulk extra dimension}
Consider a case where all the fields in the bulk are dependent on time and the bulk extra dimension ($y$) \cite{Emam2013}.
The 3- brane and the bulk cosmological constants $\Lambda$ and $\tilde{\Lambda}$ are static. 
The brane- universe contains dust and radiation with density and pressure independent of time, and it's a de-Sitter- like spacetime with $p=-\rho$. the energy stress tensors of the bulk, then will be: 
\bea
\nonumber
T_{\mu\nu}^{Bulk} &=&  G_{i\bar{j}} \partial_\mu z^i \partial_\nu z^{\bar{j}} - \frac{1}{2} g_{\mu\nu} G_{i\bar{j}} \partial_\alpha z^i \partial^\alpha z^{\bar{j}}, \\
T_{yy} &=& G_{i\bar{j}} \partial_y z^i \partial_y  z^{\bar{j}} - \frac{1}{2} g_{yy} G_{i\bar{j}} \partial_\alpha z^i \partial^\alpha z^{\bar{j}}.
\eea
Solving the Einstein's equations:
\be
\begin{aligned}
 G_{\mu\nu}+\Lambda g_{\mu\nu} =& T_{\mu\nu}^{Bulk} + T_{\mu\nu}^{3b} \\
G_{yy}+\tilde{\Lambda} g_{yy} =& T_{yy}, ~~~~~~~~ G_{yt}= T_{yt}=0,
\end{aligned}
\ee
gives:
\bea
\nonumber
G_{tt} -  \Lambda &=& \frac{1}{2} G_{i\bar{j}} ~\dot z^{i} ~ \dot z^{\bar{j}} + \frac{1}{2} b^2 G_{i\bar{j}} ~z^{i'} ~ z^{\bar{j}'} + \rho, \\ \nonumber
G_{rr}+ a^2 \Lambda & =& \frac{1}{2} a^2 G_{i\bar{j}} ~\dot z^{i} ~ \dot z^{\bar{j}} - \frac{1}{2} a^2 b^2 G_{i\bar{j}} ~z^{i'} ~ z^{\bar{j}'} + a^2 p \\
G_{yy} + b^2 \tilde{\Lambda} &=& \frac{1}{2} b^2  G_{i\bar{j}} ~\dot z^{i} ~ \dot z^{\bar{j}} + \frac{1}{2} G_{i\bar{j}} ~z^{i'} ~ z^{\bar{j}'} .
\eea
The components of the Einstein tensor are:
\bea 
\label{GG2}
\nonumber G_{tt}&=&  3\left[ {\left( {\frac{{\dot a}}{a}} \right)^2  + \left( {\frac{{\dot a}}{a}} \right)\left( {\frac{{\dot b}}{b}} \right)} \right]  ,\\ \nonumber
G_{rr}  &=& - \left( 2 \ddot a a + \dot a^2 \right)   - \frac{{a}}{b} \left( \ddot b a + 2 \dot a \dot b \right) , \\
G_{yy}  &=&  
-3  \Big(\frac{b}{a}\Big)^2 (\ddot a a + \dot a^2)  . 
\eea
These equations admit multiple branes located at various value of $y=y_I (I=1,...,N \in \mathbb{Z})$, such that $y \to \sum\limits_{I=1}^N |y-y_I|$. However we study the fields effectively at an arbitrary $y$ value or in other words as observed from our brane. For a static extra dimension, the 3-brane's Hubble parameter $H=\frac{\dot{a}}{a}$, and the bulk's scale factor $b=1$. Thus, the modified Friedmann equations become:
\bea
3 H^2 &=&  \dot z  {\dot{\bar z}} +z^\prime \bar{z}^\prime +  \rho+ \Lambda , \label{lam1} \\ 
- \Big( 2 \frac{\ddot a}{a} + H^2 \Big) &=& - \Lambda - z^\prime \bar{z}^\prime  + \dot z  {\dot{\bar z}} + p , \label{lam2} \\ 
-3 \Big(  \frac{\ddot a}{a} + H^2 \Big) &=& z^\prime \bar{z}^\prime + \dot z  {\dot{\bar z}} -  \tilde{\Lambda}. \label{lam3}
\eea 
Define $\Lambda_{eff}= z^\prime \bar{z}^\prime + \Lambda$. Where $\Lambda$ is interpreted as
the cosmological constant related to the value of the quantum mechanical vacuum energy, while $\Lambda_{eff}$ is the observed value of the dark energy density based on the brane universe rate of acceleration. While the norm of the moduli variation with respect to the bulk fifth dimension acts to reduce the first to the second. $z' \bar{z}' \equiv G_{i\bar{j}} ~z^{i'} ~ z^{\bar{j}'}$ and $\dot z  {\dot{\bar z}} \equiv G_{i\bar{j}} ~\dot z^{i} ~ \dot z^{\bar{j}} $. It is known that the scalar quantity
$G_{i\bar{j}} ~z^{i'} ~ z^{\bar{j}'}$ has a negative value.
Now substitute by Equ. (\ref{lam1}) in Equ. (\ref{lam2}), then substitute by both Equ. (\ref{lam1}), and Equ. (\ref{lam2}) into Equ. (\ref{lam3}), the equations become: 
\bea
\nonumber H^2 &=& \frac{1}{3} \Big(\Lambda_{eff} + \dot z  {\dot{\bar z}} +\rho \Big), 
\\ \nonumber \frac{\ddot a}{a} &=& \frac{1}{3} \Lambda_{eff} - \frac{2}{3} \dot z  {\dot{\bar z}}  -  \frac{1}{6} (\rho+ 3 p),
\\ \tilde{\Lambda} &=& 2 \Lambda_{eff} + z' \bar{z}'  + \frac{1}{2} (\rho- 3 p) . 
\label{fr1}
\eea
For a dynamical extra dimension: $\dot{b} \neq 0 $, 
let $\tilde{H} =\frac{\dot{b}}{b}$ that defines the Hubbel parameter of the bulk, then the modified Friedmann equations become:
\bea
\nonumber 3 \left[ H^2 + H \tilde{H} \right] &=&  \dot z  {\dot{\bar z}} + b^2 z^\prime \bar{z}^\prime +  \rho+ \Lambda , \\ \nonumber 
- \Big( 2 \frac{\ddot a}{a} + \frac{\ddot b}{b} + H^2 + H \tilde{H} \Big) &=& - \Lambda - b^2 z^\prime \bar{z}^\prime  + \dot z  {\dot{\bar z}} + p , \\ 
-3 b^2 \Big( \frac{\ddot a}{a}  + H^2 \Big) &=& z^\prime \bar{z}^\prime + b^2 \dot z  {\dot{\bar z}} - b^2 \tilde{\Lambda}.
\eea 
Let $\Lambda_{eff}= b^2 z^\prime \bar{z}^\prime + \Lambda$
\bea
\nonumber H^2 + H \tilde{H} &=& \frac{1}{3} \Big( \Lambda_{eff} + \dot z  {\dot{\bar z}} +  \rho \Big), \\ \nonumber  
\frac{\ddot a}{a} + \frac{1}{2} \frac{\ddot b}{b}  &=& \frac{1}{3} \Lambda_{eff} -\frac{2}{3}  \dot z  {\dot{\bar z}} 
-  \frac{1}{6} (\rho+ 3 p) , \\  
\tilde{\Lambda} &=& 2 \Lambda_{eff} + z' \bar{z}' -\frac{\ddot b}{b} - H \tilde{H} + \frac{1}{2} (\rho- 3 p) . 
\label{fr2}
\eea 
For a de-Sitter-like spacetime $p=-\rho$. From Equ.(\ref{fr1}), and Equ.(\ref{fr2}) we notice that the fifth extra dimension
whether a de-Sitter-like if it is static with a constant scale factor, or it can be an anti de-Sitter-like if it expands with time and 
$\frac{\ddot b}{b}$ and $H \tilde{H}$ are large enough. Also in this case the bulk scale factor $b(t)$ contributes in reducing the 3-brane's cosmological constant value. In this way the cosmological constant problem can be resolved analytically in a different way than the previous solutions have been made in this chapter. Where in this section what contributes in enhancing the brane's cosmological constant to get the observed value is the moduli's variation with respect to the fifth extra dimension.

\setcounter{equation}{0}

\chapter{The Topology of the Calabi-Yau Manifold}

In the previous chapters we have solved Einstein field equations in case of world- branes filled with dust, radiation, or both,
and dust with radiation and dark energy ($\Lambda$). These solutions shows us the time evolution of the quantity 
$G_{i\bar j} \dot z^i \dot z^{\bar j}$ \footnote{\textit {Which before we roughly called the moduli .}} at different initial conditions.
In this chapter, we will study each of the moduli $z^i$ and the metric $G_{i\bar j}$ dependence on time. 
We take the dimension of the Calabi-Yau complex structure space $\mathcal{M}_c$, $h_{(2,1)}=1$, which means 
we have a toy model with only one moduli $z$ and a single $\mathcal{M}_c$ metric component $G$. 

Although this seems a simple case, as we will see it will get us closer for a deeper understanding
of the complex structure of the CY manifold. 

Indeed there are many studies about the geometry of $\mathcal{M}_c$, like \cite{Candelas:355} and \cite{Candelas:359}, while in \cite{Belavin:2018}, \cite {Candelas:3551}, and \cite{Belavin:2017} the Calabi-Yau 3-fold is considered as a quintic threefold in the $\mathbb{P}^4$ projection space
\footnote{\em See Appendix (\ref{cy}).}. Here however we do not need to make the last assumption because we already know the time behavior of
$G_{i\bar j} \dot z^i \dot z^{\bar j}$, also the field equations of the moduli $z^i$ and its complex conjugate are well known through
$\mathcal{N}=2$ $\D=5$ supergravity, Equ. (\ref{zzz}) . 

Since all solutions that we have introduced till now for the $\N=2$ $\D=5$ supergravity are numeric, in the last of the chapter, we solve 
Friedmann- like equations (\ref{eee}) analytically in the case of $b=1$ and show how both solutions manifest $G_{i\bar j} \dot z^i \dot z^{\bar j}$ correlations to the brane scale factor $a(t)$. 
\section{The moduli field equations}
Starting by the field equations of the moduli $z^i$ and its complex conjugate  $z^{\bar{i}}$
\bea \nonumber
& (\Delta z^i) \star 1 + \Gamma^i_{jk} dz^j \wedge \star dz^k + \frac{1}{2} e^\sigma G^{i\bar{j}} \partial_{\bar{j}}  \langle d \Xi | \underset{\Lambda}{\mathbf {\Lambda}} |  \star d\Xi \rangle  =0 \\
& (\Delta z^{\bar{i}}) \star 1 + \Gamma^{\bar{i}}_{\bar{j}\bar{k}} dz^{\bar{j}} \wedge \star dz^{\bar{k}} + \frac{1}{2} e^\sigma G^{\bar{i}j} \partial_{j}  \langle d \Xi |  \underset{\Lambda}{\mathbf \Lambda} | \star d \Xi \rangle   =0 
\eea

\begin{figure}[H]
  \begin{subfigure}[t]{.5\linewidth}
    \centering
    \includegraphics[width=1.0\columnwidth]{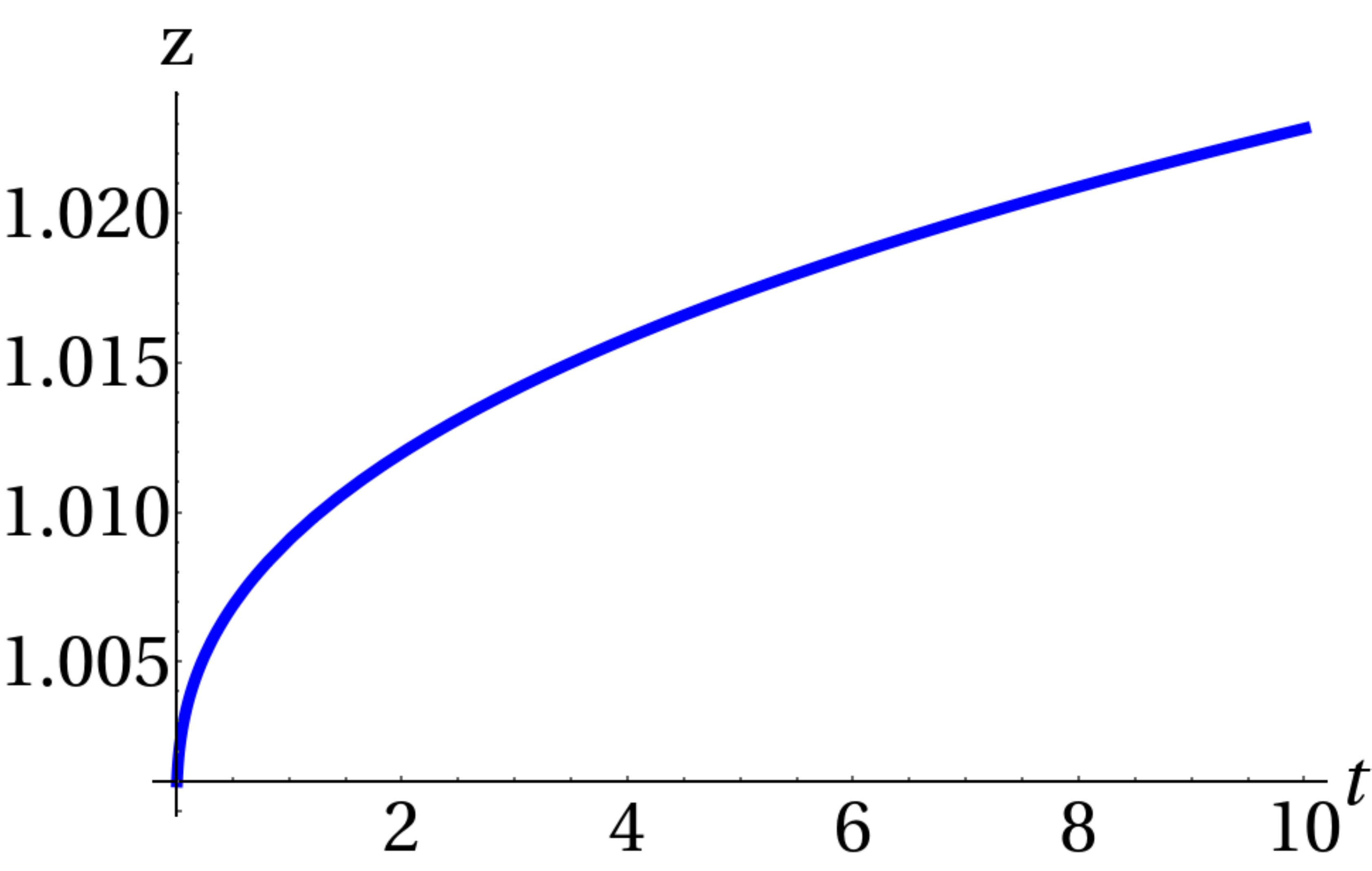}
\end{subfigure}
\qquad
\begin{subfigure}[t]{.5\linewidth}
    \centering
    \includegraphics[width=1.0\columnwidth]{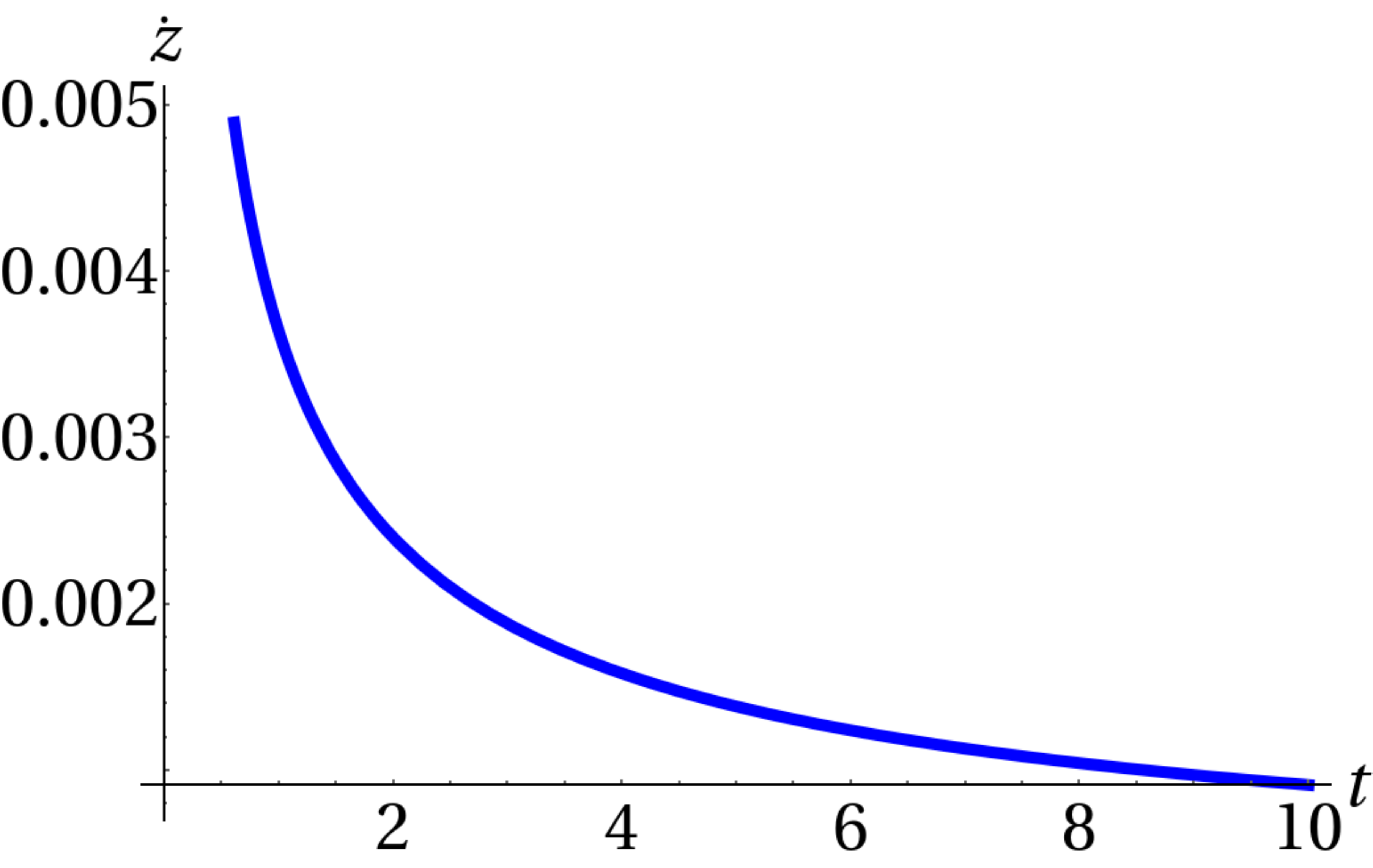}
  \end{subfigure}
\caption{(Left panel): The moduli is plotted versus time. 
(Right panel): The moduli velocity is plotted versus time. For $C =1$, and in case 
of radiation, dust and $\Lambda$ filled brane with $\Lambda=1$ and $\tilde{\Lambda}=0$. }
\label{zz}
\end{figure}

But from the BPS condition Equ. (\ref{bps}), The moduli field equations become:
\bea \nonumber
& (\Delta z^i) \star 1 + \Gamma^i_{jk} dz^j \wedge \star dz^k + G^{i\bar{j}} (\partial_{\bar{j}} G_{l\bar{k}}) dz^l \wedge \star dz^{\bar{k}}   =0 \\
& (\Delta z^{\bar{i}}) \star 1 + \Gamma^{\bar{i}}_{\bar{j}\bar{k}} dz^{\bar{j}} \wedge \star dz^{\bar{k}} + G^{\bar{i}j} (\partial_j G_{l\bar{k}}) dz^l \wedge \star dz^{\bar{k}}  =0.
\eea
Dropping the differential forms formulation, we get:
\bea \nonumber
& \nabla^2 z^i + \Gamma^i_{jk} \partial_\mu z^j \partial^\mu z^k + G^{i\bar{j}} (\partial_{\bar{j}} G_{l\bar{k}})\partial_\mu z^l \partial^\mu z^{\bar{k}}  =0 \\
& \nabla^2 z^{\bar{i}} + \Gamma^{\bar{i}}_{\bar{j}\bar{k}} \partial_\mu z^{\bar{j}} \partial^\mu z^{\bar{k}} +
G^{\bar{i}j} (\partial_{j} G_{l\bar{k}})\partial_\mu z^l \partial^\mu z^{\bar{k}}   =0.
\label{zz23}
\eea
The connections or the Christoffel symbols are related to the metric by \cite{CYM2-2}:
\be 
\Gamma^i_{jk} = G^{i\bar{p}} \partial_j G_{k\bar{p}}, ~~~~~~~~ \Gamma^{\bar{i}}_{\bar{j}\bar{k}} = G^{p\bar{i}} \partial_{\bar{j}} 
G_{\bar{k}p},
\ee
substitute in Equ. (\ref{zz23}), we get:
\bea \nonumber
& \nabla^2 z^i + G^{i\bar{p}} \partial_j G_{k\bar{p}} ~ \partial_\mu z^j \partial^\mu z^k + G^{i\bar{j}} (\partial_{\bar{j}} G_{l\bar{k}})\partial_\mu z^l \partial^\mu z^{\bar{k}}  =0 \\
& \nabla^2 z^{\bar{i}} +  G^{p\bar{i}} \partial_{\bar{j}} 
G_{\bar{k}p}~ \partial_\mu z^{\bar{j}} \partial^\mu z^{\bar{k}} +
G^{\bar{i}j} (\partial_{j} G_{l\bar{k}})\partial_\mu z^l \partial^\mu z^{\bar{k}}   =0 
\label{zz34}\eea
The moduli are independent of the 3- spatial dimensions. And consider the Hodge number $h_{2,1} =1$,  which means we have only one moduli $z$, its complex conjugate $z^*$, a single K\"{a}hler metric component $G$ and the dimension of the moduli space $\M_C$ is unity. So that 
Equ. (\ref{zz34}) simplify to:    
\bea \nonumber
& g^{tt} \ddot{z} +\frac {1}{G} (\partial_z G) ~ \dot {z}^2 + \frac {1}{G} (\partial_{z^*} G) \dot{z} \dot{z}^*  =0 \\
& g^{tt} \ddot{z}^* + \frac {1}{G^*}(\partial_{z^*} G^*) ~ \dot {z^*}^2 + \frac {1}{G^*} (\partial_z G) \dot{z} \dot{z}^* =0 
\eea
From the Robertson-Walker like metric Equ. (\ref{metricE}), the moduli field equations become:
\be
\ddot{z} +\frac {1}{G} (\partial_z G) ~ \dot {z}^2 + \frac {1}{G} (\partial_{z^*} G) \dot{z} \dot{z}^*  =0 
\label{z}
\ee
\be
\ddot{z}^* + \frac {1}{G^*}(\partial_{z^*} G^*) ~ \dot {z^*}^2 + \frac {1}{G^*} (\partial_z G) \dot{z} \dot{z}^* =0 
\label{zstar}
\ee
%
We can solve the equation of $\ddot{z}$ or $\ddot{z^*}$ with 
the numeric solution obtained for the moduli velocity norm 
$G_{i\bar{j}} \dot z^i \dot z^{\bar{j}}$ in chapters (\ref{ch4}) and (\ref{RDE}). 
Where for instance, in chapter (\ref{RDE}) EFE have been solved in case
the total density equals the dust plus radiation densities $\rho = \rho_r+ \rho_m \propto 1/a^4 +1/a^3$,
the total pressure equals to the radiation pressure $p=p_r=\rho_r/3$, $\Lambda=1$ (de Sitter space), and 
$\tilde{\Lambda}=0$. That yields the  brane's and the bulk's scale factors, and $\left| {G_{i\bar j} \dot z^i \dot z^{\bar j}} \right|$ as functions in time. Using  suitable fitting functions, we get the solution of the velocity norm of the complex structure moduli given by
\be
{G_{i\bar j} \dot z^i \dot z^{\bar j}} (t) \simeq -0.4 (t + 0.004)^{-0.9}.
\label{zsol}
\ee
The brane's scale factor varies with time exponentially $a(t) \sim e^{0.2~t}$ which means 
the brane- universe undergoes an inflationary expansion. While the bulk scale factor is given by 
$b(t) \sim 0.06 e^{0.87~t}$.
Solving the moduli field equation Equ. (\ref{z}) with Equ. (\ref{zsol}), gives
the moduli's variation with time \cite{salem2022}:
\be
z(t) \simeq C + \frac{0.001 +0.25~ t}{(1 + 250~ t)^{0.6}},
\label{z-t}
\ee
for $\dot{z}[0]=0.1$. C is the integration constant, for $z[0]=1$, $C \sim 1$. We have made a further approximation here by considering the moduli real. In Fig. (\ref{zz}- left) and (\ref{zz}- right) the moduli and the moduli velocity are plotted versus time, respectively. 

\begin{figure}[H]
  \begin{subfigure}[t]{.5\linewidth}
    \centering
    \includegraphics[width=1.0\columnwidth]{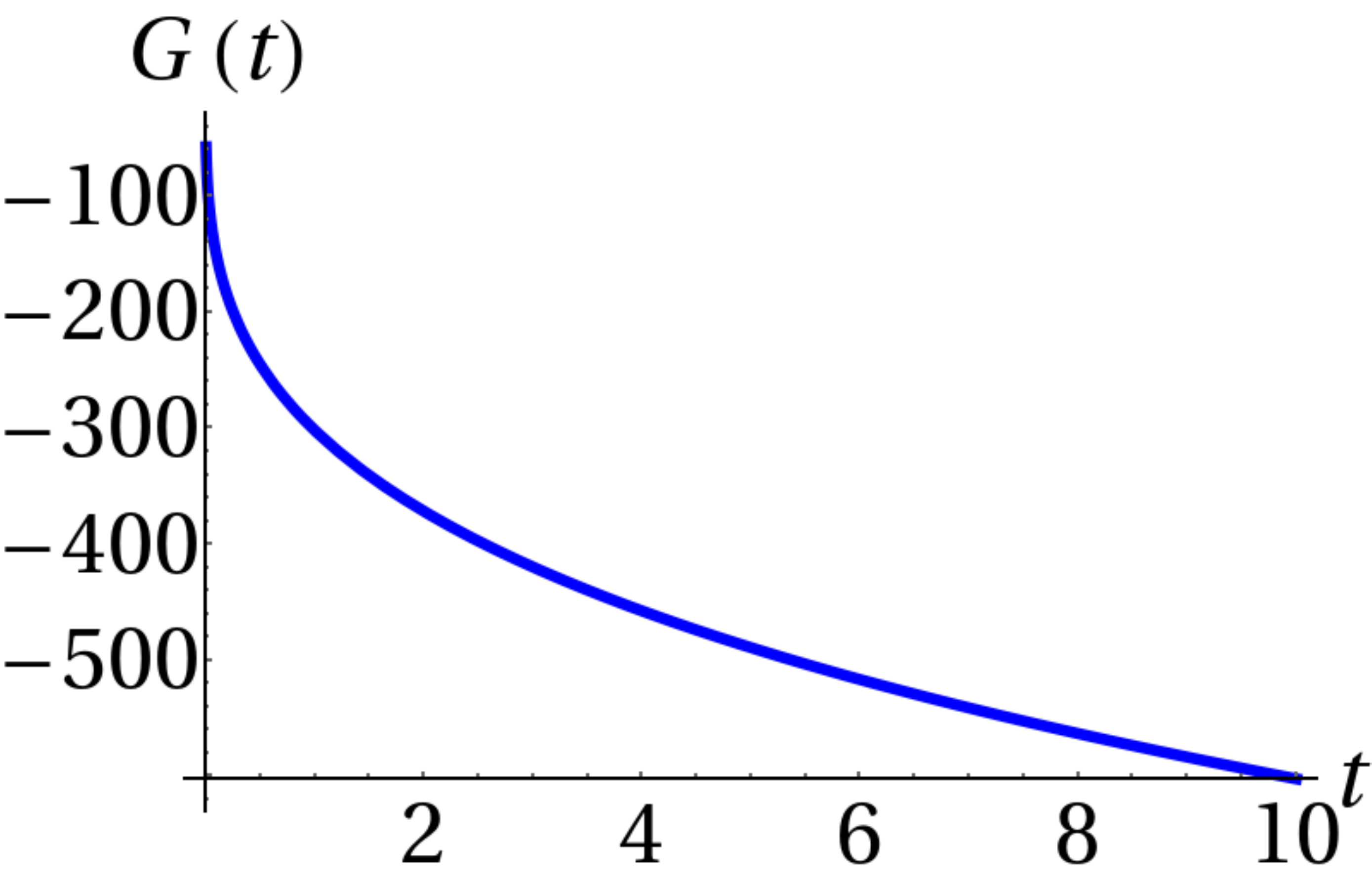}
  \end{subfigure}
\qquad
\begin{subfigure}[t]{.5\linewidth}
    \centering
    \includegraphics[width=1.0\columnwidth]{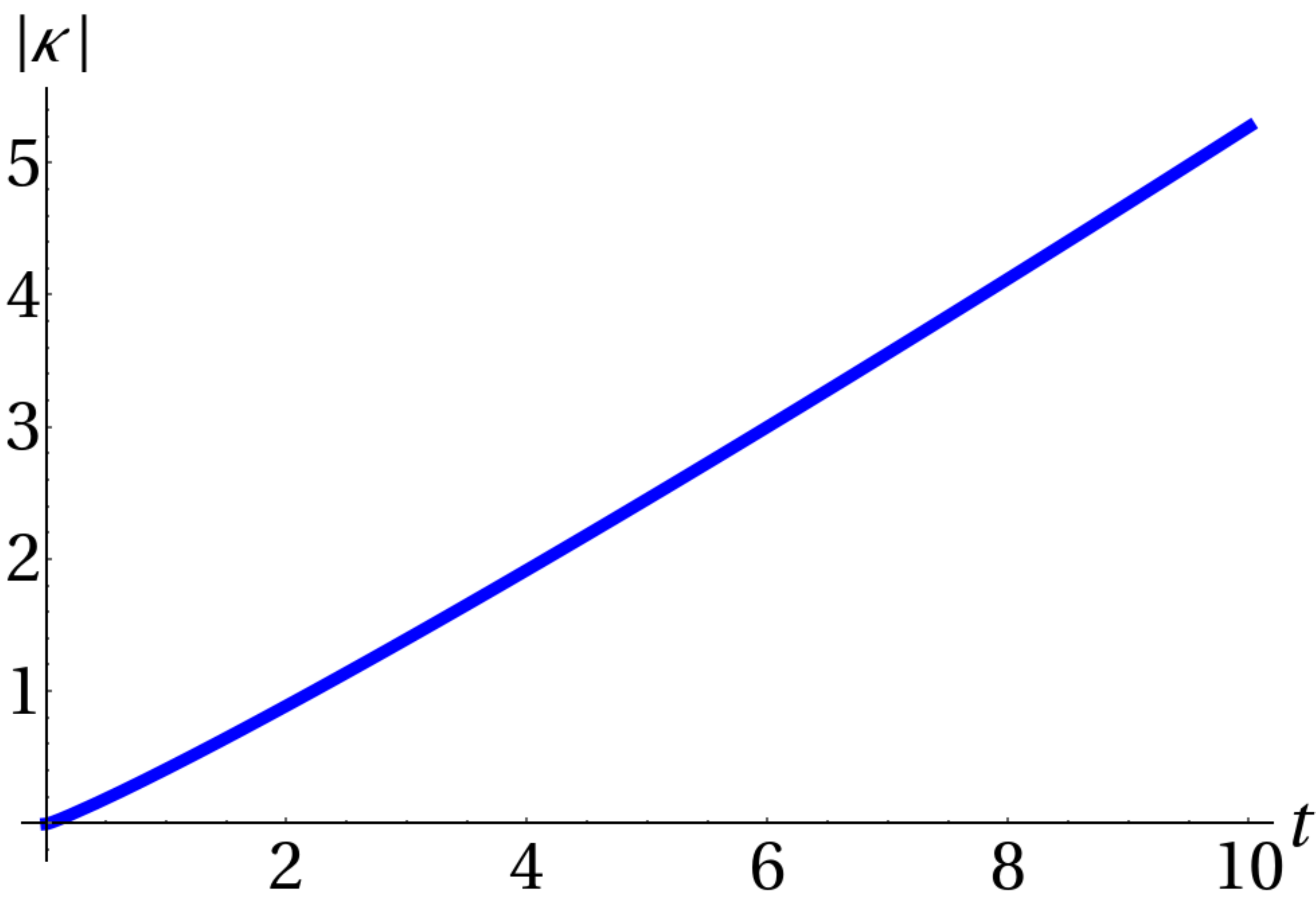}
  \end{subfigure}
\caption{(Left panel): One component of the K\"{a}hler  metric multiplied by a factor $10^{-2}$ is
plotted versus time. (Right panel):
The K\"{a}hler potential of $\M_C$ is plotted versus time.}
\label{G}
\end{figure}

\begin{figure}[H]
  \begin{subfigure}[t]{.5\linewidth}
    \centering
    \includegraphics[width=1.0\columnwidth]{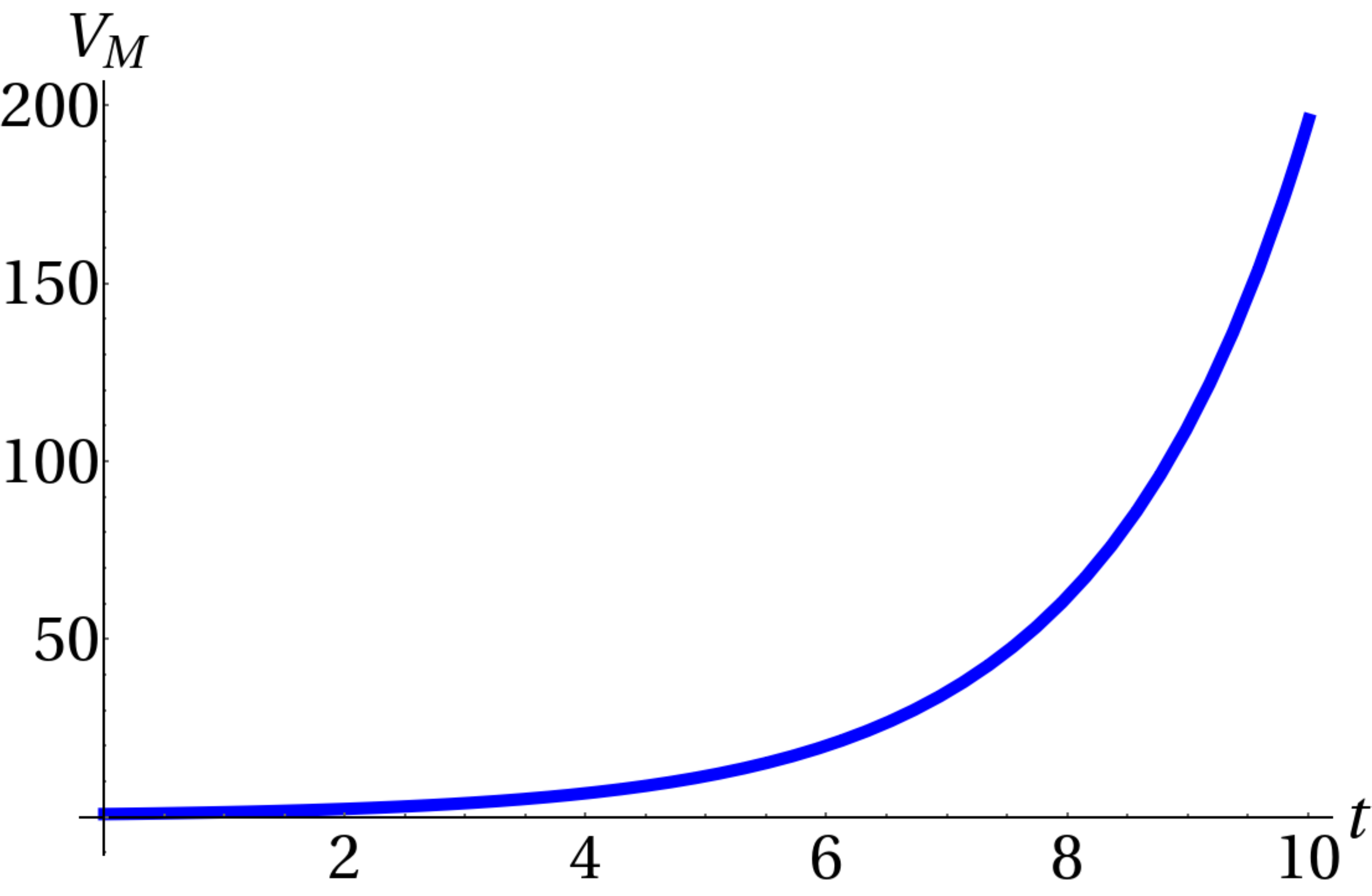}
  \end{subfigure}
\qquad
\begin{subfigure}[t]{.5\linewidth}
    \centering
    \includegraphics[width=1.0\columnwidth]{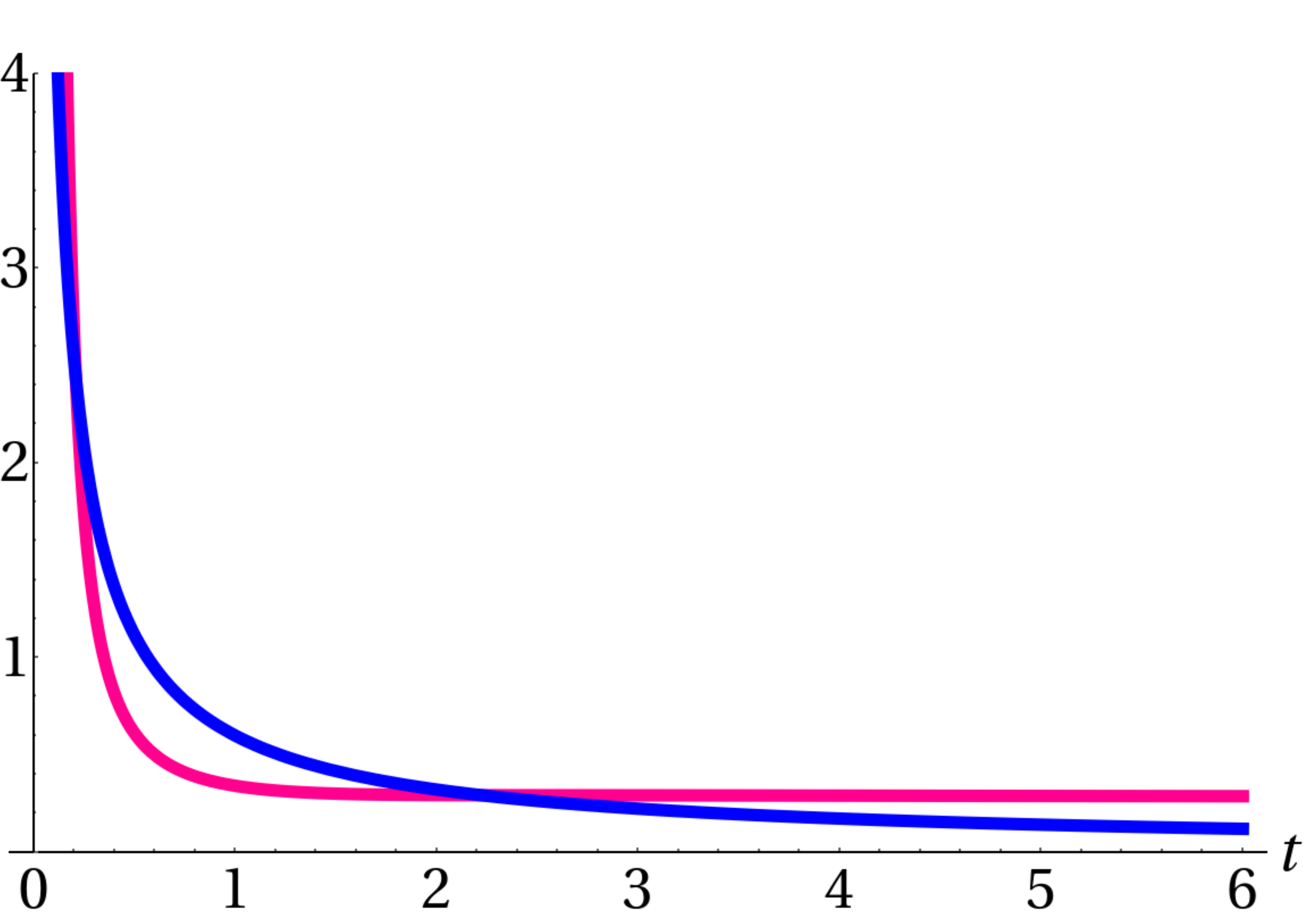}
  \end{subfigure}
\caption{(Left panel): The volume of the Calabi-Yau manifold plotted versus time.
The Hodge number $h_{2,1}=1$.
(Right panel):
Numeric and analytic $G_{i\bar{j}} \dot z^i \dot z^{\bar{j}}$ versus time in red and blue curves, respectively.
The Hodge number $h_{2,1}=1$.}
\label{vol}
\end{figure}

The K\"{a}hler metric can be directly obtained by substituting $\dot{z}$ solution in Equ. (\ref{zsol}). In Fig. (\ref{G}- left)
one component of the metric $G_{i\bar{j}}$ multiplied by a factor $10^{-2}$ is plotted versus time. 
%
We will use Equ. (\ref{pot22}) to get $\kappa$ as a function in time.
According to our approximation it can be written as:
\be 
G = \partial_z \partial_{z^*} \kappa.
\label{Gk}
\ee  
Let all fields depend on time, it becomes:
\be
G(t)  = \frac{\partial}{\partial t} \left( \frac{1}{\dot{z}} \frac{\dot{\kappa}}{\dot{z^*}} \right).
\ee
Solving for the K\"{a}hler potential, yields:
\be
\kappa (t) = - 0.298~ t- 0.0056 ~ t^2 .
\ee
Fig. (\ref{G}- right) shows the absolute value of the potential plotted versus time.
Since the volume of the CY manifold $\M$ is related to the K\"{a}hler potential by $\text{Vol}(\M)=e^{-\kappa}$
(\ref{veve}),
we have plotted in Fig. (\ref{vol}- left) the volume of the Calabi-Yau manifold versus time. As seen it increases with time.
For the sake of comparison, Fig. (\ref{vol}- right) shows $G_{i\bar{j}} \dot z^i \dot z^{\bar{j}}$ plotted versus time as it is obtained directly from the numeric solution of the field equations (\ref{EE3}) without any approximations a long a side as it is obtained when solving Equ. (\ref{zsol}) with Equ. (\ref{z-t}).

\section{Analytic solutions for the field equations}
Modeling a 3 brane- world in $\D=5$ $\N=2$ supergravity has been studied in many earlier papers. In the first \cite{Emam:2015laa} an `empty' brane-world was analytically studied, and a correlation between the dynamics of the brane, \emph{i.e.} the time evolution of the scale factors, was found. 
This was further explored in chapter (\ref{ch4}) where a couple of brane cases was considered: A dust filled brane and a radiation filled brane. 

In all cases not only was the previous correlation confirmed, leading to an expanding brane brought about by the complex structure moduli, but we have also found that the early decay of the norm of the moduli's flow velocity $G_{i\bar j} \dot z^i \dot z^{\bar j}$ directly leads to a period of rapid accelerated expansion of the brane. That  means the complex structure moduli can play the role of the inflaton. 

The studies in chapter (\ref{ch4}) were only numeric. In this section we will
solve the field equations analytically to manifest the moduli 
correlations.

Indeed $G_{i\bar{j}} \dot z_i \dot z_{\bar{j}}^*$ 
drives the dynamics of all the theory fields, whether in the bulk or on the
brane- world. This may be understood from a topological perspective , since 
$G_{i\bar{j}}$ is just the mertic of the Calabi- Yau manifold complex structure space $\M_C$ and the moduli
can be treated as the complex coordinates of $\M_C$ and its number
($h_{2,1}$) is the dimension of that space.

Anyhow we would like to show here analytic formulas for the brane- world scale factor.
Then the analytic moduli can be obtained directly from the field equations. 
This study is different than what has been done in \cite{Emam:2015laa} where 
an analytic formula for $a(t)$  has been assumed in case of a vacuous brane-world. 
To solve analytically we will take a constant bulk scale factor $b=1$, then the field equations reduce to:  
\bea
 \nonumber 3 \left( \frac{\dot a}{a} \right)^2  &=& G_{i\bar j} \dot z^i \dot z^{\bar j}  + \rho, \\ \nonumber
 2 \left( \frac{\ddot a}{a} \right) + \left(\frac{\dot a}{a} \right)^2  &=& - G_{i\bar j} \dot z^i \dot z^{\bar j}  - p,   \\
 -\left[ {\frac{{\ddot a}}{a} + \left( {\frac{{\dot a}}{a}} \right)^2 } \right] &=& -  G_{i\bar j} \dot z^i \dot z^{\bar j}, \label{FE-consb}
\eea
where we consider $\rho$ and $p$ are constants, and $\Lambda=\tilde{\Lambda}=0$. Solving these eqautions together gives:
\be
a(t)= c_2  \cosh \left[\sqrt{3}~ \sqrt{p} ~~(c_1+t)\right]^{1/3},
\ee
where $c_1$, and $c_2$ are the constants of integration. While the energy density is given by:
\be 
\rho=p+ \frac{2 \ddot{a}}{a}+\frac{4 \dot{a}^2}{a^2}.
\label{ED}
\ee

\begin{figure}[t]
    \centering
    \includegraphics[width=0.8\columnwidth]{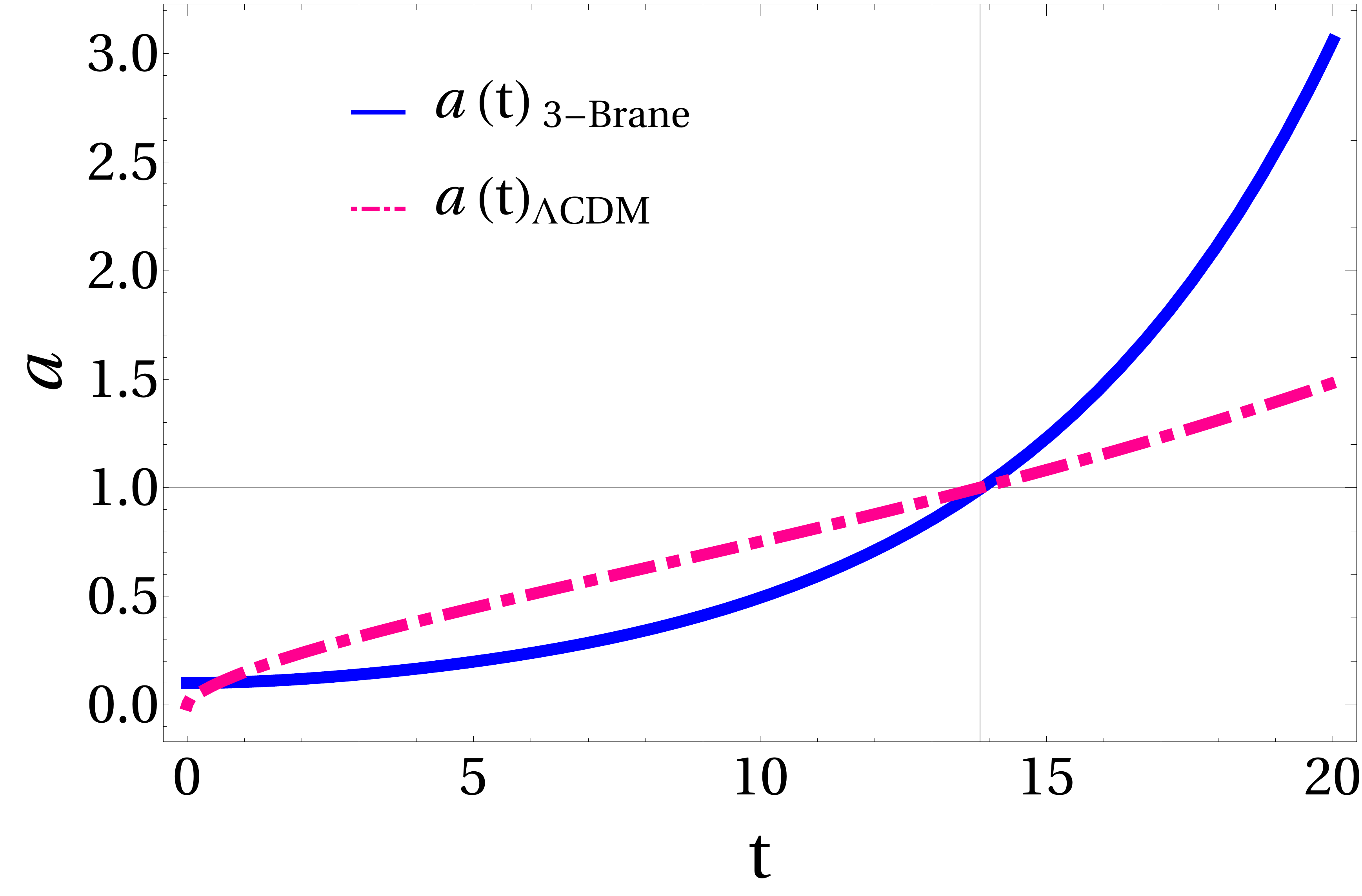}
  \caption{The blue curve: The brane's scale factor is plotted versus time
at $c_1=0$ and $c_2=0.1$ for initial conditions $a(0)=0.1$ and $\dot{a}(0)=0$. $p=0.1$. The pink dashed curve: The scale factor of $\Lambda$CDM model at $\Omega_{m0}= 0.31$, $\Omega_{r0}= 9.2 \times 10^{-5}$, and $\Omega_{\Lambda 0}=0.7$. The current universe age since the big bang is given by the vertical gird line $t_0= 13.842~ [\text{Gyr}]$, and the current value of the our universe's scale factor is given by horizontal gird line $a_0=1$. }
\label{atA}
\end{figure}

\begin{figure}[H]
  \begin{subfigure}[t]{.5\linewidth}
    \centering
    \includegraphics[width=1.0\columnwidth]{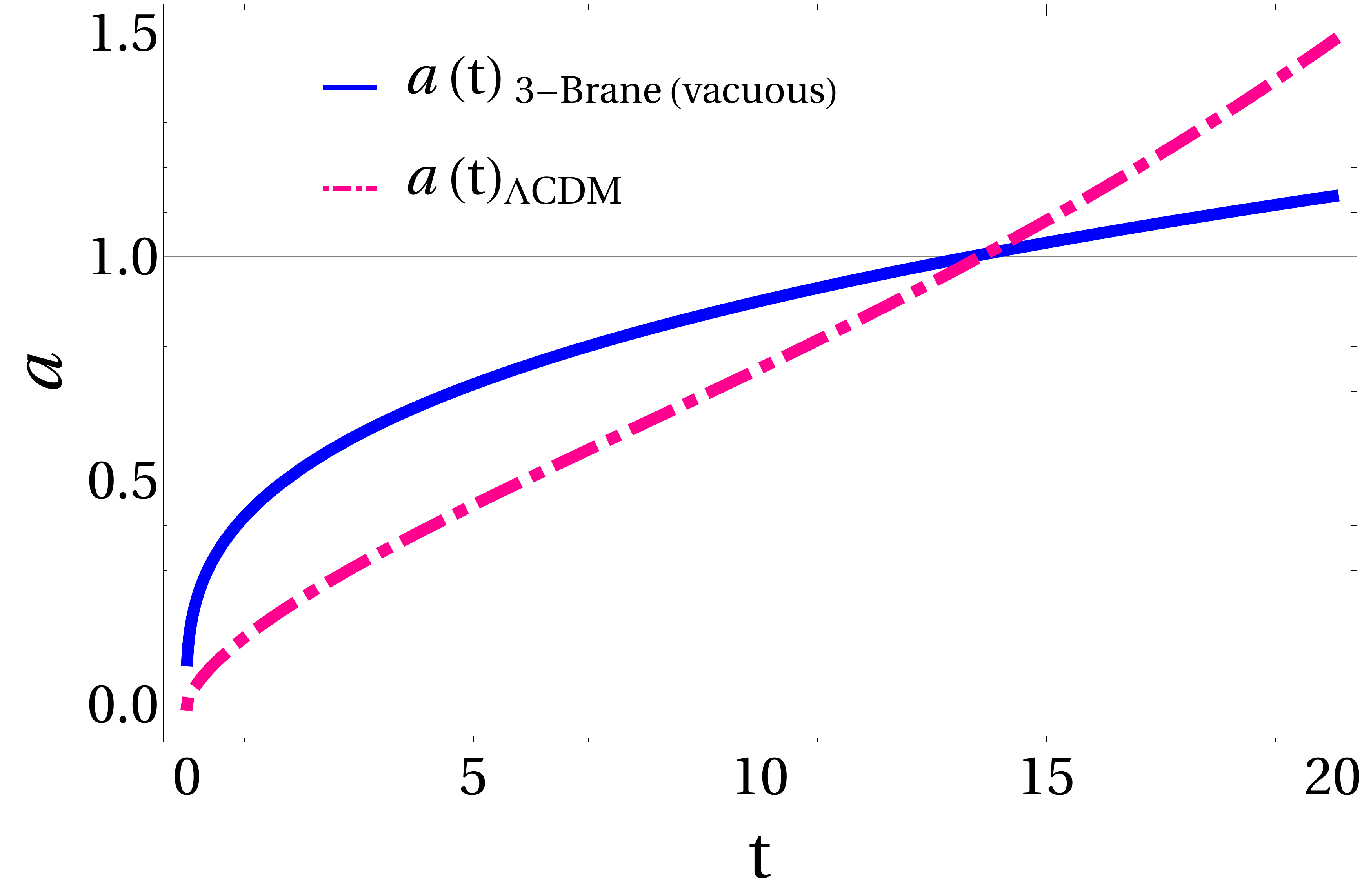}
  \end{subfigure}
\qquad
\begin{subfigure}[t]{.5\linewidth}
    \centering
    \includegraphics[width=1.0\columnwidth]{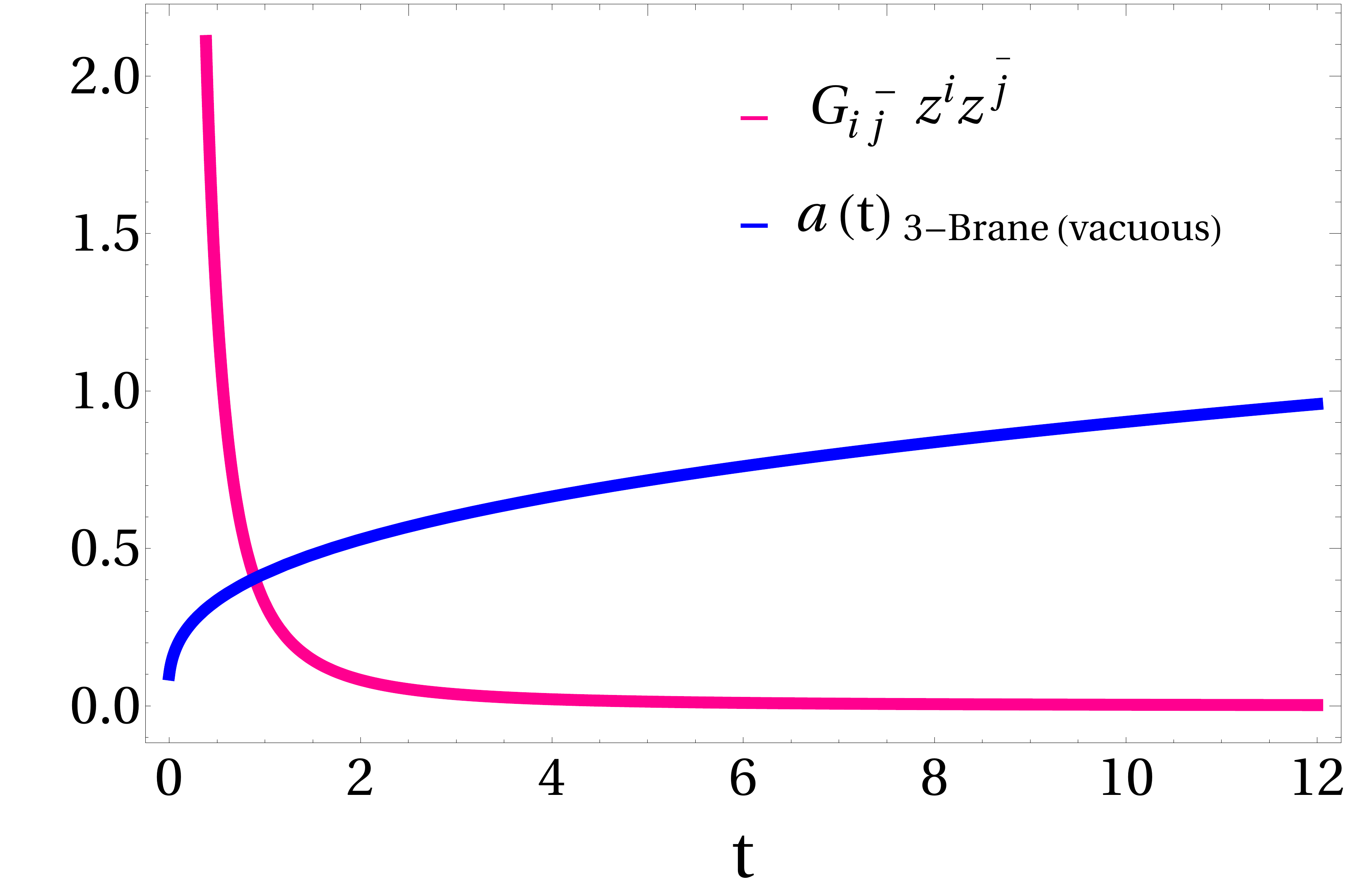}
  \end{subfigure}
\caption{(Left panel): The brane's scale factor is plotted versus time by the blue curve $\rho=p=\Lambda=0$. $c_1= -0.04,~ c_2 =  0.29 $, $a(0)=0.1,~ \dot{a}(0)=2.4 $. While $\Lambda$CDM scale factor is plotted by the pink curve. (Right panel): The moduli $G_{i\bar j} \dot z^i \dot z^{\bar j}$ and the brane's scale factor are plotted versus time.}
\label{atAv}
\end{figure}

In Fig. (\ref{atA}), the brane's scale factor is plotted versus time by the blue curve at $c_1=0$ and $c_2=0.1$ for initial conditions $a(0)=0.1$ and $\dot{a}(0)=0$. $p=0.1$, and from (\ref{ED}) $\rho \sim 3 p =0.3$ . While the scale factor of $\Lambda$CDM model \footnote{\em Lambda cold matter model: The standard model of big bang cosmology.} \cite{Planck:2018} is plotted by the dashed pink curve at current matter density parameter $\Omega_{m0}= 0.31$, current radiation density parameter $\Omega_{r0}= 9.2 \times 10^{-5}$, and $\Omega_{\Lambda 0} = 0.7$. The current value of the Hubble parameter is given by $H_0= 0.0686751 ~[\text{Gyr}^{-1}]$. The vertical grid line is at the present time of the universe $t_0= 13.842~ [\text{Gyr}]$, and the horizontal grid line is at the current value of our universe scale factor, parameterized at $a_0=1$. 
For a vacuous brane- universe $\rho=p=\Lambda=0$, the field equations become:
\bea
\nonumber 3 \left( \frac{\dot a}{a} \right)^2  &=& G_{i\bar j} \dot z^i \dot z^{\bar j},  \\
\nonumber 2 \left( \frac{\ddot a}{a} \right) + \left(\frac{\dot a}{a} \right)^2  &=& - G_{i\bar j} \dot z^i \dot z^{\bar j}, \\
 -\left[ {\frac{{\ddot a}}{a} + \left( {\frac{{\dot a}}{a}} \right)^2 } \right] &=& -  G_{i\bar j} \dot z^i \dot z^{\bar j}, \label{FE-consb2}
\eea
Solving together yields: 
\be
a(t)= c_2  (3 t - c_1)^{1/3},
\label{atv}
\ee
while the moduli $G_{i\bar j} \dot z^i \dot z^{\bar j}$ can be given by any of the three equations in (\ref{FE-consb2}). The most fitted values of $c_1$ and $c_2$ for the $\Lambda$CDM scale factor are given by $c_1= -0.04$, and $c_2 =  0.29$ at IC $a(0)=0.1$ and $\dot{a}(0)=2.4$. This shown in Fig. (\ref{atAv}-Left) where the brane's scale factor (\ref{atv}) is plotted versus time by the blue curve, while $\Lambda$CDM scale factor is plotted by the dashed pink curve. Fig. (\ref{atAv}-Right) shows the correlation between the moduli and the brane's scale factor where both are plotted versus time. 

\chapter{Conclusion}

We have studied a 3-brane embedded in higher dimensional bulk. Studying higher dimensions and brane- theories have manifested themselves in recent years as they lead to understand superstring theories, the dimensional reduction and the topological spaces where these theories exist. Brane- cosmology then has been emerged - supersymmetric, non-supersymmetric - or within the landscape of string theory \cite{sus, st} as modeling the universe as expanding branes in different evolution stages like inflation , accelerating or decelerating expansions. 

Indeed our aim is that usually the early times of the universe like inflation or radiation dominant eras are investigated solely by general relativity theories, while the current accelerating expansion era is investigated by quantum field theories like supersymmetric theories. On the other hand, there is no such distinction in the universe time evolution 
through our study framework. 

We have continued previous works began before \cite{Emam2013} and \cite{Emam:2015laa}. Where in \cite{Emam2013} a 3-brane embedded in 
$\mathcal{N}=2$ five dimensional supergravity with hypermultiplets in the bulk have been studied. All the hypermultiplets fields depend on the bulk’s fifth extra dimension y as well as time. The brane- universe considered a de-Sitter space filled only with energy ( cosmological constant $\Lambda$ ). In  \cite{Emam:2015laa} a BPS condition has been imposed on the 3-brane which was totally vacuous, while the hypermultiplets taken to be very near from the brane- world, i.e., the y- dependence vanished. It has shown in the last that the moduli of the complex structure of the underlying Calabi–Yau (CY) space act as a possible agent for the various cosmological stages of the considered brane. However that wasn’t a pretty realistic model because the brane was empty of any matter, radiation or energy. So may be questions raised that how far the results there connect to our universe. Also there was assumptions made about the brane scale factor $a(t)$ and the velocity norm of the moduli $G_{i\bar{j}} \dot{z}^i \dot{z}^{\bar{j}}$ to evolve with time similarly as our universe’s time evolution.

So we have studied a BPS 3-brane embedded in $\mathcal{N}=2$ $\mathcal{D}=5$ supergravity. The branes studied once filled with radiation ensemble the early times of our baby universe after the big bang, once filled with matter ensemble the later on times and once filled with matter, radiation and energy to ensemble the current accelerated expansion era. All these cases investigated at different initial conditions, three of which are big bang/ initial singularity- like conditions. That yields different brane- universes for whose history are similar to 
our own. 

Confirming to the previous results, we have found that velocity norm of the Calabi–Yau complex structure moduli is strongly correlated to brane scale factor $a(t)$. However we didn’t need to make any assumptions, we got our results directly by solving the Einstein’s field equations numerically. $G_{i\bar{j}} \dot{z}^i \dot{z}^{\bar{j}}$ starts large and quickly decays to zero, `causing` the expansion of the 
brane- universe. This is consistent with arguments that our universe’s big bang must have produced large, but unstable, quantities of moduli that decayed very quickly \cite{Bodeker:2006ij, Dine:2006ii}. Most solutions  have an early, and very short, period of positive acceleration for the brane’s scale factor; strongly suggesting a causality between the decaying moduli velocity norms and an early inflationary period. 

Also we show here that the moduli norm not only related to $a(t)$, but it’s also related to the bulk’s scale factor $b(t)$ which points to the dimensions of the bulk itself and related to the remaining hypermultiplet fields which exhibit interesting early time behavior, then mostly vanish asymptotically following the behavior of $G_{i\bar{j}} \dot{z}^i \dot{z}^{\bar{j}}$. Moreover, we have found that in many cases the brane-universes exhibit late time acceleration with introducing a brane cosmological constant $\Lambda$ or at vanishing $\Lambda$ and introducing an extra bulk’s cosmological constant $\tilde{\Lambda}$ . Declares that the problem of the dark energy can be interpreted by only the effects of a higher dimensional bulk. 

Some of the 3-branes solutions describe universes that show very early inflation, late time deceleration and current time accelerated expansion just exactly like our own universe . Even within these solutions we can predict various scenarios about the universe’s destiny. We studied the moduli's variation with respect to the fifth extra dimension, where the cosmological constant problem can be resolved analytically.

These results shed light on the role of the topology of the Calabi-Yau manifold. In other words what are the deep correlations that make
$G_{i\bar{j}} \dot{z}^i \dot{z}^{\bar{j}}$ derives the whole time evolution of the  brane- world and the bulk  as well as the other hypermultiplet fields. 

That made us go in the last chapter further towards the Calabi–Yau manifold topological structure to answer questions like, how the moduli fields $z^i$ vary with time ? what are the time and space dependence of the CY complex structure space shown in its metric  $G_{i\bar{j}}$ ? what about the dimensions of CY manifold? it’s related to the dilaton field which evolve in time, so does the volume of the CY manifold really depend on time? we studied analytically the moduli and the metric component G in case the dimension of the CY complex structure space $\mathcal{M}_c$, $h_{(2,1)}=1$, i.e., $i,\bar{j} =1$. For instance, in case when the Calabi-Yau manifold is considered as a quintic 3-fold, $h_{(2,1)}=101$, however here we do not need to make this assumption because we already know many information about $G_{i\bar{j}} \dot{z}^i \dot{z}^{\bar{j}}$. Then we have found analytic solutions to the the potential of the complex structure space and the volume of the Calabi–Yau manifold which as shown change with time even with a rare rate.

In summary, we outlined previous studies about 3brane- universes embedded in five dimensional bulk of 
$\mathcal{N}=2$ supergravity. Which gives a comprehensive interpretation about our universe’s cosmological aspects since its birth till the current epoch. That may has not been provided in such way before, i.e., a complete explanation of the universe’s history and future predictions.
We have taken into account the study of the topology of the CY manifold where the theory itself live. We paid an attention that may be there are other multiverses have been evolved with our own . Because there are no any reasons that the very tuned initial conditions that created our universe are special or unique. 

\appendix
\setcounter{equation}{0}


\chapter{General Relativity notes} 
\label{GR}
In this appendix, we will review some basic concepts in general relativity \cite{MUKHANOV2005}, \cite{Hobson2006}, \cite{Wald:1984}, \cite{Gary}, \cite{Frieman2015}, \cite{Terzic2008}, \cite{Bellido-lec}, \cite{Emam2018}. 

\section*{The equivalence principle}
The effect of gravity can be cancelled locally by using free falling frame.\\
\section*{The geodesic equation} 
The world line of a particle freely falling under gravity is given in general by the geodesic equation 
\be%
\frac{d^2 x^\mu}{d \tau^2} + \Gamma^\mu_{\nu\sigma} \frac{d x^\nu}{d \tau} \frac{d x^\sigma}{d\tau} = 0,
\ee%
where $\Gamma^\mu_{\nu\sigma}$ are called the Christoffel symbol, or the affine connection
\be%
\Gamma^\lambda_{\mu\nu}= \frac{1}{2} g^{\lambda\kappa} [ (\partial_\mu g_{\nu\kappa}) + (\partial_\nu g_{\mu\kappa}) - ( \partial_\kappa g_{\mu\nu}) ] . 
\ee%
\section*{The curvature tensor}
The Riemann tensor is defined by:
\be%
R^{\sigma}_{~\mu\rho\nu}=  (\partial_\rho \Gamma^\sigma_{\mu\nu} ) - (\partial_\mu \Gamma^\sigma_{\rho\nu} ) + \Gamma^\alpha_{\mu\nu} 
\Gamma^\sigma_{\alpha\rho} - \Gamma^\alpha_{\rho\nu} \Gamma^\sigma_{\alpha\mu}
\ee%
It is related to the curvature of a manifold in such that at flat region of a manifold, we may choose coordinates where the line element takes the form:
\be%
ds^2=\epsilon_1 (dX^1)^2 + \epsilon_2 (dX^2)^2 + … + \epsilon_N (d X^N)^2,
\nonumber%
\ee%
where $\epsilon_a=\pm 1$
throughout the region. In these coordinates $\Gamma^\lambda_{\mu\nu}$ and its derivatives are zero, and hence
\be%
R^{~~~~\sigma}_{\mu\rho\nu}=0
\nonumber%
\ee%
While for the manifold the line element is given by:
\be%
ds^2= g_{\mu\nu} (x) dx^\mu dx^\nu ,
\nonumber%
\ee%
and the covariant derivative of a vector field takes the form:
\be%
\nabla_\nu v_\mu = \partial_\nu v_\mu - \Gamma^\lambda_{\mu\nu} v_\lambda. 
\nonumber%
\ee%
The Ricci tensor is given by :
\be%
R_{\mu\nu} =R^{~~~~\rho}_{\mu\rho\nu},
\nonumber%
\ee%
while the Ricci scalar
\be%
R= g^{\mu\nu} R_{\mu\nu},
\nonumber%
\ee%
and the Einstein tensor
\be%
G^{;\mu\nu} \equiv R^{\mu\nu}- \frac{1}{2} g^{\mu\nu} R.
\ee%
\section*{The gravitational field equations}
In an IRF ( instantaneous rest frame ) the components of the energy–momentum tensor of a perfect fluid are given by
\be%
[T^{\mu\nu}]= \left(\begin{array}{cccc}
\rho c^2 & 0 & 0 & 0 \\
0 & p & 0 & 0 \\
0 & 0 & p & 0 \\
0 & 0 & 0 & p \\
\end{array}
\right),
\ee
so that
\be%
T^{\mu\nu}=(\rho + p/c^2) u^\mu u^\nu - p \delta^{\mu\nu} ,
\label{EMT}
\ee%
where $u^\mu=(c,0,0,0)$.
Einstein field equations are given by:
\be
G_{\mu\nu} = \kappa T_{\mu\nu}, 
\label{EFEA1}
\ee
or:
\be%
R_{\mu\nu}- \frac{1}{2} g_{\mu\nu} R = -\kappa T_{\mu\nu},\label{FE}%
\ee%
where $\kappa= 8\pi G/c^4$. Or in another form 
\be%
R_{\mu\nu}= - \kappa(T_{\mu\nu}-\frac{1}{2} T g_{\mu\nu}),
\ee%
where $T \equiv T^\mu_\mu$.
\section*{The Schwarzschild geometry}
Schwarzschild explored the metric $g_{\mu\nu}$
representing the static spherically symmetric gravitational field in the empty space surrounding some massive spherical object such as a star.
The Schwarzschild metric for the empty spacetime outside a spherical body of mass M is
\be%
ds^2 = c^2  \bigg( 1- \frac{2\mu}{ r} \bigg) dt^2- \bigg( 1 - \frac{2\mu}{r} \bigg)^{-1}
dr^2 - r^2 d\theta^2 - r^2 \sin^2 \theta d\phi^2 , 
\label{sch}
\ee%
where $\mu=GM/c^2$.
Clearly, the metric functions are infinite at $r=2\mu$ which is known as the Schwarzschild radius. 
\newpage
\section*{The Friedmann–Robertson–Walker geometry}
\vspace{-0.1em}
(The maximal symmetric 3-dimensional space)
\subsection*{The cosmological principle}
At any particular time the universe looks the same from all positions in space at a particular time and all directions in space at any point are equivalent.
Or: on the very largest scales, to high accuracy the universe is isotropic and homogenous. The line element takes the form
\be%
ds^2= c^2 dt^2 - g_{ij} dx^i dx^j ,
\nonumber%
\ee%
where the $g_{ij}$ are functions of the coordinates $(t,x^i)$. However to incorporate the time independence , the time can enter the $g_{ij}$ only through a common factor , so that the ratios of small distances are the same at all times. Hence the metric must take the form
\be%
ds^2= c^2 dt^2 - S^2(t) h_{ij} dx^i dx^j ,
\ee%
where S(t) is a time-dependent scale factor and the $h_{ij}$ are functions of the coordinates $x^i$.  
Accordingly after some calculations (\ref{sch}) becomes
\be%
ds^2= c^2 dt^2 - a^2(t) \Bigg[\frac{dr^2}{1-kr^2} + r^2 (d\theta^2  +sin^2 \theta d\phi^2)\Bigg], \label{FRW-m2}
\ee%
which's called Friedmann–Robertson–Walker (FRW) metric. a(t) is a scale function. $k$ takes the values −1, 0, or 1 depending on whether the spatial section has negative (pseudo-sphere, $a^2 < 0$ ) , zero (2- dimension flat space) or positive ( sphere, $a^2 > 0$) curvature respectively. It is also clear that the coordinates r appearing in the FRW metric are still comoving, i.e. the worldline of a galaxy, ignoring any peculiar velocity, has fixed values of $(r,\theta,\phi)$. Another more convenient form 
\be%
ds^2 = c^2 dt^2 - a^2(t) \big[d\chi^2+ S^2 (\chi) (d\theta^2+ \sin^2 \theta d\phi^2)\big]
\ee%
where the function $S(\chi) =( \sin \chi, \chi, \sinh \chi)$ if  k= ( 1,0,-1) , respectively. \\

\large{The Hubble parameter}
\be%
H(t)=\frac{\dot a(t)}{a(t)}.
\ee%

\large{The deceleration parameter}
\be%
q(t) = - \frac{\ddot a(t) a(t)}{\dot a^2(t)}.
\nonumber%
\ee%

{\large Hubble’s law}
\be%
v= H_0 d,
\label{HL}
\ee%
The galaxies appear to recede from us with a recession speed proportional to their distance from us , (the expansion of the universe- 1929).
\vspace{0.5cm}

\textbf {The cosmological field equations (FRW equations):}
Solving the gravitational field equations \eqref{FE} with the FRW metric \eqref{FRW-m2} yields: 

\be%
\nonumber%
\dot a^2 = \frac{8\pi G}{3}\rho a^2 -c^2 k,
\ee%
\be%
\ddot a = - \frac{4 \pi G}{3} \bigg(\rho + \frac{3p}{c^2}\bigg) a.
\label{fr}
\ee%
These two differential equations determine the time evolution of the scale factor a(t)( literally control the time evolution of the universe , for instances they tell the universe can‘t be static) and are known as the Friedmann–Lemaître equations.
\section*{The Continuity equation}
The energy density is conserved according to
\be%
\dot \rho = - 3 ( \rho + p ) H,
\ee%
for dust p $<< \rho $, $\dot \rho =-3 \rho H$.
\bigskip

- Each component of the cosmological fluid is modelled as a perfect fluid with an equation of state of the form
\be%
p_i = w_i \rho_i c^2, 
\ee%
where $w_i=0$ for pressure less ‘dust’, $w_i=1/3$ for radiation and $w_i=-1$ for the vacuum.
\section*{Cosmological parameters}
\begin{figure}[!t]
\centering
\includegraphics[scale=0.55]{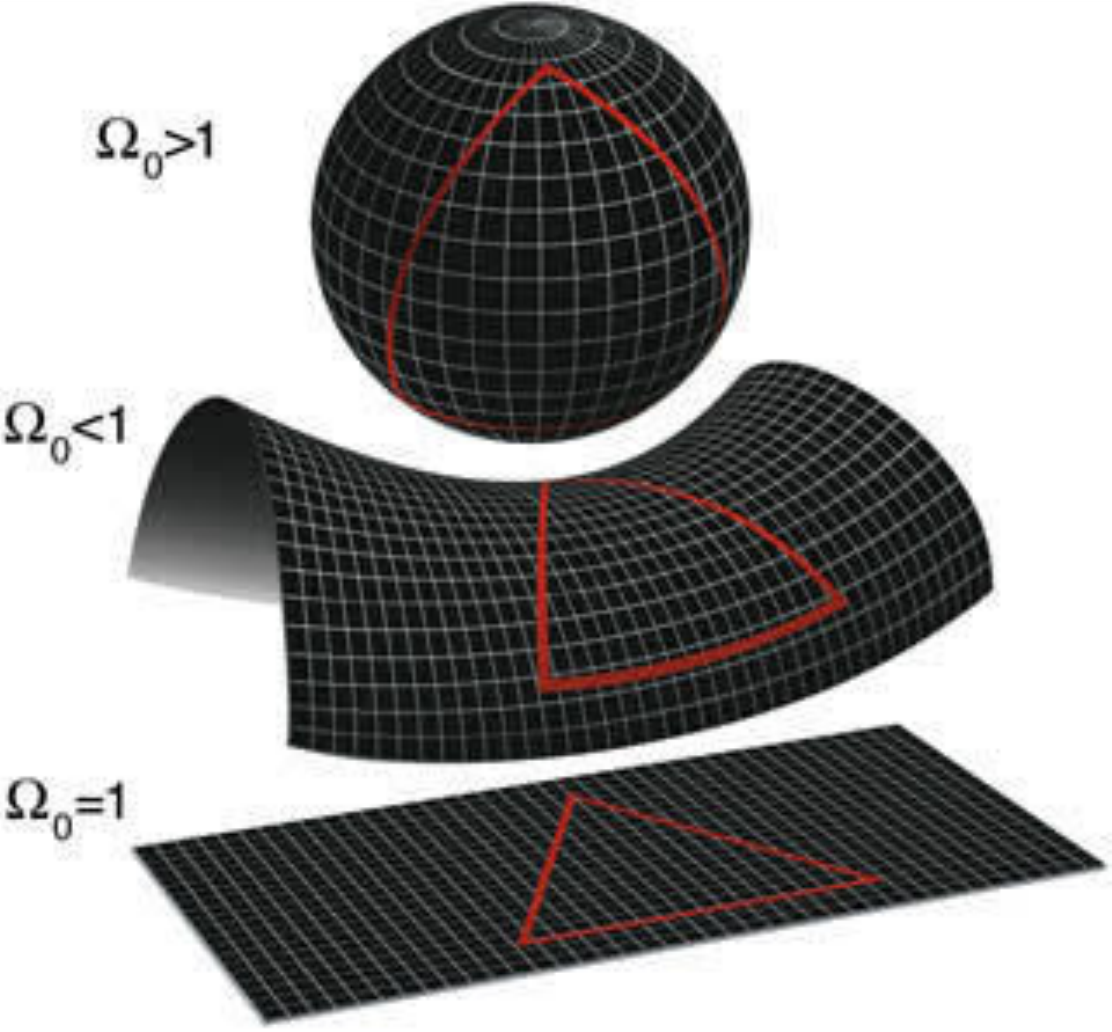}
\caption{Possible shapes of the universe depending on the density parameter $\Omega$.}
\label{dp}
\end{figure}

The density parameters or simply densities, are dimensionless parameters defined by
\be%
\Omega_i(t)\equiv \frac{8\pi G}{3H^2(t)} \rho_i(t),
\ee%
where i denotes 'm', 'r' or $'\Lambda'$, and $\rho$ is the mass density. Define the total density in the universe
\be%
\Omega= \Omega_m+ \Omega_r + \Omega_\Lambda  =1-\Omega_k,
\ee%
where $\Omega_k(t)= - \frac{c^2 k}{H^2(t)a^2(t)}$ is called the curvature density parameter. So it’s clear that $\Omega$ determine the spatial curvature of the universe, see Fig. (\ref{dp}).
\section*{A variational approach to general relativity}
The action S for a set of fields defined on some general four-dimensional spacetime manifold should take the form of an integral of the Lagrangian density, of the fields $\phi^a$  and their first ( and possibly higher) derivatives over some four-dimensional region $R$ of the spacetime %
\be%
S = \int_\mathcal{R} \mathcal{L} (\Phi^a, \partial_\mu \Phi^a, \partial_\mu \partial_\nu \Phi^a, ..) d^4 x.
\ee%
Define the field Lagrangian L
\be%
\mathcal{L} = L \sqrt{- g},
\nonumber
\ee%
where g is the determinant of the metric tensor in that coordinate system. The Euler- Lagrange (EL) equations can be found easily by the principle of least action $\delta S=0$
\be%
\frac{\partial \mathcal{L}}{\partial \Phi^a} - \partial_\mu \bigg [\frac{\partial \mathcal{L}}{\partial (\partial_\mu \Phi^a)} \bigg] = 0.
\ee%
\section*{The Einstein–Hilbert action}
\be%
S_{EH}= \int_\mathcal{R} R \sqrt{-g} ~ d^4 x.
\label{EHA}
\ee%
By demanding that $\delta S_{SH}=0$ and using the fact that the variation $\delta g^{\mu \nu}$ is arbitrary, we thus recover Einstein’s field equations in vacuo:%
\be%
G_{\mu\nu} \equiv g_{\mu\nu}-\frac{1}{2} g_{\mu\nu} R =0.
\nonumber
\ee%
In the presence of matter
\be%
S= \frac{1}{2\kappa} S_{EH} + S_M =
\int_\mathcal{R} \bigg( \frac{1}{2\kappa} \mathcal{L}_{EH}+\mathcal{L}_M \bigg)~  d^4 x,
\nonumber
\ee%
where $\kappa=8\pi G/c^4$. Let us consider the varying of the action with respect to the (inverse) metric , to obtain
\be%
\frac{1}{2\kappa} \frac{\delta \mathcal{L}_{EH}}{\delta g^{\mu\nu}} + \frac{\delta \mathcal{L}_M}{\delta g^{\mu\nu}} =0.
\nonumber%
\ee%
But 
\be%
\frac{\delta \mathcal{L}_{EH}}{\delta g^{\mu\nu}} = \sqrt{-g} G_{\mu \nu},
\nonumber%
\ee%
Define the energy - momentum tensor of the non-gravitational fields (or ‘matter’) by
\be%
T_{\mu\nu}=\frac{2}{\sqrt{-g}} \frac{\delta \mathcal{L}_M}{\delta g^{\mu\nu}}.
\nonumber%
\ee%
Then we recover the full Einstein Equations
\be%
G_{\mu\nu} = - \kappa T_{\mu\nu}.
\nonumber
\ee%
\section*{The cosmological constant \protect\footnote{\cite{BookME}.}} 
During the discovery of the general relativity by Albert Einstein in 1915, it was thought that the universe is static. While by looking at the first equation of Friedmann equations (\ref{fr}), we will find for a flat universe $k=0$, the scale factor $\left(\frac{\dot{a}}{a}\right)$ vanishes only if the universe is empty from any matter or energy $\rho=0$. In 1917 Einstein added the  
cosmological constant term to the right- hand side of the first Friedmann equation to cancel the effect of the density, and to guarantee a vanishing $\dot{a}$. The complete Einstein field equations (\ref{EFEA1}) become:
\be
G_{\mu\nu} +\Lambda g_{\mu\nu} = \kappa T_{\mu\nu}.
\ee
The cosmological constant $\Lambda$ is assumed to be a positive real number. It can be accounted as a part of the matter/energy content of the universe by moving it to the right- hand side:
\be
G_{\mu\nu}  = \kappa T_{\mu\nu}-\Lambda g_{\mu\nu}.
\ee
As we show this term generates a negative pressure, and thus acts as a `repulsive` force, counteracting the attractive gravitational force in the universe. Later, in 1929 Edwin Hubble formulated Hubble's law (\ref{HL}) which states that the universe is expanding ($\dot{a} >0$). Einstein returned and removed the cosmological constant from his equations, calling it his `greatest blunder`, because he  predicted the expansion of the universe theoretically. In 1998 the cosmological society was astonished by the discovery the speed of the expansion of the universe is indeed accelerating with time. A discovery that earned the $2011$ Nobel prize in physics 
\cite{Lambda-Art, Lambda-A}. The research teams was observing type $Ia$ supernovae \cite{supernova}, and found that over 50 distant supernova whose light was weaker than expected. 
This was a strong sign that the universe has an accelerated expansion ($\ddot{a} >0 $). 
Nowadays \footnote{The age of the universe is 14 billion years since the big bang.} the cosmological constant term is added to Einstein equations to account for this late- time accelerated expansion. However, it's not clear yet what is the nature of $\Lambda$, sometimes it's referred by dark energy, some kind of unkonwn energy who has a repulsive effect and drive the universe's acceleration. On the other hand the cosmological constant term is related to the vacuum energy, consider for instance the energy- momentum tensor of a scalar field 
\be%
T^\alpha_\beta = \frac{1}{2}\partial^\alpha \phi \partial_\beta \phi - \bigg( \frac{1}{2} \partial^\gamma \phi \partial_\gamma \phi - V(\phi) \bigg) \delta^\alpha_\beta, 
\ee%
It can be in the form of a perfect fluid  
\be%
T^{\mu\nu}=(\rho + p/c^2) u^\mu u^\nu - p \delta^{\mu\nu} ,
\label{EMT}
\ee%
by defining 
\be 
\rho \equiv \frac{1}{2} \dot \phi^2 + V(\phi), ~
p \equiv\frac{1}{2} \dot \phi^2 - V(\phi), ~
u^\alpha \equiv \partial^\alpha/ \sqrt{\partial^\gamma \phi \partial_\gamma \Phi}. 
\ee%
If the potential $V(\phi)$ has a local minimum at some point $\phi_0$ then 
\be
\rho = - p = - V(\phi_0) , 
~~~ T^\alpha_\beta =  V(\phi_0) \delta^\alpha_\beta.
\ee%
Substituting in the Einstein equations imitates a cosmological term
\be
G_{\mu\nu} - \Lambda g_{\mu\nu} = - \kappa V(\phi_0) g_{\mu\nu} ,
\ee
means that $\Lambda= \kappa V(\phi_0)$. So that
the cosmological term can always be interpreted as the contribution of
vacuum energy to the Einstein equations. The value of the theoretical vacuum energy density 
arises from the quantum field theory is given by
\be 
\rho^{(th)}_{\Lambda} = (10^{18} ~ \text{GeV})^4 ,
\ee 
while the cosmological observations imply 
\be 
\rho^{(obs)}_{\Lambda} \leq (10^{-12} ~ \text{GeV})^4.
\ee 
This huge discrepancy between the theoretical and observational values is called the cosmological constant problem.

\chapter[Differential forms]{Differential forms\protect\footnote{\cite{CYM2-2}.}} 
\label{Df}
\section{Differential forms on manifolds}
Consider a D-dimensional Riemannian/Lorentzian manifold $\mathcal{M}$ . A differential form $\omega$ or
$\omega_p$ of order p on $\mathcal{M}$, also known as a p-form, is a totally antisymmetric tensor of type (0, p).
It may be defined in terms of the differentials $dx^\mu$ , for instance
\be \nonumber \text{1- form} : \omega_{(1)}=\omega_0~ dx, \ee
\be \nonumber \text{2- form} : \omega_{(2)} = \omega_0~ dx ~ dy.\ee
But any surface has a direction upward and downward, like the flux. So define the wedge product:
\be \nonumber dx \wedge dy = - dy \wedge dx, \ee
So that:
\be \nonumber \text{3-form} : \omega_{(3)}= \omega_0~ dx \wedge dy \wedge dz, \ee
\be \nonumber \text{4-forms} : \omega_{(4)}=\omega_0~  dt \wedge dx \wedge dy \wedge dz. \ee
The differential form component should be antisymmetric, because one can define 
\be \nonumber \omega_{(2)}= \omega_{12}~ dx \wedge dy= w_{21}~ dy \wedge dx, \ee
leads to $\omega_{12}= - \omega_{21}$, in general
\be \nonumber \omega_{(i)} = \frac{1}{2}\omega_{ij} dx^i \wedge dx^j, \ee
\be \nonumber \omega_{(3)}= \frac{1}{3!} \omega_{ijk}~ dx^i \wedge dx^j \wedge dx^k, \ee
In the standard way, 1-forms themselves acting as basis 
\be \nonumber \omega_p = \frac{1}{p!} ~ \omega_{p_i ……p_p}~ dx^{\mu_i} \wedge …………\wedge dx^{\mu_p}. \ee
The so-called wedge product $\wedge$ is defined such that:
\be \nonumber dx^{\mu_i} \wedge  dx^{\mu_i}=0. \ee 
\be \nonumber dx^{\mu_i} \wedge  dx^{\mu_j}= - dx^{\mu_j} \wedge  dx^{\mu_i}. \ee 
\subsection*{The exterior derivative}
The so-called exterior derivative $d= dx^\nu~ \partial_\nu$ is defined as an operator that maps p-forms
into (p + 1)-forms for instance:
\be
d \omega_{(3)} = \frac{1}{3!} (\partial_\alpha w_{\mu\nu\rho}) ~ dx^\alpha \wedge dx^\mu \wedge dx^\nu \wedge dx^\rho = w_{(4)},\ee
in general:
\be
\lambda_{p+1}= d \omega_p  = dx^\nu~ \partial_\nu~ \omega_p =\frac{1}{p!} ~ ( \partial_\nu~ \omega_{\mu_1 ... \mu_p}) ~~ dx^\nu \wedge dx^{\mu_1} \wedge ... \wedge dx^{\mu_p}  ,
\ee
satisfying the product rule
\be
d (\omega_p \wedge \eta_r) = (d \omega_p) \wedge \eta_r + (-1)^p \omega_p \wedge (d\eta_r), 
\ee
as well as
\be 
d^2 = 0.
\label{cc}
\ee
An operator satisfying (\ref{cc}) is called nilpotent. Differential forms that can be written
as exterior derivatives of other forms, such as $\lambda = d \omega$, are called exact, while forms whose
exterior derivative vanishes, e.g. $d \lambda  = 0$, are called closed. Because of (\ref{cc}), exact forms are
always closed, while the converse is not necessarily true. For example;
\bea \nonumber A &=& A_\mu dx^\mu = A_0 dt +A_i dx_i ,\\ \nonumber
dA &=& \frac{1}{2} (\partial_\alpha A_\mu ) ~ dx^\alpha \wedge dx^\mu, \\ \nonumber &=&
(\partial_\alpha A_\mu) dx^\alpha \wedge dx_ \mu + (\partial_\mu A_\alpha) dx^\mu \wedge dx^\alpha ,\\ \nonumber F&=&( \partial_\alpha A_\mu - \partial_\mu A_\alpha) dx^\alpha \wedge dx^\mu, \\ \nonumber
dF &=& d^2 A =0.
\eea
\subsection*{Hodge operator}
Define the Hodge-duality operator $\star$, mapping p-forms into 
(D - p)- forms, for instance:
\be 
\star( dx \wedge dy) = dt \wedge dz,
\ee
\be
\star F_{\mu\nu} = F^{\mu\nu} ,  
\ee
so that
\be
\mathcal{L}_{EM} = \frac{1}{4} F \wedge \star F,
\ee
also in this language, the volume form is the Hodge dual of the identity,
\be \star 1 = \sqrt{|g|} ~ dx^1 \wedge ...\wedge dx^D. 
\ee
\section{The vielbein formulation of the general relativity}
Consider a Riemannian manifold with a line element $ds^2 = g_{\mu\nu} dx^\mu dx^\nu$ ,
where $g_{\mu\nu}$ is a gauge field, and the curvature tensor $R^{\sigma}_{~\mu\rho\nu}$ as the field strength. 
Consider this manifold is Lorentzian, which means every point on this curved manifold is Lorantz invariant, i.e., every point is locally flat. 

Take a tangent plane to every point, and define on it little vectors called beins, in which they can transform under local special relativistic transformations, and under global general coordinates transformations, because they can move on the curved manifold, so they have two indices $e_\mu^{\hat{\alpha}}$. We can define a $1$- form 
\be e^{\alpha} = e^{\hat{\alpha}}_\mu ~ dx^\mu, \ee
in this language:  
\be g_{\mu\nu} = e_\mu^{\hat{\alpha}} ~ e_\nu^{\hat{\beta}} ~ \eta_{\hat{\alpha} \hat{\beta}}, \ee
\be \nonumber ds^2 = e_\mu^{\hat{\alpha}} ~ e_\nu^{\hat{\beta}} ~\eta_{\hat{\alpha} \hat{\beta}}~ dx^\mu dx^\nu, \ee
\be ds^2 = e^{\hat\alpha} e^{\hat\beta} \eta_{\hat\alpha \hat\beta}, \ee
with
\be e^{\hat\alpha}_\mu= e^{\hat\alpha}_\nu~~ \frac{\partial x^\nu}{\partial x^\mu}= e^{\hat\beta}_\mu~~ \Lambda^{\hat \alpha}_{\hat \beta}. \ee
Define the $1$-form spin connection:
\be 
\omega^{\hat{\alpha} \hat{\beta}} = \omega^{\hat{\alpha}\hat{\beta}}_\mu dx^\mu,
\ee
\be \omega_\mu^{\hat\alpha\hat\beta}= e^{\hat\alpha \rho} (\Delta_\mu e^{\hat \beta}_\rho ) = e^{\hat\alpha\rho} (\partial_\mu e^{\hat\beta}_\rho - \Gamma^\nu_{\mu \rho} e^{\hat\beta}_\nu),
\ee
roughly speaking the Bein is the square root of the metric and the spin connection is the square of the Christoffel symbol. 
Spin connection couples spinors to the GR. Now the Ricci tensor is given by
\bea 
\nonumber R^{\hat \alpha \hat \beta} &=&
d \omega^{\hat \alpha \hat \beta} + \omega^{\hat\alpha}_{\hat\rho} \wedge \omega^{\hat\rho \hat \beta},\\ 
R^{\hat \alpha \hat \beta} &=& R^{\hat \alpha \hat \beta }_{\mu\nu} ~ dx^\mu \wedge dx^\nu, 
\eea
and the Riemann tensor:
\be
R_{\mu\nu\rho r} =
e_{\hat\alpha \mu } e_{\hat \beta \nu } R^{\hat \alpha \hat \beta}_{\rho r}.
\ee
For the torsion $\tau_{\mu\nu}^\rho = \Gamma^\rho_{\mu\nu}-\Gamma^\rho_{\nu\mu}=0,$ define the 2-form 
\be  \tau^{\hat\alpha}~= d e^{\hat\alpha} + \omega^{\hat\alpha}_{\hat r} e^{\hat r}, \ee
where 
\be \omega^{\hat\alpha}_{\hat{r}} = \eta_{\hat r \hat \rho} \omega^{\hat \alpha\hat\rho}. 
\ee
\bea \nonumber
 \tau^{\hat \alpha}&=& \tau^{\hat \alpha}_{\mu\nu} dx^\mu \wedge dx^\nu ,\\
 \tau^\rho_{\mu\nu} &=& e^\rho_{\hat \alpha} \tau^{\hat \alpha}_{\mu\nu}.
\eea
Since the gravitino has two indices $\Psi^{\mu a}$, it acts as the gauge field which has to be introduced to make local SUSY. Define  the covariant derivative that couples $\Psi$ to gravity
\be \tilde{\Delta}_\mu \Psi =
\partial_\mu \Psi+ \frac{1}{4} \omega^{\hat\alpha\hat\beta}_\mu \gamma_{\hat\alpha\hat\beta} \Psi,
\ee
which means the spinors now transform in a curved background. The EH action becomes (Palatini action)
\bea \nonumber S &=& \int d^4 x \sqrt{-|g|} R, \\
&=& \int d^4 x ~ |e| ~ e^\mu_{\hat \mu} e^\nu_{\hat \nu}~  R^{\hat \mu \hat \nu}_{\mu\nu}. 
\eea
%
%
%
%
\chapter[Topics on geometery and topology]{Topics on geometery and topology \protect\footnote{\cite{He2018, Berglund1994}.}} 
\label{cy}
\section{Calabi-Yau 3-fold as a hypersurface}
The hypersurface is a manifold of dimension $(n-1)$ embedded in an ambient space of dimension $n$, generally 
an Euclidean space or a projective space $\mathbb{CP}$. Hypersurfaces can then defined by a single implicit 
equation, at least locally (near every point) and sometimes globally.
\subsection*{The projective space}
Is attained by adding a " point at infinity " to an affine space. Hence, affine complex coordinates $(x,y)$ 
$\in \mathbb{C}^2$ is promoted to projective coordinates $(x,y,z)$ $\in \mathbb{C}^3$ with the extra identification
that $(x,y,z) \sim \lambda (x,y,z)$ for any non-zero $\lambda \in \mathbb{C}^x$. In other words, the elliptic curve is 
cubic in $\mathbb{CP}^2$ with homogenous coordinates $[x:y:z]$. 
Note that, some times the complex algebraic geometery is refered by $\mathbb{P}^n$. The Complex projective space is defined by
\bea
\nonumber \mathbb{P}^n_{[z_1:z_2:...:z_n]} :=& \mathbb{C}^{n+1}_{(x_0,x_1,...,x_n)} / \sim ~ ; 
\\ & (x_0,x_1,...,x_n) \sim \lambda (x_0,x_1,...,x_n).
\eea
Now let us see how a polynomial embedded in the above $\mathbb{P}^n$
\subsection*{The quintic}
So we can construct a 1- dimentional Calabi- Yau manifold like the torus a cubic in 3 dimensions in $\mathbb{P}^2 (n=2)$.
Note that:
\begin{itemize}
\item $1= 2-1$ : the dimension of the CY manifold is 1 less than that of the ambient space since it's defined by a single
polynomial; an example of a hypersurface.
\item $3= 2+1$ : the degree of the polynomial exceeds the dimension of the ambient projective space by 1,
and is thus equal to the number of the homogenous coordintes ; this is CY condition.  
\end{itemize}
General speaking:
\begin{description}
\item [Calabi- Yau (n-1)- fold ($\mathcal{M}$):] can be constructed as a hypersurface in $\mathbb{P}^n$, i.e.,
as a single homogenous polynomial of degree d in $n+1$ projective coordinates. Litrally:

\item [ Proposition:] A homogenous degree $d=n+1$ polynomial (in  $n+1$ projective coordinates)
as a hypersurface in $\mathbb{P}^n$ defines a Calabi-Yau (n-1)- fold. 
\end{description}

The simplest example is the quintic $Q$ : It's a CY 3-fold with degree 5 and a polynomial in $\mathbb{P}^4$.   
A generic quintic hypersurface in $\mathbb{P}^4(5)_{(z_0:...:z_4)}$ can be writtin as
\be
Q:= \sum_{i=0}^4 z_i^5 - \psi \mathop{\Pi}^4_{i=0} z_i, ~~~~ \psi \in \mathbb{C},
\ee
This is called "Fermat form" , with $\psi$ an explicit complex parameter.
\be 
\psi = h^{(2,1)} (Q) = 101,
\ee
complex space deformation parameters, while K\"{a}hler class has only one parameter $h^{(1,1)} (Q)=1$.

\backmatter

\bibliographystyle{abbrv}
\bibliography{the1}

\end{document}